\documentclass{report}

\usepackage[letterpaper,top=2cm,bottom=2cm,left=3cm,right=3cm,marginparwidth=1.75cm]{geometry}

\usepackage{amsmath}
\usepackage{graphicx}
\usepackage{amsfonts}
\usepackage{bbm}
\usepackage{multirow}
\usepackage{float}
\usepackage{cite}
\usepackage{listings}
\usepackage[table]{xcolor}
\usepackage{changepage}
\usepackage{mathtools}
\usepackage[frozencache,cachedir=.]{minted}
\usepackage{authblk}

\usepackage{caption}
\usepackage{subcaption}

\usepackage[colorlinks=true, allcolors=blue]{hyperref}

\title{Goodness of Fit Tests Based on Joint Densities of Multiple Sample Statistics}
\author{Roman Guchenko \\ \texttt{romanguchenko@yandex.ru}}
\affil{St.Petersburg, Russia}

\providecommand{\keywords}[1]
{
  \small	
  \textbf{\textit{Keywords ---}} #1
}

\newcommand{\givenSample}{x_1, \dots, x_n}
\newcommand{\abstractSampleSpace}{\mathbb{X}}
\newcommand{\unknownDistribution}{\mathcal{P}}
\newcommand{\nullHypothesisDistribution}{\mathcal{P}_0}
\newcommand{\distributionOfSamplesOfSizeN}[1]{\mathcal{P}_n (#1)}
\newcommand{\empiricalDistributionOfSamplesOfSizeN}[1]{\mathcal{P}_n^m (#1)}

\newcommand{\arbitrarySample}{x_1^*, \dots, x_n^*}

\begin{document}
\maketitle

\begin{abstract}
In this research we present various goodness of fit tests that are based on simulated confidence sets for joint densities of multiple sample statistics.
We focus on the case where null distribution parameters are considered to be known and the distribution itself is absolutely continious.
The text consists of two main parts. 

The first part is about constructing hyperrectangle confidence sets. 
It continues the ideas from \cite{AldorNoimanBrownBujaRolkeStine2013} and \cite{sailynoja2022graphical}, where the authors construct "horizontal", i.e. based on joint distribution of order statistics, and "vertical", i.e. based on joint distribution of eCDF values, confidence bounds for cumulative distribution functions.
We propose tests based on simulated hyperrectangle confidence sets for principal components of order statistics and some other combination of statistics, and demonstrate their superiority in terms of power over the tests from \cite{AldorNoimanBrownBujaRolkeStine2013} and \cite{sailynoja2022graphical} and classical tests like Kolmogorov--Smirnov, Cramer--von Mises and Andersen--Darling. We also make a comparison with Zhang tests \cite{zhang2002powerful, zhang2001powerful}, show that the proposed tests are generally on par with them, and examine for which alternative which test should be used. Principal component based tests do have interesting geometrical properties. For example, for normal null first principal component of order statistics is mean, second principal component is a linear analog of variance, and so on. We investigate these geometrical properties.

The second part is about constructing confidence sets of arbitrary shape using highest density regions framework \cite{hyndman1996computing}. The core idea is simple: we generate many samples from null hypothesis distribution, calculate different statistics, say "features" in machine learning terms, for each sample, and get the joint distribution of sample statistics for the situation when null is true. Now if the point consisting of sample statistics calculated from a given sample for which we perform goodness of fit lies in a dense region of joint distribution of sample statistics, then we should not reject the null hypothesis. If, on the other hand, the point lies in sparse region, we should reject. 
We wanted to claim this idea, but after a deeper literature search we found that the similar approach was already used in~\cite{king2020}. The difference between~\cite{king2020} and the present work is that instead of kernel density estimators to determine dense and sparse regions we use the methodology to detect highest density regions through application of k-nearest neighbors algorithm described in \cite{deliu2024alternative}. It gives us an opportunity to study higher dimensions as nearest neighbors suffer less from curse of dimensionality than kernel density estimates. 
We explore the tests that use order statistics as sample statistics and examine the distribution of empirical cumulative distribution functions, then proceed to tests based on joint distribution of sample mean, variance, skewness and kurtosis, and finally we combine Kolmogorov--Smirnov, Cramer--von Mises and Anderson--Darling tests into a single more powerful test. We demonstrate that the proposed tests are powerful against the wide range of alternatives.

In the very end of the text we present a sketch on what to do in the two sample case. We introduce permutation tests based on joint density of multiple sample statistics and briefly show how they work. We show that the permutation test based on joint distribution of sample mean, sample variance, sample skewness, and sample kurtosis for the smallest sample fairs quite well compared to the permutation test based on energy distance, which is considered to be quite powerful.

Finally, throughout the paper another interesting question is briefly discussed. Probability integral transform is extensively used in goodness of fit testing since it gives the opportunity to test if the transformed sample is from uniform distribution instead of testing if the original sample is from a given distribution. But why uniform is so special? Why we do not transform samples from arbitrary distributions to some distribution other than uniform? We show that in many situations it is worthwhile to try and transform your sample to standard normal and then apply the goodness of fit test for standard normal. Generally, if you have a powerful goodness of fit test for some distribution you may transform your sample to this distribution and apply your test. 

This research relies heavily on computations. We decided that it is appropriate to place the full code inside the text. So the text much resembles a Jupyter notebook. We believe that the code can be valuable to many readers. If not, it can be safely skipped in favor of tables, graphs and text. The main results are described in the ``Introduction'' chapter. The code is written in R language \cite{Rcore} with some parts that require performance written in C++ \cite{cpp20} and compiled with Rcpp package \cite{eddelbuettel2011rcpp, rcpppackage}.

The text still requires some polishing and will be revised, but we consider it to be quite readable already.
\end{abstract} \hspace{10pt}


\tableofcontents

\chapter{Introduction}

\section{Problem formulation}
\label{section:problem_formulation}

Given an independent identically distributed sample 
\begin{equation}
\givenSample{}; \qquad x_i \in \abstractSampleSpace{}, \, i = 1, \dots, n;
\end{equation}
from some unknown distribution $\unknownDistribution{}$ on a set $\abstractSampleSpace{}$, let us consider the problem of testing the simple hypothesis
\begin{equation}
\label{H_0_first_formulation}
H_0: \unknownDistribution{} = \nullHypothesisDistribution{};
\end{equation} 
against the most general alternative
\begin{equation}
\label{H_1_first_formulation}
H_1: \unknownDistribution{} \neq \nullHypothesisDistribution{};
\end{equation}
for some fixed null hypothesis distribution $\nullHypothesisDistribution{}$.

It is instrumental to think about the sample $\mathrm{X} = \{\givenSample{}\}$ as of a single random element from 
\begin{equation}
\distributionOfSamplesOfSizeN{\unknownDistribution{}},
\end{equation} 
the distribution of samples of size $n$ from $\unknownDistribution$ defined on $\abstractSampleSpace{}^n$. The problem of testing (\ref{H_0_first_formulation}) against (\ref{H_1_first_formulation}) then can be reformulated as a problem of testing
\begin{equation}
\label{H_0_second_formulation}
H_0: \distributionOfSamplesOfSizeN{\unknownDistribution{}} = \distributionOfSamplesOfSizeN{\nullHypothesisDistribution{}};
\end{equation}
against 
\begin{equation}
\label{H_1_second_formulation}
H_1: \distributionOfSamplesOfSizeN{\unknownDistribution{}} \neq \distributionOfSamplesOfSizeN{\nullHypothesisDistribution{}}.
\end{equation}

Let us pick $k$ arbitrary symmetric statistics 
\begin{equation}
T_j : \abstractSampleSpace{}^n \rightarrow \mathbb{R}; \qquad j = 1,\dots,k;
\end{equation}
on the basis of which we plan to decide, whether $H_0$ should be rejected or not. 
In the end, we are interested in
\begin{equation}
\mathcal{P}_{T_1, \dots, T_k}(\distributionOfSamplesOfSizeN{\nullHypothesisDistribution}),
\end{equation}
the joint distribution of
 \begin{equation}
 T_1(\mathrm{X}^*), \dots , T_k(\mathrm{X}^*),
 \end{equation}
where $\mathrm{X}^*$ is an arbitrary random element from $\distributionOfSamplesOfSizeN{\nullHypothesisDistribution}$, i.e. an arbitrary sample $\mathrm{X}^* = \{ \arbitrarySample{} \}$ of size $n$ from $\nullHypothesisDistribution{}$. If 
\begin{equation}
 T_1(\mathrm{X}), \dots , T_k(\mathrm{X})
 \end{equation}
 happens to be untypical for $\mathcal{P}_{T_1, \dots, T_k}(\distributionOfSamplesOfSizeN{\nullHypothesisDistribution})$, then we should reject hypothesis $H_0$, otherwise we should not.
 
Generally, joint distribution $\mathcal{P}_{T_1, \dots, T_k}(\distributionOfSamplesOfSizeN{\nullHypothesisDistribution})$ is a complex object and it is rarely possible to determine it in a direct way.
Luckily, the distribution $\distributionOfSamplesOfSizeN{\nullHypothesisDistribution{}}$ can be approximated with
\begin{equation} 
\empiricalDistributionOfSamplesOfSizeN{\nullHypothesisDistribution{}}, 
\end{equation}
an empirical distribution based on $m$ simulated random elements 
\begin{equation} 
\label{labelSampleOfSamples}
\mathrm{X}_1^*, \dots, \mathrm{X}_m^*
\end{equation}
from $\distributionOfSamplesOfSizeN{\nullHypothesisDistribution{}}$ ($m$ random samples of size $n$ from $\nullHypothesisDistribution$).
We will refer to (\ref{labelSampleOfSamples}) as sample of samples from $\nullHypothesisDistribution$.
Then the distribution  $\mathcal{P}_{T_1, \dots, T_k}(\distributionOfSamplesOfSizeN{\nullHypothesisDistribution})$ can be approximated with  
\begin{equation}
\mathcal{P}_{T_1, \dots, T_k}(\empiricalDistributionOfSamplesOfSizeN{\nullHypothesisDistribution}),
\end{equation}
an empirical joint distribution of $T_1, \dots, T_k$ with the support consisting of rows from the table
\begin{equation}
\label{data_table}
\begin{matrix}
T_1(\mathrm{X}_1^*) & T_2(\mathrm{X}_1^*) & \dots & T_k(\mathrm{X}_1^*) \\
T_1(\mathrm{X}_2^*) & T_2(\mathrm{X}_2^*) & \dots & T_k(\mathrm{X}_2^*) \\
\vdots & \vdots & \ddots & \vdots \\
T_1(\mathrm{X}_m^*) & T_2(\mathrm{X}_m^*) & \dots & T_k(\mathrm{X}_m^*) 
\end{matrix}
\end{equation}

Rows of the table~(\ref{data_table}) correspond to samples of size $n$ from $\mathcal{P}_0$, columns correspond to different statistics of those samples. In machine learning terms samples are called ``objects'' and statistics are called ``features''. 
The larger is the sample of samples size $m$, the better is approximation of $\mathcal{P}_{T_1, \dots, T_k}(\distributionOfSamplesOfSizeN{\nullHypothesisDistribution})$  by $\mathcal{P}_{T_1, \dots, T_k}(\empiricalDistributionOfSamplesOfSizeN{\nullHypothesisDistribution})$. In our framework $n$, the sample size, is fixed.

Marginal statistic distributions
\begin{equation}
\mathcal{P}_{T_j}(\mathcal{P}_n(\mathcal{P}_0)); \; j = 1,\dots,k,
\end{equation} 
defined as distributions of $T_j(\mathrm{X}^*)$, where $\mathrm{X}^*$ is from $\mathcal{P}_n(\mathcal{P}_0)$, are then approximated by empirical marginal statistic distributions
\begin{equation}
\mathcal{P}_{T_j}(\mathcal{P}_n^m(\mathcal{P}_0)); \; j = 1,\dots,k,
\end{equation} 
defined by columns of table~(\ref{data_table}).

If marginal distributions can be considered as absolutely continuous with densities
\begin{equation}
\label{marginal_densities}
\mathcal{D}\left[\mathcal{P}_{T_j}(\mathcal{P}_n(\mathcal{P}_0)) \right] \; \; j = 1,\dots,k,
\end{equation}
then $k$-dimensional vectors of statistics from samples of size $n$ from  $\mathcal{P}_0$ form a density 
\begin{equation}
\label{joint_density}
\mathcal{D}\left[\mathcal{P}_{T_1, \dots, T_k} (\distributionOfSamplesOfSizeN{\nullHypothesisDistribution})\right]
\end{equation}
defined on $\mathbb{R}^k$. From now on we will presume that this assumption holds.

Taking all that into account, we can reduce the problem of testing~(\ref{H_0_second_formulation}) against~(\ref{H_1_second_formulation}) to the problem of estimating density~(\ref{joint_density}) using~(\ref{data_table}) in the point $T_1(\mathrm{X}), \dots, T_k(\mathrm{X})$ and concluding if lies in a highest density region or not.

\section{Detecting highest density regions for simulated empirical joint densities of sample statistics}
\label{HDR_section}

In this text we will use two approaches to construct estimated highest density regions for the joint density~(\ref{joint_density}) of chosen samples statistics $T_1, \dots, T_k$.

\subsection{Hyperrectange confidence sets}
\label{subsection_hyperrectangle_confidence_sets}

The first simpler approach is based on simultaneous bounds for estimates of marginal densities~(\ref{marginal_densities}). It was proposed in \cite{hyndman1996computing}, and used in \cite{AldorNoimanBrownBujaRolkeStine2013} and \cite{sailynoja2022graphical} to construct goodness of fit tests based on tight confidence bounds for cumulative distribution functions. Following the mentioned works we will assume that the marginal densities are unimodal. For unimodal densities their highest density regions are somewhat close to standard confidence bounds.  

Let us fix some significance level
\begin{equation} 
\alpha \in [0,1].
\end{equation}
Now let us remember that the columns of the table (\ref{data_table}) define empirical marginal statistics distributions. Hence  
simulated confidence $\alpha \gamma w$-bounds for marginal densities can be found as
\begin{equation}
\label{marginal_confidence_bounds}
\left(L_j [\gamma],R_j [\gamma] \right) = \left( \mathrm{q} \left[ \left\{ T_j (\mathrm{X}_i^*) \right\}_{i = 1}^m, \alpha \gamma w_j / 2 \right], \; \mathrm{q} \left[ \left\{ T_j (\mathrm{X}_i^*) \right\}_{i = 1}^m, 1 - \alpha \gamma w_j / 2 \right] \right); \; j = 1,\dots,k,
\end{equation}
where $\mathrm{q} \left[ \mathbb{S}, \delta \right]$ is an $\delta$-quantile for set of real numbers $\mathbb{S}$, $\gamma$ is a fitted parameter, and $w = (w_1, \dots, w_k)$ is a vector of statistic weights that defines statistics relative importances in the test. 

To find simultaneous bounds that contain $100 (1-\alpha)$ percents of empirical joint statistics distribution, i.e. simultaneous bounds that contain $100 (1-\alpha)$ percents of rows of table~(\ref{data_table}), we need to approximately solve the equation
\begin{equation}
\label{sim_bound_eq}
\#
\big\{
\left[ T_1(\mathrm{X}_i^*), \dots, T_k (\mathrm{X}_i^*) \right] \; | \; 
T_j(\mathrm{X}_i^*) \in \left(L_j [\gamma],R_j [\gamma] \right), \; j = 1, \dots, k; \; 
i = 1, \dots, m
\big\} / m \approx 1-\alpha,
\end{equation}
where $\# \{ \dots \}$ is the number of elements of the set defined inside $\{ \dots \}$, by choosing appropriate $\gamma$.
Since
\begin{equation*} 
\left(L_j [\gamma_1],R_j [\gamma_1] \right) \subset \left(L_j [\gamma_2],R_j [\gamma_2] \right) \text{ when } \gamma_1 > \gamma_2, 
\end{equation*}
we expect equation~(\ref{sim_bound_eq}) to have a single root, which can be found, for example, with bisection method. 

Pairs of left and right bounds defined in~(\ref{marginal_confidence_bounds}) with $\gamma^*$ that is a root of~(\ref{sim_bound_eq}) form a hyperrectangle confidence set
\begin{equation}
\label{hyperrectangle_bounds_definition}
\mathbb{B} = \left\{ (z_1, \dots, z_k) \in \mathbb{R}^k \; | \; z_j \in (L_j[\gamma^*], R_j[\gamma^*]), \; j = 1,\dots,k \right\}
\end{equation}
in $\mathbb{R}^k$, hence the name of the paragraph.
One might expect this bounds to work well when the cloud of rows from table~(\ref{data_table}) has relatively simple convex form in $\mathbb{R}^k$.

In this text we will consider two unexplored aspects of the described approach. The first aspect is about the weight vector $w$ in~(\ref{marginal_confidence_bounds}). In~\cite{AldorNoimanBrownBujaRolkeStine2013} the authors construct hyperrectangle bounds for joint distribution of order statistics, in~\cite{sailynoja2022graphical} --- for joint distribution of empirical cumulative distribution function values calculated on a grid. In both papers equal weights are used. One can choose weights so that, for example, several first and several last order statistics would have the higher weights, and expect that the resulted test would be better in determining differences from the reference distribution on the edges, but in general it is not obvious how one should choose weights and what weights will lead to better power for the most alternatives. The second aspect to consider is that one can rotate the space of rows of table~(\ref{data_table}), then construct different hyperrectangle bounds such that faces of hyperrectangle would be parallel to the rotated axis hyperplanes, and expect that these bounds would have different performance in terms of power. Here it is not obvious how rotation affects power and how to choose it ``optimally''. Principal component analysis applied to the joint empirical distribution of sample statistics answers both questions: ``How to choose weights?'' and ``How to choose rotation?\footnote{Principal component analysis, of cause, can be also used to lower the dimension of studied space of statistics, but in this text we always use all the components, that is why we talk about honest space rotation.}''. Extensive experiments that we present later in the text show that principal components based approach gives much higher power than methods from~\cite{AldorNoimanBrownBujaRolkeStine2013} and~\cite{sailynoja2022graphical} for diverse range of alternatives. 

Goodness of fit tests based on hyperrectangle bounds are studied in chapter~\ref{chapter:hyperrectangle_bounds}.

\subsection{Direct joint density estimation}
\label{section:direct_joint_density_estimation}

The second approach to constructing highest density regions for goodness of fit testing is very straightforward yet computationally intensive. It is based on the ideas from~\cite{deliu2024alternative} and relies on direct joint density estimation. 

Each column of table~(\ref{data_table}) determines a marginal empirical distribution of some corresponding test statistic. The totality of rows from table~(\ref{data_table}) determines an empirical joint $H_0$ distribution of all test statistics. We assume that the joint density exists and we try to estimate this joint density with empirical joint distribution.  
In~\cite{deliu2024alternative} the authors describe the methods of highest density regions estimation for arbitrary multidimensional densities. One of the mentioned methods is based on average distance to $\lfloor \sqrt{m} \rfloor$ nearest neighbors. To estimate density value at some arbitrary point $z \in \mathbb{R}^k$ implicitly based on simulated empirical joint distribution, we 
\begin{enumerate}
\item first select a distance $d$ between points in $\mathbb{R}^k$, Euclidean for example; 
\item then calculate distances 
\begin{align}
\bigg\{ d_i (z) = d \big( z, \left[ T_1(X^*_i), \dots, T_k(X^*_i) \right] \big) \bigg\}_{i = 1}^m
\end{align}
between $z$ and all rows of table~(\ref{data_table});
\item then sort those distances so that 
\begin{align}
d_{(1)} (z) \leq d_{(2)} (z) \leq \dots \leq d_{(m)} (z);
\end{align}
$d_{(i)} (z)$ denotes the distance to $i$'th nearest neighbor for $z$;
\item and finally calculate ``sparsity'' estimate
\begin{align}
\label{eq:sparsity_estimate}
\hat{f}(z) = \frac{1}{\lfloor \sqrt{m} \rfloor} \sum_{i = 1}^{\lfloor \sqrt{m} \rfloor} d_{(i)} (z).
\end{align}
\end{enumerate}
The intuition behind formula~(\ref{eq:sparsity_estimate}) is very clear: if $z$ is a point where joint distribution is dense, then there will be a lot of points in the close neighborhood of $z$ and average distance to those nearest neighbors will be small; on the other hand, if $z$ is a point where joint distribution is sparse, then there will be less points in $z$'s close neighborhood and average distance to neighbors will be large.
On the $\lfloor \sqrt{m} \rfloor$ value we refer the reader to~\cite{deliu2024alternative}.

How do we construct a test based on the introduced ``sparsity'' estimate~(\ref{eq:sparsity_estimate})? For each support point of empirical joint distribution let us compute its sparsity:
\begin{align}
\label{eq:sparsity_empirical_distribution}
\mathbb{Q} = \left\{ \hat{f} (\left[ T_1(X^*_j), \dots, T_k(X^*_j) \right])  \right\}_{j = 1}^m = \left\{ \frac{1}{\lfloor \sqrt{m} \rfloor} \sum_{i = 1}^{\lfloor \sqrt{m} \rfloor} d_{(i)} (\left[ T_1(X^*_j), \dots, T_k(X^*_j) \right]) \right\}_{j = 1}^m.
\end{align}
The highest density region is then defined as follows:
\begin{align}
\label{eq:hdr}
\mathbb{B} = \left\{ z \in \mathbb{R}^k \; | \; \hat{f} (z) < \mathrm{q}(\mathbb{Q}, 1 - \alpha) \right\}.
\end{align}
Such region, contrary to hyperrectangle, can have an arbitrary complex form. 

To determine if the point $\left[ T_1(X), \dots, T_k(X) \right]$ is in the highest density region, we don't have to construct the region~\eqref{eq:hdr} explicitly, we can just check if 
\begin{align}
\label{eq:sparsity_check}
\hat{f} (\left[ T_1(X), \dots, T_k(X) \right])) < \mathrm{q}(\mathbb{Q}, 1 - \alpha).
\end{align}
If (\ref{eq:sparsity_check}) is satisfied, it means that $\left[ T_1(X), \dots, T_k(X) \right]$ has sparsity level like, say, $95\%$ of points in empirical joint $H_0$ distribution, so $H_0$ should not be rejected. If, on the contrary, the sparsity is too high and (\ref{eq:sparsity_check}) is not satisfied, $H_0$ should be rejected.

We note that the naive density estimation algorithm based on nearest neighbors has complexity quadratic in $m$. Luckily, some tricks exist to ease the computations. Still, the algorithm is expensive.

Goodness of fit tests based on the methodology described above are studied in chapter~\ref{chapter:joint_density_tests}. 
After a more careful literature search we found that an approach similar to ours was proposed in~\cite{king2020}. In~\cite{king2020} the authors use kernel density estimators to estimate joint density in a point. Kernel density estimators perform well in two/tree dimensional spaces, but suffer in higher dimensions. On the other hand, k-nearest neighbors approach from~\cite{deliu2024alternative} applied to goodness of fit problems in the way we propose it here gives us an opportunity to study relatively high-dimensional spaces with dozens of test statistics. When these many statistics make sense? For example, when we study the distribution of order statistics directly for some fixed $n$. Order statistics define empirical cumulative distribution functions.  So we can construct goodness of fit tests that are based on highest density regions for densities of empirical cumulative distribution functions for samples with fixed size $n$ from some given distribution. To our knowledge, these tests are novel.

\section{On estimating test's power by simulation}
\label{section:test_power}

To estimate powers of proposed goodness of fit tests that are based on confidence sets~\eqref{hyperrectangle_bounds_definition} and~\eqref{eq:hdr}, we will use standard simulation-based approach.

Let us define 
\begin{equation}
\label{data_table_general}
\mathbf{T} (\mathcal{P}, n, l) = 
\begin{bmatrix}
T_1(\mathrm{Z}_1^*) & T_2(\mathrm{Z}_1^*) & \dots & T_k(\mathrm{Z}_1^*) \\
T_1(\mathrm{Z}_2^*) & T_2(\mathrm{Z}_2^*) & \dots & T_k(\mathrm{Z}_2^*) \\
\vdots & \vdots & \ddots & \vdots \\
T_1(\mathrm{Z}_l^*) & T_2(\mathrm{Z}_l^*) & \dots & T_k(\mathrm{Z}_l^*) 
\end{bmatrix}
,
\end{equation}
where $Z_1^*, \dots, Z_l^*$ are from $\mathcal{P}_n^l(\mathcal{P})$, i.e. are $l$ samples of size $n$ from $\mathcal{P}$, and $\mathcal{P}$ is some arbitrary distribution. Then table~(\ref{data_table}) is $\mathbf{T} (\mathcal{P}_0, n, m)$ in this notation. Let us further define as
\begin{equation}
\mathbb{B} (\mathbf{T} (\mathcal{P}_0, n, m), \alpha) = \mathbb{B} (\mathrm{X}_1^*, \dots, \mathrm{X}_n^*)
\end{equation}
the confidence set that was constructed from $\mathbf{T} (\mathcal{P}_0, n, m)$ with the help of one of the methods from section~\ref{HDR_section} (or with any other method, actually).

Then to estimate $\mathbb{B} (\mathbf{T} (\mathcal{P}_0, n, m))$-test's rejection power we follow the steps: 
\begin{enumerate}
\item First of all, we choose $\mathcal{P}_1$, the specific alternative distribution, against which we are going to calculate test's power.
\item Next we simulate $Y_1^*, \dots, Y_l^*$ from $\mathcal{P}_n (\mathcal{P}_1)$ and construct
\begin{equation}
\label{eq:alternative_sample_of_samples}
\mathbf{T} (\mathcal{P}_1, n, l) = 
\begin{bmatrix}
T_1(\mathrm{Y}_1^*) & T_2(\mathrm{Y}_1^*) & \dots & T_k(\mathrm{Y}_1^*) \\
T_1(\mathrm{Y}_2^*) & T_2(\mathrm{Y}_2^*) & \dots & T_k(\mathrm{Y}_2^*) \\
\vdots & \vdots & \ddots & \vdots \\
T_1(\mathrm{Y}_l^*) & T_2(\mathrm{Y}_l^*) & \dots & T_k(\mathrm{Y}_l^*) 
\end{bmatrix}
.
\end{equation}
\item Finally, we calculate the estimate of test's power
\begin{equation}
\hat{\beta}(\mathcal{P}_1) = \# \big\{ [T_1(\mathrm{Y}_i^*), \dots, T_k(\mathrm{Y}_i^*)] \; | \; [T_1(\mathrm{Y}_i^*), \dots, T_k(\mathrm{Y}_i^*)] \notin \mathbb{B} (\mathbf{T} (\mathcal{P}_0, n, m), \alpha) ; \; i = 1,\dots,l \big\} / l.
\end{equation}
In words: the value $\hat{\beta}(\mathcal{P}_1)$ is the number of rows from table~\eqref{eq:alternative_sample_of_samples} that are out of test's confidence set $\mathbb{B} (\mathbf{T} (\mathcal{P}_0, n, m), \alpha)$.

\end{enumerate}

\noindent
If $Y_1^*, \dots, Y_l^*$ are not from $\mathcal{P}_n (\mathcal{P}_1)$ but are from $\mathcal{P}_n (\mathcal{P}_0)$, then $\hat{\alpha} = \hat\beta(\mathcal{P}_0)$ is an estimate for first type error and should approximately be $\alpha$.

We note that 
$$\hat{\beta}(\mathcal{P}_1) = \hat\beta (\mathrm{X}_1, \dots, \mathrm{X}_m, \mathrm{Y}_1, \dots, \mathrm{Y}_l),$$
i.e. the power (or type 1 error) estimate is a function of both $H_0$ sample of samples~\eqref{data_table} and $H_1$ sample of sample~\eqref{eq:alternative_sample_of_samples}. It is a random variable itself and has a distribution. In our numerical studies we compute  power (and type 1 error) estimates multiple times in a loop and then report means with standard deviations. Per each iteration we  simulate~\eqref{data_table} only once and then use the same~\eqref{data_table} with different simulated alternatives~\eqref{eq:alternative_sample_of_samples}. Technically, it is not entirely correct as it makes estimates for different alternatives dependent through the same~\eqref{data_table}, but in practice this dependence is negligible and the reuse of~\eqref{data_table} helps to reduce computation.

\section{Probability integral transform, inverse transform sampling, and how they work together}
\label{section:how:to:use:transforms:general}

The well known probability integral transform states that 
\begin{equation}
\label{eq:pit_transform}
F_0(x^*) \sim U[0,1], \text{ if } x^* \sim \mathcal{P}_0;
\end{equation}
for any absolutely continuous distribution $\mathcal{P}_0$ with cumulative distribution function $F_0$, where $U[0,1]$ is the uniform distribution on $[0,1]$. Because of that, for a given sample $\givenSample$ one can test whether the sample $F_0(x_1), \dots, F_0(x_n)$ comes from $U[0,1]$ instead of testing that $\givenSample$ comes from $\mathcal{P}_0$ directly. 
Classical tests like Kolmogorov--Smirnov, Cramer--von Mises, and Anderson--Darling all compare the empirical cumulative distribution function of probability-integral-transformed sample with cumulative distribution function of uniform distribution. In this regard we can say that these classical tests are the tests for uniformity that are used in combination with probability integral transform.

Another well known transform which is used a lot in statistical simulations\footnote{See ``inverse transform sampling''.} states that for any absolutely continuous distribution $\mathcal{P}_{0^\prime}$ with cumulative distribution function $F_{0^\prime}$ we have
\begin{align}
\label{eq:inverse_stansform}
F_{0^\prime}^{-1}\left( y \right) \sim \mathcal{P}_{0^\prime}, \text{ if } y \sim U[0,1].
\end{align}
Let us call~\eqref{eq:inverse_stansform} ``the inverse transform''.

Combining the probability integral transform~\eqref{eq:pit_transform} and inverse transform~\eqref{eq:inverse_stansform} we get
\begin{equation}
\label{equation:transformationToArbitraryDistribution}
F_{0^\prime}^{-1}\left(F_0(x^*)\right) \sim \mathcal{P}_{0^\prime}, \text{ if } x^* \sim \mathcal{P}_0.
\end{equation}
Now if we have a good goodness of fit test for distribution $\mathcal{P}_{0^\prime}$, we can use this test in combination with two-step transform~\eqref{equation:transformationToArbitraryDistribution} to check goodness of fit for arbitrary distribution $\mathcal{P}_0$. 

In this paper we show that there is value in constructing tests not only for uniform distribution but for other distributions as well.

\section{How the research went}

\begin{enumerate}
\item 
This research originated from our interest in functional bounds for empirical cumulative distribution functions and their applications in goodness-of-fit testing. We also wondered how balls in the space of empirical cumulative distribution functions might look like given different metrics.
Cumulative distribution functions are fully defined by vectors of order statistics (sorted samples). That is why a big chunk of the work is devoted to the study of joint distribution of order statistics.

\item We already knew about Kolmogorov--Smirnov functional bounds and came along paper~\cite{AldorNoimanBrownBujaRolkeStine2013}, where the authors constructed tighter and more powerful bounds compared to KS bounds.
\item From paper~\cite{AldorNoimanBrownBujaRolkeStine2013} we discovered paper~\cite{BujaRolke_CalibrationSimultaneity} (as it was cited in~\cite{AldorNoimanBrownBujaRolkeStine2013}), where the authors describe a general approach to constructing simultaneous confidence bounds for dependent variables, and paper~\cite{sailynoja2022graphical} (as it was citing~\cite{AldorNoimanBrownBujaRolkeStine2013}), where the authors study simultaneous bounds for empirical cumulative distribution function values calculated on a grid for uniform distribution.
\item Results from papers~\cite{AldorNoimanBrownBujaRolkeStine2013} and~\cite{sailynoja2022graphical} looked interesting, moreover~\cite{sailynoja2022graphical} was pretty recent, so we decided to check if anything can be done in this direction by asking and answering a series of simple questions. What if we just apply principal component analysis to simulated dataset of order statistics or simulated dataset of eCDF values, and then proceed with methodology from~\cite{BujaRolke_CalibrationSimultaneity, AldorNoimanBrownBujaRolkeStine2013, sailynoja2022graphical}? Will the bounds for principal components be better in power than the bounds for original set of statistics? Will there be a good interpretation for principal component bounds? Can we still stay graphical?
\item
We constructed new tests based on hyperrectangle bounds (approach described in section~\ref{subsection_hyperrectangle_confidence_sets}) for principal components of order statistics for samples from standard normal and standard uniform distributions (see sections~\ref{section:pc_test_normal} and~\ref{section:hyperrectabgle_order_stats_uniform_full} of chapter~\ref{chapter:hyperrectangle_bounds}).
\item It turned out that principal component tests work quite spectacularly for normal and uniform distributions in terms of power if compared to tests from~\cite{AldorNoimanBrownBujaRolkeStine2013} and~\cite{sailynoja2022graphical}. The geometry of the process is as follows. Both original and principal component based tests try to put the same cloud of points into a multi-dimensional box, so that the box would contain, say, 95 percent of points from the cloud. They differ only in rotation of the said box and in the weights that determine lengths of the box edges. For some distributions like normal or uniform principal components introduce a felicitous rotation that helps to construct a box that can separate null hypothesis points from alternative points well. For other distributions like beta(0.5,0.5) direct principal component based tests don't work that well.
\item For continuous data probability integral transform is often used to reduce arbitrary goodness of fit problems to uniformity checks. But by further plugging the PIT-transformed sample into inverse of some CDF we can reduce any goodness of fit problem to any other goodness of fit problem. So we can check goodness of fit for normal with a test tailored for uniform, goodness of fit for uniform with a test for normal, goodness of fit for beta with a test for gamma, and so on. While this trick is definitely known since we found it for example in~\cite{berkowitz2001testing}, it looks like it is not used widely. Using the described transformation paired with normal or uniform principal component tests we acquire powerful goodness of fit tests for arbitrary distributions.
\item Now to the principal components interpretation. In the order-statistics-based test for standard normal first principal component stands for sample mean, second component stands for linear analog of sample variance, third component looks at the difference between distribution center and distribution tails, and so on. In short, for normal-order-statistics-based case principal components are very interpretable! For uniform-order-statistics-based case the components have periodic sine/cosine symmetric/asymmetric structure. We were pleased to find such a well-formed and clear structure in a random table object. Then we were less pleased to find out that the principal component behavior for order statistics vectors was at least partly known from the empirical stochastic process theory. ChatGPT helped us to look through the literature. Chapter 8 of~\cite{ramsay2005functional}, describing the connection between principal component analysis for functional data and Karhunen--Loève expansion\footnote{Original works~\cite{karhunen1947,loeve1948} are in German and French, so we looked through more modern textbooks \cite{loeve1977probability,adler2007random} in English.} of stochastic processes into sums of covariance operator eigenfunctions, became a good general entry point for us. Then we proceeded with chapters 2 and 3 of~\cite{david2003order} with information about joint distributions of order statistics and order statistics covariance matrices.
From~\cite{shorack1986empirical} we learned that order statistics for uniform can be approximated with Brownian bridge, and from~\cite{deheuvels2006kl} --- that Brownian bridge can be expanded into the sum of sines via Karhunen--Loève expansion, hence the sine principal component structure we got in our uniform goodness of fit tests.
As for the normal order statistics case, we were not able to find papers or books that discuss ``first principal component of normal order statistics is proportional to mean, second --- is some linear analog of variance'', so we claim it as observation of our own.

\item 
For the covariance matrix of uniform order statistics we were able to find a compact analytical representation in~\cite{david2003order}. Using this representation we derived analytically ``loadings'' and component weights to use in our principal component based test. To our knowledge, these loadings and weights are novel.

\item 
During our numerical studies we noticed that simultaneous bounds for order statistics introduced in~\cite{AldorNoimanBrownBujaRolkeStine2013} give the same power when testing uniformity directly and testing uniformity via the bounds for normal distribution in combination with two-step transform~\eqref{equation:transformationToArbitraryDistribution}. We proved the proposition about invariance of these bounds. Later we found that the similar result is mentioned in \cite{weine2023qqconf} using a bit different language of ``equal local levels''.

\item Since principal component based tests worked very well for normal and uniform distributions, we decided to compare them not only with tests from~\cite{AldorNoimanBrownBujaRolkeStine2013} and~\cite{sailynoja2022graphical}, but also with classical Kolmogorov--Smirnov~\cite{Kolmogorov1933}, Cramer--von Mieses~\cite{Cramer1928,vonMises1931,vonMises1939_English} and Andersen--Darling~\cite{AndersonDarling1952,AndersonDarling1954} tests. New tests showed significantly better performance for most alternatives.

\item 
Then we searched for more recent and more performant tests and came along Zhang tests~\cite{zhang2002powerful, zhang2001powerful}. After an extensive numerical study we concluded that our tests are on par with Zhang tests: for some alternatives our tests are better, for some they are worse, but overall performance is close.

\item
Among other recent works on goodness of fit testing we found \cite{covington2025powerful} and \cite{Desgagne2026}. These works look interesting, but we did not compare our tests with tests from them. 

\item We also came along the paper \cite{novoa2023ocench}, where the authors constructed non-convex hulls of joint densities of several statistics for purposes of anomaly detection. At first their approach looked appropriate for goodness of fit, but it was not obvious how to adapt it to formal hypothesis testing with given significance level. 

\item 
When we construct hyperrectangle bounds, we actually try to put the most dense part of the joint statistics density into a multidimensional box. 
The next logical step is working with this joint density directly and finding the dense region more accurately. We were not aware of the concept of highest density regions, so we wasted some time and actually reinwented it for two-dimensional case using kernel density estimators. The method worked fine for the test based on joint distribution of two distances, but kernel density estimates did not work well in higher dimensions. We tried to perform a test based on simultaneous bounds for all pairs of order statistics, but this test lacked power. At some point we came along the paper~\cite{deliu2024alternative}, from which we learned about~\cite{hyndman1996computing}, the source paper about highest density regions. From~\cite{deliu2024alternative} we also learned about k-nearest neighbor method to estimate density in higher dimensions. With k-nearest neighbors technique we were finally able to construct the tests based on highest density regions for joint distribution of order statistics --- the tests we wanted in the first place (see sections~\ref{sec:hdr_uniform_order_stat_test} and~\ref{sec:hdr_normal_order_stats_test} of chapter~\ref{chapter:joint_density_tests}). The test based on HDR for uniform order statistics is pretty balanced and has good power, but surprisingly the test based on hyperrectangle bounds for principal components still looks better. During the next round of checking the novelty of the achieved results, we searched through the literature and found the paper~\cite{king2020}, where the authors propose very similar HDR-based goodness of fit methodology. The difference with present work is that to estimate the joint density in a point in~\cite{king2020} the authors use kernel density estimators and here we use k-nearest neighbors. The nearest neighbors approach gives us the opportunity to explore joint densities in much higher dimensions. As for the concrete tests, to our understanding, the proposed tests based of HDRs of order statistics, tests based on mean/variance/skewness/kurtosis, and tests based on multiple distances are new. 

\item
It is known that joint density of order statistics from uniform samples is uniform in a simplex. If joint density is uniform, isn't it pointless to search for the highest density region? We demonstrate that the approach based on nearest neighbors described in section~\ref{section:direct_joint_density_estimation} is still able to produce meaningful results: instead of highest density region it finds a bound region of the simplex, where the local density estimates are the lowest. 
 
\end{enumerate}

\chapter{Hyperrectangle bounds}
\label{chapter:hyperrectangle_bounds}

In this chapter
\begin{enumerate}
\item 
we introduce tests based on hyperrectangle bounds (subsection~\ref{subsection_hyperrectangle_confidence_sets}) for principal components of normal (section~\ref{section:pc_test_normal}) and uniform (section~\ref{section:hyperrectabgle_order_stats_uniform_full}) order statistics,
\item describe their geometrical properties,
\item
compare them with tests based on hyperrectangle bounds for order statistics from~\cite{AldorNoimanBrownBujaRolkeStine2013} (sections~\ref{section:pc_test_normal} and~\ref{section:hyperrectabgle_order_stats_uniform_full}), 
\item 
tests based on hyperrectangle bounds for eCDF values from~\cite{sailynoja2022graphical} (section~\ref{sec:2022verticals}),  
\item classical Kolmogorov--Smirnov, Cramer--vor Mises, and Anderson--Darling tests (sections~\ref{section:classical_tests_normal} and~\ref{section:classical_tests_uniform}),
\item and Zhang $Z_K$, $Z_A$, $Z_C$ tests (sections~\ref{section:zhang:normal} and~\ref{sec:zhang_uniform}).
\end{enumerate}
In addition 
\begin{enumerate}
\item we introduce tests based on hyperrectangle bounds for order statistics and all pairwise distances between order statistics (sections~\ref{section:os_pairwise_dist_normal},~\ref{section:os_pairwise_dist_uniform}, and~\ref{section:os_pairwise_dist_uniform_via_normal}),
\item discuss how to apply two-step transform (section~\ref{section:how:to:use:transforms:general}) to check uniformity via tests based on normal distribution (sections~\ref{section:uniform:via:normal} and~\ref{section:os_pairwise_dist_uniform_via_normal}),
\item discuss the invariance of bounds based on order statistics (section~\ref{section:invariance}).
\end{enumerate}

\section{Goodness of fit test based on hyperrectangle bounds for joint density of principal components of order statistics: samples from standard normal distribution [pc, \texorpdfstring{$N(0,1)$}{N(0,1)}]}
\label{section:pc_test_normal}

We start the whole research by checking a simple idea: what if instead of hyperrectangle bounds for order statistics of standard normal distribution constructed in~\cite{AldorNoimanBrownBujaRolkeStine2013}, we construct hyperrectangle bounds for principal components of that order statistics? Will the power of such test be better? 

In terms of geometry there can be at least two ways of interpreting the results from~\cite{AldorNoimanBrownBujaRolkeStine2013}. The authors themselves say that they construct strict functional bounds that contain cumulative distribution function with a specified high probability, say 95 percents. Another way of looking at it is that they construct a multi-dimensional box or hyperrectangle in $n$-dimensional space that contains the 95 percent of joint density of order statistics. The faces of that hyperrectangle are parallel to coordinate hyperplanes, each hyperplane corresponding to marginal distribution of some order statistic.

When we apply principal components to order statistics we just rotate the space where their joint density lives.
After that we again can place the 95 percents of joint density into a rotated hyperrectangle using a variation of numerical method from~\cite{AldorNoimanBrownBujaRolkeStine2013} applied to principal components of order statistics. The faces of this new hyperrectangle will be parallel to hyperplanes that correspond to those principal components.

In~\cite{AldorNoimanBrownBujaRolkeStine2013} a hyperectangle is defined by $n$ confidence intervals fitted simultaneously, each interval corresponding to a marginal distribution of some order statistic. All intervals have the same weight during the fitting process. Principal components, on the other hand, come with the natural ``importance'' for each component based on its variance, so it is logical to use this ``importance'' during simultaneous fitting process to make edges of the rotated hyperrectangle that correspond to principal components with higher ``importance'' smaller, which in tern should make the test ``look more attentively'' at ``important'' components. 

The section is organized as follows: first we repeat the results from~\cite{AldorNoimanBrownBujaRolkeStine2013}, then we introduce our principal component test and compare. Subsections that contain the code with functions that will be reused in subsequent parts of the text are marked with ``\textasteriskcentered''~sign.

Again, the text contains a lot of \texttt{R} and C++ code and can be thought as a long Jupyter notebook.

Now, let us proceed to the code. 

\subsection{Sample size and number of samples \textasteriskcentered}
\label{section:sample_size_num_samples}

Let us define standard values for the sample size $n$ and for the number of samples $m$ that will be used during the simulation studies throughout the whole chapter:
\begin{minted}[mathescape, linenos]{r}
n <- 10        # sample size
m <- 1000000   # number of samples
\end{minted}
We focus on the small $n$ situation since we want to be able to do plotting and it is problematic in high dimension. 
We will study the asymptotics later.

\subsection{Functions to generate order statistics (sorted samples) from a given distribution \textasteriskcentered}
\label{section:functions_to_generate_order_stats}

To be able to construct confidence sets and power estimates we need a function to calculate $\mathbf{T} (\mathcal{P}, n, l) $ that was defined in~(\ref{data_table_general}). In all our studies for simplicity we set
$$l = m.$$
In this section we consider the case when $T_1, \dots, T_k$ are order statistics for samples of size $n$, so 
$$k = n.$$
The table~(\ref{data_table_general}) then looks as follows: 

\begin{equation}
\label{sorted_samples_table}
\mathbf{T} (\mathcal{P}, n, m) = 
\begin{bmatrix}
z_{1,(1)} & z_{1,(2)} & \dots & z_{1,(n)} \\
z_{2,(1)} & z_{2,(2)} & \dots & z_{2,(n)}  \\
\vdots & \vdots & \ddots & \vdots \\
z_{m,(1)} & z_{m,(2)}  & \dots & z_{m,(n)}  
\end{bmatrix}
,
\end{equation}
where $z_{i,(j)}$ denotes $j$'th order statistic of $i$'th sample from a given distribution $\mathcal{P}$.

A function that generates a table like~(\ref{sorted_samples_table}) which rows are sorted samples from a given distribution can be written in base \textsf{R} as follows:

\begin{minted}[mathescape, linenos]{r}
get.sorted.samples <- function(
    sample.generation.function,  # like rnorm or runif
    n, m
)
{
    samples <- sample.generation.function(n * m)  # generate a large vector
    dim(samples) <- c(m, n)                       # turn it into a matrix
    t(apply(samples, 1, sort))                    # apply sort to each row, transpose 
}
\end{minted}
Here \texttt{m} and \texttt{n} are resulting table's dimensions, and \texttt{sample.generation.function} is a function that given integer size returns an \textsf{R}'s \texttt{vector} of that size with i.i.d. values from some distribution. The final result is  \textsf{R}'s \texttt{matrix} object of size $m \times n$.

We'll be needing to generate a lot of sorted samples in the upcoming experiments, so table generation operation should be fast. In \texttt{get.sorted.samples} the \texttt{t(apply(samples, 1, sort))} is quite slow. Since R is an interpreted language, it has a significant overhead related to indirect loop in \texttt{apply}. Moreover, \texttt{apply} makes a full copy of \texttt{samples} along the way. To compensate, we rewrite this part with C++ in a way that avoids full coping, compile it, and integrate the resulted low level binary code back into \texttt{R} with \texttt{Rcpp} package:

\begin{minted}[mathescape, linenos, texcomments]{r}
library(Rcpp)  # see \cite{rcpppackage}
cppFunction('
void rowSortWithBuffer(NumericMatrix x) {
    int m = x.nrow();
    int n = x.ncol();
    std::vector<double> buf(n);                   // define a row buffer

    for (int i = 0; i < m; i++) {                 // loop over matrix rows
        for (int j = 0; j < n; j++) {             // write a row into the buffer
            buf[j] = x(i, j);
        }

        std::sort(buf.begin(), buf.end());        // sort the buffer

        for (int j = 0; j < n; j++) {             // write sorted buffer back to row
            x(i, j) = buf[j];
        }
    }
}
')
\end{minted}

In \textsf{R}, \texttt{matrix} objects are stored in memory as \texttt{vector}s column by column, so we can not sort \texttt{matrix} rows in place with stdsort since the elements of a row are not adjacent to each other. That is why instead we copy each row into a row buffer, then sort this buffer, and then copy the row back.

The updated table generation code then looks as follows:

\begin{minted}[mathescape, linenos]{r}
get.sorted.samples.cpp <- function(
    sample.generation.function,  # like rnorm or runif
    n, m
)
{
    samples <- sample.generation.function(n * m)  # generate a large vector
    dim(samples) <- c(m, n)                       # turn it into a matrix
    rowSortWithBuffer(samples)                    # sort matrix "in place" row by row
    samples                                       # return sorted matrix
}

\end{minted}

Sure we could have done the generating process even faster than it is in \texttt{get.sorted.samples.cpp function}. For example, we could have stored table~(\ref{sorted_samples_table}) in transposed form which would have given us the opportrunity to sort each row in place, or, maybe, we could have done sample generation and sorting inside a single loop, but we think that the current variant is already good enough and it will serve us well in experiments to follow. 

\subsection{Generation of order statistics for samples from standard normal distribution}
\label{section:standard_normal_order_stats}

Now we generate $m$ sorted samples of size $n$ from standard normal distribution like this:
\begin{minted}[mathescape, linenos]{r}
sorted.samples.std.normal <- get.sorted.samples.cpp(rnorm, n, m)
\end{minted}
Throughout section \ref{section:pc_test_normal} standard normal will be our $H_0$ distribution.

\subsection{Histograms of order statistics for samples from standard normal distribution}
Let us plot histograms for columns of \texttt{sorted.samples.std.normal} that contain empirical distributions of order statistics for normal samples of size 10:
\begin{minted}[mathescape, linenos]{r}
par(mfcol = c(5, 2))
for(i in 1:n)
    hist(
        sorted.samples.std.normal[,i], xlim = c(-6, 6), 
        main = paste0(i," order statistic"), xlab = "x"
    )
\end{minted}

\begin{figure}[ht]
\centering
\includegraphics[width=12cm]{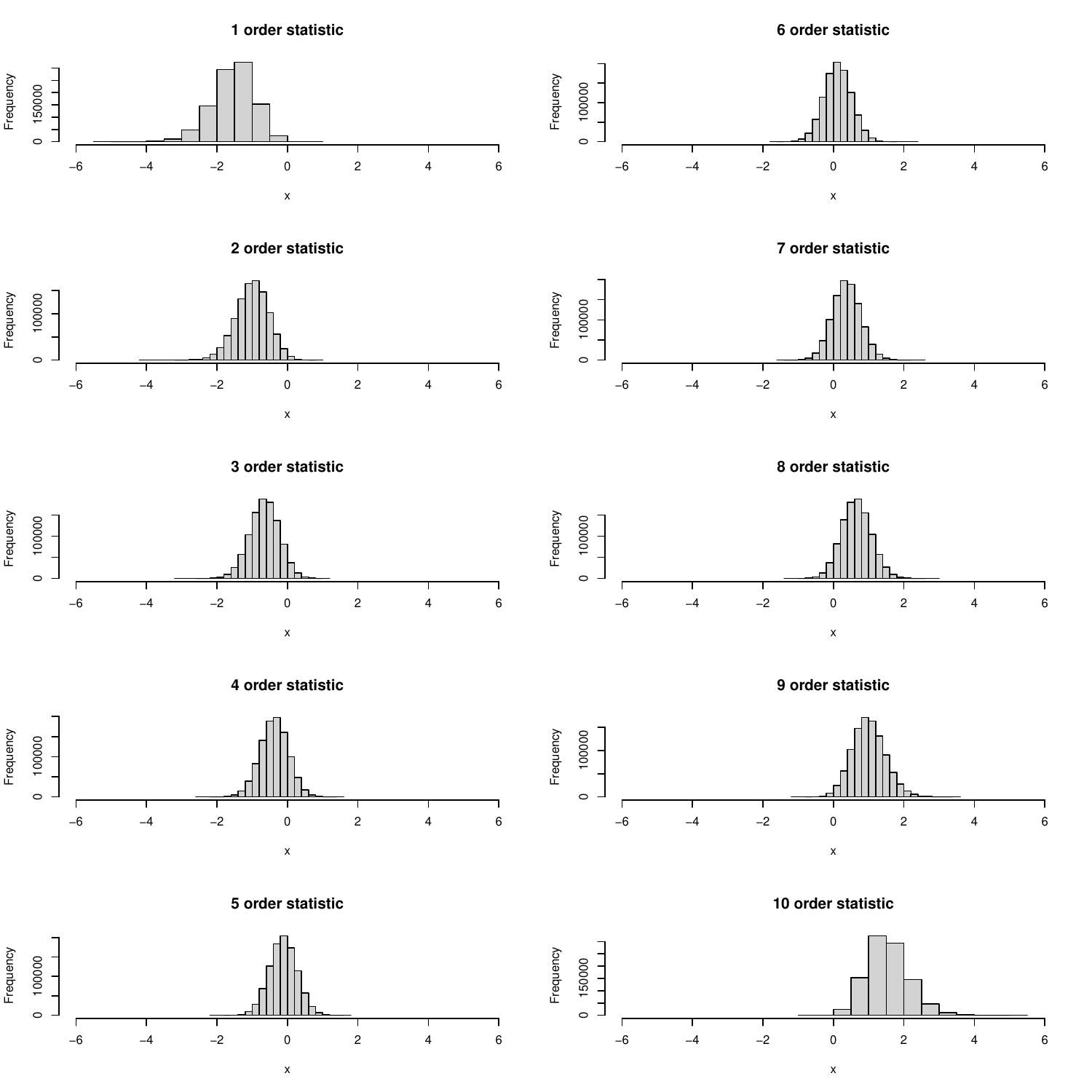}
\caption{Histograms for empirical distributions of order statistics for samples from standard normal distribution of size 10.}
\label{figure:normal_ordered_sample_hists}
\end{figure}

\noindent
The plot is presented on figure~\ref{figure:normal_ordered_sample_hists}.
We see that empirical distributions of order statistics are unimodal, so we can safely use confidence intervals to determine one dimensional highest density regions\footnote{We note that confidence intervals and highest density regions for nonsymmetric one-dimensional densities may not coincide, but they are close. The point is that since all order statistics distributions are unimodal, we don't need a more complex method, standard confidence intervals are fine.}.

\subsection{Function to calculate hyperrectangle bounds \textasteriskcentered}
\label{section:calculate.hyperrectangle.bounds}

Let us proceed to implementing a function to calculate hyperrectangle bounds defined in~(\ref{hyperrectangle_bounds_definition}) using the algorithm described in subsection~\ref{subsection_hyperrectangle_confidence_sets}. The algorithm requires to compute quantiles over the same arrays of data multiple times, and \texttt{R}'s standard \texttt{quantile} function does not expect sorted data, so it is sorting over and over again each time it is run even if the input is already sorted. To overcome this we first implement the analog of \texttt{quantile} function that expects sorted \texttt{vector} as input and does not sort:

\begin{minted}[mathescape, linenos]{r}
quantile_sorted <- function(x, p) {  # same as base quantile(...) with type=7 except for sorting
    n <- length(x)
    h <- (n - 1) * p + 1
    i <- floor(h)
    frac <- h - i
    x[i] + frac * (x[i + 1] - x[i])
}
\end{minted}

\noindent
Then we implement a function that solves equation~(\ref{sim_bound_eq}) via bisection method:

\begin{minted}[mathescape, linenos, texcomments]{r}
calculate.hyperrectangle.bounds <- function(
    alpha,         # significance level
    gamma.left,    # left  bound for bisection method
    gamma.right,   # right bound for bisection method
    w,             # the vector of weights; should satisfy: length(w) = ncol(stat.matrix) 
    stat.matrix    # statistics data matrix of type (\ref{data_table})
)
{
    m                <- nrow(stat.matrix)  # infer number of rows    from data
    k                <- ncol(stat.matrix)  # infer number of columns from data
    left.bounds      <- numeric(k)         # allocate memory for left  bounds of hyperrectangle
    right.bounds     <- numeric(k)         # allocate memory for right bounds of hyperrectangle
    check.sim.bounds <- matrix(0, m, k)    # allocate memory for rejection checks

    stat.matrix.sorted.cols <- apply(stat.matrix, 2, sort)  # sort each column of stat.matrix
    
    while(gamma.right - gamma.left > 0.00001)  # bisection loop
    {
        gamma <- (gamma.right + gamma.left) / 2  # interval center in bisection method
        
        for(i in 1:k){  # loop over stat.matrix columns to compute bounds and rejections
            left.bounds[i]  <- quantile_sorted(stat.matrix.sorted.cols[,i],   gamma*w[i]*alpha/2)
            right.bounds[i] <- quantile_sorted(stat.matrix.sorted.cols[,i], 1-gamma*w[i]*alpha/2)
            check.sim.bounds[,i] <- 
                (stat.matrix[,i] >=  left.bounds[i]) & 
                (stat.matrix[,i] <= right.bounds[i])
        }
      
        sim.check <- rowSums(check.sim.bounds) == k  # for all rows look if all checks passed
        positive.checks <- sum(sim.check) / m        # proportion of rows with all checks passed

        if(positive.checks < 1 - alpha){  # choose left or right subinterval in bisection method
            gamma.right <- gamma
        } else {
            gamma.left  <- gamma
        }
        
        print(gamma)     # print gamma to monitor progress
        flush.console()  # refresh the console for progress to appear
    }
    list(                                    # function returns this:
        left.bounds     = left.bounds,       # left  bounds for hyperrectangle
        right.bounds    = right.bounds,      # right bounds for hyperrectangle
        positive.checks = positive.checks,   # estimated 1 - alpha as sanity check
        gamma           = gamma              # resulted optimal gamma
    )
}
\end{minted}

\noindent
The function returns \texttt{R}'s \texttt{list} with hyperrectangle bounds as defined in~(\ref{hyperrectangle_bounds_definition}) (left and right bounds for each coordinate axis), estimated left part of~(\ref{sim_bound_eq}) that should be $\approx 1 - \alpha$, and optimal $\gamma^*$, the solution of~(\ref{sim_bound_eq}).

\subsection{Calculate hyperectangle bounds for normal distribution}
\label{section:hyperrectangle_bounds_normal}

Now let us use the function defined in the previous section to determine hyperrectangle bounds for order statistics of standard normal distribution\footnote{Remember two interpretations of simultaneous bounds for order statistics: functional bounds for empirical cumulative distribution function and hyperrectangle bounds for vectors of order statistics.} and repeat the result from~\cite{AldorNoimanBrownBujaRolkeStine2013}:

\begin{minted}[mathescape, linenos, texcomments]{r}
res.hyperrectangle.bounds.std.normal <- calculate.hyperrectangle.bounds(  # subsection \ref{section:calculate.hyperrectangle.bounds}
    alpha = 0.05,
    gamma.left  = 0,
    gamma.right = 1,
    w = rep(1, n),            # for now we use equal weights
    sorted.samples.std.normal # was defined in subsection \ref{section:standard_normal_order_stats}
)
\end{minted}

\noindent
To check optimal $\gamma^*$ we write:

\begin{minted}[mathescape, linenos]{r}
res.hyperrectangle.bounds.std.normal$gamma
\end{minted}

\begin{verbatim}
0.147636413574219
\end{verbatim}

\noindent
To check the left part of equation~(\ref{sim_bound_eq}):

\begin{minted}[mathescape, linenos]{r}
res.hyperrectangle.bounds.std.normal$positive.checks
\end{minted}

\begin{verbatim}
0.950008
\end{verbatim}

\noindent
It is very close to $0.95 = 1 - \alpha$.

\noindent
Finally, to print hyperrectangle bounds~(\ref{hyperrectangle_bounds_definition}) we do:

\begin{minted}[mathescape, linenos]{r}
for(i in 1:n) 
    print(c(
        res.hyperrectangle.bounds.std.normal$left.bounds[i], 
        res.hyperrectangle.bounds.std.normal$right.bounds[i]
    ))
\end{minted}

\begin{verbatim}
[1] -3.3796029 -0.1769341
[1] -2.3523762  0.1512951
[1] -1.8390947  0.4232829
[1] -1.4728362  0.6653704
[1] -1.1743844  0.9160947
[1] -0.909935   1.179191
[1] -0.6679801  1.4719381
[1] -0.4190711  1.8332817
[1] -0.1530355  2.3522358
[1]  0.1837448  3.3701641
\end{verbatim}

\noindent
Well, the bounds look symmetric, minimum is around $-3$, maximum is around $3$. To check them properly we need plots.

\subsection{Function to plot order statistics bounds \textasteriskcentered}

So let us do the plotting. We start with writing a function to plot hyperrectangle bounds for order statistics of standard normal samples computed in subsection~\ref{section:hyperrectangle_bounds_normal} as functional bounds for empirical cumulative distribution function in the way it was done in~\cite{AldorNoimanBrownBujaRolkeStine2013}:  

\begin{minted}[mathescape, linenos, texcomments]{r}
plot.order.statistics.bounds <- function(
    sorted.samples.matrix,  # this function expects sorted samples matrix, no other stats allowed
    x.grid,                 # grid for x axis to use in the plot
    theoretical.cdf,        # cdf function like pnorm or punif
    hyperrectangle.bounds   # result of calculate.hyperrectangle.bounds function, see subsec. \ref{section:calculate.hyperrectangle.bounds}
)
{
    n <- ncol(sorted.samples.matrix)  # infer number of order statistics from data
    
    order.stat.mins <- apply(sorted.samples.matrix, 2, min)  # min for each order statistic
    order.stat.maxs <- apply(sorted.samples.matrix, 2, max)  # max for each order statistic
    
    plot(  # plot of theoretical CDF on the provided grid
        x.grid, 
        theoretical.cdf(x.grid), 
        type = 'l', 
        lwd = 10,
        xlab = "x",
        ylab = "F(x)"
    )
    for(i in 1:n)  # loop over order statistics
    {
        rect(  # tight bounds from \cite{AldorNoimanBrownBujaRolkeStine2013} for each statistic
            xleft   = hyperrectangle.bounds$left.bounds[i], 
            xright  = hyperrectangle.bounds$right.bounds[i], 
            ybottom = (i - 1) / n, 
            ytop    = i / n, 
            border  = "red"
        )
        rect(  # absolute minimums and maximums for order statistics seen during simulation
            xleft   = order.stat.mins[i], 
            xright  = order.stat.maxs[i], 
            ybottom = (i - 1) / n, 
            ytop    = i / n
        )
    }
}
\end{minted}

\subsection{Plotting order statistics bounds for normal distribution}

Ok, now we are ready to plot simultaneous bounds for order statistics of standard normal distribution that contain $100(1-\alpha)$ percents of empirical cumulative distribution functions:

\begin{minted}[mathescape, linenos, texcomments]{r}
plot.order.statistics.bounds(
    sorted.samples.matrix = sorted.samples.std.normal,             # defined in subsection \ref{section:standard_normal_order_stats}
    x.grid                = seq(-6, 6, 0.01),                      # to cover very rare outliers 
    theoretical.cdf       = pnorm,                                 # normal CDF
    hyperrectangle.bounds = res.hyperrectangle.bounds.std.normal   # defined in subsection \ref{section:hyperrectangle_bounds_normal}
)
\end{minted}

\begin{figure}[H]
\centering
\includegraphics[width=12cm]{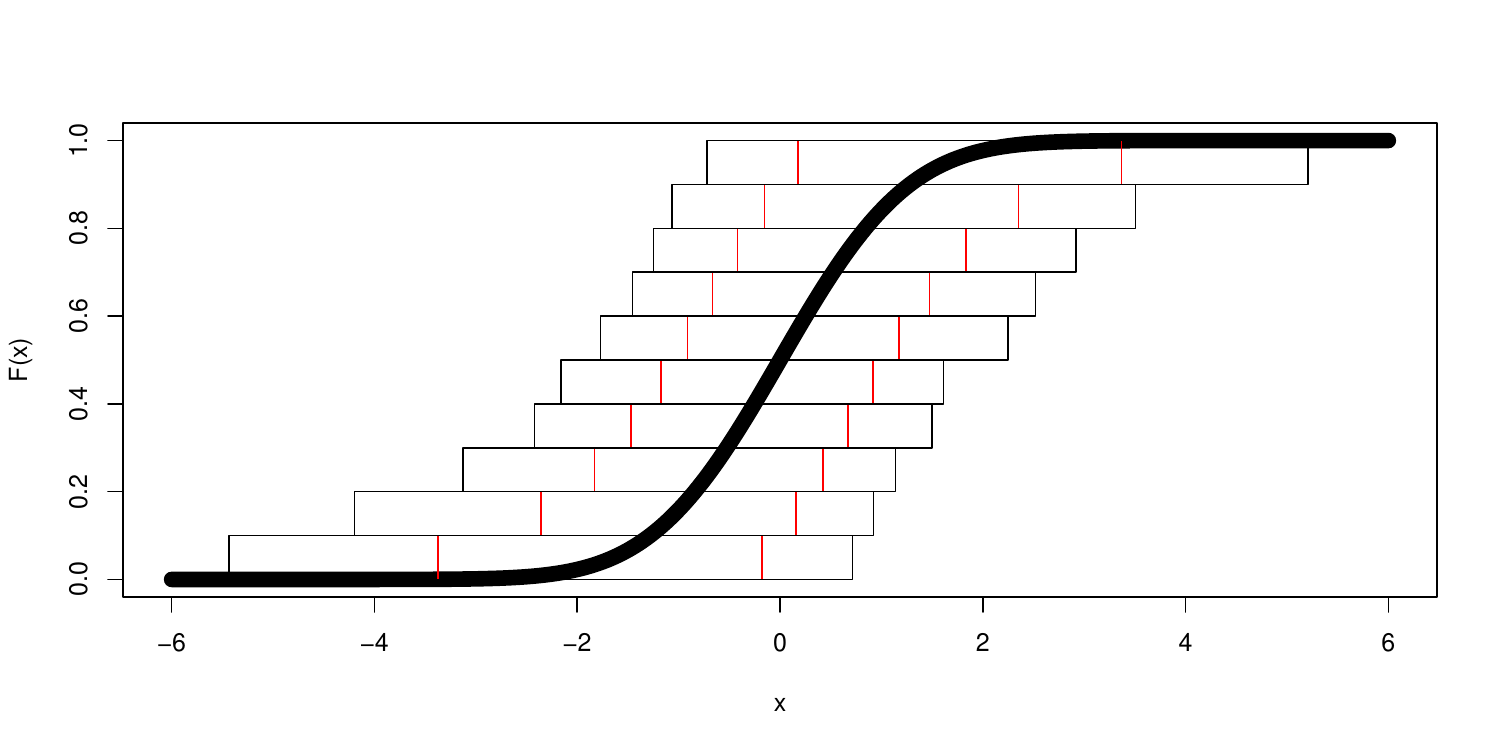}
\caption{Red - numerically estimated functional $0.95$-confidence bounds for empirical cumulative distribution function of the sample from standard normal distribution of size 10 based on simultaneous bounds for simulated standard normal order statistics. Black - minimums and maximums for each order statistics among 1000000 simulated samples.}
\end{figure}

While these bounds are visually significantly tighter compared to Kolmogorov--Smirnov bounds (see~\cite{AldorNoimanBrownBujaRolkeStine2013}), they, as we will see later in the text, lack power against many important alternatives. Why is that? Because not every possible empirical cumulative distribution function that fits inside these bounds is a probable standard normal cumulative distribution function. Moreover, some probable standard normal cumulative distribution functions may lie outside of these bounds.

Can we remain graphical and construct drawable bounds that do have better power than bounds from~\cite{AldorNoimanBrownBujaRolkeStine2013}? As we will see, principal component based tests are the positive answer to this question. But to be interpretable bound-containing plots for principal component based tests should be drawn in different space. 

\subsection{Function to plot 2d projections of multidimensional dataset along with hyperrectangle bounds: contour plots \textasteriskcentered}
\label{section:2d.projections.contour}

We can not directly draw $n$-dimensional vectors of order statistics and corresponding confidence hyperrectangle for them, but we can draw 2d projections of these objects for all possible pairs of coordinate axes. Let us write a function that does it:

\begin{minted}[mathescape, linenos, texcomments]{r}
library(MASS)  # see \cite{mass2002}
plot.2d.projections.contour <- function(
    stat.matrix,           # statistics data matrix of type (\ref{data_table_general})
    hyperrectangle.bounds  # result of calculate.hyperrectangle.bounds, see subsection \ref{section:calculate.hyperrectangle.bounds}
)
{
    upper.panel<-function(x, y, ...){ # used for plots above diagonal
        i <- parent.frame(2)$i    # infer first  axis index from context
        j <- parent.frame(2)$j    # infer second axis index from context
        res <- MASS::kde2d(x, y)  # construct 2d kernel density estimate
        contour(res$x, res$y, res$z, add = TRUE)  # do contour plot based on kde
        rect(  # draw confidence rectangle for corresponding axes
            xleft   = hyperrectangle.bounds$left.bounds[j], 
            xright  = hyperrectangle.bounds$right.bounds[j], 
            ybottom = hyperrectangle.bounds$left.bounds[i], 
            ytop    = hyperrectangle.bounds$right.bounds[i]
        )
    }
    lower.panel <- function(x, y, digits = 2, cex.cor = 4, ...) {  # below diagonal
        usr <- par("usr"); on.exit(par(usr=usr))
        par(usr = c(0, 1, 0, 1))
        r <- abs(cor(x, y))  # compute correlation between columns of stat.matrix
        txt <- format(c(r, r), digits = digits)[1]  # some hack for proper formatting
        text(0.5, 0.5, txt, cex = cex.cor)          # print correlation 
    }
    pairs(  
        stat.matrix, 
        lower.panel = lower.panel, 
        upper.panel = upper.panel, 
        cex.labels  = 5, 
        cex.axis    = 2, 
        row1attop = FALSE
    )
}
\end{minted}

This function produces an upper-triangle matrix of kernel density estimate plots for all possible 2d projections of joint density of \texttt{stat.matrix} columns as well as corresponding rectangular 2d projections of given hyperrectangle confidence bounds. The lower triangle is populated with absolute values of correlations between pairs of \texttt{stat.matrix} columns. 

To compute kernel density estimates we use \texttt{kde2d} function from package \texttt{MASS}. The package contains implementations for methods from the book~\cite{mass2002}, and the name of the package itself is an abbreviation for the book title, ``Modern applied statistics with S''.

\subsection{Plotting 2d projections of sorted samples from normal distribution with corresponding hyperrectangle bounds: contour plots}
\label{section:2d.projections.contour.normal}

Now let us plot a matrix of 2d projections for empirical joint density of order statistics for standard normal distribution and corresponding projections of hyperrectangle bounds that contain $95$ percents of that joint density.

\begin{minted}[mathescape, linenos, texcomments]{r}
plot.2d.projections.contour(             # defined in subsection \ref{section:2d.projections.contour}
    sorted.samples.std.normal,           # defined in subsection \ref{section:standard_normal_order_stats}
    res.hyperectangle.bounds.std.normal  # defined in subsection \ref{section:hyperrectangle_bounds_normal}
)
\end{minted}

\begin{figure}[ht]
\centering
\includegraphics[width=15cm]{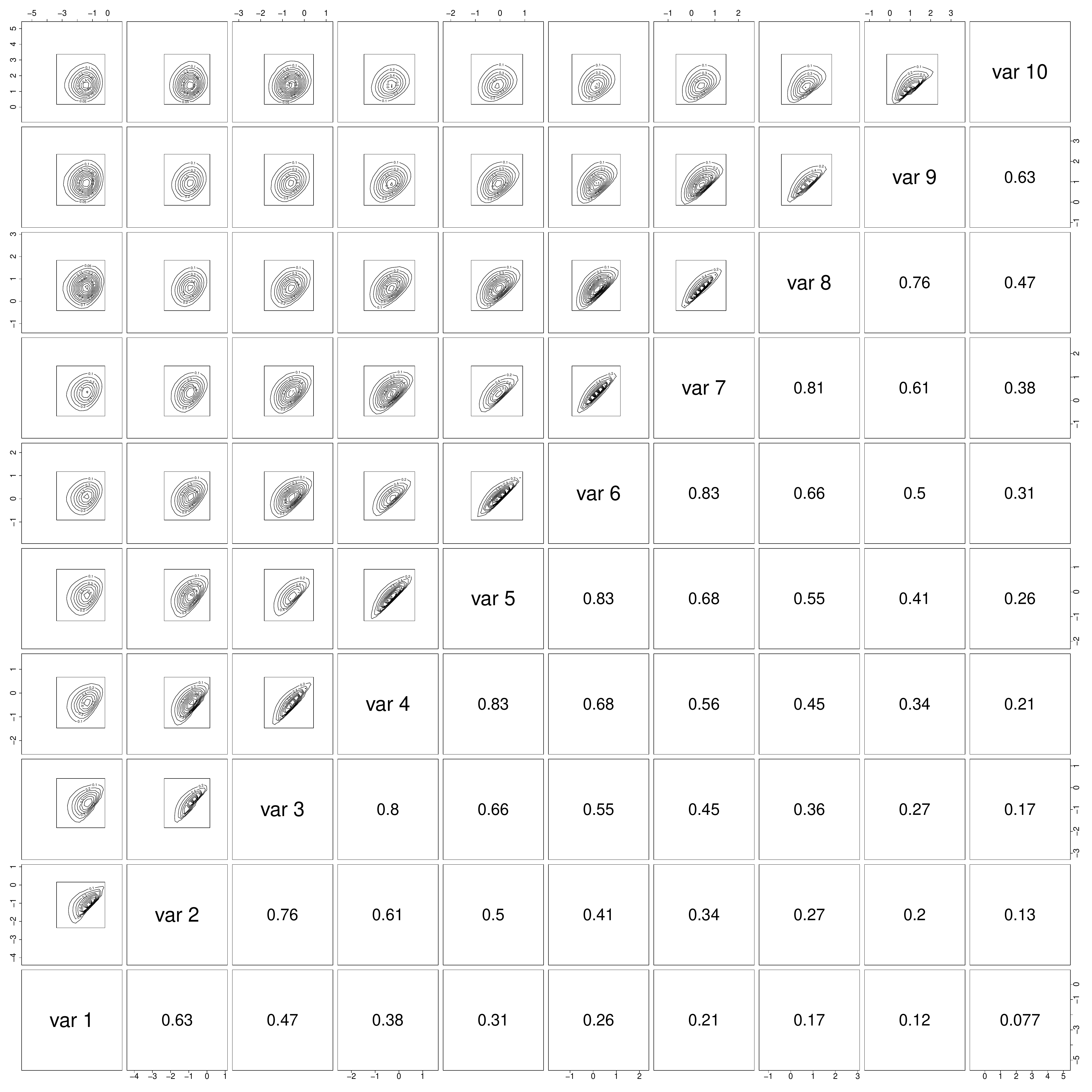}
\caption{Pairwise contour plots for 2d projections of empirical joint density of order statistics for standard normal with pairwise correlations of order statistics.}
\label{figure:contour_order_samples_normal}
\end{figure}

The plot is presented on figure~\ref{figure:contour_order_samples_normal}. First of all, we see that all contours are inside confidence rectangles which is expected and tells us that most of the joint density really lies inside the constructed hyperrectangle.  

We notice that correlations between adjacent middle order statistics are the highest, adjacent order statistics on the edges are slightly less correlated than in the middle, and correlations decrease with the increase of the distance between the indices of order statistics. Thus, 5'th and 6'th order statistics are highly correlated, 4'th and 8'th are much less correlated, and 1'st and 10'th are almost uncorrelated. 

We also notice that for pairs of close order statistics there is a lot of space inside 2d rectangular projections of confidence hyperrectangle where joint density for this pair of statistics is sparse. Examine var~5 vs var~6 plot, for example. Most of the points in empirical joint density for this pair are concentrated around $x = y$ line of the plot. Sure, nothing will be below this line since 6'th order statistic is always greater than 5'th order statistic, so the lower right part of confidence rectangle will always be blank. That is fine. But it looks like upper left part of confidence rectangle does not have a lot of points too. And it appears to be a frequent situation among the 2d projection plots. So it looks like when we try to put joint density of standard normal order statistics into a hyperrectangle, we end up with a confidence set that has a lot of regions where this joint density is sparse. That potentially can lead to low power against some alternatives. 

\subsection{Function to plot 2d projections of multidimentional dataset via hexbin package \textasteriskcentered}
\label{section:2d.projections.hexplom}

Now let us do the same plots but with different visualization tools. Contour plots presented in subsection~\ref{section:2d.projections.contour.normal} give us an idea on where the most points in projected 2d densities are situated, but we would also like to know where the low density regions actually end. For this purpose we will use lightly customized hexbin plots from \texttt{hexbin} package~\cite{hexbinpackage}. Hexbin plots divide the drawing space into hexes and then color each hex depending on the number of points caught by that hex. We end up with hex-based heat map that serves the same purpose in 2d situation as histogram surves in 1d situation. Now to the actual function:

\begin{minted}[mathescape, linenos, texcomments]{r}
library(hexbin)   # see \cite{hexbinpackage}
library(lattice)  # see \cite{latticebook}
library(grid)     # part of base R

plot.2d.projections.hexplom <- function(
    stat.matrix,           # statistics data matrix of type (\ref{data_table_general})
    hyperrectangle.bounds  # result of calculate.hyperrectangle.bounds function
)
{
    custom.panel <- function(x, y, i, j, ...) { 
        panel.hexplom(x, y, ...)  # draw a standard hexplom panel 
        grid.rect(                # draw a 2d projection of hyperrectangle on top of it 
            (hyperrectangle.bounds$right.bounds[j] + hyperrectangle.bounds$left.bounds[j]) / 2, 
            (hyperrectangle.bounds$right.bounds[i] + hyperrectangle.bounds$left.bounds[i]) / 2, 
            width  = hyperrectangle.bounds$right.bounds[j] - hyperrectangle.bounds$left.bounds[j], 
            height = hyperrectangle.bounds$right.bounds[i] - hyperrectangle.bounds$left.bounds[i],
            default.units = "native",          
            gp = gpar(col = "black", fill = NA, lwd = 4)   
        ) 
    }
    hexplom(
        stat.matrix,
        panel = custom.panel,  # redefine default panel function 
        varname.cex = 4,       # change variable name font size
        axis.text.cex = 2,     # change axes font size
        varnames = sapply(1:ncol(stat.matrix), function(x) paste("var", x))  # redefine var names
    )
}
\end{minted}

Package \texttt{hexbin} uses \texttt{lattice} visualization system~\cite{latticebook} that is suited well for multivariate data and works quite fast. To modify hexbin plots we also use \texttt{grid} package, which is part of \texttt{Rcore}~\cite{Rcore}.

\subsection{Plot of 2d projections of sorted samples from normal distribution via hexbin}
\label{section:2d.projections.hexbin.normal}

Now let us do exactly the same plot as in subsection~\ref{section:2d.projections.contour.normal}, but with hexbin-based function:

\begin{minted}[mathescape, linenos, texcomments]{r}
plot.2d.projections.hexplom(              # defined in subsection \ref{section:2d.projections.hexplom}
    sorted.samples.std.normal,            # defined in subsection \ref{section:standard_normal_order_stats}
    res.hyperrectangle.bounds.std.normal  # defined in subsection \ref{section:hyperrectangle_bounds_normal}
)
\end{minted}

\begin{figure}[ht]
\centering
\includegraphics[width=15cm]{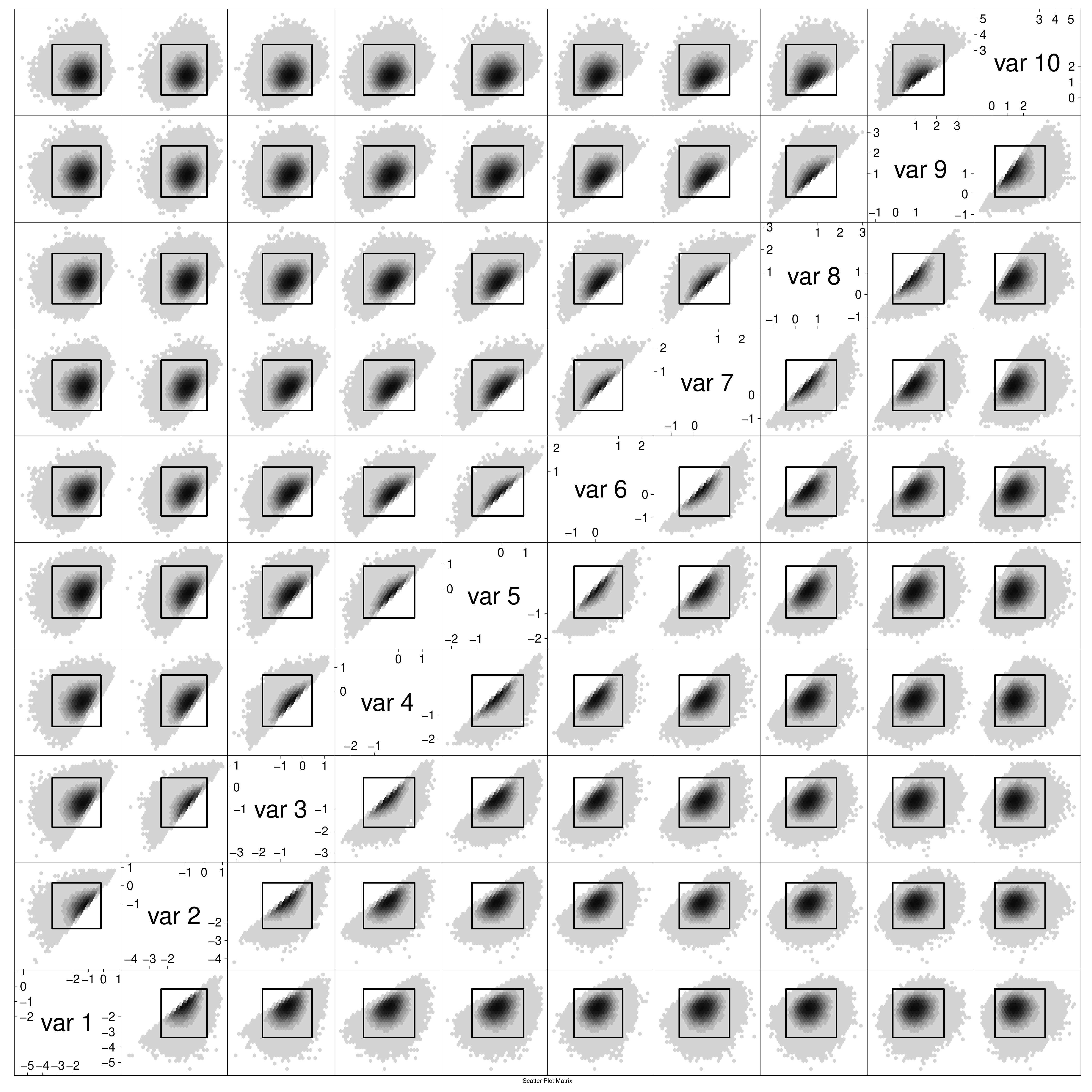}
\caption{Pairwise hexbin plots for 2d projections of empirical joint density of order statistics for standard normal samples of size 10.}
\label{figure:2d_projections_order_stats_normal}
\end{figure}

The plots are presented on figure~\ref{figure:2d_projections_order_stats_normal}.
On these plots grey corresponds to lower density levels and black corresponds to higher density levels. They give a better picture for grey low density regions and their absolute bounds than plots from subsection~\ref{section:2d.projections.contour.normal}, where contours cover mostly high density regions. 

\subsection{Calculate principal components for ordered samples from normal distribution}
\label{section:principal_components_normal}

As we have seen on the plots in subsections~\ref{section:2d.projections.contour.normal} and~\ref{section:2d.projections.hexbin.normal}, hyperrectangle bounds for joint distribution of normal order statistics contain a lot of regions where joint density is sparse, which is in term potentially bad for tests based on those bounds.

But what if we rotate the space and then construct hyperrectangle bounds for rotated data? We may end up with better bounds that do have less sparse density regions inside them. The natural thing to try here is principal component analysis applied to a matrix of simulated sorted samples:

\begin{minted}[mathescape, linenos]{r}
sorted.samples.std.normal.p.comp <- princomp(sorted.samples.std.normal)
sorted.samples.std.normal.p.comp
sorted.samples.std.normal.p.comp$loadings
\end{minted}

\begin{verbatim}
Call:
princomp(x = sorted.samples.std.normal)

Standard deviations:
   Comp.1    Comp.2    Comp.3    Comp.4    Comp.5    Comp.6    Comp.7    Comp.8 
1.0005157 0.6757698 0.4973716 0.3746707 0.2918326 0.2351403 0.1949415 0.1650635 
   Comp.9   Comp.10 
0.1427509 0.1249043 

 10  variables and  1000000 observations.

Loadings:
      Comp.1 Comp.2 Comp.3 Comp.4 Comp.5 Comp.6 Comp.7 Comp.8 Comp.9 Comp.10
 [1,]  0.317  0.535  0.565  0.439  0.274  0.147                             
 [2,]  0.317  0.365  0.108 -0.265 -0.490 -0.502 -0.376 -0.210               
 [3,]  0.316  0.243 -0.121 -0.383 -0.237  0.157  0.470  0.501  0.331  0.131 
 [4,]  0.316  0.140 -0.247 -0.287  0.119  0.403  0.169 -0.319 -0.545 -0.363 
 [5,]  0.316        -0.304 -0.104  0.334  0.195 -0.330 -0.317  0.295  0.590 
 [6,]  0.316        -0.304  0.104  0.335 -0.190 -0.330  0.319  0.293 -0.591 
 [7,]  0.316 -0.141 -0.249  0.287  0.118 -0.403  0.165  0.319 -0.544  0.366 
 [8,]  0.316 -0.244 -0.123  0.381 -0.239 -0.162  0.466 -0.503  0.330 -0.133 
 [9,]  0.316 -0.364  0.109  0.262 -0.492  0.504 -0.371  0.211               
[10,]  0.316 -0.535  0.565 -0.436  0.277 -0.148                             

               Comp.1 Comp.2 Comp.3 Comp.4 Comp.5 Comp.6 Comp.7 Comp.8 Comp.9
SS loadings       1.0    1.0    1.0    1.0    1.0    1.0    1.0    1.0    1.0
Proportion Var    0.1    0.1    0.1    0.1    0.1    0.1    0.1    0.1    0.1
Cumulative Var    0.1    0.2    0.3    0.4    0.5    0.6    0.7    0.8    0.9
               Comp.10
SS loadings        1.0
Proportion Var     0.1
Cumulative Var     1.0
\end{verbatim}

To perform principal component analysis we use \texttt{princomp} function from base \texttt{R}. We are interested in \texttt{loadings} matrix that gives us coefficients to apply to the original data for rotating and getting principal components, and in vector of standard deviations for principal components that gives us component ``importances''. 

Each principal component is a linear combination of order statistics, and coefficients of this linear combination are given by corresponding column from \texttt{loadings} matrix. 

We see that 
\begin{enumerate}
\item first principal component is responsible for location and is proportional to sample mean, 
\item second principal component is responsible for scale and is some kind of linear analog of variance,
\item third, fifth, seventh and ninth components are symmetric and measure, roughly speaking, the differences between sample middle and sample tails, performing together the function of kurtosis,
\item fourth, sixth, eighth and tenth components are asymmetric, performing together the function of asymmetry.  
\end{enumerate}
So we got very interpretable principal components!

\subsection{Histograms, principal components, normal distribution}

Now let us plot histograms of principal components we've got in subsection~\ref{section:principal_components_normal}:

\begin{minted}[mathescape, linenos]{r}
par(mfcol = c(5, 2))
for(i in 1:n)
    hist(
        sorted.samples.std.normal.p.comp$scores[,i],
        main = paste0(i," order statistic principal component"), xlab = "x"
    )
\end{minted}

\begin{figure}[ht]
\centering
\includegraphics[width=15cm]{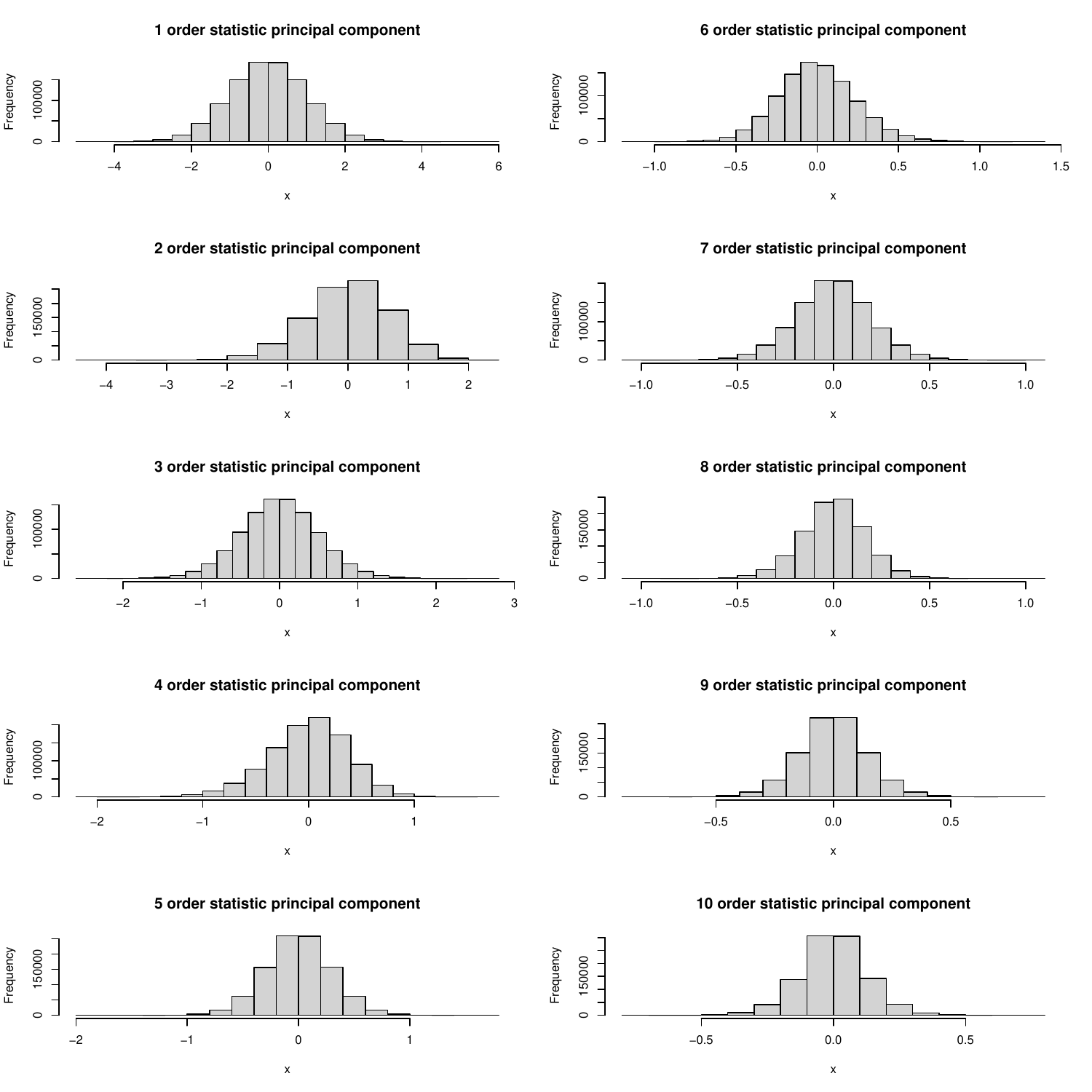}
\caption{Histograms for empirical distributions of principal components of order statistics for samples from standard normal distribution of size 10.}
\label{figure:hists_os_pc_normal}
\end{figure}

\noindent
The histograms are presented on figure~\ref{figure:hists_os_pc_normal}.
We see that empirical distributions of principal components are unimodal, so we can once again use confidence intervals instead of one-dimensional highest density regions. 

\subsection{Calculate hyperrectangle bounds for principal components of ordered samples from normal distribution; equal weights}
\label{section:hyperrectangle.bounds.std.normal.pc.1}

Principal component analysis gives natural ``importances'' for components based on their variances, and it is logical to use them inside \texttt{calculate.hyperrectangle.bounds} as weights. As we will see later, this variance-weights approach produces a better test than equal-weights approach, but we will nonetheless start with equal-weights. 

In this subsection let us calculate hyperrectangle bounds for principal components of standard normal order statistics using equal weights for components:

\begin{minted}[mathescape, linenos, texcomments]{r}
res.hyperrectangle.bounds.std.normal.pc.1 <- calculate.hyperrectangle.bounds(  # subsection \ref{section:calculate.hyperrectangle.bounds}
    alpha = 0.05,
    gamma.left  = 0,
    gamma.right = 2,
    w = rep(1, n),                           # equal weights for principal components
    sorted.samples.std.normal.p.comp$scores  # prin.comp. of normal order stats; subsec. \ref{section:principal_components_normal}
)
\end{minted}

\noindent
We look at optimal $\gamma^*$, the solution of~(\ref{sim_bound_eq}):

\begin{minted}[mathescape, linenos]{r}
res.hyperrectangle.bounds.std.normal.pc.1$gamma
\end{minted}

\begin{verbatim}
0.114158630371094
\end{verbatim}

\noindent
Then we check the proportion of rows of matrix \texttt{sorted.samples.std.normal.p.comp\$scores} that are inside of hyperrectangle bounds \texttt{res.hyperrectangle.bounds.std.normal.pc.1}:

\begin{minted}[mathescape, linenos]{r}
res.hyperrectangle.bounds.std.normal.pc.1$positive.checks
\end{minted}

\begin{verbatim}
0.950015
\end{verbatim}

\noindent
Finally, we print hyperrectangle bounds for each principal component:

\begin{minted}[mathescape, linenos]{r}
for(i in 1:n) 
    print(c(
        res.hyperrectangle.bounds.std.normal.pc.1$left.bounds[i], 
        res.hyperrectangle.bounds.std.normal.pc.1$right.bounds[i]
    ))
\end{minted}

\begin{verbatim}
[1] -2.761522  2.760003
[1] -2.040742  1.634704
[1] -1.428760  1.434087
[1] -1.2102316  0.9576026
[1] -0.8432864  0.8402371
[1] -0.6320974  0.7212072
[1] -0.5746237  0.5726034
[1] -0.5136518  0.4653171
[1] -0.4240698  0.4236289
[1] -0.3939270  0.3735682
\end{verbatim}

\newpage
\subsection{Plot of 2d projections of principal components of sorted samples from normal distribution; equal weights}
\label{section:2d.projections.contour.pc1}

Ok, now let us plot a matrix of 2d projections for empirical joint density of principal components of standard normal order statistics and corresponding projections of hyperrectangle bounds constructed in subsection~\ref{section:hyperrectangle.bounds.std.normal.pc.1} with equal weight for each principal component: 

\begin{minted}[mathescape, linenos, texcomments]{r}
plot.2d.projections.contour(                   # defined in subsection \ref{section:2d.projections.contour}
    sorted.samples.std.normal.p.comp$scores,   # defined in subsection \ref{section:principal_components_normal}
    res.hyperrectangle.bounds.std.normal.pc.1  # defined in subsection \ref{section:hyperrectangle.bounds.std.normal.pc.1}
)
\end{minted}

\begin{figure}[H]
\centering
\includegraphics[width=15cm]{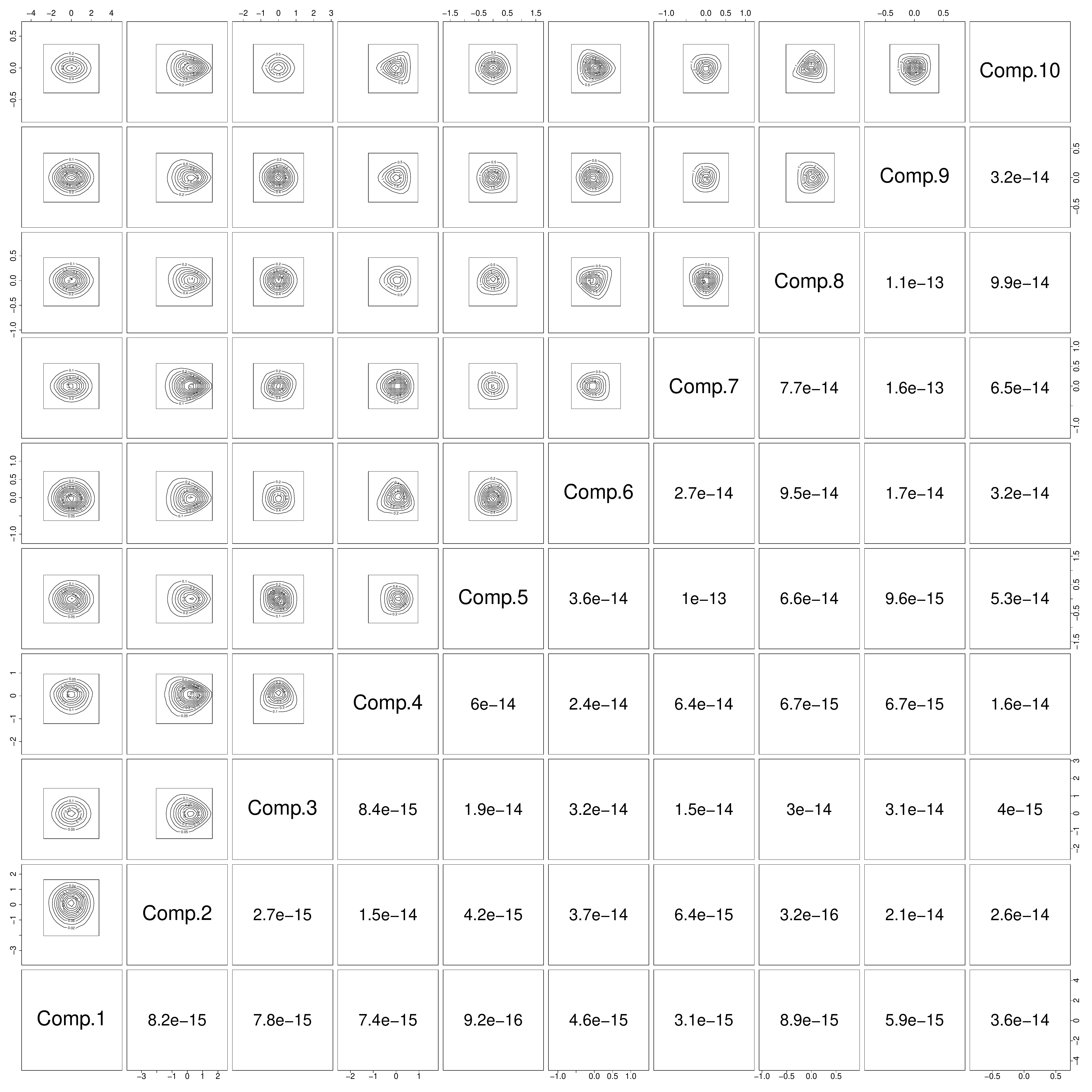}
\caption{Pairwise contour plots for 2d projections of empirical joint density of principal components of standard normal order statistics. Confidence rectangles are constructed using equal weights for principal components. Principal components are uncorrelated by construction, which we see in lower triangle.}
\label{figure:contour_pc1_order_stat_normal}
\end{figure}

We see that contours of projected densities for principal components from figure~\ref{figure:contour_pc1_order_stat_normal} are ``more round'' compared to contours of projected densities for unrotated order statistics from figure~\ref{figure:contour_order_samples_normal}. We also see that contours of principal components cover greater areas inside confidence rectangles compared to contours of order statistics which potentially should be good for hyperrectangle's power.  

\newpage
\subsection{Plot of 2d projections of principal components of sorted samples from normal distribution via hexplom; equal weights}
\label{section:2d.projections.hexplom.pc1}

Next let us do the same plot as in the previous section, but with the help of \texttt{hexbin} package:

\begin{minted}[mathescape, linenos, texcomments]{r}
plot.2d.projections.hexplom(                   # defined in subsection \ref{section:2d.projections.hexplom}
    sorted.samples.std.normal.p.comp$scores,   # defined in subsection \ref{section:principal_components_normal}
    res.hyperrectangle.bounds.std.normal.pc.1  # defined in subsection \ref{section:hyperrectangle.bounds.std.normal.pc.1}
)

\end{minted}

\begin{figure}[H]
\centering
\includegraphics[width=15cm]{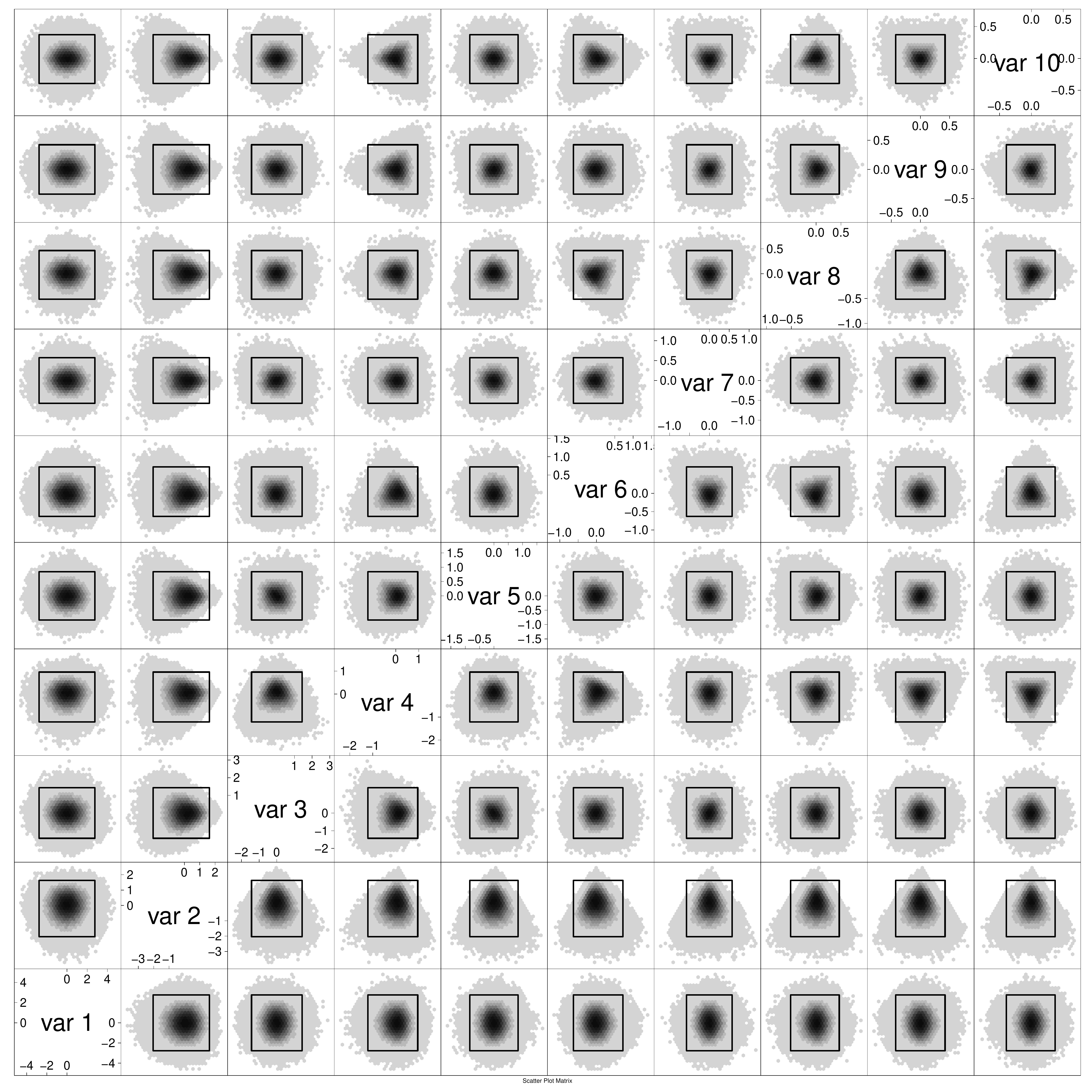}
\caption{Pairwise hexbin plots for 2d projections of empirical joint density of principal components of order statistics for standard normal samples of size 10. Confidence rectangles are constructed using equal weights for principal components.}
\label{figure:hexbin_pc1_order_stat_normal}
\end{figure}

The data is essentially the same as the data that was used to plot figure~\ref{figure:2d_projections_order_stats_normal}, only the rotation is different.
Compared to figure~\ref{figure:contour_pc1_order_stat_normal}, on figure~\ref{figure:hexbin_pc1_order_stat_normal} we can observe the grey low-density projected bounds of the 10-dimensional cloud of principal components. 

\newpage
\subsection{Calculate hyperrectangle bounds for principal components of ordered samples from normal distribution; principal component based weights}
\label{section:hyperrectangle.bounds.std.normal.pc.2}

In subsection~\ref{section:hyperrectangle.bounds.std.normal.pc.1} we've constructed hyperrectangle bounds for principal components of standard normal order statistics using the equal weights for the components. In this subsection we will construct hyperrectangle bounds for components using variance-based weights.
Then we will plot projections like in subsections \ref{section:2d.projections.contour.pc1} and \ref{section:2d.projections.hexplom.pc1}.
Now to the code:

\begin{minted}[mathescape, linenos, texcomments]{r}
w_ <- sorted.samples.std.normal.p.comp$sdev^2 / sum(sorted.samples.std.normal.p.comp$sdev^2)

res.hyperrectangle.bounds.std.normal.pc.2 <- calculate.hyperrectangle.bounds(  # subsection \ref{section:calculate.hyperrectangle.bounds}
    alpha = 0.05,
    gamma.left  = 0,
    gamma.right = 2,
    w = w_,  # principal component weights based on their relative variance
    sorted.samples.std.normal.p.comp$scores  # prin.comp. of normal order stats; subsec. \ref{section:principal_components_normal}
)
\end{minted}

\noindent
Let us print variance-based weights for principal component:

\begin{minted}[mathescape, linenos]{r}
w_
\end{minted}

\begin{verbatim}
[1] 0.479619436 0.218799596 0.118525262 0.067258693 0.040805299 0.026491289
[7] 0.018207804 0.013054224 0.009763526 0.007474872
\end{verbatim}

\noindent
We see that most of variability in data is covered by first four principal components. We will still use all of them nonetheless.

Optimal $\gamma^*$ is:

\begin{minted}[mathescape, linenos]{r}
res.hyperrectangle.bounds.std.normal.pc.2$gamma
\end{minted}

\begin{verbatim}
1.05715179443359
\end{verbatim}

\noindent
Proportion of positive checks is fine:

\begin{minted}[mathescape, linenos]{r}
res.hyperrectangle.bounds.std.normal.pc.2$positive.checks
\end{minted}

\begin{verbatim}
0.95
\end{verbatim}

\noindent
Hyperrectangle bounds for variance-based weights look as follows:

\begin{minted}[mathescape, linenos]{r}
for(i in 1:n) 
    print(c(
        res.hyperrectangle.bounds.std.normal.pc.2$left.bounds[i], 
        res.hyperrectangle.bounds.std.normal.pc.2$right.bounds[i]
    ))
\end{minted}

\begin{verbatim}
[1] -2.240466   2.232921
[1] -1.849241   1.522729
[1] -1.414755   1.417066
[1] -1.289917   1.007060
[1] -0.9457654  0.9429225
[1] -0.7330497  0.8552919
[1] -0.7068574  0.7018197
[1] -0.657535   0.594390
[1] -0.5629872  0.5559923
[1] -0.5374102  0.4979408
\end{verbatim}

\noindent
Compared to the bounds from subsection~\ref{section:hyperrectangle.bounds.std.normal.pc.1}, these bounds are tighter for first and looser for last principal components.

\newpage
\subsection{Plot of 2d projections of principal components of sorted samples from normal distribution; principal component based weights}

In this subsection we do the plots like in subsection~\ref{section:2d.projections.contour.pc1}, but for hyperrectangle bounds constructed with variance-based weights:

\begin{minted}[mathescape, linenos, texcomments]{r}
plot.2d.projections.contour(                   # defined in subsection \ref{section:2d.projections.contour}
    sorted.samples.std.normal.p.comp$scores,   # defined in subsection \ref{section:principal_components_normal}
    res.hyperrectangle.bounds.std.normal.pc.2  # defined in subsection \ref{section:hyperrectangle.bounds.std.normal.pc.2}
)
\end{minted}

\begin{figure}[H]
\centering
\includegraphics[width=15cm]{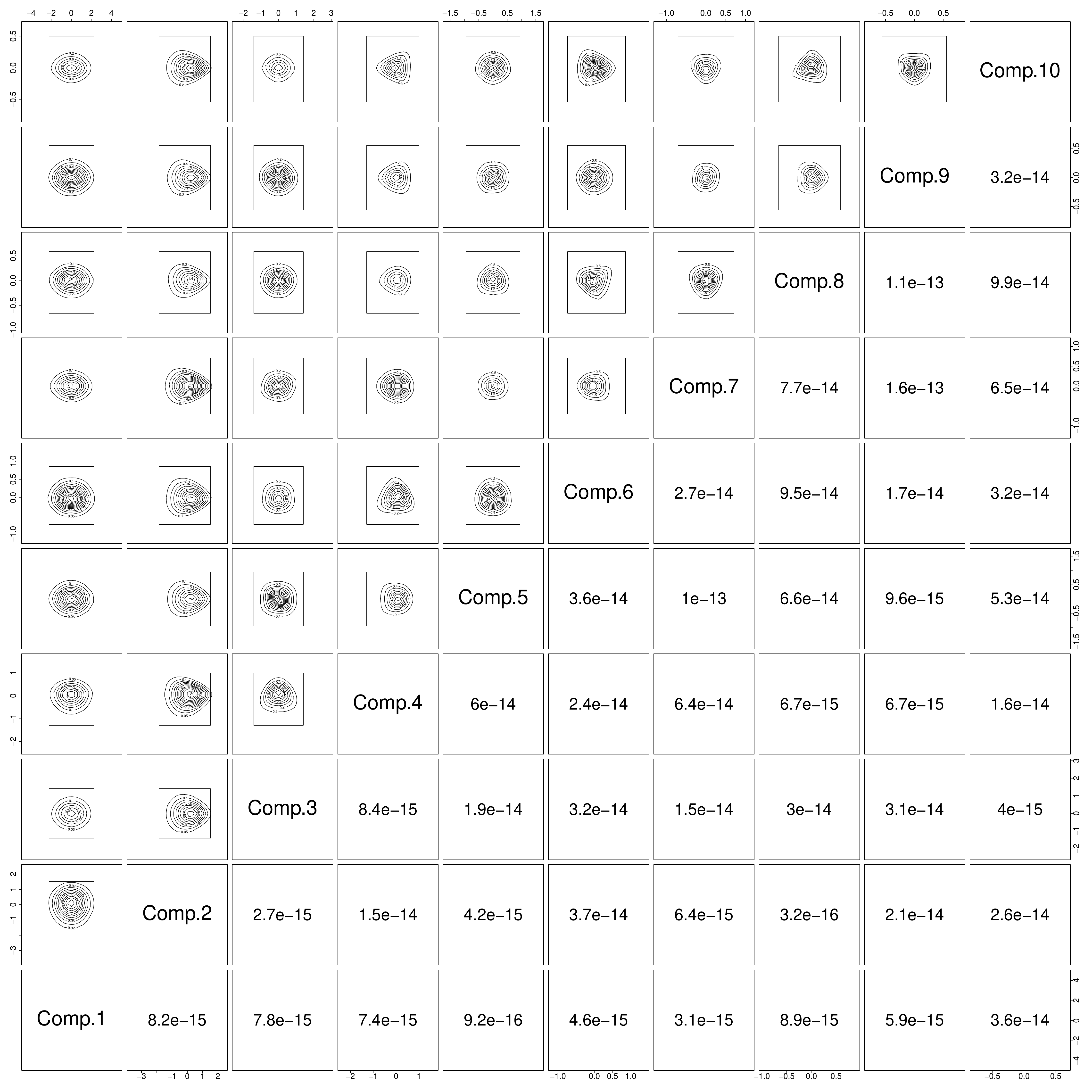}
\caption{Like figure \ref{figure:contour_pc1_order_stat_normal}, but hyperrectangle bounds are variance-based.}
\label{figure:contour_pc2_order_stat_normal}
\end{figure}

\newpage
\subsection{Plot of 2d projections of principal components of sorted samples from normal distribution via hexplom; principal component based weights}

And in this subsection we do the plots like in subsection~\ref{section:hyperrectangle.bounds.std.normal.pc.1}, but for variance-based hyperrectangle bounds:

\begin{minted}[mathescape, linenos, texcomments]{r}
plot.2d.projections.hexplom(                   # defined in subsection \ref{section:2d.projections.hexplom}
    sorted.samples.std.normal.p.comp$scores,   # defined in subsection \ref{section:principal_components_normal}
    res.hyperrectangle.bounds.std.normal.pc.2  # defined in subsection \ref{section:hyperrectangle.bounds.std.normal.pc.2}
)
\end{minted}

\begin{figure}[H]
\centering
\includegraphics[width=15cm]{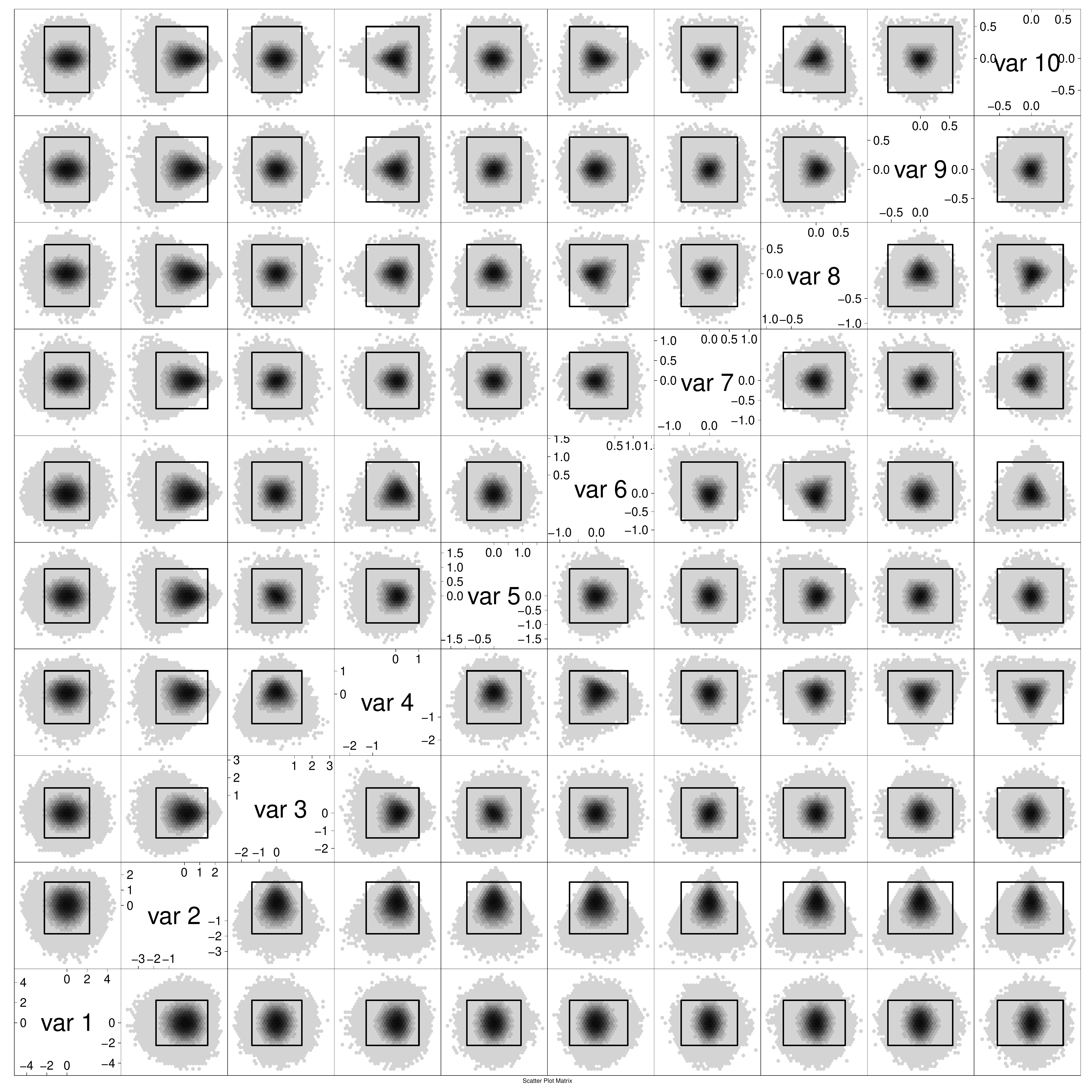}
\caption{Like figure \ref{figure:hexbin_pc1_order_stat_normal}, but hyperrectangle bounds are variance-based.}
\label{figure:hexbin_pc2_order_stat_normal}
\end{figure}

For \texttt{var 1} vs \texttt{var 2} plot we see that the confidence rectangle is significantly closer to the most dense black-painted area than the corresponding confidence rectangle on figure~\ref{figure:hexbin_pc1_order_stat_normal}. For \texttt{var 9} vs \texttt{var 10} plot we see the opposite: much looser confidence rectangle that is significantly farther from the most dense black area if compared to the corresponding rectangle on figure~\ref{figure:hexbin_pc1_order_stat_normal}.

\newpage
\subsection{Function to check hyperrectangle bounds \textasteriskcentered}
\label{section:check.hyperrectangle.bounds}

Ok, we did some plotting and now we somewhat understand the geometry of joint distribution of order statistics from standard normal samples, rotated or not. 

It is time to finally compare the actual powers of direct simultaneous confidence bounds for order statistics, hyperrectangle equal-weighted bounds for principal components  of those order statistics, and hyperrectangle variance-weighted bounds.

To do that, let us first write a general function that given empirical joint distribution of some statistics and some hyperrectangle bounds tells us how many points of empirical distribution lie outside of this bounds (see section~\ref{section:test_power} for notes on power estimation):

\begin{minted}[mathescape, linenos, texcomments]{r}
check.hyperrectangle.bounds <- function(
    stat.matrix,           # statistics data matrix of type (\ref{data_table_general}) 
    hyperrectangle.bounds  # result of calculate.hyperrectangle.bounds, see subsection \ref{section:calculate.hyperrectangle.bounds}
)
{
    m <- nrow(stat.matrix)           # infer number of rows    from stat.matrix
    k <- ncol(stat.matrix)           # infer number of columns from stat.matrix
    check.bounds <- matrix(0, m, k)  # allocate memory for rejection checks
    for(i in 1:k)  # loop over stat.matrix columns
    {
        check.bounds[,i] <-  # check the bounds in column in vectorized manner
            (stat.matrix[,i] < hyperrectangle.bounds$left.bounds[i] ) | 
            (stat.matrix[,i] > hyperrectangle.bounds$right.bounds[i])
    }

    check <- apply(check.bounds, 1, sum) == 0  # number of stat.matrix rows inside the bounds 
    1 - sum(check) / m  # proportion of stat.matrix rows outside the bounds
}
\end{minted}

Having the function, we check if constructed confidence bounds of three mentioned types provide the desired type-1 error $\alpha = 0.05$ defined before. To do that check, we simulate another sample of 10-sized samples from standard normal ($H_0$ distribution), apply \texttt{check.hyperrectangle.bounds} function to that sample with hyperrectangle bounds for order statistics from subsection~\ref{section:hyperrectangle_bounds_normal} (that were constructed using the different sample of samples from the same $H_0$ distribution), then apply principal component coefficients we got in subsection~\ref{section:principal_components_normal} to simulated sample of samples, then apply \texttt{check.hyperrectangle.bounds} two more times: to principal-component-transformed data with equal-weighted hyperrectangle bounds for principal components from subsection~\ref{section:hyperrectangle.bounds.std.normal.pc.1}, and to principal-component-transformed data with variance-weighted hyperrectangle bounds from subsection~\ref{section:hyperrectangle.bounds.std.normal.pc.2} (both equal-weighted and variance-weighted bounds were constructed on principal components of another sample of samples from $H_0$, that's why we expect to get a proper type-1 error):

\begin{minted}[mathescape, linenos, texcomments]{r}
sorted.samples.H0 <- get.sorted.samples.cpp(  # defined in \ref{section:functions_to_generate_order_stats}
    function(x) rnorm(x, 0, 1),  # standard normal; H0 distribution
    n, m                         # defined in \ref{section:sample_size_num_samples}
)
check.hyperrectangle.bounds(
    sorted.samples.H0, 
    res.hyperrectangle.bounds.std.normal  # defined in subsection \ref{section:hyperrectangle_bounds_normal}
)
p.comp.H0 <- predict(  # uses loadings from given princomp object to compute PCs for new data
    sorted.samples.std.normal.p.comp,  # defined in subsection \ref{section:principal_components_normal}
    sorted.samples.H0
)
check.hyperrectangle.bounds(
    p.comp.H0, 
    res.hyperrectangle.bounds.std.normal.pc.1  # defined in subsection \ref{section:hyperrectangle.bounds.std.normal.pc.1}
)
check.hyperrectangle.bounds(
    p.comp.H0, 
    res.hyperrectangle.bounds.std.normal.pc.2  # defined in subsection \ref{section:hyperrectangle.bounds.std.normal.pc.2}
)
\end{minted}

\begin{verbatim}
0.049859
0.0498150000000001
0.049759
\end{verbatim}

We see that all three bounds (simultaneous for order stats, equal-weighted for PCs of order stats, and variance-weighted for PCs of order stats) provide the expected type-1 error. It is important that we constructed hyperrectangle bounds using one sample of samples from standard normal and performed bounds check with another sample of samples from standard normal.

We can use the same \texttt{check.hyperrectangle.bounds} function to calculate power. To do that we need to simulate sample of 10-sized samples from any distribution other than standard normal ($H_1$ distribution) and perform the same chain of actions described in the previous paragraph. It will give us estimates of power against the chosen distribution for each of the three bounds:

\begin{minted}[mathescape, linenos]{r}
sorted.samples.H1 <- get.sorted.samples.cpp(
    function(x) rnorm(x, 0, 0.2),  # normal with sd = 0.2; H1 distribution
    n, m)
check.hyperrectangle.bounds(sorted.samples.H1, res.hyperrectangle.bounds.std.normal)
p.comp.H1 <- predict(sorted.samples.std.normal.p.comp, sorted.samples.H1)
check.hyperrectangle.bounds(p.comp.H1, res.hyperrectangle.bounds.std.normal.pc.1)
check.hyperrectangle.bounds(p.comp.H1, res.hyperrectangle.bounds.std.normal.pc.2)
\end{minted}

\begin{verbatim}
0.256675
0.999974
1
\end{verbatim}

We see that at least against $N(0,0.2^2)$ alternative both principal-component-based tests show greater power than order-statistics-based test from~\cite{AldorNoimanBrownBujaRolkeStine2013}. Later on we will do a very detailed power study, but for now let us proceed with plots that will illustrate the geometry of how these tests reject.

Pay attention: we do not perform PCA on sample of samples from alternative distribution, we use \texttt{loadings} coefficients that we got from PCA performed on sample of samples from null distribution.

\subsection{Function to plot \texorpdfstring{$H_0$}{H0} vs \texorpdfstring{$H_1$}{H1} \textasteriskcentered}
\label{section:H0vsH1_ggplot}

Figures
\ref{figure:contour_order_samples_normal},
\ref{figure:2d_projections_order_stats_normal};
\ref{figure:contour_pc1_order_stat_normal},
\ref{figure:hexbin_pc1_order_stat_normal};
\ref{figure:contour_pc2_order_stat_normal},
\ref{figure:hexbin_pc2_order_stat_normal}
only illustrate the geometry of $H_0$ distribution and corresponding $H_0$ bounds.
In this section we will write a function to plot given $H_1$ distribution along with given $H_0$ distribution and the bounds. Package \texttt{hexbin} is fast but is not very well suited for plotting multiple datasets on the same \texttt{hexplom} grid, so we have to use less performant but more flexible \texttt{ggplot2} \cite{ggplot2_hadley} based libraries:

\begin{minted}[mathescape, linenos, texcomments]{r}
library(GGally)   # see \cite{GGally_package}; for ggpairs
library(ggrastr)  # see \cite{ggrastr_package}; to reduce plot size by rasterizing some elements

ggplot.H0.H1.hexplot.pairs <- function(
    stat.matrix.H0,             # statistics data matrix of type \eqref{data_table}
    stat.matrix.H1,             # statistics data matrix of type \eqref{data_table_general}
    res.hyperrectangle.bounds,  # result of calculate.hyperrectangle.bounds, see subsec. \ref{section:calculate.hyperrectangle.bounds}
    col.nums,                   # column indices to use in plots 
    row.sample.size = 10000     # number of rows to sample from each stat.matrix for plots
)
{
    row.sample.size.H0 <- min(row.sample.size, nrow(stat.matrix.H0))  # adjust sizes
    row.sample.size.H1 <- min(row.sample.size, nrow(stat.matrix.H1))  # for safe sampling
    
    df = rbind(  # construct a data.frame to use with ggplot functions
        cbind(as.data.frame(stat.matrix.H0[1:row.sample.size.H0,]), group='H_0'),
        cbind(as.data.frame(stat.matrix.H1[1:row.sample.size.H1,]), group='H_1')
    )

    custom_hex <- function(data, mapping, bins = 25, ...) {  # hexbin panel
        vars <- names(data)[col.nums]  # variable names
        
        x_var <- rlang::as_label(mapping$x)  # get x variable name from context
        y_var <- rlang::as_label(mapping$y)  # get y variable name from context
        
        i <- match(x_var, vars)  # infer index from x variable name
        j <- match(y_var, vars)  # infer index from y variable name
        
        ggplot(data = data, mapping = mapping) +  # general plot canvas
        rasterize(geom_hex(bins = bins, ...), dpi = 50) +  # rasterized hexbin plot
        geom_rect(  # projections of proper hyperrectangle bounds
            xmin = res.hyperrectangle.bounds$left.bounds[i], 
            xmax = res.hyperrectangle.bounds$right.bounds[i],
            ymin = res.hyperrectangle.bounds$left.bounds[j], 
            ymax = res.hyperrectangle.bounds$right.bounds[j],
            fill = NA,
            color = "red",
            alpha = 0.5,
            inherit.aes = FALSE   
        ) +
        theme_minimal()
    }

    flip_rows_ggpairs <- function(ggpairs_plot) {  # to align with previous figures
        n <- length(ggpairs_plot$plots)  
        p <- ggpairs_plot$nrow  
        
        plot_matrix <- matrix(1:n, nrow = p, ncol = p, byrow = TRUE)
        plot_matrix_reversed <- plot_matrix[p:1, ]
        
        ggpairs_plot$plots <- ggpairs_plot$plots[as.vector(t(plot_matrix_reversed))]
        ggpairs_plot$yAxisLabels <- rev(ggpairs_plot$yAxisLabels)
        
        ggpairs_plot
    }

    flip_rows_ggpairs(ggpairs(  # ggpairs with reordered pairwise plots
        df,
        columns = col.nums,
        lower = list(continuous = wrap(custom_hex, bins = 25)),
        upper = list(continuous = wrap(custom_hex, bins = 25)),
        aes(color = group, alpha = 0.9)
    ) + theme_minimal() + theme(
        axis.text  = element_text(size = 20),  
        axis.title = element_text(size = 20),  
        strip.text = element_text(size = 20),
        panel.grid = element_blank()
    ))
}
\end{minted}
We note that \texttt{ggplot2} can't handle hexbin plots for unsampled \texttt{stat.matrix.H0} and \texttt{stat.matrix.H1}, so we only use 10000 rows from each. It should be enough for illustration purposes.

Now we can proceed to plots with standard normal null and various alternatives.

\subsection{Plots for \texorpdfstring{$H_0: N(0,1)$}{H0:N(0,1)} vs \texorpdfstring{$H_1: N(1,1)$}{H1:N(1,1)} case}

We start by checking the geometry in case of location shift. We execute the same code as in the end of subsection~\ref{section:check.hyperrectangle.bounds}, but with $N(1,1)$ sample generating function:

\begin{minted}[mathescape, linenos]{r}
sorted.samples.H1 <- get.sorted.samples.cpp(function(x) rnorm(x, 1, 1), n, m)
check.hyperrectangle.bounds(sorted.samples.H1, res.hyperrectangle.bounds.std.normal)
p.comp.H1 <- predict(sorted.samples.std.normal.p.comp, sorted.samples.H1)
check.hyperrectangle.bounds(p.comp.H1, res.hyperrectangle.bounds.std.normal.pc.1)
check.hyperrectangle.bounds(p.comp.H1, res.hyperrectangle.bounds.std.normal.pc.2)
\end{minted}

\begin{verbatim}
0.786175
0.669873
0.827209
\end{verbatim}

We see that variance-based bounds for PCs of order statistics are a bit better and equal-weighted bounds for PCs are worse than bounds for order statistics. As we will see a bit later, for normal null variance-based bounds are universally better than equal-weight bounds, so we will not do plots for equal-weights case.

\newpage

Ok, we start with illustration based on \texttt{hexbin} package. We only plot $H_1$ distribution and $H_0$ bounds, since \texttt{hexplom} function does not support plotting several datasets. On the good side, \texttt{hexplom} is able to handle the whole $H_1$ distribution and plot detailed grey low density regions for it.

Order statistics for $N(1,1)$ along with bounds for $N(0,1)$ for samples of size 10 look as follows:

\begin{minted}[mathescape, linenos, texcomments]{r}
plot.2d.projections.hexplom(              # defined in subsection \ref{section:2d.projections.hexplom}
    sorted.samples.H1,                    # defined in this subsection
    res.hyperrectangle.bounds.std.normal  # defined in subsection \ref{section:hyperrectangle_bounds_normal}
)
\end{minted}

\begin{figure}[H]
\centering
\includegraphics[width=15cm]{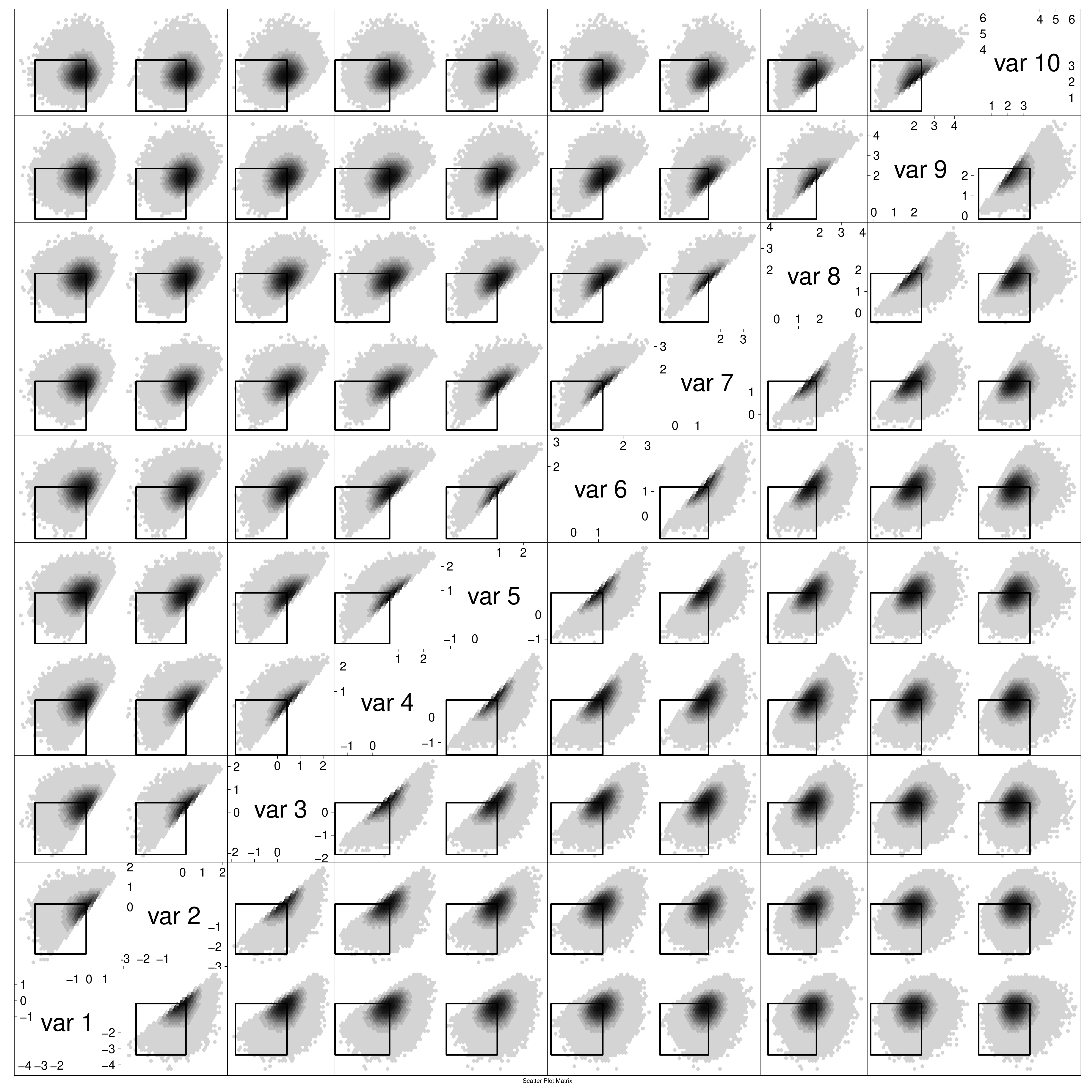}
\caption{Order statistics from $N(1,1)$, hyperectangle bounds for $N(0,1)$.}
\label{figure:N01vsN11hexplom}
\end{figure}

\newpage

\begin{minted}[mathescape, linenos, texcomments]{r}
ggplot.H0.H1.hexplot.pairs(               # defined in subsection \ref{section:H0vsH1_ggplot}
    sorted.samples.std.normal,            # defined in subsection \ref{section:standard_normal_order_stats}
    sorted.samples.H1,                    # defined in this subsection
    res.hyperrectangle.bounds.std.normal, # defined in subsection \ref{section:hyperrectangle_bounds_normal}
    col.nums=1:10
)
\end{minted}

\begin{figure}[H]
\centering
\includegraphics[width=15cm]{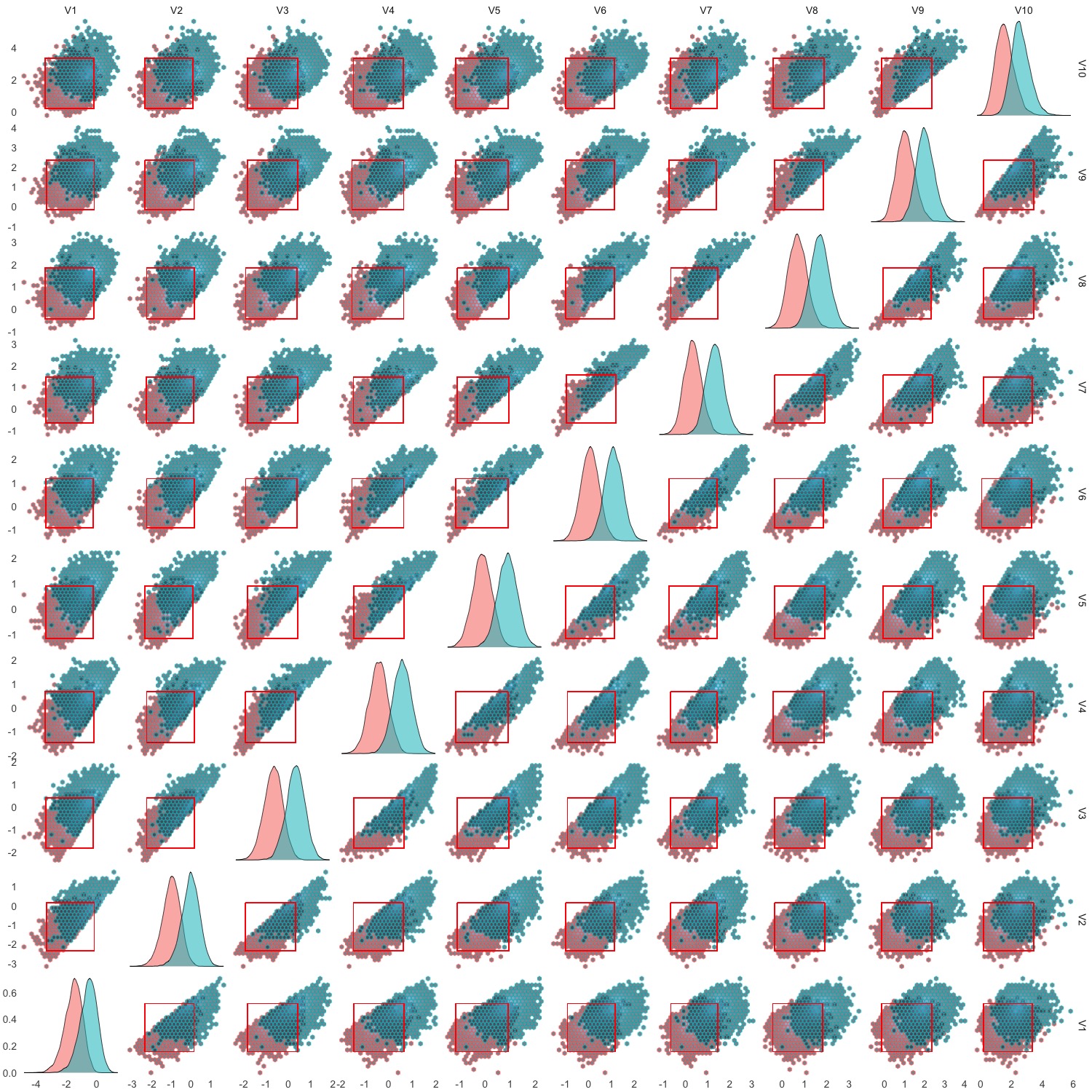}
\caption{Order statistics from $N(0,1)$ (red hexes) and $N(1,1)$ (blue hexes), hyperrectangle bounds for $N(0,1)$ (red rectangles).}
\label{figure:N01vsN11ggplot}
\end{figure}

We see that on figures~\ref{figure:N01vsN11hexplom} and~\ref{figure:N01vsN11ggplot} all order statistics from $N(1,1)$ are equally shifted against order statistics from $N(0,1)$.

\newpage

Now to the plots for principal components of standard normal order statistics. Plot based on \texttt{hexplom} is as follows: 

\begin{minted}[mathescape, linenos, texcomments]{r}
plot.2d.projections.hexplom(                   # defined in subsection \ref{section:2d.projections.hexplom}
    p.comp.H1,                                 # defined in this subsection
    res.hyperrectangle.bounds.std.normal.pc.2  # defined in subsection \ref{section:hyperrectangle.bounds.std.normal.pc.2}
)
\end{minted}

\begin{figure}[H]
\centering
\includegraphics[width=15cm]{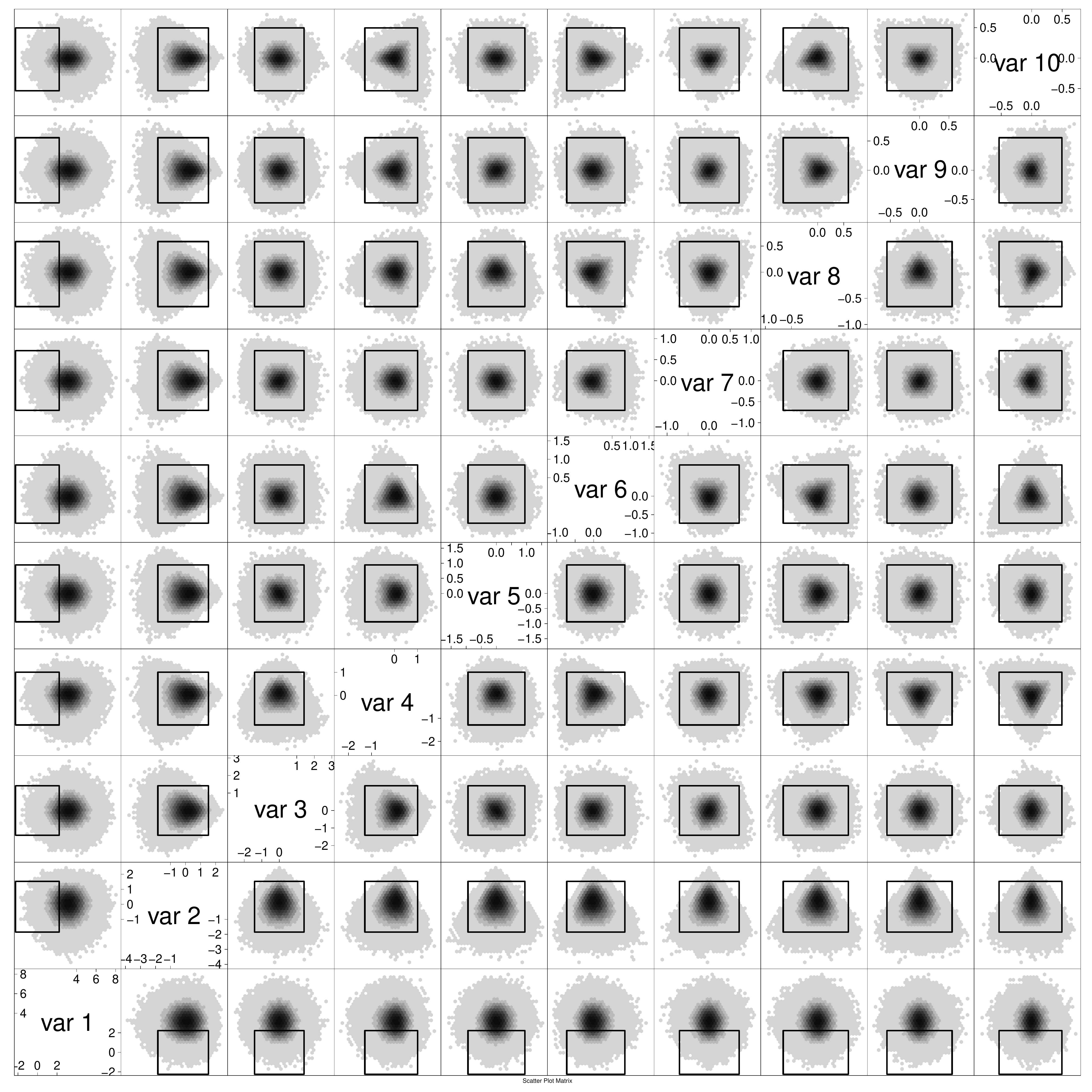}
\caption{Principal components of order statistics from $N(1,1)$, hyperrectangle bounds for principal components of order statistics from N(0,1).}
\label{figure:NN01vsNN11pc_hexplom}
\end{figure}

\newpage

\begin{minted}[mathescape, linenos, texcomments]{r}
ggplot.H0.H1.hexplot.pairs(                     # defined in subsection \ref{section:H0vsH1_ggplot}
    sorted.samples.std.normal.p.comp$scores,    # defined in subsection \ref{section:principal_components_normal}
    p.comp.H1,                                  # defined in this subsection
    res.hyperrectangle.bounds.std.normal.pc.2,  # defined in subsection \ref{section:hyperrectangle.bounds.std.normal.pc.2}
    col.nums = 1:10
)
\end{minted}

\begin{figure}[H]
\centering
\includegraphics[width=15cm]{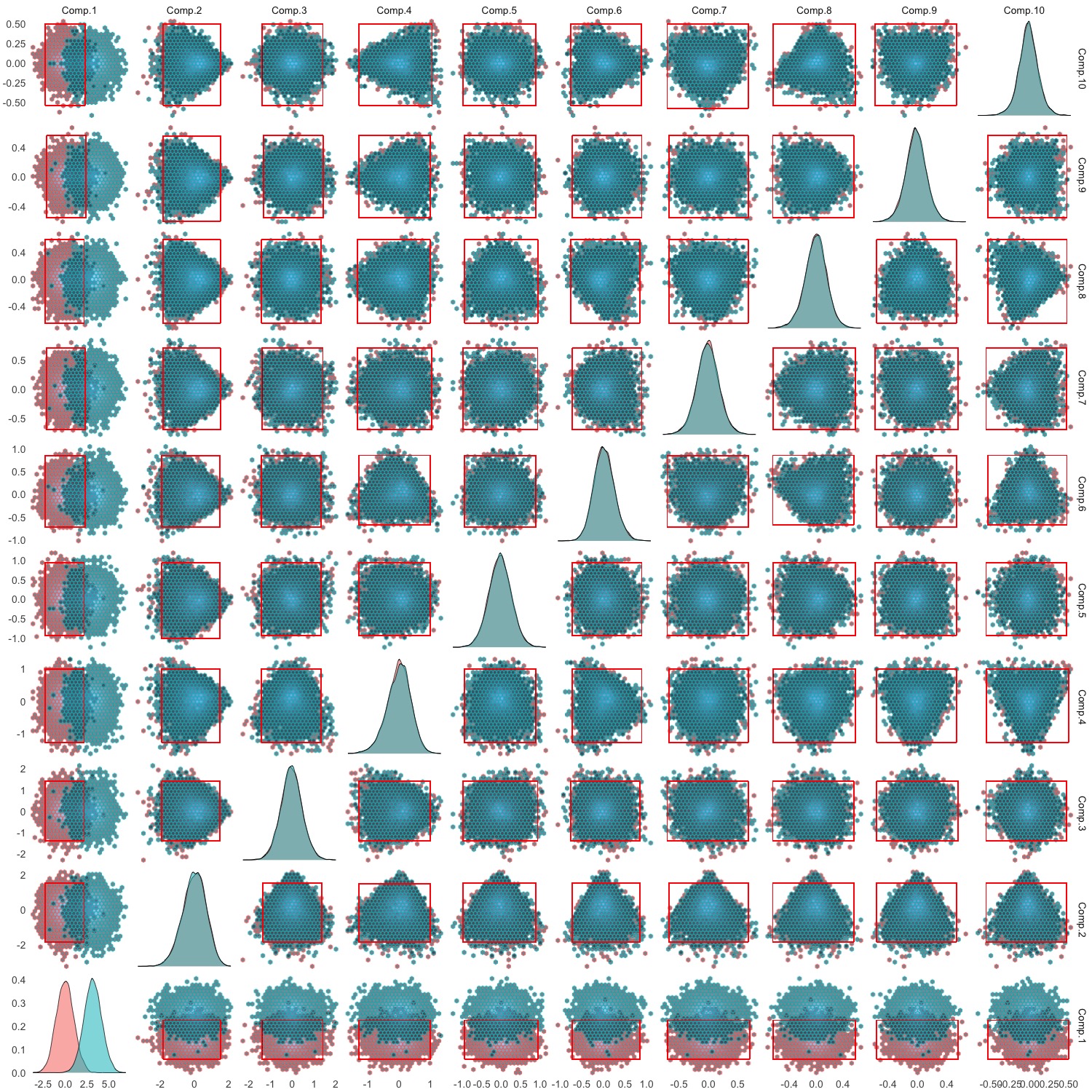}
\caption{Principal components of order statistics from $N(0,1)$ (red hexes) and $N(1,1)$ (blue hexes), hyperrectangle bounds for $N(0,1)$ (red rectangles).}
\label{figure:N01vsN11pc_gg}
\end{figure}

On figures~\ref{figure:NN01vsNN11pc_hexplom} and~\ref{figure:N01vsN11pc_gg} we see that only first principal component for $N(1,1)$ that is proportional to sample mean is shifted against the first principal component for $N(0,1)$. All other principal components stay the same, which is expected given the principal component structure we examined in subsection~\ref{section:principal_components_normal}, but is delightful nonetheless.

\newpage

\subsection{Plots for \texorpdfstring{$H_0: N(0,1)$}{H0:N(0,1)} vs \texorpdfstring{$H_1: N(0,1.5^2)$}{H1:N(0,1.5**2)} case}
\label{section:N01vsN015}

We examined the power and the geometry for location shit, now we will look at scale shift. The powers for $N(0,1)$ vs $N(0,1.5^2)$ case are as follows:

\begin{minted}[mathescape, linenos]{r}
sorted.samples.H1 <- get.sorted.samples.cpp(function(x) rnorm(x, 0, 1.5), n, m)
check.hyperrectangle.bounds(sorted.samples.H1, res.hyperrectangle.bounds.std.normal)
p.comp.H1 <- predict(sorted.samples.std.normal.p.comp, sorted.samples.H1)
check.hyperrectangle.bounds(p.comp.H1, res.hyperrectangle.bounds.std.normal.pc.1)
check.hyperrectangle.bounds(p.comp.H1, res.hyperrectangle.bounds.std.normal.pc.2)
\end{minted}

\begin{verbatim}
0.449476
0.446975
0.464596
\end{verbatim}

\noindent
We see that the powers are not that different.

Now to the plots:

\begin{minted}[mathescape, linenos, texcomments]{r}
plot.2d.projections.hexplom(              # defined in subsection \ref{section:2d.projections.hexplom}
    sorted.samples.H1,                    # defined in this subsection
    res.hyperrectangle.bounds.std.normal  # defined in subsection \ref{section:hyperrectangle_bounds_normal}
)
\end{minted}

\begin{figure}[H]
\centering
\includegraphics[width=15cm]{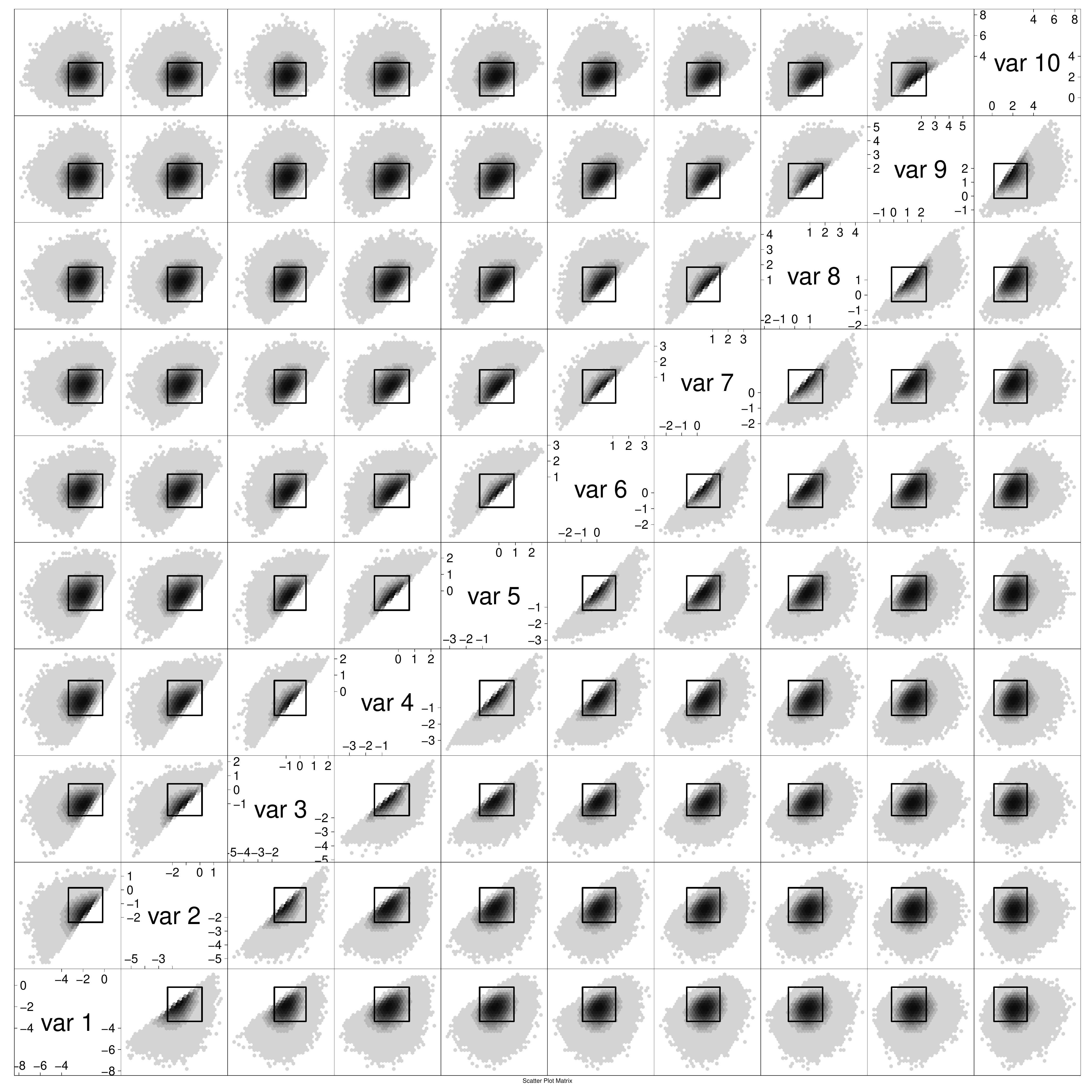}
\caption{Order statistics from $N(0,1.5^2)$, hyperrectangle bounds for $N(0,1)$.}
\label{figure:N01vsN015_hexplom}
\end{figure}

\newpage

\begin{minted}[mathescape, linenos, texcomments]{r}
ggplot.H0.H1.hexplot.pairs(                # defined in subsection \ref{section:H0vsH1_ggplot}
    sorted.samples.std.normal,             # defined in subsection \ref{section:standard_normal_order_stats}
    sorted.samples.H1,                     # defined in this subsection 
    res.hyperrectangle.bounds.std.normal,  # defined in subsection \ref{section:hyperrectangle_bounds_normal}
    col.nums=1:10
)
\end{minted}

\begin{figure}[H]
\centering
\includegraphics[width=15cm]{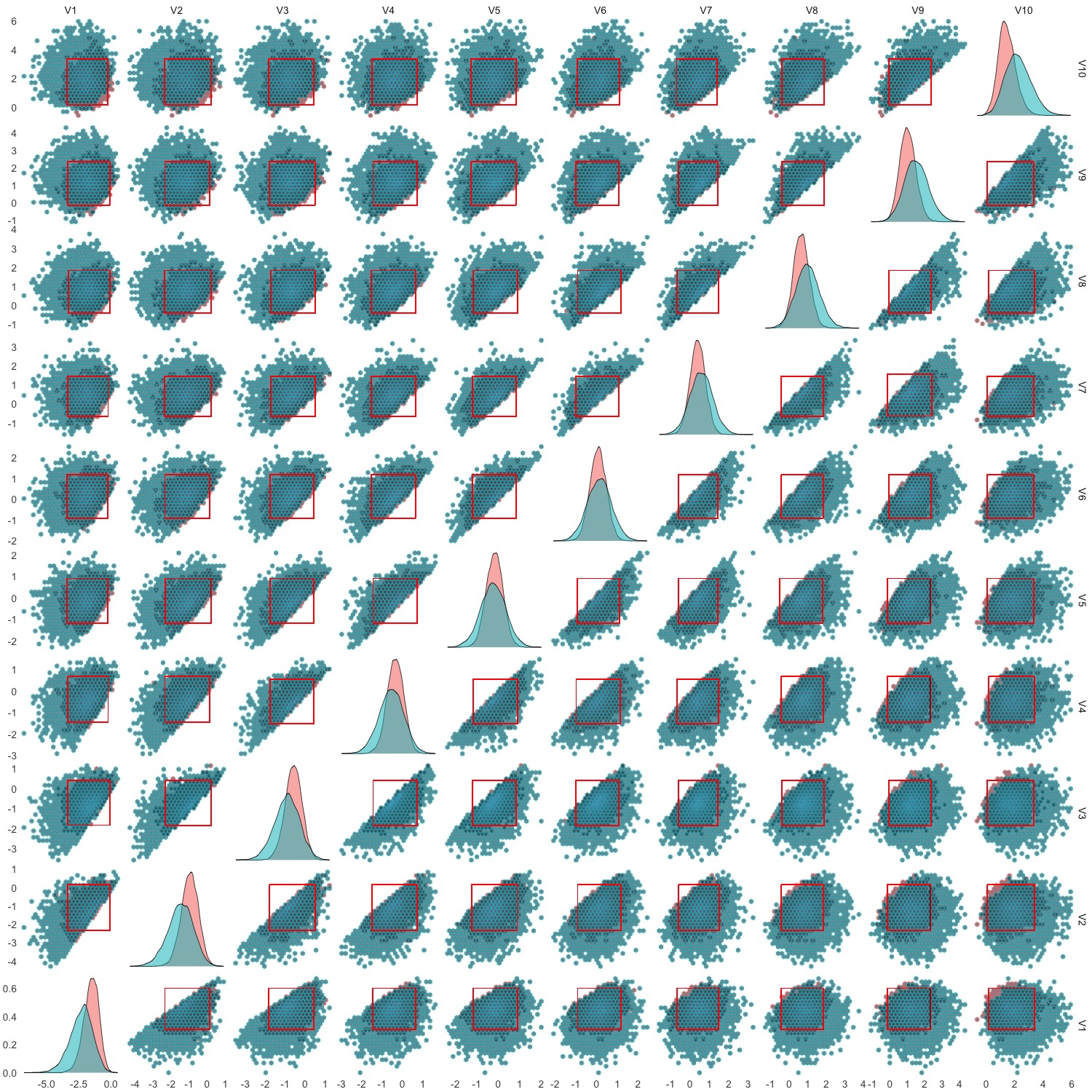}
\caption{Order statistics from $N(0,1)$ (red hexes) and $N(0,1.5^2)$ (blue hexes), hyperrectangle bounds for $N(0,1)$ (red rectangles).}
\label{figure:N01vsN015_gg}
\end{figure}

On figures~\ref{figure:N01vsN015_hexplom} and~\ref{figure:N01vsN015_gg} we see that all marginal distributions of order statistics from $N(0,1.5^2)$ have larger scale than corresponding marginal distributions of order statistics from $N(0,1)$. First order statistic of $N(0,1.5^2)$ is significantly shifted to the left relative to first order statistic of $N(0,1)$, for second to fifth order statistics the shift shrinks with each subsequent index, for sixth to tenth the shift grows and goes to the right.

\newpage

Now to the plots of principal components of order statistics. Plot based on \texttt{hexbin} package:

\begin{minted}[mathescape, linenos, texcomments]{r}
plot.2d.projections.hexplom(                   # defined in subsection \ref{section:2d.projections.hexplom}
    p.comp.H1,                                 # defined in this subsection
    res.hyperrectangle.bounds.std.normal.pc.2  # defined in subsection \ref{section:hyperrectangle.bounds.std.normal.pc.2}
)
\end{minted}

\begin{figure}[H]
\centering
\includegraphics[width=15cm]{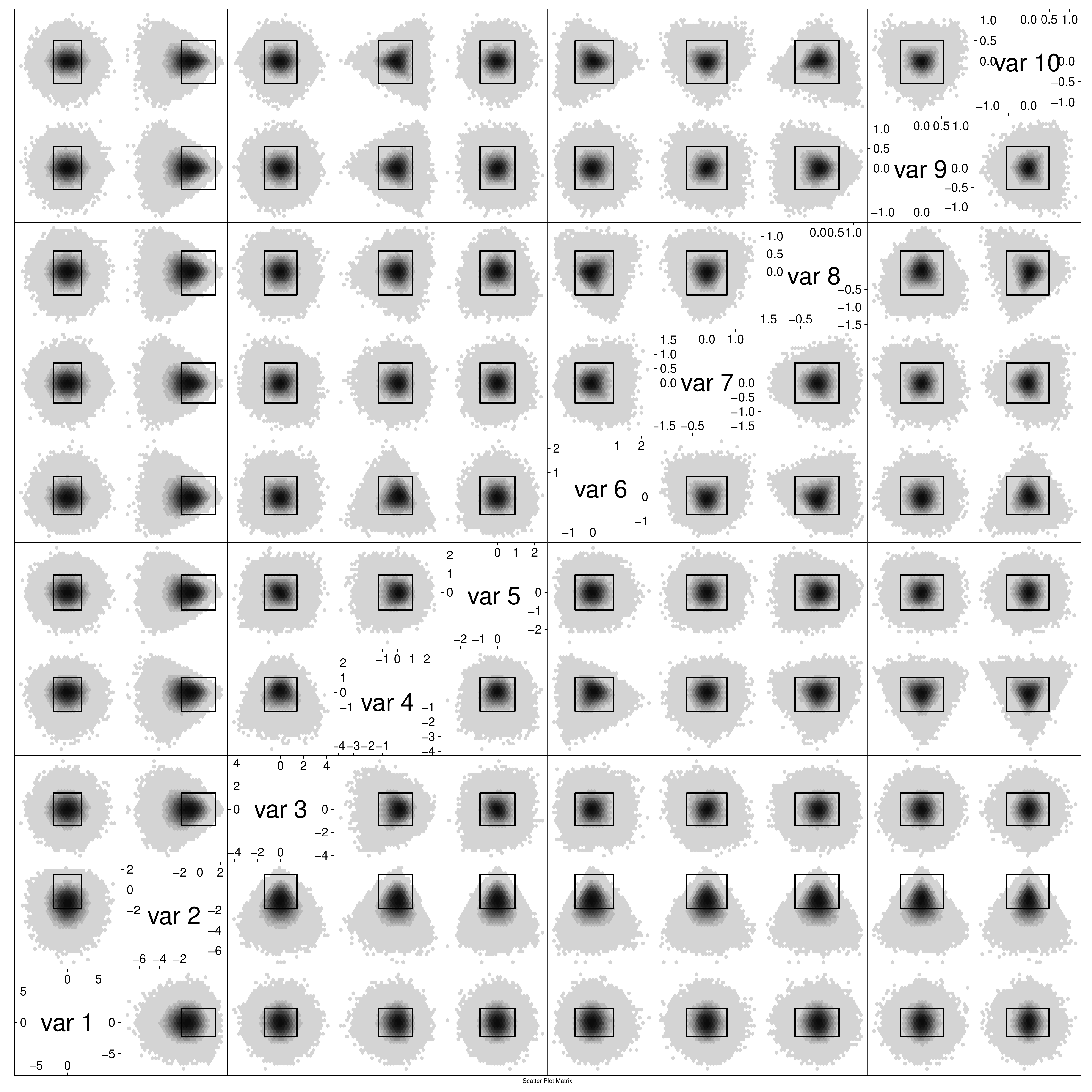}
\caption{Principal components of order statistics from $N(0,1.5^2)$, hyperectangle bounds for principal components of order statistics from $N(0,1)$.}
\label{figure:N01vsN015pc_hexbin}
\end{figure}

\newpage

Plot based on \texttt{ggplot2} package:

\begin{minted}[mathescape, linenos, texcomments]{r}
ggplot.H0.H1.hexplot.pairs(                     # defined in subsection \ref{section:H0vsH1_ggplot}
    sorted.samples.std.normal.p.comp$scores,    # defined in subsection \ref{section:principal_components_normal}
    p.comp.H1,                                  # defined in this subsection
    res.hyperrectangle.bounds.std.normal.pc.2,  # defined in subsection \ref{section:hyperrectangle.bounds.std.normal.pc.2}
    col.nums = 1:10
)
\end{minted}

\begin{figure}[H]
\centering
\includegraphics[width=15cm]{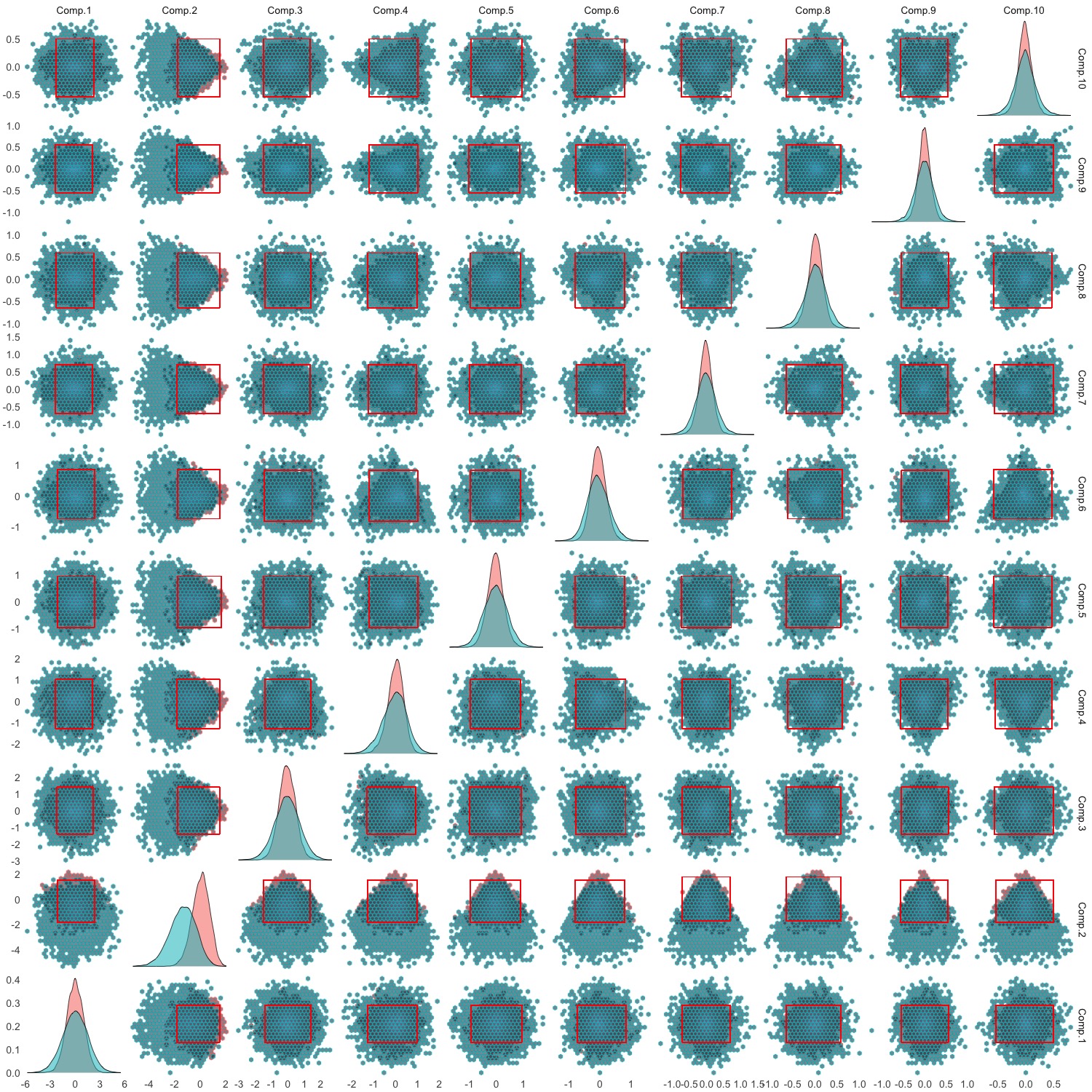}
\caption{Principal components of order statistics from $N(0,1)$ (red hexes) and from $N(0,1.5^2)$ (blue hexes), hyperrectangle bounds for principal components of order statistics from $N(0,1)$ (red rectangles).}
\label{figure:N01vsN015pc_gg}
\end{figure}

On figures~\ref{figure:N01vsN015pc_hexbin} and~\ref{figure:N01vsN015pc_gg} we see that all marginal distributions for principal components of order statistics from $N(0,1.5^2)$ have larger scale than corresponding marginal distributions for components of order statistics from $N(0,1)$.  
We also see that for all principal components except the second one, marginal distributions for $N(0,1)$ and $N(0,1.5^2)$ do have the same centers. For the second principal component marginal distribution for $N(0,1.5^2)$ is significantly shifted to the left relative to marginal distribution for $N(0,1)$. We conclude that the second principal component plays a decisive role in rejection decision when alternative sample distribution differs from null distribution in scale. 

\subsection{Plots for \texorpdfstring{$H_0: N(0,1)$}{H0:N(0,1)} vs \texorpdfstring{$H_1: N(0,0.6^2)$}{H1:N(0,0.6**2)} case}

In subsection~\ref{section:N01vsN015} we examined the case where scale of alternative distribution is larger than scale of null distribution. In this section we look at geometry for situation when alternative's scale is lower than null's. The powers for $N(0,0.6^2)$ alternative are as follows:

\begin{minted}[mathescape, linenos]{r}
sorted.samples.H1 <- get.sorted.samples.cpp(function(x) rnorm(x, 0, 0.6), n, m)
check.hyperrectangle.bounds(sorted.samples.H1, res.hyperrectangle.bounds.std.normal)
p.comp.H1 <- predict(sorted.samples.std.normal.p.comp, sorted.samples.H1)
check.hyperrectangle.bounds(p.comp.H1, res.hyperrectangle.bounds.std.normal.pc.1)
check.hyperrectangle.bounds(p.comp.H1, res.hyperrectangle.bounds.std.normal.pc.2)
\end{minted}

\begin{verbatim}
0.019184
0.100494
0.167011
\end{verbatim}

\noindent
We see that for this case hyperrectangle bounds for order stats do have near zero power. Let us find out why.

\begin{minted}[mathescape, linenos]{r}
plot.2d.projections.hexplom(sorted.samples.H1, res.hyperrectangle.bounds.std.normal)
\end{minted}

\begin{figure}[H]
\centering
\includegraphics[width=15cm]{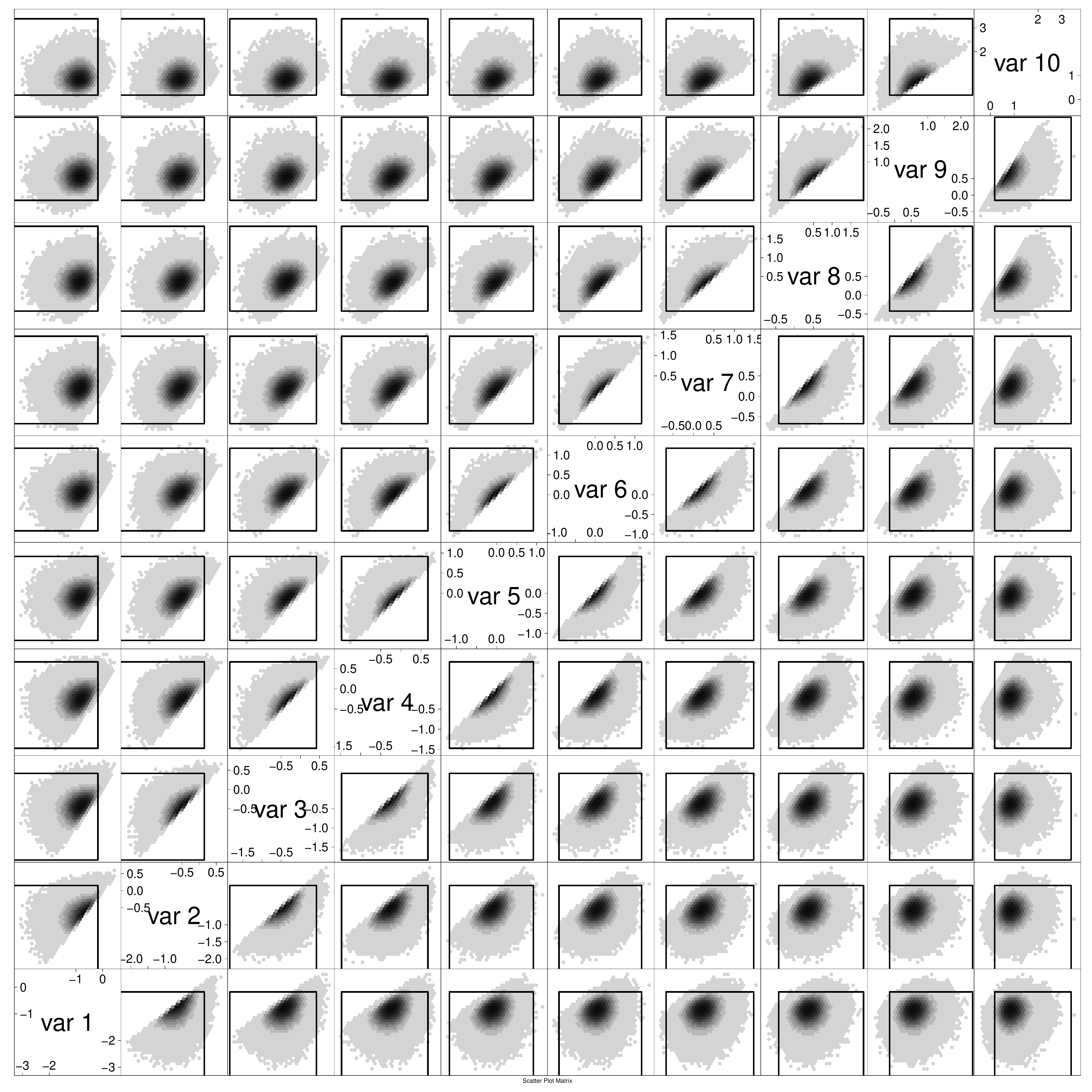}
\caption{Order statistics for $N(0,0.6^2)$, hyperrectangle bounds for $N(0,1)$.}
\label{figure:N01vsN006_hexbin}
\end{figure}

\newpage

\begin{minted}[mathescape, linenos]{r}
ggplot.H0.H1.hexplot.pairs(
    sorted.samples.std.normal, sorted.samples.H1,
    res.hyperrectangle.bounds.std.normal, 
    col.nums=1:10
)
\end{minted}

\begin{figure}[H]
\centering
\includegraphics[width=15cm]{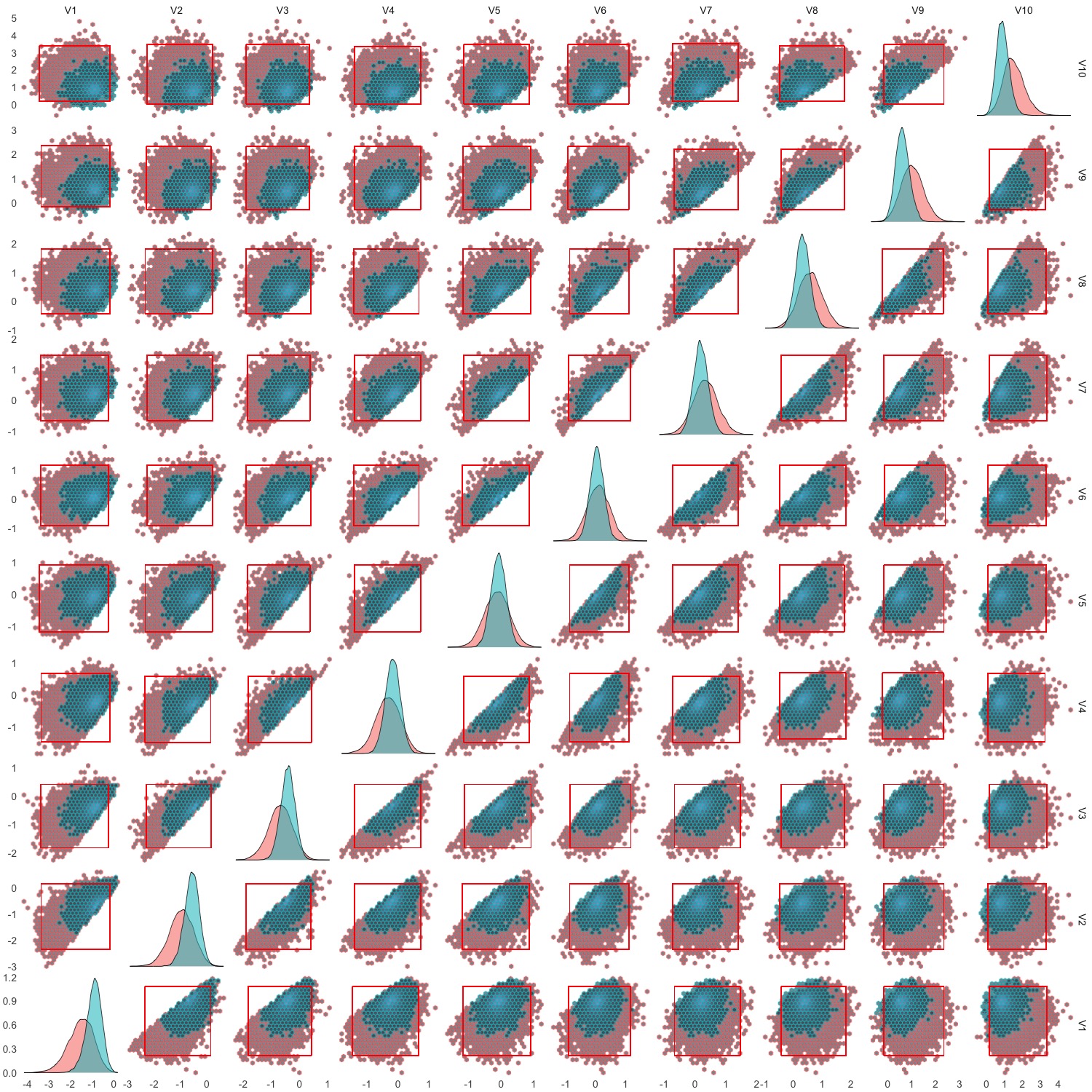}
\caption{Order statistics from $N(0, 1)$ (red hexes) and $N(0,0.6^2)$ (blue hexes), hyperectangle bounds for N(0,1) (red rectangles)}
\end{figure}

We see that almost all blue hexes are inside red rectangles! That explains near zero power of hyperrectangle bounds for order statistics.

\newpage

Now to the plots for principal components of order statistics:

\begin{minted}[mathescape, linenos]{r}
plot.2d.projections.hexplom(p.comp.H1, res.hyperrectangle.bounds.std.normal.pc.2)
\end{minted}

\begin{figure}[H]
\centering
\includegraphics[width=15cm]{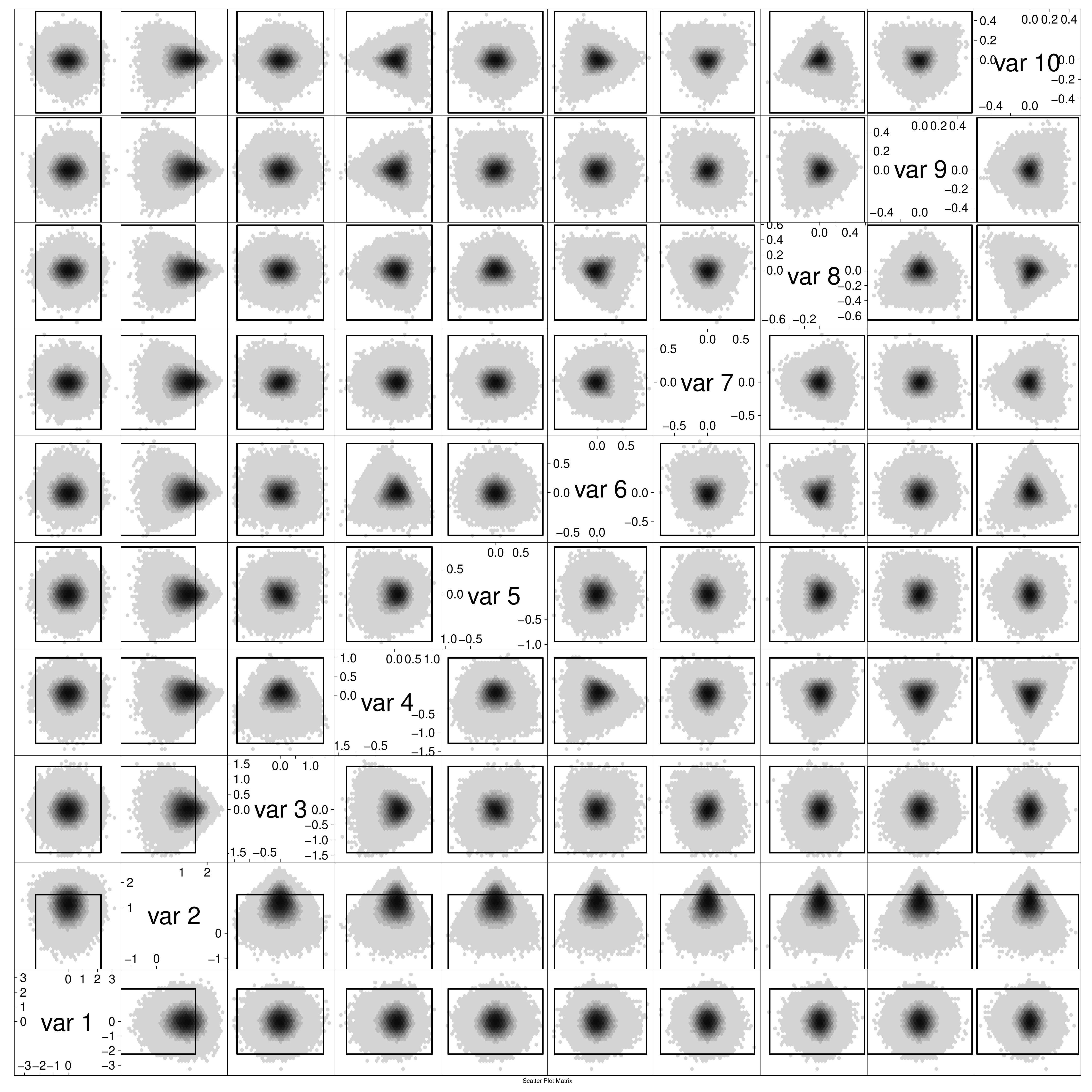}
\caption{Principal components of order statistics for $N(0,0.6^2)$, PC-based hyperrectangle bounds for $N(0,1)$.}
\end{figure}

\newpage

\begin{minted}[mathescape, linenos]{r}
ggplot.H0.H1.hexplot.pairs(
    sorted.samples.std.normal.p.comp$scores, p.comp.H1, 
    res.hyperrectangle.bounds.std.normal.pc.2, col.nums = 1:10)
\end{minted}

\begin{figure}[H]
\centering
\includegraphics[width=15cm]{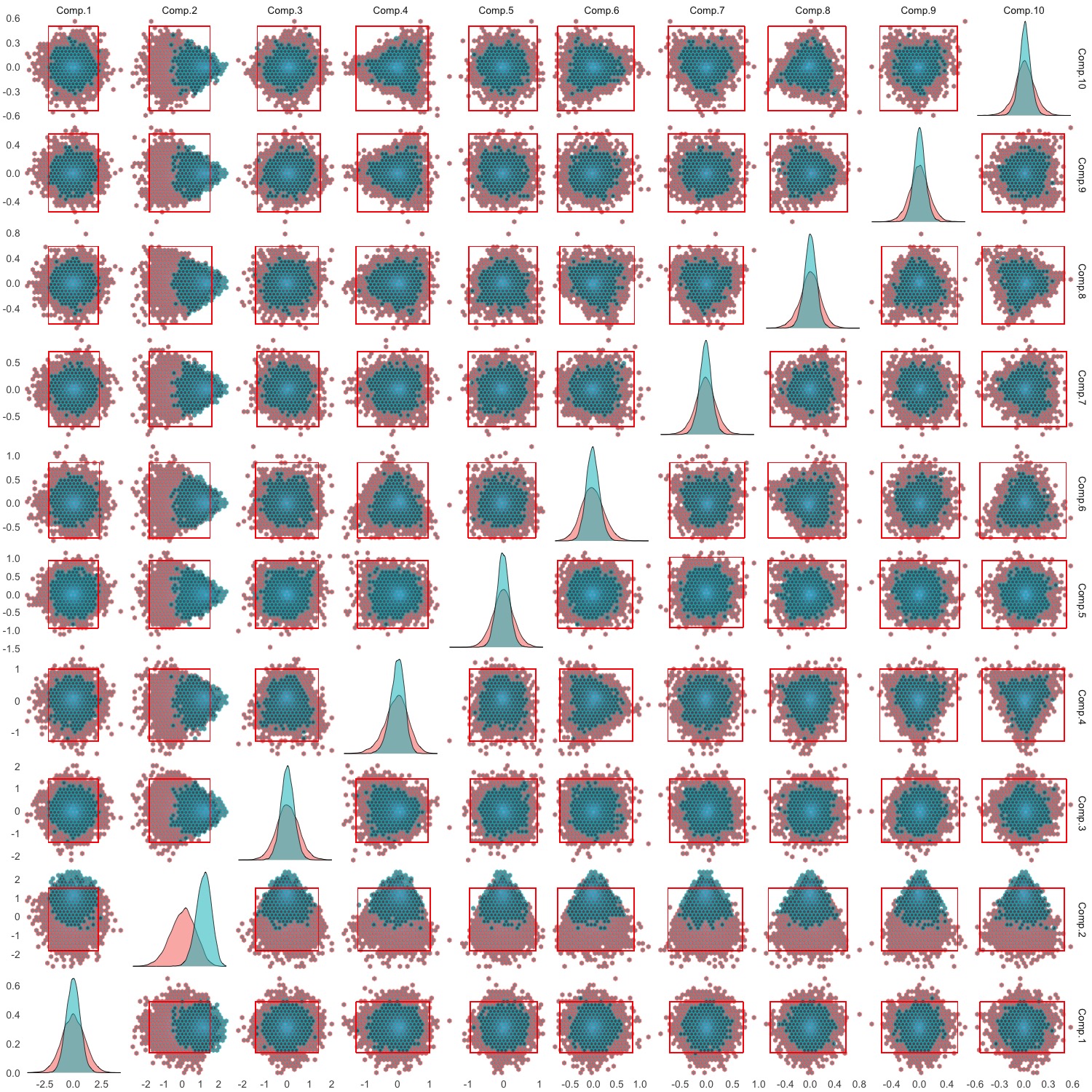}
\caption{Principal components of order statistics from $N(0,1)$ (red hexes) and from $N(0,0.6^2)$ (blue hexes), hyperrectangle bounds for principal components of order statistics from $N(0,1)$ (red rectangles).}
\end{figure}

We see that only on pairwise plots associated with second principal component (9th row and 2nd column) blue hexes go outside of red confidence rectangles. For all other pairwise plots blue hexes stay inside bounds. Marginal density of second principal component for $N(0,0.6^2)$ is shifted to the right relative to marginal density of PCs for $N(0,1)$ (remember that on figure~\ref{figure:N01vsN015pc_gg} for $N(0,1.5^2)$ marginal density was shifted to the left). We confirm again that second principal component is the most important one when we compare null with the alternative that is different in scale.

\newpage
\subsection{Plots for \texorpdfstring{$H_0: N(0,1)$}{H0:N(0,1)} vs \texorpdfstring{$H_1: \mathrm{Gamma}(1,1) - 1$}{H1:Gamma(1,1)-1} case}

Let $\mathrm{Gamma}(a, b)$ denote gamma distribution with shape $a$ and scale $b$. 
In this subsection we will compare standard normal null and gamma alternative.
We note that
\begin{equation} 
\label{eq:gamma_mean_var}
\text{if } \xi \sim \mathrm{Gamma}(a, 1 / \sqrt{a}) \text{ and } \eta = \xi  - a / \sqrt{a}, \text{ then } \mathbf{E} \eta  = 0 \text{ and } \mathbf{D} \eta  = 1;
\end{equation}
where $\mathbf{E}[.]$ and $\mathbf{D}[.]$ stand for random variable mean and variance. We think that $\eta$ should work as a good example alternative since it should be harder for a test tailored for standard normal to detect the difference when alternative's mean and variance are the same as for standard normal.

For this subsection we choose $a = 1$ (other shapes will be checked later in the text) and generate 10-sized samples from $\mathrm{Gamma}(1,1) - 1$:

\begin{minted}[mathescape, linenos]{r}
shape <- 1
sorted.samples.H1 <- get.sorted.samples.cpp(
    function(x) rgamma(x, shape = shape, scale = 1 / sqrt(shape)) - shape / sqrt(shape), 
    n, m)
check.hyperrectangle.bounds(sorted.samples.H1, res.hyperrectangle.bounds.std.normal)
p.comp.H1 <- predict(sorted.samples.std.normal.p.comp, sorted.samples.H1)
check.hyperrectangle.bounds(p.comp.H1, res.hyperrectangle.bounds.std.normal.pc.1)
check.hyperrectangle.bounds(p.comp.H1, res.hyperrectangle.bounds.std.normal.pc.2)
\end{minted}

\begin{verbatim}
0.20199
0.250448
0.273291
\end{verbatim}
\noindent
We see that principal component based tests have better power here.

\newpage
Now to the plots. For order stats we have:

\begin{minted}[mathescape, linenos]{r}
ggplot.H0.H1.hexplot.pairs(
    sorted.samples.std.normal, sorted.samples.H1,
    res.hyperrectangle.bounds.std.normal, 
    col.nums=1:10
)
\end{minted}

\begin{figure}[H]
\centering
\includegraphics[width=15cm]{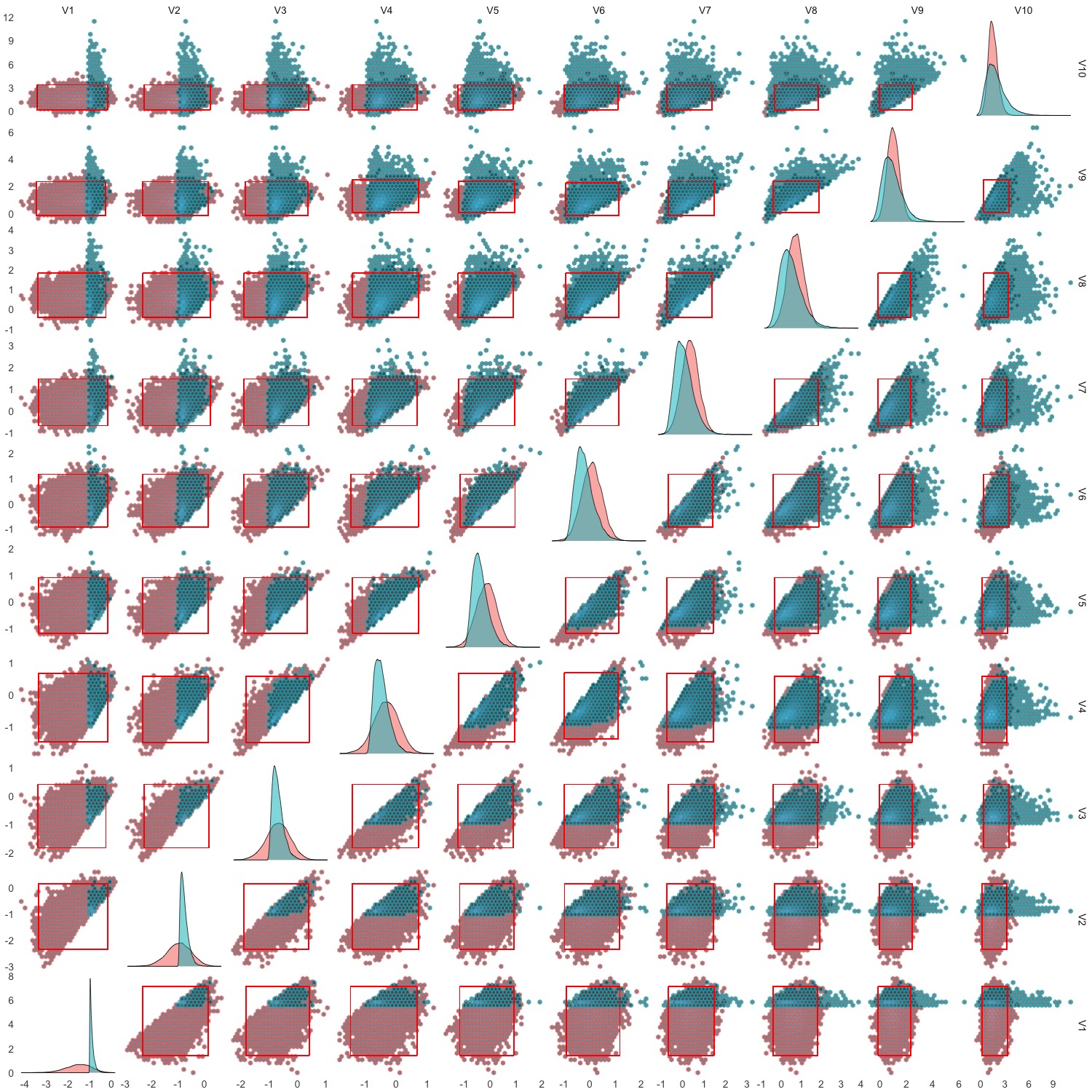}
\caption{Order statistics; $N(0,1)$ (red hexes), $\mathrm{Gamma}(1,1)-1$ (blue hexes); hyperrectangle bounds for $N(0,1)$ (red rectangles).}
\end{figure}

Here we see that blue hexes lie outside of confidence rectangles mostly for pairwise plots associated with 6th to 10th order statistics.

\newpage

And for principal components of order stats we have:

\begin{minted}[mathescape, linenos]{r}
ggplot.H0.H1.hexplot.pairs(
    sorted.samples.std.normal.p.comp$scores, p.comp.H1, 
    res.hyperrectangle.bounds.std.normal.pc.2, col.nums = 1:10)
\end{minted}

\begin{figure}[H]
\centering
\includegraphics[width=15cm]{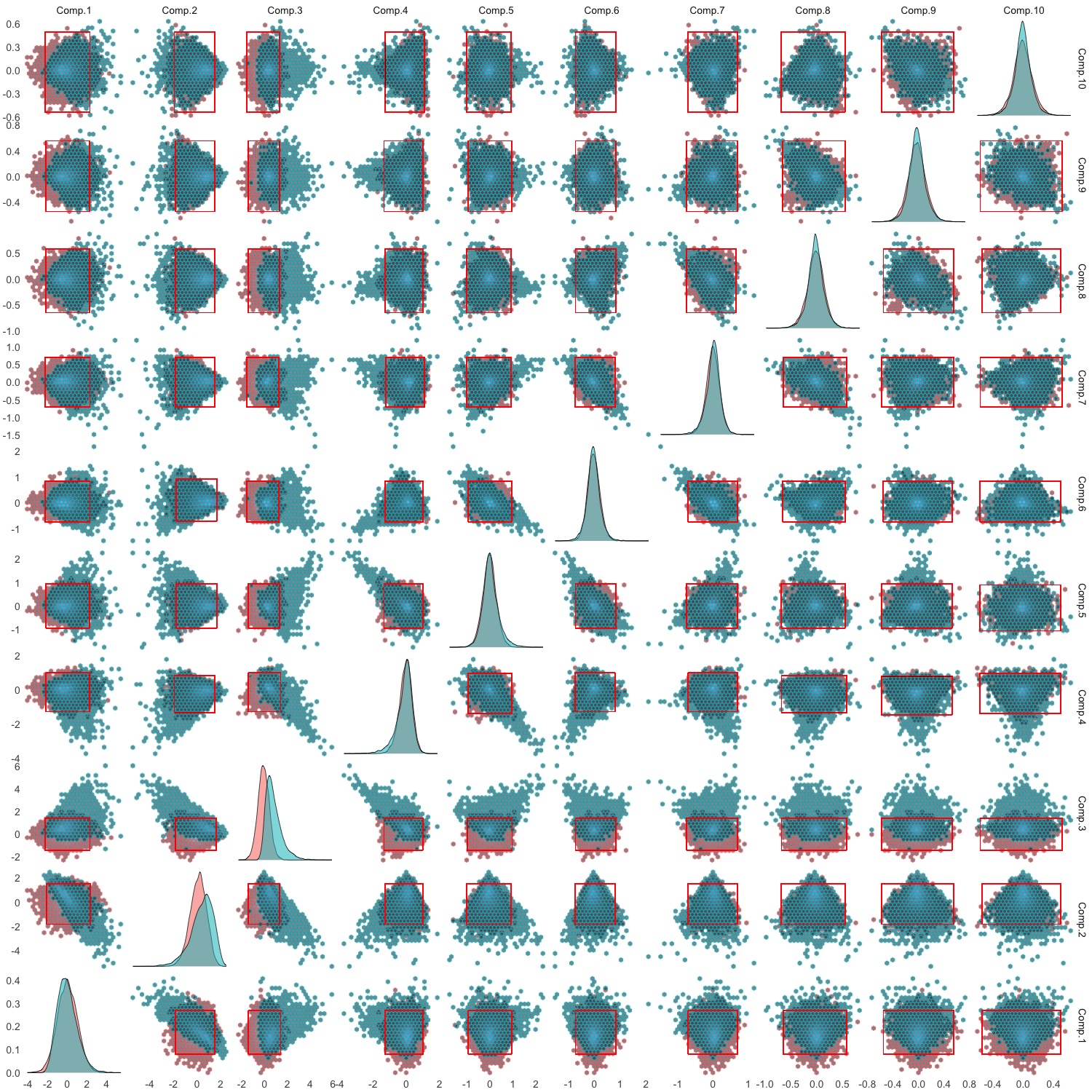}
\caption{Principal components of order statistics; $N(0,1)$ (red hexes), $\mathrm{Gamma}(1,1)-1$ (blue hexes); hyperrectangle bounds for $N(0,1)$ (red rectangles).}
\end{figure}

And here it looks like most of blue hexes that lie outside the bounds are associated with first five principal components. 

\newpage
\subsection{Plots for \texorpdfstring{$H_0: N(0,1)$}{H0:N(0,1)} vs \texorpdfstring{$H_1: \mathrm{Cauchy}(0,0.2)$}{H1:Cauchy(0,0.2)} case}

In this subsection we examine the case of heavy tailed Cauchy distribution:

\begin{minted}[mathescape, linenos]{r}
sorted.samples.H1 <- get.sorted.samples.cpp(function(x) rcauchy(x, 0, 0.2), n, m)
check.hyperrectangle.bounds(sorted.samples.H1, res.hyperrectangle.bounds.std.normal)
p.comp.H1 <- predict(sorted.samples.std.normal.p.comp, sorted.samples.H1)
check.hyperrectangle.bounds(p.comp.H1, res.hyperrectangle.bounds.std.normal.pc.1)
check.hyperrectangle.bounds(p.comp.H1, res.hyperrectangle.bounds.std.normal.pc.2)
\end{minted}

\begin{verbatim}
0.393696
0.621951
0.66153
\end{verbatim}

\noindent
We see that the power is significantly better for principal component based bounds.

\begin{minted}[mathescape, linenos]{r}
ggplot.H0.H1.hexplot.pairs(sorted.samples.std.normal, sorted.samples.H1,
                           res.hyperrectangle.bounds.std.normal, col.nums=1:10)
\end{minted}

\begin{figure}[H]
\centering
\includegraphics[width=15cm]{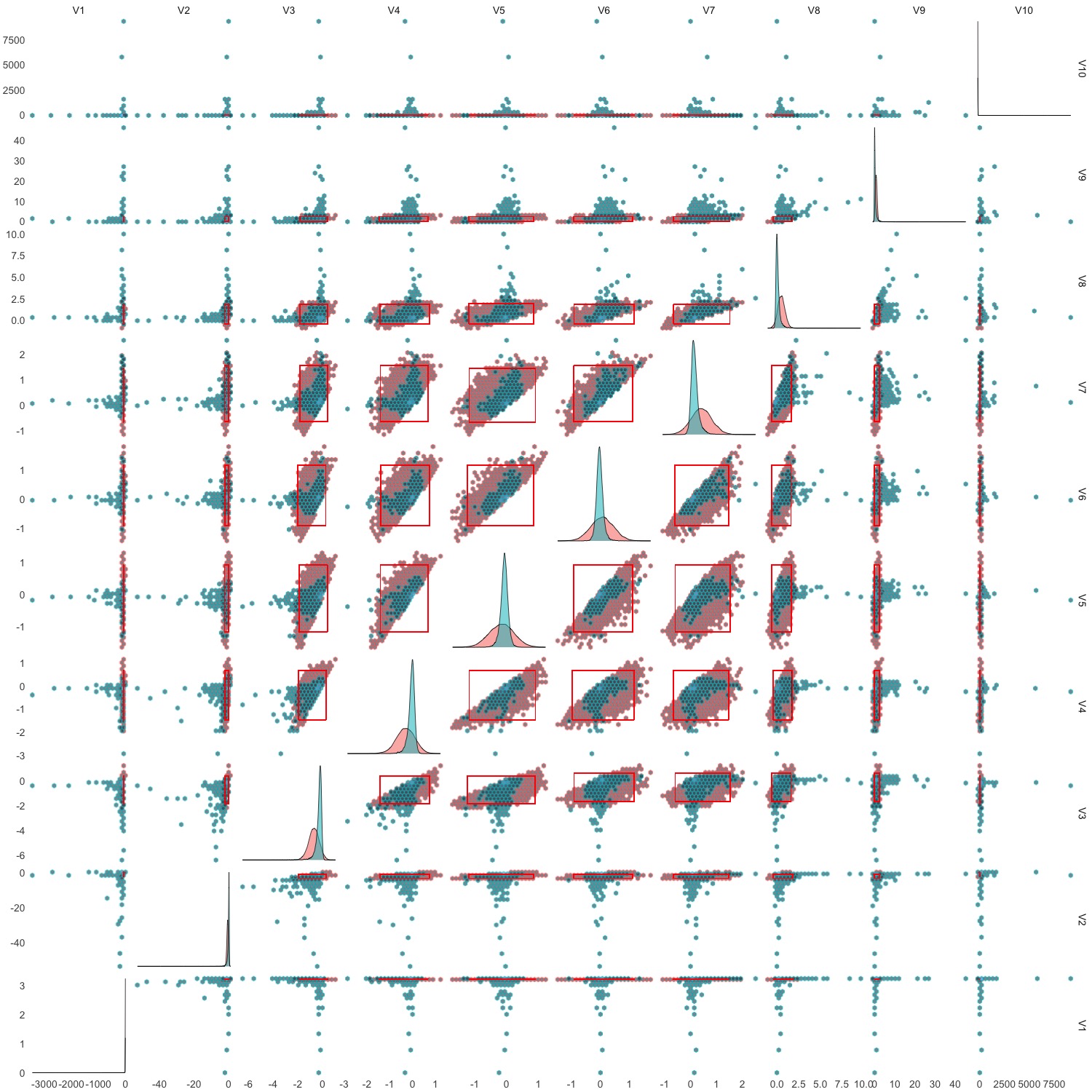}
\caption{Order statistics; $N(0,1)$ (red hexes), $\mathrm{Cauchy}(0,0.2)$ (blue hexes); hyperrectangle bounds for $N(0,1)$ (red rectangles).}
\end{figure}

We see that on plots associated with 4th to 7th order stats blue hexes mostly lie inside the bounds. For 1st, 2nd, 3rd, 8th, 9th and 10th order stats that are on the tails we see the most rejections.

\newpage

\begin{minted}[mathescape, linenos]{r}
ggplot.H0.H1.hexplot.pairs(
    sorted.samples.std.normal.p.comp$scores,  
    p.comp.H1, 
    res.hyperrectangle.bounds.std.normal.pc.2,    
    col.nums = 1:10
)
\end{minted}

\begin{figure}[H]
\centering
\includegraphics[width=15cm]{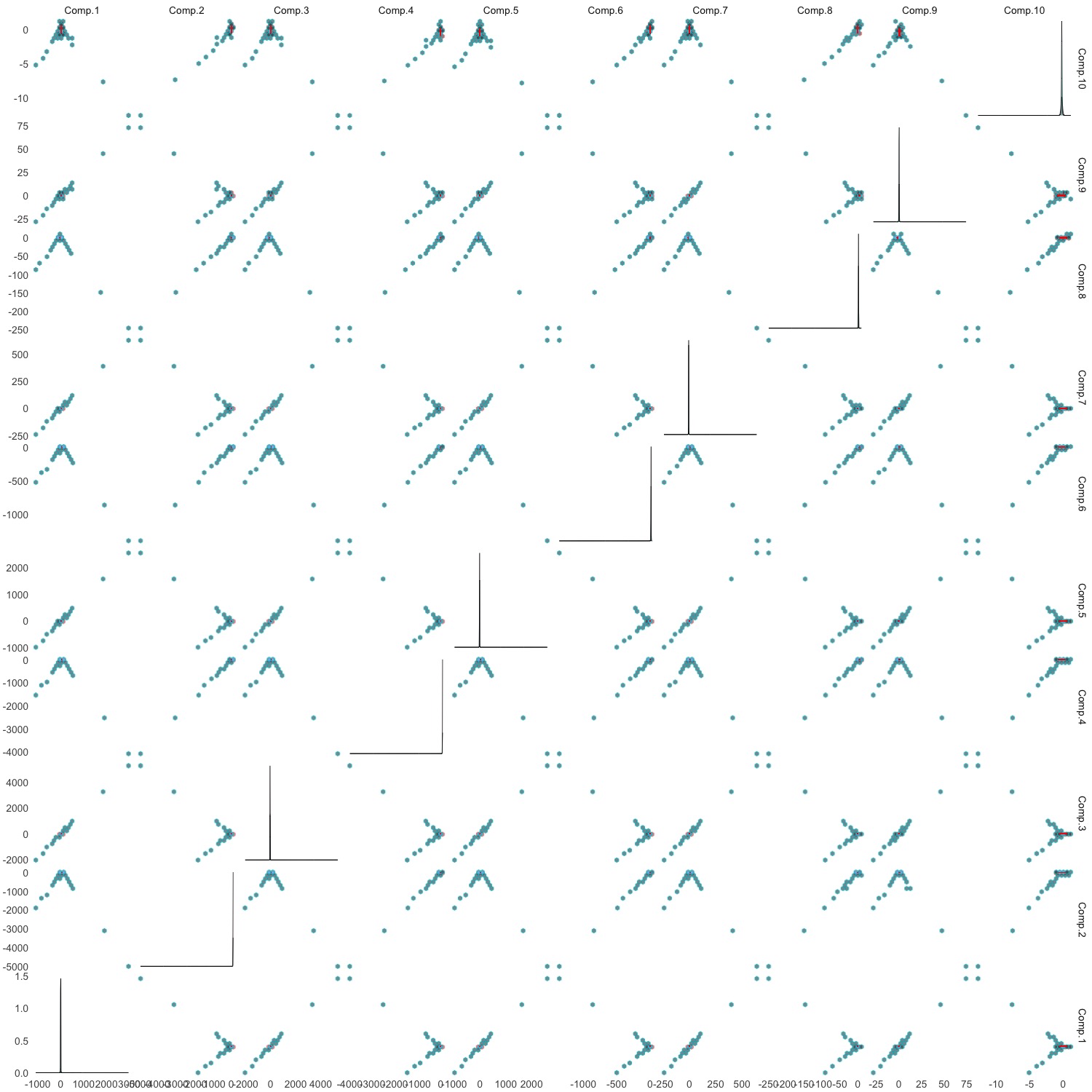}
\caption{Principal components of order statistics; $N(0,1)$ (red hexes), $\mathrm{Cauchy}(0,0.2)$ (blue hexes); hyperrectangle bounds for $N(0,1)$ (red rectangles).}
\end{figure}

On this plot we see that there are a lot of wild blue outliers that lie outside the bounds for each principal component. 

\newpage
\subsection{Plots for \texorpdfstring{$H_0: N(0,1)$}{H0:N(0,1)} vs \texorpdfstring{$H_1: \mathrm{t}(4)$}{H1:t(4)} case}

The last example will be about Student distribution, another heavy tailed in our list:

\begin{minted}[mathescape, linenos]{r}
sorted.samples.H1 <- get.sorted.samples.cpp(function(x) rt(x, df=4), n, m)
check.hyperrectangle.bounds(sorted.samples.H1, res.hyperrectangle.bounds.std.normal)
p.comp.H1 <- predict(sorted.samples.std.normal.p.comp, sorted.samples.H1)
check.hyperrectangle.bounds(p.comp.H1, res.hyperrectangle.bounds.std.normal.pc.1)
check.hyperrectangle.bounds(p.comp.H1, res.hyperrectangle.bounds.std.normal.pc.2)
\end{minted}

\begin{verbatim}
0.326463
0.308857
0.317131
\end{verbatim}

\begin{minted}[mathescape, linenos]{r}
ggplot.H0.H1.hexplot.pairs(sorted.samples.std.normal, sorted.samples.H1,
                           res.hyperrectangle.bounds.std.normal, col.nums=1:10)
\end{minted}

\begin{figure}[H]
\centering
\includegraphics[width=15cm]{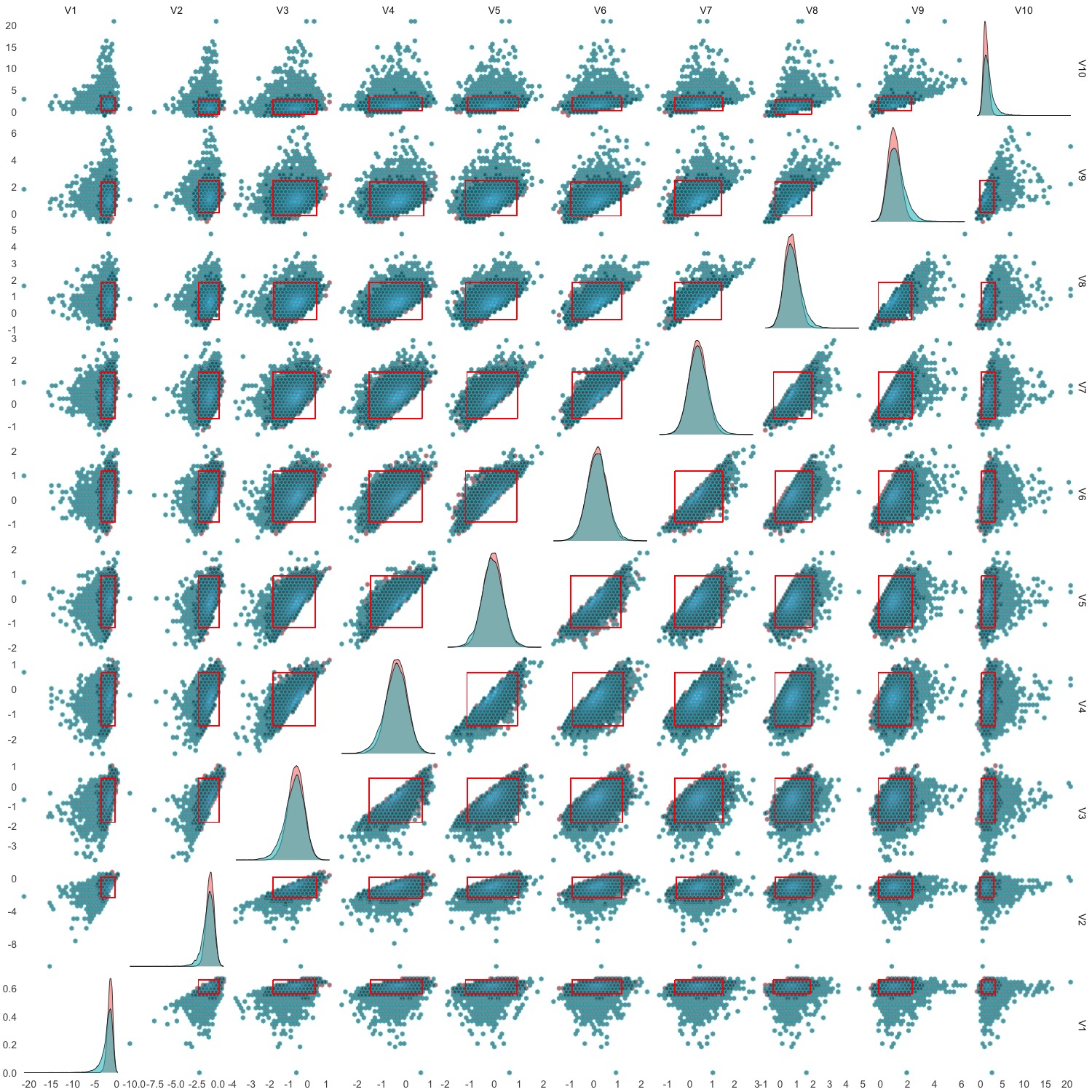}
\caption{Order statistics; $N(0,1)$ (red hexes), $\mathrm{t}(4)$ (blue hexes); hyperrectangle bounds for $N(0,1)$ (red rectangles).}
\end{figure}

Here again it looks like tail-associated order statistics play more significant role than center ones when we try to distinguish between $N(0,1)$ and $\mathrm{t}(4)$.

\newpage

\begin{minted}[mathescape, linenos]{r}
ggplot.H0.H1.hexplot.pairs(
    sorted.samples.std.normal.p.comp$scores, 
    p.comp.H1, 
    res.hyperrectangle.bounds.std.normal.pc.2, 
    col.nums = 1:10
)
\end{minted}

\begin{figure}[H]
\centering
\includegraphics[width=15cm]{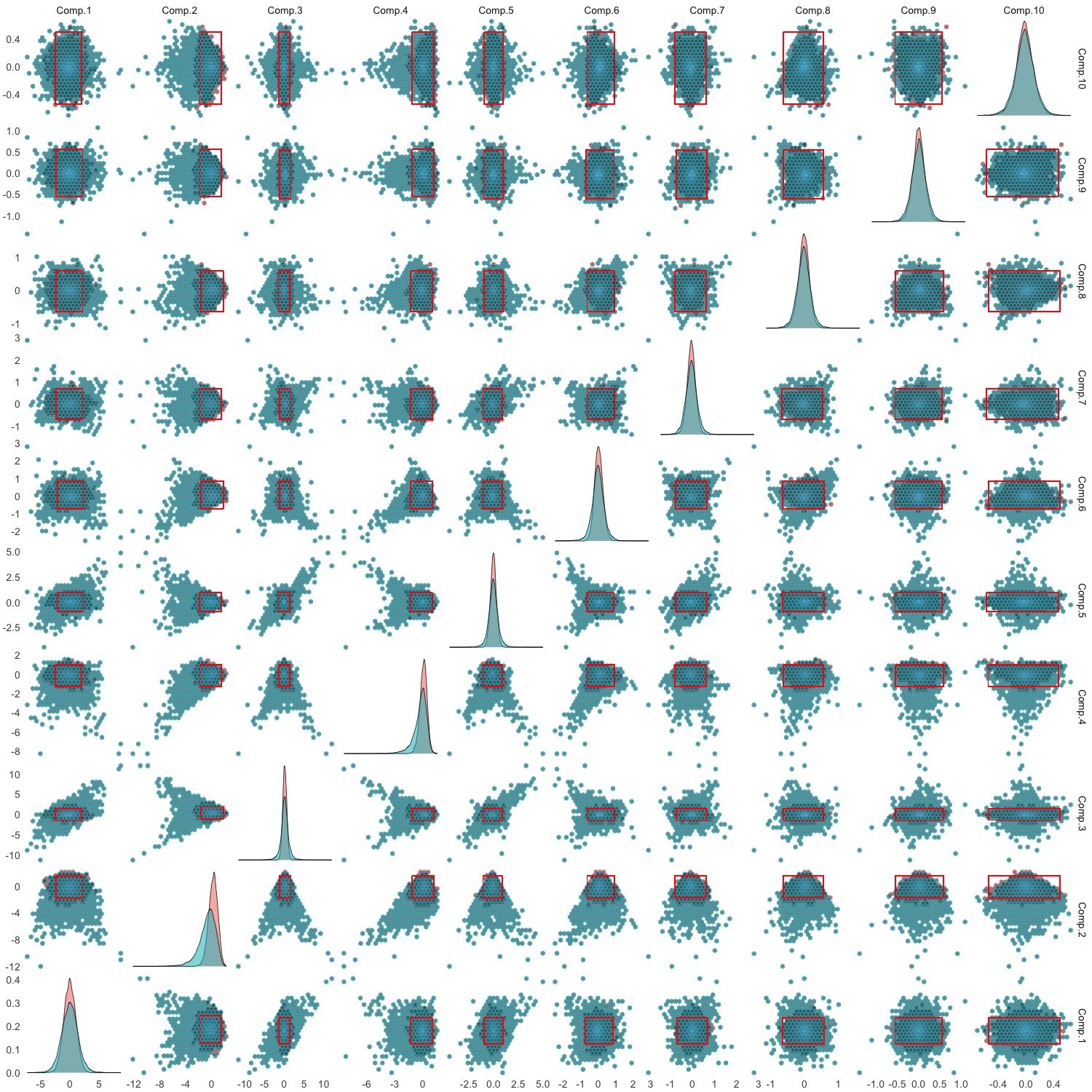}
\caption{Principal components of order statistics; $N(0,1)$ (red hexes), $\mathrm{t}(4)$ (blue hexes); hyperrectangle bounds for $N(0,1)$ (red rectangles).}
\label{figure:normal:vs:t}
\end{figure}

For this plot we don't have special comments.

Student distribution is the only example where we've seen that the variance-weighted principal component based test has slightly worth power compared to order statistics based test.

\newpage

\subsection{Function to calculate the number of rejections \textasteriskcentered}
\label{section:calc_num_rej1}

Ok, we've drawn some plots and calculated some powers for several isolated alternatives. Now we want to do detailed power analysis for order-stats-based, equal-weighted principal-component-based and variance-weighted principal-component-based tests. 

In section \ref{section:test_power} we already discussed that power estimate is a random variable that depends on both sample of samples from null hypothesis distribution and sample of samples from alternative hypothesis distribution. Because of that it is desirable not to stop after getting just a single power estimate but instead to simulate sample of samples from null and sample of samples from alternative multiple times to get empirical distribution for power estimate and then report, for example, mean and variance for this distribution.

So we write:

\begin{minted}[mathescape, linenos, texcomments]{r}
calc.rejections.sm.pc.bounds <- function(
    sample.generation.function.H0,   # single function to generate samples from H0 distribution
    sample.generation.functions.H1,  # list of functions to generate samples for multiple H1s
    get.stat.matrix,  # function to generate sample of samples and construct table of type (\ref{data_table_general}) 
    n,  # sample size for each sample in sample of samples  
    m,  # sample of sample size
    alpha,  # significance level for hyperrectangle bounds
    h,  # how many times we generate sample of samples for H0 and each H1
    gamma.left.sm = 0, gamma.right.sm = 10, # range for bisection method; stat.matrix test
    w.sm = rep(1, n),  # custom weights for test based on stat.matrix
    gamma.left.pc = 0, gamma.right.pc = 10, # range for bisection method; PCs of stat.matrix test
    w.pc = rep(1, n)   # custom weights for test based on principal components of stat.matrix
)
{
    s <- length(sample.generation.functions.H1)  # number of alternatives
    res.sm    <- matrix(0, h, s)  # allocate space for powers of stat.matrix test
    res.pc.w1 <- matrix(0, h, s)  # allocate space for powers of PC test with custom weights 
    res.pc.w2 <- matrix(0, h, s)  # allocate space for powers of PC test with variance weights
    for(i in 1:h)  # we generate sample of samples for H0 and each H1 multiple times
    {
        print(paste("iteration", i))
        flush.console()
        stat.matrix.H0 <- get.stat.matrix(sample.generation.function.H0, n, m)  # of type (\ref{data_table})
        print("calculate hyperrectangle.bounds.H0.sm")
        flush.console()
        res.h.bounds.H0.sm <- calculate.hyperrectangle.bounds(  # based on stat.matrix cols 
            alpha           = alpha,
            gamma.left      = gamma.left.sm,
            gamma.right     = gamma.right.sm,
            w               = w.sm,  # custom weights for stat.matrix
            stat.matrix     = stat.matrix.H0
        )
        stat.matrix.H0.p.comp <- princomp(stat.matrix.H0)  # of type (\ref{data_table}), but with PCs as features
        print("calculate hyperrectangle.bounds.H0.pc.w1")
        flush.console()
        res.h.bounds.H0.pc.w1 <- calculate.hyperrectangle.bounds(
            alpha           = alpha,
            gamma.left      = gamma.left.pc,
            gamma.right     = gamma.right.pc,
            w               = w.pc,  # custom weights for principal components
            stat.matrix     = stat.matrix.H0.p.comp$scores
        )
        print("calculate hyperrectangle.bounds.H0.pc.w2")
        flush.console()
        res.h.bounds.H0.pc.w2 <- calculate.hyperrectangle.bounds(
            alpha           = alpha,
            gamma.left      = gamma.left.pc,
            gamma.right     = gamma.right.pc,
            w               = stat.matrix.H0.p.comp$sdev^2 / sum(stat.matrix.H0.p.comp$sdev^2),  # var-based
            stat.matrix     = stat.matrix.H0.p.comp$scores
        )

        for(j in 1:s){  # loop over alternatives
            print(paste("H1", j))
            flush.console()
            stat.matrix.H1 <- get.stat.matrix(sample.generation.functions.H1[[j]], n, m)  # of type (\ref{data_table_general}) 
            res.sm[i,j]    <- check.hyperrectangle.bounds(stat.matrix.H1, res.h.bounds.H0.sm   )
            p.comp.H1      <- predict(stat.matrix.H0.p.comp, stat.matrix.H1)  # of type (\ref{data_table_general}), but with PCs
            res.pc.w1[i,j] <- check.hyperrectangle.bounds(p.comp.H1,      res.h.bounds.H0.pc.w1)
            res.pc.w2[i,j] <- check.hyperrectangle.bounds(p.comp.H1,      res.h.bounds.H0.pc.w2)
        }
        
        print(paste("iteration", i, "done"))
        flush.console()
    }
    list(  # matrices with power estimates; each col is a power distribution for H0 vs some H1 
        res.sm    = res.sm,
        res.pc.w1 = res.pc.w1,
        res.pc.w2 = res.pc.w2
    )
}
\end{minted}

This function supports general statistics matrices of type~(\ref{data_table_general}), not only order statistics matrices. We will later use this function to construct other tests that are not based on order statistics. Also, this function supports custom weights for raw \texttt{stat.matrix} and its principal components. In this text we don't use custom weights. Also we could have added as a parameter the number of principal components to consider, but we did not do it since in all our examples we use all principal components.

We note that in \texttt{for(i in 1:h)\{...\}} loop we generate $H_0$ sample of samples only once for each iteration and then use it inside the inner \texttt{for(j in 1:s)\{...\}} loop many times to calculate power estimates for $H_1$s, so the distributions of powers for $H_1$s that we get are correlated. We also note that for our purposes it is not a problem.

\subsection{Calculate the number of rejections, normal \texorpdfstring{$H_0$}{H0}, normal \texorpdfstring{$H_1$}{H1}}
\label{section:rejections:normal:pc}

We are finally ready to perform the previously announced extensive power study for order-statistics-based, equal-weighted PC-based and variance-weighted PC-based tests!

We start with powers for ``standard normal vs other normals'' case: 

\begin{minted}[mathescape, linenos]{r}
res <- calc.rejections.sm.pc.bounds(
    sample.generation.function.H0  = rnorm,
    sample.generation.functions.H1 = c(
        lapply(c(0.05, 0.1, 0.15), function(sd)   function(nsamples) rnorm(nsamples, 0,    sd)), 
        lapply(seq(0.2, 0.9, 0.1), function(sd)   function(nsamples) rnorm(nsamples, 0,    sd)),
        rnorm,
        lapply(seq(1.1, 2.4, 0.1), function(sd)   function(nsamples) rnorm(nsamples, 0,    sd)),
        lapply(seq(2.5,   4, 0.5), function(sd)   function(nsamples) rnorm(nsamples, 0,    sd)),
        lapply(seq(0.1,   2, 0.1), function(mean) function(nsamples) rnorm(nsamples, mean, 1 ))
    ),
    get.stat.matrix = get.sorted.samples.cpp,
    n = n, m = m,
    alpha = 0.05,
    h = 30
)
\end{minted}

List \texttt{res} contains 3 matrices with power estimates: \texttt{res.sm} is for order statistics based test, \texttt{res.pc.w1} is for equal-weighted PC-based test, and \texttt{res.pc.w2} is for variance-weighted PC-based test. Each matrix has 30 rows and 50 columns, each column corresponds to the empirical distribution of power for ``standard normal vs some other normal''. For each column of each matrix we calculate sample mean and sample standard deviation:

\begin{minted}[mathescape, linenos]{r}
create.rejection.table <- function(res.rejections)
    list(
        sm.means    = round(apply(res.rejections$res.sm,    2, mean), 4),
        pc.w1.means = round(apply(res.rejections$res.pc.w1, 2, mean), 4),
        pc.w2.means = round(apply(res.rejections$res.pc.w2, 2, mean), 4),
        sm.sds    = formatC(apply(res.rejections$res.sm,    2, sd  ), format = "e", digits = 1),
        pc.w1.sds = formatC(apply(res.rejections$res.pc.w1, 2, sd  ), format = "e", digits = 1),
        pc.w2.sds = formatC(apply(res.rejections$res.pc.w2, 2, sd  ), format = "e", digits = 1)
    )
\end{minted}

Then we report everything in the tables:

\begin{minted}[mathescape, linenos]{r}
create.rejection.table(res)
\end{minted}

\begin{table}[H]
\centering
\begin{tabular}{ |l|l||>{\columncolor[gray]{0.9}}p{1cm}|p{1.1cm}||>{\columncolor[gray]{0.9}}p{1cm}|p{1.25cm}||>{\columncolor[gray]{0.9}}p{1cm}|p{1.25cm}| } 
 \hline
 $H_0$ & $H_1$ & \multicolumn{2}{c||}{OS} & \multicolumn{2}{c||}{PC1} & \multicolumn{2}{c|}{PC2} \\
 \hline
 &  & mean & sd & mean & sd & mean & sd \\
 \hline
 \hline

$\mathcal{N}(0, 1)$ & $\mathcal{N}(0, 0.05^2)$ & 1 & 1.8e-06 & 1 & 0.0e+00 & 1 & 0.0e+00\\
$\mathcal{N}(0, 1)$ & $\mathcal{N}(0, 0.1^2)$ & 0.91 & 6.7e-03 & 1 & 0.0e+00 & 1 & 0.0e+00\\
$\mathcal{N}(0, 1)$ & $\mathcal{N}(0, 0.15^2)$ & 0.5112 & 1.2e-02 & 1 & 0.0e+00 & 1 & 0.0e+00\\
$\mathcal{N}(0, 1)$ & $\mathcal{N}(0, 0.2^2)$ & 0.2486 & 7.5e-03 & 1 & 6.6e-06 & 1 & 1.3e-06\\
$\mathcal{N}(0, 1)$ & $\mathcal{N}(0, 0.3^2)$ & 0.08 & 2.4e-03 & 0.9425 & 1.6e-03 & 0.9806 & 4.5e-04\\
$\mathcal{N}(0, 1)$ & $\mathcal{N}(0, 0.4^2)$ & 0.0386 & 9.5e-04 & 0.5911 & 4.2e-03 & 0.736 & 2.7e-03\\
$\mathcal{N}(0, 1)$ & $\mathcal{N}(0, 0.5^2)$ & 0.0243 & 5.6e-04 & 0.2577 & 3.2e-03 & 0.3807 & 2.8e-03\\
$\mathcal{N}(0, 1)$ & $\mathcal{N}(0, 0.6^2)$ & 0.0185 & 3.6e-04 & 0.0994 & 1.4e-03 & 0.165 & 1.7e-03\\
$\mathcal{N}(0, 1)$ & $\mathcal{N}(0, 0.7^2)$ & 0.0166 & 2.9e-04 & 0.0393 & 6.7e-04 & 0.0697 & 7.3e-04\\
$\mathcal{N}(0, 1)$ & $\mathcal{N}(0, 0.8^2)$ & 0.0184 & 2.3e-04 & 0.0215 & 3.3e-04 & 0.0354 & 3.6e-04\\
$\mathcal{N}(0, 1)$ & $\mathcal{N}(0, 0.9^2)$ & 0.0272 & 2.1e-04 & 0.0263 & 1.8e-04 & 0.0316 & 2.2e-04\\
\hline
$\mathcal{N}(0, 1)$ & $\mathcal{N}(0, 1^2)$ & 0.05 & 3.1e-04 & 0.05 & 3.1e-04 & 0.05 & 3.3e-04\\
\hline
$\mathcal{N}(0, 1)$ & $\mathcal{N}(0, 1.1^2)$ & 0.0939 & 4.3e-04 & 0.0943 & 3.6e-04 & 0.0921 & 4.2e-04\\
$\mathcal{N}(0, 1)$ & $\mathcal{N}(0, 1.2^2)$ & 0.1614 & 6.4e-04 & 0.1605 & 5.4e-04 & 0.1603 & 5.3e-04\\
$\mathcal{N}(0, 1)$ & $\mathcal{N}(0, 1.3^2)$ & 0.2491 & 8.4e-04 & 0.2465 & 6.5e-04 & 0.2517 & 8.0e-04\\
$\mathcal{N}(0, 1)$ & $\mathcal{N}(0, 1.4^2)$ & 0.3488 & 9.0e-04 & 0.345 & 7.1e-04 & 0.357 & 7.7e-04\\
$\mathcal{N}(0, 1)$ & $\mathcal{N}(0, 1.5^2)$ & 0.4515 & 1.1e-03 & 0.4478 & 8.6e-04 & 0.4655 & 1.2e-03\\
$\mathcal{N}(0, 1)$ & $\mathcal{N}(0, 1.6^2)$ & 0.5491 & 9.6e-04 & 0.5461 & 9.8e-04 & 0.5677 & 9.1e-04\\
$\mathcal{N}(0, 1)$ & $\mathcal{N}(0, 1.7^2)$ & 0.6368 & 9.9e-04 & 0.635 & 7.9e-04 & 0.6579 & 1.0e-03\\
$\mathcal{N}(0, 1)$ & $\mathcal{N}(0, 1.8^2)$ & 0.7122 & 8.6e-04 & 0.7115 & 9.6e-04 & 0.7339 & 9.4e-04\\
$\mathcal{N}(0, 1)$ & $\mathcal{N}(0, 1.9^2)$ & 0.7748 & 8.0e-04 & 0.7751 & 6.9e-04 & 0.7956 & 6.5e-04\\
$\mathcal{N}(0, 1)$ & $\mathcal{N}(0, 2^2)$ & 0.8251 & 6.2e-04 & 0.826 & 6.3e-04 & 0.844 & 5.7e-04\\
$\mathcal{N}(0, 1)$ & $\mathcal{N}(0, 2.1^2)$ & 0.8652 & 5.1e-04 & 0.8665 & 5.8e-04 & 0.8819 & 4.7e-04\\
$\mathcal{N}(0, 1)$ & $\mathcal{N}(0, 2.2^2)$ & 0.8964 & 4.9e-04 & 0.898 & 5.4e-04 & 0.9108 & 4.9e-04\\
$\mathcal{N}(0, 1)$ & $\mathcal{N}(0, 2.3^2)$ & 0.9206 & 4.0e-04 & 0.9221 & 3.6e-04 & 0.9326 & 3.6e-04\\
$\mathcal{N}(0, 1)$ & $\mathcal{N}(0, 2.4^2)$ & 0.9392 & 3.2e-04 & 0.9406 & 3.2e-04 & 0.9492 & 3.1e-04\\
$\mathcal{N}(0, 1)$ & $\mathcal{N}(0, 2.5^2)$ & 0.9534 & 2.1e-04 & 0.9547 & 2.9e-04 & 0.9616 & 2.5e-04\\
$\mathcal{N}(0, 1)$ & $\mathcal{N}(0, 3^2)$ & 0.9873 & 1.1e-04 & 0.9879 & 1.1e-04 & 0.9901 & 8.8e-05\\
$\mathcal{N}(0, 1)$ & $\mathcal{N}(0, 3.5^2)$ & 0.9963 & 6.7e-05 & 0.9965 & 6.4e-05 & 0.9972 & 6.2e-05\\
$\mathcal{N}(0, 1)$ & $\mathcal{N}(0, 4^2)$ & 0.9988 & 3.4e-05 & 0.9989 & 4.1e-05 & 0.9991 & 3.3e-05\\

\hline
\end{tabular}
\caption{$H_0$ --- standard normal, $H_1$ --- other normals that differ in scale.}
\label{table:H0stdnormH1normscale}
\end{table}

In table~\ref{table:H0stdnormH1normscale} ``OS'' stands for the order statistics based test from paper~\cite{AldorNoimanBrownBujaRolkeStine2013}, ``PC1'' stands for the newly proposed test based on principal components of order statistics with equal weights, and ``PC2'' stands for the newly proposed test based on principal components of order statistics with weights proportional to components variances. We will use these notations in the upcoming tables as well.

First two columns of the table describe $H_0$ and $H_1$. In 3'rd, 5'th and 7'th columns we print means of empirical power estimates based on $30$ simulation runs. In 4'th, 6'th and 8'th columns --- corresponding standard deviations. During each run for both $H_0$ and $H_1$ we simulate $m = 1000000$ samples of size $n = 10$. 

In table~\ref{table:H0stdnormH1normscale} we investigate $H_0: N(0,1)$ vs $H_1:N(0, \sigma^2)$ case for different values of $\sigma$. We see that for $\sigma <  1$  ``PC1'' and ``PC2'', the proposed principal component based tests, significantly outperform ``OS'', the order statistics based test from~\cite{AldorNoimanBrownBujaRolkeStine2013}. For $\sigma > 1$ ``PC2'' slightly outperforms two other tests, except for $\sigma = 1.1$ and $\sigma=1.2$ cases, where it is behind in the third decimal place, which is insignificant. Overall, we can conclude that ``PC2'' is the best here.

We also do simulation for $\sigma = 1$ case. In this situation simulated $H_0$ and $H_1$ distributions are the same, so instead of mean power estimates in 3'rd, 5'th and 7'th columns we get type 1 error estimates. We see that all the estimates are equal to $0.05$. That is a good sanity check and means that the constructed confidence sets work as expected. 

\begin{table}[H]
\centering
\begin{tabular}{ |l|l||>{\columncolor[gray]{0.9}}p{1cm}|p{1.1cm}||>{\columncolor[gray]{0.9}}p{1cm}|p{1.1cm}||>{\columncolor[gray]{0.9}}p{1cm}|p{1.1cm}| } 
 \hline
 $H_0$ & $H_1$ & \multicolumn{2}{c||}{OS} & \multicolumn{2}{c||}{PC1} & \multicolumn{2}{c|}{PC2} \\
 \hline
 &  & mean & sd & mean & sd & mean & sd \\

 \hline
 \hline
 $\mathcal{N}(0, 1)$ & $\mathcal{N}(0, 1)$ & 0.05 & 3.1e-04 & 0.05 & 3.1e-04 & 0.05 & 3.3e-04\\
 \hline
$\mathcal{N}(0, 1)$ & $\mathcal{N}(0.1, 1)$ & 0.058 & 3.8e-04 & 0.0524 & 3.5e-04 & 0.0573 & 3.8e-04\\
$\mathcal{N}(0, 1)$ & $\mathcal{N}(0.2, 1)$ & 0.0824 & 5.6e-04 & 0.0606 & 3.7e-04 & 0.0802 & 4.7e-04\\
$\mathcal{N}(0, 1)$ & $\mathcal{N}(0.3, 1)$ & 0.1249 & 6.0e-04 & 0.0777 & 5.6e-04 & 0.1222 & 5.1e-04\\
$\mathcal{N}(0, 1)$ & $\mathcal{N}(0.4, 1)$ & 0.1872 & 8.0e-04 & 0.1084 & 8.6e-04 & 0.1867 & 9.4e-04\\
$\mathcal{N}(0, 1)$ & $\mathcal{N}(0.5, 1)$ & 0.2688 & 9.8e-04 & 0.1575 & 1.4e-03 & 0.2747 & 1.1e-03\\
$\mathcal{N}(0, 1)$ & $\mathcal{N}(0.6, 1)$ & 0.3671 & 1.0e-03 & 0.2288 & 1.8e-03 & 0.3829 & 1.3e-03\\
$\mathcal{N}(0, 1)$ & $\mathcal{N}(0.7, 1)$ & 0.4767 & 1.3e-03 & 0.3224 & 2.3e-03 & 0.5035 & 1.5e-03\\
$\mathcal{N}(0, 1)$ & $\mathcal{N}(0.8, 1)$ & 0.5887 & 1.1e-03 & 0.4335 & 2.3e-03 & 0.6249 & 1.3e-03\\
$\mathcal{N}(0, 1)$ & $\mathcal{N}(0.9, 1)$ & 0.6946 & 1.1e-03 & 0.5531 & 2.6e-03 & 0.7354 & 1.1e-03\\
$\mathcal{N}(0, 1)$ & $\mathcal{N}(1, 1)$ & 0.7868 & 1.0e-03 & 0.6698 & 2.4e-03 & 0.827 & 9.2e-04\\
$\mathcal{N}(0, 1)$ & $\mathcal{N}(1.1, 1)$ & 0.8606 & 7.5e-04 & 0.7728 & 1.9e-03 & 0.8954 & 7.2e-04\\
$\mathcal{N}(0, 1)$ & $\mathcal{N}(1.2, 1)$ & 0.9151 & 4.6e-04 & 0.8553 & 1.5e-03 & 0.9419 & 3.2e-04\\
$\mathcal{N}(0, 1)$ & $\mathcal{N}(1.3, 1)$ & 0.9517 & 3.1e-04 & 0.9147 & 1.1e-03 & 0.9702 & 2.9e-04\\
$\mathcal{N}(0, 1)$ & $\mathcal{N}(1.4, 1)$ & 0.9747 & 1.9e-04 & 0.954 & 6.6e-04 & 0.9861 & 1.6e-04\\
$\mathcal{N}(0, 1)$ & $\mathcal{N}(1.5, 1)$ & 0.9877 & 1.2e-04 & 0.9771 & 4.1e-04 & 0.994 & 8.3e-05\\
$\mathcal{N}(0, 1)$ & $\mathcal{N}(1.6, 1)$ & 0.9945 & 9.7e-05 & 0.9896 & 2.3e-04 & 0.9977 & 5.6e-05\\
$\mathcal{N}(0, 1)$ & $\mathcal{N}(1.7, 1)$ & 0.9977 & 5.4e-05 & 0.9957 & 9.8e-05 & 0.9992 & 2.8e-05\\
$\mathcal{N}(0, 1)$ & $\mathcal{N}(1.8, 1)$ & 0.9991 & 3.3e-05 & 0.9984 & 5.2e-05 & 0.9997 & 1.6e-05\\
$\mathcal{N}(0, 1)$ & $\mathcal{N}(1.9, 1)$ & 0.9997 & 2.0e-05 & 0.9994 & 2.8e-05 & 0.9999 & 1.0e-05\\
$\mathcal{N}(0, 1)$ & $\mathcal{N}(2, 1)$ & 0.9999 & 8.1e-06 & 0.9998 & 1.3e-05 & 1 & 4.5e-06\\
 
  \hline
\end{tabular}
\caption{$H_0$ --- standard normal, $H_1$ --- other normals that differ in location.}
\label{table:H0stdnormH1normlocation}
\end{table}

In table~\ref{table:H0stdnormH1normlocation} we investigate $H_0: N(0,1)$ vs $H_1:N(\mu, 1)$ case for different values of $\mu$. We see that ``PC1'' lags behind two other tests when we check for difference in location. Generally, ``PC2'' slightly outperforms ``OS'' except for $\mu = 0.1$, $\mu = 0.2$ and $\mu = 0.3$ cases where ``OS'' is insignificantly better (again in the third decimal). ``PC2'' looks the best here as well.

\subsection{Calculate the number of rejections, normal \texorpdfstring{$H_0$}{H0}, Cauchy \texorpdfstring{$H_1$}{H1}}
\label{section:pc:normal:cauchy}

Let us move from normal alternatives to Cauchy alternatives. Cauchy distribution is still bell-shaped, but has heavier tails compared to normal. We fix Cauchy's location parameter to be $0$ and research power for different values of scale:

\begin{minted}[mathescape, linenos]{r}
res.cauchy <- calc.rejections.sm.pc.bounds(
    sample.generation.function.H0  = rnorm,
    sample.generation.functions.H1 = c(
        lapply(
            seq(0.01, 0.09, 0.01), 
            function(scale) 
                function(nsamples) 
                    rcauchy(nsamples, 0, scale)
        ),
        lapply(
            seq(0.1,  2,    0.1 ), 
            function(scale) 
                function(nsamples) 
                    rcauchy(nsamples, 0, scale)
        )
    ),
    get.stat.matrix = get.sorted.samples.cpp,
    n = n, m = m,
    alpha = 0.05,
    h = 30
)
\end{minted}

\begin{minted}[mathescape, linenos]{r}
create.rejection.table(res.cauchy)
\end{minted}

\begin{table}[H]
\centering
\begin{tabular}{ |l|l||>{\columncolor[gray]{0.9}}p{1cm}|p{1.1cm}||>{\columncolor[gray]{0.9}}p{1cm}|p{1.1cm}||>{\columncolor[gray]{0.9}}p{1cm}|p{1.1cm}| } 
 \hline
 $H_0$ & $H_1$ & \multicolumn{2}{c||}{OS} & \multicolumn{2}{c||}{PC1} & \multicolumn{2}{c|}{PC2} \\
 \hline
 &  & mean & sd & mean & sd & mean & sd \\
 \hline
 \hline
 
$\mathcal{N}(0, 1)$ & $\mathrm{Cauchy}(0, 0.01)$ & 0.9781 & 3.9e-04 & 0.9927 & 1.4e-04 & 0.9958 & 9.5e-05\\
$\mathcal{N}(0, 1)$ & $\mathrm{Cauchy}(0, 0.02)$ & 0.925 & 1.3e-03 & 0.9804 & 2.1e-04 & 0.987 & 1.7e-04\\
$\mathcal{N}(0, 1)$ & $\mathrm{Cauchy}(0, 0.03)$ & 0.8558 & 2.1e-03 & 0.9642 & 3.7e-04 & 0.9746 & 2.7e-04\\
$\mathcal{N}(0, 1)$ & $\mathrm{Cauchy}(0, 0.04)$ & 0.7814 & 3.0e-03 & 0.9451 & 4.5e-04 & 0.9594 & 3.3e-04\\
$\mathcal{N}(0, 1)$ & $\mathrm{Cauchy}(0, 0.05)$ & 0.7086 & 3.6e-03 & 0.9236 & 5.6e-04 & 0.9418 & 3.3e-04\\
$\mathcal{N}(0, 1)$ & $\mathrm{Cauchy}(0, 0.06)$ & 0.6419 & 3.9e-03 & 0.9006 & 6.9e-04 & 0.9226 & 4.9e-04\\
$\mathcal{N}(0, 1)$ & $\mathrm{Cauchy}(0, 0.07)$ & 0.5833 & 4.1e-03 & 0.8765 & 7.7e-04 & 0.9021 & 5.6e-04\\
$\mathcal{N}(0, 1)$ & $\mathrm{Cauchy}(0, 0.08)$ & 0.5333 & 4.2e-03 & 0.8516 & 7.9e-04 & 0.8806 & 5.2e-04\\
$\mathcal{N}(0, 1)$ & $\mathrm{Cauchy}(0, 0.09)$ & 0.4922 & 3.9e-03 & 0.8268 & 9.1e-04 & 0.8589 & 7.1e-04\\
$\mathcal{N}(0, 1)$ & $\mathrm{Cauchy}(0, 0.1)$ & 0.4591 & 3.7e-03 & 0.8024 & 9.2e-04 & 0.8369 & 7.3e-04\\
$\mathcal{N}(0, 1)$ & $\mathrm{Cauchy}(0, 0.2)$ & 0.3911 & 1.4e-03 & 0.6205 & 1.1e-03 & 0.6597 & 8.0e-04\\
$\mathcal{N}(0, 1)$ & $\mathrm{Cauchy}(0, 0.3)$ & 0.4734 & 7.1e-04 & 0.5749 & 8.0e-04 & 0.5983 & 7.3e-04\\
$\mathcal{N}(0, 1)$ & $\mathrm{Cauchy}(0, 0.4)$ & 0.5666 & 6.5e-04 & 0.6058 & 7.8e-04 & 0.6161 & 7.7e-04\\
$\mathcal{N}(0, 1)$ & $\mathrm{Cauchy}(0, 0.5)$ & 0.6501 & 4.6e-04 & 0.6615 & 5.3e-04 & 0.6661 & 5.8e-04\\
$\mathcal{N}(0, 1)$ & $\mathrm{Cauchy}(0, 0.6)$ & 0.7208 & 5.0e-04 & 0.7197 & 5.4e-04 & 0.7228 & 5.5e-04\\
$\mathcal{N}(0, 1)$ & $\mathrm{Cauchy}(0, 0.7)$ & 0.779 & 5.4e-04 & 0.7727 & 6.2e-04 & 0.7759 & 6.1e-04\\
$\mathcal{N}(0, 1)$ & $\mathrm{Cauchy}(0, 0.8)$ & 0.8265 & 4.8e-04 & 0.8185 & 4.7e-04 & 0.8221 & 4.9e-04\\
$\mathcal{N}(0, 1)$ & $\mathrm{Cauchy}(0, 0.9)$ & 0.8648 & 3.6e-04 & 0.8569 & 3.5e-04 & 0.8608 & 3.7e-04\\
$\mathcal{N}(0, 1)$ & $\mathrm{Cauchy}(0, 1)$ & 0.895 & 3.1e-04 & 0.8881 & 3.3e-04 & 0.8921 & 3.3e-04\\
$\mathcal{N}(0, 1)$ & $\mathrm{Cauchy}(0, 1.1)$ & 0.919 & 2.2e-04 & 0.9133 & 2.6e-04 & 0.9171 & 3.0e-04\\
$\mathcal{N}(0, 1)$ & $\mathrm{Cauchy}(0, 1.2)$ & 0.9376 & 2.4e-04 & 0.9331 & 2.5e-04 & 0.9366 & 2.4e-04\\
$\mathcal{N}(0, 1)$ & $\mathrm{Cauchy}(0, 1.3)$ & 0.9521 & 1.7e-04 & 0.9487 & 2.0e-04 & 0.9518 & 2.0e-04\\
$\mathcal{N}(0, 1)$ & $\mathrm{Cauchy}(0, 1.4)$ & 0.9632 & 1.6e-04 & 0.9606 & 2.2e-04 & 0.9634 & 1.9e-04\\
$\mathcal{N}(0, 1)$ & $\mathrm{Cauchy}(0, 1.5)$ & 0.9718 & 1.7e-04 & 0.9699 & 1.9e-04 & 0.9722 & 1.7e-04\\
$\mathcal{N}(0, 1)$ & $\mathrm{Cauchy}(0, 1.6)$ & 0.9784 & 1.5e-04 & 0.9771 & 1.5e-04 & 0.979 & 1.5e-04\\
$\mathcal{N}(0, 1)$ & $\mathrm{Cauchy}(0, 1.7)$ & 0.9834 & 9.3e-05 & 0.9825 & 1.1e-04 & 0.9841 & 9.6e-05\\
$\mathcal{N}(0, 1)$ & $\mathrm{Cauchy}(0, 1.8)$ & 0.9873 & 1.5e-04 & 0.9866 & 1.4e-04 & 0.9879 & 1.2e-04\\
$\mathcal{N}(0, 1)$ & $\mathrm{Cauchy}(0, 1.9)$ & 0.9902 & 1.2e-04 & 0.9898 & 1.0e-04 & 0.9908 & 8.6e-05\\
$\mathcal{N}(0, 1)$ & $\mathrm{Cauchy}(0, 2)$ & 0.9925 & 1.0e-04 & 0.9921 & 1.1e-04 & 0.993 & 9.1e-05\\
 
   \hline
\end{tabular}
\caption{$H_0$ --- standard normal, $H_1$ --- Cauchy distributions with different scales.}
\label{table:H0stdnormH1cauchyscale}
\end{table}

In table~\ref{table:H0stdnormH1cauchyscale} we see that for scales $\geq 0.5$ all tests do have almost the same power. For scales between $0.01$ and $0.5$ new tests are significantly better and ``PC2'' test is the best.

\subsection{Calculate the number of rejections, normal \texorpdfstring{$H_0$}{H0}, gamma \texorpdfstring{$H_1$}{H1}}
\label{section:rejections:pc:gamma}

Now let us check the power for gamma alternatives. Following~(\ref{eq:gamma_mean_var}) we construct gamma alternative distributions so that they have zero mean and unit variance.
We denote standardized gamma generating function as
\begin{equation}
\label{eq:g:definition}
\mathrm{G}(a) = \mathrm{Gamma}(a, 1 / \sqrt{a}) - a / \sqrt{a}; \; a > 0 .
\end{equation} 
For small values of $a$, $G(a)$ distributions are severely skewed, while for large values of $a$ they are very normal-like:
\begin{minted}[mathescape, linenos]{r}
par(mfrow = c(2, 3))  
for(shape in c(0.1, 0.5, 1, 10, 100, 1000)) 
    hist(
        (rgamma(1000000, shape = shape, rate = 1) - shape) / sqrt(shape), 
        main = paste("a =", shape), 
        xlab = 'x'
    )
\end{minted}
\begin{figure}[H]
\centering
\includegraphics[width=15cm]{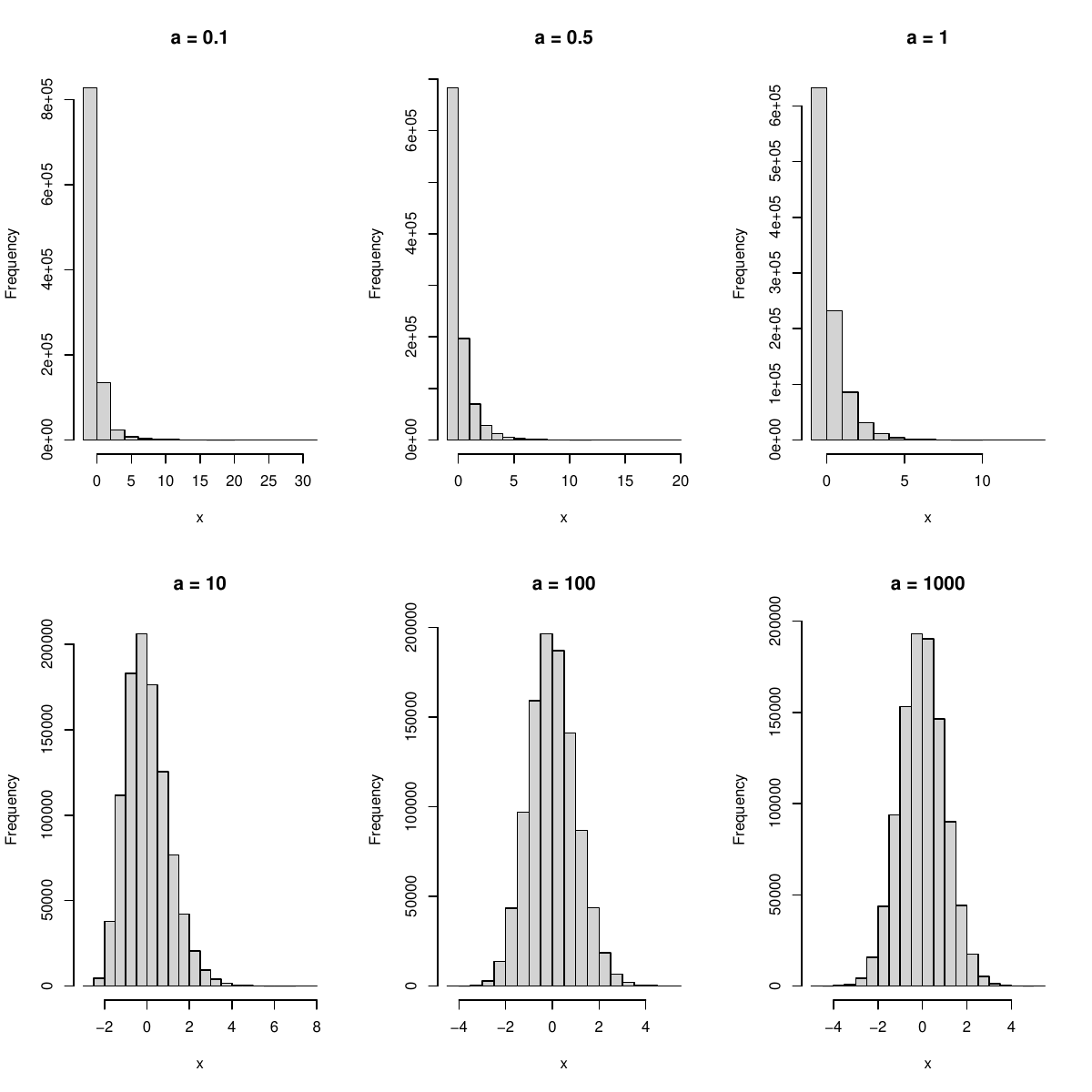}
\caption{$\mathrm{G}(a)$ distributions for different values of $a$.}
\end{figure}

\noindent
Now let us proceed to the powers:

\begin{minted}[mathescape, linenos]{r}
res.gamma <- calc.rejections.sm.pc.bounds(
    sample.generation.function.H0  = rnorm,
    sample.generation.functions.H1 = 
        lapply(
            c(seq(0.1, 0.9, 0.1), 1:5, 10, 100, 1000), 
            function(shape) 
                function(nsamples) 
                    (rgamma(nsamples, shape = shape, rate = 1) - shape) / sqrt(shape)
        ),
    get.stat.matrix = get.sorted.samples.cpp,
    n = n, m = m,
    alpha = 0.05,
    h = 30
)
\end{minted}

\begin{table}[H]
\centering
\begin{tabular}{ |l|l||>{\columncolor[gray]{0.9}}p{1cm}|p{1.1cm}||>{\columncolor[gray]{0.9}}p{1cm}|p{1.1cm}||>{\columncolor[gray]{0.9}}p{1cm}|p{1.1cm}| } 
 \hline
 $H_0$ & $H_1$ & \multicolumn{2}{c||}{OS} & \multicolumn{2}{c||}{PC1} & \multicolumn{2}{c|}{PC2} \\
 \hline
 &  & mean & sd & mean & sd & mean & sd \\
 \hline
 \hline
 
 $\mathcal{N}(0, 1)$ & $\mathrm{G}(0.1)$ & 0.5275 & 1.9e-03 & 0.8512 & 1.4e-03 & 0.8884 & 1.1e-03\\
$\mathcal{N}(0, 1)$ & $\mathrm{G}(0.2)$ & 0.3899 & 1.6e-03 & 0.7028 & 1.8e-03 & 0.7508 & 1.4e-03\\
$\mathcal{N}(0, 1)$ & $\mathrm{G}(0.3)$ & 0.3327 & 1.3e-03 & 0.5839 & 1.9e-03 & 0.6337 & 1.5e-03\\
$\mathcal{N}(0, 1)$ & $\mathrm{G}(0.4)$ & 0.2947 & 1.1e-03 & 0.4931 & 1.8e-03 & 0.5406 & 1.4e-03\\
$\mathcal{N}(0, 1)$ & $\mathrm{G}(0.5)$ & 0.2678 & 8.8e-04 & 0.4242 & 1.9e-03 & 0.4675 & 1.7e-03\\
$\mathcal{N}(0, 1)$ & $\mathrm{G}(0.6)$ & 0.2479 & 7.0e-04 & 0.3713 & 1.6e-03 & 0.41 & 1.5e-03\\
$\mathcal{N}(0, 1)$ & $\mathrm{G}(0.7)$ & 0.2326 & 8.5e-04 & 0.3303 & 1.6e-03 & 0.3644 & 1.5e-03\\
$\mathcal{N}(0, 1)$ & $\mathrm{G}(0.8)$ & 0.2201 & 7.0e-04 & 0.2979 & 1.5e-03 & 0.3278 & 1.4e-03\\
$\mathcal{N}(0, 1)$ & $\mathrm{G}(0.9)$ & 0.2095 & 8.2e-04 & 0.2718 & 1.3e-03 & 0.2979 & 1.2e-03\\
$\mathcal{N}(0, 1)$ & $\mathrm{G}(1)$ & 0.2004 & 7.4e-04 & 0.2504 & 1.4e-03 & 0.2732 & 1.3e-03\\
$\mathcal{N}(0, 1)$ & $\mathrm{G}(2)$ & 0.1501 & 5.6e-04 & 0.1525 & 7.7e-04 & 0.1594 & 7.3e-04\\
$\mathcal{N}(0, 1)$ & $\mathrm{G}(3)$ & 0.1269 & 5.8e-04 & 0.1195 & 6.4e-04 & 0.123 & 6.0e-04\\
$\mathcal{N}(0, 1)$ & $\mathrm{G}(4)$ & 0.1131 & 4.8e-04 & 0.103 & 4.9e-04 & 0.1051 & 4.6e-04\\
$\mathcal{N}(0, 1)$ & $\mathrm{G}(5)$ & 0.1037 & 3.9e-04 & 0.0929 & 4.7e-04 & 0.0944 & 4.5e-04\\
$\mathcal{N}(0, 1)$ & $\mathrm{G}(10)$ & 0.0815 & 4.1e-04 & 0.0723 & 4.2e-04 & 0.0729 & 4.4e-04\\
$\mathcal{N}(0, 1)$ & $\mathrm{G}(100)$ & 0.0539 & 2.4e-04 & 0.0524 & 2.7e-04 & 0.0525 & 2.7e-04\\
$\mathcal{N}(0, 1)$ & $\mathrm{G}(1000)$ & 0.0504 & 3.4e-04 & 0.0502 & 2.7e-04 & 0.0502 & 3.3e-04\\

   \hline
\end{tabular}
\caption{$H_0$ --- standard normal, $H_1$ --- standardized gamma distributions with different $a$'s.}
\label{table:H0stdnormH1gamma}
\end{table}

\noindent
In table~\ref{table:H0stdnormH1gamma} we see that new tests can capture differences in skew much better than test from~\cite{AldorNoimanBrownBujaRolkeStine2013}. We also see that the test from~\cite{AldorNoimanBrownBujaRolkeStine2013} has slightly better power for $a >= 3$, but the difference is very small. We conclude that ``PC2'' test is again the best overall.

\subsection{Calculate the number of rejections, normal \texorpdfstring{$H_0$}{H0}, Student \texorpdfstring{$H_1$}{H1}}
\label{section:rejections:pc:normal:student}

Finally, let us calculate power for Student alternatives, another fat-tailed family of distributions:

\begin{minted}[mathescape, linenos]{r}
res.t <- calc.rejections.sm.pc.bounds(
    sample.generation.function.H0  = rnorm,
    sample.generation.functions.H1 = 
        lapply(seq(1,5), function(df) function(nsamples) rt(nsamples, df = df)),
    get.stat.matrix = get.sorted.samples.cpp,
    n = n, m = m,
    alpha = 0.05,
    h = 30
)
\end{minted}

\begin{table}[H]
\centering
\begin{tabular}{ |l|l||>{\columncolor[gray]{0.9}}p{1cm}|p{1.1cm}||>{\columncolor[gray]{0.9}}p{1cm}|p{1.1cm}||>{\columncolor[gray]{0.9}}p{1cm}|p{1.1cm}| } 
 \hline
 $H_0$ & $H_1$ & \multicolumn{2}{c||}{OS} & \multicolumn{2}{c||}{PC1} & \multicolumn{2}{c|}{PC2} \\
 \hline
 &  & mean & sd & mean & sd & mean & sd \\
 \hline
 \hline
 
$\mathcal{N}(0, 1)$ & $\mathrm{t}(1)$ & 0.8951 & 3.2e-04 & 0.8882 & 2.8e-04 & 0.8923 & 3.1e-04\\
$\mathcal{N}(0, 1)$ & $\mathrm{t}(2)$ & 0.6232 & 8.8e-04 & 0.6052 & 7.4e-04 & 0.6146 & 7.5e-04\\
$\mathcal{N}(0, 1)$ & $\mathrm{t}(3)$ & 0.4372 & 8.6e-04 & 0.4177 & 7.3e-04 & 0.4271 & 6.9e-04\\
$\mathcal{N}(0, 1)$ & $\mathrm{t}(4)$ & 0.3264 & 5.6e-04 & 0.3089 & 6.8e-04 & 0.3169 & 5.9e-04\\
$\mathcal{N}(0, 1)$ & $\mathrm{t}(5)$ & 0.2579 & 6.5e-04 & 0.2428 & 7.7e-04 & 0.2495 & 8.1e-04\\

   \hline
\end{tabular}
\caption{$H_0$ --- standard normal, $H_1$ --- Student distributions with different degrees of freedom.}
\label{table:H0stdnormH1student}
\end{table}

\noindent
In table~\ref{table:H0stdnormH1student} we see that for Student alternatives all tests perform more or less equally well.

\subsection{Asymptotics check, \texorpdfstring{$n = 50$}{n=50}}
\label{section:asymptotics:n50}

For small sample case when $n = 10$ we performed an extensive power study for standard normal $H_0$ and different $H_1$ alternatives. We found that new ``PC2'' test significantly outperforms the test from~\cite{AldorNoimanBrownBujaRolkeStine2013} in many scenarios, and when the latter test is better, the former test is not far behind. But what about new test's asymptotics? Let us have a quick look on what is going on with power when we increase $n$. We start the asymptotics study with $n = 50$:

\begin{minted}[mathescape, linenos]{r}
res.n50 <- calc.rejections.sm.pc.bounds(
    sample.generation.function.H0  = rnorm,
    sample.generation.functions.H1 = 
        list(
            rnorm,
            function(nsamples) runif(nsamples, -sqrt(3), sqrt(3)),
            function(nsamples) rnorm(nsamples, 0.2, 1),
            function(nsamples) rnorm(nsamples, 0, 0.8),
            function(nsamples) rnorm(nsamples, 0, 1.2),
            function(nsamples) rt(nsamples, df = 5),
            function(nsamples, shape = 5) 
                (rgamma(nsamples, shape = shape, rate = 1) - shape) / sqrt(shape)
        ),
    get.stat.matrix = get.sorted.samples.cpp,
    n = 50, m = m,
    alpha = 0.05,
    h = 30
)
\end{minted}

\begin{table}[H]
\centering
\begin{tabular}{ |l|l||>{\columncolor[gray]{0.9}}p{1cm}|p{1.1cm}||>{\columncolor[gray]{0.9}}p{1cm}|p{1.1cm}||>{\columncolor[gray]{0.9}}p{1cm}|p{1.1cm}| } 
 \hline
 $H_0$ & $H_1$ & \multicolumn{2}{c||}{OS} & \multicolumn{2}{c||}{PC1} & \multicolumn{2}{c|}{PC2} \\
 \hline
 &  & mean & sd & mean & sd & mean & sd \\
 \hline
 \hline
 
 $\mathcal{N}(0, 1)$ & $\mathcal{N}(0, 1)$ & 0.05 & 3.5e-04 & 0.0501 & 3.1e-04 & 0.05 & 3.9e-04\\
$\mathcal{N}(0, 1)$ & $\mathrm{U}(-\sqrt{3}, \sqrt{3})$ & 0.146 & 7.6e-04 & 0.4609 & 5.1e-03 & 0.6056 & 2.7e-03\\
$\mathcal{N}(0, 1)$ & $\mathcal{N}(0.2, 1)$ & 0.2073 & 8.5e-04 & 0.0799 & 8.9e-04 & 0.2071 & 1.2e-03\\
$\mathcal{N}(0, 1)$ & $\mathcal{N}(0, 0.8)$ & 0.1209 & 1.1e-03 & 0.0843 & 1.8e-03 & 0.272 & 1.8e-03\\
$\mathcal{N}(0, 1)$ & $\mathcal{N}(0, 1.2)$ & 0.3235 & 1.0e-03 & 0.2779 & 1.5e-03 & 0.36 & 1.4e-03\\
$\mathcal{N}(0, 1)$ & t(5) & 0.5852 & 1.0e-03 & 0.5098 & 1.3e-03 & 0.5906 & 7.8e-04\\
$\mathcal{N}(0, 1)$ & $\mathrm{G}(5)$ & 0.2375 & 1.1e-03 & 0.2406 & 2.7e-03 & 0.3798 & 2.0e-03\\
 
    \hline
\end{tabular}
\caption{Asymptotics, $n = 50$.}
\label{table:H0stdnormal:n50}
\end{table}

\noindent
In table~\ref{table:H0stdnormal:n50} we've added one more alternative: $\mathrm{U}(-\sqrt{3}, \sqrt{3})$ --- uniform distribution on $[-\sqrt{3},\sqrt{3}]$. It was pointless to look at this alternative for $n = 10$ as for such small samples it was indistinguishable from standard normal. For $n = 50$ we see that all three tests can reject uniform $H_1$ and that ``PC2'' test is significantly better at that. We also see that
\begin{enumerate}
\item for normal location shift 
there is almost no difference between ``OS'' and ``PC2''; 
\item ``PC2'' is significantly better for normal scales that are $< 1$ and slightly better for scales that are $> 1$ than ``OS'';
\item performance of ``OS'' and ``PC2'' for Student alternative is almost the same;
\item for gamma alternative ``PC2'' is better than ``OS''.
\end{enumerate}

\subsection{Asymptotics check, \texorpdfstring{$n = 100$}{n=100}}

Let us continue asymptotics study with examining the $n = 100$ case:

\begin{minted}[mathescape, linenos]{r}
res.n100 <- calc.rejections.sm.pc.bounds(
    sample.generation.function.H0  = rnorm,
    sample.generation.functions.H1 = 
        list(
            rnorm,
            function(nsamples) runif(nsamples, -sqrt(3), sqrt(3)),
            function(nsamples) rnorm(nsamples, 0.2, 1),
            function(nsamples) rnorm(nsamples, 0, 0.8),
            function(nsamples) rnorm(nsamples, 0, 1.2),
            function(nsamples) rt(nsamples, df = 5),
            function(nsamples, shape = 5) 
                (rgamma(nsamples, shape = shape, rate = 1) - shape) / sqrt(shape)
        ),
    get.stat.matrix = get.sorted.samples.cpp,
    n = 100, m = m,
    alpha = 0.05,
    h = 30
)
\end{minted}

\begin{table}[H]
\centering
\begin{tabular}{ |l|l||>{\columncolor[gray]{0.9}}p{1cm}|p{1.1cm}||>{\columncolor[gray]{0.9}}p{1cm}|p{1.1cm}||>{\columncolor[gray]{0.9}}p{1cm}|p{1.1cm}| } 
 \hline
 $H_0$ & $H_1$ & \multicolumn{2}{c||}{OS} & \multicolumn{2}{c||}{PC1} & \multicolumn{2}{c|}{PC2} \\
 \hline
 &  & mean & sd & mean & sd & mean & sd \\
 \hline
 \hline
 
 $\mathcal{N}(0, 1)$ & $\mathcal{N}(0, 1)$ & 0.0501 & 2.5e-04 & 0.0502 & 2.9e-04 & 0.0501 & 3.5e-04\\
$\mathcal{N}(0, 1)$ & $\mathrm{U}(-\sqrt{3}, \sqrt{3})$ & 0.2611 & 1.7e-03 & 0.9126 & 3.2e-03 & 0.9783 & 4.3e-04\\
$\mathcal{N}(0, 1)$ & $\mathcal{N}(0.2, 1)$ & 0.3639 & 1.4e-03 & 0.119 & 1.9e-03 & 0.3845 & 1.2e-03\\
$\mathcal{N}(0, 1)$ & $\mathcal{N}(0, 0.8)$ & 0.3353 & 2.2e-03 & 0.2685 & 7.6e-03 & 0.6435 & 2.2e-03\\
$\mathcal{N}(0, 1)$ & $\mathcal{N}(0, 1.2)$ & 0.5048 & 1.4e-03 & 0.4089 & 2.4e-03 & 0.5954 & 1.2e-03\\
$\mathcal{N}(0, 1)$ & t(5) & 0.8006 & 8.9e-04 & 0.7014 & 1.3e-03 & 0.8125 & 5.6e-04\\
$\mathcal{N}(0, 1)$ & $\mathrm{G}(5)$ & 0.4322 & 1.8e-03 & 0.4897 & 5.7e-03 & 0.7335 & 2.0e-03\\
 
     \hline
\end{tabular}
\caption{Asymptotics, $n = 100$.}
\label{table:H0stdnormal:n100}
\end{table}

\noindent
Overall conclusions from table~\ref{table:H0stdnormal:n100} are the same as conclusions listed in the end of subsection~\ref{section:asymptotics:n50}.

\subsection{How to use normality test to check goodness of fit for any given distribution}
\label{section:how:to:use:normal:for:arbitrary}

Let us discuss how to use normality tests to check goodness of fit for arbitrary distribution. Suppose we have a sample
\begin{align*}
x_1, \dots, x_n \sim F
\end{align*} 
from some absolutely continuous distribution with cumulative probability distribution function $F$. If hypothesis 
\begin{align*}
H_0: F = F_0
\end{align*} 
is true, then
\begin{align}
F_0(x_1), \dots, F_0(x_n) &\sim \mathrm{U}(0,1), \label{eq:transform1}\\
F^{-1}_{N(0,1)} ( F_0(x_1) ), \dots, F^{-1}_{N(0,1)} ( F_0(x_n) ) &\sim N(0,1). \label{eq:transform2}
\end{align}
In other words, a sample from any given distribution can be transformed to another sample with two consequent transformations, probability integral transform and inverse CDF transform. This transformed sample can then be checked for normality. If normality hypothesis is rejected,  the original $H_0$ hypothesis should be rejected as well, since, if $H_0$ were true, the transformed sample would be normally distributed. 

\subsection{Conclusion for the section}

Practically, there is no situation where one should prefer the OS test from~\cite{AldorNoimanBrownBujaRolkeStine2013} over the proposed ``PC2'' test, since ``PC2'' test is either more powerful or almost as powerful as OS test for large amount of investigated alternatives (see tables~\ref{table:H0stdnormH1normscale}--\ref{table:H0stdnormal:n100}). ``PC2'' test still has geometrical interpretation (see figures~\ref{figure:N01vsN11hexplom}--\ref{figure:normal:vs:t} and comments about interpretation of principal components in section~\ref{section:principal_components_normal}), even if it is not as straightforward as geometrical interpretation of OS tests (that gives graphical eCDF bounds via simultaneous bounds for order statistics).

\newpage

\section{Classical goodness of fit tests (Kolmogorov--Smirnov, Cra\-mer--von Mises, Anderson--Darling): samples from standard normal distribution [cl, \texorpdfstring{$N(0,1)$}{N(0,1)}]}
\label{section:classical_tests_normal}

In this section we provide the power analysis for Kolmogorov--Smirnov~\cite{Kolmogorov1933}, Cramer--von Mises~\cite{Cramer1928,vonMises1931,vonMises1939_English} and Anderson--Darling~\cite{AndersonDarling1952,AndersonDarling1954} classical tests using the same $H_1$ alternatives and sample sizes as in section~\ref{section:pc_test_normal}. We then compare the powers of classical tests and newly proposed ``PC2'' test from section~\ref{section:pc_test_normal} and show that ``PC2'' test is substantially better.

Kolmogorov--Smirnov, Cramer--von Mises and Anderson--Darling tests are based on the following distances between empirical cumulative distribution function $F_n$ and theoretical $H_0$ cumulative distribution function $F_0$:
\begin{align}
D_{KS}(F_n, F_0) &= \sup_{x\in\mathbb{R}} \left|F_n(x)-F_0(x)\right| ; \label{eq:KSdist1} \\
D_{CvM}(F_n, F_0) &=
n \int_{-\infty}^{\infty}
\left(F_n(x)-F_0(x)\right)^2
\, dF_0(x) ;  \label{eq:CvMdist1} \\
D_{AD}(F_n, F_0) &=
n \int_{-\infty}^{\infty}
\frac{\left(F_n(x)-F_0(x)\right)^2}
{F_0(x)\bigl(1-F_0(x)\bigr)}
\, dF_0(x) . \label{eq:ADdist1}
\end{align}
The computational formulas for (\ref{eq:KSdist1}), (\ref{eq:CvMdist1}), (\ref{eq:ADdist1}) are as follows:
\begin{align}
D_{KS}(F_n, F_0) &= \max_{1\le i\le n}\max\left\{\frac{i}{n}-F_0\!\left(x_{(i)}\right),\;F_0\!\left(x_{(i)}\right)-\frac{i-1}{n}\right\}; \label{eq:KSdist2} \\
D_{CvM}(F_n, F_0) &=
\frac{1}{12n}
+
\sum_{i=1}^{n}
\left(
\frac{2i-1}{2n}
-
F_0\!\left(x_{(i)}\right)
\right)^2; \label{eq:CvMdist2} \\
D_{AD}(F_n, F_0) &=
-n
-
\sum_{i=1}^{n}
\frac{2i-1}{n}
\left[
\log F_0\!\left(x_{(i)}\right)
+
\log\!\left(1-F_0\!\left(x_{(n+1-i)}\right)\right)
\right], \label{eq:ADdist2}
\end{align}
where $x_{(i)}$ stands for i'th order statistics from the input sample. Formulas~(\ref{eq:KSdist2}), (\ref{eq:CvMdist2}), (\ref{eq:ADdist2}) are derived from~(\ref{eq:KSdist1}), (\ref{eq:CvMdist1}), (\ref{eq:ADdist1}) using the fact that empirical cumulative distribution function is a step function with jumps at order statistics combined with probability integral transform $u_i = F_0(x_{(i)})$, which converts the integral functionals into sums over ordered uniform variables (see, for example, \cite{AndersonDarling1952, Stephens1974, ShorackWellner1986}).

\subsection{Functions to calculate classical distances \textasteriskcentered}
\label{section:functions:classical:distances}

Let us implement formulas~(\ref{eq:KSdist2}), (\ref{eq:CvMdist2}), (\ref{eq:ADdist2}) straightforwardly using base \texttt{R}:

\begin{minted}[mathescape, linenos, texcomments]{r}
distance.KS  <- function(s, F_0, n)  # implementation of (\ref{eq:KSdist2})  for sorted sample s
    max( pmax((1:n) / n - F_0(s), F_0(s) - (0:(n-1)) / n) )
distance.CvM <- function(s, F_0, n)  # implementation of (\ref{eq:CvMdist2}) for sorted sample s
    1 / (12 * n) + sum( (2 * (1:n) - 1) / (2 * n) - F_0(s) )**2
distance.AD  <- function(s, F_0, n)  # implementation of (\ref{eq:ADdist2}) for sorted sample s
    -n - sum( (2 * (1:n) - 1) / n * (log(F_0(s)) + log(1 - F_0(rev(s)))) )
\end{minted}
Here each distance function expects a sorted sample, a theoretical cumulative distribution function $F_0$, and a sample size $n$.

To get $H_0$ distance distributions we may apply \texttt{distance.KS}, \texttt{distance.CvM}, and \texttt{distance.AD} functions to \texttt{sorted.samples.std.normal} \texttt{R}'s matrix that was defined in section~\ref{section:standard_normal_order_stats}:

\begin{minted}[mathescape, linenos]{r}
F_0 <- pnorm  # cdf of standard normal
distance.distribution.KS.H0  <- apply(
    sorted.samples.std.normal, 1, function(samp) distance.KS( samp, F_0, n))
distance.distribution.CvM.H0 <- apply(
    sorted.samples.std.normal, 1, function(samp) distance.CvM(samp, F_0, n))
distance.distribution.AD.H0  <- apply(
    sorted.samples.std.normal, 1, function(samp) distance.AD( samp, F_0, n))
\end{minted}
Note that $F_0$ here is cumulative distribution function of standard normal and $n$ was defined in section~\ref{section:sample_size_num_samples}.

For simulations to follow we will need to calculate KS, CvM and AD distances for $H_0$ and $H_1$ distributions a lot, so we write a function in C++ to do it faster than in base \texttt{R}:

\begin{minted}[mathescape, linenos]{r}
cppFunction("
NumericVector classicalDistances(
    NumericMatrix sorted_samples,   // matrix with sorted samples in rows
    Function F_0                    // cdf for H_0
) {
    int m = sorted_samples.nrow();  // infer number of samples from sorted_samples matrix
    int n = sorted_samples.ncol();  // infer sample size       from sorted_samples matrix

    NumericMatrix out(m, 3);        // allocate memory for the final result

    double CvM_constant = double(1) / (12 * n);    // calculate constant for CvM distance

    for (int i = 0; i < m; i++) {   // loop over rows of sorted_samples matrix 
        NumericVector sorted_sample = sorted_samples(i, _);  // get specific sample
        NumericVector F_0_values = F_0(sorted_sample);  // calculate F_0 for this sample

        double KS_dist_tmp  = 0.0;  // initiate KS  distance for sample
        double CvM_dist_tmp = 0.0;  // initiate CvM distance for sample
        double AD_dist_tmp  = 0.0;  // initiate AS  distance for sample

        for (int j = 0; j < n; j++) {  // loop over sample order statistics to calc dists
            double t1_KS = double(j + 1) / n - F_0_values[j];
            double t2_KS = F_0_values[j] - double(j) / n;
            double t3_KS = t1_KS > t2_KS ? t1_KS : t2_KS;
            if (t3_KS > KS_dist_tmp) KS_dist_tmp = t3_KS;

            CvM_dist_tmp += double(2 * (j + 1) - 1) / (2 * n) - F_0_values[j];

            double t_AD = std::log(F_0_values[j]) + std::log(1 - F_0_values[n - j - 1]);
            AD_dist_tmp += double(2 * (j + 1) - 1) / n * t_AD;
        }

        out(i, 0) = KS_dist_tmp;                                 // 1st res col is for KS
        out(i, 1) = CvM_constant + CvM_dist_tmp * CvM_dist_tmp;  // 2nd res col is for CvM
        out(i, 2) = -n - AD_dist_tmp;                            // 3rd res col is for AD
    }

    return out;
}
")
\end{minted}

\noindent
This code

\begin{minted}[mathescape, linenos]{r}
distance.distribution.KS.H0[1:5]
distance.distribution.CvM.H0[1:5]
distance.distribution.AD.H0[1:5]
\end{minted}

\noindent
will give the same values as this code

\begin{minted}[mathescape, linenos]{r}
classicalDistances(sorted.samples.std.normal, pnorm)[1:5,]
\end{minted}

\subsection{Function to calculate rejections for classical tests \textasteriskcentered}

Now let us write a function to calculate number of rejections for classical tests in the similar way we did it in subsection~\ref{section:calc_num_rej1}:

\begin{minted}[mathescape, linenos]{r}
calc.rejections.classical.tests <- function(
    sample.generation.function.H0,   # single function to generate samples from H0 distribution
    sample.generation.functions.H1,  # list of functions to generate samples for multiple H1s
    F_0,    # cdf for H0 distribution
    alpha,  # significance level 
    n,      # sample size for each sample in sample of samples
    m,      # sample of sample size
    h       # how many times we generate sample of samples for H0 and each H1
)
{
    s <- length(sample.generation.functions.H1)  # number of alternatives
    res.KS  <- matrix(0, h, s)  # allocate space for powers of KS  test
    res.CvM <- matrix(0, h, s)  # allocate space for powers of CvM test
    res.AD  <- matrix(0, h, s)  # allocate space for powers of AD test
    for(i in 1:h)  # we generate sample of samples for H0 and each H1 multiple times
    {
        print(paste("iteration", i))
        flush.console()
        
        # generate sorted samples from H0 distribution:
        sorted.samples.H0 <- get.sorted.samples.cpp(sample.generation.function.H0, n, m)
        
        print("calculate bounds")
        flush.console()
        
        # calculate distance distributions between F_0 and H0 ecdfs:
        classical.distance.distributions.H0  <- classicalDistances(sorted.samples.H0, F_0)
        # calculate bounds for each distance distribution:
        bound.KS.H0  <- quantile(classical.distance.distributions.H0[,1], 1 - alpha)
        bound.CvM.H0 <- quantile(classical.distance.distributions.H0[,2], 1 - alpha)        
        bound.AD.H0  <- quantile(classical.distance.distributions.H0[,3], 1 - alpha)
                                              
        for(j in 1:s){  # loop over alternatives
            print(paste("H1", j))
            flush.console()

            # generate sorted samples from H1 distribution: 
            sorted.samples.H1 <- get.sorted.samples.cpp(sample.generation.functions.H1[[j]], n, m)
            # calculate distance distributions between F_0 and H1 ecdfs:
            classical.distance.distributions.H1  <- classicalDistances(sorted.samples.H1, F_0)
            
            # calculate number of rejections for each distance:                                       
            res.KS[i,j]  <- sum(classical.distance.distributions.H1[,1] >= bound.KS.H0 ) / m
            res.CvM[i,j] <- sum(classical.distance.distributions.H1[,2] >= bound.CvM.H0) / m
            res.AD[i,j]  <- sum(classical.distance.distributions.H1[,3] >= bound.AD.H0 ) / m
        }
        print(paste("iteration", i, "done"))
        flush.console()
    }
    list(  # matrices with power estimates; each col is a power distribution for H0 vs some H1
        res.KS  = res.KS,
        res.CvM = res.CvM,
        res.AD  = res.AD
    )
}
\end{minted}

\subsection{Calculate rejections for classical tests: normal \texorpdfstring{$H_0$}{H0}, normal \texorpdfstring{$H_1$}{H1}}
\label{section:rejections:normal:normal:classic}

Now we are ready to perform power analysis for classical tests. We start with standard normal $H_0$ and other normal alternatives: 

\begin{minted}[mathescape, linenos]{r}
res <- calc.rejections.classical.tests(
    sample.generation.function.H0  = rnorm,
    sample.generation.functions.H1 = c(
        lapply(c(0.05, 0.1, 0.15), function(sd)   function(nsamples) rnorm(nsamples, 0,    sd)), 
        lapply(seq(0.2, 0.9, 0.1), function(sd)   function(nsamples) rnorm(nsamples, 0,    sd)),
        rnorm,
        lapply(seq(1.1, 2.4, 0.1), function(sd)   function(nsamples) rnorm(nsamples, 0,    sd)),
        lapply(seq(2.5,   4, 0.5), function(sd)   function(nsamples) rnorm(nsamples, 0,    sd)),
        lapply(seq(0.1,   2, 0.1), function(mean) function(nsamples) rnorm(nsamples, mean, 1 ))
    ),
    F_0   = pnorm,
    alpha = 0.05,
    h     = 30, n = n, m = m
)
\end{minted}

\noindent
The function to prepare the final tables is as follows: 

\begin{minted}[mathescape, linenos]{r}
create.rejection.table.classic <- function(res.rejections)
    list(
        KS.means  = round(apply(res.rejections$res.KS,  2, mean), 4),
        CvM.means = round(apply(res.rejections$res.CvM, 2, mean), 4),
        AD.means  = round(apply(res.rejections$res.AD,  2, mean), 4),
        KS.sds  = formatC(apply(res.rejections$res.KS,  2, sd  ), format = "e", digits = 1),
        CvM.sds = formatC(apply(res.rejections$res.CvM, 2, sd  ), format = "e", digits = 1),
        AD.sds  = formatC(apply(res.rejections$res.AD,  2, sd  ), format = "e", digits = 1)
    )
\end{minted}

\noindent
Finally, to the tables:

\begin{minted}[mathescape, linenos]{r}
create.rejection.table.classic(res)
\end{minted}

\noindent
Table for $H_0: N(0,1)$ vs $H_1: N(0, \sigma^2)$ is as follows:

\begin{table}[H]
\centering
\begin{tabular}{ |l|l||>{\columncolor[gray]{0.9}}p{1cm}|p{1.25cm}||>{\columncolor[gray]{0.9}}p{1cm}|p{1.25cm}||>{\columncolor[gray]{0.9}}p{1cm}|p{1.25cm}| } 
 \hline
  $H_0$ & $H_1$ & \multicolumn{2}{c||}{KS} & \multicolumn{2}{c||}{CvM} & \multicolumn{2}{c|}{AD} \\
 \hline
 &  & mean & sd & mean & sd & mean & sd \\
 \hline
 \hline
 
 $\mathcal{N}(0, 1)$ & $\mathcal{N}(0, 0.05^2)$ & 1 & 0.0e+00 & 0 & 0.0e+00 & 1 & 0.0e+00\\
$\mathcal{N}(0, 1)$ & $\mathcal{N}(0, 0.1^2)$ & 0.9902 & 2.9e-04 & 0 & 0.0e+00 & 0.9996 & 3.5e-05\\
$\mathcal{N}(0, 1)$ & $\mathcal{N}(0, 0.15^2)$ & 0.7834 & 2.3e-03 & 0 & 0.0e+00 & 0.8744 & 3.3e-03\\
$\mathcal{N}(0, 1)$ & $\mathcal{N}(0, 0.2^2)$ & 0.4689 & 2.5e-03 & 0 & 0.0e+00 & 0.4724 & 5.4e-03\\
$\mathcal{N}(0, 1)$ & $\mathcal{N}(0, 0.3^2)$ & 0.1643 & 1.1e-03 & 0 & 5.5e-07 & 0.0814 & 1.6e-03\\
$\mathcal{N}(0, 1)$ & $\mathcal{N}(0, 0.4^2)$ & 0.0789 & 6.0e-04 & 1e-04 & 9.3e-06 & 0.0201 & 4.2e-04\\
$\mathcal{N}(0, 1)$ & $\mathcal{N}(0, 0.5^2)$ & 0.0504 & 4.2e-04 & 0.0013 & 4.1e-05 & 0.0104 & 1.8e-04\\
$\mathcal{N}(0, 1)$ & $\mathcal{N}(0, 0.6^2)$ & 0.0396 & 3.0e-04 & 0.0056 & 8.6e-05 & 0.01 & 1.5e-04\\
$\mathcal{N}(0, 1)$ & $\mathcal{N}(0, 0.7^2)$ & 0.0366 & 2.5e-04 & 0.0136 & 1.3e-04 & 0.0133 & 1.4e-04\\
$\mathcal{N}(0, 1)$ & $\mathcal{N}(0, 0.8^2)$ & 0.0377 & 2.4e-04 & 0.0245 & 2.2e-04 & 0.0201 & 2.3e-04\\
$\mathcal{N}(0, 1)$ & $\mathcal{N}(0, 0.9^2)$ & 0.0424 & 2.7e-04 & 0.037 & 2.4e-04 & 0.0317 & 2.7e-04\\
\hline
$\mathcal{N}(0, 1)$ & $\mathcal{N}(0, 1^2)$ & 0.05 & 2.4e-04 & 0.05 & 2.7e-04 & 0.05 & 2.8e-04\\
\hline
$\mathcal{N}(0, 1)$ & $\mathcal{N}(0, 1.1^2)$ & 0.0606 & 2.9e-04 & 0.063 & 3.5e-04 & 0.0785 & 3.8e-04\\
$\mathcal{N}(0, 1)$ & $\mathcal{N}(0, 1.2^2)$ & 0.0741 & 4.3e-04 & 0.0753 & 3.6e-04 & 0.1201 & 5.8e-04\\
$\mathcal{N}(0, 1)$ & $\mathcal{N}(0, 1.3^2)$ & 0.0902 & 4.1e-04 & 0.0869 & 4.6e-04 & 0.1767 & 7.3e-04\\
$\mathcal{N}(0, 1)$ & $\mathcal{N}(0, 1.4^2)$ & 0.1086 & 4.5e-04 & 0.0974 & 3.8e-04 & 0.2471 & 7.6e-04\\
$\mathcal{N}(0, 1)$ & $\mathcal{N}(0, 1.5^2)$ & 0.1291 & 4.3e-04 & 0.1074 & 3.9e-04 & 0.3278 & 8.7e-04\\
$\mathcal{N}(0, 1)$ & $\mathcal{N}(0, 1.6^2)$ & 0.151 & 5.2e-04 & 0.1163 & 4.5e-04 & 0.4133 & 9.9e-04\\
$\mathcal{N}(0, 1)$ & $\mathcal{N}(0, 1.7^2)$ & 0.174 & 6.6e-04 & 0.1245 & 5.7e-04 & 0.4985 & 8.3e-04\\
$\mathcal{N}(0, 1)$ & $\mathcal{N}(0, 1.8^2)$ & 0.1977 & 5.5e-04 & 0.132 & 4.9e-04 & 0.5786 & 8.9e-04\\
$\mathcal{N}(0, 1)$ & $\mathcal{N}(0, 1.9^2)$ & 0.222 & 6.9e-04 & 0.1389 & 5.0e-04 & 0.6517 & 9.3e-04\\
$\mathcal{N}(0, 1)$ & $\mathcal{N}(0, 2^2)$ & 0.2462 & 7.4e-04 & 0.1451 & 4.4e-04 & 0.7154 & 8.6e-04\\
$\mathcal{N}(0, 1)$ & $\mathcal{N}(0, 2.1^2)$ & 0.2704 & 6.3e-04 & 0.1508 & 5.1e-04 & 0.7697 & 7.7e-04\\
$\mathcal{N}(0, 1)$ & $\mathcal{N}(0, 2.2^2)$ & 0.2942 & 5.8e-04 & 0.156 & 5.2e-04 & 0.815 & 6.4e-04\\
$\mathcal{N}(0, 1)$ & $\mathcal{N}(0, 2.3^2)$ & 0.3179 & 6.6e-04 & 0.161 & 4.8e-04 & 0.8523 & 5.2e-04\\
$\mathcal{N}(0, 1)$ & $\mathcal{N}(0, 2.4^2)$ & 0.3409 & 7.3e-04 & 0.1655 & 5.4e-04 & 0.8826 & 4.0e-04\\
$\mathcal{N}(0, 1)$ & $\mathcal{N}(0, 2.5^2)$ & 0.3631 & 7.1e-04 & 0.1697 & 5.7e-04 & 0.9069 & 4.6e-04\\
$\mathcal{N}(0, 1)$ & $\mathcal{N}(0, 3^2)$ & 0.4645 & 8.0e-04 & 0.187 & 5.9e-04 & 0.971 & 2.0e-04\\
$\mathcal{N}(0, 1)$ & $\mathcal{N}(0, 3.5^2)$ & 0.5474 & 7.0e-04 & 0.1998 & 6.3e-04 & 0.9906 & 9.2e-05\\
$\mathcal{N}(0, 1)$ & $\mathcal{N}(0, 4^2)$ & 0.6148 & 8.0e-04 & 0.2102 & 6.2e-04 & 0.9967 & 5.2e-05\\
 
    \hline
\end{tabular}
\caption{Classical tests. $H_0$ --- standard normal, $H_1$ --- other normals that differ in scale.}
\label{table:H0stdnormal:H1othernormals_scale:classical}
\end{table}

\noindent
Table~\ref{table:H0stdnormal:H1othernormals_scale:classical} should be compared with table~\ref{table:H0stdnormH1normscale}. We are mainly interested in checking classical tests against ``PC2'' test since ``PC2'' is the clear winner in the previous group of tests from table~~\ref{table:H0stdnormH1normscale}. We see that for $\sigma < 1$ KS test and AD test powers are similar. They both are far behind ``PC2'' test power. CvM test is very weak in this case. For $\sigma > 1$ AD test is a winner among classical tests. It is still noticably behind ``PC2'' test though.

Table for $H_0: N(0,1)$ vs $H_1: N(\mu, 1)$ is as follows:

\begin{table}[H]
\centering
\begin{tabular}{ |l|l||>{\columncolor[gray]{0.9}}p{1cm}|p{1.1cm}||>{\columncolor[gray]{0.9}}p{1cm}|p{1.1cm}||>{\columncolor[gray]{0.9}}p{1cm}|p{1.1cm}| } 
 \hline
  $H_0$ & $H_1$ & \multicolumn{2}{c||}{KS} & \multicolumn{2}{c||}{CvM} & \multicolumn{2}{c|}{AD} \\
 \hline
 &  & mean & sd & mean & sd & mean & sd \\
 \hline
 \hline 
 
 $\mathcal{N}(0, 1)$ & $\mathcal{N}(0, 1^2)$ & 0.05 & 2.4e-04 & 0.05 & 2.7e-04 & 0.05 & 2.8e-04\\
 
  \hline 
 
 $\mathcal{N}(0, 1)$ & $\mathcal{N}(0.1, 1)$ & 0.0584 & 3.4e-04 & 0.0609 & 3.5e-04 & 0.0607 & 3.2e-04\\
$\mathcal{N}(0, 1)$ & $\mathcal{N}(0.2, 1)$ & 0.0844 & 4.7e-04 & 0.094 & 4.0e-04 & 0.0935 & 4.7e-04\\
$\mathcal{N}(0, 1)$ & $\mathcal{N}(0.3, 1)$ & 0.129 & 5.5e-04 & 0.1509 & 6.6e-04 & 0.1504 & 6.4e-04\\
$\mathcal{N}(0, 1)$ & $\mathcal{N}(0.4, 1)$ & 0.1924 & 5.8e-04 & 0.2316 & 7.2e-04 & 0.2312 & 7.5e-04\\
$\mathcal{N}(0, 1)$ & $\mathcal{N}(0.5, 1)$ & 0.2741 & 7.6e-04 & 0.3332 & 7.3e-04 & 0.3337 & 8.3e-04\\
$\mathcal{N}(0, 1)$ & $\mathcal{N}(0.6, 1)$ & 0.3705 & 9.1e-04 & 0.4494 & 8.0e-04 & 0.4515 & 9.5e-04\\
$\mathcal{N}(0, 1)$ & $\mathcal{N}(0.7, 1)$ & 0.4764 & 8.8e-04 & 0.57 & 7.8e-04 & 0.574 & 8.6e-04\\
$\mathcal{N}(0, 1)$ & $\mathcal{N}(0.8, 1)$ & 0.584 & 7.6e-04 & 0.6841 & 8.2e-04 & 0.6898 & 9.3e-04\\
$\mathcal{N}(0, 1)$ & $\mathcal{N}(0.9, 1)$ & 0.6853 & 7.8e-04 & 0.7825 & 7.3e-04 & 0.7894 & 7.1e-04\\
$\mathcal{N}(0, 1)$ & $\mathcal{N}(1, 1)$ & 0.7746 & 7.0e-04 & 0.8607 & 6.4e-04 & 0.8674 & 6.0e-04\\
$\mathcal{N}(0, 1)$ & $\mathcal{N}(1.1, 1)$ & 0.8473 & 4.8e-04 & 0.9168 & 4.5e-04 & 0.9228 & 4.2e-04\\
$\mathcal{N}(0, 1)$ & $\mathcal{N}(1.2, 1)$ & 0.9024 & 4.0e-04 & 0.9538 & 3.0e-04 & 0.9586 & 2.9e-04\\
$\mathcal{N}(0, 1)$ & $\mathcal{N}(1.3, 1)$ & 0.9413 & 3.1e-04 & 0.9763 & 1.7e-04 & 0.9795 & 1.4e-04\\
$\mathcal{N}(0, 1)$ & $\mathcal{N}(1.4, 1)$ & 0.9668 & 2.1e-04 & 0.9887 & 9.1e-05 & 0.9907 & 1.1e-04\\
$\mathcal{N}(0, 1)$ & $\mathcal{N}(1.5, 1)$ & 0.9824 & 1.6e-04 & 0.995 & 7.9e-05 & 0.9962 & 6.0e-05\\
$\mathcal{N}(0, 1)$ & $\mathcal{N}(1.6, 1)$ & 0.9913 & 1.0e-04 & 0.998 & 5.0e-05 & 0.9985 & 4.0e-05\\
$\mathcal{N}(0, 1)$ & $\mathcal{N}(1.7, 1)$ & 0.9959 & 6.8e-05 & 0.9992 & 2.5e-05 & 0.9995 & 2.2e-05\\
$\mathcal{N}(0, 1)$ & $\mathcal{N}(1.8, 1)$ & 0.9982 & 3.7e-05 & 0.9997 & 1.8e-05 & 0.9998 & 1.2e-05\\
$\mathcal{N}(0, 1)$ & $\mathcal{N}(1.9, 1)$ & 0.9993 & 2.1e-05 & 0.9999 & 1.0e-05 & 1 & 7.1e-06\\
$\mathcal{N}(0, 1)$ & $\mathcal{N}(2, 1)$ & 0.9997 & 1.3e-05 & 1 & 4.5e-06 & 1 & 4.3e-06\\
 
     \hline
\end{tabular}
\caption{Classical tests. $H_0$ --- standard normal, $H_1$ --- other normals that differ in location.}
\label{table:H0stdnormal:H1othernormals_location:classical}
\end{table}

\noindent
Table~\ref{table:H0stdnormal:H1othernormals_location:classical} should be compared to table~\ref{table:H0stdnormH1normlocation}. We see that CvM test and AD test are a bit better than ``PC2'' test in determining the difference in location.

\subsection{Calculate rejections for classical tests: normal \texorpdfstring{$H_0$}{H0}, Cauchy \texorpdfstring{$H_1$}{H1}}
\label{section:classical_tests:cauchy}

Now let us mimic subsection~\ref{section:pc:normal:cauchy}, but for classical tests:

\begin{minted}[mathescape, linenos]{r}
res.cauchy <- calc.rejections.classical.tests(
    sample.generation.function.H0  = rnorm,
    sample.generation.functions.H1 = c(
        lapply(
            seq(0.01, 0.09, 0.01), 
            function(scale) 
                function(nsamples) 
                    rcauchy(nsamples, 0, scale)
        ),
        lapply(
            seq(0.1,  2,    0.1 ), 
            function(scale) 
                function(nsamples) 
                    rcauchy(nsamples, 0, scale)
        )
    ),
    F_0   = pnorm,
    alpha = 0.05,
    h     = 30, n = n, m = m
)
\end{minted}

\begin{minted}[mathescape, linenos]{r}
create.rejection.table.classic(res.cauchy)
\end{minted}

\begin{table}[H]
\centering
\begin{tabular}{ |l|l||>{\columncolor[gray]{0.9}}p{1cm}|p{1.1cm}||>{\columncolor[gray]{0.9}}p{1cm}|p{1.25cm}||>{\columncolor[gray]{0.9}}p{1cm}|p{1.1cm}| } 
 \hline
  $H_0$ & $H_1$ & \multicolumn{2}{c||}{KS} & \multicolumn{2}{c||}{CvM} & \multicolumn{2}{c|}{AD} \\
 \hline
 &  & mean & sd & mean & sd & mean & sd \\
 \hline
 \hline
 
 $\mathcal{N}(0, 1)$ & $\mathrm{Cauchy}(0, 0.01)$ & 0.9845 & 1.5e-04 & 0 & 0.0e+00 & 0.9894 & 1.8e-04\\
$\mathcal{N}(0, 1)$ & $\mathrm{Cauchy}(0, 0.02)$ & 0.9447 & 3.1e-04 & 0 & 6.8e-07 & 0.949 & 5.5e-04\\
$\mathcal{N}(0, 1)$ & $\mathrm{Cauchy}(0, 0.03)$ & 0.8891 & 6.9e-04 & 0 & 1.9e-06 & 0.8823 & 1.2e-03\\
$\mathcal{N}(0, 1)$ & $\mathrm{Cauchy}(0, 0.04)$ & 0.8246 & 7.6e-04 & 0 & 4.0e-06 & 0.8016 & 1.4e-03\\
$\mathcal{N}(0, 1)$ & $\mathrm{Cauchy}(0, 0.05)$ & 0.7559 & 1.1e-03 & 0 & 5.0e-06 & 0.7167 & 1.9e-03\\
$\mathcal{N}(0, 1)$ & $\mathrm{Cauchy}(0, 0.06)$ & 0.6867 & 1.3e-03 & 1e-04 & 6.1e-06 & 0.635 & 2.2e-03\\
$\mathcal{N}(0, 1)$ & $\mathrm{Cauchy}(0, 0.07)$ & 0.6199 & 1.2e-03 & 1e-04 & 1.2e-05 & 0.5608 & 2.1e-03\\
$\mathcal{N}(0, 1)$ & $\mathrm{Cauchy}(0, 0.08)$ & 0.5568 & 1.5e-03 & 2e-04 & 1.2e-05 & 0.4955 & 2.0e-03\\
$\mathcal{N}(0, 1)$ & $\mathrm{Cauchy}(0, 0.09)$ & 0.4985 & 1.5e-03 & 3e-04 & 1.7e-05 & 0.44 & 2.2e-03\\
$\mathcal{N}(0, 1)$ & $\mathrm{Cauchy}(0, 0.1)$ & 0.4452 & 1.5e-03 & 5e-04 & 2.1e-05 & 0.3939 & 1.9e-03\\
$\mathcal{N}(0, 1)$ & $\mathrm{Cauchy}(0, 0.2)$ & 0.1524 & 7.7e-04 & 0.0053 & 8.7e-05 & 0.2645 & 6.9e-04\\
$\mathcal{N}(0, 1)$ & $\mathrm{Cauchy}(0, 0.3)$ & 0.0715 & 5.8e-04 & 0.0165 & 1.5e-04 & 0.3293 & 6.4e-04\\
$\mathcal{N}(0, 1)$ & $\mathrm{Cauchy}(0, 0.4)$ & 0.0482 & 3.6e-04 & 0.0317 & 2.5e-04 & 0.4149 & 7.0e-04\\
$\mathcal{N}(0, 1)$ & $\mathrm{Cauchy}(0, 0.5)$ & 0.0445 & 2.7e-04 & 0.048 & 3.0e-04 & 0.4995 & 4.3e-04\\
$\mathcal{N}(0, 1)$ & $\mathrm{Cauchy}(0, 0.6)$ & 0.0507 & 3.0e-04 & 0.0638 & 2.6e-04 & 0.5785 & 6.5e-04\\
$\mathcal{N}(0, 1)$ & $\mathrm{Cauchy}(0, 0.7)$ & 0.0634 & 3.4e-04 & 0.0786 & 3.9e-04 & 0.6502 & 4.7e-04\\
$\mathcal{N}(0, 1)$ & $\mathrm{Cauchy}(0, 0.8)$ & 0.0811 & 3.6e-04 & 0.092 & 3.7e-04 & 0.7134 & 5.3e-04\\
$\mathcal{N}(0, 1)$ & $\mathrm{Cauchy}(0, 0.9)$ & 0.1025 & 5.1e-04 & 0.1039 & 4.8e-04 & 0.7678 & 3.8e-04\\
$\mathcal{N}(0, 1)$ & $\mathrm{Cauchy}(0, 1)$ & 0.127 & 5.9e-04 & 0.1149 & 4.7e-04 & 0.8137 & 3.3e-04\\
$\mathcal{N}(0, 1)$ & $\mathrm{Cauchy}(0, 1.1)$ & 0.1534 & 5.5e-04 & 0.1247 & 5.2e-04 & 0.8518 & 4.2e-04\\
$\mathcal{N}(0, 1)$ & $\mathrm{Cauchy}(0, 1.2)$ & 0.1813 & 6.4e-04 & 0.1331 & 4.6e-04 & 0.8827 & 4.1e-04\\
$\mathcal{N}(0, 1)$ & $\mathrm{Cauchy}(0, 1.3)$ & 0.2098 & 5.9e-04 & 0.141 & 5.1e-04 & 0.9078 & 3.0e-04\\
$\mathcal{N}(0, 1)$ & $\mathrm{Cauchy}(0, 1.4)$ & 0.2387 & 5.9e-04 & 0.1481 & 3.7e-04 & 0.9276 & 2.8e-04\\
$\mathcal{N}(0, 1)$ & $\mathrm{Cauchy}(0, 1.5)$ & 0.2677 & 6.3e-04 & 0.1545 & 4.8e-04 & 0.9434 & 2.9e-04\\
$\mathcal{N}(0, 1)$ & $\mathrm{Cauchy}(0, 1.6)$ & 0.2961 & 5.8e-04 & 0.1601 & 5.3e-04 & 0.9559 & 1.9e-04\\
$\mathcal{N}(0, 1)$ & $\mathrm{Cauchy}(0, 1.7)$ & 0.324 & 8.5e-04 & 0.1657 & 5.8e-04 & 0.9656 & 1.9e-04\\
$\mathcal{N}(0, 1)$ & $\mathrm{Cauchy}(0, 1.8)$ & 0.3512 & 7.4e-04 & 0.1704 & 4.9e-04 & 0.9732 & 1.5e-04\\
$\mathcal{N}(0, 1)$ & $\mathrm{Cauchy}(0, 1.9)$ & 0.3776 & 7.0e-04 & 0.175 & 5.4e-04 & 0.9791 & 1.4e-04\\
$\mathcal{N}(0, 1)$ & $\mathrm{Cauchy}(0, 2)$ & 0.4029 & 7.5e-04 & 0.1792 & 5.0e-04 & 0.9836 & 1.4e-04\\
 
      \hline
\end{tabular}
\caption{Classical tests. $H_0$ --- standard normal, $H_1$ --- Cauchy with different scales.}
\label{table:H0stdnormal:H1cauchy_scale:classical}
\end{table}

\noindent
Table~\ref{table:H0stdnormal:H1cauchy_scale:classical} should be compared with table~\ref{table:H0stdnormH1cauchyscale}. We see that CvM test does not work well at all for Cauchy alternatives. AD test looks better than KS test overall. KS test has local power minimum of $0.0445$ at scale = 0.5, AD test --- $0.2645$ at scale = 0.2. ``PC2'' test has local minimum of $0.5983$ at 0.3. We see that ``PC2'' test is universally better in terms of power at any point of scale grid for Cauchy alternatives.

\subsection{Calculate rejections for classical tests: normal \texorpdfstring{$H_0$}{H0}, gamma \texorpdfstring{$H_1$}{H1}}
\label{section:rejections:classic:normal:gamma}

Here we repeat power simulations from subsection~\ref{section:rejections:pc:gamma} about gamma alternatives, but for classical tests:

\begin{minted}[mathescape, linenos]{r}
res.gamma <- calc.rejections.classical.tests(
    sample.generation.function.H0  = rnorm,
    sample.generation.functions.H1 = 
        lapply(
            c(seq(0.1,0.9,0.1),1:5,10,100,1000), 
            function(shape) 
                function(nsamples) 
                    (rgamma(nsamples, shape = shape, rate = 1) - shape) / sqrt(shape)
        ),
    F_0   = pnorm,
    alpha = 0.05,
    h     = 30, n = n, m = m 
)
\end{minted}

\begin{table}[H]
\centering
\begin{tabular}{ |l|l||>{\columncolor[gray]{0.9}}p{1cm}|p{1.1cm}||>{\columncolor[gray]{0.9}}p{1cm}|p{1.1cm}||>{\columncolor[gray]{0.9}}p{1cm}|p{1.1cm}| } 
 \hline
  $H_0$ & $H_1$ & \multicolumn{2}{c||}{KS} & \multicolumn{2}{c||}{CvM} & \multicolumn{2}{c|}{AD} \\
 \hline
 &  & mean & sd & mean & sd & mean & sd \\
 \hline
 \hline
 
 $\mathcal{N}(0, 1)$ & $\mathrm{G}(0.1)$ & 0.5276 & 1.3e-03 & 4e-04 & 1.9e-05 & 0.6366 & 1.5e-03\\
$\mathcal{N}(0, 1)$ & $\mathrm{G}(0.2)$ & 0.3561 & 8.2e-04 & 0.0011 & 2.9e-05 & 0.4072 & 1.2e-03\\
$\mathcal{N}(0, 1)$ & $\mathrm{G}(0.3)$ & 0.2643 & 7.6e-04 & 0.0129 & 1.6e-04 & 0.2923 & 1.2e-03\\
$\mathcal{N}(0, 1)$ & $\mathrm{G}(0.4)$ & 0.2148 & 8.0e-04 & 0.0234 & 2.2e-04 & 0.2283 & 1.1e-03\\
$\mathcal{N}(0, 1)$ & $\mathrm{G}(0.5)$ & 0.1837 & 6.0e-04 & 0.0298 & 2.9e-04 & 0.1883 & 1.0e-03\\
$\mathcal{N}(0, 1)$ & $\mathrm{G}(0.6)$ & 0.1621 & 6.8e-04 & 0.0342 & 2.9e-04 & 0.162 & 6.9e-04\\
$\mathcal{N}(0, 1)$ & $\mathrm{G}(0.7)$ & 0.1463 & 5.2e-04 & 0.0375 & 2.7e-04 & 0.1435 & 6.5e-04\\
$\mathcal{N}(0, 1)$ & $\mathrm{G}(0.8)$ & 0.134 & 5.1e-04 & 0.0399 & 2.6e-04 & 0.1297 & 6.7e-04\\
$\mathcal{N}(0, 1)$ & $\mathrm{G}(0.9)$ & 0.1244 & 4.5e-04 & 0.0418 & 3.4e-04 & 0.1193 & 6.2e-04\\
$\mathcal{N}(0, 1)$ & $\mathrm{G}(1)$ & 0.1166 & 6.8e-04 & 0.0432 & 3.1e-04 & 0.1111 & 6.6e-04\\
$\mathcal{N}(0, 1)$ & $\mathrm{G}(2)$ & 0.0826 & 3.5e-04 & 0.0486 & 2.9e-04 & 0.0773 & 4.5e-04\\
$\mathcal{N}(0, 1)$ & $\mathrm{G}(3)$ & 0.0715 & 4.8e-04 & 0.0498 & 3.3e-04 & 0.0674 & 4.4e-04\\
$\mathcal{N}(0, 1)$ & $\mathrm{G}(4)$ & 0.066 & 3.9e-04 & 0.0501 & 3.3e-04 & 0.0627 & 4.0e-04\\
$\mathcal{N}(0, 1)$ & $\mathrm{G}(5)$ & 0.0627 & 3.1e-04 & 0.0502 & 2.4e-04 & 0.06 & 3.4e-04\\
$\mathcal{N}(0, 1)$ & $\mathrm{G}(10)$ & 0.0564 & 3.3e-04 & 0.0503 & 3.2e-04 & 0.0549 & 3.5e-04\\
$\mathcal{N}(0, 1)$ & $\mathrm{G}(100)$ & 0.0506 & 4.0e-04 & 0.05 & 2.8e-04 & 0.0505 & 3.7e-04\\
$\mathcal{N}(0, 1)$ & $\mathrm{G}(1000)$ & 0.0501 & 3.4e-04 & 0.05 & 2.5e-04 & 0.0501 & 3.5e-04\\
 
       \hline
\end{tabular}
\caption{Classical tests. $H_0$ --- standard normal, $H_1$ --- standardized gamma with different skews.}
\label{table:H0stdnormal:H1gamma_scale:classical}
\end{table}

\noindent
Table~\ref{table:H0stdnormal:H1gamma_scale:classical} should be compared with table~\ref{table:H0stdnormH1gamma}. Remember that $\mathrm{G}(a)$ was defined in~(\ref{eq:g:definition}). We see that CvM test does not work for gamma alternatives. AD test works a bit better than KS test. All classical tests are worse than ``PC2'' test from table~\ref{table:H0stdnormH1gamma} for standardized gamma alternatives.

\subsection{Calculate rejections for classical tests: normal \texorpdfstring{$H_0$}{H0}, Student \texorpdfstring{$H_1$}{H1}}
\label{section:rejections:classical:normal:student}

This subsection mimics subsection~\ref{section:rejections:pc:normal:student} about Student alternatives, but for classical tests:

\begin{minted}[mathescape, linenos]{r}
res.t <- calc.rejections.classical.tests(
    sample.generation.function.H0  = rnorm,
    sample.generation.functions.H1 = 
        lapply(seq(1,5), function(df) function(nsamples) rt(nsamples, df = df)),
    F_0   = pnorm,
    alpha = 0.05,
    h     = 30, n = n, m = m 
)
\end{minted}

\begin{table}[H]
\centering
\begin{tabular}{ |l|l||>{\columncolor[gray]{0.9}}p{1cm}|p{1.1cm}||>{\columncolor[gray]{0.9}}p{1cm}|p{1.1cm}||>{\columncolor[gray]{0.9}}p{1cm}|p{1.1cm}| } 
 \hline
  $H_0$ & $H_1$ & \multicolumn{2}{c||}{KS} & \multicolumn{2}{c||}{CvM} & \multicolumn{2}{c|}{AD} \\
 \hline
 &  & mean & sd & mean & sd & mean & sd \\
 \hline
 \hline
 
$\mathcal{N}(0, 1)$ & $\mathrm{t}(1)$ & 0.1271 & 4.5e-04 & 0.1148 & 4.6e-04 & 0.8137 & 3.8e-04\\
$\mathcal{N}(0, 1)$ & $\mathrm{t}(2)$ & 0.0791 & 4.2e-04 & 0.0865 & 4.1e-04 & 0.4491 & 9.5e-04\\
$\mathcal{N}(0, 1)$ & $\mathrm{t}(3)$ & 0.0671 & 3.8e-04 & 0.0752 & 3.6e-04 & 0.2664 & 8.4e-04\\
$\mathcal{N}(0, 1)$ & $\mathrm{t}(4)$ & 0.0618 & 3.3e-04 & 0.0691 & 3.2e-04 & 0.1819 & 6.2e-04\\
$\mathcal{N}(0, 1)$ & $\mathrm{t}(5)$ & 0.0591 & 3.4e-04 & 0.0654 & 4.0e-04 & 0.139 & 6.2e-04\\
 
        \hline
\end{tabular}
\caption{Classical tests. $H_0$ --- standard normal, $H_1$ --- Student alternatives with different degrees of freedom.}
\label{table:H0stdnormal:H1student:classical}
\end{table}

\noindent
Table~\ref{table:H0stdnormal:H1student:classical} should be compared with table~\ref{table:H0stdnormH1student}. We see that KS and CvM tests don't work well for Student alternatives. AD test works slightly worse than ``PC2'' test.

\subsection{Asymptotics check for classical tests, \texorpdfstring{$n = 50$}{n=50}}
\label{section:asymptotics:classical:n50}

Let us check asymptotics for classical tests like we did it for PC-based tests in subsection~\ref{section:asymptotics:n50}:

\begin{minted}[mathescape, linenos]{r}
res.n50 <- calc.rejections.classical.tests(
    sample.generation.function.H0  = rnorm,
    sample.generation.functions.H1 = 
        list(
            rnorm,
            function(nsamples) runif(nsamples, -sqrt(3), sqrt(3)),
            function(nsamples) rnorm(nsamples, 0.2, 1),
            function(nsamples) rnorm(nsamples, 0, 0.8),
            function(nsamples) rnorm(nsamples, 0, 1.2),
            function(nsamples) rt(nsamples, df = 5),
            function(nsamples, shape = 5) 
                (rgamma(nsamples, shape = shape, rate = 1) - shape) / sqrt(shape)
        ),
    F_0   = pnorm,
    alpha = 0.05,
    h     = 30, n = 50, m = m 
)
\end{minted}

\begin{table}[H]
\centering
\begin{tabular}{ |l|l||>{\columncolor[gray]{0.9}}p{1cm}|p{1.1cm}||>{\columncolor[gray]{0.9}}p{1cm}|p{1.1cm}||>{\columncolor[gray]{0.9}}p{1cm}|p{1.1cm}| } 
 \hline
  $H_0$ & $H_1$ & \multicolumn{2}{c||}{KS} & \multicolumn{2}{c||}{CvM} & \multicolumn{2}{c|}{AD} \\
 \hline
 &  & mean & sd & mean & sd & mean & sd \\
 \hline
 \hline
 
 $\mathcal{N}(0, 1)$ & $\mathcal{N}(0, 1)$ & 0.05 & 3.4e-04 & 0.05 & 3.0e-04 & 0.05 & 3.6e-04\\
$\mathcal{N}(0, 1)$ & $\mathrm{U}(-\sqrt{3}, \sqrt{3})$ & 0.1553 & 8.2e-04 & 0.069 & 4.1e-04 & 0.1455 & 7.6e-04\\
$\mathcal{N}(0, 1)$ & $\mathcal{N}(0.2, 1)$ & 0.2284 & 6.2e-04 & 0.2812 & 7.7e-04 & 0.2757 & 7.3e-04\\
$\mathcal{N}(0, 1)$ & $\mathcal{N}(0, 0.8)$ & 0.0772 & 5.1e-04 & 0.025 & 2.1e-04 & 0.0737 & 5.6e-04\\
$\mathcal{N}(0, 1)$ & $\mathcal{N}(0, 1.2)$ & 0.1059 & 5.2e-04 & 0.0745 & 3.2e-04 & 0.2091 & 7.8e-04\\
$\mathcal{N}(0, 1)$ & t(5) & 0.064 & 4.1e-04 & 0.0649 & 3.1e-04 & 0.2046 & 7.6e-04\\
$\mathcal{N}(0, 1)$ & $\mathrm{G}(5)$ & 0.124 & 5.9e-04 & 0.0731 & 3.1e-04 & 0.1165 & 4.9e-04\\
 
         \hline
\end{tabular}
\caption{Classical tests asymptotics, n = 50.}
\label{table:H0stdnormal:n50:classic}
\end{table}

\noindent
Table~\ref{table:H0stdnormal:n50:classic} should be compared with table~\ref{table:H0stdnormal:n50}. We see that ``PC2'' test works much better then classical tests for "normal vs uniform" case, CvM and AD tests work better for "normal vs shifted normal" case, ``PC2'' is better for "normal vs scaled normal", "normal vs Student" and "normal vs gamma".

\subsection{Asymptotics check for classical tests, \texorpdfstring{$n = 100$}{n=100}}

Finally, let us look at power of classical tests for larger samples:

\begin{minted}[mathescape, linenos]{r}
res.n100 <- calc.rejections.classical.tests(
    sample.generation.function.H0  = rnorm,
    sample.generation.functions.H1 = 
        list(
            rnorm,
            function(nsamples) runif(nsamples, -sqrt(3), sqrt(3)),
            function(nsamples) rnorm(nsamples, 0.2, 1),
            function(nsamples) rnorm(nsamples, 0, 0.8),
            function(nsamples) rnorm(nsamples, 0, 1.2),
            function(nsamples) rt(nsamples, df = 5),
            function(nsamples, shape = 5) 
                (rgamma(nsamples, shape = shape, rate = 1) - shape) / sqrt(shape)
        ),
    F_0   = pnorm,
    alpha = 0.05,
    h     = 30, n = 100, m = m 
)
\end{minted}

\begin{table}[H]
\centering
\begin{tabular}{ |l|l||>{\columncolor[gray]{0.9}}p{1cm}|p{1.1cm}||>{\columncolor[gray]{0.9}}p{1cm}|p{1.1cm}||>{\columncolor[gray]{0.9}}p{1cm}|p{1.1cm}| } 
 \hline
  $H_0$ & $H_1$ & \multicolumn{2}{c||}{KS} & \multicolumn{2}{c||}{CvM} & \multicolumn{2}{c|}{AD} \\
 \hline
 &  & mean & sd & mean & sd & mean & sd \\
 \hline
 \hline
 
 $\mathcal{N}(0, 1)$ & $\mathcal{N}(0, 1)$ & 0.05 & 2.4e-04 & 0.0499 & 3.3e-04 & 0.05 & 3.4e-04\\
$\mathcal{N}(0, 1)$ & $\mathrm{U}(-\sqrt{3}, \sqrt{3})$ & 0.2602 & 8.5e-04 & 0.0688 & 3.9e-04 & 0.2894 & 1.0e-03\\
$\mathcal{N}(0, 1)$ & $\mathcal{N}(0.2, 1)$ & 0.4054 & 6.9e-04 & 0.4969 & 8.4e-04 & 0.4884 & 8.1e-04\\
$\mathcal{N}(0, 1)$ & $\mathcal{N}(0, 0.8)$ & 0.148 & 7.1e-04 & 0.025 & 2.2e-04 & 0.232 & 1.1e-03\\
$\mathcal{N}(0, 1)$ & $\mathcal{N}(0, 1.2)$ & 0.1548 & 5.4e-04 & 0.0743 & 3.8e-04 & 0.349 & 9.9e-04\\
$\mathcal{N}(0, 1)$ & t(5) & 0.0716 & 3.2e-04 & 0.0648 & 3.5e-04 & 0.3058 & 7.7e-04\\
$\mathcal{N}(0, 1)$ & $\mathrm{G}(5)$ & 0.2047 & 5.3e-04 & 0.1022 & 4.4e-04 & 0.2163 & 8.0e-04\\
 
          \hline
\end{tabular}
\caption{Classical tests asymptotics, n = 100.}
\label{table:H0stdnormal:n100:classic}
\end{table}

\noindent
Table~\ref{table:H0stdnormal:n100:classic} should be compared with table~\ref{table:H0stdnormal:n100}. All conclusions are the same as in the previous paragraph. We can just observe power dynamics.

\subsection{Conclusions}

Classical tests are better than proposed ``PC2'' test only when checking "standard normal vs shifted normal" situation. In all other situations ``PC2'' test has better power.

\newpage

\section{Zhang likelihood-based goodness of fit tests (\texorpdfstring{$Z_K, Z_A, Z_C$}{Zk,Za,Zc}): samples from standard normal distribution [zh, \texorpdfstring{$N(0,1)$}{N(0,1)}]}
\label{section:zhang:normal}

In section~\ref{section:pc_test_normal} we've shown that the proposed PC2 test based on principal components of order statistics outperforms  the test based on order statistics from~\cite{AldorNoimanBrownBujaRolkeStine2013} for normal null vs various alternatives and is never significantly worse than this test. In section~\ref{section:classical_tests_normal} we've additionally shown, that PC2 tests is more powerful than classical KS, CvM, and AD tests except when checking for normal location shift. In this section we will compare our test with more recent and powerful tests. 

Likelihood-ratio-based goodness-of-fit tests proposed by Zhang in~\cite{zhang2001powerful, zhang2002powerful} posses top tier power and often outperform classical tests, as shown in multiple simulation studies (see, for example, \cite{Romao2010, rolke2020supplementalstudiessimultaneousgoodnessoffit, Rolke2024}). If we denote likelihood ratio for $F_0$ and $F_n$ as 
\begin{align}
G_t^2 = 2n \left[
F_n(t)\log\left(\frac{F_n(t)}{F_0(t)}\right)
+ \left(1 - F_n(t)\right)\log\left(\frac{1 - F_n(t)}{1 - F_0(t)}\right)
\right],
\end{align}
then statistics for Zhang tests are 
\begin{align}
Z_K &= \sup_{t \in \mathbb{R}} G_t^2; \label{eq:Z_K_1} \\
Z_A &=
\int_{-\infty}^{\infty}
G_t^2 \,
\frac{dF_n(t)}{F_n(t)\left(1 - F_n(t)\right)}; \label{eq:Z_A_1} \\
Z_C &=
\int_{-\infty}^{\infty}
G_t^2 \,
\frac{dF_0(t)}{F_0(t)\left(1 - F_0(t)\right)}. \label{eq:Z_C_1}
\end{align}
After several transformations of~(\ref{eq:Z_K_1}), (\ref{eq:Z_A_1}), (\ref{eq:Z_C_1}) and by applying some assumptions (see~\cite{zhang2002powerful}), one can get the final computation formulas:
\begin{align}
Z_K &=
\max_{1 \le i \le n}
\left[
\left(i - \frac{1}{2}\right)
\log\left(\frac{i - \frac{1}{2}}{n\,F_0(X_{(i)})}\right)
+
\left(n - i + \frac{1}{2}\right)
\log\left(\frac{n - i + \frac{1}{2}}{n\left(1 - F_0(X_{(i)})\right)}\right)
\right] ;  \label{eq:Z_K_2}\\
Z_A &=
-\sum_{i=1}^{n}
\left[
\frac{\log\left(F_0(X_{(i)})\right)}{n - i + \frac{1}{2}}
+
\frac{\log\left(1 - F_0(X_{(i)})\right)}{i - \frac{1}{2}}
\right] ; \label{eq:Z_A_2} \\
Z_C &=
\sum_{i=1}^{n}
\left[
\log\left(
\frac{F_0(X_{(i)})^{-1} - 1}
{\left\{n - \frac{1}{2}\right\} / \left\{ i - \frac{3}{4} \right\} - 1}
\right)
\right]^2 . \label{eq:Z_C_2}
\end{align}

\subsection{Functions to calculate Zhang distances \textasteriskcentered}

Once again we start with implementing test statistics, this time~(\ref{eq:Z_K_2}), (\ref{eq:Z_A_2}),  (\ref{eq:Z_C_2}), using plain \texttt{R}:

\begin{minted}[mathescape, linenos, texcomments]{r}
distance.Zk <- function(s, F_0, n)  # implementation of (\ref{eq:Z_K_2})
    max((1:n-0.5)*log((1:n-0.5)/(n*F_0(s)))+(n-1:n+0.5)*log((n-1:n+0.5)/(n*(1-F_0(s)))))
distance.Za <- function(s, F_0, n)  # implementation of (\ref{eq:Z_A_2})
    -sum(log(F_0(s))/(n - 1:n + 0.5) + log(1 - F_0(s))/(1:n - 0.5))
distance.Zc <- function(s, F_0, n)  # implementation of (\ref{eq:Z_C_2})
    sum(log((1/F_0(s)-1)/((n-0.5)/(1:n-0.75)-1))^2)
\end{minted}
We note that the functions above expect sorted samples as their first argument.

In this section $H_0$ will be again standard normal:
\begin{minted}[mathescape, linenos]{r}
F_0 <- pnorm
\end{minted}
We can get distribution of each Zhang statistic like this:
\begin{minted}[mathescape, linenos]{r}
distance.distribution.Zk.H0 <- apply(
    sorted.samples.std.normal, 1, function(samp) distance.Zk(samp, F_0, n))
distance.distribution.Za.H0 <- apply(
    sorted.samples.std.normal, 1, function(samp) distance.Za(samp, F_0, n))
distance.distribution.Zc.H0 <- apply(
    sorted.samples.std.normal, 1, function(samp) distance.Zc(samp, F_0, n))
\end{minted}
To calculate this distributions faster, we write a C++ function using \texttt{Rcpp}:
\begin{minted}[mathescape, linenos]{r}
cppFunction("
NumericMatrix ZhangDistances(
    NumericMatrix sorted_samples,     // matrix with sorted samples in rows
    Function F_0                      // cdf for H_0
) {
    int m = sorted_samples.nrow();    // infer number of samples from sorted_samples matrix
    int n = sorted_samples.ncol();    // infer sample size       from sorted_samples matrix

    NumericMatrix out(m, 3);          // allocate memory for the final result

    for (int i = 0; i < m; i++) {   // loop over sorted samples
        NumericVector sorted_sample = sorted_samples(i, _);  // get specific sample
        NumericVector F_0_values = F_0(sorted_sample);  // calculate F_0 for this sample

        double Zk_dist_tmp = 0.0;     // initiate Z_K statistic for sample
        double Za_dist_tmp = 0.0;     // initiate Z_A statistic for sample
        double Zc_dist_tmp = 0.0;     // initiate Z_C statistic for sample

        for (int j = 0; j < n; j++) {  // loop over sample's order stats to calc Z_K, Z_A, Z_C
            double t1 = j + 0.5;
            double t2 = n - j - 0.5;

            double t_Zk = t1 * std::log(t1 / (n *      F_0_values[j])) + 
                          t2 * std::log(t2 / (n * (1 - F_0_values[j])));
            if (t_Zk > Zk_dist_tmp) Zk_dist_tmp = t_Zk;

            Za_dist_tmp += std::log(    F_0_values[j]) / t2 + 
                           std::log(1 - F_0_values[j]) / t1;

            double t_Zc = std::log((1 / F_0_values[j] - 1) / ((n - 0.5) / (j + 0.25) - 1));
            Zc_dist_tmp += t_Zc * t_Zc;
        }

        out(i, 0) =  Zk_dist_tmp;     // 1st res col is for Z_K
        out(i, 1) = -Za_dist_tmp;     // 2nd res col is for Z_A
        out(i, 2) =  Zc_dist_tmp;     // 3rd res col is for Z_C
    }

    return out;
}
")
\end{minted}
We can compare C++ function result 
\begin{minted}[mathescape, linenos]{r}
ZhangDistances(sorted.samples.std.normal, F_0)[1:10,]
\end{minted}
with plain \texttt{R} functions 
\begin{minted}[mathescape, linenos]{r}
distance.distribution.Zk.H0[1:10]
distance.distribution.Za.H0[1:10]
distance.distribution.Zc.H0[1:10]
\end{minted}
The values are the same.

\subsection{Function to calculate rejections for Zhang tests \textasteriskcentered}
Let us write a function to calculate number of rejections for Zhang tests:
\begin{minted}[mathescape, linenos]{r}
calc.rejections.Zhang.tests <- function(
    sample.generation.function.H0,   # single function to generate samples from H0 distribution
    sample.generation.functions.H1,  # list of functions to generate samples for multiple H1s
    F_0,    # cdf for H0 distribution
    alpha,  # significance level
    h,      # how many times we generate sample of samples for H0 and each H1
    n,      # sample size for each sample in sample of samples
    m       # sample of sample size
)
{
    s <- length(sample.generation.functions.H1)  # number of alternatives
    res.Zk <- matrix(0, h, s)  # allocate space for powers of Z_K test
    res.Za <- matrix(0, h, s)  # allocate space for powers of Z_A test
    res.Zc <- matrix(0, h, s)  # allocate space for powers of Z_C test
    for(i in 1:h)  # we generate sample of samples for H0 and each H1 multiple times
    {
        print(paste("iteration", i))
        flush.console()
        
        # generate sorted samples from H0 distribution:
        sorted.samples.H0 <- get.sorted.samples.cpp(sample.generation.function.H0, n, m)
        
        print("calculate bounds")
        flush.console()
        
        # calculate distance distributions between F_0 and H0 ecdfs:
        Zhang.distance.distributions.H0  <- ZhangDistances(sorted.samples.H0, F_0)
        # calculate bounds for each distance distribution:
        bound.Zk.H0 <- quantile(Zhang.distance.distributions.H0[,1], 1 - alpha)
        bound.Za.H0 <- quantile(Zhang.distance.distributions.H0[,2], 1 - alpha)        
        bound.Zc.H0 <- quantile(Zhang.distance.distributions.H0[,3], 1 - alpha)
                                              
        for(j in 1:s){  # loop over alternatives
            print(paste("H1", j))
            flush.console()

            # generate sorted samples from H1 distribution:
            sorted.samples.H1 <- get.sorted.samples.cpp(sample.generation.functions.H1[[j]], n, m)
            # calculate distance distributions between F_0 and H1 ecdfs:
            Zhang.distance.distributions.H1  <- ZhangDistances(sorted.samples.H1, F_0)
            
            # calculate number of rejections for each distance:                                       
            res.Zk[i,j] <- sum(Zhang.distance.distributions.H1[,1] >= bound.Zk.H0) / m
            res.Za[i,j] <- sum(Zhang.distance.distributions.H1[,2] >= bound.Za.H0) / m
            res.Zc[i,j] <- sum(Zhang.distance.distributions.H1[,3] >= bound.Zc.H0) / m
        }
        print(paste("iteration", i, "done"))
        flush.console()
    }
    list(  # matrices with power estimates; each col is a power distribution for H0 vs some H1
        res.Zk = res.Zk,
        res.Za = res.Za,
        res.Zc = res.Zc
    )
}
\end{minted}

\subsection{Calculate rejections for Zhang tests: normal \texorpdfstring{$H_0$}{H0}, normal \texorpdfstring{$H_1$}{H1}}

This subsection is like subsection \ref{section:rejections:normal:pc} and subsection \ref{section:rejections:normal:normal:classic}, but for Zhang tests. We investigate the power of $Z_K$, $Z_A$, and $Z_C$ tests for ``standard normal null vs other normal alternatives'' case:

\begin{minted}[mathescape, linenos]{r}
res <- calc.rejections.Zhang.tests(
    sample.generation.function.H0  = rnorm,
    sample.generation.functions.H1 = c(
        lapply(c(0.05, 0.1, 0.15), function(sd)   function(nsamples) rnorm(nsamples, 0,    sd)), 
        lapply(seq(0.2, 0.9, 0.1), function(sd)   function(nsamples) rnorm(nsamples, 0,    sd)),
        rnorm,
        lapply(seq(1.1, 2.4, 0.1), function(sd)   function(nsamples) rnorm(nsamples, 0,    sd)),
        lapply(seq(2.5,   4, 0.5), function(sd)   function(nsamples) rnorm(nsamples, 0,    sd)),
        lapply(seq(0.1,   2, 0.1), function(mean) function(nsamples) rnorm(nsamples, mean, 1 ))
    ),
    F_0   = pnorm,
    alpha = 0.05,
    h     = 30,
    n = n, m = m
)
\end{minted}
Here is the table creation function for Zhang tests:
\begin{minted}[mathescape, linenos]{r}
create.rejection.table.Zhang <- function(res.rejections)
    list(
        Zk.means = round(apply(res.rejections$res.Zk, 2, mean), 4),
        Za.means = round(apply(res.rejections$res.Za, 2, mean), 4),
        Zc.means = round(apply(res.rejections$res.Zc, 2, mean), 4),
        Zk.sds = formatC(apply(res.rejections$res.Zk, 2, sd  ), format = "e", digits = 1),
        Za.sds = formatC(apply(res.rejections$res.Za, 2, sd  ), format = "e", digits = 1),
        Zc.sds = formatC(apply(res.rejections$res.Zc, 2, sd  ), format = "e", digits = 1)
    )
\end{minted}

\noindent
The table for $H_0 : N(0,1)$ vs $H_1 : N(0, \sigma^2)$ is as follows:
\begin{minted}[mathescape, linenos]{r}
create.rejection.table.Zhang(res)
\end{minted}

\begin{table}[H]
\centering
\begin{tabular}{ |l|l||>{\columncolor[gray]{0.9}}p{1cm}|p{1.25cm}||>{\columncolor[gray]{0.9}}p{1cm}|p{1.25cm}||>{\columncolor[gray]{0.9}}p{1cm}|p{1.25cm}| } 
 \hline
   $H_0$ & $H_1$ & \multicolumn{2}{c||}{$Z_K$} & \multicolumn{2}{c||}{$Z_A$} & \multicolumn{2}{c|}{$Z_C$} \\
 \hline
 &  & mean & sd & mean & sd & mean & sd \\
 \hline
 \hline
 
 $\mathcal{N}(0, 1)$ & $\mathrm{N}(0,0.05^2)$ & 1 & 0.0e+00 & 1 & 0.0e+00 & 1 & 0.0e+00\\
$\mathcal{N}(0, 1)$ & $\mathrm{N}(0,0.1^2)$ & 0.9978 & 7.2e-05 & 1 & 0.0e+00 & 1 & 0.0e+00\\
$\mathcal{N}(0, 1)$ & $\mathrm{N}(0,0.15^2)$ & 0.8811 & 1.6e-03 & 1 & 0.0e+00 & 1 & 1.2e-06\\
$\mathcal{N}(0, 1)$ & $\mathrm{N}(0,0.2^2)$ & 0.5941 & 2.5e-03 & 0.9998 & 1.3e-05 & 0.9974 & 1.0e-04\\
$\mathcal{N}(0, 1)$ & $\mathrm{N}(0,0.3^2)$ & 0.2205 & 1.4e-03 & 0.9085 & 9.2e-04 & 0.763 & 2.8e-03\\
$\mathcal{N}(0, 1)$ & $\mathrm{N}(0,0.4^2)$ & 0.0991 & 6.1e-04 & 0.5462 & 1.7e-03 & 0.3365 & 2.5e-03\\
$\mathcal{N}(0, 1)$ & $\mathrm{N}(0,0.5^2)$ & 0.0563 & 4.1e-04 & 0.2569 & 1.1e-03 & 0.1264 & 1.2e-03\\
$\mathcal{N}(0, 1)$ & $\mathrm{N}(0,0.6^2)$ & 0.0386 & 3.1e-04 & 0.1227 & 6.4e-04 & 0.0525 & 5.7e-04\\
$\mathcal{N}(0, 1)$ & $\mathrm{N}(0,0.7^2)$ & 0.031 & 2.3e-04 & 0.0688 & 4.2e-04 & 0.0279 & 3.3e-04\\
$\mathcal{N}(0, 1)$ & $\mathrm{N}(0,0.8^2)$ & 0.0295 & 1.9e-04 & 0.0485 & 2.5e-04 & 0.022 & 2.4e-04\\
$\mathcal{N}(0, 1)$ & $\mathrm{N}(0,0.9^2)$ & 0.0347 & 2.5e-04 & 0.0438 & 2.6e-04 & 0.0282 & 2.5e-04\\
\hline
$\mathcal{N}(0, 1)$ & $\mathrm{N}(0,1^2)$ & 0.05 & 2.8e-04 & 0.05 & 2.9e-04 & 0.05 & 3.2e-04\\
\hline
$\mathcal{N}(0, 1)$ & $\mathrm{N}(0,1.1^2)$ & 0.08 & 3.6e-04 & 0.0699 & 3.6e-04 & 0.0944 & 3.9e-04\\
$\mathcal{N}(0, 1)$ & $\mathrm{N}(0,1.2^2)$ & 0.128 & 4.9e-04 & 0.1093 & 4.6e-04 & 0.165 & 5.0e-04\\
$\mathcal{N}(0, 1)$ & $\mathrm{N}(0,1.3^2)$ & 0.1939 & 5.7e-04 & 0.1718 & 6.0e-04 & 0.258 & 6.2e-04\\
$\mathcal{N}(0, 1)$ & $\mathrm{N}(0,1.4^2)$ & 0.2738 & 6.4e-04 & 0.2546 & 7.8e-04 & 0.3639 & 6.0e-04\\
$\mathcal{N}(0, 1)$ & $\mathrm{N}(0,1.5^2)$ & 0.3622 & 8.3e-04 & 0.351 & 8.2e-04 & 0.4723 & 7.0e-04\\
$\mathcal{N}(0, 1)$ & $\mathrm{N}(0,1.6^2)$ & 0.453 & 7.5e-04 & 0.4516 & 9.6e-04 & 0.5739 & 7.6e-04\\
$\mathcal{N}(0, 1)$ & $\mathrm{N}(0,1.7^2)$ & 0.5401 & 6.6e-04 & 0.5479 & 9.1e-04 & 0.6631 & 6.4e-04\\
$\mathcal{N}(0, 1)$ & $\mathrm{N}(0,1.8^2)$ & 0.6201 & 7.1e-04 & 0.6349 & 7.6e-04 & 0.7382 & 4.7e-04\\
$\mathcal{N}(0, 1)$ & $\mathrm{N}(0,1.9^2)$ & 0.6905 & 5.1e-04 & 0.7098 & 6.0e-04 & 0.7986 & 5.8e-04\\
$\mathcal{N}(0, 1)$ & $\mathrm{N}(0,2^2)$ & 0.7508 & 5.7e-04 & 0.7721 & 6.3e-04 & 0.8467 & 4.9e-04\\
$\mathcal{N}(0, 1)$ & $\mathrm{N}(0,2.1^2)$ & 0.801 & 5.3e-04 & 0.8225 & 5.6e-04 & 0.8836 & 3.9e-04\\
$\mathcal{N}(0, 1)$ & $\mathrm{N}(0,2.2^2)$ & 0.8423 & 4.7e-04 & 0.8629 & 5.0e-04 & 0.9122 & 3.0e-04\\
$\mathcal{N}(0, 1)$ & $\mathrm{N}(0,2.3^2)$ & 0.8754 & 4.6e-04 & 0.8941 & 4.2e-04 & 0.9336 & 2.6e-04\\
$\mathcal{N}(0, 1)$ & $\mathrm{N}(0,2.4^2)$ & 0.902 & 3.8e-04 & 0.9185 & 2.8e-04 & 0.9499 & 2.3e-04\\
$\mathcal{N}(0, 1)$ & $\mathrm{N}(0,2.5^2)$ & 0.9229 & 3.2e-04 & 0.9373 & 2.7e-04 & 0.9621 & 2.2e-04\\
$\mathcal{N}(0, 1)$ & $\mathrm{N}(0,3^2)$ & 0.9769 & 1.8e-04 & 0.9828 & 1.5e-04 & 0.9902 & 8.6e-05\\
$\mathcal{N}(0, 1)$ & $\mathrm{N}(0,3.5^2)$ & 0.9927 & 7.4e-05 & 0.9949 & 6.3e-05 & 0.9972 & 4.7e-05\\
$\mathcal{N}(0, 1)$ & $\mathrm{N}(0,4^2)$ & 0.9975 & 4.6e-05 & 0.9983 & 4.4e-05 & 0.9991 & 3.3e-05\\

      \hline
\end{tabular}
\caption{Zhang tests. $H_0$ --- standard normal, $H_1$ --- other normals that differ in scale.}
\label{table:H0stdnormal:H1othernormals_scale:zhang}
\end{table}

\noindent
Table~\ref{table:H0stdnormal:H1othernormals_scale:zhang} should be compared with tables~\ref{table:H0stdnormH1normscale} and~\ref{table:H0stdnormal:H1othernormals_scale:classical}. 
We see that for $\sigma < 1$ $Z_C$ and $Z_A$ tests perform pretty well, much better than classical tests from table~\ref{table:H0stdnormal:H1othernormals_scale:classical}, but not as well as PC2 test from table~\ref{table:H0stdnormH1normscale}. $Z_A$ is better than $Z_C$, $Z_K$ is behind $Z_C$. 
For $\sigma > 1$ $Z_C$ is universally better than $Z_A$, $Z_K$ is slightly better than $Z_A$ for smaller values of $\sigma$ and slightly worse for larger values of $\sigma$. $Z_C$ and PC2 tests are very close to each other when $\sigma > 1$. Overall, we can conclude that PC2 test is the best for checking  normal scale when compared to classical and Zhang tests. We note that in this particular experiment $n = 10$. We will check larger sample sizes later.

The table for $H_0 : N(0,1)$ vs $H_1 : N(\mu, 1)$ is as follows:

\begin{table}[H]
\centering
\begin{tabular}{ |l|l||>{\columncolor[gray]{0.9}}p{1cm}|p{1.1cm}||>{\columncolor[gray]{0.9}}p{1cm}|p{1.1cm}||>{\columncolor[gray]{0.9}}p{1cm}|p{1.1cm}| } 
 \hline
   $H_0$ & $H_1$ & \multicolumn{2}{c||}{$Z_K$} & \multicolumn{2}{c||}{$Z_A$} & \multicolumn{2}{c|}{$Z_C$} \\
 \hline
 &  & mean & sd & mean & sd & mean & sd \\
 \hline
 \hline
 
 $\mathcal{N}(0, 1)$ & $\mathrm{N}(0.1, 1)$ & 0.0586 & 4.0e-04 & 0.0597 & 3.5e-04 & 0.0591 & 3.7e-04\\
$\mathcal{N}(0, 1)$ & $\mathrm{N}(0.2, 1)$ & 0.0849 & 3.9e-04 & 0.0892 & 3.6e-04 & 0.0872 & 4.1e-04\\
$\mathcal{N}(0, 1)$ & $\mathrm{N}(0.3, 1)$ & 0.1302 & 4.8e-04 & 0.1405 & 5.5e-04 & 0.1364 & 5.7e-04\\
$\mathcal{N}(0, 1)$ & $\mathrm{N}(0.4, 1)$ & 0.196 & 5.7e-04 & 0.2148 & 6.1e-04 & 0.2085 & 7.2e-04\\
$\mathcal{N}(0, 1)$ & $\mathrm{N}(0.5, 1)$ & 0.2809 & 7.1e-04 & 0.3105 & 7.3e-04 & 0.3026 & 8.1e-04\\
$\mathcal{N}(0, 1)$ & $\mathrm{N}(0.6, 1)$ & 0.382 & 7.5e-04 & 0.4229 & 7.6e-04 & 0.4142 & 1.0e-03\\
$\mathcal{N}(0, 1)$ & $\mathrm{N}(0.7, 1)$ & 0.4925 & 8.6e-04 & 0.5423 & 7.0e-04 & 0.5339 & 7.7e-04\\
$\mathcal{N}(0, 1)$ & $\mathrm{N}(0.8, 1)$ & 0.6043 & 7.6e-04 & 0.6588 & 7.1e-04 & 0.6516 & 7.9e-04\\
$\mathcal{N}(0, 1)$ & $\mathrm{N}(0.9, 1)$ & 0.7084 & 8.2e-04 & 0.762 & 6.5e-04 & 0.7565 & 8.0e-04\\
$\mathcal{N}(0, 1)$ & $\mathrm{N}(1, 1)$ & 0.7979 & 5.5e-04 & 0.8456 & 5.0e-04 & 0.8419 & 4.9e-04\\
$\mathcal{N}(0, 1)$ & $\mathrm{N}(1.1, 1)$ & 0.869 & 3.9e-04 & 0.9072 & 3.7e-04 & 0.9049 & 4.2e-04\\
$\mathcal{N}(0, 1)$ & $\mathrm{N}(1.2, 1)$ & 0.9207 & 3.8e-04 & 0.9485 & 3.2e-04 & 0.9473 & 3.4e-04\\
$\mathcal{N}(0, 1)$ & $\mathrm{N}(1.3, 1)$ & 0.9554 & 2.6e-04 & 0.9737 & 1.6e-04 & 0.9731 & 1.5e-04\\
$\mathcal{N}(0, 1)$ & $\mathrm{N}(1.4, 1)$ & 0.9767 & 2.2e-04 & 0.9877 & 1.5e-04 & 0.9874 & 1.6e-04\\
$\mathcal{N}(0, 1)$ & $\mathrm{N}(1.5, 1)$ & 0.9887 & 1.4e-04 & 0.9947 & 7.2e-05 & 0.9946 & 6.6e-05\\
$\mathcal{N}(0, 1)$ & $\mathrm{N}(1.6, 1)$ & 0.9949 & 7.3e-05 & 0.9979 & 4.6e-05 & 0.9979 & 4.4e-05\\
$\mathcal{N}(0, 1)$ & $\mathrm{N}(1.7, 1)$ & 0.9979 & 3.9e-05 & 0.9993 & 2.2e-05 & 0.9992 & 2.3e-05\\
$\mathcal{N}(0, 1)$ & $\mathrm{N}(1.8, 1)$ & 0.9992 & 2.8e-05 & 0.9998 & 1.7e-05 & 0.9998 & 1.6e-05\\
$\mathcal{N}(0, 1)$ & $\mathrm{N}(1.9, 1)$ & 0.9997 & 1.6e-05 & 0.9999 & 7.8e-06 & 0.9999 & 8.6e-06\\
$\mathcal{N}(0, 1)$ & $\mathrm{N}(2, 1)$ & 0.9999 & 1.0e-05 & 1 & 5.4e-06 & 1 & 5.1e-06\\
 
       \hline
\end{tabular}
\caption{Zhang tests. $H_0$ --- standard normal, $H_1$ --- other normals that differ in location.}
\label{table:H0stdnormal:H1othernormals_location:zhang}
\end{table}

\noindent
Table~\ref{table:H0stdnormal:H1othernormals_location:zhang} should be compared to tables~\ref{table:H0stdnormH1normlocation} and~\ref{table:H0stdnormal:H1othernormals_location:classical}. We see that $Z_A$ and $Z_C$ are slightly better at detecting location shift than PC2 test, $Z_K$ is worse than PC2. CvM and AD are better than $Z_A$ and $Z_C$. We can conclude that AD is the best among checked tests for normal location shifts.

\subsection{Calculate rejections for Zhang tests: normal \texorpdfstring{$H_0$}{H0},  Cauchy \texorpdfstring{$H_1$}{H1}}

This subsection is like subsections~\ref{section:pc:normal:cauchy} and~\ref{section:classical_tests:cauchy}, but for Zhang tests:

\begin{minted}[mathescape, linenos]{r}
res.cauchy.Zhang <- calc.rejections.Zhang.tests(
    sample.generation.function.H0  = rnorm,
    sample.generation.functions.H1 = c(
        lapply(
            seq(0.01, 0.09, 0.01), 
            function(scale) 
                function(nsamples) 
                    rcauchy(nsamples, 0, scale)
        ),
        lapply(
            seq(0.1,  2,    0.1 ), 
            function(scale) 
                function(nsamples) 
                    rcauchy(nsamples, 0, scale)
        )
    ),
    F_0   = pnorm,
    alpha = 0.05,
    h     = 30,
    n = n, m = m
)
\end{minted}

\begin{table}[H]
\centering
\begin{tabular}{ |l|l||>{\columncolor[gray]{0.9}}p{1cm}|p{1.1cm}||>{\columncolor[gray]{0.9}}p{1cm}|p{1.1cm}||>{\columncolor[gray]{0.9}}p{1cm}|p{1.1cm}| } 
 \hline
   $H_0$ & $H_1$ & \multicolumn{2}{c||}{$Z_K$} & \multicolumn{2}{c||}{$Z_A$} & \multicolumn{2}{c|}{$Z_C$} \\
 \hline
 &  & mean & sd & mean & sd & mean & sd \\
 \hline
 \hline
 
 $\mathcal{N}(0, 1)$ & $\mathrm{Cauchy}(0, 0.01)$ & 0.9891 & 1.2e-04 & 0.9987 & 4.2e-05 & 0.9967 & 7.3e-05\\
$\mathcal{N}(0, 1)$ & $\mathrm{Cauchy}(0, 0.02)$ & 0.961 & 2.8e-04 & 0.9941 & 8.4e-05 & 0.9848 & 2.1e-04\\
$\mathcal{N}(0, 1)$ & $\mathrm{Cauchy}(0, 0.03)$ & 0.9215 & 4.8e-04 & 0.9857 & 1.4e-04 & 0.9639 & 4.3e-04\\
$\mathcal{N}(0, 1)$ & $\mathrm{Cauchy}(0, 0.04)$ & 0.875 & 6.5e-04 & 0.9734 & 1.9e-04 & 0.9358 & 5.0e-04\\
$\mathcal{N}(0, 1)$ & $\mathrm{Cauchy}(0, 0.05)$ & 0.8254 & 9.3e-04 & 0.9576 & 3.1e-04 & 0.9029 & 8.0e-04\\
$\mathcal{N}(0, 1)$ & $\mathrm{Cauchy}(0, 0.06)$ & 0.7754 & 1.0e-03 & 0.9387 & 3.7e-04 & 0.8675 & 9.6e-04\\
$\mathcal{N}(0, 1)$ & $\mathrm{Cauchy}(0, 0.07)$ & 0.7266 & 1.1e-03 & 0.9171 & 4.5e-04 & 0.831 & 1.1e-03\\
$\mathcal{N}(0, 1)$ & $\mathrm{Cauchy}(0, 0.08)$ & 0.6809 & 1.3e-03 & 0.8936 & 5.5e-04 & 0.7949 & 1.3e-03\\
$\mathcal{N}(0, 1)$ & $\mathrm{Cauchy}(0, 0.09)$ & 0.6388 & 1.5e-03 & 0.8688 & 6.0e-04 & 0.7602 & 1.2e-03\\
$\mathcal{N}(0, 1)$ & $\mathrm{Cauchy}(0, 0.1)$ & 0.6011 & 1.2e-03 & 0.8433 & 6.8e-04 & 0.7273 & 1.3e-03\\
$\mathcal{N}(0, 1)$ & $\mathrm{Cauchy}(0, 0.2)$ & 0.4293 & 8.0e-04 & 0.6297 & 8.9e-04 & 0.5388 & 1.1e-03\\
$\mathcal{N}(0, 1)$ & $\mathrm{Cauchy}(0, 0.3)$ & 0.454 & 7.3e-04 & 0.5609 & 7.7e-04 & 0.5376 & 8.3e-04\\
$\mathcal{N}(0, 1)$ & $\mathrm{Cauchy}(0, 0.4)$ & 0.5251 & 5.4e-04 & 0.5812 & 6.3e-04 & 0.6007 & 6.2e-04\\
$\mathcal{N}(0, 1)$ & $\mathrm{Cauchy}(0, 0.5)$ & 0.6016 & 5.7e-04 & 0.6339 & 6.4e-04 & 0.6735 & 5.7e-04\\
$\mathcal{N}(0, 1)$ & $\mathrm{Cauchy}(0, 0.6)$ & 0.6717 & 5.7e-04 & 0.6925 & 5.4e-04 & 0.739 & 4.7e-04\\
$\mathcal{N}(0, 1)$ & $\mathrm{Cauchy}(0, 0.7)$ & 0.7327 & 4.6e-04 & 0.7474 & 3.9e-04 & 0.794 & 3.5e-04\\
$\mathcal{N}(0, 1)$ & $\mathrm{Cauchy}(0, 0.8)$ & 0.7847 & 3.8e-04 & 0.7958 & 3.9e-04 & 0.839 & 2.9e-04\\
$\mathcal{N}(0, 1)$ & $\mathrm{Cauchy}(0, 0.9)$ & 0.8279 & 4.7e-04 & 0.8368 & 4.8e-04 & 0.8751 & 4.4e-04\\
$\mathcal{N}(0, 1)$ & $\mathrm{Cauchy}(0, 1)$ & 0.8632 & 3.2e-04 & 0.8708 & 3.4e-04 & 0.9036 & 2.6e-04\\
$\mathcal{N}(0, 1)$ & $\mathrm{Cauchy}(0, 1.1)$ & 0.892 & 3.2e-04 & 0.8985 & 3.0e-04 & 0.926 & 2.7e-04\\
$\mathcal{N}(0, 1)$ & $\mathrm{Cauchy}(0, 1.2)$ & 0.9151 & 2.9e-04 & 0.9207 & 2.6e-04 & 0.9434 & 2.2e-04\\
$\mathcal{N}(0, 1)$ & $\mathrm{Cauchy}(0, 1.3)$ & 0.9334 & 3.0e-04 & 0.9383 & 2.9e-04 & 0.9568 & 2.1e-04\\
$\mathcal{N}(0, 1)$ & $\mathrm{Cauchy}(0, 1.4)$ & 0.9479 & 2.6e-04 & 0.9522 & 2.4e-04 & 0.9671 & 1.7e-04\\
$\mathcal{N}(0, 1)$ & $\mathrm{Cauchy}(0, 1.5)$ & 0.9593 & 2.1e-04 & 0.963 & 2.0e-04 & 0.9749 & 1.8e-04\\
$\mathcal{N}(0, 1)$ & $\mathrm{Cauchy}(0, 1.6)$ & 0.9682 & 1.6e-04 & 0.9714 & 1.6e-04 & 0.981 & 1.3e-04\\
$\mathcal{N}(0, 1)$ & $\mathrm{Cauchy}(0, 1.7)$ & 0.9752 & 2.1e-04 & 0.9779 & 1.9e-04 & 0.9855 & 1.5e-04\\
$\mathcal{N}(0, 1)$ & $\mathrm{Cauchy}(0, 1.8)$ & 0.9806 & 1.1e-04 & 0.983 & 1.1e-04 & 0.9889 & 9.4e-05\\
$\mathcal{N}(0, 1)$ & $\mathrm{Cauchy}(0, 1.9)$ & 0.9849 & 1.5e-04 & 0.9869 & 1.2e-04 & 0.9916 & 9.6e-05\\
$\mathcal{N}(0, 1)$ & $\mathrm{Cauchy}(0, 2)$ & 0.9882 & 1.0e-04 & 0.9898 & 8.7e-05 & 0.9936 & 7.9e-05\\
 
       \hline
\end{tabular}
\caption{Zhang tests. $H_0$ --- standard normal, $H_1$ --- Cauchy with different scales.}
\label{table:H0stdnormal:H1cauchy:zhang}
\end{table}

\noindent
Table~\ref{table:H0stdnormal:H1cauchy:zhang} should be compared with tables~\ref{table:H0stdnormH1cauchyscale} and~\ref{table:H0stdnormal:H1cauchy_scale:classical}. We see that $Z_A$ and $Z_C$ are universally better than $Z_K$. Both $Z_A$ and $Z_C$ have local minima at $0.3$, $0.5609$ and $0.5376$ respectively. PC2 test also has minimum at 0.3, $0.5983$. For scale $< 0.3$ PC2 is slightly better than $Z_C$, significantly better than $Z_K$, and slightly worse than $Z_A$ for scale $\leq 0.1$. For scale $> 0.3$ PC2 is slightly better than $Z_A$, better than $Z_K$ and slightly worse than $Z_C$ for scale $\geq 0.5$. All in all, PC2 is very close to $Z_A$ and $Z_C$ with higher minimum power at tested parameter range. Among classical tests AD is the most powerful with local minimum of $0.2645$ at $0.2$. Around this minimum it is significantly behind $Z_K$, $Z_A$, $Z_C$ and PC2. 

\subsection{Calculate rejections for Zhang tests: normal \texorpdfstring{$H_0$}{H0}, gamma \texorpdfstring{$H_1$}{H1}}

This subsection is like subsections \ref{section:rejections:pc:gamma} and \ref{section:rejections:classic:normal:gamma}, but for Zhang tests:

\begin{minted}[mathescape, linenos]{r}
res.gamma.Zhang <- calc.rejections.Zhang.tests(
    sample.generation.function.H0  = rnorm,
    sample.generation.functions.H1 = 
        lapply(
            c(seq(0.1, 0.9, 0.1), 1:5, 10, 100, 1000), 
            function(shape) 
                function(nsamples) 
                    (rgamma(nsamples, shape = shape, rate = 1) - shape) / sqrt(shape)
        ),
    F_0   = pnorm,
    alpha = 0.05,
    h     = 30,
    n = n, m = m
)
\end{minted}

\begin{table}[H]
\centering
\begin{tabular}{ |l|l||>{\columncolor[gray]{0.9}}p{1cm}|p{1.1cm}||>{\columncolor[gray]{0.9}}p{1cm}|p{1.1cm}||>{\columncolor[gray]{0.9}}p{1cm}|p{1.1cm}| } 
 \hline
   $H_0$ & $H_1$ & \multicolumn{2}{c||}{$Z_K$} & \multicolumn{2}{c||}{$Z_A$} & \multicolumn{2}{c|}{$Z_C$} \\
 \hline
 &  & mean & sd & mean & sd & mean & sd \\
 \hline
 \hline
 
 $\mathcal{N}(0, 1)$ & $\mathrm{G}(0.1)$ & 0.5422 & 7.4e-04 & 0.9307 & 7.2e-04 & 0.7915 & 1.2e-03\\
$\mathcal{N}(0, 1)$ & $\mathrm{G}(0.2)$ & 0.3956 & 9.9e-04 & 0.6975 & 1.1e-03 & 0.6033 & 1.4e-03\\
$\mathcal{N}(0, 1)$ & $\mathrm{G}(0.3)$ & 0.3235 & 8.3e-04 & 0.5296 & 1.1e-03 & 0.4881 & 1.1e-03\\
$\mathcal{N}(0, 1)$ & $\mathrm{G}(0.4)$ & 0.277 & 8.2e-04 & 0.4211 & 1.2e-03 & 0.4132 & 1.2e-03\\
$\mathcal{N}(0, 1)$ & $\mathrm{G}(0.5)$ & 0.2457 & 7.0e-04 & 0.349 & 8.7e-04 & 0.3612 & 1.1e-03\\
$\mathcal{N}(0, 1)$ & $\mathrm{G}(0.6)$ & 0.2233 & 6.2e-04 & 0.2991 & 7.0e-04 & 0.3235 & 9.1e-04\\
$\mathcal{N}(0, 1)$ & $\mathrm{G}(0.7)$ & 0.2061 & 6.4e-04 & 0.2627 & 6.6e-04 & 0.2948 & 8.0e-04\\
$\mathcal{N}(0, 1)$ & $\mathrm{G}(0.8)$ & 0.192 & 5.9e-04 & 0.2352 & 6.5e-04 & 0.2722 & 7.0e-04\\
$\mathcal{N}(0, 1)$ & $\mathrm{G}(0.9)$ & 0.1808 & 5.7e-04 & 0.2139 & 6.0e-04 & 0.2541 & 6.9e-04\\
$\mathcal{N}(0, 1)$ & $\mathrm{G}(1)$ & 0.1715 & 6.7e-04 & 0.1973 & 6.6e-04 & 0.2393 & 8.9e-04\\
$\mathcal{N}(0, 1)$ & $\mathrm{G}(2)$ & 0.1222 & 4.9e-04 & 0.1233 & 4.6e-04 & 0.1652 & 5.0e-04\\
$\mathcal{N}(0, 1)$ & $\mathrm{G}(3)$ & 0.1021 & 5.0e-04 & 0.0992 & 4.7e-04 & 0.1359 & 4.7e-04\\
$\mathcal{N}(0, 1)$ & $\mathrm{G}(4)$ & 0.091 & 4.9e-04 & 0.0872 & 4.2e-04 & 0.1194 & 4.6e-04\\
$\mathcal{N}(0, 1)$ & $\mathrm{G}(5)$ & 0.0838 & 4.5e-04 & 0.08 & 3.5e-04 & 0.1086 & 3.8e-04\\
$\mathcal{N}(0, 1)$ & $\mathrm{G}(10)$ & 0.0682 & 3.4e-04 & 0.0654 & 2.6e-04 & 0.0836 & 3.5e-04\\
$\mathcal{N}(0, 1)$ & $\mathrm{G}(100)$ & 0.0519 & 2.8e-04 & 0.0514 & 2.5e-04 & 0.0538 & 3.4e-04\\
$\mathcal{N}(0, 1)$ & $\mathrm{G}(1000)$ & 0.0501 & 2.5e-04 & 0.05 & 3.0e-04 & 0.0503 & 3.1e-04\\

      \hline
\end{tabular}
\caption{Zhang tests. $H_0$ --- standard normal, $H_1$ --- gamma with different skews.}
\label{table:H0stdnormal:H1gamma:zhang}
\end{table}

\noindent
Table \ref{table:H0stdnormal:H1gamma:zhang} should be compared with tables \ref{table:H0stdnormH1gamma} and \ref{table:H0stdnormal:H1gamma_scale:classical}. We see that $Z_A$, $Z_C$ and PC2 are better than $Z_K$. $Z_A$ is better than $Z_C$ for $a \leq 0.4$ and worse than $Z_C$ for $a > 0.4$. PC2 is worse than $Z_A$ only for $a = 0.1$, for $a > 0.1$ PC2 is better than $Z_A$. PC2 is better than $Z_C$ for $a \leq 1$, and worse than $Z_C$ for $a > 1$. All in all, there is no clear winner here since $Z_A$, $Z_C$ and PC2 excel at different values of parameter $a$. $Z_K$ is again behind other Zhang tests and PC2 test. It has power comparable to KS and AD tests.

\subsection{Calculate rejections for Zhang tests: normal \texorpdfstring{$H_0$}{H0}, Student \texorpdfstring{$H_1$}{H1}}

This subsection is like subsections \ref{section:rejections:pc:normal:student} and \ref{section:rejections:classical:normal:student}, but for Zhang tests:

\begin{minted}[mathescape, linenos]{r}
res.t.Zhang <- calc.rejections.Zhang.tests(
    sample.generation.function.H0  = rnorm,
    sample.generation.functions.H1 = 
        lapply(seq(1,5), function(df) function(nsamples) rt(nsamples, df = df)),
    F_0   = pnorm,
    alpha = 0.05,
    h     = 30,
    n = n, m = m
)
\end{minted}

\begin{table}[H]
\centering
\begin{tabular}{ |l|l||>{\columncolor[gray]{0.9}}p{1cm}|p{1.1cm}||>{\columncolor[gray]{0.9}}p{1cm}|p{1.1cm}||>{\columncolor[gray]{0.9}}p{1cm}|p{1.1cm}| } 
 \hline
   $H_0$ & $H_1$ & \multicolumn{2}{c||}{$Z_K$} & \multicolumn{2}{c||}{$Z_A$} & \multicolumn{2}{c|}{$Z_C$} \\
 \hline
 &  & mean & sd & mean & sd & mean & sd \\
 \hline
 \hline
 
$\mathcal{N}(0, 1)$ & $\mathrm{t}(1)$ & 0.8632 & 4.1e-04 & 0.8708 & 3.0e-04 & 0.9035 & 2.4e-04\\
$\mathcal{N}(0, 1)$ & $\mathrm{t}(2)$ & 0.5461 & 6.9e-04 & 0.5605 & 6.9e-04 & 0.6452 & 6.3e-04\\
$\mathcal{N}(0, 1)$ & $\mathrm{t}(3)$ & 0.3538 & 6.7e-04 & 0.3667 & 5.7e-04 & 0.462 & 6.1e-04\\
$\mathcal{N}(0, 1)$ & $\mathrm{t}(4)$ & 0.2503 & 6.3e-04 & 0.2605 & 6.0e-04 & 0.35 & 6.4e-04\\
$\mathcal{N}(0, 1)$ & $\mathrm{t}(5)$ & 0.1915 & 5.0e-04 & 0.1993 & 5.6e-04 & 0.2794 & 5.5e-04\\
 
       \hline
\end{tabular}
\caption{Zhang tests. $H_0$ --- standard normal, $H_1$ --- Student with different degrees of freedom.}
\label{table:H0stdnormal:H1student:zhang}
\end{table}

\noindent
Table \ref{table:H0stdnormal:H1student:zhang} should be compared with tables \ref{table:H0stdnormH1student} and \ref{table:H0stdnormal:H1student:classical}. PC2 test is better than $Z_K$ and $Z_A$, and worse than $Z_C$. All Zhang tests are better than AD, the best performing test among classical tests.

\subsection{Asymptotics check for Zhang tests, \texorpdfstring{$n = 50$}{n=50}}

This subsection is like subsections \ref{section:asymptotics:n50} and \ref{section:asymptotics:classical:n50}, but for Zhang tests:

\begin{minted}[mathescape, linenos]{r}
res.n50.Zhang <- calc.rejections.Zhang.tests(
    sample.generation.function.H0  = rnorm,
    sample.generation.functions.H1 = 
        list(
            rnorm,
            function(nsamples) runif(nsamples, -sqrt(3), sqrt(3)),
            function(nsamples) rnorm(nsamples, 0.2, 1),
            function(nsamples) rnorm(nsamples, 0, 0.8),
            function(nsamples) rnorm(nsamples, 0, 1.2),
            function(nsamples) rt(nsamples, df = 5),
            function(nsamples, shape = 5) 
                (rgamma(nsamples, shape = shape, rate = 1) - shape) / sqrt(shape)
        ),
    F_0   = pnorm,
    alpha = 0.05,
    h     = 30,
    n = 50, m = m
)
\end{minted}

\begin{table}[H]
\centering
\begin{tabular}{ |l|l||>{\columncolor[gray]{0.9}}p{1cm}|p{1.1cm}||>{\columncolor[gray]{0.9}}p{1cm}|p{1.1cm}||>{\columncolor[gray]{0.9}}p{1cm}|p{1.1cm}| } 
 \hline
   $H_0$ & $H_1$ & \multicolumn{2}{c||}{$Z_K$} & \multicolumn{2}{c||}{$Z_A$} & \multicolumn{2}{c|}{$Z_C$} \\
 \hline
 &  & mean & sd & mean & sd & mean & sd \\
 \hline
 \hline
 
$\mathcal{N}(0, 1)$ & $\mathcal{N}(0, 1)$ & 0.05 & 3.2e-04 & 0.05 & 2.9e-04 & 0.0499 & 3.1e-04\\
$\mathcal{N}(0, 1)$ & $\mathrm{U}(-\sqrt{3}, \sqrt{3})$ & 0.1513 & 6.3e-04 & 0.1033 & 8.9e-04 & 0.1236 & 1.0e-03\\
$\mathcal{N}(0, 1)$ & $\mathcal{N}(0.2, 1)$ & 0.2156 & 6.1e-04 & 0.2242 & 6.9e-04 & 0.2235 & 6.8e-04\\
$\mathcal{N}(0, 1)$ & $\mathcal{N}(0, 0.8)$ & 0.1581 & 6.5e-04 & 0.3607 & 9.3e-04 & 0.2564 & 1.1e-03\\
$\mathcal{N}(0, 1)$ & $\mathcal{N}(0, 1.2)$ & 0.2777 & 8.5e-04 & 0.2279 & 8.7e-04 & 0.3428 & 7.2e-04\\
$\mathcal{N}(0, 1)$ & t(5) & 0.4751 & 8.1e-04 & 0.5067 & 8.9e-04 & 0.6467 & 6.3e-04\\
$\mathcal{N}(0, 1)$ & $\mathrm{G}(5)$ & 0.1936 & 7.1e-04 & 0.3034 & 1.0e-03 & 0.3263 & 8.8e-04\\
 
        \hline
\end{tabular}
\caption{Zhang tests asymptotics, $n = 50$.}
\label{table:zhang:asymptotics:n50}
\end{table}

\noindent
Table \ref{table:zhang:asymptotics:n50} should be compared to tables \ref{table:H0stdnormal:n50} and \ref{table:H0stdnormal:n50:classic}. We see that:
\begin{enumerate}
\item for uniform alternative $Z_K$ is the best among Zhang tests, and is much (4 times) worse than PC2 test;
\item for normal location shift Zhang tests are behind CvM and AD, but slightly better than PC2;
\item for normal scale, PC2 is slightly better than $Z_C$ and significantly better than $Z_K$; for scales $> 1$ PC2 is better than $Z_A$ and for scales $< 1$ PC2 is worse;
\item for Student alternative PC2 is better than $Z_A$ and $Z_K$, and worse than $Z_C$;   
\item for gamma alternative PC2 is better than all Zhang tests. 
\end{enumerate}

\subsection{Asymptotics check for Zhang tests, \texorpdfstring{$n = 100$}{n=100}}

Now to the same alternatives for larger samples:

\begin{minted}[mathescape, linenos]{r}
res.n100.Zhang <- calc.rejections.Zhang.tests(
    sample.generation.function.H0  = rnorm,
    sample.generation.functions.H1 = 
        list(
            rnorm,
            function(nsamples) runif(nsamples, -sqrt(3), sqrt(3)),
            function(nsamples) rnorm(nsamples, 0.2, 1),
            function(nsamples) rnorm(nsamples, 0, 0.8),
            function(nsamples) rnorm(nsamples, 0, 1.2),
            function(nsamples) rt(nsamples, df = 5),
            function(nsamples, shape = 5) 
                (rgamma(nsamples, shape = shape, rate = 1) - shape) / sqrt(shape)
        ),
    F_0   = pnorm,
    alpha = 0.05,
    h     = 30,
    n = 100, m = m
)
\end{minted}

\begin{table}[H]
\centering
\begin{tabular}{ |l|l||>{\columncolor[gray]{0.9}}p{1cm}|p{1.1cm}||>{\columncolor[gray]{0.9}}p{1cm}|p{1.1cm}||>{\columncolor[gray]{0.9}}p{1cm}|p{1.1cm}| } 
 \hline
   $H_0$ & $H_1$ & \multicolumn{2}{c||}{$Z_K$} & \multicolumn{2}{c||}{$Z_A$} & \multicolumn{2}{c|}{$Z_C$} \\
 \hline
 &  & mean & sd & mean & sd & mean & sd \\
 \hline
 \hline
 
 $\mathcal{N}(0, 1)$ & $\mathcal{N}(0, 1)$ & 0.0501 & 3.5e-04 & 0.0501 & 2.9e-04 & 0.0501 & 2.7e-04\\
$\mathcal{N}(0, 1)$ & $\mathrm{U}(-\sqrt{3}, \sqrt{3})$ & 0.2702 & 9.1e-04 & 1 & 5.7e-07 & 0.8537 & 2.0e-03\\
$\mathcal{N}(0, 1)$ & $\mathcal{N}(0.2, 1)$ & 0.3754 & 9.6e-04 & 0.3908 & 8.1e-04 & 0.3943 & 9.2e-04\\
$\mathcal{N}(0, 1)$ & $\mathcal{N}(0, 0.8)$ & 0.3907 & 1.1e-03 & 0.6935 & 7.6e-04 & 0.6106 & 1.1e-03\\
$\mathcal{N}(0, 1)$ & $\mathcal{N}(0, 1.2)$ & 0.4627 & 9.5e-04 & 0.4365 & 7.4e-04 & 0.5512 & 7.2e-04\\
$\mathcal{N}(0, 1)$ & t(5) & 0.718 & 7.3e-04 & 0.7666 & 6.0e-04 & 0.8567 & 4.4e-04\\
$\mathcal{N}(0, 1)$ & $\mathrm{G}(5)$ & 0.4123 & 1.2e-03 & 0.7701 & 8.4e-04 & 0.6439 & 8.9e-04\\
 
         \hline
\end{tabular}
\caption{Zhang tests asymptotics, $n = 100$.}
\label{table:zhang:asymptotics:n100}
\end{table}

\noindent
We see that for $n = 100$ $Z_A$ and $Z_C$ tests become powerful for uniform alternative. Moreover, $Z_A$ becomes more powerful than PC2 for gamma alternative. Everything else stays the same as for the $n = 50$ case.

\subsection{Conclusions}

We see that there is no clear "winner" between PC2 and Zhang tests, but we can surely state that PC2 is on par with $Z_A$ and $Z_C$.

\newpage

\section{Goodness of fit test based on hyperrectangle bounds for joint density of principal components of order statistics: samples from standard uniform distribution [pc, \texorpdfstring{$\mathrm{U}(0,1)$}{U(0,1)}]}
\label{section:hyperrectabgle_order_stats_uniform_full}

In section \ref{section:pc_test_normal} we introduced the test based on principal components of standard normal order statistics and showed how it can be used to check goodness of fit for any distribution. In this section we will do the same for the test based on principal components of standard uniform order statistics. 

In classical and Zhang tests probability integral transform is used as a part of these tests: in essence, we first apply the transform and then check if the transformed sample's eCDF is near uniform CDF according to some distance. Probability integral transform can also be used with the uniform test proposed in this section to make it applicable to any continuous distribution.

\subsection{Generation of order statistics for samples from standard uniform distribution}
\label{section:standard_uniform_order_stats}

We start with generating $m$ sorted samples of size $n$ from standard uniform distribution:
\begin{minted}[mathescape, linenos]{r}
sorted.samples.uniform <- get.sorted.samples.cpp(runif, n, m)
\end{minted}
Remember that $m$ and $n$ were defined in subsection~\ref{section:sample_size_num_samples}, \texttt{get.sorted.samples.cpp} function --- in subsection~\ref{section:functions_to_generate_order_stats}. The resulted table \texttt{sorted.samples.uniform} of test statistics is of~(\ref{sorted_samples_table}) form.

Throughout section~\ref{section:hyperrectabgle_order_stats_uniform_full} standard uniform will be our $H_0$ distribution.

\subsection{Histograms of ordered statistics from uniform distribution}

Now we plot histograms for columns of \texttt{sorted.samples.uniform} table that contain empirical distributions of order statistics for uniform samples of size $n = 10$:
\begin{minted}[mathescape, linenos]{r}
par(mfcol = c(5, 2))
for(i in 1:n)
    hist(
        sorted.samples.uniform[,i], xlim = c(0,1), 
        main = paste0(i," order statistic"), xlab = "x"
    )
\end{minted}

\begin{figure}[ht]
\centering
\includegraphics[width=12cm]{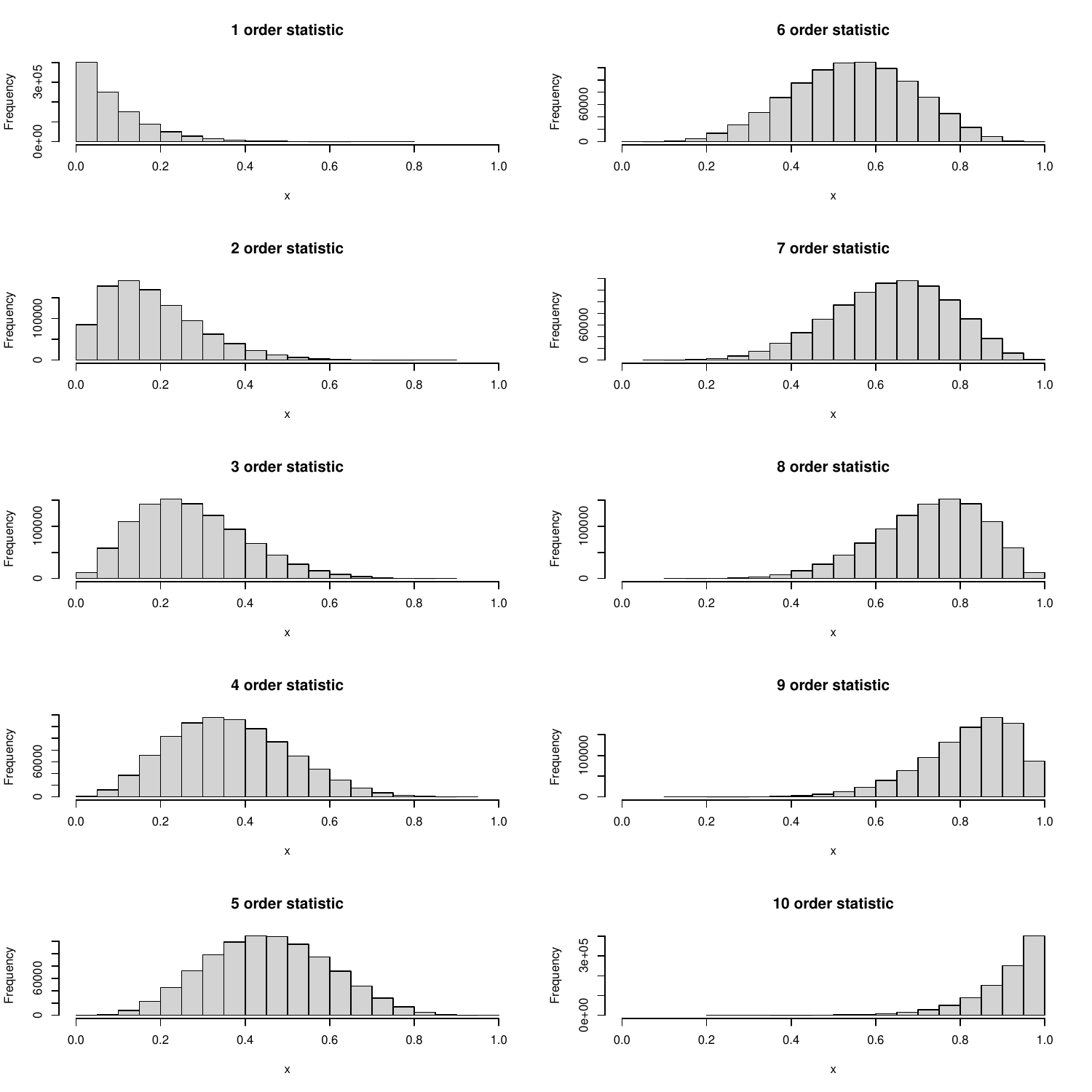}
\caption{Histograms for empirical distributions of order statistics for samples from standard uniform distribution of size $10$.}
\label{figure:uniform_ordered_sample_hists}
\end{figure}

The histograms are presented on figure~\ref{figure:uniform_ordered_sample_hists}.
We see that distributions of order statistics are again unimodal. 
This means we can safely use confidence intervals instead of one-dimensional highest density regions.
We also see that 1st order statistic has mode near 0, and 10th order statistic has mode near 1. Thus maybe for 1st and 10th order statistics we could better use one-side confidence intervals. 

\subsection{Function to calculate hyperrectangle bounds with interval type \textasteriskcentered}
\label{section:hr:bounds:mixed:type}

Function \texttt{calculate.hyperrectangle.bounds} from subsection~\ref{section:calculate.hyperrectangle.bounds} can only handle two-sided intervals. Let us write another function that can handle one-sided intervals as well:

\begin{minted}[mathescape, linenos, texcomments]{r}
calculate.hyperrectangle.bounds.with.interval.type <- function(
    alpha,        # significance level
    gamma.left,   # left bound for bisection method
    gamma.right,  # right bound for bisection method
    w,            # the vector of weights; should satisfy: length(w) = ncol(stat.matrix)
    stat.matrix,  # statistics data matrix of type (\ref{data_table})
    interval.types = rep("two sides", length(w)),  # types of confidence intervals; same size as w
    default.left.bounds  = NULL,  # custom left  bound for each interval; same size as w 
    default.right.bounds = NULL   # custom right bound for each interval; same size as w
)
{
    m                <- nrow(stat.matrix)  # infer number of    rows from data
    k                <- ncol(stat.matrix)  # infer number of columns from data
    left.bounds      <- numeric(k)         # allocate memory for left  bounds of hyperrectangle
    right.bounds     <- numeric(k)         # allocate memory for right bounds of hyperrectangle
    check.sim.bounds <- matrix(0, m, k)    # allocate memory for rejection checks

    stat.matrix.sorted.cols <- apply(stat.matrix, 2, sort)  # sort each column of stat.matrix
    
    while(gamma.right - gamma.left > 0.00001)  # bisection loop
    {
        gamma <- (gamma.right + gamma.left) / 2  # interval center in bisection method
        for(i in 1:k){  # loop over stat.matrix columns to compute bounds and rejections
            if(interval.types[i] == "two sides") 
            {
                left.bounds[i]  <- quantile_sorted(stat.matrix.sorted.cols[,i],     gamma*w[i]*alpha/2)
                right.bounds[i] <- quantile_sorted(stat.matrix.sorted.cols[,i], 1 - gamma*w[i]*alpha/2)
            } else if(interval.types[i] == "left side") {
                left.bounds[i]  <- quantile_sorted(stat.matrix.sorted.cols[,i],     gamma*w[i]*alpha  )
                right.bounds[i] <- default.right.bounds[i]
            } else if(interval.types[i] == "right side") {
                left.bounds[i]  <- default.left.bounds[i]
                right.bounds[i] <- quantile_sorted(stat.matrix.sorted.cols[,i], 1 - gamma*w[i]*alpha  )
            }
            check.sim.bounds[,i] <- 
                (stat.matrix[,i] >= left.bounds[i]) & 
                (stat.matrix[,i] <= right.bounds[i])
        }
      
        sim.check <- rowSums(check.sim.bounds) == k  # for all rows look if all checks passed
        positive.checks <- sum(sim.check) / m        # proportion of rows with all checks passed

        if(positive.checks < 1 - alpha){  # choose left or right subinterval in bisection method
            gamma.right <- gamma
        } else {
            gamma.left <- gamma
        }
        
        print(gamma)     # print gamma to monitor progress
        flush.console()  # refresh the console for progress to appear
    }
    list(                                   # function returns this:
        left.bounds     = left.bounds,      # left bounds for hyperrectangle
        right.bounds    = right.bounds,     # right bounds for hyperrectangle
        positive.checks = positive.checks,  # estimated 1 - alpha as sanity check
        gamma           = gamma             # resulted optimal gamma
    )
}
\end{minted}

This function has 3 additional parameters if compared to \texttt{calculate.hyperrectangle.bounds}: 
\begin{enumerate}
\item \texttt{interval.types}, that is a character \texttt{vector} of the same size as \texttt{w}, where each value can be \texttt{"two sides"}, \texttt{"left side" or \texttt{"right side"}};
\item \texttt{default.left.bounds} and \texttt{default.right.bounds}, that are numeric vectors of size \texttt{w} giving corresponding custom bounds for  one-sided intervals.
\end{enumerate}

\subsection{Calculate hyperectangle bounds for uniform distribution: different interval types}
\label{section:calc_bounds_uniform}

Let us apply the function from previous subsection to get two types of hyperrectangle bounds. First type uses two-sided confidence intervals for all order statistics:

\begin{minted}[mathescape, linenos]{r}
res.hr.bounds.uniform.two.side.bounds <- calculate.hyperrectangle.bounds.with.interval.type(
    alpha = 0.05,
    gamma.left  = 0,
    gamma.right = 1,
    rep(1, n),
    sorted.samples.uniform
)
\end{minted}
The resulted bound are like this:
\begin{minted}[mathescape, linenos]{r}
for(i in 1:n) 
    print(c(
        res.hr.bounds.uniform.two.side.bounds$left.bounds[i], 
        res.hr.bounds.uniform.two.side.bounds$right.bounds[i]
    ))
\end{minted}

\begin{verbatim}
[1] 0.000373162 0.429293036
[1] 0.009405357 0.560171456
[1] 0.03315181 0.66237723
[1] 0.07033786 0.74770595
[1] 0.1197788 0.8202572
[1] 0.180793 0.879739
[1] 0.2531332 0.9294063
[1] 0.3377238 0.9669189
[1] 0.4390001 0.9907563
[1] 0.5719531 0.9996312
\end{verbatim}

Second hyperrectangle type uses right sided confidence interval for 1st order statistic with left bound fixed at 0, left sided confidence interval for 10th order statistic with right side fixed at 1, and two-sided confidence intervals for all other order statistics:
\begin{minted}[mathescape, linenos]{r}
res.hr.bounds.uniform.mixed.bounds <- calculate.hyperrectangle.bounds.with.interval.type(
    alpha = 0.05,
    gamma.left  = 0,
    gamma.right = 1,
    rep(1, n),
    sorted.samples.uniform,
    interval.types = c(rep("right side", 1), rep("two sides", 8), rep("left side", 1)),
    default.left.bounds  = rep(0, n),
    default.right.bounds = rep(1, n)
)
\end{minted}
The bounds are as follows:
\begin{minted}[mathescape, linenos]{r}
for(i in 1:n) 
    print(c(
        res.hr.bounds.uniform.mixed.bounds$left.bounds[i], 
        res.hr.bounds.uniform.mixed.bounds$right.bounds[i]
    ))
\end{minted}

\begin{verbatim}
[1] 0.0000000   0.3883262
[1] 0.009283598 0.559746340
[1] 0.03315469  0.66325239
[1] 0.07065085  0.74805713
[1] 0.1200487   0.8195192
[1] 0.1803404   0.8803111
[1] 0.2527674   0.9295968
[1] 0.3383371   0.9668086
[1] 0.4402455   0.9906281
[1] 0.6128585   1.0000000
\end{verbatim}

\subsection{Plot confidence bounds for mixed bounds types}

Now let us plot simultaneous bounds for order statistics of standard uniform distribution. We plot mixed bounds from the previous subsection:

\begin{minted}[mathescape, linenos, texcomments]{r}
plot.order.statistics.bounds(
    sorted.samples.matrix = sorted.samples.uniform,             # defined in subsection \ref{section:standard_uniform_order_stats}
    x.grid                = seq(0, 1, 0.01),                    
    theoretical.cdf       = punif,                              # uniform CDF
    hyperrectangle.bounds = res.hr.bounds.uniform.mixed.bounds  # defined in subsection \ref{section:calc_bounds_uniform}
)
\end{minted}

\begin{figure}[H]
\centering
\includegraphics[width=12cm]{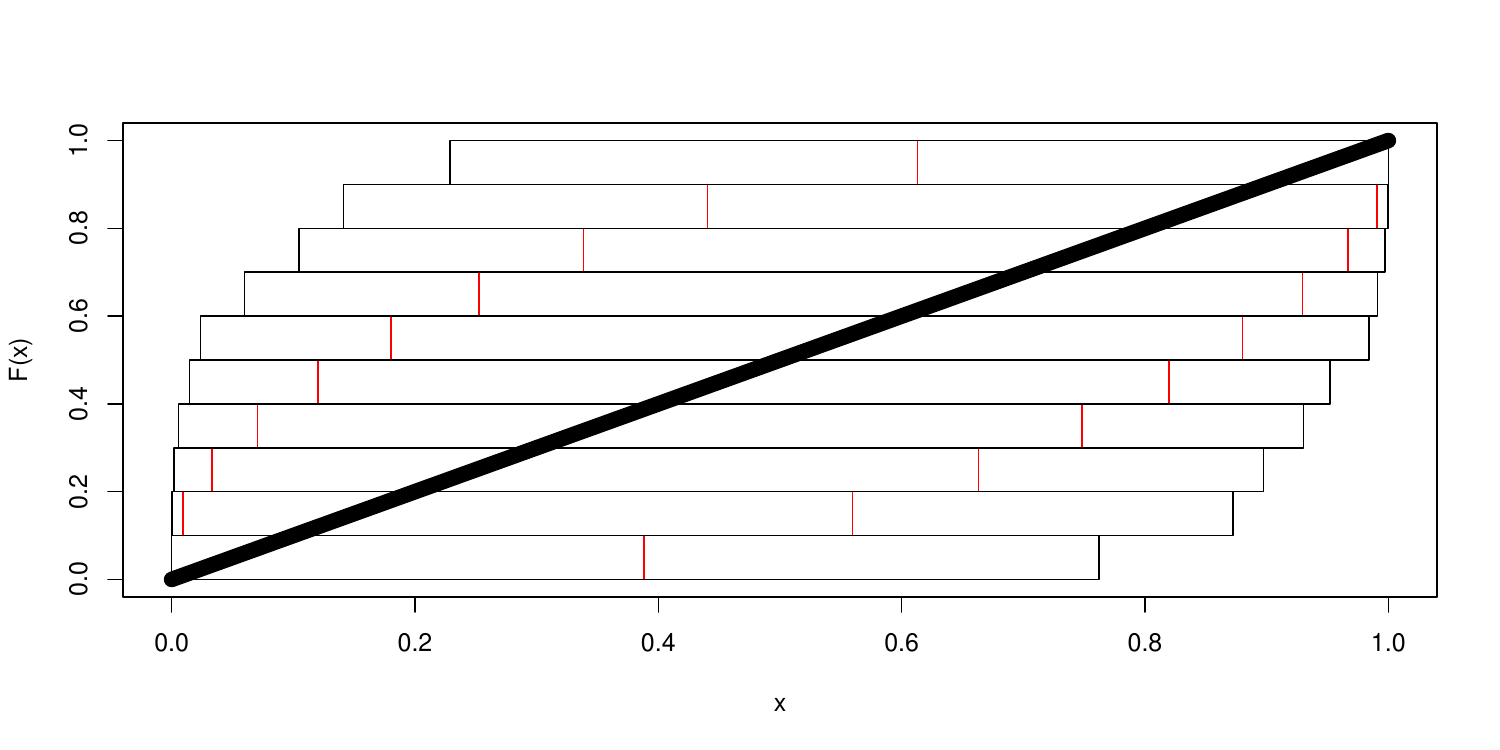}
\caption{Red --- estimated 0.95-confidence bound for empirical cumulative distribution function of a sample from uniform distribution of size 10; black --- minimum and maximum values for order statistics among 1000000 generated samples.}
\end{figure}

\newpage
\subsection{Plotting 2d projections of sorted samples from uniform distribution with corresponding mixed type hyperrectangle bounds: contour plots}
\label{section:2dplots:uniform}

Let us plot a matrix of 2d projections for empirical joint density of order statistics for standard uniform distribution and corresponding projections of mixed type hyperrectangle bounds that contain 95 percents of that joint density:

\begin{minted}[mathescape, linenos, texcomments]{r}
plot.2d.projections.contour(            # defined in subsection \ref{section:2d.projections.contour}
    sorted.samples.uniform,             # defined in subsection \ref{section:standard_uniform_order_stats}
    res.hr.bounds.uniform.mixed.bounds  # defined in subsection \ref{section:calc_bounds_uniform}
)
\end{minted}

\begin{figure}[H]
\centering
\includegraphics[width=12cm]{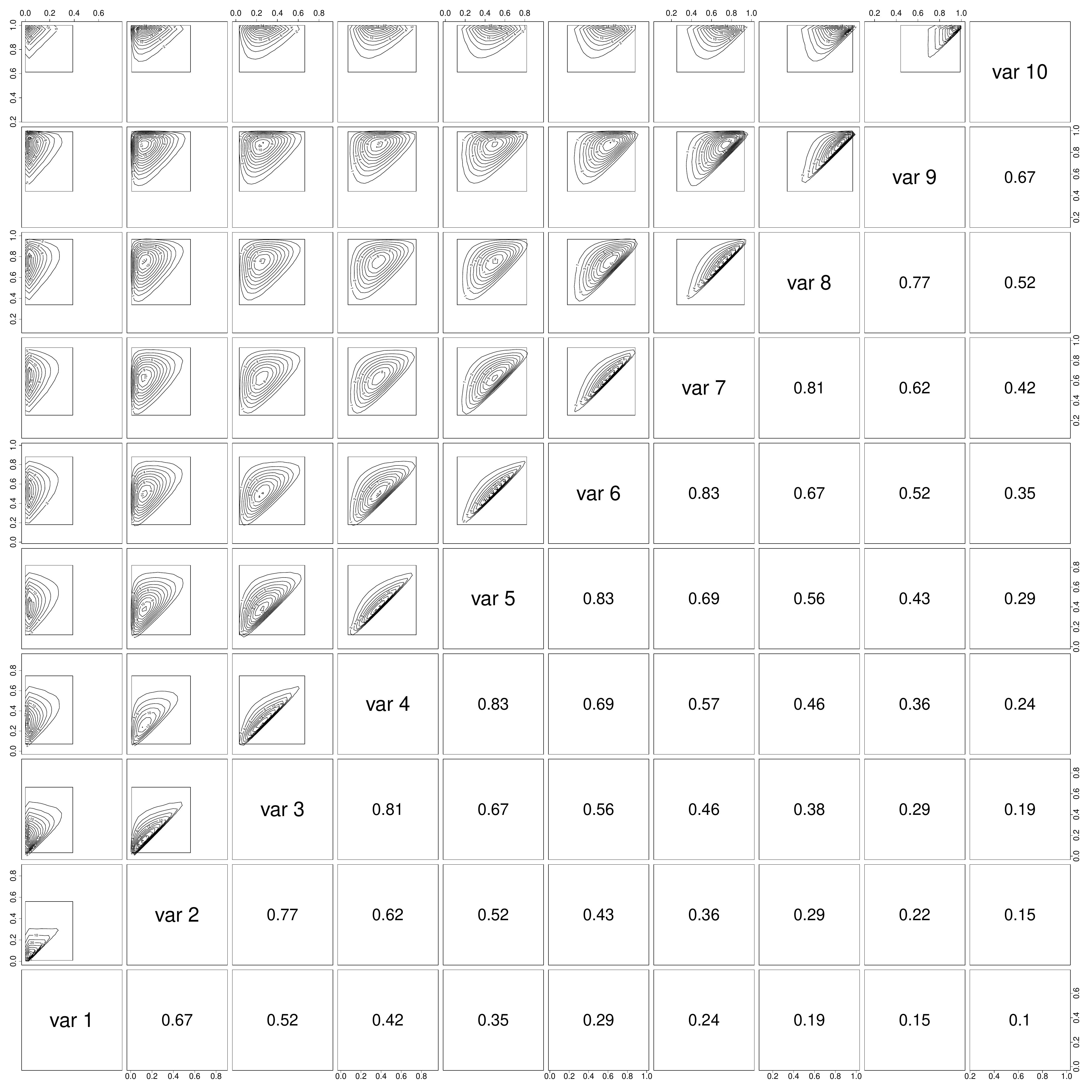}
\caption{Pairwise contour plots for 2d projections of empirical joint density of order statistics for standard uniform with pairwise correlations of order statistics.}
\label{figure:contours:uniform:os}
\end{figure}

Most of the conclusions are the same as for figure~\ref{figure:contour_order_samples_normal} that illustrated standard normal samples of size 10. We again see that all contours are inside confidence rectangles. Correlations are very similar: the largest are for adjacent order statistics, and they decrease with the increase of the distance between indices of order statistics. For uniformly distributed samples, the contours tend to appear more triangular than circular, unlike the contours observed for normally distributed samples. For pairs of consequent order statistics, contours are flat around the diagonal and there is a lot of space inside of confidence rectangles that is not covered by contours.

\newpage
\subsection{Plot 2d projections, principal components, ordered samples, uniform distribution, hexplom}
\label{section:2dplots:uniform:hexplom}

Now let us do exactly the same plot, but with hexbin function:
\begin{minted}[mathescape, linenos, texcomments]{r}
plot.2d.projections.hexplom(            # defined in subsection \ref{section:2d.projections.hexplom}
    sorted.samples.uniform,             # defined in subsection \ref{section:standard_uniform_order_stats}
    res.hr.bounds.uniform.mixed.bounds  # defined in subsection \ref{section:calc_bounds_uniform}
)
\end{minted}

\begin{figure}[H]
\centering
\includegraphics[width=12cm]{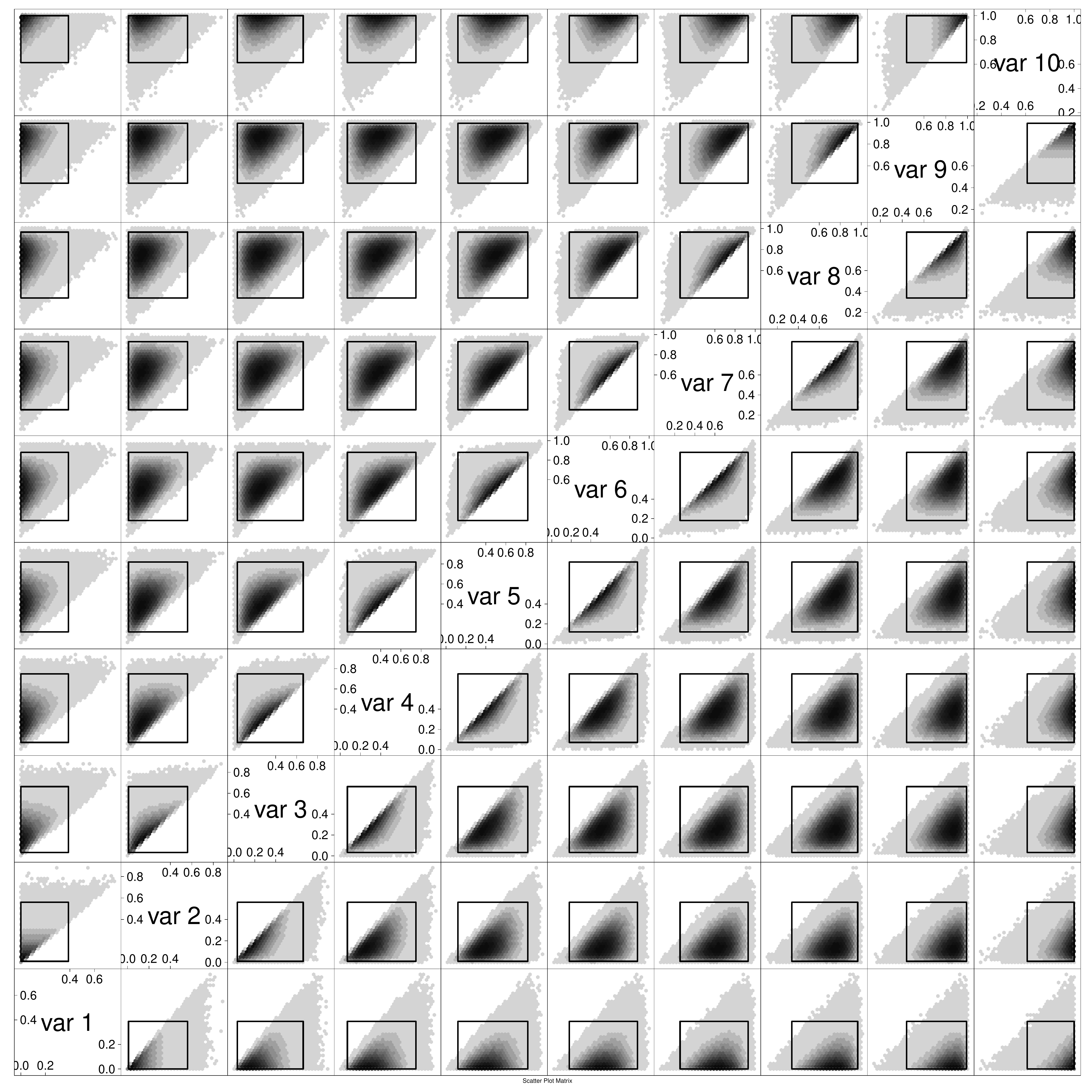}
\caption{Pairwise hexbin plots for 2d projections of empirical joint density of order statistics for standard uniform samples of size 10.}
\end{figure}

\noindent
We see that all 2d projections are triangular. That is because order statistics of uniform samples live in a simplex
\begin{align}
\label{eq:simplex}
\Delta_n
=
\left\{
(x_1,\dots,x_n)\in \mathbb{R}^n :
0 \le x_1 \le x_2 \le \cdots \le x_n \le 1
\right\}.
\end{align}
Moreover, joint distribution of uniform order statistics is uniform in the simplex $\Delta_n$.
We'll discuss the geometry of uniform order statistics in more detail in the next chapter.

\newpage
\subsection{Calculate principal components for ordered samples from uniform distribution}
\label{section:pc:samples:from:unidorm}

On plots in subsections \ref{section:2dplots:uniform} and \ref{section:2dplots:uniform:hexplom} we see again that there is a lot of free space inside confidence hyperrectangle projections. Let us try space rotation from principal component analysis to fix that:

\begin{minted}[mathescape, linenos]{r}
sorted.samples.uniform.p.comp <- princomp(sorted.samples.uniform)
sorted.samples.uniform.p.comp
sorted.samples.uniform.p.comp$loadings
\end{minted}

\begin{verbatim}
Call:
princomp(x = sorted.samples.uniform)

Standard deviations:
    Comp.1     Comp.2     Comp.3     Comp.4     Comp.5     Comp.6     Comp.7 
0.30559848 0.15432658 0.10475906 0.08056415 0.06645039 0.05756201 0.05168339 
    Comp.8     Comp.9    Comp.10 
0.04781334 0.04533968 0.04397556 

 10  variables and  1000000 observations.

Loadings:
      Comp.1 Comp.2 Comp.3 Comp.4 Comp.5 Comp.6 Comp.7 Comp.8 Comp.9 Comp.10
 [1,]  0.120  0.231  0.324  0.389  0.421  0.423  0.380  0.326  0.234  0.121 
 [2,]  0.230  0.387  0.423  0.323  0.117 -0.121 -0.316 -0.424 -0.390 -0.232 
 [3,]  0.322  0.422  0.230 -0.122 -0.388 -0.391 -0.121  0.227  0.420  0.323 
 [4,]  0.388  0.322 -0.121 -0.422 -0.228  0.235  0.420  0.127 -0.321 -0.387 
 [5,]  0.422  0.121 -0.387 -0.229  0.323  0.321 -0.229 -0.392  0.120  0.421 
 [6,]  0.422 -0.120 -0.387  0.231  0.323 -0.325 -0.233  0.386  0.118 -0.420 
 [7,]  0.388 -0.322 -0.120  0.422 -0.231 -0.226  0.427 -0.117 -0.321  0.387 
 [8,]  0.322 -0.422  0.230  0.118 -0.389  0.385 -0.121 -0.230  0.424 -0.324 
 [9,]  0.231 -0.388  0.422 -0.321  0.119  0.120 -0.327  0.418 -0.388  0.233 
[10,]  0.120 -0.231  0.323 -0.388  0.423 -0.420  0.393 -0.319  0.228 -0.121 

               Comp.1 Comp.2 Comp.3 Comp.4 Comp.5 Comp.6 Comp.7 Comp.8 Comp.9
SS loadings       1.0    1.0    1.0    1.0    1.0    1.0    1.0    1.0    1.0
Proportion Var    0.1    0.1    0.1    0.1    0.1    0.1    0.1    0.1    0.1
Cumulative Var    0.1    0.2    0.3    0.4    0.5    0.6    0.7    0.8    0.9
               Comp.10
SS loadings        1.0
Proportion Var     0.1
Cumulative Var     1.0
\end{verbatim}

\noindent
For the case of order statistics from uniform distribution we can derive the \texttt{loadings} matrix analytically\footnote{Thanks goes to ChatGPT \cite{chatgpt2026} for pointing this out.}. This matrix is based on singular value decomposition of uniform order statistics covariance matrix. Let us denote uniform order statistics as 
$U_{(1)} \leq \dots \leq U_{(n)}$. Then covariance matrix for them is as follows (see~\cite{david2003order}):
\begin{align}
\Sigma_{ij}
=
\operatorname{Cov}(U_{(i)},U_{(j)})
=
\frac{\min(i,j)\,\bigl(n-\max(i,j)+1\bigr)}
{(n+1)^2(n+2)}; \\
\Sigma
=
\frac{1}{(n+1)^2(n+2)}
\begin{pmatrix}
n & n-1 & n-2 & \cdots & 1\\
n-1 & 2(n-1) & 2(n-2) & \cdots & 2\\
n-2 & 2(n-2) & 3(n-2) & \cdots & 3\\
\vdots & \vdots & \vdots & \ddots & \vdots\\
1 & 2 & 3 & \cdots & n
\end{pmatrix}.
\end{align}
The inverse of $\Sigma$ has a simple tree-diagonal form:
\begin{align}
\Sigma^{-1}
=
(n+1)(n+2)
\begin{pmatrix}
2 & -1 & 0 & \cdots & 0\\
-1 & 2 & -1 & \ddots & \vdots\\
0 & -1 & 2 & \ddots & 0\\
\vdots & \ddots & \ddots & \ddots & -1\\
0 & \cdots & 0 & -1 & 2
\end{pmatrix}.
\end{align}
Singular value decomposition for $\Sigma^{-1}$ is as follows: 
\begin{align}
\Sigma^{-1} = V \Lambda^{-1} V^\top, 
\end{align}
where
\begin{align}
\label{eq:uniform:order:stats:pc:sine}
V_{ij}
=
\sqrt{\frac{2}{n+1}}\,
\sin\!\left(\frac{\pi i j}{n+1}\right),
\qquad i,j=1,\dots,n,
\end{align}
and
\begin{align}
\Lambda=\operatorname{diag}(\lambda_1,\dots,\lambda_n),
\end{align}
with
\begin{align}
\label{eq:weights:unif}
\lambda_i
=
\frac{1}
{4(n+1)(n+2)
\sin^2\!\left(\frac{\pi i}{2(n+1)}\right)},
\qquad k=1,\dots,n.
\end{align}
Finally, $\Sigma$ does have the same eigenvectors $V$ as $\Sigma^{-1}$:
\begin{align}
\Sigma = V \Lambda V^\top.
\end{align}
Matrix $V$ is the \texttt{loadings} matrix and $\sqrt{\lambda_i}$ are ``standard deviations'' from \texttt{princomp} result.

Let us do the checks by programming the derived formulas directly:
\begin{minted}[mathescape, linenos]{r}
V <- outer( 1:n, 1:n, function(i, j) sqrt(2/(n+1))*sin(pi*i*j/(n+1)) )
V
\end{minted}

\begin{verbatim}
           [,1]       [,2]       [,3]       [,4]       [,5]       [,6]       [,7]
 [1,] 0.1201312  0.2305300  0.3222527  0.3878684  0.4220613  0.4220613  0.3878684
 [2,] 0.2305300  0.3878684  0.4220613  0.3222527  0.1201312 -0.1201312 -0.3222527
 [3,] 0.3222527  0.4220613  0.2305300 -0.1201312 -0.3878684 -0.3878684 -0.1201312
 [4,] 0.3878684  0.3222527 -0.1201312 -0.4220613 -0.2305300  0.2305300  0.4220613
 [5,] 0.4220613  0.1201312 -0.3878684 -0.2305300  0.3222527  0.3222527 -0.2305300
 [6,] 0.4220613 -0.1201312 -0.3878684  0.2305300  0.3222527 -0.3222527 -0.2305300
 [7,] 0.3878684 -0.3222527 -0.1201312  0.4220613 -0.2305300 -0.2305300  0.4220613
 [8,] 0.3222527 -0.4220613  0.2305300  0.1201312 -0.3878684  0.3878684 -0.1201312
 [9,] 0.2305300 -0.3878684  0.4220613 -0.3222527  0.1201312  0.1201312 -0.3222527
[10,] 0.1201312 -0.2305300  0.3222527 -0.3878684  0.4220613 -0.4220613  0.3878684
            [,8]       [,9]      [,10]
 [1,]  0.3222527  0.2305300  0.1201312
 [2,] -0.4220613 -0.3878684 -0.2305300
 [3,]  0.2305300  0.4220613  0.3222527
 [4,]  0.1201312 -0.3222527 -0.3878684
 [5,] -0.3878684  0.1201312  0.4220613
 [6,]  0.3878684  0.1201312 -0.4220613
 [7,] -0.1201312 -0.3222527  0.3878684
 [8,] -0.2305300  0.4220613 -0.3222527
 [9,]  0.4220613 -0.3878684  0.2305300
[10,] -0.3222527  0.2305300 -0.1201312
\end{verbatim}

\begin{minted}[mathescape, linenos]{r}
lambda <- sapply( 1:n, function(i) 1/(4*(n+1)*(n+2)*sin(pi*i/(2*(n+1)))^2) )
sqrt(lambda)
\end{minted}

\begin{verbatim}
[1] 0.30579674 0.15447066 0.10476129 0.08049598 0.06645598 0.05758444 0.05173163
[8] 0.04784288 0.04535668 0.04396693
\end{verbatim}

\noindent
We see that numerical and analytical results coincide. So the principal components of order statistics for uniform samples are sine waves defined by formula~(\ref{eq:uniform:order:stats:pc:sine}). They don't have such nice interpretation as principal components of normal order statistics. Potentially we can use formulas~(\ref{eq:uniform:order:stats:pc:sine}) and~(\ref{eq:weights:unif})
to determine space rotation and principal component weights in our algorithm for confidence hyperrectangle construction to save some machine time by skipping numerical SVD of \texttt{sorted.samples.uniform} matrix, but we still need this matrix to solve equation~(\ref{sim_bound_eq}) numerically, so optimizing SVD step will not give us a big speedup. That is why we'll proceed with numerical SVD as before.

\subsection{Print histograms of principal components}

Let us plot histograms of principal components from previous section:

\begin{minted}[mathescape, linenos]{r}
par(mfcol = c(5, 2))
for(i in 1:n)
    hist(
        sorted.samples.uniform.p.comp$scores[,i],
        main = paste0(i," order statistic"), xlab = "x"
    )
\end{minted}

\begin{figure}[H]
\centering
\includegraphics[width=12cm]{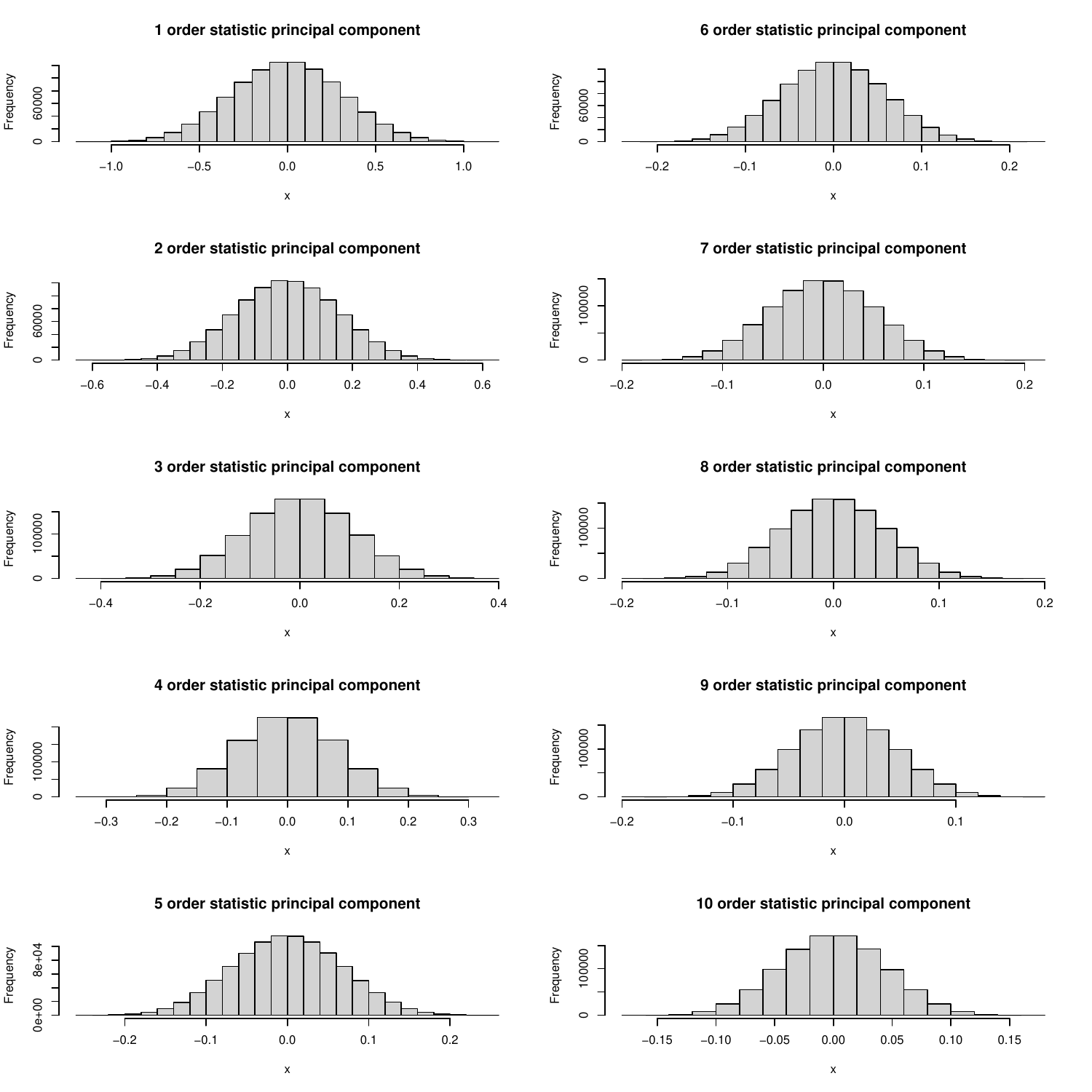}
\caption{Histograms for empirical distributions of principal components of order statistics for samples from standard uniform distribution of size 10.}
\end{figure}

\noindent
Empirical distributions of principal components are again unimodal, so confidence intervals instead of highest density regions are fine. No modes are near the border, so only two-sided confidence intervals should be used for all order statistics.

\newpage
\subsection{Calculate hyperrectangle bounds, principal components, uniform distribution}
\label{section:hr:bounds:uniform:pc}

It is time to construct hyperrectangle bounds for principal components of uniform order statistics:

\begin{minted}[mathescape, linenos]{r}
res.hyperrectangle.bounds.uniform.pc <- calculate.hyperrectangle.bounds(
    alpha = 0.05,
    gamma.left  = 0,
    gamma.right = 2,
    w = sorted.samples.uniform.p.comp$sdev^2 / sum(sorted.samples.uniform.p.comp$sdev^2),
    sorted.samples.uniform.p.comp$scores
)
\end{minted}
The two-sided variance-based bounds are as follows:

\begin{minted}[mathescape, linenos]{r}
for(i in 1:n) 
    print(c(
        res.hyperrectangle.bounds.uniform.pc$left.bounds[i], 
        res.hyperrectangle.bounds.uniform.pc$right.bounds[i]
    ))
\end{minted}

\begin{verbatim}
[1] -0.6412571  0.6428143
[1] -0.3904547  0.3896175
[1] -0.2855091  0.2865183
[1] -0.2295019  0.2297811
[1] -0.1955495  0.1959322
[1] -0.1722380  0.1728755
[1] -0.1560432  0.1586393
[1] -0.1468864  0.1468978
[1] -0.1409762  0.1392950
[1] -0.1358782  0.1361864
\end{verbatim}

\newpage
\subsection{2d projections, principal components, ordered samples, uniform distribution, contour plots}

Now to the contour plots of principal component: 

\begin{minted}[mathescape, linenos, texcomments]{r}
plot.2d.projections.contour(               # defined in subsection \ref{section:2d.projections.contour}
    sorted.samples.uniform.p.comp$scores,  # defined in subsection \ref{section:pc:samples:from:unidorm}
    res.hyperrectangle.bounds.uniform.pc   # defined in subsection \ref{section:hr:bounds:uniform:pc}
)
\end{minted}

\begin{figure}[H]
\centering
\includegraphics[width=12cm]{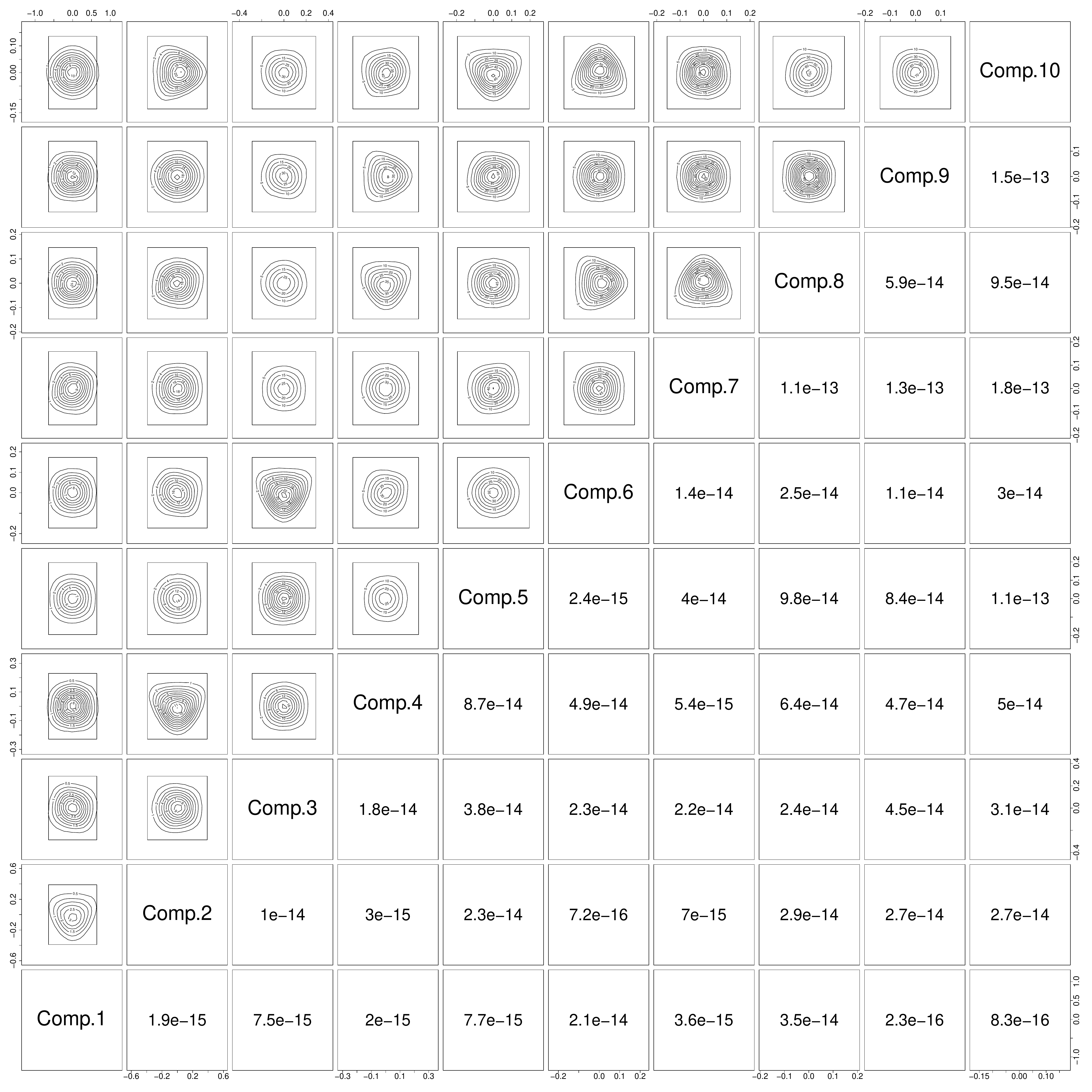}
\caption{Pairwise contour plots for 2d projections of empirical joint density of principal components of standard uniform order statistics. Confidence rectangles are constructed using variance-based weights for principal components. Principal components are uncorrelated by construction, which we see in lower triangle.}
\end{figure}

\noindent
We see that the contours are circular or smoothed triangular, they consume larger percent of area inside confidence rectangles than contours on figure~\ref{figure:contours:uniform:os}. Bounds for pairs of more important principal components are noticeably tighter than bounds for less important components, since we use variance-based weights.

\newpage
\subsection{2d projections, principal components, ordered samples, uniform distribution, hexplom}

And here are the same plots made with the help of \texttt{hexbin} package:
\begin{minted}[mathescape, linenos, texcomments]{r}
plot.2d.projections.hexplom(               # defined in subsection \ref{section:2d.projections.hexplom}
    sorted.samples.uniform.p.comp$scores,  # defined in subsection \ref{section:pc:samples:from:unidorm}
    res.hyperrectangle.bounds.uniform.pc   # defined in subsection \ref{section:hr:bounds:uniform:pc}
)
\end{minted}

\begin{figure}[H]
\centering
\includegraphics[width=12cm]{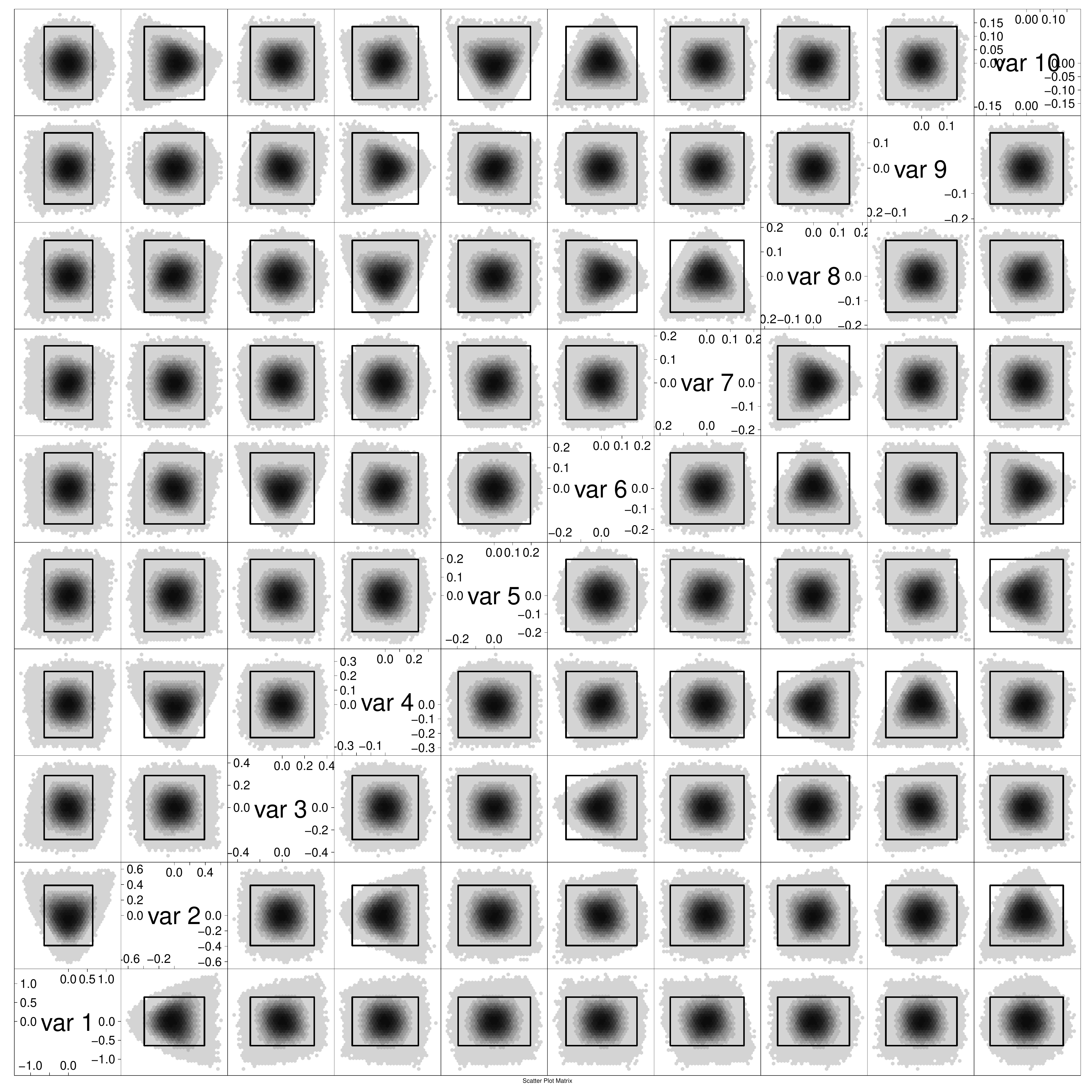}
\caption{Pairwise hexbin plots for 2d projections of empirical joint density of principal components of order statistics for standard uniform samples of size 10. Confidence rectangles are constructed using variance-based weights for principal components.}
\end{figure}

\noindent
High density regions are marked with black, low density regions are marked with grey.

\subsection{Function to calculate rejections; with interval types \textasteriskcentered}

It would be interesting to compare powers of two types of hyperrectangle bounds for uniform order statistics: one based entirely on two-sided confidence intervals and one that uses one-sided confidence intervals for first and last order statistics. We constructed the mentioned two types of bounds in subsection~\ref{section:calc_bounds_uniform}. In this subsection we modify \texttt{calc.rejections.sm.pc.bounds} function from subsection~\ref{section:calc_num_rej1} to handle one-sided confidence intervals via adoption of function for hyperrectangle bounds computation from subsection~\ref{section:hr:bounds:mixed:type}. The new rejection counting code is as follows:

\begin{minted}[mathescape, linenos, texcomments]{r}
calc.rejections.sm.pc.bounds.with.interval.types <- function(
    sample.generation.function.H0,   # single function to generate samples from H0 distribution
    sample.generation.functions.H1,  # list of functions to generate samples for multiple H1s
    get.stat.matrix,  # function to generate sample of samples and construct table of type (\ref{data_table_general}) 
    alpha,  # significance level for hyperrectangle bounds
    n,      # sample size for each sample in sample of samples 
    m,      # sample of sample size
    h,      # how many times we generate sample of samples for H0 and each H1
    gamma.left.sm = 0, gamma.right.sm = 10,  # range for bisection method; stat.matrix test
    w.sm = rep(1, n),  # custom weights for test based on stat.matrix
    gamma.left.pc = 0, gamma.right.pc = 10,  # range for bisection method; PCs of stat.matrix test
    w.pc = rep(1, n),  # custom weights for test based on principal components of stat.matrix
    interval.types.sm.mt = rep("two sides", length(w.sm)),  # interval types for sm.mt test
    default.left.bounds.sm.mt  = NULL,    # custom left  bounds for conf.ints in sm.mt test
    default.right.bounds.sm.mt = NULL,    # custom right bounds for conf.ints in sm.mt test
    interval.types.pc = rep("two sides", length(w.pc)),     # interval types for pc test
    default.left.bounds.pc  = NULL,       # custom left  bounds for conf.ints in pc test
    default.right.bounds.pc = NULL        # custom right bounds for conf.ints in pc test
)
{
    s <- length(sample.generation.functions.H1)  # number of alternatives
    res.sm    <- matrix(0, h, s)  # allocate space for powers of stat.matrix test
    res.sm.mt <- matrix(0, h, s)  # allocate space for powers of sm test with mixed type bounds
    res.pc.w1 <- matrix(0, h, s)  # allocate space for powers of PC test with custom weights 
    res.pc.w2 <- matrix(0, h, s)  # allocate space for powers of PC test with variance weights
    for(i in 1:h)  # we generate sample of samples for H0 and each H1 multiple times
    {
        print(paste("iteration", i))
        flush.console()
        stat.matrix.H0 <- get.stat.matrix(sample.generation.function.H0, n, m) # of type (\ref{data_table})
        print("calculate hyperrectangle.bounds.H0.sm")
        flush.console()
        # Bounds based on stat.matrix cols: 
        res.h.bounds.H0.sm    <- calculate.hyperrectangle.bounds.with.interval.type(
            alpha                = alpha,
            gamma.left           = gamma.left.sm,
            gamma.right          = gamma.right.sm,
            w                    = w.sm,  # custom weights for stat.matrix
            stat.matrix          = stat.matrix.H0
        )
        print("calculate hyperrectangle.bounds.H0.sm.mt")
        flush.console()
        # Bounds based on stat.matrix cols, mixed confidence interval types:
        res.h.bounds.H0.sm.mt <- calculate.hyperrectangle.bounds.with.interval.type(
            alpha                = alpha,
            gamma.left           = gamma.left.sm,
            gamma.right          = gamma.right.sm,
            w                    = w.sm,
            stat.matrix          = stat.matrix.H0,
            interval.types       = interval.types.sm.mt,
            default.left.bounds  = default.left.bounds.sm.mt,
            default.right.bounds = default.right.bounds.sm.mt
        )
        stat.matrix.H0.p.comp <- princomp(stat.matrix.H0)  # of type (\ref{data_table}), but with PCs as features
        print("calculate hyperrectangle.bounds.H0.pc.w1")
        flush.console()
        # Bounds based on principal components, custom weights:
        res.h.bounds.H0.pc.w1 <- calculate.hyperrectangle.bounds.with.interval.type(
            alpha                = alpha,
            gamma.left           = gamma.left.pc,
            gamma.right          = gamma.right.pc,
            w                    = w.pc,  # custom weights for principal components
            stat.matrix          = stat.matrix.H0.p.comp$scores,
            interval.types       = interval.types.pc,
            default.left.bounds  = default.left.bounds.pc,
            default.right.bounds = default.right.bounds.pc
        )
        print("calculate hyperrectangle.bounds.H0.pc.w2")
        flush.console()
        # Bounds based on principal components, variance based weights:
        res.h.bounds.H0.pc.w2 <- calculate.hyperrectangle.bounds.with.interval.type(
            alpha                = alpha,
            gamma.left           = gamma.left.pc,
            gamma.right          = gamma.right.pc,
            w                    = stat.matrix.H0.p.comp$sdev^2 / sum(stat.matrix.H0.p.comp$sdev^2), #var based
            stat.matrix          = stat.matrix.H0.p.comp$scores,
            interval.types       = interval.types.pc,
            default.left.bounds  = default.left.bounds.pc,
            default.right.bounds = default.right.bounds.pc
        )

        for(j in 1:s){  # loop over alternatives
            print(paste("H1", j))
            flush.console()
            stat.matrix.H1 <- get.stat.matrix(sample.generation.functions.H1[[j]], n, m)  # of type (\ref{data_table_general}) 
            res.sm[i,j]    <- check.hyperrectangle.bounds(stat.matrix.H1, res.h.bounds.H0.sm   )
            res.sm.mt[i,j] <- check.hyperrectangle.bounds(stat.matrix.H1, res.h.bounds.H0.sm.mt)
            p.comp.H1      <- predict(stat.matrix.H0.p.comp, stat.matrix.H1)  # of type (\ref{data_table_general}), but with PCs
            res.pc.w1[i,j] <- check.hyperrectangle.bounds(p.comp.H1,      res.h.bounds.H0.pc.w1)
            res.pc.w2[i,j] <- check.hyperrectangle.bounds(p.comp.H1,      res.h.bounds.H0.pc.w2)
        }
        print(paste("iteration", i, "done"))
        flush.console()
    }
    list(  # matrices with power estimates; each col is a power distribution for H0 vs some H1 
        res.sm    = res.sm,
        res.sm.mt = res.sm.mt,
        res.pc.w1 = res.pc.w1,
        res.pc.w2 = res.pc.w2
    )
}
\end{minted}

\noindent
We also modify the code for the final table preparation accordingly:

\begin{minted}[mathescape, linenos]{r}
create.rejection.table.mt <- function(res.rejections)
    list(
        sm.means    = round(apply(res.rejections$res.sm,    2, mean), 4),
        sm.mt.means = round(apply(res.rejections$res.sm.mt, 2, mean), 4),
        pc.w1.means = round(apply(res.rejections$res.pc.w1, 2, mean), 4),
        pc.w2.means = round(apply(res.rejections$res.pc.w2, 2, mean), 4),
        sm.sds    = formatC(apply(res.rejections$res.sm,    2, sd  ), format = "e", digits = 1),
        sm.mt.sds = formatC(apply(res.rejections$res.sm.mt, 2, sd  ), format = "e", digits = 1),
        pc.w1.sds = formatC(apply(res.rejections$res.pc.w1, 2, sd  ), format = "e", digits = 1),
        pc.w2.sds = formatC(apply(res.rejections$res.pc.w2, 2, sd  ), format = "e", digits = 1)
    )
\end{minted}

\noindent
We remind that
\begin{enumerate}
\item \texttt{sm} prefix is used for tests based on \texttt{stat.matrix}; 
\item \texttt{pc} prefix is used for tests based on principal components of \texttt{stat.matrix}; 
\item \texttt{pc.w1} is a test that uses equal weights for principal components;
\item \texttt{pc.w2} is a test that uses variance-based weights for principal components.
\end{enumerate}
The new \texttt{mt} prefix stands for ``mixed type'' and is used in the name of \texttt{sm.mt} test that can use custom bounds for confidence intervals. 

In the next section we will use this ``mixed type'' intervals logic to compute bounds for uniform order statistics, where the left bound for first order statistic is fixed at $0$, and the right bound for last order statistic is fixed at $1$.

\newpage
\subsection{Calculate rejections, uniform \texorpdfstring{$H_0$}{H0}, beta \texorpdfstring{$H_1$}{H1}}
\label{section:rejections:uniform:vs:beta}

In this subsection we are going to estimate powers and type 1 errors for 4 tests based on uniform order statistics. Null hypothesis $H_0 : \mathcal{P} = \mathrm{U}(0,1)$ is fixed, we need to decide on alternatives. The authors of the paper~\cite{sailynoja2022graphical}, which we will discuss in more detail later in the text, following~\cite{Stephens1974} and~\cite{MarhuendaMoralesPardo2005} suggest to test against the following families of distributions:
\begin{align}
A_a:\quad F_{1,a}(x)
&= 1-(1-x)^a,
\; 0 \le x \le 1,
\label{eq:Aa}\\[1ex]
B_a:\quad F_{2,a}(x)
&=
\begin{cases}
2^{a-1}x^a, & 0 \le x \le 0.5, \\
1-2^{a-1}(1-x)^a, & 0.5 \le x \le 1,
\end{cases}
\label{eq:Ba}\\[1ex]
C_a:\quad F_{3,a}(x)
&=
\begin{cases}
0.5-2^{a-1}(0.5-x)^a, & 0 \le x \le 0.5, \\
0.5+2^{a-1}(x-0.5)^a, & 0.5 \le x \le 1,
\end{cases}
\label{eq:Ca}
\end{align}
where $a > 0$. These families cover different situations: more points in the center of the support interval, more points on both edges, or near one of the edges.

Instead of families~\eqref{eq:Aa}--\eqref{eq:Ca} in subsequent numerical studies we will use beta distributions. Beta distributions were tested against uniform, for example, in~\cite{zhang2002powerful,zhang2001powerful}. While not being entirely equal to~\eqref{eq:Aa}--\eqref{eq:Ca}, beta family covers very similar situations and even more. We start our study with the following types of alternatives:

\begin{figure}[H]
\centering
\includegraphics[width=12cm]{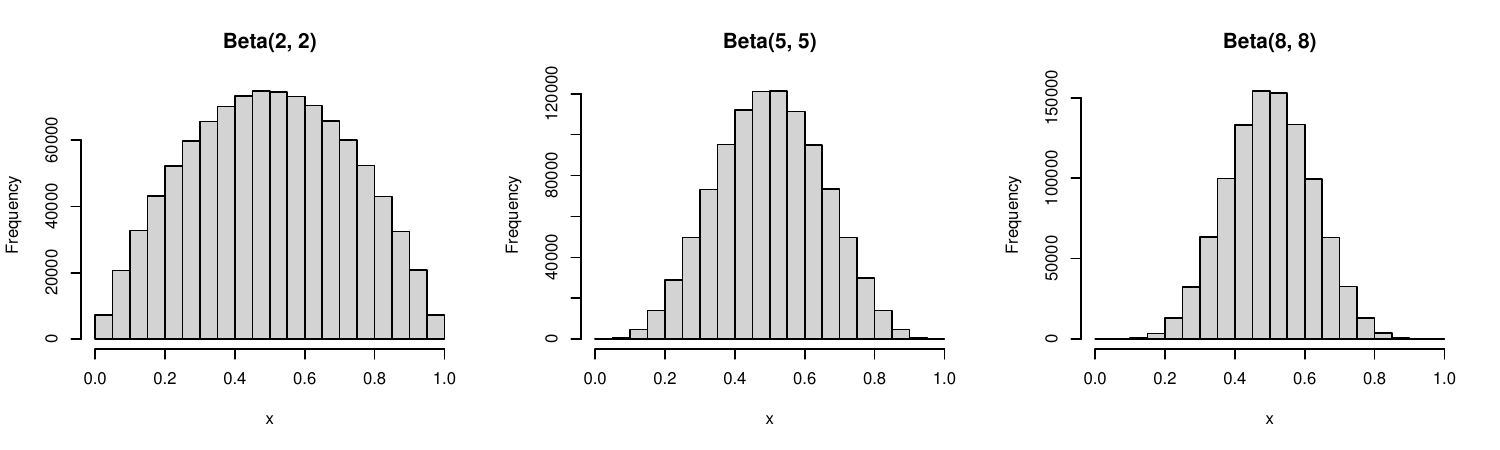}
\caption{$\mathrm{Beta}(a,a)$, $a > 1$; is symmetric,  has more points in the center;}
\label{figure:beta:aa}
\end{figure}

\begin{figure}[H]
\centering
\includegraphics[width=12cm]{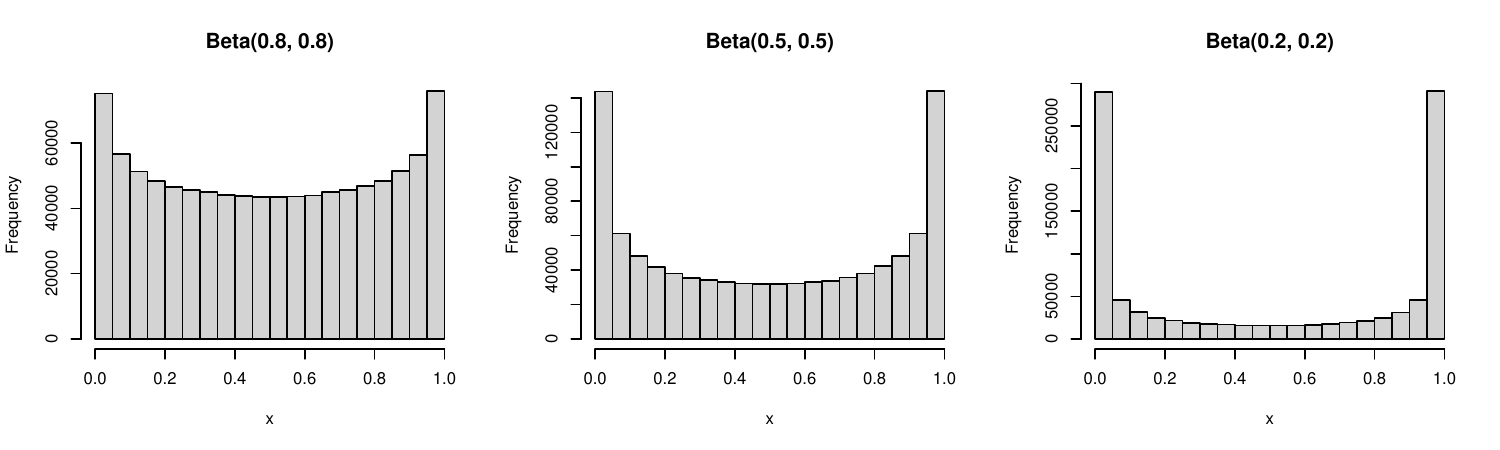}
\caption{$\mathrm{Beta}(a,a)$, $0 < a < 1$; is symmetric, has more points near the edges; }
\label{figure:beta:0a0a}
\end{figure}

\begin{figure}[H]
\centering
\includegraphics[width=12cm]{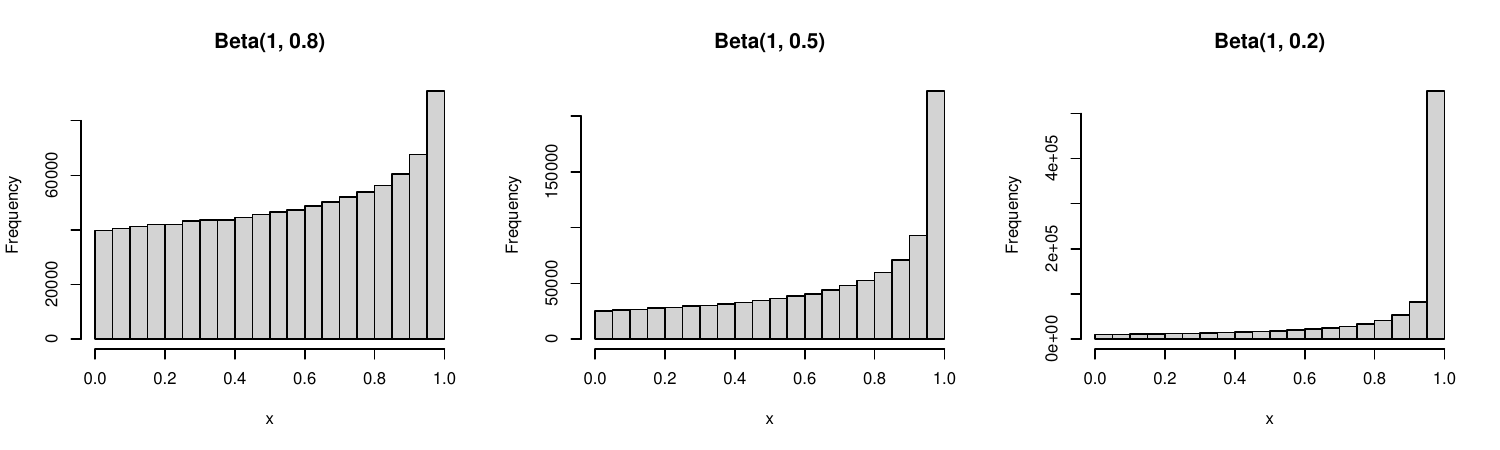}
\caption{$\mathrm{Beta(1,a)}$, $0 < a < 1$; has significant spike near the right edge;}
\label{figure:beta:10a}
\end{figure}

\begin{figure}[H]
\centering
\includegraphics[width=12cm]{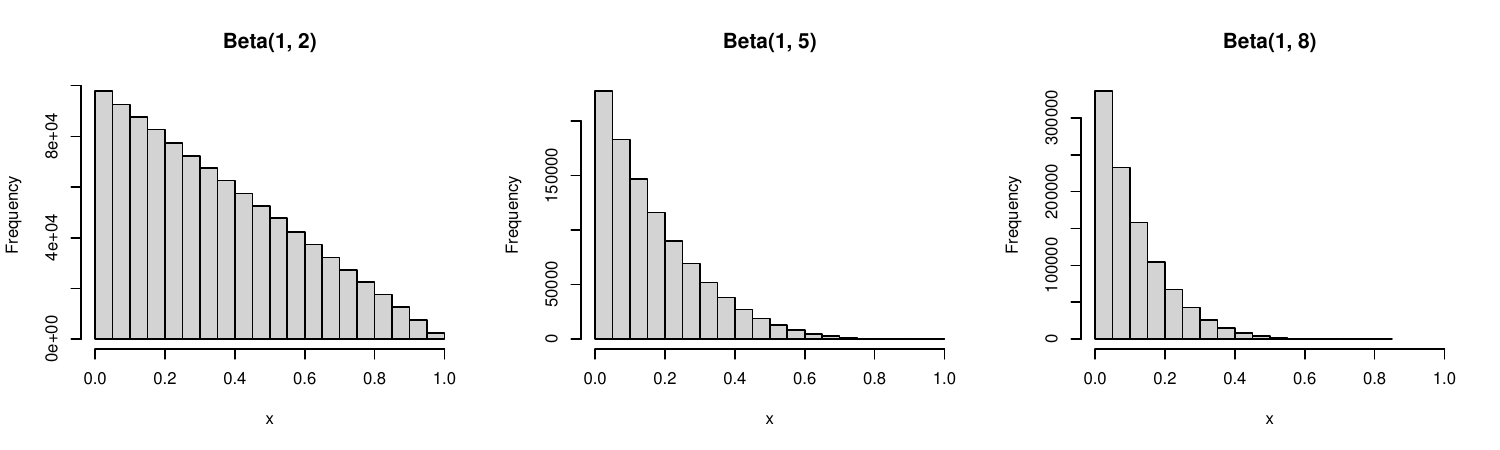}
\caption{$\mathrm{Beta(1,a)}$, $a > 1$; has sloping near the left edge.}
\label{figure:beta:1a}
\end{figure}

\noindent
All histograms were done on beta samples of size $1e6$.

Finally, let us proceed with calculating the powers of tests based on uniform order statistics:
\begin{minted}[mathescape, linenos, texcomments]{r}
res.uniform <- calc.rejections.sm.pc.bounds.with.interval.types(
    sample.generation.function.H0  = runif,   # null samples are from uniform
    sample.generation.functions.H1 = c(       # alternative samples are beta; see figures \ref{figure:beta:aa}-\ref{figure:beta:1a}
        lapply(1:10, function(x) function(nsamples) rbeta(nsamples, x, x)),
        lapply(rev(seq(0.1, 1.0, 0.1)), function(x) function(nsamples) rbeta(nsamples, x, x)),
        lapply(rev(seq(0.1, 1.0, 0.1)), function(x) function(nsamples) rbeta(nsamples, 1, x)),
        lapply(1:10, function(x) function(nsamples) rbeta(nsamples, 1, x))
    ),
    get.stat.matrix = get.sorted.samples.cpp, # we sort H0/H1 samples to get matrices of type (\ref{data_table_general})
    alpha = 0.05,                             # significance level
    n = n, m = m,                             # these are defined in subsection \ref{section:sample_size_num_samples}
    h = 30,                                   # how many full simulation cycles
    interval.types.sm.mt = c(rep("right side", 1), rep("two sides", 8), rep("left side", 1)),  
    default.left.bounds.sm.mt  = rep(0, n),
    default.right.bounds.sm.mt = rep(1, n)    # see subsection \ref{section:calc_bounds_uniform}
)
\end{minted}
The tables are like this:
\begin{minted}[mathescape, linenos]{r}
create.rejection.table.mt(res.uniform)
\end{minted}

\begin{table}[H]
\centering
\begin{tabular}{ |l|l||>{\columncolor[gray]{0.9}}p{1cm}|p{1.1cm}||>{\columncolor[gray]{0.9}}p{1cm}|p{1.1cm}||>{\columncolor[gray]{0.9}}p{1cm}|p{1.1cm}|| >{\columncolor[gray]{0.9}}p{1cm}|p{1.1cm}| } 
 \hline
  $H_0$ & $H_1$ & \multicolumn{2}{c||}{OS(U)} &  \multicolumn{2}{c||}{OS.MT(U)} & \multicolumn{2}{c||}{PC1(U)} & \multicolumn{2}{c|}{PC2(U)} \\
 \hline
 &  & mean & sd & mean & sd & mean & sd & mean & sd \\
 \hline
 \hline
 
$\mathrm{U}(0, 1)$ & $\mathrm{Beta}(1, 1)$ & 0.05 & 3.1e-04 & 0.05 & 3.1e-04 & 0.05 & 3.4e-04 & 0.05 & 3.5e-04\\
\hline
$\mathrm{U}(0, 1)$ & $\mathrm{Beta}(2, 2)$ & 0.0169 & 2.5e-04 & 0.0371 & 4.4e-04 & 0.0467 & 5.5e-04 & 0.0502 & 5.5e-04\\
$\mathrm{U}(0, 1)$ & $\mathrm{Beta}(3, 3)$ & 0.0219 & 3.7e-04 & 0.0602 & 7.5e-04 & 0.1251 & 1.4e-03 & 0.1458 & 1.1e-03\\
$\mathrm{U}(0, 1)$ & $\mathrm{Beta}(4, 4)$ & 0.0289 & 5.4e-04 & 0.09 & 1.3e-03 & 0.2451 & 2.3e-03 & 0.284 & 2.0e-03\\
$\mathrm{U}(0, 1)$ & $\mathrm{Beta}(5, 5)$ & 0.0373 & 7.2e-04 & 0.1245 & 1.7e-03 & 0.3817 & 3.0e-03 & 0.4329 & 2.4e-03\\
$\mathrm{U}(0, 1)$ & $\mathrm{Beta}(6, 6)$ & 0.0467 & 9.6e-04 & 0.1627 & 2.2e-03 & 0.5144 & 3.1e-03 & 0.5704 & 2.4e-03\\
$\mathrm{U}(0, 1)$ & $\mathrm{Beta}(7, 7)$ & 0.0571 & 1.2e-03 & 0.2039 & 2.6e-03 & 0.6311 & 3.2e-03 & 0.6858 & 2.2e-03\\
$\mathrm{U}(0, 1)$ & $\mathrm{Beta}(8, 8)$ & 0.0682 & 1.5e-03 & 0.2469 & 3.4e-03 & 0.7272 & 2.9e-03 & 0.7763 & 2.0e-03\\
$\mathrm{U}(0, 1)$ & $\mathrm{Beta}(9, 9)$ & 0.0803 & 1.8e-03 & 0.291 & 3.6e-03 & 0.8031 & 2.5e-03 & 0.8447 & 1.7e-03\\
$\mathrm{U}(0, 1)$ & $\mathrm{Beta}(10, 10)$ & 0.0932 & 2.1e-03 & 0.3358 & 4.1e-03 & 0.8603 & 2.1e-03 & 0.894 & 1.4e-03\\

 \hline

\end{tabular}
\caption{Tests based on uniform order statistics; $\mathrm{U}(0,1)$ vs $\mathrm{Beta}(a,a)$; $a \geq 1$; alternatives have modes in the center, see figure~\ref{figure:beta:aa}.}
\label{table:beta:aa:uniform:os}
\end{table}

\noindent
In table~\ref{table:beta:aa:uniform:os} the powers of 4 tests based on uniform order statistics against beta alternatives of type from figure~\ref{figure:beta:aa} are presented. Notations for tests are as follows: 
\begin{enumerate}
\item ``OS(U)'' stands for the direct order statistics based test described in~\cite{AldorNoimanBrownBujaRolkeStine2013}, but computed for uniform order statistics instead of normal ones;
\item ``OS.MT(U)'' stands for the direct uniform order statistics test, but with one-sided intervals for first and last order statistics;
\item ``PC1(U)'' stands for the test based on principal components of uniform order statistics with equal weights for principal components;
\item ``PC2(U)'' stands for the test based on principal components of uniform order statistics with variance-based weights for principal components.
\end{enumerate}
First row of the table gives the type 1 error and is presented as a sanity check. We see that type 1 error coincides with given significance level \texttt{alpha}. In ``mean'' and ``sd'' columns there are means and standard deviations for powers and type 1 errors aggregated across \texttt{h} runs. 

We also see that ``OS(U)'' test has very low power against beta alternative that concentrates points in the center, ``OS.MT(U)'' has better power, and ``PC1(U)''/``PC2(U)'' have much better power with ``PC2(U)'' being the best. Maximum value of power for ``OS(U)'' is $0.0932$, which means that this test does not work at all for presented alternatives. Maximum for ``OS.MT(U)'' is $0.3358$, 3 times larger than for ``OS(U)'', which means that the idea to try one-sided bounds for edge order statistics has its merit. ``PC1(U)'' and ``PC2(U)'' have maximums of $0.8603$ and $0.894$, which are very good.

\begin{table}[H]
\centering
\begin{tabular}{ |l|l||>{\columncolor[gray]{0.9}}p{1cm}|p{1.1cm}||>{\columncolor[gray]{0.9}}p{1cm}|p{1.1cm}||>{\columncolor[gray]{0.9}}p{1cm}|p{1.1cm}|| >{\columncolor[gray]{0.9}}p{1cm}|p{1.1cm}| } 
 \hline
  $H_0$ & $H_1$ & \multicolumn{2}{c||}{OS(U)} &  \multicolumn{2}{c||}{OS.MT(U)} & \multicolumn{2}{c||}{PC1(U)} & \multicolumn{2}{c|}{PC2(U)} \\
 \hline
 &  & mean & sd & mean & sd & mean & sd & mean & sd \\

 \hline
 \hline
 
$\mathrm{U}(0, 1)$ & $\mathrm{Beta}(1, 1)$ & 0.05 & 2.9e-04 & 0.05 & 3.3e-04 & 0.05 & 3.1e-04 & 0.0499 & 2.7e-04\\
\hline
$\mathrm{U}(0, 1)$ & $\mathrm{Beta}(0.9, 0.9)$ & 0.0754 & 4.3e-04 & 0.0686 & 4.1e-04 & 0.0629 & 3.4e-04 & 0.0642 & 3.7e-04\\
$\mathrm{U}(0, 1)$ & $\mathrm{Beta}(0.8, 0.8)$ & 0.1189 & 5.1e-04 & 0.1012 & 5.5e-04 & 0.0831 & 4.8e-04 & 0.0866 & 3.2e-04\\
$\mathrm{U}(0, 1)$ & $\mathrm{Beta}(0.7, 0.7)$ & 0.192 & 6.1e-04 & 0.157 & 6.1e-04 & 0.1155 & 6.1e-04 & 0.1223 & 3.9e-04\\
$\mathrm{U}(0, 1)$ & $\mathrm{Beta}(0.6, 0.6)$ & 0.3081 & 8.8e-04 & 0.2484 & 8.1e-04 & 0.1676 & 6.8e-04 & 0.1797 & 6.3e-04\\
$\mathrm{U}(0, 1)$ & $\mathrm{Beta}(0.5, 0.5)$ & 0.4769 & 1.0e-03 & 0.3889 & 9.6e-04 & 0.2517 & 1.0e-03 & 0.271 & 1.0e-03\\
$\mathrm{U}(0, 1)$ & $\mathrm{Beta}(0.4, 0.4)$ & 0.6857 & 9.6e-04 & 0.5817 & 9.3e-04 & 0.3829 & 1.1e-03 & 0.4105 & 1.1e-03\\
$\mathrm{U}(0, 1)$ & $\mathrm{Beta}(0.3, 0.3)$ & 0.8785 & 4.7e-04 & 0.7951 & 6.8e-04 & 0.5706 & 1.4e-03 & 0.603 & 1.2e-03\\
$\mathrm{U}(0, 1)$ & $\mathrm{Beta}(0.2, 0.2)$ & 0.9818 & 1.9e-04 & 0.9519 & 2.9e-04 & 0.7932 & 9.4e-04 & 0.8198 & 9.0e-04\\
$\mathrm{U}(0, 1)$ & $\mathrm{Beta}(0.1, 0.1)$ & 0.9998 & 1.4e-05 & 0.9985 & 3.5e-05 & 0.9635 & 3.2e-04 & 0.9714 & 2.5e-04\\
 
  \hline

\end{tabular}
\caption{Tests based on uniform order statistics; $\mathrm{U}(0,1)$ vs $\mathrm{Beta}(a,a)$; $0 < a \leq 1$; alternatives have modes on both sides, see figure~\ref{figure:beta:0a0a}.}
\label{table:beta:0a0a:uniform:os}
\end{table}

\noindent
In table~\ref{table:beta:0a0a:uniform:os} we see the opposite ranking. For symmetric beta alternatives that are concentrated on the edges of the $[0,1]$ interval ``OS(U)'' test shows the best power, ``OS.MT(U)'' is slightly behind, then goes ``PC2(U)'', and ``PC1(U)'' is the last. Though the differences between all 4 tests are not as drastic as in table~\ref{table:beta:aa:uniform:os}. Maximum powers for all 4 tests are around 1, which means that all 4 tests are able to detect the difference between uniform null and both-edge-concentarted alternatives, but power curve for ``OS(U)'' is noticeably above power curve for ``PC2(U)''.

\begin{table}[H]
\centering
\begin{tabular}{ |l|l||>{\columncolor[gray]{0.9}}p{1cm}|p{1.1cm}||>{\columncolor[gray]{0.9}}p{1cm}|p{1.1cm}||>{\columncolor[gray]{0.9}}p{1cm}|p{1.1cm}|| >{\columncolor[gray]{0.9}}p{1cm}|p{1.1cm}| } 
 \hline
 $H_0$ & $H_1$ & \multicolumn{2}{c||}{OS(U)} &  \multicolumn{2}{c||}{OS.MT(U)} & \multicolumn{2}{c||}{PC1(U)} & \multicolumn{2}{c|}{PC2(U)} \\
 \hline
 &  & mean & sd & mean & sd & mean & sd & mean & sd \\
 
 \hline
 \hline

$\mathrm{U}(0, 1)$ & $\mathrm{Beta}(1, 1)$ & 0.05 & 3.5e-04 & 0.05 & 3.0e-04 & 0.0501 & 2.8e-04 & 0.0501 & 2.8e-04\\
\hline
$\mathrm{U}(0, 1)$ & $\mathrm{Beta}(1, 0.9)$ & 0.0694 & 3.0e-04 & 0.0663 & 2.8e-04 & 0.0583 & 3.5e-04 & 0.0636 & 3.3e-04\\
$\mathrm{U}(0, 1)$ & $\mathrm{Beta}(1, 0.8)$ & 0.117 & 6.2e-04 & 0.1091 & 5.4e-04 & 0.0756 & 4.2e-04 & 0.0988 & 4.7e-04\\
$\mathrm{U}(0, 1)$ & $\mathrm{Beta}(1, 0.7)$ & 0.2094 & 8.0e-04 & 0.1935 & 7.7e-04 & 0.11 & 5.8e-04 & 0.1668 & 6.5e-04\\
$\mathrm{U}(0, 1)$ & $\mathrm{Beta}(1, 0.6)$ & 0.3611 & 1.0e-03 & 0.3345 & 9.2e-04 & 0.1761 & 7.2e-04 & 0.2798 & 7.3e-04\\
$\mathrm{U}(0, 1)$ & $\mathrm{Beta}(1, 0.5)$ & 0.5686 & 9.6e-04 & 0.5329 & 8.2e-04 & 0.2939 & 1.3e-03 & 0.4445 & 9.0e-04\\
$\mathrm{U}(0, 1)$ & $\mathrm{Beta}(1, 0.4)$ & 0.7867 & 7.8e-04 & 0.7525 & 6.7e-04 & 0.4781 & 1.4e-03 & 0.647 & 8.7e-04\\
$\mathrm{U}(0, 1)$ & $\mathrm{Beta}(1, 0.3)$ & 0.9394 & 3.5e-04 & 0.9205 & 4.4e-04 & 0.7091 & 1.5e-03 & 0.8398 & 6.4e-04\\
$\mathrm{U}(0, 1)$ & $\mathrm{Beta}(1, 0.2)$ & 0.9943 & 7.3e-05 & 0.9905 & 9.7e-05 & 0.909 & 6.7e-04 & 0.9621 & 2.7e-04\\
$\mathrm{U}(0, 1)$ & $\mathrm{Beta}(1, 0.1)$ & 1 & 5.4e-06 & 0.9999 & 8.2e-06 & 0.9934 & 9.7e-05 & 0.9982 & 3.9e-05\\ 

\hline

\end{tabular}
\caption{Tests based on uniform order statistics; $\mathrm{U}(0,1)$ vs $\mathrm{Beta}(1,a)$; $0 < a \leq 1$; alternatives have sharp modes on the right, see figure~\ref{figure:beta:10a}.}
\label{table:beta:10a:uniform:os}
\end{table}

\noindent
The rankings for the tests in table~\ref{table:beta:10a:uniform:os} are the same as the rankings in table~\ref{table:beta:0a0a:uniform:os}: ``OS(U)'', ``OS.MT(U)'', ``PC2(U)'', ``PC1(U)''. Again, for all 4 tests maximum power is close to 1. The difference between power curves for ``OS(U)'' and ``PC2(U)'' is lower then the same difference in table~\ref{table:beta:10a:uniform:os}.

\begin{table}[H]
\centering
\begin{tabular}{ |l|l||>{\columncolor[gray]{0.9}}p{1cm}|p{1.1cm}||>{\columncolor[gray]{0.9}}p{1cm}|p{1.1cm}||>{\columncolor[gray]{0.9}}p{1cm}|p{1.1cm}|| >{\columncolor[gray]{0.9}}p{1cm}|p{1.1cm}| } 
 \hline
 $H_0$ & $H_1$ & \multicolumn{2}{c||}{OS(U)} &  \multicolumn{2}{c||}{OS.MT(U)} & \multicolumn{2}{c||}{PC1(U)} & \multicolumn{2}{c|}{PC2(U)} \\
 \hline
 &  & mean & sd & mean & sd & mean & sd & mean & sd \\

 \hline
 \hline

$\mathrm{U}(0, 1)$ & $\mathrm{Beta}(1, 1)$ & 0.0501 & 2.5e-04 & 0.05 & 2.3e-04 & 0.05 & 2.8e-04 & 0.0501 & 2.8e-04\\
\hline
$\mathrm{U}(0, 1)$ & $\mathrm{Beta}(1, 2)$ & 0.3018 & 1.3e-03 & 0.3412 & 1.1e-03 & 0.1705 & 1.3e-03 & 0.3544 & 1.1e-03\\
$\mathrm{U}(0, 1)$ & $\mathrm{Beta}(1, 3)$ & 0.7181 & 1.6e-03 & 0.7622 & 1.2e-03 & 0.5326 & 2.5e-03 & 0.7886 & 1.1e-03\\
$\mathrm{U}(0, 1)$ & $\mathrm{Beta}(1, 4)$ & 0.9273 & 7.9e-04 & 0.9457 & 5.1e-04 & 0.8267 & 1.7e-03 & 0.9592 & 3.3e-04\\
$\mathrm{U}(0, 1)$ & $\mathrm{Beta}(1, 5)$ & 0.986 & 2.2e-04 & 0.9907 & 1.5e-04 & 0.9528 & 8.0e-04 & 0.9944 & 9.4e-05\\
$\mathrm{U}(0, 1)$ & $\mathrm{Beta}(1, 6)$ & 0.9978 & 6.8e-05 & 0.9987 & 4.3e-05 & 0.9898 & 2.6e-04 & 0.9994 & 2.4e-05\\
$\mathrm{U}(0, 1)$ & $\mathrm{Beta}(1, 7)$ & 0.9997 & 1.7e-05 & 0.9998 & 1.4e-05 & 0.9981 & 7.0e-05 & 0.9999 & 7.8e-06\\
$\mathrm{U}(0, 1)$ & $\mathrm{Beta}(1, 8)$ & 1 & 7.2e-06 & 1 & 3.9e-06 & 0.9997 & 1.7e-05 & 1 & 2.0e-06\\
$\mathrm{U}(0, 1)$ & $\mathrm{Beta}(1, 9)$ & 1 & 2.1e-06 & 1 & 1.2e-06 & 1 & 5.9e-06 & 1 & 5.2e-07\\
$\mathrm{U}(0, 1)$ & $\mathrm{Beta}(1, 10)$ & 1 & 5.8e-07 & 1 & 3.5e-07 & 1 & 2.1e-06 & 1 & 0.0e+00\\
 
      \hline
\end{tabular}
\caption{Tests based on uniform order statistics; $\mathrm{U}(0,1)$ vs $\mathrm{Beta}(1,a)$; $a \geq 1$; alternatives have sloping modes on the left, see figure~\ref{figure:beta:1a}.}
\label{table:beta:1a:uniform:os}
\end{table}

\noindent
In table~\ref{table:beta:1a:uniform:os} ``PC2(U)'' has the best power, ``OS.MT(U)'' and ``OS(U)'' are slightly behind, ``PC1(U)'' is the last. The overall difference between ``PC2(U)'' and ``OS.MT(U)'' is very insignificant.

Intermediate conclusions: 
\begin{enumerate}
\item there is no situation where ``PC1(U)'' is better than ``PC2(U)'' so far;
\item ``OS.MT(U)'' is slightly more balanced than ``OS(U)'';
\item ``PC2(U)'' is the most balanced test among the 4 studied tests, since it performs well for all checked alternatives; ``OS(U)'' is better for $\mathrm{Beta}(a,a)$ and $\mathrm{Beta}(1,a)$ alternatives when $0 < a < 1$, but it does not work at all for $\mathrm{Beta(a,a)}$, $a > 1$ alternative.
\end{enumerate}

\noindent
Let us continue our study by adding even more beta alternatives:

\begin{figure}[H]
\centering
\includegraphics[width=12cm]{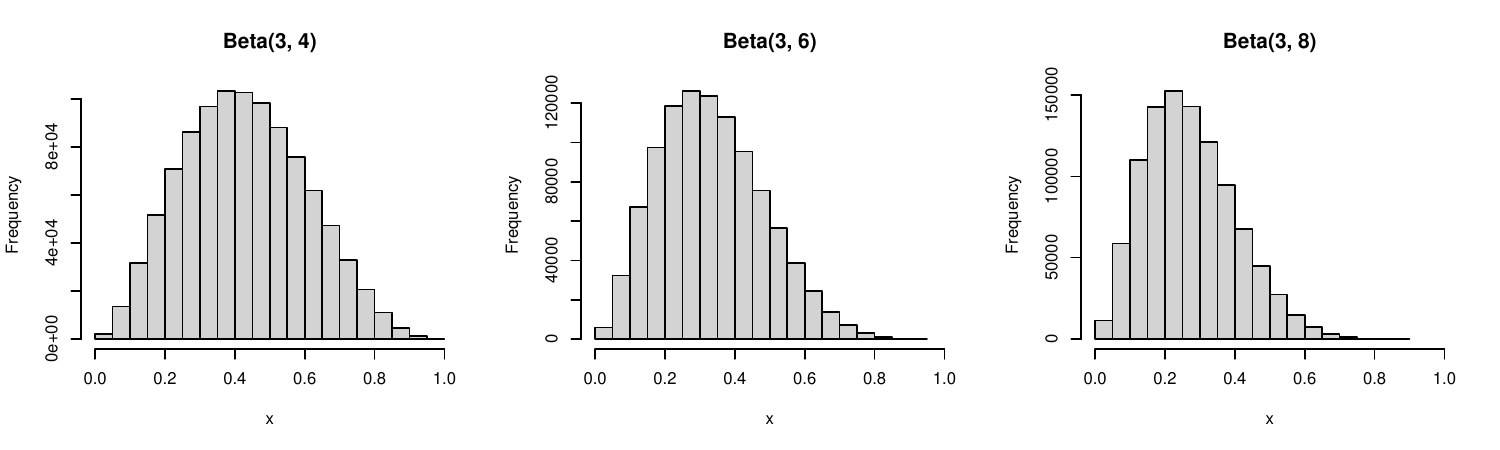}
\caption{$\mathrm{Beta}(a,b)$, $a, b > 1$, $a \neq b$; asymmetric with more points near the center;}
\label{figure:beta:3a}
\end{figure}

\begin{figure}[H]
\centering
\includegraphics[width=12cm]{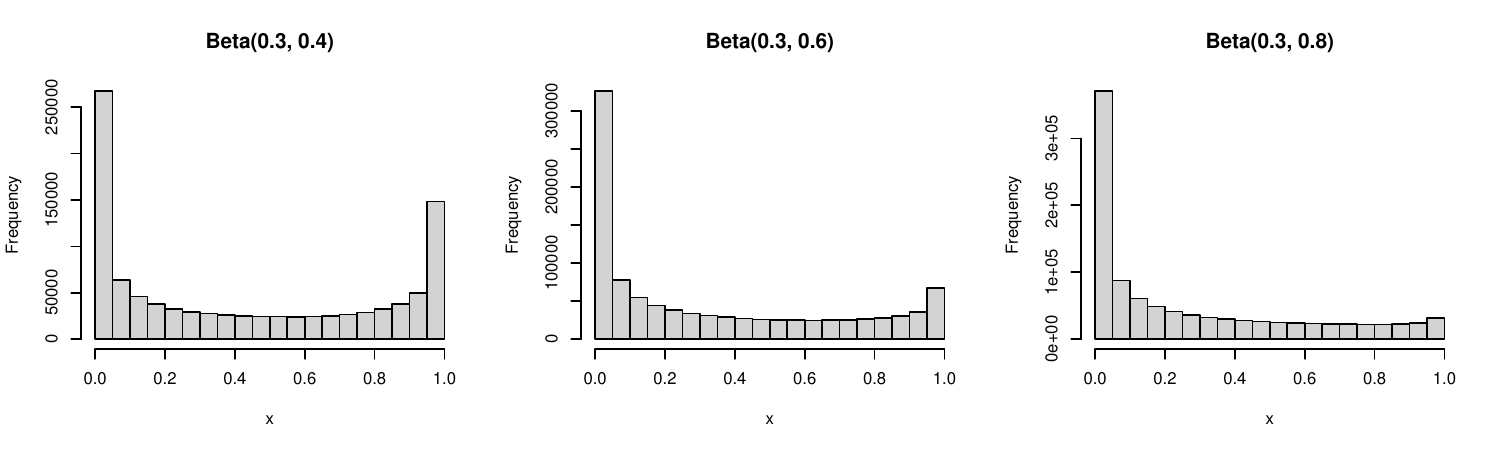}
\caption{$\mathrm{Beta}(a,b)$, $0 < a, b < 1$, $a \neq b$; asymmetric with more points near the edges;}
\label{figure:beta:030a}
\end{figure}

\begin{figure}[H]
\centering
\includegraphics[width=12cm]{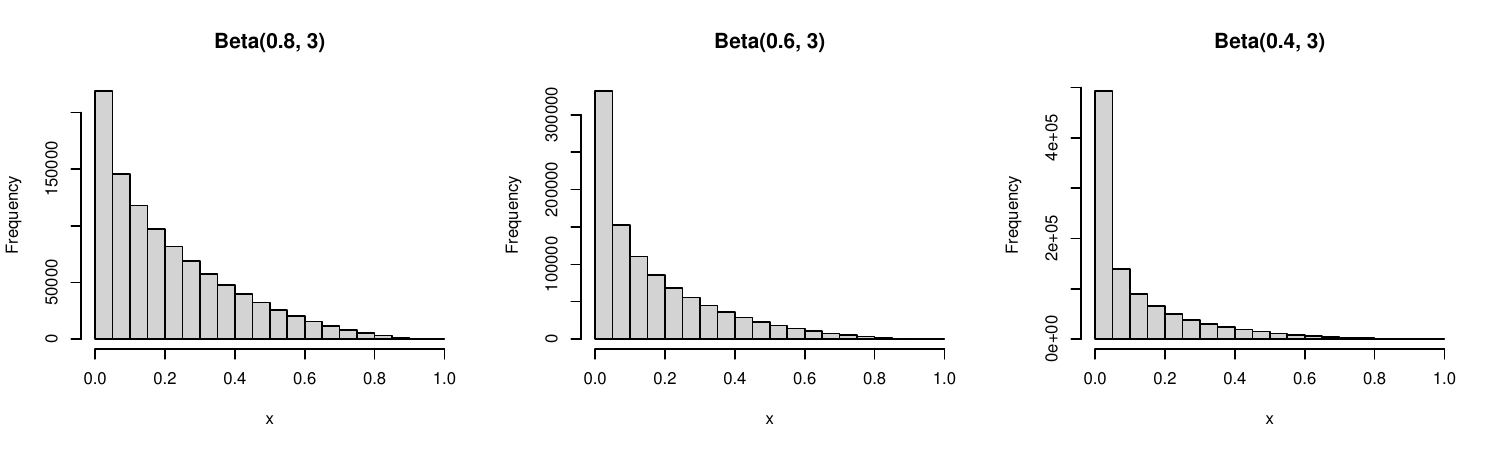}
\caption{$\mathrm{Beta}(a,b)$, $0 < a < 1$, $b > 1$; asymmetric with a long slope and a spike near one of the edges.}
\label{figure:beta:0a3}
\end{figure}

\noindent
We introduce the following grid for parameters of beta distribution:
\begin{minted}[mathescape, linenos]{r}
beta.1d.grid <- c(seq(0.1, 0.9, 0.1), 1:10)
beta.parameters.grid <- expand.grid(V1 = beta.1d.grid, V2 = beta.1d.grid)
beta.parameters.grid <- beta.parameters.grid[beta.parameters.grid$V1 <= beta.parameters.grid$V2,]
\end{minted}
The corresponding set of alternatives is as follows:
\begin{align}
\label{eq:beta_grid}
\big\{ \mathrm{Beta}(a, b) \; | \; a, b \in \left\{ 0.1, 0.2, \dots, 0.9, 1, 2, \dots, 10 \right\}, a \leq b \big\}.
\end{align}
$\mathrm{Beta}(a^*,b^*)$ has the same power as $\mathrm{Beta}(b^*,a^*)$ when tested against $\mathrm{U}(0,1)$ due to symmetry, so we only need to check betas with $a \leq b$.

Let us compute power estimates for the introduced beta grid:

\begin{minted}[mathescape, linenos, texcomments]{r}
res.uniform.beta.grid <- calc.rejections.sm.pc.bounds.with.interval.types(
    sample.generation.function.H0  = runif,
    sample.generation.functions.H1 = 
        apply(
            beta.parameters.grid,  # see formula \eqref{eq:beta_grid}
            1, 
            function(par) function(nsamples) rbeta(nsamples, par[1], par[2])
        ),
    get.stat.matrix = get.sorted.samples.cpp,
    n = n, m = m,
    alpha = 0.05,
    h = 10,
    interval.types.sm.mt = c(rep("right side", 1), rep("two sides", 8), rep("left side", 1)),
    default.left.bounds.sm.mt  = rep(0, n),
    default.right.bounds.sm.mt = rep(1, n)
)
\end{minted}

\noindent
The table creation code is as follows:
\begin{minted}[mathescape, linenos]{r}
res.uniform.beta.grid.table <- create.rejection.table.mt(res.uniform.beta.grid)
res.uniform.beta.grid.table
\end{minted}
This table is too large to be printed in the text as an ordinary table. 

Instead, let us write a function to present each ``mean'' column of this table with a tile plot: 
\begin{minted}[mathescape, linenos]{r}
plot.powers.on.grid <- function(grid, rejection.means)
{
    df = data.frame(V1 = factor(grid$V1), V2 = factor(grid$V2), val = rejection.means)
    ggplot(df, aes(V1, V2, fill = val)) +
        geom_tile() +
        scale_fill_gradient(low = "white", high = "blue", limits = c(0,1)) +
        coord_fixed()
}
\end{minted}

\newpage

\paragraph{Tile plot for ``OS(U)'': beta parameters grid}

\begin{minted}[mathescape, linenos]{r}
plot.powers.on.grid(beta.parameters.grid, res.uniform.beta.grid.table$sm.means)
\end{minted}

\begin{figure}[H]
\centering
\includegraphics[width=16cm]{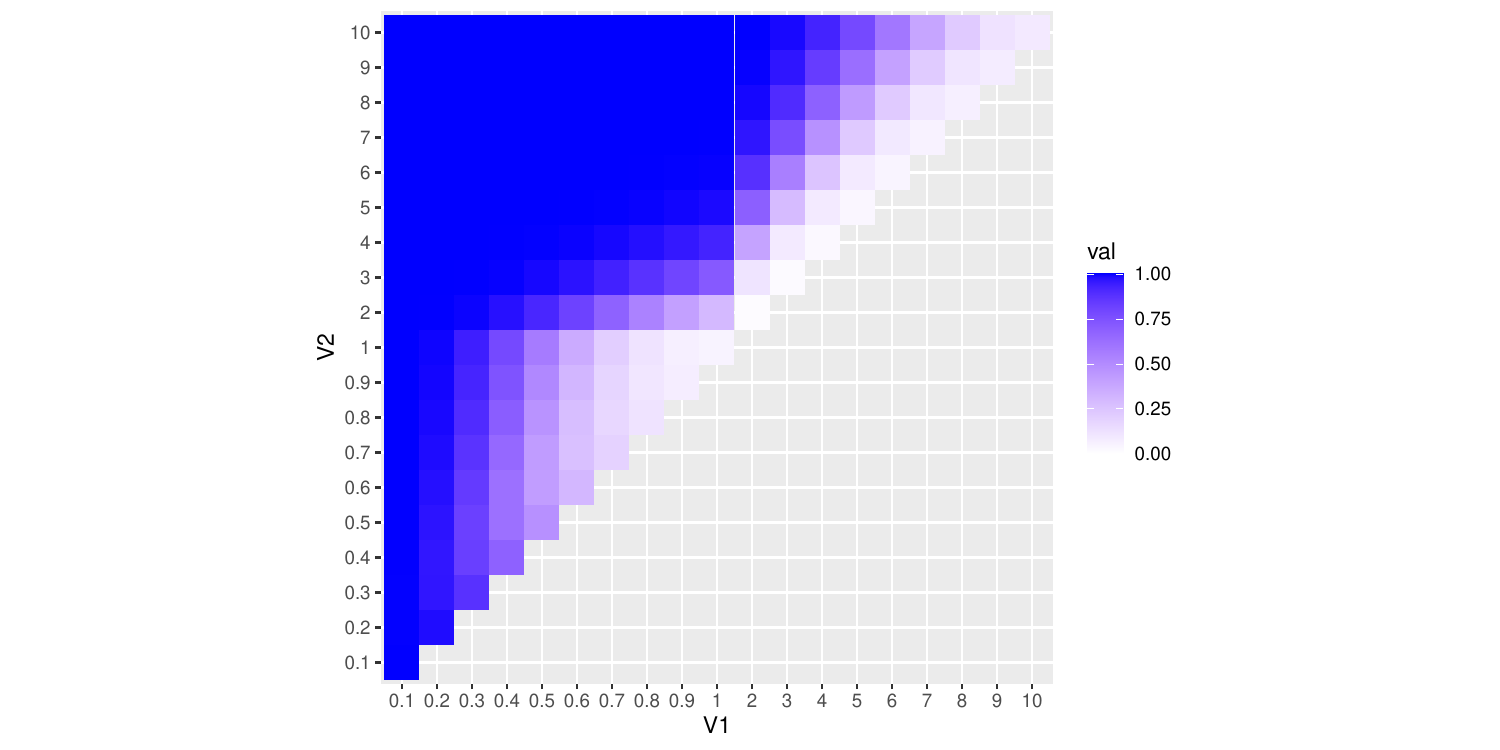}
\caption{``OS(U)'' test powers on beta grid of alternatives~\eqref{eq:beta_grid}.}
\label{figure:beta_grid:osu}
\end{figure}

\paragraph{Tile plot for ``OS.MT(U)'': beta parameters grid}

\begin{minted}[mathescape, linenos]{r}
plot.powers.on.grid(beta.parameters.grid, res.uniform.beta.grid.table$sm.mt.means)
\end{minted}

\begin{figure}[H]
\centering
\includegraphics[width=16cm]{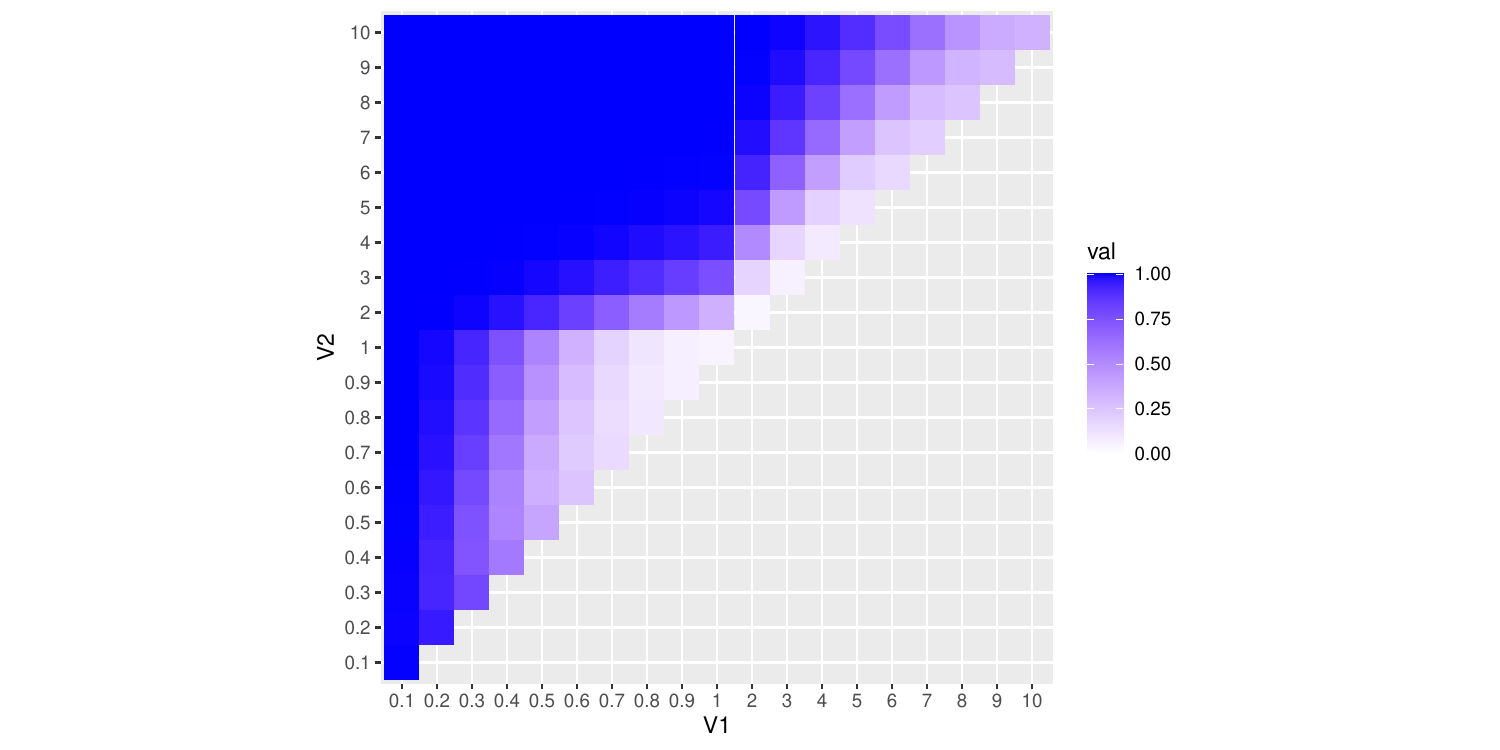}
\caption{``OS.MT(U)'' test powers on beta grid of alternatives~\eqref{eq:beta_grid}.}
\label{figure:beta_grid:osmtu}
\end{figure}

\newpage

\paragraph{Tile plot for ``PC1(U)'': beta parameters grid}

\begin{minted}[mathescape, linenos]{r}
plot.powers.on.grid(beta.parameters.grid, res.uniform.beta.grid.table$pc.w1.means)
\end{minted}

\begin{figure}[H]
\centering
\includegraphics[width=16cm]{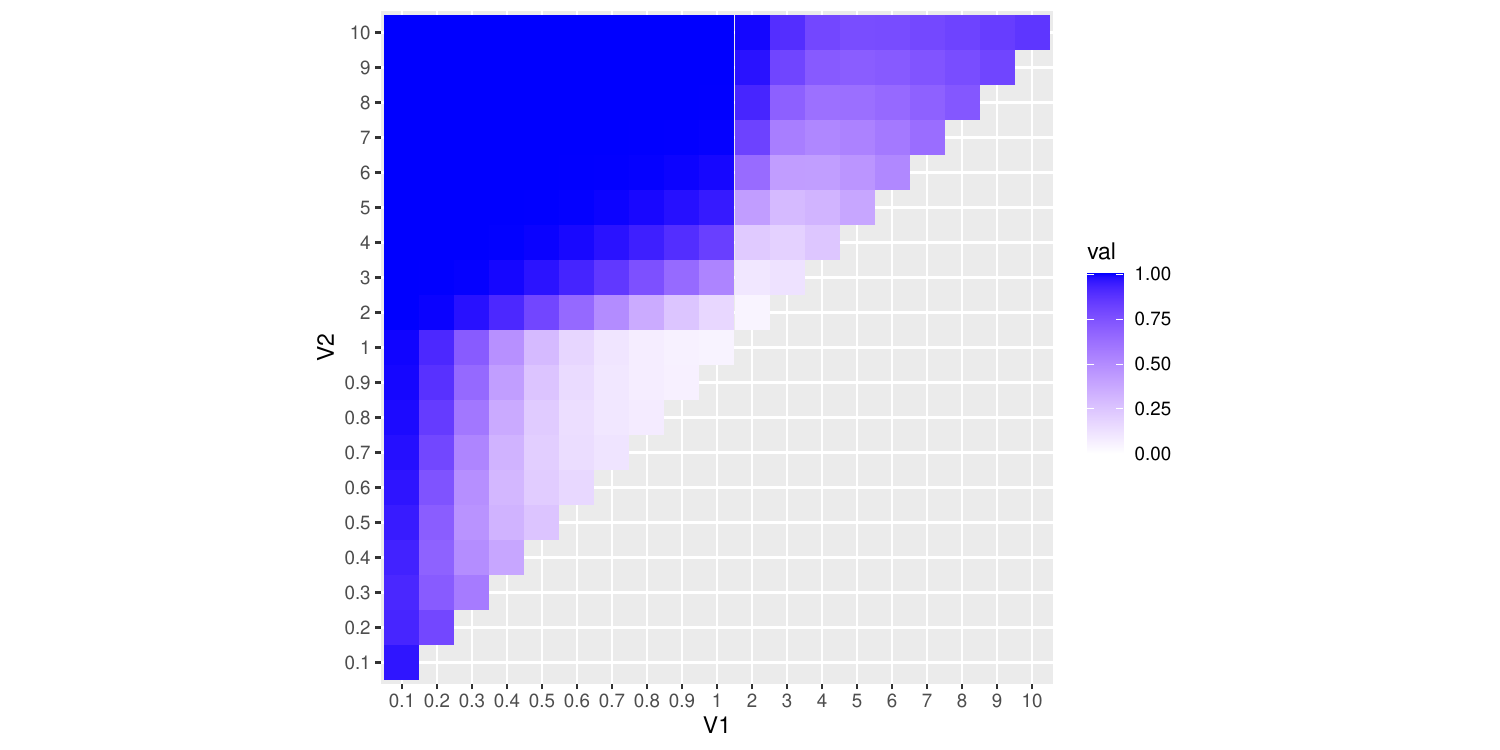}
\caption{``PC1(U)'' test powers on beta grid of alternatives~\eqref{eq:beta_grid}.}
\label{figure:beta_grid:pc1u}
\end{figure}

\paragraph{Tile plot for ``PC2(U)'': beta parameters grid}

\begin{minted}[mathescape, linenos]{r}
plot.powers.on.grid(beta.parameters.grid, res.uniform.beta.grid.table$pc.w2.means)
\end{minted}

\begin{figure}[H]
\centering
\includegraphics[width=16cm]{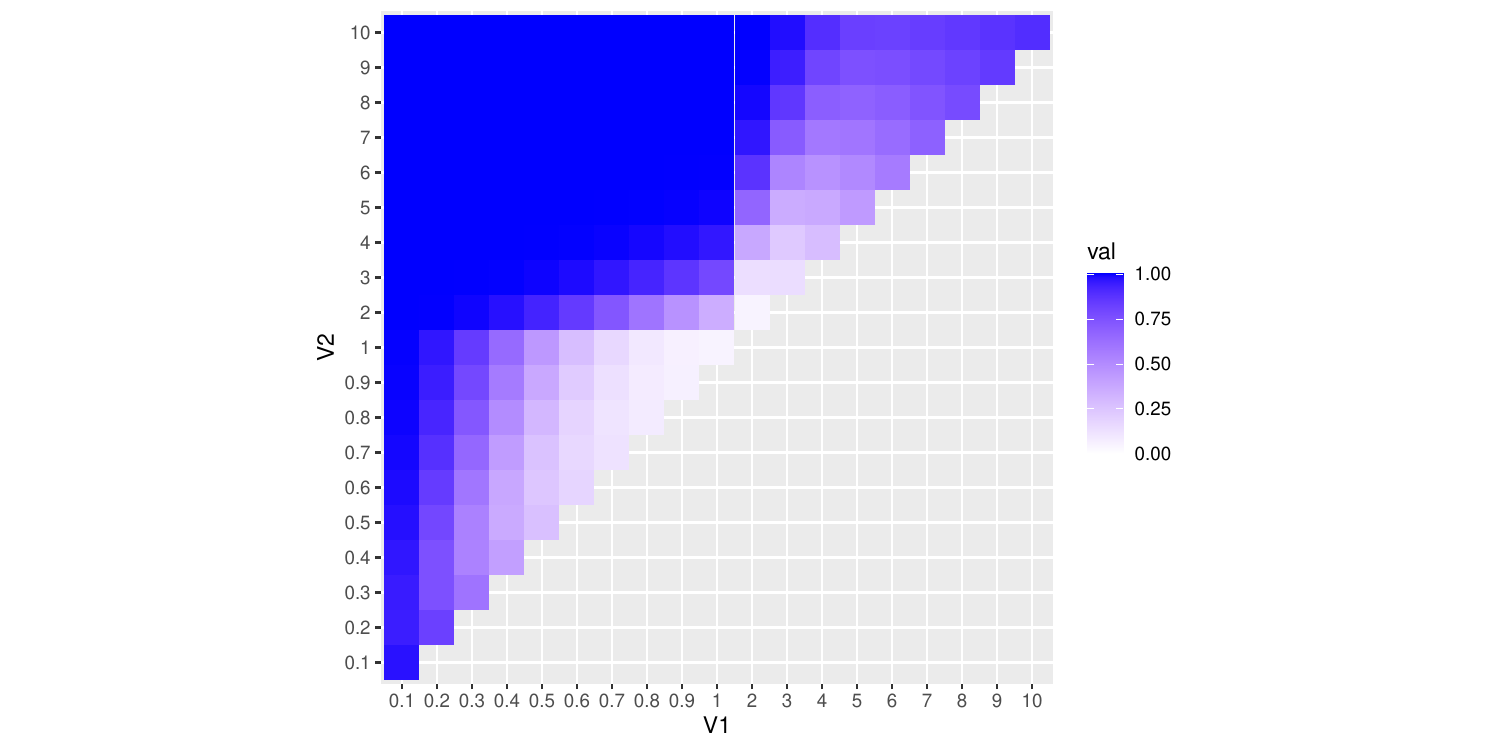}
\caption{``PC2(U)'' test powers on beta grid of alternatives~\eqref{eq:beta_grid}.}
\label{figure:beta_grid:pc2u}
\end{figure}

\newpage

\subsubsection{Comments to the tile plots}

Figures~\ref{figure:beta_grid:osu}--\ref{figure:beta_grid:pc2u} illustrate the powers of the uniform-order-statistics-based tests for uniform null against different beta alternatives defined in~\eqref{eq:beta_grid}. The parts of the plots under diagonal are left blank due to symmetry of beta alternatives mentioned under formula~\eqref{eq:beta_grid}. Point (1,1) on all the plots corresponds to uniform $H_1$ distribution ($\mathrm{Beta}(1,1)$) and presents type 1 error instead of power.

On figure~\ref{figure:beta_grid:osu} we again see that ``OS(U)'' test does not work well for $\mathrm{Beta}(a,a)$, $a > 1$ alternatives (figure~\ref{figure:beta:aa}): the corresponding tiles on upper part of the plot's diagonal are white. We also see that for $\mathrm{Beta}(a, b)$, $a, b > 1$, $a \neq b$ alternatives (figure~\ref{figure:beta:3a}), the larger is the difference between $a$ and $b$, i.e. the density's asymmetry, the larger is the test's power: see tiles in the upper right triangle of figure~\ref{figure:beta_grid:osu}. For $\mathrm{Beta}(a, b)$, $0 < a \leq 1$, $b > 1$ (figures~\ref{figure:beta:0a3} and~\ref{figure:beta:1a}) the test performs very well: see the upper left rectangle of the tile plot. For $\mathrm{Beta}(a,b)$, $0 < a,b \leq 1$ (figures~\ref{figure:beta:030a},~\ref{figure:beta:0a0a}, and~\ref{figure:beta:10a}) the power depends mostly on the distance between 1 and $\min(a,b)$: see lower left triangle of the plot. 

On figure~\ref{figure:beta_grid:osmtu} we see that ``OS.MT(U)'' test performs a bit better for $\mathrm{Beta}(a, b)$, $a, b > 1$ alternatives (figures~\ref{figure:beta:aa} and~\ref{figure:beta:3a}; upper right triangle on figure~\ref{figure:beta_grid:osmtu}), and slightly worse for $\mathrm{Beta}(a,b)$, $0 < a,b \leq 1$ alternatives (figures~\ref{figure:beta:030a},~\ref{figure:beta:0a0a}, and~\ref{figure:beta:10a}; lower left triangle on figure~\ref{figure:beta_grid:osmtu}), if compared to ``OS(U)'' test. The power for $\mathrm{Beta}(a, b)$, $0 < a \leq 1$, $b > 1$ (figures~\ref{figure:beta:0a3} and~\ref{figure:beta:1a}; upper left rectangle on figure~\ref{figure:beta_grid:osmtu}) looks the same. We note again that``OS.MT(U)'' test better captures the difference for symmetric centered alternatives than ``OS(U)'' test.

Figure~\ref{figure:beta_grid:pc1u} looks like a weaker variant of figure~\ref{figure:beta_grid:pc2u}. There is no beta alternative for which ``PC1(U)'' test performs better than ``PC2(U)'' test.

``PC2(U)'' test presented by figure~\ref{figure:beta_grid:pc2u}, if compared to ``OS(U)'' and ``OS.MT(U)'' tests, performs even better for $\mathrm{Beta}(a, b)$, $a, b > 1$ alternatives, and even worse for $\mathrm{Beta}(a,b)$, $0 < a,b \leq 1$ alternatives. We can say that ``OS.MT(U)'' test is the more balanced version of ``OS(U)'' test, and ``PC2(U)'' is more balanced than ``OS.MT(U)''. ``Balanced'' test here is a test that performs relatively well in all checked situations. An interesting observation: ``PC2(U)'' has better power for point (10,10) than for point (5,10), which means that this test handles symmetric alternatives concentrated in the center (like on figure~\ref{figure:beta:aa}) better than asymmetric alternatives concentrated near the center (like on figure~\ref{figure:beta:3a}). This is different from ``OS(U)'' and ``OS.MT(U)'' that handle  symmetry worse than asymmetry for $a, b > 1$.

\subsection{Conclusions for the section}

In section~\ref{section:hyperrectabgle_order_stats_uniform_full} we introduced the new PC2(U) test based on principal components of standard uniform order statistics.

During the simulation studies we saw that PC2(U) test is not universally better than OS(U) test. In the next section we will show that the test based on standard normal order statistics is almost universally better than OS(U) test when checking against beta alternatives.

\newpage

\section{Testing uniformity via normal distribution}
\label{section:uniform:via:normal}

In section~\ref{section:hyperrectabgle_order_stats_uniform_full} we constructed tests based on uniform order statistics and checked their powers for ``uniform null vs beta alternative'' case. We were able to construct a more ``balanced'' test based on principal components of uniform order statistics (``PC2(U)''), but it was not universally better than the test constructed on raw uniform order statistics using the approach from~\cite{AldorNoimanBrownBujaRolkeStine2013} (``OS(U)'').

In section \ref{section:pc_test_normal} we constructed tests based on normal order statistics. 
In sections~\ref{section:pc_test_normal},~\ref{section:classical_tests_normal}, and  \ref{section:zhang:normal} we showed that for ``normal null'' case the test based on principal components of normal order statistics (``PC2'') is powerful, is significantly better than the test based on raw order statistics (``OS'') and classical tests (KS, CvM, AD), and is on par with Zhang tests ($Z_K$, $Z_A$, $Z_C$).

In section \ref{section:how:to:use:transforms:general} and subsection \ref{section:how:to:use:normal:for:arbitrary} we discussed, how normal null test can be used to check goodness of fit for any other arbitrary null distribution (see transformations~\eqref{eq:transform1} and~\eqref{eq:transform2}). In this section we will use tests based on normal order statistics to check goodness of fit for uniform null. 

To handle \texttt{Inf} values that arise during transformations~\eqref{eq:transform1} and~\eqref{eq:transform2} we introduce the following function:
\begin{minted}[mathescape, linenos]{r}
handle.infs <- function(x)
{
    x[x ==  Inf] <-  100000
    x[x == -Inf] <- -100000
    x
}
\end{minted}
We start with checking beta alternatives plotted on figures~\ref{figure:beta:aa}-\ref{figure:beta:1a}.
Rejection simulation is as follows:
\begin{minted}[mathescape, linenos, texcomments]{r}
res.uniform.via.normal <- calc.rejections.sm.pc.bounds(
    sample.generation.function.H0  = rnorm, # we test for uniform via normal
    sample.generation.functions.H1 = c(     # we use \eqref{eq:transform1} and \eqref{eq:transform2} to transform to normal
        lapply(1:10, function(x) function(nsamples) qnorm(punif(rbeta(nsamples, x, x)))),
        lapply(
            rev(seq(0.1, 1.0, 0.1)), 
            function(x) function(nsamples) handle.infs(qnorm(punif(rbeta(nsamples, x, x))))
        ),
        lapply(
            rev(seq(0.1, 1.0, 0.1)), 
            function(x) function(nsamples) handle.infs(qnorm(punif(rbeta(nsamples, 1, x))))
        ),
        lapply(1:10, function(x) function(nsamples) qnorm(punif(rbeta(nsamples, 1, x))))
    ),
    get.stat.matrix = get.sorted.samples.cpp,
    n = n, m = m,
    alpha = 0.05,
    h = 30
)
\end{minted}
We note that
\begin{enumerate} 
\item despite \texttt{sample.generation.function.H0  = rnorm}, we check goodness of fit for uniform null, not normal;
\item \texttt{qnorm(punif(rbeta(...)))} is implementation of 
$$F^{-1}_{N(0,1)} ( F_{U(0,1)}(x_1) ), \dots, F^{-1}_{N(0,1)} ( F_{U(0,1)}(x_n) ),$$ 
where $x_1, \dots, x_n \sim \mathrm{Beta}(a,b)$; see formula \eqref{eq:transform2};
\item $F_{U(0,1)}(x)$ is identity function for $x \in [0,1]$, so for beta samples we could use \texttt{qnorm(rbeta(...))} instead of \texttt{qnorm(punif(rbeta(...)))};
\item if $x_1, \dots, x_n \sim U(0,1)$, then 
$$F^{-1}_{N(0,1)} ( F_{U(0,1)}(x_1) ) = F^{-1}_{N(0,1)} (x_1), \dots, F^{-1}_{N(0,1)} ( F_{U(0,1)}(x_n) ) = F^{-1}_{N(0,1)} (x_n)\sim N(0,1);$$
that is how we can check the type 1 error;
\item we can use the same transformations to check goodness of fit for any other distribution via normal based test.
\end{enumerate}

\noindent
Rejection tables are as follows:
\begin{minted}[mathescape, linenos]{r}
create.rejection.table(res.uniform.via.normal)
\end{minted}

\begin{table}[H]
\centering
\begin{tabular}{ |l|l||>{\columncolor[gray]{0.9}}p{1cm}|p{1.1cm}||>{\columncolor[gray]{0.9}}p{1cm}|p{1.1cm}||>{\columncolor[gray]{0.9}}p{1cm}|p{1.1cm}| } 
 \hline
 $H_0$ & $H_1$ & \multicolumn{2}{c||}{OS}  & \multicolumn{2}{c||}{PC1} & \multicolumn{2}{c|}{PC2} \\
 \hline
 &  & mean & sd & mean & sd & mean & sd \\
 \hline
 \hline
 
 $\mathrm{U}(0, 1)$ & $\mathrm{Beta}(1, 1)$ & 0.05 & 3.4e-04 & 0.05 & 3.3e-04 & 0.05 & 2.9e-04\\
$\mathrm{U}(0, 1)$ & $\mathrm{Beta}(2, 2)$ & 0.0169 & 2.1e-04 & 0.0525 & 7.8e-04 & 0.0918 & 9.4e-04\\
$\mathrm{U}(0, 1)$ & $\mathrm{Beta}(3, 3)$ & 0.0219 & 4.4e-04 & 0.1869 & 1.9e-03 & 0.2889 & 1.9e-03\\
$\mathrm{U}(0, 1)$ & $\mathrm{Beta}(4, 4)$ & 0.0291 & 6.9e-04 & 0.3751 & 3.2e-03 & 0.5189 & 2.5e-03\\
$\mathrm{U}(0, 1)$ & $\mathrm{Beta}(5, 5)$ & 0.0375 & 1.0e-03 & 0.562 & 3.5e-03 & 0.7093 & 2.4e-03\\
$\mathrm{U}(0, 1)$ & $\mathrm{Beta}(6, 6)$ & 0.0468 & 1.2e-03 & 0.7151 & 3.2e-03 & 0.8387 & 1.8e-03\\
$\mathrm{U}(0, 1)$ & $\mathrm{Beta}(7, 7)$ & 0.0573 & 1.6e-03 & 0.8252 & 2.7e-03 & 0.9163 & 1.2e-03\\
$\mathrm{U}(0, 1)$ & $\mathrm{Beta}(8, 8)$ & 0.0686 & 1.9e-03 & 0.8977 & 1.8e-03 & 0.9587 & 6.5e-04\\
$\mathrm{U}(0, 1)$ & $\mathrm{Beta}(9, 9)$ & 0.0807 & 2.4e-03 & 0.9423 & 1.2e-03 & 0.9804 & 4.3e-04\\
$\mathrm{U}(0, 1)$ & $\mathrm{Beta}(10, 10)$ & 0.0935 & 2.7e-03 & 0.9685 & 7.7e-04 & 0.991 & 2.2e-04\\
\hline
\end{tabular}
\caption{Tests based on normal order statistics; $\mathrm{U}(0,1)$ vs $\mathrm{Beta}(a,a)$; $a \geq 1$; alternatives have modes in the center, see figure~\ref{figure:beta:aa}.}
\label{table:beta:aa:normal:os}
\end{table}

\noindent
Table~\ref{table:beta:aa:normal:os} should be compared to table~\ref{table:beta:aa:uniform:os}. We see that the powers of ``OS(U)'' and ``OS'' tests are the same, up to the simulation error. Later we discuss why. ``PC1'' and ``PC2'' perform better than both ``PC1(U)'' and ``PC2(U)''. ``PC2'' is the best overall with significant advantage, and it almost reaches 1 for $a = 9,10$.

\begin{table}[H]
\centering
\begin{tabular}{ |l|l||>{\columncolor[gray]{0.9}}p{1cm}|p{1.1cm}||>{\columncolor[gray]{0.9}}p{1cm}|p{1.1cm}||>{\columncolor[gray]{0.9}}p{1cm}|p{1.1cm}| } 
 \hline
 $H_0$ & $H_1$ & \multicolumn{2}{c||}{OS}  & \multicolumn{2}{c||}{PC1} & \multicolumn{2}{c|}{PC2} \\
 \hline
 &  & mean & sd & mean & sd & mean & sd \\
 
 \hline
 \hline
 
 $\mathrm{U}(0, 1)$ & $\mathrm{Beta}(1, 1)$ & 0.05 & 2.6e-04 & 0.05 & 3.5e-04 & 0.05 & 3.2e-04\\
$\mathrm{U}(0, 1)$ & $\mathrm{Beta}(0.9, 0.9)$ & 0.0753 & 3.6e-04 & 0.0759 & 4.4e-04 & 0.0739 & 3.5e-04\\
$\mathrm{U}(0, 1)$ & $\mathrm{Beta}(0.8, 0.8)$ & 0.1189 & 4.4e-04 & 0.1196 & 4.2e-04 & 0.1172 & 4.9e-04\\
$\mathrm{U}(0, 1)$ & $\mathrm{Beta}(0.7, 0.7)$ & 0.1919 & 6.7e-04 & 0.1916 & 6.5e-04 & 0.1925 & 6.3e-04\\
$\mathrm{U}(0, 1)$ & $\mathrm{Beta}(0.6, 0.6)$ & 0.308 & 7.6e-04 & 0.3068 & 9.8e-04 & 0.3157 & 9.7e-04\\
$\mathrm{U}(0, 1)$ & $\mathrm{Beta}(0.5, 0.5)$ & 0.477 & 6.1e-04 & 0.4767 & 1.2e-03 & 0.4958 & 1.1e-03\\
$\mathrm{U}(0, 1)$ & $\mathrm{Beta}(0.4, 0.4)$ & 0.6855 & 6.2e-04 & 0.6888 & 1.3e-03 & 0.7123 & 9.1e-04\\
$\mathrm{U}(0, 1)$ & $\mathrm{Beta}(0.3, 0.3)$ & 0.8787 & 4.3e-04 & 0.8832 & 7.0e-04 & 0.8985 & 4.7e-04\\
$\mathrm{U}(0, 1)$ & $\mathrm{Beta}(0.2, 0.2)$ & 0.9819 & 1.9e-04 & 0.9835 & 1.4e-04 & 0.9868 & 1.4e-04\\
$\mathrm{U}(0, 1)$ & $\mathrm{Beta}(0.1, 0.1)$ & 0.9998 & 1.5e-05 & 0.9998 & 1.7e-05 & 0.9999 & 1.1e-05\\

\hline
\end{tabular}
\caption{Tests based on normal order statistics; $\mathrm{U}(0,1)$ vs $\mathrm{Beta}(a,a)$; $0 < a \leq 1$; alternatives have modes on both sides, see figure~\ref{figure:beta:0a0a}.}
\label{table:beta:0a0a:normal:os}
\end{table}

\noindent
Table~\ref{table:beta:0a0a:normal:os} should be compared to table~\ref{table:beta:0a0a:uniform:os}. In table~\ref{table:beta:0a0a:uniform:os} ``OS(U)'' test was the best; significantly better than ``PC2(U)'' test. We see that ``OS(U)'' powers coincide with ``OS'' powers again. ``PC2'' test is slightly better than ``OS'' and ``PC1'' tests for $a \leq 0.6$ and is almost equal to ``OS'' and ``PC1'' tests for $0.6 < a \leq 1$. Overall, all tests in table~\ref{table:beta:0a0a:normal:os} perform very similar.

\begin{table}[H]
\centering
\begin{tabular}{ |l|l||>{\columncolor[gray]{0.9}}p{1cm}|p{1.1cm}||>{\columncolor[gray]{0.9}}p{1cm}|p{1.1cm}||>{\columncolor[gray]{0.9}}p{1cm}|p{1.1cm}| } 
 \hline
 $H_0$ & $H_1$ & \multicolumn{2}{c||}{OS}  & \multicolumn{2}{c||}{PC1} & \multicolumn{2}{c|}{PC2} \\
 \hline
 &  & mean & sd & mean & sd & mean & sd \\

 \hline
 \hline
 
 $\mathrm{U}(0, 1)$ & $\mathrm{Beta}(1, 1)$ & 0.05 & 3.2e-04 & 0.05 & 2.7e-04 & 0.05 & 3.1e-04\\
$\mathrm{U}(0, 1)$ & $\mathrm{Beta}(1, 0.9)$ & 0.0693 & 2.9e-04 & 0.0643 & 3.4e-04 & 0.0678 & 4.5e-04\\
$\mathrm{U}(0, 1)$ & $\mathrm{Beta}(1, 0.8)$ & 0.1169 & 5.4e-04 & 0.0932 & 4.3e-04 & 0.1123 & 5.6e-04\\
$\mathrm{U}(0, 1)$ & $\mathrm{Beta}(1, 0.7)$ & 0.2091 & 7.5e-04 & 0.1511 & 7.8e-04 & 0.2002 & 8.4e-04\\
$\mathrm{U}(0, 1)$ & $\mathrm{Beta}(1, 0.6)$ & 0.361 & 8.3e-04 & 0.2605 & 1.3e-03 & 0.3486 & 9.7e-04\\
$\mathrm{U}(0, 1)$ & $\mathrm{Beta}(1, 0.5)$ & 0.5683 & 9.5e-04 & 0.4423 & 1.2e-03 & 0.5561 & 1.0e-03\\
$\mathrm{U}(0, 1)$ & $\mathrm{Beta}(1, 0.4)$ & 0.7864 & 6.1e-04 & 0.681 & 1.5e-03 & 0.7784 & 7.5e-04\\
$\mathrm{U}(0, 1)$ & $\mathrm{Beta}(1, 0.3)$ & 0.9393 & 3.3e-04 & 0.8919 & 6.8e-04 & 0.9365 & 3.5e-04\\
$\mathrm{U}(0, 1)$ & $\mathrm{Beta}(1, 0.2)$ & 0.9942 & 9.0e-05 & 0.9874 & 1.5e-04 & 0.994 & 9.5e-05\\
$\mathrm{U}(0, 1)$ & $\mathrm{Beta}(1, 0.1)$ & 1 & 6.1e-06 & 0.9999 & 1.2e-05 & 1 & 5.7e-06\\

\hline
\end{tabular}
\caption{Tests based on normal order statistics; $\mathrm{U}(0,1)$ vs $\mathrm{Beta}(1,a)$; $0 < a \leq 1$; alternatives have sharp modes on the right, see figure~\ref{figure:beta:10a}.}
\label{table:beta:10a:normal:os}
\end{table}

\noindent
Table~\ref{table:beta:10a:normal:os} should be compared to table~\ref{table:beta:10a:uniform:os}. In table~\ref{table:beta:10a:uniform:os} ``OS(U)'' test was the best; noticeably better than ``PC2(U)''. ``OS(U)'' and ``OS'' are the same. ``PC1'' is worse than ``PC2''. ``PC2'' performs almost as well as ``OS''. 

\begin{table}[H]
\centering
\begin{tabular}{ |l|l||>{\columncolor[gray]{0.9}}p{1cm}|p{1.1cm}||>{\columncolor[gray]{0.9}}p{1cm}|p{1.1cm}||>{\columncolor[gray]{0.9}}p{1cm}|p{1.1cm}| } 
 \hline
 $H_0$ & $H_1$ & \multicolumn{2}{c||}{OS}  & \multicolumn{2}{c||}{PC1} & \multicolumn{2}{c|}{PC2} \\
 \hline
 &  & mean & sd & mean & sd & mean & sd \\
 
 \hline
 \hline

$\mathrm{U}(0, 1)$ & $\mathrm{Beta}(1, 1)$ & 0.05 & 2.7e-04 & 0.0498 & 2.3e-04 & 0.0499 & 2.9e-04\\
$\mathrm{U}(0, 1)$ & $\mathrm{Beta}(1, 2)$ & 0.3018 & 1.4e-03 & 0.1368 & 1.4e-03 & 0.3102 & 1.4e-03\\
$\mathrm{U}(0, 1)$ & $\mathrm{Beta}(1, 3)$ & 0.7181 & 1.7e-03 & 0.4667 & 3.4e-03 & 0.7363 & 1.7e-03\\
$\mathrm{U}(0, 1)$ & $\mathrm{Beta}(1, 4)$ & 0.9272 & 8.4e-04 & 0.7699 & 2.7e-03 & 0.937 & 6.7e-04\\
$\mathrm{U}(0, 1)$ & $\mathrm{Beta}(1, 5)$ & 0.986 & 2.3e-04 & 0.9239 & 1.3e-03 & 0.9889 & 1.9e-04\\
$\mathrm{U}(0, 1)$ & $\mathrm{Beta}(1, 6)$ & 0.9978 & 6.3e-05 & 0.9792 & 5.5e-04 & 0.9984 & 4.2e-05\\
$\mathrm{U}(0, 1)$ & $\mathrm{Beta}(1, 7)$ & 0.9997 & 2.1e-05 & 0.9951 & 1.7e-04 & 0.9998 & 1.4e-05\\
$\mathrm{U}(0, 1)$ & $\mathrm{Beta}(1, 8)$ & 1 & 6.4e-06 & 0.999 & 5.6e-05 & 1 & 4.4e-06\\
$\mathrm{U}(0, 1)$ & $\mathrm{Beta}(1, 9)$ & 1 & 1.5e-06 & 0.9998 & 1.7e-05 & 1 & 1.4e-06\\
$\mathrm{U}(0, 1)$ & $\mathrm{Beta}(1, 10)$ & 1 & 6.7e-07 & 1 & 6.2e-06 & 1 & 6.0e-07\\

        \hline
\end{tabular}
\caption{Tests based on normal order statistics; $\mathrm{U}(0,1)$ vs $\mathrm{Beta}(1,a)$; $a \geq 1$; alternatives have sloping modes on the left, see figure~\ref{figure:beta:1a}.}
\label{table:beta:1a:normal:os}
\end{table}

\noindent
Table~\ref{table:beta:1a:normal:os} should be compared to table~\ref{table:beta:1a:uniform:os}. ``OS(U)'' and ``OS'' are the same. ``PC1'' is worse than ``PC2''. ``PC2'' is slightly better than ``OS'', but they are almost equal. ``PC2(U)'' from table~\ref{table:beta:1a:uniform:os} is a bit better than ``PC2'', but not much better.

Overall, we can say that so far ``PC2'' test is universally better than ``OS'' test. 

We continue our study with constructing power estimates for other beta alternatives listed in~\eqref{eq:beta_grid}:

\begin{minted}[mathescape, linenos, texcomments]{r}
res.uniform.via.normal.beta.grid <- calc.rejections.sm.pc.bounds(
    sample.generation.function.H0  = rnorm,
    sample.generation.functions.H1 = 
        apply(
            beta.parameters.grid,  # see formula \eqref{eq:beta_grid}
            1, 
            function(par) function(nsamples) handle.infs(qnorm(punif(rbeta(nsamples, par[1], par[2]))))
        ),
    get.stat.matrix = get.sorted.samples.cpp,
    n = n, m = m,
    alpha = 0.05,
    h = 10
)
\end{minted}
Table construction code is as follows:
\begin{minted}[mathescape, linenos]{r}
res.uniform.via.normal.beta.grid.table <- create.rejection.table(res.uniform.via.normal.beta.grid)
res.uniform.via.normal.beta.grid.table
\end{minted}
As in subsection~\ref{section:rejections:uniform:vs:beta}, instead of full table we print tile plots for each ``mean'' column of it.

\newpage

\paragraph{Tile plot for ``OS'': beta parameters grid}

\begin{minted}[mathescape, linenos, texcomments]{r}
plot.powers.on.grid(        # defined in subsection \ref{section:rejections:uniform:vs:beta} after formula \eqref{eq:beta_grid}
    beta.parameters.grid,   # defined just before formula \eqref{eq:beta_grid}
    res.uniform.via.normal.beta.grid.table$sm.means
)
\end{minted}

\begin{figure}[H]
\centering
\includegraphics[width=16cm]{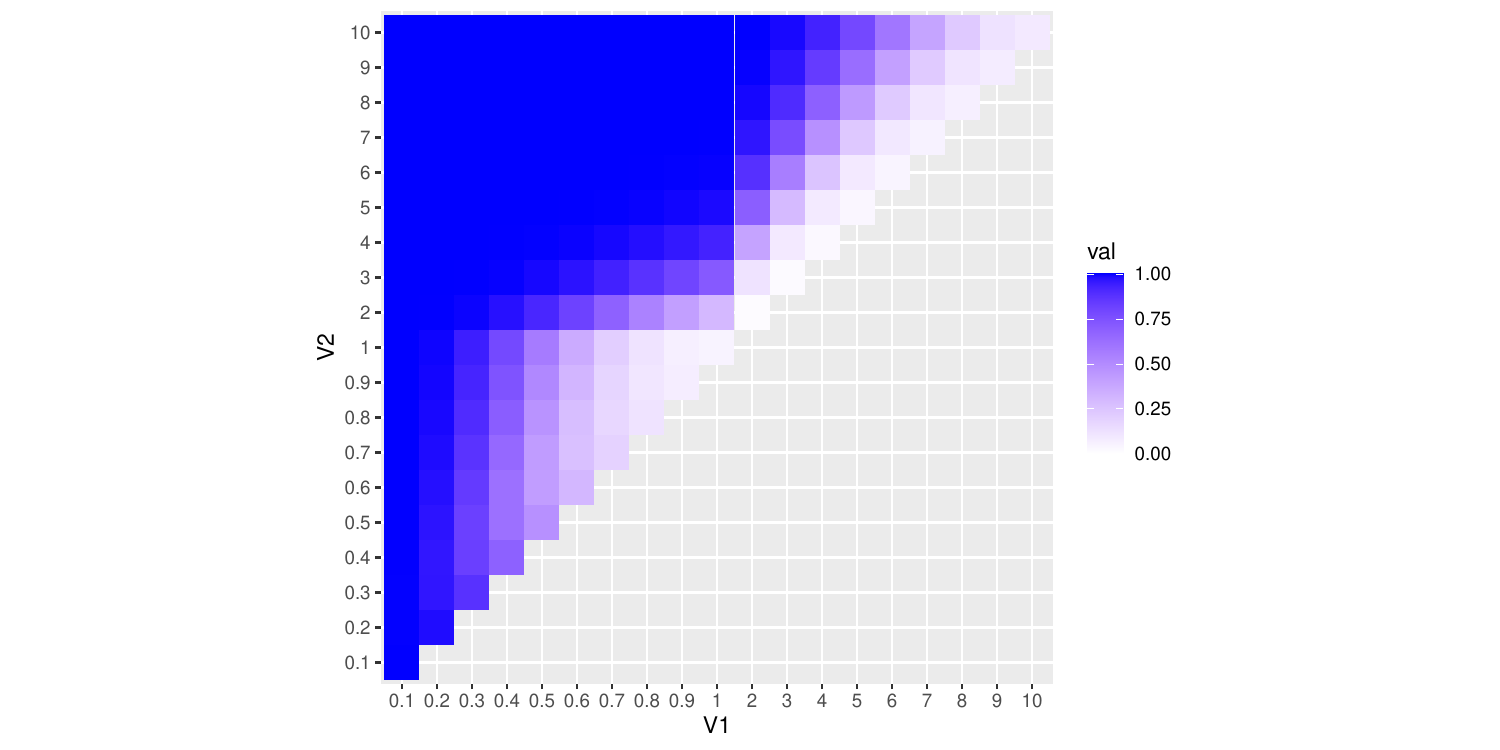}
\caption{``OS'' test powers on beta grid of alternatives~\eqref{eq:beta_grid}.}
\label{figure:beta_grid:os}
\end{figure}

\paragraph{Tile plot for ``PC1'': beta parameters grid}

\begin{minted}[mathescape, linenos]{r}
plot.powers.on.grid(beta.parameters.grid, res.uniform.via.normal.beta.grid.table$pc.w1.means)
\end{minted}

\begin{figure}[H]
\centering
\includegraphics[width=16cm]{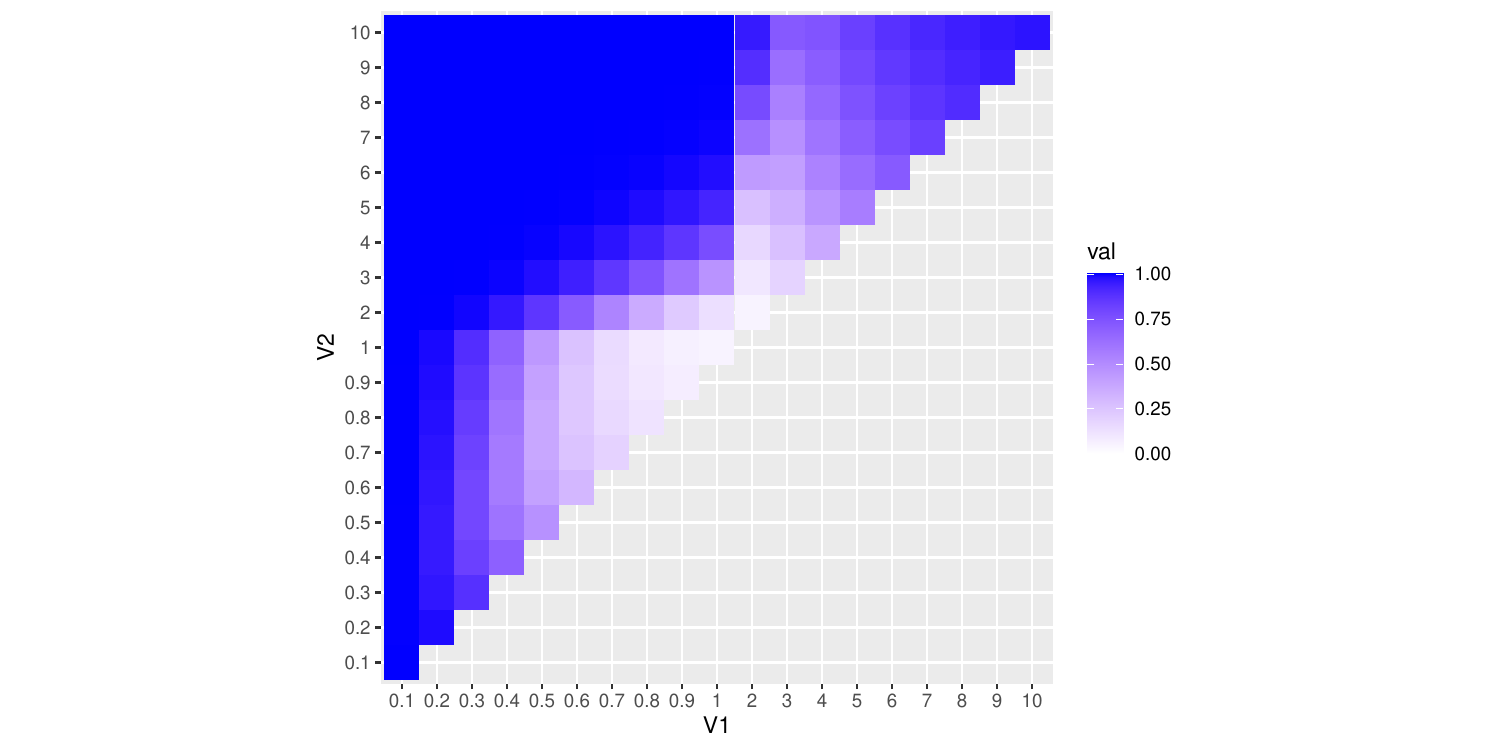}
\caption{``PC1'' test powers on beta grid of alternatives~\eqref{eq:beta_grid}.}
\label{figure:beta_grid:pc1}
\end{figure}

\newpage

\paragraph{Tile plot for ``PC2'': beta parameters grid}

\begin{minted}[mathescape, linenos]{r}
plot.powers.on.grid(beta.parameters.grid, res.uniform.via.normal.beta.grid.table$pc.w2.means)
\end{minted}

\begin{figure}[H]
\centering
\includegraphics[width=16cm]{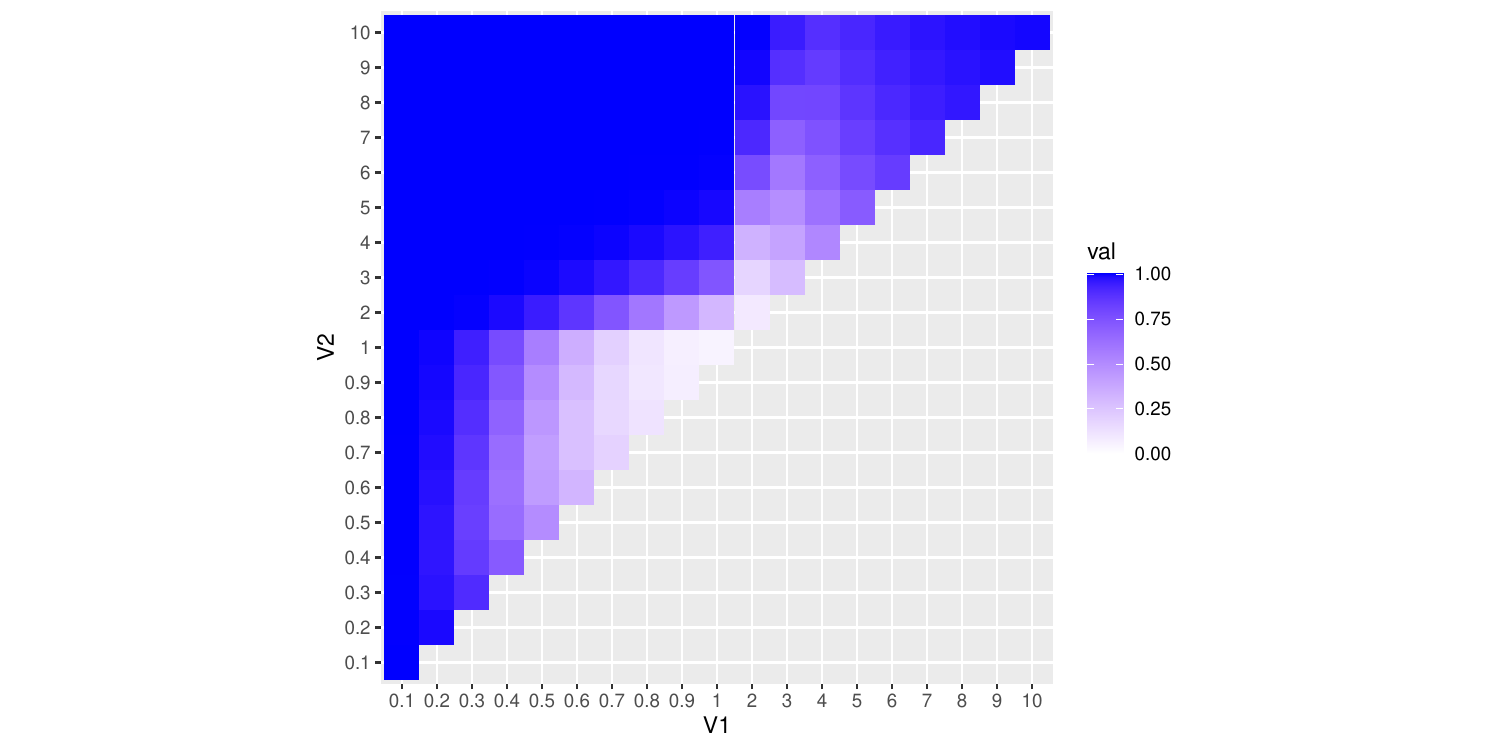}
\caption{``PC2'' test powers on beta grid of alternatives~\eqref{eq:beta_grid}.}
\label{figure:beta_grid:pc2}
\end{figure}

\subsubsection{Comments to the tile plots}

Figures~\ref{figure:beta_grid:os}--\ref{figure:beta_grid:pc2} should be compared to figures~\ref{figure:beta_grid:osu}--\ref{figure:beta_grid:pc2u}. We see that figures~\ref{figure:beta_grid:os} and~\ref{figure:beta_grid:osu} that correspond to ``OS'' and ``OS(U)'' tests are essentially the same. We discuss it in the next section. We also see that there is no situation, where ``PC1'' test (figure~\ref{figure:beta_grid:pc1}) is better than ``PC2'' test (figure~\ref{figure:beta_grid:pc2}), so we focus all our attention at analyzing ``PC2''.

Tiles for $\mathrm{Beta}(a,b)$ and $\mathrm{Beta}(b,a)$, $0 < a \leq 1$, $0 < b \leq 10$ alternatives (figures~\ref{figure:beta:0a0a}--\ref{figure:beta:1a}, \ref{figure:beta:030a}, \ref{figure:beta:0a3}) look very similar for ``OS'' and ``PC2''. The main difference is for $\mathrm{Beta}(a, b)$, $a, b > 1$ (figures~\ref{figure:beta:aa} and~\ref{figure:beta:3a}). We see that ``PC2'' test performs extremely well for symmetric alternatives concentrated in the center of the $[0,1]$ interval like (7,7) or (10,10). For ``near symmetric'' alternatives like (6,9) or (5,8) it is still very good. For more asymmetric alternatives like (3,6) or (2,5) ``PC2'' looses some power, but it is still comparable with ``OS''.

All in all, we can't say that ``PC2'' test is universally better than ``OS'', but judging by the plots ``PC2'' is better or not worse than ``OS'' in most situations.

We also note that ``PC2'' test handles alternative in the lower left triangle better than ``PC2(U)'' test.

\newpage
\section{Invariance of bounds based on order statistics}
\label{section:invariance}

In section~\ref{section:uniform:via:normal} we already mentioned that ``OS(U)'' and ``OS'' tests have the same power estimates up to the simulation error (see ``OS(U)'' and ``OS'' columns in tables \ref{table:beta:aa:uniform:os}--\ref{table:beta:1a:normal:os} as well as figures~\ref{figure:beta_grid:osu} and~\ref{figure:beta_grid:os}). In this section we will show that these tests are indeed the same.

The chain of thought is as follows:
\begin{enumerate}
\item Suppose we have a sample from uniform distribution:
$u_1, \dots, u_n \sim U(0,1)$. 
\item Consider order statistics for this sample:
$u_{(1)}, \dots, u_{(n)} $.
\item Denote distributions of
$u_{(i)}$, $i = 1, \dots, n$, 
as $\mathcal{P}_{(i)}(U)$.
\item
Consider $F$, a cumulative distribution function of some absolutely continuous distribution $\mathcal{P}(F)$.
\item 
Then $x_{(i)} = F^{-1}(u_{(i)})$, $i = 1,\dots,n$, are order statistics for $\mathcal{P}(F)$.
\item 
Denote distributions of $x_{(i)}$, $i = 1, \dots, n$, as $\mathcal{P}_{(i)}(F)$.
\item
Then $\mathrm{q}_{\alpha}(\mathcal{P}_{(i)}(F)) = F^{-1}(q_{\alpha}(\mathcal{P}_{(i)}(U)))$, where $q_{\alpha}(\dots)$ is an $\alpha$-level quantile of a corresponding distribution.
\item
From all the above it follows that equations
\begin{align}
\label{eq:invariance1}
P\left(q_{l,1}\left(\mathcal{P}_{(1)}(U)\right) \leq u_{(1)} \leq q_{r,n}(\mathcal{P}_{(1)}(U)) \land \dots \land q_{l,n}\left(\mathcal{P}_{(n)}(U)\right) \leq u_{(n)} \leq q_{r,n}(\mathcal{P}_{(n)}(U)) \right) = 1 - \alpha
\end{align}
and 
\begin{align}
\label{eq:invariance2}
P\left(q_{l,1}\left(\mathcal{P}_{(1)}(F)\right) \leq x_{(1)} \leq q_{r,1}(\mathcal{P}_{(1)}(F)) \land \dots \land q_{l,n}\left(\mathcal{P}_{(n)}(F)\right) \leq x_{(n)} \leq q_{r,n}(\mathcal{P}_{(n)}(F)) \right) = 1 - \alpha
\end{align}
are equivalent. In~\eqref{eq:invariance1} and~\eqref{eq:invariance2}: $q_{l,i}(\dots) = q_{\alpha\gamma w_i / 2} (\dots)$, $q_{r,i} (\dots) = q_{1-\alpha\gamma w_i / 2} (\dots)$, and $w_i$ are optional order statistic weighs.
\item 
From the equivalence of~\eqref{eq:invariance1} and~\eqref{eq:invariance2} it follows that if
\begin{align}
\left(L_i [\gamma] (U),R_i [\gamma] (U) \right), \; i = 1, \dots,n;
\end{align}
are the bounds of confidence hyperrectangle constructed for uniform order statistics, then
\begin{align} 
\left(F^{-1}(L_i [\gamma] (U)), F^{-1} (R_i [\gamma] (U)) \right), \; i = 1, \dots,n;
\end{align}
are the bounds of confidence hyperrectangle constructed for order statistics of distribution $\mathcal{P}(F)$.
\end{enumerate}
In short, if $F$ is absolutely continuous, $F^{-1}$ maps order statistics of uniform distribution into order statistics of $\mathcal{P}(F)$, and bounds for uniform order statistics into bounds for order statistics of $\mathcal{P}(F)$.

After the literature review we found that this fact is actually known: see~\cite{weine2023qqconf}, where it is described with a bit different notations and language (``Equal Local Levels''). 

\newpage
\section{Classical goodness of fit tests: samples from standard uniform distribution [cl, \texorpdfstring{$U(0,1)$}{U(0,1)}]}
\label{section:classical_tests_uniform}

Let us now look at the powers of classical tests for ``uniform null vs beta alternative'' situation. We start again with checking beta alternatives plotted on figures~\ref{figure:beta:aa}-\ref{figure:beta:1a}:

\begin{minted}[mathescape, linenos]{r}
res.unif.classic <- calc.rejections.classical.tests(
    sample.generation.function.H0  = runif,
    sample.generation.functions.H1 = c(
        lapply(1:10, function(x) function(nsamples) rbeta(nsamples, x, x)),
        lapply(rev(seq(0.1, 1.0, 0.1)), function(x) function(nsamples) rbeta(nsamples, x, x)),
        lapply(rev(seq(0.1, 1.0, 0.1)), function(x) function(nsamples) rbeta(nsamples, 1, x)),
        lapply(1:10, function(x) function(nsamples) rbeta(nsamples, 1, x))
    ),
    F_0   = punif,
    alpha = 0.05,
    h     = 30,
    n = n, m = m
)
\end{minted}
The tables are as follows:
\begin{minted}[mathescape, linenos]{r}
create.rejection.table.classic(res.unif.classic)
\end{minted}

\begin{table}[H]
\centering
\begin{tabular}{ |l|l||>{\columncolor[gray]{0.9}}p{1cm}|p{1.1cm}||>{\columncolor[gray]{0.9}}p{1cm}|p{1.1cm}||>{\columncolor[gray]{0.9}}p{1cm}|p{1.1cm}| } 
 \hline
 $H_0$ & $H_1$ & \multicolumn{2}{c||}{KS}  & \multicolumn{2}{c||}{CvM} & \multicolumn{2}{c|}{AD} \\
 \hline
 &  & mean & sd & mean & sd & mean & sd \\
 \hline
 \hline

$\mathrm{U}(0, 1)$ & $\mathrm{Beta}(1, 1)$ & 0.0501 & 3.4e-04 & 0.05 & 3.6e-04 & 0.05 & 3.8e-04\\
$\mathrm{U}(0, 1)$ & $\mathrm{Beta}(2, 2)$ & 0.037 & 3.4e-04 & 0.0107 & 1.0e-04 & 0.0119 & 1.5e-04\\
$\mathrm{U}(0, 1)$ & $\mathrm{Beta}(3, 3)$ & 0.0456 & 3.6e-04 & 0.0024 & 5.2e-05 & 0.0097 & 1.4e-04\\
$\mathrm{U}(0, 1)$ & $\mathrm{Beta}(4, 4)$ & 0.0593 & 5.2e-04 & 6e-04 & 2.3e-05 & 0.0126 & 2.1e-04\\
$\mathrm{U}(0, 1)$ & $\mathrm{Beta}(5, 5)$ & 0.0761 & 6.6e-04 & 1e-04 & 1.1e-05 & 0.0189 & 3.5e-04\\
$\mathrm{U}(0, 1)$ & $\mathrm{Beta}(6, 6)$ & 0.0953 & 7.7e-04 & 0 & 5.4e-06 & 0.0286 & 5.2e-04\\
$\mathrm{U}(0, 1)$ & $\mathrm{Beta}(7, 7)$ & 0.1167 & 8.0e-04 & 0 & 2.6e-06 & 0.0423 & 7.9e-04\\
$\mathrm{U}(0, 1)$ & $\mathrm{Beta}(8, 8)$ & 0.1399 & 1.1e-03 & 0 & 1.4e-06 & 0.0598 & 1.1e-03\\
$\mathrm{U}(0, 1)$ & $\mathrm{Beta}(9, 9)$ & 0.1647 & 1.2e-03 & 0 & 6.8e-07 & 0.0816 & 1.4e-03\\
$\mathrm{U}(0, 1)$ & $\mathrm{Beta}(10, 10)$ & 0.1907 & 1.5e-03 & 0 & 3.5e-07 & 0.107 & 1.9e-03\\

\hline
\end{tabular}
\caption{Classical tests; $\mathrm{U}(0,1)$ vs $\mathrm{Beta}(a,a)$; $a \geq 1$; alternatives have modes in the center, see figure~\ref{figure:beta:aa}.}
\label{table:beta:aa:classical}
\end{table}

\noindent
Table~\ref{table:beta:aa:classical} should be compared to tables~\ref{table:beta:aa:normal:os} and~\ref{table:beta:aa:uniform:os}. We see that CvM test does not have power at all, AD test has power comparable to ``OS'' test, KS test is twice as powerful as ``OS'' test. In short, classical tests does not perform for symmetric centered alternatives.

\begin{table}[H]
\centering
\begin{tabular}{ |l|l||>{\columncolor[gray]{0.9}}p{1cm}|p{1.1cm}||>{\columncolor[gray]{0.9}}p{1cm}|p{1.1cm}||>{\columncolor[gray]{0.9}}p{1cm}|p{1.1cm}| } 
 \hline
 $H_0$ & $H_1$ & \multicolumn{2}{c||}{KS}  & \multicolumn{2}{c||}{CvM} & \multicolumn{2}{c|}{AD} \\
 \hline
 &  & mean & sd & mean & sd & mean & sd \\

\hline
\hline
 
$\mathrm{U}(0, 1)$ & $\mathrm{Beta}(1, 1)$ & 0.05 & 3.1e-04 & 0.05 & 3.2e-04 & 0.0499 & 3.5e-04\\
$\mathrm{U}(0, 1)$ & $\mathrm{Beta}(0.9, 0.9)$ & 0.0568 & 3.1e-04 & 0.0586 & 3.1e-04 & 0.0672 & 3.4e-04\\
$\mathrm{U}(0, 1)$ & $\mathrm{Beta}(0.8, 0.8)$ & 0.0667 & 2.9e-04 & 0.0687 & 2.8e-04 & 0.0951 & 3.9e-04\\
$\mathrm{U}(0, 1)$ & $\mathrm{Beta}(0.7, 0.7)$ & 0.0818 & 4.0e-04 & 0.0808 & 3.3e-04 & 0.1423 & 5.2e-04\\
$\mathrm{U}(0, 1)$ & $\mathrm{Beta}(0.6, 0.6)$ & 0.105 & 4.6e-04 & 0.0952 & 4.3e-04 & 0.2232 & 7.6e-04\\
$\mathrm{U}(0, 1)$ & $\mathrm{Beta}(0.5, 0.5)$ & 0.1421 & 4.7e-04 & 0.1123 & 4.3e-04 & 0.3586 & 8.8e-04\\
$\mathrm{U}(0, 1)$ & $\mathrm{Beta}(0.4, 0.4)$ & 0.2026 & 6.4e-04 & 0.1325 & 4.7e-04 & 0.5616 & 1.0e-03\\
$\mathrm{U}(0, 1)$ & $\mathrm{Beta}(0.3, 0.3)$ & 0.3037 & 6.7e-04 & 0.1571 & 5.7e-04 & 0.7983 & 7.3e-04\\
$\mathrm{U}(0, 1)$ & $\mathrm{Beta}(0.2, 0.2)$ & 0.4738 & 7.6e-04 & 0.1872 & 5.8e-04 & 0.962 & 2.1e-04\\
$\mathrm{U}(0, 1)$ & $\mathrm{Beta}(0.1, 0.1)$ & 0.7452 & 5.7e-04 & 0.2309 & 4.5e-04 & 0.9994 & 2.1e-05\\

          \hline
\end{tabular}
\caption{Classical tests; $\mathrm{U}(0,1)$ vs $\mathrm{Beta}(a,a)$; $0 < a \leq 1$; alternatives have modes on both sides, see figure~\ref{figure:beta:0a0a}.}
\label{table:beta:0a0a:classical}
\end{table}

\noindent
Table~\ref{table:beta:0a0a:classical} should be compared to tables~\ref{table:beta:0a0a:normal:os} and~\ref{table:beta:0a0a:uniform:os}. We see that CvM test has low power, KS test has moderate power (but still low), AD test works better than ``PC2(U)'' test, but worse than ``PC2'' and ``OS'' tests. 

\begin{table}[H]
\centering
\begin{tabular}{ |l|l||>{\columncolor[gray]{0.9}}p{1cm}|p{1.1cm}||>{\columncolor[gray]{0.9}}p{1cm}|p{1.1cm}||>{\columncolor[gray]{0.9}}p{1cm}|p{1.1cm}| } 
 \hline
 $H_0$ & $H_1$ & \multicolumn{2}{c||}{KS}  & \multicolumn{2}{c||}{CvM} & \multicolumn{2}{c|}{AD} \\
 \hline
 &  & mean & sd & mean & sd & mean & sd \\

 \hline
 \hline
 
$\mathrm{U}(0, 1)$ & $\mathrm{Beta}(1, 1)$ & 0.05 & 2.7e-04 & 0.05 & 3.6e-04 & 0.05 & 3.2e-04\\
$\mathrm{U}(0, 1)$ & $\mathrm{Beta}(1, 0.9)$ & 0.0604 & 3.6e-04 & 0.0638 & 2.9e-04 & 0.0679 & 3.0e-04\\
$\mathrm{U}(0, 1)$ & $\mathrm{Beta}(1, 0.8)$ & 0.0908 & 4.5e-04 & 0.1028 & 5.3e-04 & 0.1161 & 5.1e-04\\
$\mathrm{U}(0, 1)$ & $\mathrm{Beta}(1, 0.7)$ & 0.1502 & 6.3e-04 & 0.1774 & 6.1e-04 & 0.2103 & 6.6e-04\\
$\mathrm{U}(0, 1)$ & $\mathrm{Beta}(1, 0.6)$ & 0.2505 & 7.7e-04 & 0.2975 & 6.6e-04 & 0.3633 & 7.8e-04\\
$\mathrm{U}(0, 1)$ & $\mathrm{Beta}(1, 0.5)$ & 0.402 & 8.7e-04 & 0.4659 & 8.8e-04 & 0.5702 & 9.2e-04\\
$\mathrm{U}(0, 1)$ & $\mathrm{Beta}(1, 0.4)$ & 0.599 & 9.6e-04 & 0.6639 & 7.9e-04 & 0.7857 & 6.2e-04\\
$\mathrm{U}(0, 1)$ & $\mathrm{Beta}(1, 0.3)$ & 0.8043 & 5.7e-04 & 0.8471 & 5.8e-04 & 0.9379 & 3.4e-04\\
$\mathrm{U}(0, 1)$ & $\mathrm{Beta}(1, 0.2)$ & 0.9497 & 2.8e-04 & 0.962 & 2.7e-04 & 0.9938 & 7.9e-05\\
$\mathrm{U}(0, 1)$ & $\mathrm{Beta}(1, 0.1)$ & 0.9975 & 4.4e-05 & 0.9978 & 4.1e-05 & 1 & 6.3e-06\\

          \hline
\end{tabular}
\caption{Classical tests; $\mathrm{U}(0,1)$ vs $\mathrm{Beta}(1,a)$; $0 < a \leq 1$; alternatives have sharp modes on the right, see figure~\ref{figure:beta:10a}.}
\label{table:beta:10a:classical}
\end{table}

\noindent
Table~\ref{table:beta:10a:classical} should be compared to tables~\ref{table:beta:10a:normal:os} and~\ref{table:beta:10a:uniform:os}. We see that all classical tests do have good powers. AD test is the best among classical tests and is comparable to ``PC2'' and ``OS'' tests.

\begin{table}[H]
\centering
\begin{tabular}{ |l|l||>{\columncolor[gray]{0.9}}p{1cm}|p{1.1cm}||>{\columncolor[gray]{0.9}}p{1cm}|p{1.22cm}||>{\columncolor[gray]{0.9}}p{1cm}|p{1.22cm}| } 
 \hline
 $H_0$ & $H_1$ & \multicolumn{2}{c||}{KS}  & \multicolumn{2}{c||}{CvM} & \multicolumn{2}{c|}{AD} \\
 \hline
 &  & mean & sd & mean & sd & mean & sd \\
 
 \hline
 \hline

$\mathrm{U}(0, 1)$ & $\mathrm{Beta}(1, 1)$ & 0.05 & 2.6e-04 & 0.05 & 3.4e-04 & 0.0501 & 3.5e-04\\
$\mathrm{U}(0, 1)$ & $\mathrm{Beta}(1, 2)$ & 0.3844 & 9.3e-04 & 0.4516 & 9.6e-04 & 0.4099 & 9.6e-04\\
$\mathrm{U}(0, 1)$ & $\mathrm{Beta}(1, 3)$ & 0.8054 & 6.6e-04 & 0.8758 & 5.3e-04 & 0.8441 & 6.8e-04\\
$\mathrm{U}(0, 1)$ & $\mathrm{Beta}(1, 4)$ & 0.9618 & 2.5e-04 & 0.985 & 1.5e-04 & 0.9776 & 2.0e-04\\
$\mathrm{U}(0, 1)$ & $\mathrm{Beta}(1, 5)$ & 0.9944 & 8.8e-05 & 0.9988 & 3.7e-05 & 0.9979 & 5.2e-05\\
$\mathrm{U}(0, 1)$ & $\mathrm{Beta}(1, 6)$ & 0.9993 & 2.9e-05 & 0.9999 & 1.3e-05 & 0.9998 & 1.3e-05\\
$\mathrm{U}(0, 1)$ & $\mathrm{Beta}(1, 7)$ & 0.9999 & 7.4e-06 & 1 & 1.9e-06 & 1 & 2.7e-06\\
$\mathrm{U}(0, 1)$ & $\mathrm{Beta}(1, 8)$ & 1 & 2.0e-06 & 1 & 5.0e-07 & 1 & 6.7e-07\\
$\mathrm{U}(0, 1)$ & $\mathrm{Beta}(1, 9)$ & 1 & 8.1e-07 & 1 & 0.0e+00 & 1 & 0.0e+00\\
$\mathrm{U}(0, 1)$ & $\mathrm{Beta}(1, 10)$ & 1 & 2.5e-07 & 1 & 0.0e+00 & 1 & 0.0e+00\\

          \hline
\end{tabular}
\caption{Classical tests; $\mathrm{U}(0,1)$ vs $\mathrm{Beta}(1,a)$; $a \geq 1$; alternatives have sloping modes on the left, see figure~\ref{figure:beta:1a}.}
\label{table:beta:1a:classical}
\end{table}

\noindent
Table~\ref{table:beta:1a:classical} should be compared to tables~\ref{table:beta:1a:normal:os} and~\ref{table:beta:1a:uniform:os}. We see that all classical tests outperform ``OS'' and ``PC2'' tests. The difference is not drastic, but still. CvM test shines here.

Now to the plots on beta grid~\eqref{eq:beta_grid}:

\begin{minted}[mathescape, linenos, texcomments]{r}
res.unif.beta.grid.classic<- calc.rejections.classical.tests(
    sample.generation.function.H0  = runif,
    sample.generation.functions.H1 = apply(
        beta.parameters.grid,  # see formula \eqref{eq:beta_grid} 
        1, 
        function(par) function(nsamples) rbeta(nsamples, par[1], par[2])
    ),
    F_0   = punif,
    alpha = 0.05,
    h     = 10,
    n = n, m = m
)
\end{minted}

\begin{minted}[mathescape, linenos]{r}
res.unif.beta.grid.classic.table <- create.rejection.table.classic(res.unif.beta.grid.classic)
\end{minted}

\newpage
\paragraph{Tile plot for ``KS'': beta parameters grid}

\begin{minted}[mathescape, linenos, texcomments]{r}
plot.powers.on.grid(        # defined in subsection \ref{section:rejections:uniform:vs:beta} after formula \eqref{eq:beta_grid}
    beta.parameters.grid,   # defined just before formula \eqref{eq:beta_grid}
    res.unif.beta.grid.classic.table$KS.means
)
\end{minted}

\begin{figure}[H]
\centering
\includegraphics[width=16cm]{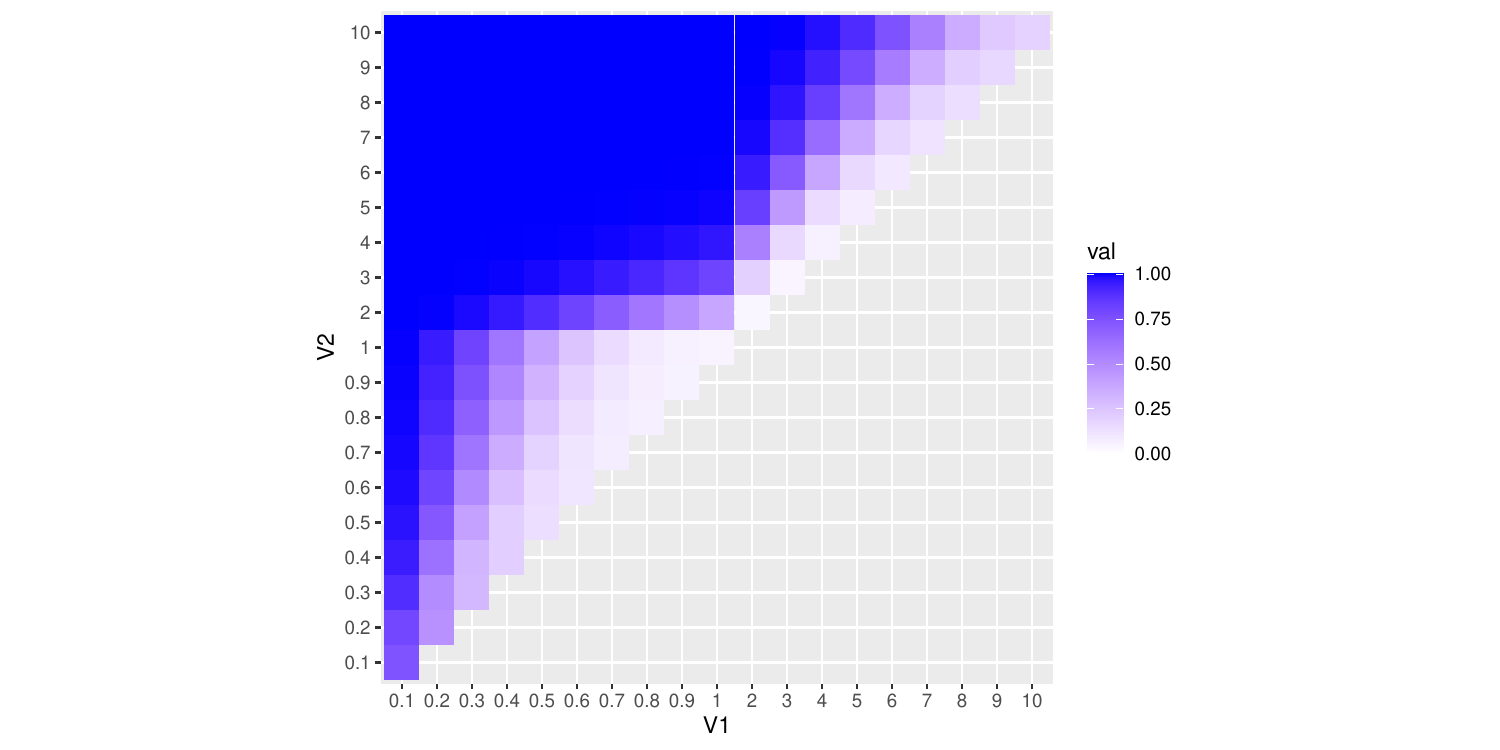}
\caption{``KS'' test powers on beta grid of alternatives~\eqref{eq:beta_grid}.}
\label{figure:beta_grid:ks}
\end{figure}

\paragraph{Tile plot for ``CvM'': beta parameters grid}

\begin{minted}[mathescape, linenos]{r}
plot.powers.on.grid(beta.parameters.grid, res.unif.beta.grid.classic.table$CvM.means)
\end{minted}

\begin{figure}[H]
\centering
\includegraphics[width=16cm]{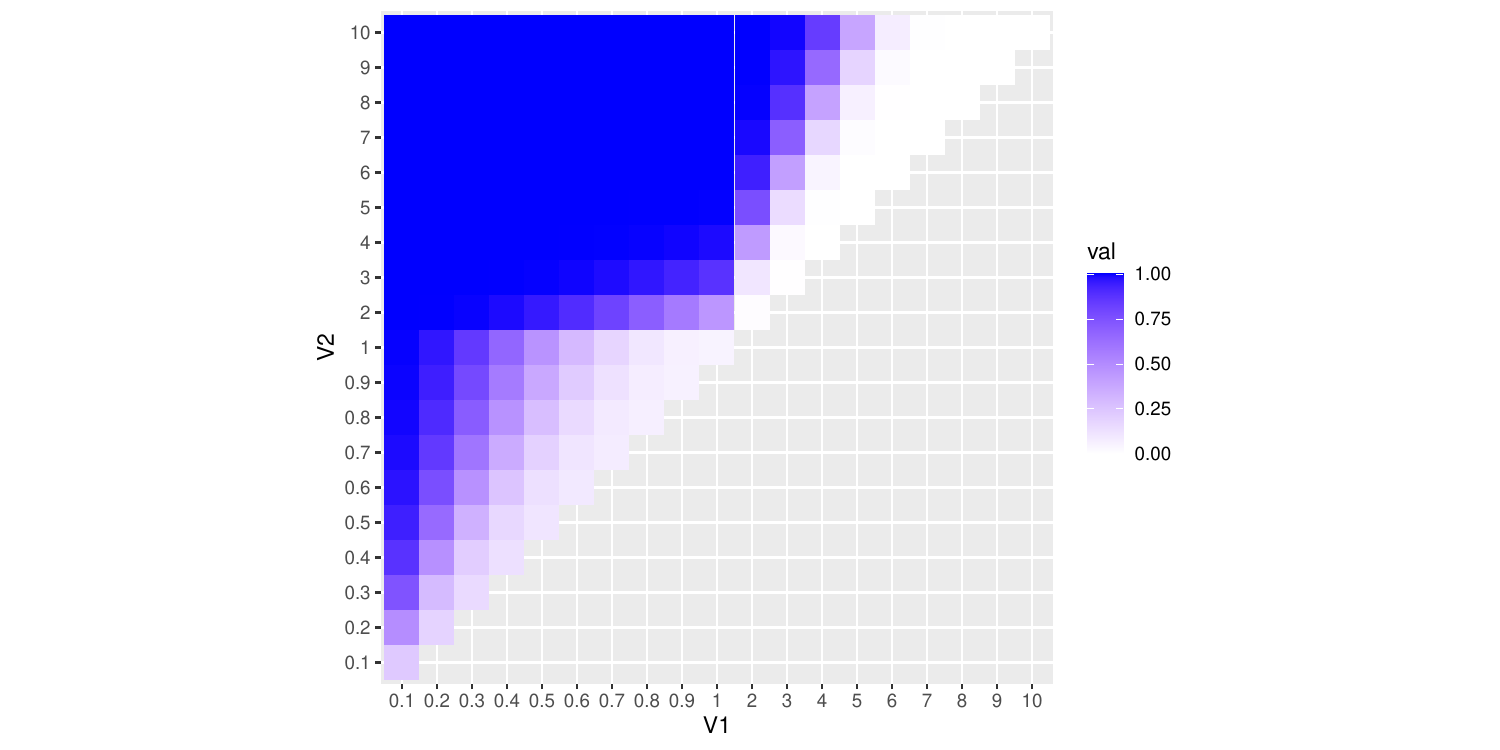}
\caption{``CvM'' test powers on beta grid of alternatives~\eqref{eq:beta_grid}.}
\label{figure:beta_grid:cvm}
\end{figure}

\paragraph{Tile plot for ``AD'': beta parameters grid}

\begin{minted}[mathescape, linenos]{r}
plot.powers.on.grid(beta.parameters.grid, res.unif.beta.grid.classic.table$AD.means)
\end{minted}

\begin{figure}[H]
\centering
\includegraphics[width=16cm]{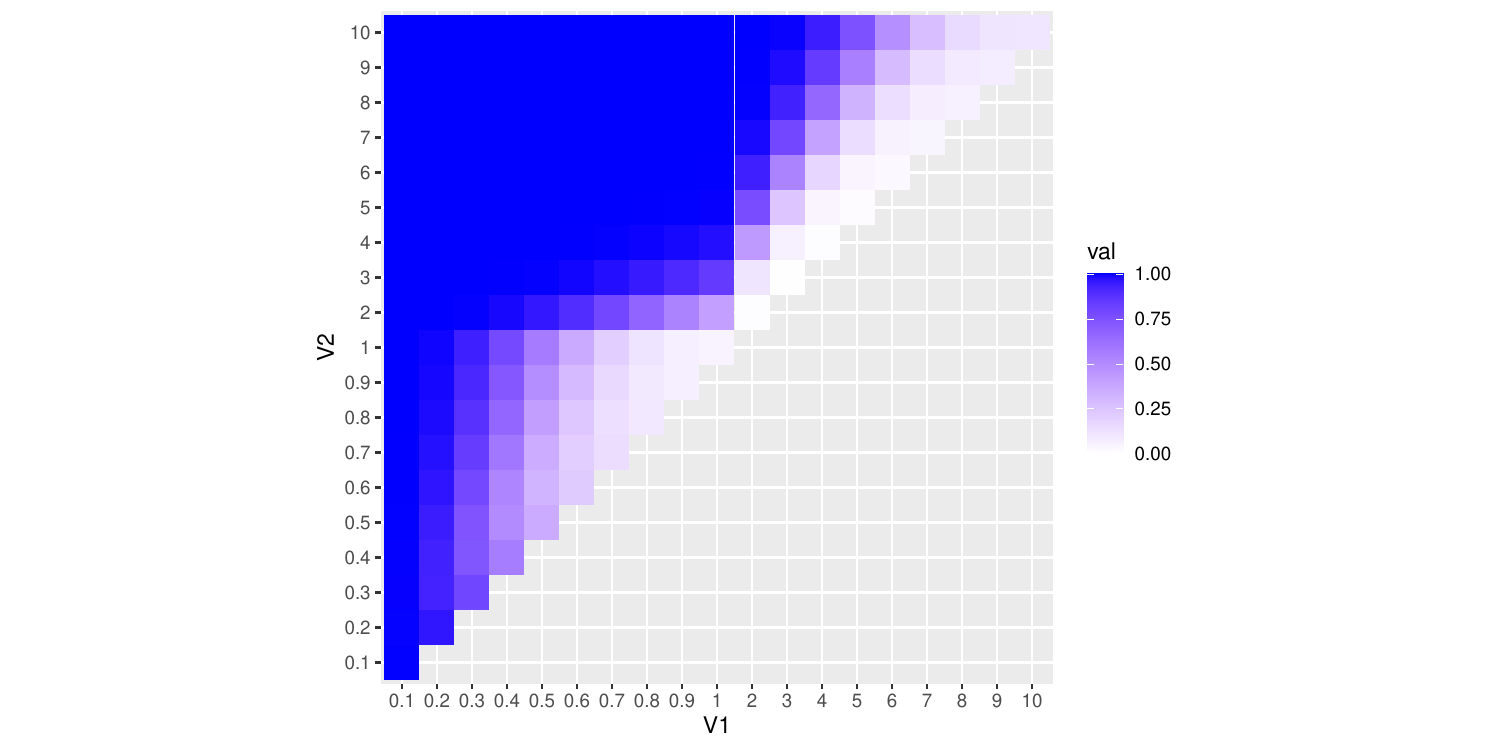}
\caption{``AD'' test powers on beta grid of alternatives~\eqref{eq:beta_grid}.}
\label{figure:beta_grid:ad}
\end{figure}

\noindent
Figures~\ref{figure:beta_grid:ks}--\ref{figure:beta_grid:ad} should be compared to figures~\ref{figure:beta_grid:osu}--\ref{figure:beta_grid:pc2u} (uniform order statistics based tests), and figures~\ref{figure:beta_grid:os}--\ref{figure:beta_grid:pc2} (normal order statistics based tests). KS test (figure~\ref{figure:beta_grid:ks}) is the most balanced among classical tests, but lacks power, CvM test (figure~\ref{figure:beta_grid:cvm}) mostly does not look well, especially for $\mathrm{Beta}(a,b)$, $a, b > 1$ alternatives (upper left triangle). AD test (figure~\ref{figure:beta_grid:ad}) looks very similar to ``OS'' test (figure~\ref{figure:beta_grid:os}). In summary, classical tests are weaker than ``PC2'' test on beta grid.

\newpage
\section{Zhang tests (\texorpdfstring{$Z_K, Z_A, Z_C$}{Zk,Za,Zc}): samples from standard uniform distribution [zh, \texorpdfstring{$U(0,1)$}{U(0,1)}]}
\label{sec:zhang_uniform}

Zhang tests for beta alternatives plotted on figures~\ref{figure:beta:aa}-\ref{figure:beta:1a} are as follows:

\begin{minted}[mathescape, linenos, texcomments]{r}
res.unif.Zhang <- calc.rejections.Zhang.tests(
    sample.generation.function.H0  = runif,
    sample.generation.functions.H1 = c(
        lapply(1:10, function(x) function(nsamples) rbeta(nsamples, x, x)),
        lapply(rev(seq(0.1, 1.0, 0.1)), function(x) function(nsamples) rbeta(nsamples, x, x)),
        lapply(rev(seq(0.1, 1.0, 0.1)), function(x) function(nsamples) rbeta(nsamples, 1, x)),
        lapply(1:10, function(x) function(nsamples) rbeta(nsamples, 1, x))
    ),
    F_0   = punif,
    alpha = 0.05,
    h     = 30,
    n = n, m = m
)
\end{minted}
The tables are:
\begin{minted}[mathescape, linenos]{r}
create.rejection.table.Zhang(res.unif.Zhang)
\end{minted}

\begin{table}[H]
\centering
\begin{tabular}{ |l|l||>{\columncolor[gray]{0.9}}p{1cm}|p{1.1cm}||>{\columncolor[gray]{0.9}}p{1cm}|p{1.1cm}||>{\columncolor[gray]{0.9}}p{1cm}|p{1.1cm}| } 
 \hline
 $H_0$ & $H_1$ & \multicolumn{2}{c||}{$Z_K$}  & \multicolumn{2}{c||}{$Z_A$} & \multicolumn{2}{c|}{$Z_C$} \\
 \hline
 &  & mean & sd & mean & sd & mean & sd \\
 \hline
 \hline

$\mathrm{U}(0, 1)$ & $\mathrm{Beta}(1,1)$ & 0.05 & 3.8e-04 & 0.05 & 2.8e-04 & 0.05 & 2.7e-04\\
$\mathrm{U}(0, 1)$ & $\mathrm{Beta}(2,2)$ & 0.0327 & 2.7e-04 & 0.0809 & 5.6e-04 & 0.0329 & 3.4e-04\\
$\mathrm{U}(0, 1)$ & $\mathrm{Beta}(3,3)$ & 0.0487 & 3.9e-04 & 0.1966 & 9.3e-04 & 0.0913 & 8.7e-04\\
$\mathrm{U}(0, 1)$ & $\mathrm{Beta}(4,4)$ & 0.0699 & 6.4e-04 & 0.3553 & 1.3e-03 & 0.1897 & 1.4e-03\\
$\mathrm{U}(0, 1)$ & $\mathrm{Beta}(5,5)$ & 0.095 & 9.4e-04 & 0.5187 & 1.6e-03 & 0.3134 & 2.2e-03\\
$\mathrm{U}(0, 1)$ & $\mathrm{Beta}(6,6)$ & 0.1232 & 1.1e-03 & 0.6614 & 1.6e-03 & 0.4439 & 2.4e-03\\
$\mathrm{U}(0, 1)$ & $\mathrm{Beta}(7,7)$ & 0.154 & 1.3e-03 & 0.7725 & 1.3e-03 & 0.567 & 2.3e-03\\
$\mathrm{U}(0, 1)$ & $\mathrm{Beta}(8,8)$ & 0.1867 & 1.8e-03 & 0.8528 & 1.1e-03 & 0.6739 & 2.4e-03\\
$\mathrm{U}(0, 1)$ & $\mathrm{Beta}(9,9)$ & 0.2211 & 1.9e-03 & 0.9077 & 7.1e-04 & 0.7616 & 2.0e-03\\
$\mathrm{U}(0, 1)$ & $\mathrm{Beta}(10,10)$ & 0.2566 & 2.1e-03 & 0.9435 & 6.4e-04 & 0.8298 & 1.9e-03\\

\hline
\end{tabular}
\caption{Zhang tests; $\mathrm{U}(0,1)$ vs $\mathrm{Beta}(a,a)$; $a \geq 1$; alternatives have modes in the center, see figure~\ref{figure:beta:aa}.}
\label{table:beta:aa:zhang:os}
\end{table}

\noindent
Table~\ref{table:beta:aa:zhang:os} should be compared to tables~\ref{table:beta:aa:classical},~\ref{table:beta:aa:normal:os}, and~\ref{table:beta:aa:uniform:os}. We see that $Z_K$ test is not much better than KS test, $Z_C$ is slightly worse than ``PC2(U)'', $Z_A$ is slightly better than ``PC2(U)'' and worse than ``PC2''. So Zhang tests can't beat ``PC2'' here.

\begin{table}[H]
\centering
\begin{tabular}{ |l|l||>{\columncolor[gray]{0.9}}p{1cm}|p{1.1cm}||>{\columncolor[gray]{0.9}}p{1cm}|p{1.1cm}||>{\columncolor[gray]{0.9}}p{1cm}|p{1.1cm}| } 
 \hline
 $H_0$ & $H_1$ & \multicolumn{2}{c||}{$Z_K$}  & \multicolumn{2}{c||}{$Z_A$} & \multicolumn{2}{c|}{$Z_C$} \\
 \hline
 &  & mean & sd & mean & sd & mean & sd \\
 \hline
 \hline
 
 $\mathrm{U}(0, 1)$ & $\mathrm{Beta}(1,1)$ & 0.05 & 3.1e-04 & 0.05 & 2.5e-04 & 0.05 & 3.4e-04\\
$\mathrm{U}(0, 1)$ & $\mathrm{Beta}(0.9,0.9)$ & 0.0676 & 3.7e-04 & 0.0607 & 3.1e-04 & 0.0751 & 4.4e-04\\
$\mathrm{U}(0, 1)$ & $\mathrm{Beta}(0.8,0.8)$ & 0.0986 & 4.2e-04 & 0.0832 & 3.8e-04 & 0.1195 & 5.0e-04\\
$\mathrm{U}(0, 1)$ & $\mathrm{Beta}(0.7,0.7)$ & 0.1524 & 5.8e-04 & 0.129 & 5.5e-04 & 0.1954 & 5.9e-04\\
$\mathrm{U}(0, 1)$ & $\mathrm{Beta}(0.6,0.6)$ & 0.2436 & 5.9e-04 & 0.2183 & 6.1e-04 & 0.3184 & 6.4e-04\\
$\mathrm{U}(0, 1)$ & $\mathrm{Beta}(0.5,0.5)$ & 0.3894 & 1.0e-03 & 0.3753 & 9.3e-04 & 0.4972 & 7.8e-04\\
$\mathrm{U}(0, 1)$ & $\mathrm{Beta}(0.4,0.4)$ & 0.5941 & 8.9e-04 & 0.6033 & 9.2e-04 & 0.7115 & 6.4e-04\\
$\mathrm{U}(0, 1)$ & $\mathrm{Beta}(0.3,0.3)$ & 0.8188 & 6.2e-04 & 0.8404 & 4.7e-04 & 0.897 & 3.5e-04\\
$\mathrm{U}(0, 1)$ & $\mathrm{Beta}(0.2,0.2)$ & 0.9669 & 2.2e-04 & 0.9758 & 2.0e-04 & 0.9864 & 1.3e-04\\
$\mathrm{U}(0, 1)$ & $\mathrm{Beta}(0.1,0.1)$ & 0.9995 & 1.9e-05 & 0.9997 & 1.6e-05 & 0.9999 & 1.3e-05\\
 
 \hline
\end{tabular}
\caption{Zhang tests; $\mathrm{U}(0,1)$ vs $\mathrm{Beta}(a,a)$; $0 < a \leq 1$; alternatives have modes on both sides, see figure~\ref{figure:beta:0a0a}.}
\label{table:beta:0a0a:zhang:os}
\end{table}

\noindent
Table~\ref{table:beta:0a0a:zhang:os} should be compared to tables~\ref{table:beta:0a0a:classical},~\ref{table:beta:0a0a:normal:os}, and~\ref{table:beta:0a0a:uniform:os}. $Z_K$ and $Z_A$ are similar to ``OS.MT(U)'', $Z_C$ is more powerful and comparable with ``OS'' and ``PC2''.

\begin{table}[H]
\centering
\begin{tabular}{ |l|l||>{\columncolor[gray]{0.9}}p{1cm}|p{1.1cm}||>{\columncolor[gray]{0.9}}p{1cm}|p{1.1cm}||>{\columncolor[gray]{0.9}}p{1cm}|p{1.1cm}| } 
 \hline
 $H_0$ & $H_1$ & \multicolumn{2}{c||}{$Z_K$}  & \multicolumn{2}{c||}{$Z_A$} & \multicolumn{2}{c|}{$Z_C$} \\
 \hline
 &  & mean & sd & mean & sd & mean & sd \\
 \hline
 \hline
 
 $\mathrm{U}(0, 1)$ & $\mathrm{Beta}(1,1)$ & 0.0499 & 3.1e-04 & 0.05 & 3.4e-04 & 0.05 & 2.8e-04\\
$\mathrm{U}(0, 1)$ & $\mathrm{Beta}(1,0.9)$ & 0.0659 & 3.8e-04 & 0.0636 & 3.7e-04 & 0.0705 & 3.2e-04\\
$\mathrm{U}(0, 1)$ & $\mathrm{Beta}(1,0.8)$ & 0.1084 & 4.2e-04 & 0.1053 & 3.7e-04 & 0.1229 & 3.8e-04\\
$\mathrm{U}(0, 1)$ & $\mathrm{Beta}(1,0.7)$ & 0.1928 & 7.5e-04 & 0.1915 & 6.4e-04 & 0.2246 & 6.2e-04\\
$\mathrm{U}(0, 1)$ & $\mathrm{Beta}(1,0.6)$ & 0.335 & 7.3e-04 & 0.3395 & 6.0e-04 & 0.3886 & 6.4e-04\\
$\mathrm{U}(0, 1)$ & $\mathrm{Beta}(1,0.5)$ & 0.5362 & 8.5e-04 & 0.5491 & 9.9e-04 & 0.6039 & 6.7e-04\\
$\mathrm{U}(0, 1)$ & $\mathrm{Beta}(1,0.4)$ & 0.7589 & 6.9e-04 & 0.7751 & 6.7e-04 & 0.8158 & 4.6e-04\\
$\mathrm{U}(0, 1)$ & $\mathrm{Beta}(1,0.3)$ & 0.9267 & 3.6e-04 & 0.9362 & 3.2e-04 & 0.9521 & 2.4e-04\\
$\mathrm{U}(0, 1)$ & $\mathrm{Beta}(1,0.2)$ & 0.9924 & 7.9e-05 & 0.994 & 9.0e-05 & 0.9959 & 6.6e-05\\
$\mathrm{U}(0, 1)$ & $\mathrm{Beta}(1,0.1)$ & 1 & 7.7e-06 & 1 & 6.4e-06 & 1 & 4.5e-06\\
 
  \hline
\end{tabular}
\caption{Zhang tests; $\mathrm{U}(0,1)$ vs $\mathrm{Beta}(1,a)$; $0 < a \leq 1$; alternatives have sharp modes on the right, see figure~\ref{figure:beta:10a}.}
\label{table:beta:10a:zhang:os}
\end{table}

\noindent
Table~\ref{table:beta:10a:zhang:os} should be compared to tables~\ref{table:beta:10a:classical},~\ref{table:beta:10a:normal:os}, and~\ref{table:beta:10a:uniform:os}. $Z_K$ and $Z_A$ are slightly behind ``PC2'' and ``OS'', $Z_C$ is a bit more powerful than ``PC2'' and ``OS''.

\begin{table}[H]
\centering
\begin{tabular}{ |l|l||>{\columncolor[gray]{0.9}}p{1cm}|p{1.1cm}||>{\columncolor[gray]{0.9}}p{1cm}|p{1.1cm}||>{\columncolor[gray]{0.9}}p{1cm}|p{1.1cm}| } 
 \hline
 $H_0$ & $H_1$ & \multicolumn{2}{c||}{$Z_K$}  & \multicolumn{2}{c||}{$Z_A$} & \multicolumn{2}{c|}{$Z_C$} \\
 \hline
 &  & mean & sd & mean & sd & mean & sd \\
 \hline
 \hline
 
 $\mathrm{U}(0, 1)$ & $\mathrm{Beta}(1,1)$ & 0.05 & 3.5e-04 & 0.05 & 3.3e-04 & 0.0501 & 3.4e-04\\
$\mathrm{U}(0, 1)$ & $\mathrm{Beta}(1,2)$ & 0.3498 & 1.0e-03 & 0.4399 & 8.8e-04 & 0.3637 & 9.7e-04\\
$\mathrm{U}(0, 1)$ & $\mathrm{Beta}(1,3)$ & 0.7724 & 9.6e-04 & 0.8684 & 6.7e-04 & 0.806 & 9.1e-04\\
$\mathrm{U}(0, 1)$ & $\mathrm{Beta}(1,4)$ & 0.9501 & 4.1e-04 & 0.9835 & 1.7e-04 & 0.9672 & 3.2e-04\\
$\mathrm{U}(0, 1)$ & $\mathrm{Beta}(1,5)$ & 0.9919 & 1.1e-04 & 0.9986 & 4.0e-05 & 0.9963 & 7.6e-05\\
$\mathrm{U}(0, 1)$ & $\mathrm{Beta}(1,6)$ & 0.9989 & 4.0e-05 & 0.9999 & 8.0e-06 & 0.9997 & 2.0e-05\\
$\mathrm{U}(0, 1)$ & $\mathrm{Beta}(1,7)$ & 0.9999 & 8.1e-06 & 1 & 2.1e-06 & 1 & 6.2e-06\\
$\mathrm{U}(0, 1)$ & $\mathrm{Beta}(1,8)$ & 1 & 4.1e-06 & 1 & 4.1e-07 & 1 & 1.4e-06\\
$\mathrm{U}(0, 1)$ & $\mathrm{Beta}(1,9)$ & 1 & 8.5e-07 & 1 & 0.0e+00 & 1 & 2.5e-07\\
$\mathrm{U}(0, 1)$ & $\mathrm{Beta}(1,10)$ & 1 & 3.1e-07 & 1 & 0.0e+00 & 1 & 1.8e-07\\
 
   \hline
\end{tabular}
\caption{Zhang tests; $\mathrm{U}(0,1)$ vs $\mathrm{Beta}(1,a)$; $a \geq 1$; alternatives have sloping modes on the left, see figure~\ref{figure:beta:1a}.}
\label{table:beta:1a:zhang:os}
\end{table}
 
 \noindent
Table~\ref{table:beta:1a:zhang:os} should be compared to tables~\ref{table:beta:1a:classical},~\ref{table:beta:1a:normal:os}, and~\ref{table:beta:1a:uniform:os}. All Zhang tests are a bit better than ``OS'' and ``PC2'' tests. $Z_A$ is the best.

So far $Z_A$ and $Z_C$ look powerful. Let us construct the power plots on the extended beta parameters grid:

\begin{minted}[mathescape, linenos, texcomments]{r}
res.unif.beta.grid.Zhang <- calc.rejections.Zhang.tests(
    sample.generation.function.H0  = runif,
    sample.generation.functions.H1 = apply(
        beta.parameters.grid,  # see formula \eqref{eq:beta_grid}
        1, 
        function(par) function(nsamples) rbeta(nsamples, par[1], par[2])
    ),
    F_0   = punif,
    alpha = 0.05,
    h     = 10,
    n = n, m = m
)
\end{minted}
The table for the plots is as follows:
\begin{minted}[mathescape, linenos]{r}
res.unif.beta.grid.Zhang.table <- create.rejection.table.Zhang(res.unif.beta.grid.Zhang)
\end{minted}

\newpage

\paragraph{Tile plot for $Z_K$: beta parameters grid}

\begin{minted}[mathescape, linenos, texcomments]{r}
plot.powers.on.grid(        # defined in subsection \ref{section:rejections:uniform:vs:beta} after formula \eqref{eq:beta_grid}
    beta.parameters.grid,   # defined just before formula \eqref{eq:beta_grid}
    res.unif.beta.grid.Zhang.table$Zk.means
)
\end{minted}

\begin{figure}[H]
\centering
\includegraphics[width=16cm]{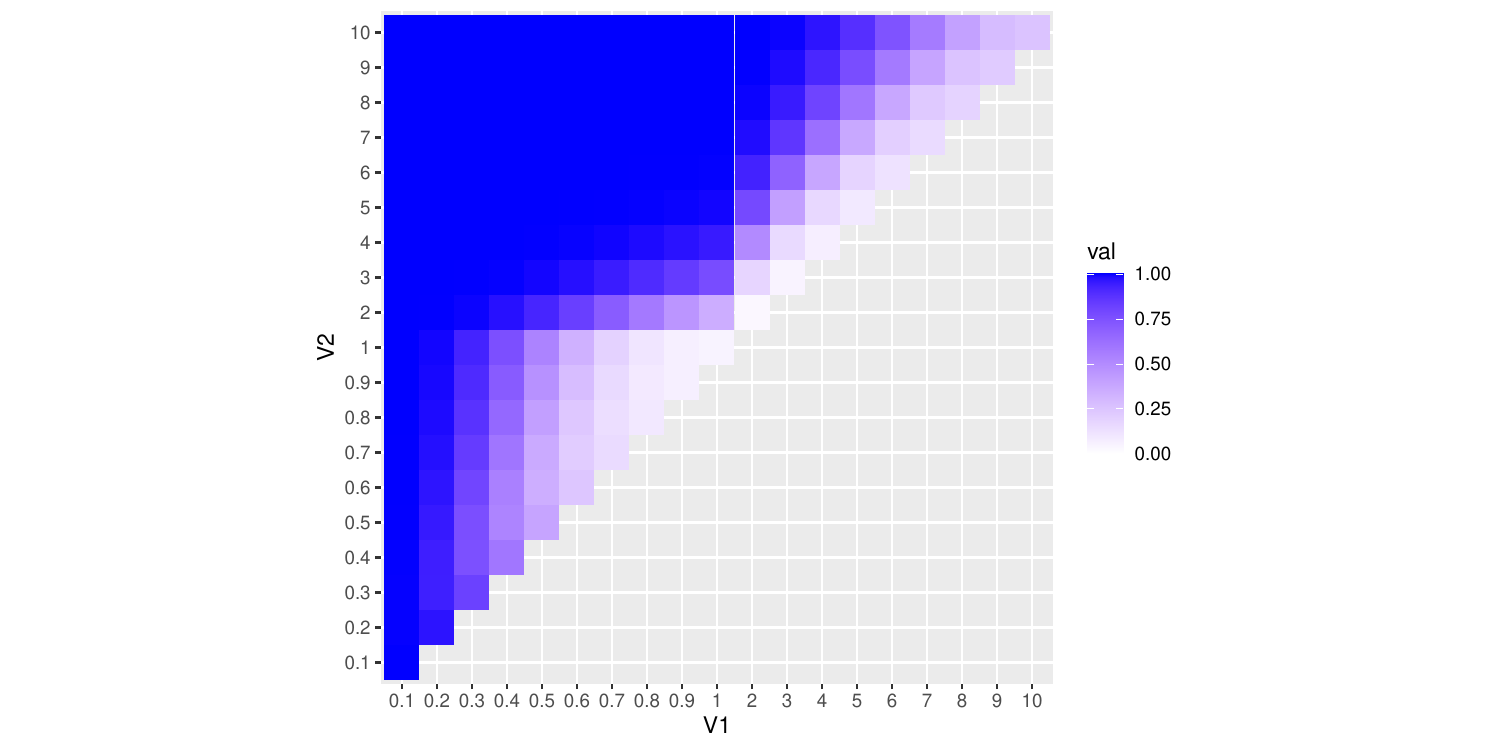}
\caption{$Z_K$ test powers on beta grid of alternatives~\eqref{eq:beta_grid}.}
\label{figure:beta_grid:zk}
\end{figure}

\paragraph{Tile plot for $Z_A$: beta parameters grid}

\begin{minted}[mathescape, linenos]{r}
plot.powers.on.grid(beta.parameters.grid, res.unif.beta.grid.Zhang.table$Za.means)
\end{minted}

\begin{figure}[H]
\centering
\includegraphics[width=16cm]{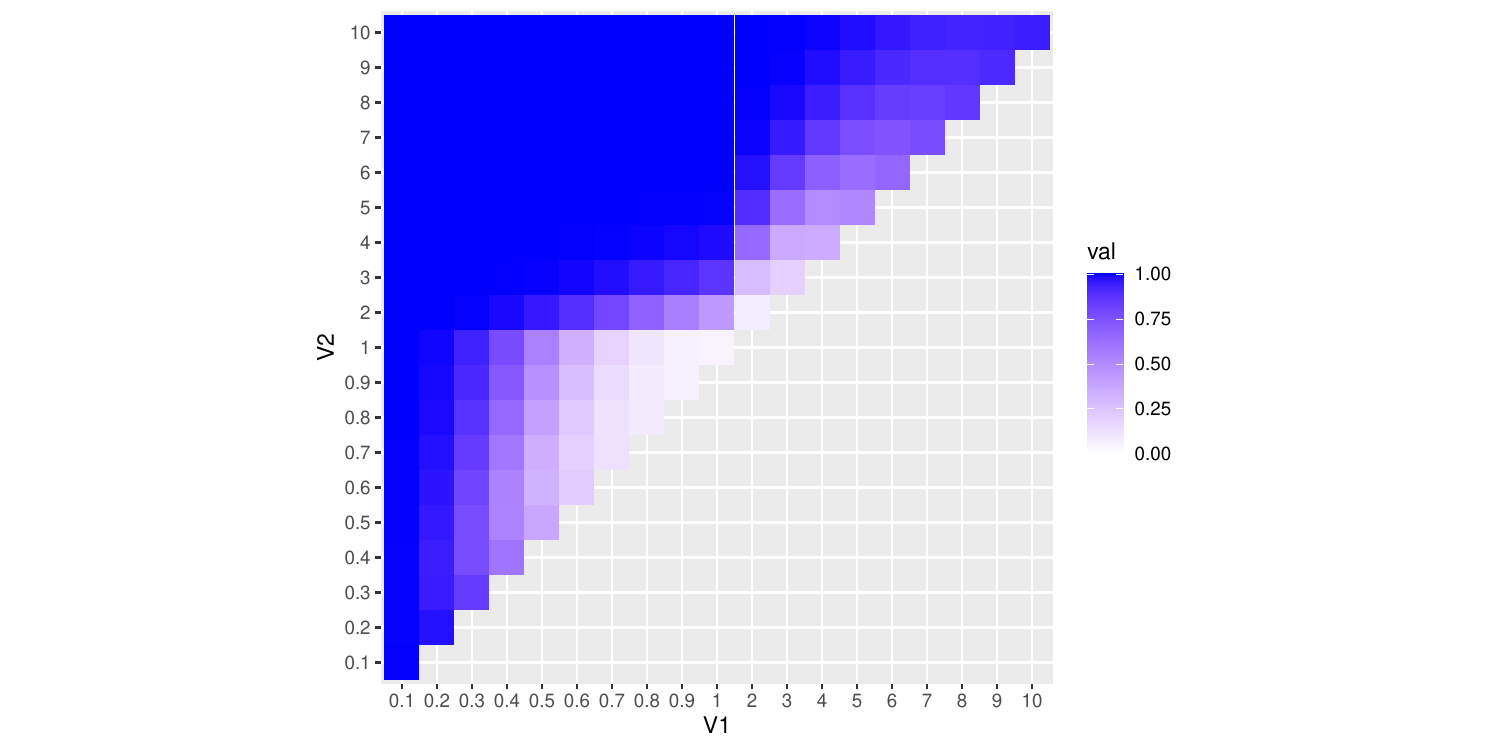}
\caption{$Z_A$ test powers on beta grid of alternatives~\eqref{eq:beta_grid}.}
\label{figure:beta_grid:za}
\end{figure}

\newpage

\paragraph{Tile plot for $Z_C$: beta parameters grid}

\begin{minted}[mathescape, linenos]{r}
plot.powers.on.grid(beta.parameters.grid, res.unif.beta.grid.Zhang.table$Zc.means)
\end{minted}

\begin{figure}[H]
\centering
\includegraphics[width=16cm]{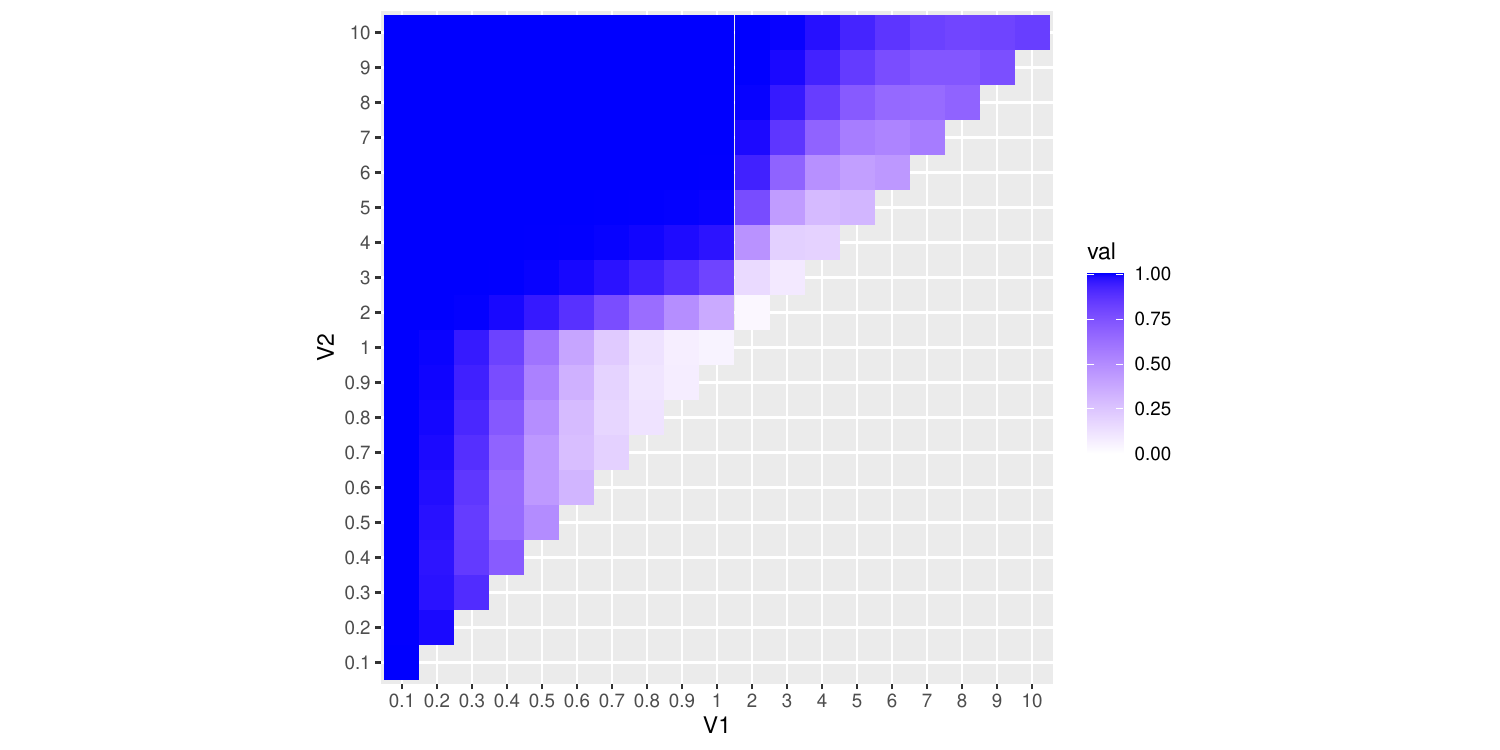}
\caption{$Z_C$ test powers on beta grid of alternatives~\eqref{eq:beta_grid}.}
\label{figure:beta_grid:zc}
\end{figure}

\noindent
Figures~\ref{figure:beta_grid:zk}--\ref{figure:beta_grid:zc}  should be compared to figures~\ref{figure:beta_grid:osu}--\ref{figure:beta_grid:pc2u} (uniform order statistics based tests), figures~\ref{figure:beta_grid:os}--\ref{figure:beta_grid:pc2} (normal order statistics based tests), and figures~\ref{figure:beta_grid:ks}--\ref{figure:beta_grid:ad} (classical tests). We see that figure~\ref{figure:beta_grid:zk} for $Z_K$ test resembles figure~\ref{figure:beta_grid:osmtu} for ``OS.MT(U)'' test. Figure~\ref{figure:beta_grid:za} for $Z_A$ test looks prominent with strong and pre\-dict\-able performance for $\mathrm{Beta}(a,b)$, $a,b > 1$ alternatives (upper left triangle of the plot). Figure~\ref{figure:beta_grid:zc} for $Z_C$ test
looks less impressive than the figure for $Z_A$ test. 

All in all, $Z_C$ and $Z_A$ tests are powerful in ``uniform vs beta'' case, they are comparable to the proposed ``PC2'' and ``PC2(U)'' tests. ``PC2'' and ``PC2(U)'' tests perform better for symmetric $\mathrm{Beta}(a,a)$, $a > 1$ alternatives (especially, ``PC2'' test), while $Z_C$ and $Z_A$ better handle asymmetric situation $\mathrm{Beta}(a,b)$, $a,b > 1$, $a \neq b$ (especially, $Z_A$ test). For other beta alternatives these tests perform more or less the same. All these tests are more preferable compared to ``OS'' graphical test from~\cite{AldorNoimanBrownBujaRolkeStine2013}. 

\newpage
\section{Tests based on simultaneous bounds for eCDF values on a grid}
\label{sec:2022verticals}

In this section we are going to recreate tests from~\cite{sailynoja2022graphical} and introduce new tests based on principal components of eCDF values computed on a grid.

\subsection{Function to compute empirical cumulative distribution function values on a grid \textasteriskcentered}
To compute eCDF values given some \texttt{x.grid} we could use the following pure \texttt{R} code:
\begin{minted}[mathescape, linenos]{r}
get.ecdfs.on.grid <- function(sample.generation.function) {
    samples <- sample.generation.function(n * m)
    dim(samples) <- c(m, n)
    t(apply(samples, 1, function(samp) ecdf(samp)(x.grid)))
}
\end{minted}
To do the same job faster, we rewrite critical parts in C++:
\begin{minted}[mathescape, linenos]{r}
cppFunction('
NumericMatrix ecdfs_on_grid_cpp(NumericMatrix samples, NumericVector x_grid) 
{
    int m = samples.nrow();             // infer number of samples 
    int n = samples.ncol();             // infer sample size 
    int x_grid_size = x_grid.size();    // infer grid size

    NumericMatrix out(m, x_grid_size);  // reserve memory for the final result

    std::vector<double> buf(n);         // reserve memory for the buffer

    for(int i = 0; i < m; i++) {        // iterate over the samples

        for(int j = 0; j < n; j++) {    // write sample in the buffer
            buf[j] = samples(i, j);
        }

        std::sort(buf.begin(), buf.end());  // sort the sample in the buffer

        int f = 0; 

        for(int g = 0; g < x_grid_size; g++) {  // compute ecdf values on grid
            double x = x_grid[g];
            while(f < n && buf[f] <= x) f++;
            out(i, g) = double(f) / n;
        }
    }
    
    return out;
}
')
\end{minted}
We fix the grid:
\begin{minted}[mathescape, linenos]{r}
x.grid <- seq(0.05, 0.95, 0.05)
\end{minted}
And the final function to get statistics is as follows:
\begin{minted}[mathescape, linenos]{r}
get.ecdfs.on.grid.cpp <- function(sample.generation.function, n, m){
    samples <- sample.generation.function(n * m)
    dim(samples) <- c(m, n)
    ecdfs_on_grid_cpp(samples, x.grid)
}
\end{minted}

\subsection{Calculate rejections: uniform $H_0$, beta $H_1$}
Let us use new \texttt{get.ecdfs.on.grid.cpp} function in combination with \texttt{calc.rejections.sm.pc.bounds}:
\begin{minted}[mathescape, linenos]{r}
res.uniform.vs.beta.F <- calc.rejections.sm.pc.bounds(
    sample.generation.function.H0  = runif,
    sample.generation.functions.H1 = c(
        lapply(1:10, function(x) function(nsamples) rbeta(nsamples, x, x)),
        lapply(rev(seq(0.1, 1.0, 0.1)), function(x) function(nsamples) rbeta(nsamples, x, x)),
        lapply(rev(seq(0.1, 1.0, 0.1)), function(x) function(nsamples) rbeta(nsamples, 1, x)),
        lapply(1:10, function(x) function(nsamples) rbeta(nsamples, 1, x))
    ),
    get.stat.matrix = get.ecdfs.on.grid.cpp,
    alpha = 0.05,
    n = n, m = m,
    h = 10,
    w.sm = rep(1, length(x.grid)),
    w.pc = rep(1, length(x.grid))
)
\end{minted}
Now, to the tables:
\begin{minted}[mathescape, linenos]{r}
res.uniform.vs.beta.F.table <- create.rejection.table(res.uniform.vs.beta.F)
res.uniform.vs.beta.F.table
\end{minted}

\begin{table}[H]
\centering
\begin{tabular}{ |l|l||>{\columncolor[gray]{0.9}}p{1cm}|p{1.1cm}||>{\columncolor[gray]{0.9}}p{1cm}|p{1.1cm}||>{\columncolor[gray]{0.9}}p{1cm}|p{1.1cm}| } 
 \hline
 $H_0$ & $H_1$ & \multicolumn{2}{c||}{ECDFV}  & \multicolumn{2}{c||}{ECDFV.PC1} & \multicolumn{2}{c|}{ECDFV.PC2} \\
 \hline
 &  & mean & sd & mean & sd & mean & sd \\
 \hline
 \hline
 
$\mathrm{U}(0, 1)$ & $\mathrm{Beta}(1, 1)$ & 0.0491 & 2.1e-03 & 0.0501 & 3.0e-04 & 0.05 & 3.1e-04\\
$\mathrm{U}(0, 1)$ & $\mathrm{Beta}(2, 2)$ & 0.0314 & 1.1e-03 & 0.0842 & 8.7e-04 & 0.0934 & 7.9e-04\\
$\mathrm{U}(0, 1)$ & $\mathrm{Beta}(3, 3)$ & 0.0463 & 1.0e-03 & 0.1823 & 2.5e-03 & 0.2609 & 1.9e-03\\
$\mathrm{U}(0, 1)$ & $\mathrm{Beta}(4, 4)$ & 0.0667 & 9.4e-04 & 0.3349 & 4.2e-03 & 0.4764 & 2.5e-03\\
$\mathrm{U}(0, 1)$ & $\mathrm{Beta}(5, 5)$ & 0.0906 & 9.2e-04 & 0.5029 & 5.2e-03 & 0.6673 & 2.5e-03\\
$\mathrm{U}(0, 1)$ & $\mathrm{Beta}(6, 6)$ & 0.1176 & 9.8e-04 & 0.6532 & 5.1e-03 & 0.8052 & 2.0e-03\\
$\mathrm{U}(0, 1)$ & $\mathrm{Beta}(7, 7)$ & 0.1471 & 7.5e-04 & 0.771 & 4.4e-03 & 0.8927 & 1.4e-03\\
$\mathrm{U}(0, 1)$ & $\mathrm{Beta}(8, 8)$ & 0.1787 & 8.0e-04 & 0.8554 & 3.4e-03 & 0.9437 & 8.3e-04\\
$\mathrm{U}(0, 1)$ & $\mathrm{Beta}(9, 9)$ & 0.2117 & 7.5e-04 & 0.9119 & 2.5e-03 & 0.9715 & 5.4e-04\\
$\mathrm{U}(0, 1)$ & $\mathrm{Beta}(10, 10)$ & 0.2458 & 6.6e-04 & 0.9478 & 1.8e-03 & 0.986 & 3.2e-04\\

           \hline
\end{tabular}
\caption{ECDF values based tests; $\mathrm{U}(0,1)$ vs $\mathrm{Beta}(a,a)$; $a \geq 1$; alternatives have modes in the center, see figure~\ref{figure:beta:aa}.}
\label{table:beta:aa:ecdf:values}
\end{table}

\noindent
Table~\ref{table:beta:aa:ecdf:values} should be compared to tables~\ref{table:beta:aa:zhang:os},~\ref{table:beta:aa:classical},~\ref{table:beta:aa:normal:os}, and~\ref{table:beta:aa:uniform:os}.
In this table ``ECDFV'' denotes the test from~\cite{sailynoja2022graphical} based on eCDF values on a grid, ``ECDFV.PC1'' --- principal component based test with equal weights, and ``ECDFV.PC2'' --- principal component based test with variance dependent weights. We see that ``ECDFV'' test lacks power for symmetric alternatives concentrated in the center; ``ECDFV.PC1'' and ``ECDFV.PC2'' have very good power. ``ECDFV.PC2'' is almost as good as ``PC2''.

\begin{table}[H]
\centering
\begin{tabular}{ |l|l||>{\columncolor[gray]{0.9}}p{1cm}|p{1.1cm}||>{\columncolor[gray]{0.9}}p{1cm}|p{1.1cm}||>{\columncolor[gray]{0.9}}p{1cm}|p{1.1cm}| } 
 \hline
 $H_0$ & $H_1$ & \multicolumn{2}{c||}{ECDFV}  & \multicolumn{2}{c||}{ECDFV.PC1} & \multicolumn{2}{c|}{ECDFV.PC2} \\
 \hline
 &  & mean & sd & mean & sd & mean & sd \\
 
 \hline
 \hline
 
$\mathrm{U}(0, 1)$ & $\mathrm{Beta}(1, 1)$ & 0.0491 & 2.1e-03 & 0.0501 & 2.4e-04 & 0.05 & 2.5e-04\\
$\mathrm{U}(0, 1)$ & $\mathrm{Beta}(0.9, 0.9)$ & 0.0613 & 2.4e-03 & 0.0511 & 2.9e-04 & 0.0566 & 3.7e-04\\
$\mathrm{U}(0, 1)$ & $\mathrm{Beta}(0.8, 0.8)$ & 0.0801 & 2.4e-03 & 0.0552 & 3.3e-04 & 0.0684 & 4.3e-04\\
$\mathrm{U}(0, 1)$ & $\mathrm{Beta}(0.7, 0.7)$ & 0.1087 & 2.8e-03 & 0.0655 & 5.2e-04 & 0.0889 & 5.2e-04\\
$\mathrm{U}(0, 1)$ & $\mathrm{Beta}(0.6, 0.6)$ & 0.1534 & 2.9e-03 & 0.0883 & 8.8e-04 & 0.1246 & 8.3e-04\\
$\mathrm{U}(0, 1)$ & $\mathrm{Beta}(0.5, 0.5)$ & 0.223 & 2.9e-03 & 0.1354 & 1.5e-03 & 0.1876 & 1.3e-03\\
$\mathrm{U}(0, 1)$ & $\mathrm{Beta}(0.4, 0.4)$ & 0.3311 & 2.6e-03 & 0.2282 & 2.8e-03 & 0.2961 & 2.1e-03\\
$\mathrm{U}(0, 1)$ & $\mathrm{Beta}(0.3, 0.3)$ & 0.4944 & 1.9e-03 & 0.3968 & 4.0e-03 & 0.4717 & 2.5e-03\\
$\mathrm{U}(0, 1)$ & $\mathrm{Beta}(0.2, 0.2)$ & 0.7174 & 9.9e-04 & 0.6573 & 4.5e-03 & 0.7145 & 2.8e-03\\
$\mathrm{U}(0, 1)$ & $\mathrm{Beta}(0.1, 0.1)$ & 0.9373 & 3.4e-04 & 0.9251 & 1.8e-03 & 0.9405 & 1.4e-03\\

           \hline
\end{tabular}
\caption{ECDF values based tests; $\mathrm{U}(0,1)$ vs $\mathrm{Beta}(a,a)$; $0 < a \leq 1$; alternatives have modes on both sides, see figure~\ref{figure:beta:0a0a}.}
\label{table:beta:0a0a:ecdf:values}
\end{table}

\noindent
Table~\ref{table:beta:0a0a:ecdf:values} should be compared to tables~\ref{table:beta:0a0a:zhang:os},~\ref{table:beta:0a0a:classical},~\ref{table:beta:0a0a:normal:os}, and~\ref{table:beta:0a0a:uniform:os}. All three tests have relatively ok performance, but not the top-notch. They are a bit worse than ``PC1(U)''.

\begin{table}[H]
\centering
\begin{tabular}{ |l|l||>{\columncolor[gray]{0.9}}p{1cm}|p{1.1cm}||>{\columncolor[gray]{0.9}}p{1cm}|p{1.1cm}||>{\columncolor[gray]{0.9}}p{1cm}|p{1.1cm}| } 
 \hline
 $H_0$ & $H_1$ & \multicolumn{2}{c||}{ECDFV}  & \multicolumn{2}{c||}{ECDFV.PC1} & \multicolumn{2}{c|}{ECDFV.PC2} \\
 \hline
 &  & mean & sd & mean & sd & mean & sd \\
 
 \hline
 \hline
 
$\mathrm{U}(0, 1)$ & $\mathrm{Beta}(1, 1)$ & 0.0484 & 1.9e-03 & 0.05 & 3.4e-04 & 0.05 & 3.3e-04\\
$\mathrm{U}(0, 1)$ & $\mathrm{Beta}(1, 0.9)$ & 0.0617 & 2.8e-03 & 0.0515 & 3.9e-04 & 0.0593 & 3.8e-04\\
$\mathrm{U}(0, 1)$ & $\mathrm{Beta}(1, 0.8)$ & 0.0976 & 4.3e-03 & 0.0579 & 6.5e-04 & 0.0868 & 5.2e-04\\
$\mathrm{U}(0, 1)$ & $\mathrm{Beta}(1, 0.7)$ & 0.1674 & 6.2e-03 & 0.075 & 1.2e-03 & 0.1421 & 7.8e-04\\
$\mathrm{U}(0, 1)$ & $\mathrm{Beta}(1, 0.6)$ & 0.2844 & 8.0e-03 & 0.1148 & 2.0e-03 & 0.237 & 1.2e-03\\
$\mathrm{U}(0, 1)$ & $\mathrm{Beta}(1, 0.5)$ & 0.4554 & 8.6e-03 & 0.1975 & 3.3e-03 & 0.3815 & 1.5e-03\\
$\mathrm{U}(0, 1)$ & $\mathrm{Beta}(1, 0.4)$ & 0.6644 & 7.0e-03 & 0.3488 & 4.8e-03 & 0.5723 & 1.5e-03\\
$\mathrm{U}(0, 1)$ & $\mathrm{Beta}(1, 0.3)$ & 0.8585 & 3.9e-03 & 0.5792 & 5.4e-03 & 0.7776 & 1.2e-03\\
$\mathrm{U}(0, 1)$ & $\mathrm{Beta}(1, 0.2)$ & 0.9721 & 9.9e-04 & 0.8335 & 3.7e-03 & 0.9338 & 6.2e-04\\
$\mathrm{U}(0, 1)$ & $\mathrm{Beta}(1, 0.1)$ & 0.9992 & 4.4e-05 & 0.9824 & 7.1e-04 & 0.9952 & 1.0e-04\\

           \hline
\end{tabular}
\caption{ECDF values based tests; $\mathrm{U}(0,1)$ vs $\mathrm{Beta}(1,a)$; $0 < a \leq 1$; alternatives have sharp modes on the right, see figure~\ref{figure:beta:10a}.}
\label{table:beta:10a:ecdf:values}
\end{table}

\noindent
Table~\ref{table:beta:10a:ecdf:values} should be compared to tables~\ref{table:beta:10a:zhang:os},~\ref{table:beta:10a:classical},~\ref{table:beta:10a:normal:os}, and~\ref{table:beta:10a:uniform:os}.
``ECDFV'' is comparable with ``PC2(U)'', ``ECDFV.PC2'' is a bit worse than KS, ``ECDFV.PC1'' lags behind but still almost reaches 1 for $\mathrm{Beta}(1,0.1)$.

\begin{table}[H]
\centering
\begin{tabular}{ |l|l||>{\columncolor[gray]{0.9}}p{1cm}|p{1.1cm}||>{\columncolor[gray]{0.9}}p{1cm}|p{1.1cm}||>{\columncolor[gray]{0.9}}p{1cm}|p{1.1cm}| } 
 \hline
 $H_0$ & $H_1$ & \multicolumn{2}{c||}{ECDFV}  & \multicolumn{2}{c||}{ECDFV.PC1} & \multicolumn{2}{c|}{ECDFV.PC2} \\
 \hline
 &  & mean & sd & mean & sd & mean & sd \\

 \hline
 \hline

$\mathrm{U}(0, 1)$ & $\mathrm{Beta}(1, 1)$ & 0.0485 & 1.9e-03 & 0.0499 & 2.6e-04 & 0.05 & 3.0e-04\\
$\mathrm{U}(0, 1)$ & $\mathrm{Beta}(1, 2)$ & 0.3535 & 1.1e-02 & 0.1349 & 1.8e-03 & 0.3762 & 1.3e-03\\
$\mathrm{U}(0, 1)$ & $\mathrm{Beta}(1, 3)$ & 0.7751 & 9.4e-03 & 0.4541 & 4.3e-03 & 0.8219 & 9.9e-04\\
$\mathrm{U}(0, 1)$ & $\mathrm{Beta}(1, 4)$ & 0.9511 & 3.4e-03 & 0.7777 & 3.7e-03 & 0.9729 & 2.4e-04\\
$\mathrm{U}(0, 1)$ & $\mathrm{Beta}(1, 5)$ & 0.9921 & 7.2e-04 & 0.9362 & 1.6e-03 & 0.9973 & 4.9e-05\\
$\mathrm{U}(0, 1)$ & $\mathrm{Beta}(1, 6)$ & 0.999 & 1.4e-04 & 0.9859 & 5.0e-04 & 0.9998 & 1.1e-05\\
$\mathrm{U}(0, 1)$ & $\mathrm{Beta}(1, 7)$ & 0.9999 & 2.2e-05 & 0.9974 & 1.3e-04 & 1 & 4.4e-06\\
$\mathrm{U}(0, 1)$ & $\mathrm{Beta}(1, 8)$ & 1 & 4.0e-06 & 0.9996 & 2.3e-05 & 1 & 6.8e-07\\
$\mathrm{U}(0, 1)$ & $\mathrm{Beta}(1, 9)$ & 1 & 1.1e-06 & 0.9999 & 8.4e-06 & 1 & 0.0e+00\\
$\mathrm{U}(0, 1)$ & $\mathrm{Beta}(1, 10)$ & 1 & 4.6e-07 & 1 & 2.9e-06 & 1 & 0.0e+00\\
 
           \hline
\end{tabular}
\caption{ECDF values based tests; $\mathrm{U}(0,1)$ vs $\mathrm{Beta}(1,a)$; $a \geq 1$; alternatives have sloping modes on the left, see figure~\ref{figure:beta:1a}.}
\label{table:beta:1a:ecdf:values}
\end{table}

\noindent
Table~\ref{table:beta:1a:ecdf:values} should be compared to tables~\ref{table:beta:1a:zhang:os},~\ref{table:beta:1a:classical},~\ref{table:beta:1a:normal:os}, and~\ref{table:beta:1a:uniform:os}. ``ECDFV.PC2'' is a bit behind AD, ``ECDFV'' --- a bit behind KS, ``ECDFV.PC1'' is again behind.

Now to the beta grid:
\begin{minted}[mathescape, linenos, texcomments]{r}
res.uniform.vs.beta.2dgrid.F <- calc.rejections.sm.pc.bounds(
    sample.generation.function.H0  = runif,
    sample.generation.functions.H1 = apply(
        beta.parameters.grid,  # see formula \eqref{eq:beta_grid} 
        1, 
        function(par) function(nsamples) rbeta(nsamples, par[1], par[2])
    ),
    get.stat.matrix = get.ecdfs.on.grid.cpp,
    alpha = 0.05,
    n = n, m = m,
    h = 10,
    w.sm = rep(1, length(x.grid)),
    w.pc = rep(1, length(x.grid))
)
\end{minted}
Tables for the plots are as follows:
\begin{minted}[mathescape, linenos, texcomments]{r}
res.uniform.vs.beta.2dgrid.F.table <- create.rejection.table(res.uniform.vs.beta.2dgrid.F)
res.uniform.vs.beta.2dgrid.F.table
\end{minted}
Finally, to the plots:

\paragraph{Tile plot for ``ECDFV'': beta parameters grid}

\begin{minted}[mathescape, linenos]{r}
plot.powers.on.grid(beta.parameters.grid, res.uniform.vs.beta.2dgrid.F.table$sm.means)
\end{minted}

\begin{figure}[H]
\centering
\includegraphics[width=16cm]{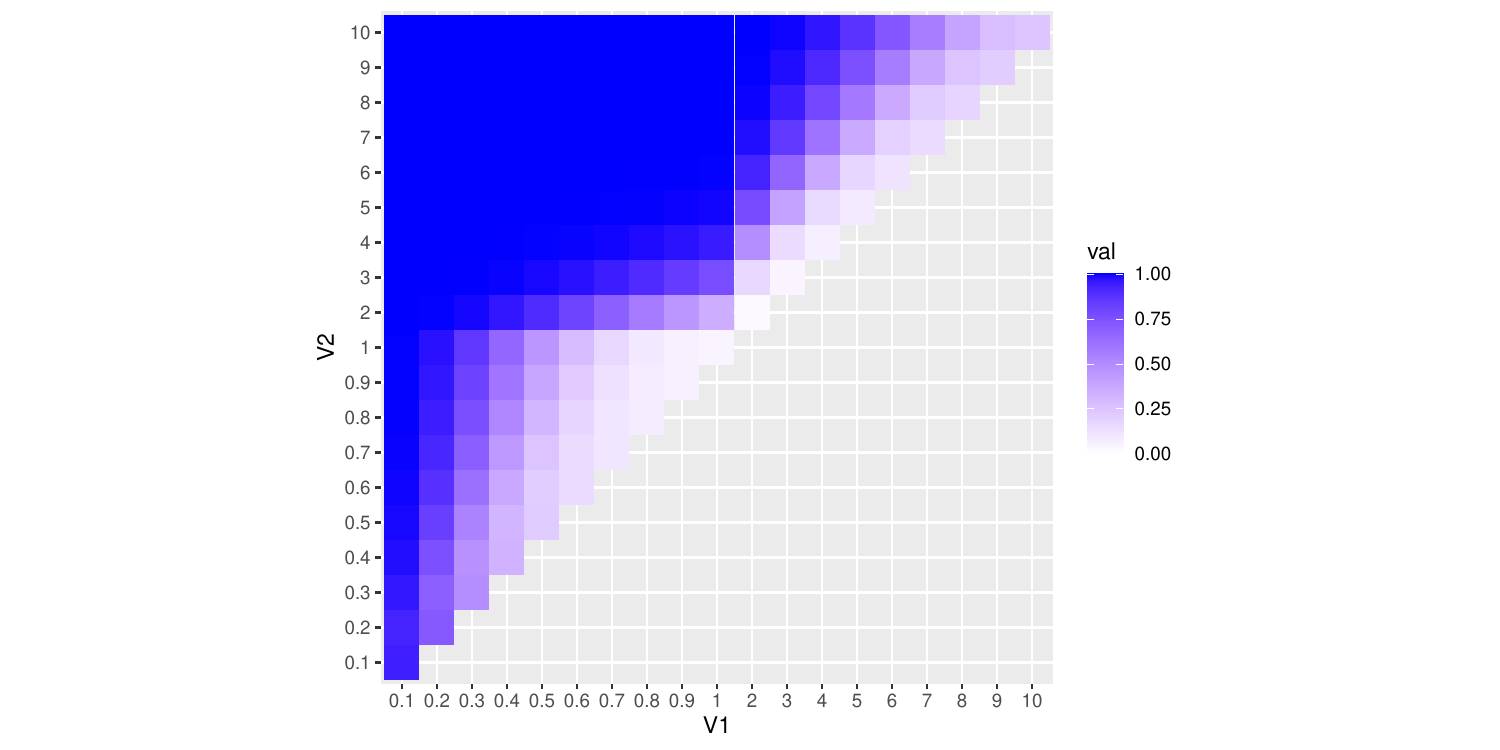}
\caption{``ECDFV'' test powers on beta grid of alternatives~\eqref{eq:beta_grid}.}
\label{figure:beta_grid:ecdfv}
\end{figure}

\paragraph{Tile plot for ``ECDFV.PC1'': beta parameters grid}

\begin{minted}[mathescape, linenos]{r}
plot.powers.on.grid(beta.parameters.grid, res.uniform.vs.beta.2dgrid.F.table$pc.w1.means)
\end{minted}

\begin{figure}[H]
\centering
\includegraphics[width=16cm]{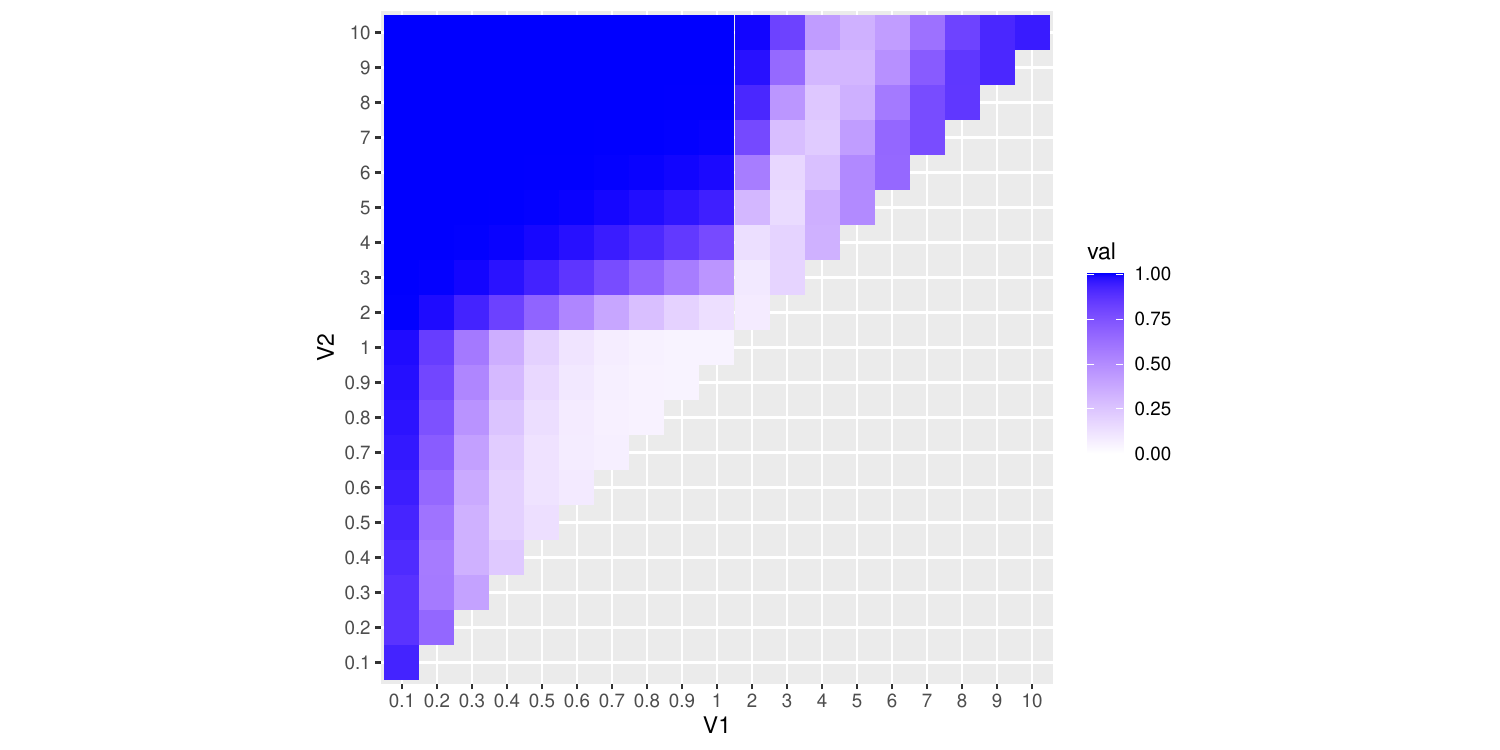}
\caption{``ECDFV.PC1'' test powers on beta grid of alternatives~\eqref{eq:beta_grid}.}
\label{figure:beta_grid:ecdfv:pc1}
\end{figure}

\newpage
\paragraph{Tile plot for ``ECDFV.PC2'': beta parameters grid}

\begin{minted}[mathescape, linenos]{r}
plot.powers.on.grid(beta.parameters.grid, res.uniform.vs.beta.2dgrid.F.table$pc.w2.means)
\end{minted}

\begin{figure}[H]
\centering
\includegraphics[width=16cm]{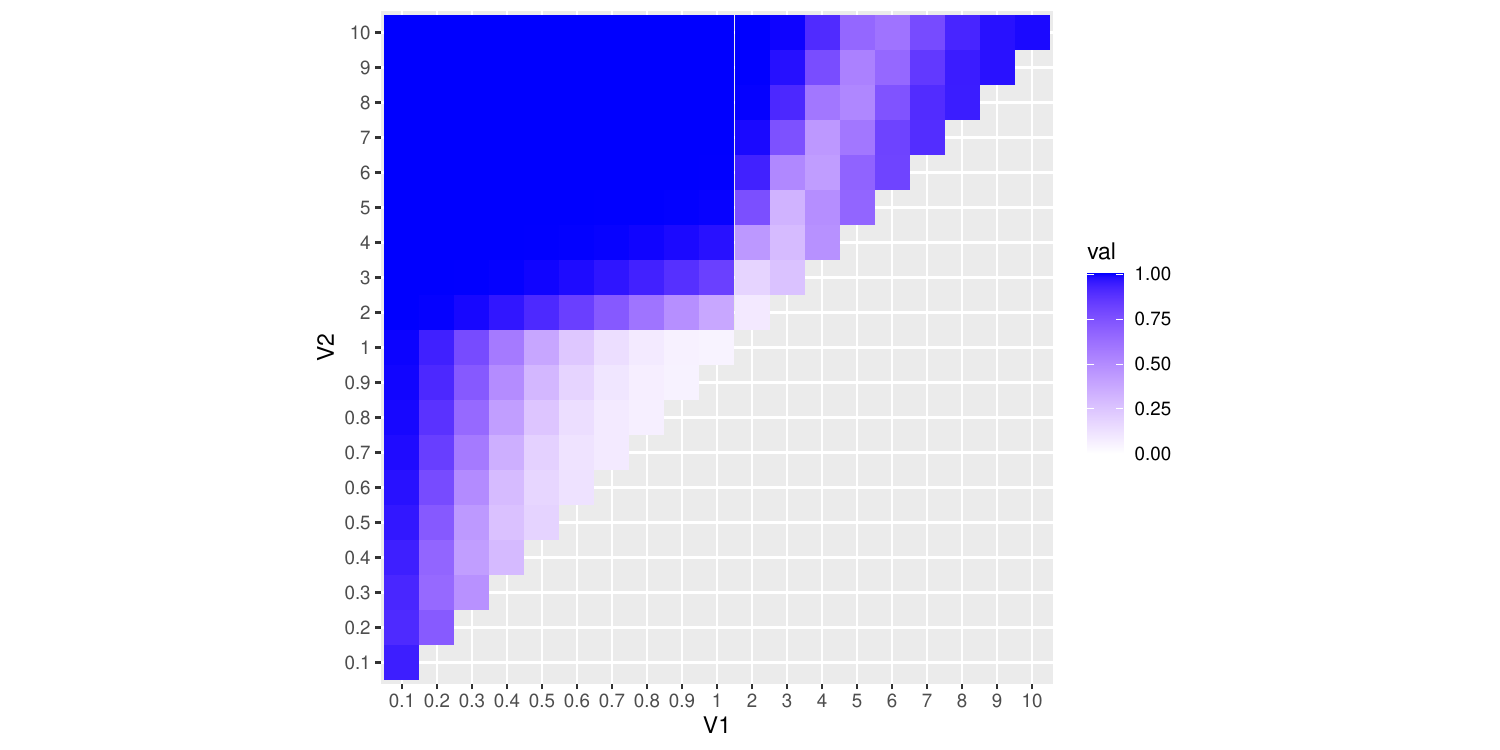}
\caption{``ECDFV.PC2'' test powers on beta grid of alternatives~\eqref{eq:beta_grid}.}
\label{figure:beta_grid:ecdfv:pc2}
\end{figure}

\noindent
Figures~\ref{figure:beta_grid:ecdfv}--\ref{figure:beta_grid:ecdfv:pc2} should be compared to figures~\ref{figure:beta_grid:osu}--\ref{figure:beta_grid:pc2u} (uniform order statistics based tests), figures~\ref{figure:beta_grid:os}--\ref{figure:beta_grid:pc2} (normal order statistics based tests), figures~\ref{figure:beta_grid:ks}--\ref{figure:beta_grid:ad} (classical tests), and figures~\ref{figure:beta_grid:zk}--\ref{figure:beta_grid:zc} (Zhang tests). Well, figures~\ref{figure:beta_grid:ecdfv:pc1} and~\ref{figure:beta_grid:ecdfv:pc2} for ``ECDFV.PC1'' and ``ECDFV.PC2'' tests don't look prominent. They suffer the same effect as figures \ref{figure:beta_grid:pc1} and \ref{figure:beta_grid:pc2} for ``PC1'' and ``PC2'' tests, but in a more severe way: they work well for symmetric alternatives concentrated in the center, and lack power for asymmetric centered alternatives. Perhaps, denser grid could have helped, but we did not try it. Figure~\ref{figure:beta_grid:ecdfv} for ``ECDFV'' test looks similar to figure~\ref{figure:beta_grid:ks} for KS test.

\newpage
\section{Order statistics and pairwise distances between order statistics; normal distribution case}
\label{section:os_pairwise_dist_normal}

In this section we introduce a new test that is based on hyperrectangle bounds for not only order statistics, but also for all pairwise distances between order statistics. For such ``features'', since they are collinear, it is pointless to try principal component analysis. That is why we only construct  hyperrectangle bounds without rotation. In this section we use standard normal order statistics. In the tables and figures to follow the test is denoted as ``OS+PWD''\footnote{Which stands for ``order statistics + pairwise distances''.}. 

\subsection{Function to calculate rejections for a single test \textasteriskcentered}
We'll need a new function to calculate rejections for only one test in a loop:
\begin{minted}[mathescape, linenos]{r}
calc.rejections.sm.bounds <- function(
    sample.generation.function.H0,
    sample.generation.functions.H1,
    get.stat.matrix,
    n, m,
    alpha,
    h,
    gamma.left.sm = 0, gamma.right.sm = 10, w.sm = rep(1, n)
)
{
    s <- length(sample.generation.functions.H1)
    res.sm    <- matrix(0, h, s)
    for(i in 1:h)
    {
        print(paste("iteration", i))
        flush.console()
        stat.matrix.H0 <- get.stat.matrix(sample.generation.function.H0, n, m)
        print("calculate hyperrectangle.bounds.H0.sm")
        flush.console()
        res.hyperrectangle.bounds.H0.sm <- calculate.hyperrectangle.bounds(
            alpha       = alpha,
            gamma.left  = gamma.left.sm,
            gamma.right = gamma.right.sm,
            w           = w.sm,
            stat.matrix = stat.matrix.H0
        )

        for(j in 1:s){
            print(paste("H1", j))
            flush.console()
            stat.matrix.H1 <- get.stat.matrix(sample.generation.functions.H1[[j]], n, m)
            res.sm[i,j]    <- check.hyperrectangle.bounds(stat.matrix.H1, res.hyperrectangle.bounds.H0.sm)
        }
        
        print(paste("iteration", i, "done"))
        flush.console()
    }
    list(
        res.sm = res.sm
    )
}
\end{minted}
Function to prepare power tables is as follows:

\begin{minted}[mathescape, linenos]{r}
create.rejection.table.sm <- function(res.rejections)
    list(
        sm.means = round(apply(res.rejections$res.sm, 2, mean), 4),
        sm.sds = formatC(apply(res.rejections$res.sm, 2, sd  ), format = "e", digits = 1)
    )
\end{minted}

\subsection{Function to compute pairwise distances between order statistics \textasteriskcentered}
Let us also write a C++ function to create vectors of pairwise distances efficiently:
\begin{minted}[mathescape, linenos]{r}
cppFunction('
NumericMatrix ordered_samples_with_os_diffs(const NumericMatrix& x) {
  const int m = x.nrow();
  const int n = x.ncol();
  const int ndist = n * (n - 1) / 2;
  const int out_cols = n + ndist;

  NumericMatrix out(m, out_cols);

  for (int i = 0; i < m; ++i) {
    for (int j = 0; j < n; ++j) {
      out(i, j) = x(i, j);
    }

    int k = n;
    for (int a = 0; a < n - 1; ++a) {
      const double xa = x(i, a);
      for (int b = a + 1; b < n; ++b) {
        out(i, k++) = x(i, b) - xa;
      }
    }
  }

  return out;
}
')
\end{minted}
To compare the result of C++ code and raw \texttt{R} code we do:
\begin{minted}[mathescape, linenos]{r}
sorted.samples <- get.sorted.samples.cpp(rnorm, n, m)
t(apply(sorted.samples, 1, function(x) c(x, as.vector(dist(x))) ))[1:10,]
ordered_samples_with_os_diffs(sorted.samples)[1:10,]
\end{minted}
They are the same. 

\subsection{Calculate rejections: normal $H_0$, normal $H_1$}
Now let us calculate rejections for ``standard normal vs other normals'' situation: 
\begin{minted}[mathescape, linenos]{r}
res <- calc.rejections.sm.bounds(
    sample.generation.function.H0  = rnorm,
    sample.generation.functions.H1 = c(
        lapply(c(0.05, 0.1, 0.15), function(sd)   function(nsamples) rnorm(nsamples, 0,    sd)), 
        lapply(seq(0.2, 0.9, 0.1), function(sd)   function(nsamples) rnorm(nsamples, 0,    sd)),
        rnorm,
        lapply(seq(1.1, 2.4, 0.1), function(sd)   function(nsamples) rnorm(nsamples, 0,    sd)),
        lapply(seq(2.5,   4, 0.5), function(sd)   function(nsamples) rnorm(nsamples, 0,    sd)),
        lapply(seq(0.1,   2, 0.1), function(mean) function(nsamples) rnorm(nsamples, mean, 1 ))
    ),
    get.stat.matrix = function(sample.generation.function, n, m){
        sorted.samples <- get.sorted.samples.cpp(sample.generation.function, n, m)
        ordered_samples_with_os_diffs(sorted.samples)
    },
    n = n, m = m,
    alpha = 0.05,
    h = 10,
    w.sm = rep(1, n + n * (n - 1) / 2)
)
\end{minted}
We note that there are $n$ + $n (n - 1) / 2$ statistics (55 for $n = 10$), since we have n order statistics and $n (n - 1) / 2$ pairwise distances.

The tables for normal alternatives are:
\begin{minted}[mathescape, linenos]{r}
res.table <- create.rejection.table.sm(res)
\end{minted}

\begin{table}[H]
\centering

\begin{minipage}[t]{.48\textwidth}
\vspace{0pt}
\centering

 \begin{tabular}{ |l|p{1.5cm}||>{\columncolor[gray]{0.9}}p{1cm}|p{1.1cm}| } 
 \hline
  $H_0$ & $H_1$ & \multicolumn{2}{c|}{OS+PWD}  \\
  \hline
 &  & mean & sd  \\
 
 \hline
 \hline
 
 $\mathcal{N}(0, 1)$ & $\mathrm{N}(0,0.05^2)$ & 1 & 0.0e+00\\
$\mathcal{N}(0, 1)$ & $\mathrm{N}(0,0.1^2)$ & 1 & 0.0e+00\\
$\mathcal{N}(0, 1)$ & $\mathrm{N}(0,0.15^2)$ & 1 & 7.9e-07\\
$\mathcal{N}(0, 1)$ & $\mathrm{N}(0,0.2^2)$ & 0.9988 & 7.4e-05\\
$\mathcal{N}(0, 1)$ & $\mathrm{N}(0,0.3^2)$ & 0.8626 & 3.0e-03\\
$\mathcal{N}(0, 1)$ & $\mathrm{N}(0,0.4^2)$ & 0.5049 & 3.3e-03\\
$\mathcal{N}(0, 1)$ & $\mathrm{N}(0,0.5^2)$ & 0.2528 & 2.1e-03\\
$\mathcal{N}(0, 1)$ & $\mathrm{N}(0,0.6^2)$ & 0.1319 & 8.8e-04\\
$\mathcal{N}(0, 1)$ & $\mathrm{N}(0,0.7^2)$ & 0.077 & 7.4e-04\\
$\mathcal{N}(0, 1)$ & $\mathrm{N}(0,0.8^2)$ & 0.0513 & 5.2e-04\\
$\mathcal{N}(0, 1)$ & $\mathrm{N}(0,0.9^2)$ & 0.0423 & 3.2e-04\\
$\mathcal{N}(0, 1)$ & $\mathrm{N}(0,1^2)$ & 0.05 & 2.3e-04\\
$\mathcal{N}(0, 1)$ & $\mathrm{N}(0,1.1^2)$ & 0.0803 & 3.9e-04\\
$\mathcal{N}(0, 1)$ & $\mathrm{N}(0,1.2^2)$ & 0.1377 & 4.3e-04\\
$\mathcal{N}(0, 1)$ & $\mathrm{N}(0,1.3^2)$ & 0.2202 & 5.8e-04\\
$\mathcal{N}(0, 1)$ & $\mathrm{N}(0,1.4^2)$ & 0.3205 & 6.6e-04\\
$\mathcal{N}(0, 1)$ & $\mathrm{N}(0,1.5^2)$ & 0.4269 & 8.8e-04\\
$\mathcal{N}(0, 1)$ & $\mathrm{N}(0,1.6^2)$ & 0.5294 & 5.7e-04\\
$\mathcal{N}(0, 1)$ & $\mathrm{N}(0,1.7^2)$ & 0.6224 & 5.5e-04\\
$\mathcal{N}(0, 1)$ & $\mathrm{N}(0,1.8^2)$ & 0.7023 & 4.7e-04\\
$\mathcal{N}(0, 1)$ & $\mathrm{N}(0,1.9^2)$ & 0.7683 & 6.1e-04\\
$\mathcal{N}(0, 1)$ & $\mathrm{N}(0,2^2)$ & 0.8211 & 4.7e-04\\
$\mathcal{N}(0, 1)$ & $\mathrm{N}(0,2.1^2)$ & 0.8632 & 4.1e-04\\
$\mathcal{N}(0, 1)$ & $\mathrm{N}(0,2.2^2)$ & 0.8956 & 3.7e-04\\
$\mathcal{N}(0, 1)$ & $\mathrm{N}(0,2.3^2)$ & 0.9205 & 2.9e-04\\
$\mathcal{N}(0, 1)$ & $\mathrm{N}(0,2.4^2)$ & 0.9394 & 2.0e-04\\
$\mathcal{N}(0, 1)$ & $\mathrm{N}(0,2.5^2)$ & 0.9539 & 2.0e-04\\
$\mathcal{N}(0, 1)$ & $\mathrm{N}(0,3^2)$ & 0.9877 & 8.3e-05\\
$\mathcal{N}(0, 1)$ & $\mathrm{N}(0,3.5^2)$ & 0.9964 & 4.8e-05\\
$\mathcal{N}(0, 1)$ & $\mathrm{N}(0,4^2)$ & 0.9989 & 2.9e-05\\
 
 \hline
 \end{tabular}
 \caption{$H_0$ --- standard normal, $H_1$ --- other normals that differ in scale.}
 \label{table:ospwd:normal:sd}
 
 \end{minipage}
\hfill
\begin{minipage}[t]{.48\textwidth}
\vspace{120pt}
\centering

 \begin{tabular}{ |l|p{1.2cm}||>{\columncolor[gray]{0.9}}p{1cm}|p{1.1cm}| } 
 \hline
  $H_0$ & $H_1$ & \multicolumn{2}{c|}{OS+PWD}  \\
  \hline
 &  & mean & sd  \\
 \hline
 \hline
 
 $\mathcal{N}(0, 1)$ & $\mathrm{N}(0.1, 1)$ & 0.0527 & 2.5e-04\\
$\mathcal{N}(0, 1)$ & $\mathrm{N}(0.2, 1)$ & 0.0611 & 4.8e-04\\
$\mathcal{N}(0, 1)$ & $\mathrm{N}(0.3, 1)$ & 0.0778 & 4.1e-04\\
$\mathcal{N}(0, 1)$ & $\mathrm{N}(0.4, 1)$ & 0.1059 & 7.2e-04\\
$\mathcal{N}(0, 1)$ & $\mathrm{N}(0.5, 1)$ & 0.1493 & 8.4e-04\\
$\mathcal{N}(0, 1)$ & $\mathrm{N}(0.6, 1)$ & 0.2102 & 1.0e-03\\
$\mathcal{N}(0, 1)$ & $\mathrm{N}(0.7, 1)$ & 0.2899 & 1.2e-03\\
$\mathcal{N}(0, 1)$ & $\mathrm{N}(0.8, 1)$ & 0.3857 & 1.5e-03\\
$\mathcal{N}(0, 1)$ & $\mathrm{N}(0.9, 1)$ & 0.4922 & 1.8e-03\\
$\mathcal{N}(0, 1)$ & $\mathrm{N}(1, 1)$ & 0.6011 & 1.4e-03\\
$\mathcal{N}(0, 1)$ & $\mathrm{N}(1.1, 1)$ & 0.7046 & 1.3e-03\\
$\mathcal{N}(0, 1)$ & $\mathrm{N}(1.2, 1)$ & 0.794 & 1.0e-03\\
$\mathcal{N}(0, 1)$ & $\mathrm{N}(1.3, 1)$ & 0.8657 & 7.9e-04\\
$\mathcal{N}(0, 1)$ & $\mathrm{N}(1.4, 1)$ & 0.9182 & 5.9e-04\\
$\mathcal{N}(0, 1)$ & $\mathrm{N}(1.5, 1)$ & 0.9536 & 3.7e-04\\
$\mathcal{N}(0, 1)$ & $\mathrm{N}(1.6, 1)$ & 0.9757 & 2.3e-04\\
$\mathcal{N}(0, 1)$ & $\mathrm{N}(1.7, 1)$ & 0.9882 & 1.4e-04\\
$\mathcal{N}(0, 1)$ & $\mathrm{N}(1.8, 1)$ & 0.9947 & 9.4e-05\\
$\mathcal{N}(0, 1)$ & $\mathrm{N}(1.9, 1)$ & 0.9978 & 6.0e-05\\
$\mathcal{N}(0, 1)$ & $\mathrm{N}(2, 1)$ & 0.9992 & 1.7e-05\\

\hline
\end{tabular}
\caption{$H_0$ --- standard normal, $H_1$ --- other normals that differ in location.}
\label{table:ospwd:normal:mean}

\end{minipage}

\end{table}

\noindent
In table~\ref{table:ospwd:normal:sd}, where $N(0,1)$ is tested against $N(0,\sigma^2)$, we see that ``OS+PWD'' test has decent power for $\sigma < 1$, which is comparable to $Z_A$ test. For $\sigma > 1$ new test's power is between $Z_A$ and $Z_C$. In table~\ref{table:ospwd:normal:mean}, where $N(0,1)$ is tested against $N(\mu, 1)$, we see that ``OS+PWD'' test has very average power.

\subsection{Calculate rejections: normal $H_0$, Cauchy/gamma/Student $H_1$}
 
 For Cauchy/gamma/Student alternatives the results are as follows:
\begin{minted}[mathescape, linenos]{r}
res.os.diff.cauchy.gamma.t <- calc.rejections.sm.bounds(
    sample.generation.function.H0  = rnorm,
    sample.generation.functions.H1 = c(
        # cauchy
        lapply(seq(0.01, 0.09, 0.01), function(scale) function(nsamples) rcauchy(nsamples, 0, scale)),
        lapply(seq(0.1,  2,    0.1 ), function(scale) function(nsamples) rcauchy(nsamples, 0, scale)),
        # gamma
        lapply(
            c(seq(0.1,0.9,0.1),1:5,10,100,1000), 
            function(shape) 
                function(nsamples) (rgamma(nsamples, shape = shape, rate = 1) - shape) / sqrt(shape)
        ),
        # t
        lapply(seq(1,5), function(df) function(nsamples) rt(nsamples, df = df))
    ),
    get.stat.matrix = function(sample.generation.function, n, m){
        sorted.samples <- get.sorted.samples.cpp(sample.generation.function, n, m)
        ordered_samples_with_os_diffs(sorted.samples)
    },
    n = n, m = m,
    alpha = 0.05,
    h = 10,
    w.sm = rep(1, n + n * (n - 1) / 2)
)
res.os.diff.cauchy.gamma.t.table <- create.rejection.table.sm(res.os.diff.cauchy.gamma.t)
\end{minted}

\begin{table}[H]
\centering

\begin{minipage}[t]{.48\textwidth}
\vspace{0pt}
\centering

 \begin{tabular}{ |l|p{2.3cm}||>{\columncolor[gray]{0.9}}p{1cm}|p{1.1cm}| } 
 \hline
  $H_0$ & $H_1$ & \multicolumn{2}{c|}{OS+PWD}  \\
  \hline
 &  & mean & sd  \\
 \hline
 \hline
 
 $\mathcal{N}(0, 1)$ & $\mathrm{Cauchy}(0, 0.01)$ & 1 & 3.2e-07\\
$\mathcal{N}(0, 1)$ & $\mathrm{Cauchy}(0, 0.02)$ & 1 & 5.2e-06\\
$\mathcal{N}(0, 1)$ & $\mathrm{Cauchy}(0, 0.03)$ & 0.9998 & 1.3e-05\\
$\mathcal{N}(0, 1)$ & $\mathrm{Cauchy}(0, 0.04)$ & 0.999 & 4.2e-05\\
$\mathcal{N}(0, 1)$ & $\mathrm{Cauchy}(0, 0.05)$ & 0.997 & 9.2e-05\\
$\mathcal{N}(0, 1)$ & $\mathrm{Cauchy}(0, 0.06)$ & 0.9928 & 1.4e-04\\
$\mathcal{N}(0, 1)$ & $\mathrm{Cauchy}(0, 0.07)$ & 0.986 & 2.3e-04\\
$\mathcal{N}(0, 1)$ & $\mathrm{Cauchy}(0, 0.08)$ & 0.976 & 4.0e-04\\
$\mathcal{N}(0, 1)$ & $\mathrm{Cauchy}(0, 0.09)$ & 0.9629 & 4.5e-04\\
$\mathcal{N}(0, 1)$ & $\mathrm{Cauchy}(0, 0.1)$ & 0.9466 & 4.5e-04\\
$\mathcal{N}(0, 1)$ & $\mathrm{Cauchy}(0, 0.2)$ & 0.7233 & 1.5e-03\\
$\mathcal{N}(0, 1)$ & $\mathrm{Cauchy}(0, 0.3)$ & 0.6108 & 1.2e-03\\
$\mathcal{N}(0, 1)$ & $\mathrm{Cauchy}(0, 0.4)$ & 0.614 & 1.2e-03\\
$\mathcal{N}(0, 1)$ & $\mathrm{Cauchy}(0, 0.5)$ & 0.6604 & 7.0e-04\\
$\mathcal{N}(0, 1)$ & $\mathrm{Cauchy}(0, 0.6)$ & 0.7169 & 5.2e-04\\
$\mathcal{N}(0, 1)$ & $\mathrm{Cauchy}(0, 0.7)$ & 0.7698 & 4.8e-04\\
$\mathcal{N}(0, 1)$ & $\mathrm{Cauchy}(0, 0.8)$ & 0.8163 & 3.9e-04\\
$\mathcal{N}(0, 1)$ & $\mathrm{Cauchy}(0, 0.9)$ & 0.8549 & 4.6e-04\\
$\mathcal{N}(0, 1)$ & $\mathrm{Cauchy}(0, 1)$ & 0.8868 & 3.7e-04\\
$\mathcal{N}(0, 1)$ & $\mathrm{Cauchy}(0, 1.1)$ & 0.9121 & 2.8e-04\\
$\mathcal{N}(0, 1)$ & $\mathrm{Cauchy}(0, 1.2)$ & 0.9323 & 2.3e-04\\
$\mathcal{N}(0, 1)$ & $\mathrm{Cauchy}(0, 1.3)$ & 0.948 & 2.5e-04\\
$\mathcal{N}(0, 1)$ & $\mathrm{Cauchy}(0, 1.4)$ & 0.9601 & 2.4e-04\\
$\mathcal{N}(0, 1)$ & $\mathrm{Cauchy}(0, 1.5)$ & 0.9696 & 1.3e-04\\
$\mathcal{N}(0, 1)$ & $\mathrm{Cauchy}(0, 1.6)$ & 0.9768 & 2.2e-04\\
$\mathcal{N}(0, 1)$ & $\mathrm{Cauchy}(0, 1.7)$ & 0.9823 & 1.2e-04\\
$\mathcal{N}(0, 1)$ & $\mathrm{Cauchy}(0, 1.8)$ & 0.9865 & 1.2e-04\\
$\mathcal{N}(0, 1)$ & $\mathrm{Cauchy}(0, 1.9)$ & 0.9896 & 7.4e-05\\
$\mathcal{N}(0, 1)$ & $\mathrm{Cauchy}(0, 2)$ & 0.9921 & 7.1e-05\\
 
 \hline
\end{tabular}
\caption{$H_0$ --- standard normal, $H_1$ --- Cauchy distributions with different scale.}
\label{table:ospwd:cauchy}

\end{minipage}
\hfill
\begin{minipage}[t]{.48\textwidth}
\vspace{0pt}
\centering

 \begin{tabular}{ |l|p{1.2cm}||>{\columncolor[gray]{0.9}}p{1cm}|p{1.1cm}| } 
 \hline
  $H_0$ & $H_1$ & \multicolumn{2}{c|}{OS+PWD}  \\
  \hline
 &  & mean & sd  \\
 \hline
 \hline
 
 $\mathcal{N}(0, 1)$ & $\mathrm{G}(0.1)$ & 0.9997 & 2.1e-05\\
$\mathcal{N}(0, 1)$ & $\mathrm{G}(0.2)$ & 0.9873 & 1.2e-04\\
$\mathcal{N}(0, 1)$ & $\mathrm{G}(0.3)$ & 0.9308 & 3.6e-04\\
$\mathcal{N}(0, 1)$ & $\mathrm{G}(0.4)$ & 0.8272 & 7.8e-04\\
$\mathcal{N}(0, 1)$ & $\mathrm{G}(0.5)$ & 0.7029 & 1.1e-03\\
$\mathcal{N}(0, 1)$ & $\mathrm{G}(0.6)$ & 0.5843 & 1.1e-03\\
$\mathcal{N}(0, 1)$ & $\mathrm{G}(0.7)$ & 0.484 & 9.6e-04\\
$\mathcal{N}(0, 1)$ & $\mathrm{G}(0.8)$ & 0.4053 & 8.3e-04\\
$\mathcal{N}(0, 1)$ & $\mathrm{G}(0.9)$ & 0.3456 & 8.3e-04\\
$\mathcal{N}(0, 1)$ & $\mathrm{G}(1)$ & 0.301 & 1.2e-03\\
$\mathcal{N}(0, 1)$ & $\mathrm{G}(2)$ & 0.1593 & 5.7e-04\\
$\mathcal{N}(0, 1)$ & $\mathrm{G}(3)$ & 0.1275 & 4.3e-04\\
$\mathcal{N}(0, 1)$ & $\mathrm{G}(4)$ & 0.1114 & 4.7e-04\\
$\mathcal{N}(0, 1)$ & $\mathrm{G}(5)$ & 0.1011 & 4.6e-04\\
$\mathcal{N}(0, 1)$ & $\mathrm{G}(10)$ & 0.0786 & 3.8e-04\\
$\mathcal{N}(0, 1)$ & $\mathrm{G}(100)$ & 0.0534 & 3.7e-04\\
$\mathcal{N}(0, 1)$ & $\mathrm{G}(1000)$ & 0.0504 & 1.5e-04\\
 
  \hline
\end{tabular}
\caption{$H_0$ --- standard normal, $H_1$ --- gamma with different skews.}
\label{table:ospwd:gamma}

\vspace{21pt}

 \begin{tabular}{ |l|p{1.2cm}||>{\columncolor[gray]{0.9}}p{1cm}|p{1.1cm}| } 
 \hline
  $H_0$ & $H_1$ & \multicolumn{2}{c|}{OS+PWD}  \\
  \hline
 &  & mean & sd  \\
 \hline
 \hline
 
 $\mathcal{N}(0, 1)$ & $\mathrm{t}(1)$ & 0.8866 & 2.5e-04\\
$\mathcal{N}(0, 1)$ & $\mathrm{t}(2)$ & 0.5993 & 6.1e-04\\
$\mathcal{N}(0, 1)$ & $\mathrm{t}(3)$ & 0.4084 & 6.5e-04\\
$\mathcal{N}(0, 1)$ & $\mathrm{t}(4)$ & 0.2977 & 7.1e-04\\
$\mathcal{N}(0, 1)$ & $\mathrm{t}(5)$ & 0.2313 & 8.8e-04\\

  \hline
\end{tabular}
\caption{$H_0$ --- standard normal, $H_1$ --- Student with different degrees of freedom.}
\label{table:ospwd:student}

\end{minipage}

\end{table}

\noindent
In table~\ref{table:ospwd:cauchy} we see that ``OS+PWD'' has excellent performance for Cauchy alternatives that is even better than performance of ``PC2'' test. In table~\ref{table:ospwd:gamma} we also see that ``OS+PWD'' has the best performance for gamma alternatives among all the tests reviewed in this text so far. In table~\ref{table:ospwd:student} we see that new test's performance for Student alternatives is comparable to ``PC2''.

\subsection{Conclusions}
In this section we constructed a pretty powerful test based on standard normal order statistics and pairwise distances between these order statistics. With this we demonstrated that there is merit in trying different ``features'' beside order statistics. We note that the described test makes sense mostly for moderately small values of $n$, since number of statistics grows quadratically with $n$.  

\newpage
\section{Order statistics and pairwise distances between order statistics; uniform distribution case}
\label{section:os_pairwise_dist_uniform}

Now let us briefly construct the test based on uniform order statistics and the differences between them. We'll call this test ``OS(U)+PWD''\footnote{Which stands for ``order statistics (uniform) + pairwise distances''.}. Skipping the tables for peculiar beta alternatives, we'll fast forward to more general beta grid plots:

\begin{minted}[mathescape, linenos, texcomments]{r}
res.os.diff.beta.2dgrid <- calc.rejections.sm.bounds(
    sample.generation.function.H0  = runif,
    sample.generation.functions.H1 = apply(
        beta.parameters.grid,  # see formula \eqref{eq:beta_grid} 
        1, 
        function(par) function(nsamples) rbeta(nsamples, par[1], par[2])
    ),
    get.stat.matrix = function(sample.generation.function, n, m){
        sorted.samples <- get.sorted.samples.cpp(sample.generation.function, n, m)
        ordered_samples_with_os_diffs(sorted.samples)
    },
    n = n, m = m,
    alpha = 0.05,
    h = 10,
    w.sm = rep(1, n + n * (n - 1) / 2)
)

res.os.diff.beta.2dgrid.table <- create.rejection.table.sm(res.os.diff.beta.2dgrid)

plot.powers.on.grid(beta.parameters.grid, res.os.diff.beta.2dgrid.table$sm.means)
\end{minted}

\begin{figure}[H]
\centering
\includegraphics[width=16cm]{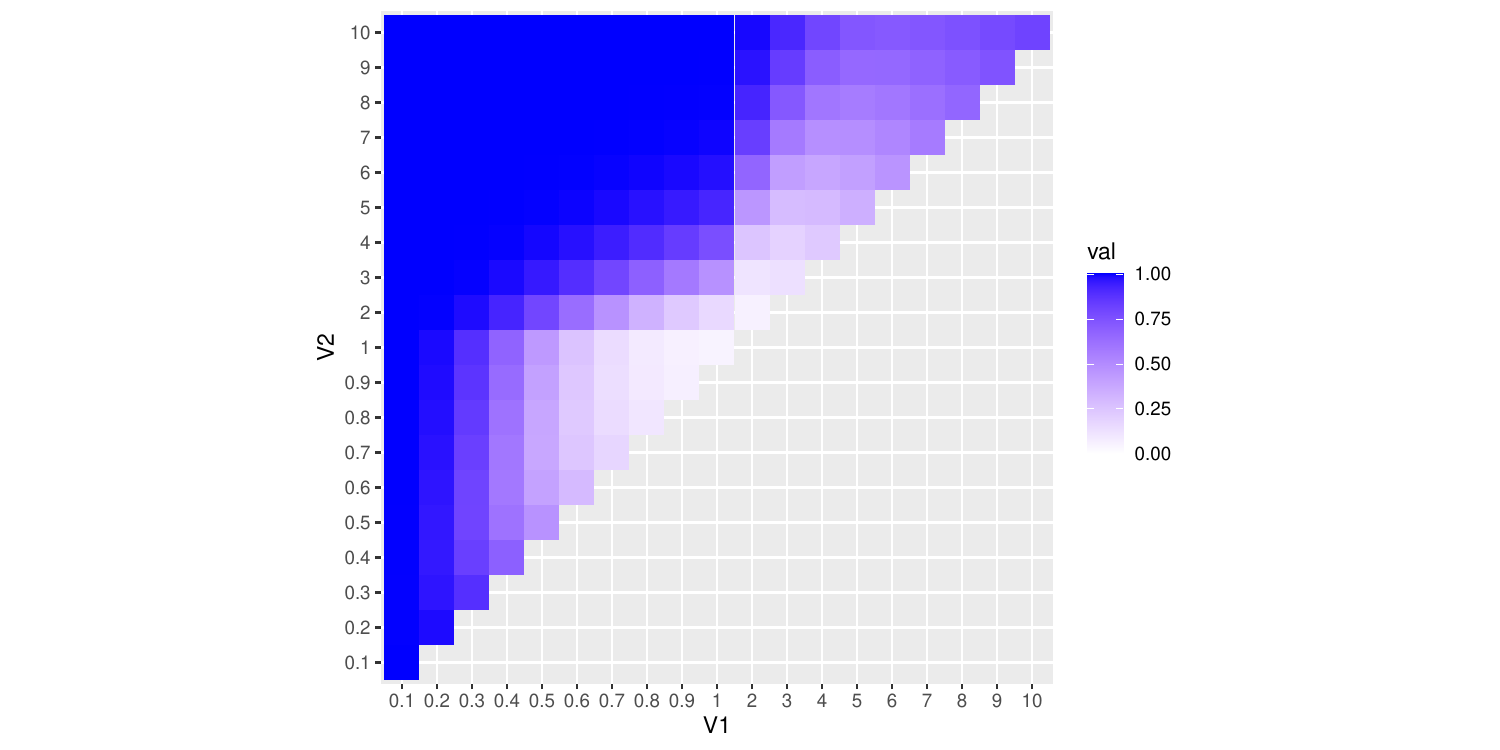}
\caption{OS(U)+PWD}
\label{figure:beta:grid:osupwd}
\end{figure}

\noindent
On figure~\ref{figure:beta:grid:osupwd} we see that ``OS(U)+PWD'' test looks better than ``OS(U)'' test for symmetric $\mathrm{Beta}(a,a)$, $a > 1$ alternatives, and for other alternatives it looks more or less the same as ``OS(U)'' test. This means that adding pairwise distances as additional statistics to the original uniform order statistics based test fixes its weak point.

\newpage
\section{Order statistics and pairwise distances between order statistics; uniform via normal}
\label{section:os_pairwise_dist_uniform_via_normal}

Let us also try the normal order statistics based ``OS+PWD'' test from section~\ref{section:os_pairwise_dist_normal} in situation with uniform $H_0$ and beta $H_1$: 

\begin{minted}[mathescape, linenos, texcomments]{r}
res.os.diff.unif.via.norm.beta.2dgrid <- calc.rejections.sm.bounds(
    sample.generation.function.H0  = rnorm,
    sample.generation.functions.H1 = apply(
        beta.parameters.grid,  # see formula \eqref{eq:beta_grid}
        1, 
        function(par) function(nsamples) handle.infs(qnorm(rbeta(nsamples, par[1], par[2])))
    ),
    get.stat.matrix = function(sample.generation.function, n, m){
        sorted.samples <- get.sorted.samples.cpp(sample.generation.function, n, m)
        ordered_samples_with_os_diffs(sorted.samples)
    },
    n = n, m = m,
    alpha = 0.05,
    h = 10,
    w.sm = rep(1, n + n * (n - 1) / 2)
)

res.os.diff.unif.via.norm.beta.2dgrid.table <- create.rejection.table.sm(res.os.diff.unif.via.norm.beta.2dgrid)

plot.powers.on.grid(beta.parameters.grid, res.os.diff.unif.via.norm.beta.2dgrid$sm.means)
\end{minted}

\begin{figure}[H]
\centering
\includegraphics[width=16cm]{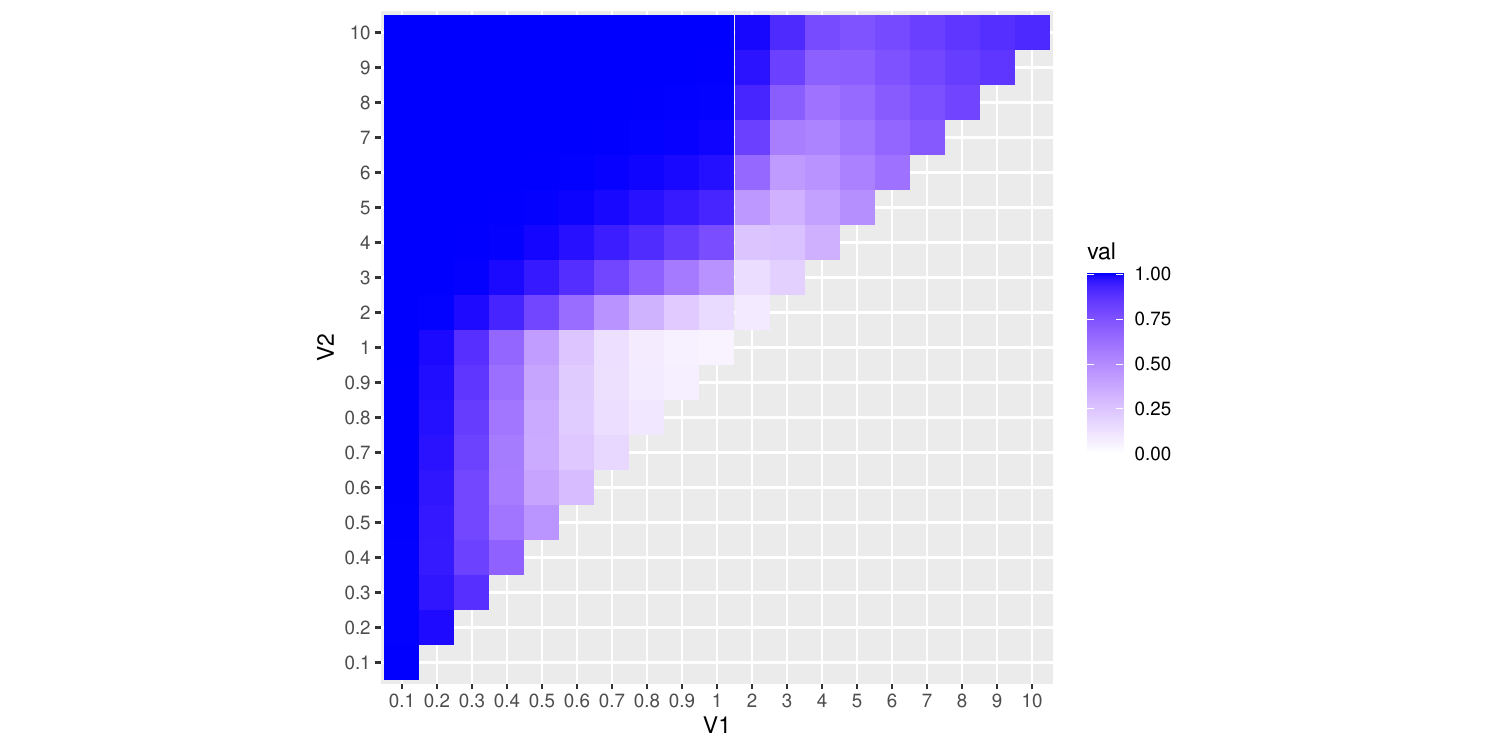}
\caption{OS+PWD}
\label{figure:beta:grid:ospwd}
\end{figure}

\noindent
On figure~\ref{figure:beta:grid:ospwd} we see that ``OS+PWD'' test has even better power for symmetric $\mathrm{Beta}(a,a)$, $a > 1$ alternatives compared to ``OS(U)+PWD'' test (figure~\ref{figure:beta:grid:osupwd}). Everything else looks the same. For asymmetric $\mathrm{Beta}(a,b)$, $a,b > 1$, $a \neq b$ alternatives test's power is less than for symmetric alternatives. We conclude that this ``better for symmetric than asymmetric'' behavior is a general characteristic of ``uniform via normal'' approach.

\chapter{Joint density tests}
\label{chapter:joint_density_tests}

In this chapter we implement the approach that was described in subsection~\ref{section:direct_joint_density_estimation}.

\section{When hyperrectange bounds are bad: joint distribution of two distances example}

Remember \texttt{distance.distribution.KS.H0} and \texttt{distance.distribution.AD.H0}, the \texttt{R} vectors with Kolmogo\-rov--Smirnov and Anderson--Darling distances between standard normal CDF and simulated eCDFs for standard normal samples of size 10. They were defined in  section~\ref{section:functions:classical:distances}. For ease of narration let us redefine them here as follows:

\begin{minted}[mathescape, linenos, texcomments]{r}
n <- 10       # sample size
m <- 1000000  # number of samples

sorted.samples.std.normal <- get.sorted.samples.cpp(rnorm, n, m)  # see section \ref{section:functions_to_generate_order_stats}

distance.KS  <- function(s, F_0, n)
    max( pmax((1:n) / n - F_0(s), F_0(s) - (0:(n-1)) / n) )
distance.AD  <- function(s, F_0, n) 
    -n - sum( (2 * (1:n) - 1) / n * (log(F_0(s)) + log(1 - F_0(rev(s)))))
    
distance.distribution.KS.H0 <- 
    apply(sorted.samples.std.normal, 1, function(samp) distance.KS(samp, pnorm, n))
distance.distribution.AD.H0 <- 
    apply(sorted.samples.std.normal, 1, function(samp) distance.AD(samp, pnorm, n))
\end{minted}

\noindent
In this section we will study tests based on joint distribution of KS and AD distances. We will store the empirical joint distribution as 

\begin{minted}[mathescape, linenos, texcomments]{r}
distances.KS.AD <- data.frame(
    x = distance.distribution.KS.H0,
    y = distance.distribution.AD.H0
)
\end{minted}

\noindent
We then construct two types of rectangle bounds. The first type uses two-sided confidence intervals:

\begin{minted}[mathescape, linenos, texcomments]{r}
res.hyperrectangle.bounds.KS.AD <- calculate.hyperrectangle.bounds(  # defined in sec. \ref{section:calculate.hyperrectangle.bounds}
    alpha = 0.05,                     # confidence level
    gamma.left = 0, gamma.right = 2,  # hyperparameter search bounds
    w = rep(1, 2),                    # weights of both statistics are equal
    distances.KS.AD                   # joint distribution of KS and AD distances
)
\end{minted}
The bounds are as follows:
\begin{minted}[mathescape, linenos, texcomments]{r}
res.hyperrectangle.bounds.KS.AD
\end{minted}

\begin{verbatim}
$left.bounds    0.134369464567924   0.224633646311647
$right.bounds   0.463745801729359   3.437268425475
\end{verbatim}

\noindent
The second type of rectangular bounds uses right-sided confidence intervals with both left bounds fixed at $0$:

\begin{minted}[mathescape, linenos, texcomments]{r}
res.hyperrectangle.bounds.KS.AD.rsi <- calculate.hyperrectangle.bounds.with.interval.type( # sec. \ref{section:hr:bounds:mixed:type}
    alpha = 0.05,                           # confidence level
    gamma.left = 0, gamma.right = 2,        # hyperparameter search bounds
    w = rep(1, 2),                          # weights of both statistics are equal
    distances.KS.AD,                        # joint distribution of KS and AD distances
    interval.types = rep("right side", 2),  # both intervals are right-sided 
    default.left.bounds  = rep(0, 2),       # left bounds for both intervals is 0
    default.right.bounds = NULL             # right bounds are computed
)
\end{minted}
These bounds are as follows:
\begin{minted}[mathescape, linenos, texcomments]{r}
res.hyperrectangle.bounds.KS.AD.rsi
\end{minted}
\begin{verbatim}
$left.bounds    0                   0
$right.bounds   0.426337672998619   2.78160532094437
\end{verbatim}

\noindent
To plot confidence rectangles we define the following function:
\begin{minted}[mathescape, linenos, texcomments]{r}
plot.confidence.rectangle <- function(rectangle.bounds, lty)
    grid.rect(                
        (rectangle.bounds$right.bounds[1] + rectangle.bounds$left.bounds[1]) / 2, 
        (rectangle.bounds$right.bounds[2] + rectangle.bounds$left.bounds[2]) / 2, 
        width  = rectangle.bounds$right.bounds[1] - rectangle.bounds$left.bounds[1], 
        height = rectangle.bounds$right.bounds[2] - rectangle.bounds$left.bounds[2],
        default.units = "native",          
        gp = gpar(col = "black", fill = NA, lwd = 1, lty = lty)   
    ) 
\end{minted}

\noindent
We also define a custom panel function to plot data along with two confidence rectangles computed earlier:

\begin{minted}[mathescape, linenos, texcomments]{r}
panel.with.KS.AD.bounds <- function(x, y, ...) {
    panel.hexbinplot(x, y, ...)
    plot.confidence.rectangle(res.hyperrectangle.bounds.KS.AD    , 1)
    plot.confidence.rectangle(res.hyperrectangle.bounds.KS.AD.rsi, 2)
}
\end{minted}

\noindent
Now we are ready to plot joint distribution of KS and AD distances and two types of confidence rectangles for it:

\begin{minted}[mathescape, linenos, texcomments]{r}
hexbinplot(
    y ~ x,
    data  = distances.KS.AD,
    xbins = 50,
    panel = panel.with.KS.AD.bounds,
    xlab = "KS distance", ylab = "AD distance",
    xlim = c(-0.1, 1),    ylim = c(-0.5, 12) # xlim = c(-0.1, 0.6), ylim = c(-0.5, 4) - zoomed options
)
\end{minted}

\begin{figure}[h]
\centering
\begin{minipage}{.48\textwidth}
  \centering
  \includegraphics[width=.9\linewidth]{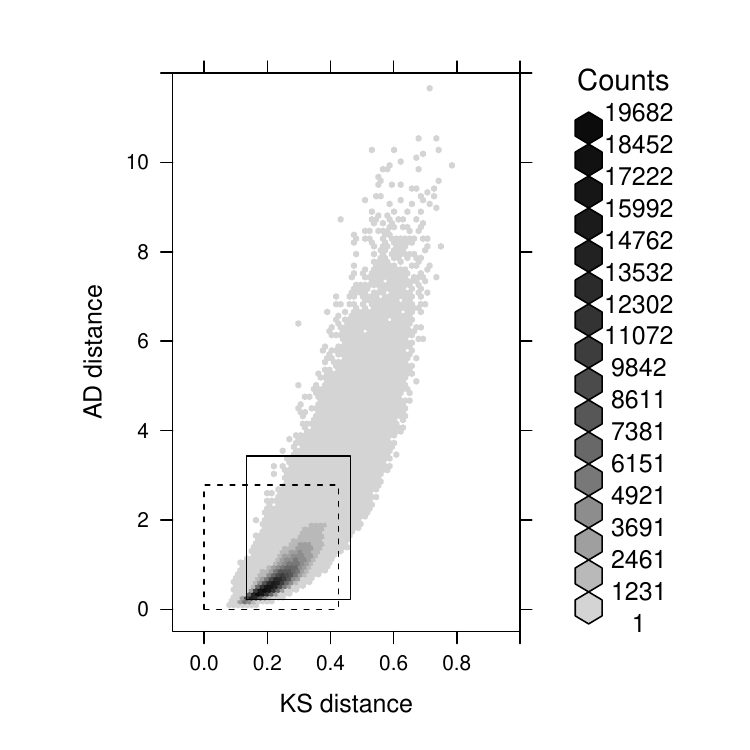}
  \captionof{figure}{Joint distribution of KS and AD distances to normal CDF for normal samples of size $n = 10$.}
  \label{figure:KS_AD_joint}
\end{minipage}
\hfill
\begin{minipage}{.48\textwidth}
  \centering
  \includegraphics[width=.9\linewidth]{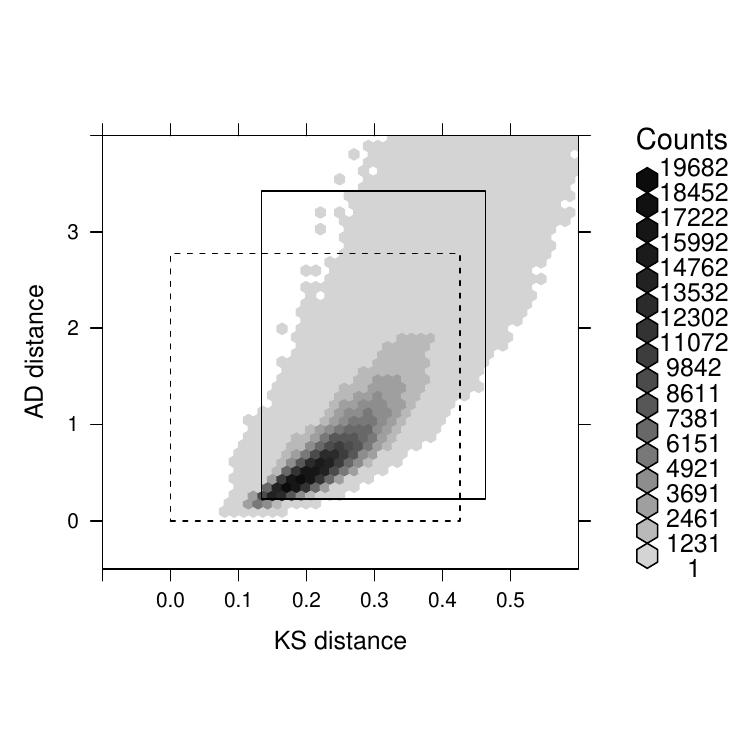}
  \captionof{figure}{Joint distribution of KS and AD distances to normal CDF for normal samples of size $n = 10$. Zoomed-in plot.}
  \label{figure:KS_AD_joint_zoomed}
\end{minipage}
\end{figure}

\noindent
The generated plots are presented on figure~\ref{figure:KS_AD_joint} (uncommented \texttt{xlim} and \texttt{ylim}) and figure~\ref{figure:KS_AD_joint_zoomed} (commented \texttt{xlim} and \texttt{ylim}; zoomed-in region). Black hexagons correspond to the areas with the highest concentration of empirical joint distribution points, lighter hexagons correspond to sparser areas. White areas don't contain points at all. Confidence rectangle based on two-side intervals is plotted with a solid line, rectangle based on right-side intervals --- with a dashed line.

On these plots we see that
\begin{enumerate}
\item joint distribution of KS and AD distances has a long grey sparse train (see figure~\ref{figure:KS_AD_joint}); 
\item there is a lot of white space inside both confidence rectangles, especially inside the one that starts at point $(0,0)$ (the dashed one);
\item black hexagons with highest density of points are very close to the solid rectangle's left bottom corner boundary, some darker areas are even outside of this rectangle.  
\end{enumerate}
In this situation rectangular bounds look suboptimal. We will try rotation trick from previous chapter a bit later, but it seems that in this particular case it will not help much.

Now let us construct a highest density region with a method described in section~\ref{section:direct_joint_density_estimation}. First of all, to apply k-nearest neighbors algorithm we need to restrict ourselves with a smaller number of distance pairs:
\begin{minted}[mathescape, linenos, texcomments]{r}
statistics.H0 <- distances.KS.AD[1:100000,]
\end{minted}
It is needed to reduce computation, since general nearest neighbors algorithm's computational cost depends quadratically on the number of compared points. 

\begin{figure}[h]
\centering
\begin{minipage}{.48\textwidth}
  \centering
  \includegraphics[width=.9\linewidth]{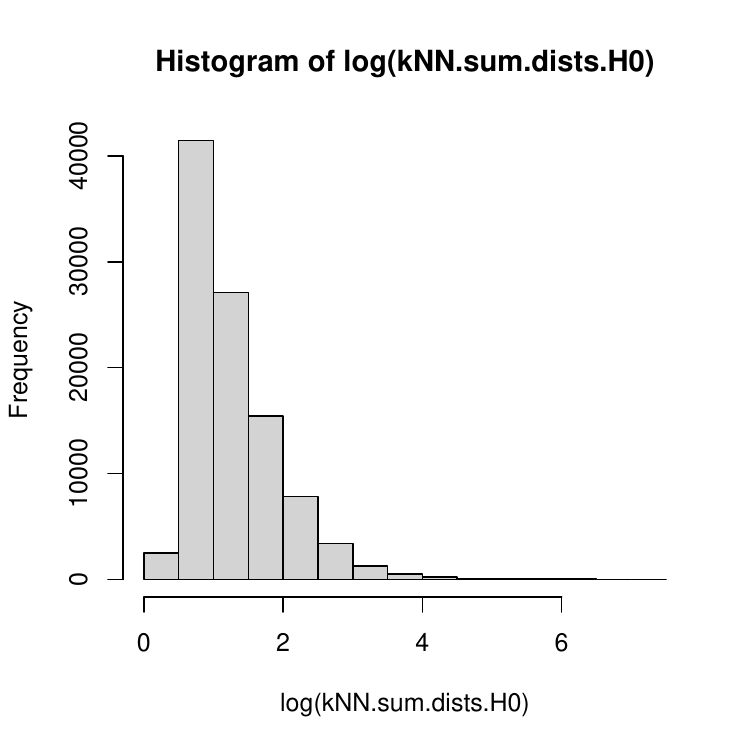}
  \captionof{figure}{Histogram for log values of $\mathbb{Q}$ for joint distribution of KS and AD distances.}
  \label{figure:hist_Q_KS_AD}
\end{minipage}
\hfill
\begin{minipage}{.48\textwidth}
  \centering
  \includegraphics[width=.9\linewidth]{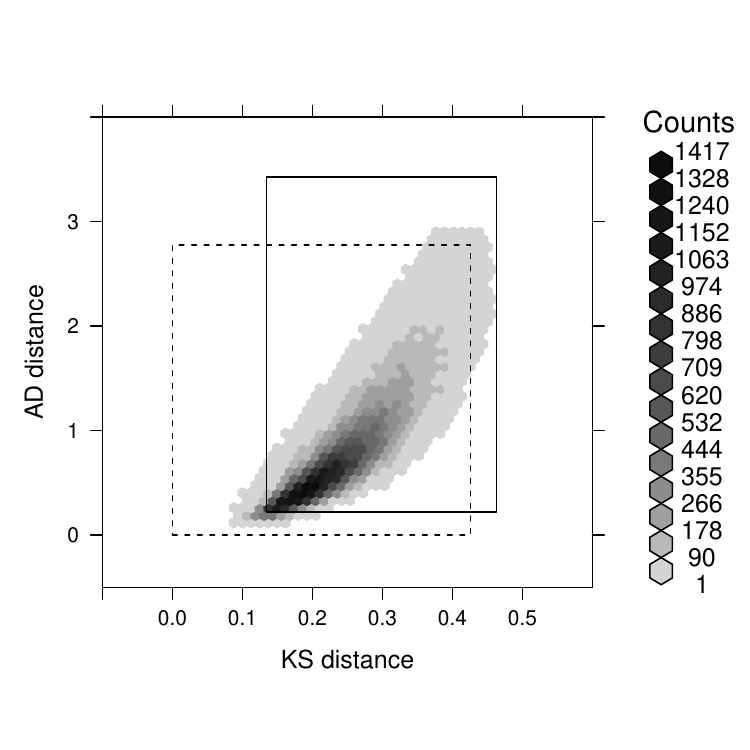}
  \captionof{figure}{Highest density region compared with two confidence rectangles.}
  \label{figure:hdr_with_2_conf_rect}
\end{minipage}
\end{figure}

Nearest neighbors algorithm has one hyperparameter: k --- the number of nearest neighbors to search for a given point. Unfortunately, $k$ was previously reserved to denote the number of statistics in our multi-statistic tests (see section~\ref{section:problem_formulation}). We use different fonts for this two parameters\footnote{But in the \texttt{R} code they still look the same.}. Hope this will not lead to much confusion. To determine the value for k, we use the recipe from~\cite{deliu2024alternative}: $\text{k} = \lfloor \sqrt{m} \rfloor$, where $m$ is the joint distribution sample size:

\begin{minted}[mathescape, linenos, texcomments]{r}
k = floor(sqrt(100000))
k
\end{minted}

\begin{verbatim}
316
\end{verbatim}

\noindent
To compute distances to k-nearest neighbors we will use the \texttt{FNN} library:
\begin{minted}[mathescape, linenos, texcomments]{r}
library(FNN)
\end{minted}
We will discuss this library and the existing alternatives in more detail later in the text. For now we just do this:

\begin{minted}[mathescape, linenos, texcomments]{r}
mat.H0 <- knnx.dist(data = statistics.H0, query = statistics.H0, k = k)
kNN.sum.dists.H0 <- rowSums(mat.H0)
\end{minted}
For each row of matrix \texttt{query} function \texttt{knnx.dist} returns distances to \texttt{k} nearest neighbors of this row from matrix \texttt{data} and puts these distances in corresponding row of retult matrix \texttt{mat.H0} (hence it will be $100000 \times 316$ numeric \texttt{R} matrix). So here for each point of joint distribution \texttt{statistics.H0} we calculate distances to this point's k-nearest neighbors in the joint distribution.
Then we sum up distances to nearest neighbors in each row of matrix \texttt{mat.H0} and get ``empirical distribution of sparsity'' \texttt{kNN.sum.dists.H0} (numeric \texttt{R} vector of size 100000) defined as $\mathbb{Q}$ in \eqref{eq:sparsity_empirical_distribution}. Then we compute $\mathrm{q}$ from the inequality~\eqref{eq:sparsity_check}:

\begin{minted}[mathescape, linenos, texcomments]{r}
q <- quantile(kNN.sum.dists.H0, 0.95)
q
\end{minted}

\begin{verbatim}
95%: 13.0944152566381
\end{verbatim}

\noindent
Distribution \texttt{kNN.sum.dists.H0} has long right tail. We first compute the logarithm of $\mathrm{q}$:

\begin{minted}[mathescape, linenos, texcomments]{r}
log(q)
\end{minted}

\begin{verbatim}
95%: 2.57218582302132
\end{verbatim}

\noindent
Then we plot a histogram for the log values of empirical distribution $\mathbb{Q}$:

\begin{minted}[mathescape, linenos, texcomments]{r}
hist(log(kNN.sum.dists.H0))
\end{minted}
This plot is presented on figure~\ref{figure:hist_Q_KS_AD}.

Finally, let us plot the announced highest density region. 
To do this, from the points of joint density that were used to construct $\mathbb{Q}$, we exclude those points that do not satisfy inequality~\eqref{eq:sparsity_check} and then plot the rest:

\begin{minted}[mathescape, linenos, texcomments]{r}
hexbinplot(
    y ~ x,
    data = distances.KS.AD[1:100000,][kNN.sum.dists.H0 < q,],  # see inequality \eqref{eq:sparsity_check}
    xbins = 30,
    panel = panel.with.KS.AD.bounds,  # we again add both confidence rectangles for comparison
    xlab = "KS distance", ylab = "AD distance",
    xlim = c(-0.1, 0.6),  ylim = c(-0.5, 4)      # zoomed-in graph like figure \ref{figure:KS_AD_joint_zoomed}
)
\end{minted}
This plot is presented on figure~\ref{figure:hdr_with_2_conf_rect}.

\noindent
On figure~\ref{figure:hdr_with_2_conf_rect} we see the estimated highest density region for joint density of KS and AD distances, confidence rectangle based on two-side confidence intervals (solid line), and confidence rectangle based on one-side confidence intervals (dashed line). 
The plot for the full density with the same $x$ and $y$ regions is on the figure~\ref{figure:KS_AD_joint_zoomed}. 

What can we say from the figure~\ref{figure:hdr_with_2_conf_rect}? 
Well, all three confidence sets contain 95\% of joint density, but they look quite different from each other.
Let us measure the power for goodness of fit tests based on each of the presented confidence sets. 

To do that, we first simulate sample of samples from some $H_1$ distribution ($N(0,1.5^2)$ in this particular case) and then compute joint distribution of KS and AD distances between standard normal CDF and eCDFs for samples of size $n=10$ from $H_1$ distribution:

\begin{minted}[mathescape, linenos, texcomments]{r}
sorted.samples.H1 <- get.sorted.samples.cpp(function(x) rnorm(x, 0, 1.5), n, 100000)
distance.distribution.KS.H1 <- 
    apply(sorted.samples.H1, 1, function(samp) distance.KS(samp, pnorm, n))
distance.distribution.AD.H1 <- 
    apply(sorted.samples.H1, 1, function(samp) distance.AD(samp, pnorm, n))
\end{minted}
We then store the joint distribution in a \texttt{data.frame}:
\begin{minted}[mathescape, linenos, texcomments]{r}
statistics.H1 <- data.frame(
    x = distance.distribution.KS.H1,
    y = distance.distribution.AD.H1
)
\end{minted}
Then we check the percent of samples, statistics for which lay outside of particular confidence set.  

For confidence rectangle based on two-side confidence intervals we have this:
\begin{minted}[mathescape, linenos, texcomments]{r}
check.hyperrectangle.bounds(statistics.H1, res.hyperrectangle.bounds.KS.AD)
\end{minted}

\begin{verbatim}
0.20009
\end{verbatim}
For confidence rectangle based on one-side confidence intervals we have this:
\begin{minted}[mathescape, linenos, texcomments]{r}
check.hyperrectangle.bounds(statistics.H1, res.hyperrectangle.bounds.KS.AD.rsi)
\end{minted}

\begin{verbatim}
0.28524
\end{verbatim}
For highest density region we have this:
\begin{minted}[mathescape, linenos, texcomments]{r}
mat.H1 <- knnx.dist(data = statistics.H0, query = statistics.H1, k = k)
kNN.sum.dists.H1 <- rowSums(mat.H1)
sum(kNN.sum.dists.H1 > q) / 100000
\end{minted}

\begin{verbatim}
0.41102
\end{verbatim}
We compute power estimate for highest density region as follows: for each point in joint distribution \texttt{statistics.H1} we find distances to its k-nearest neighbors in distribution \texttt{statistics.H0} and then compute the sum of these distances to measure the point's sparsity. If sparsity is too high, we reject. In the end, we calculate the number of rejections to get the power estimate. This procedure leads to the same result as if for each point of \texttt{statistics.H1} we checked if the point was outside HDR from figure~\ref{figure:hdr_with_2_conf_rect}.

We see that for this particular $H_1$ highest density region works better than both rectangle bounds. It is interesting to compare the result for joint KS--AD density test to the results for raw KS and AD tests from table~\ref{table:H0stdnormal:H1othernormals_scale:classical}. We see that for $N(0,1.5^2)$ KS test has average estimated power of $0.1291$ and AD test --- $0.3278$, so the KS--AD joint density test looks promising. 

In later sections we will present and evaluate the test based not on 2, but on 3 classical distances. 

Now let us try to rotate the axes via principal component approach and construct hyperrectangles for these rotated axes. Firstly, we do PCA as follows:

\begin{minted}[mathescape, linenos, texcomments]{r}
statistics.H0.p.comp <- princomp(statistics.H0)
statistics.H0.p.comp
statistics.H0.p.comp$loadings
\end{minted}

\begin{verbatim}
Call:
princomp(x = statistics.H0)

Standard deviations:
    Comp.1     Comp.2 
0.77113534 0.04083815 

 2  variables and  100000 observations.

Loadings:
  Comp.1 Comp.2
x         0.996
y  0.996       

               Comp.1 Comp.2
SS loadings       1.0    1.0
Proportion Var    0.5    0.5
Cumulative Var    0.5    1.0
\end{verbatim}
Then we construct equal-weighted hyperrectangle bounds for principal components:
\begin{minted}[mathescape, linenos, texcomments]{r}
res.hyperrectangle.bounds.KS.AD.pc1 <- calculate.hyperrectangle.bounds(  
    alpha = 0.05,
    gamma.left  = 0,
    gamma.right = 10,
    w = rep(1,2),  
    statistics.H0.p.comp$scores
)
res.hyperrectangle.bounds.KS.AD.pc1
\end{minted}

\begin{verbatim}
$left.bounds   -0.792721191117782   -0.0849965974476037
$right.bounds   2.62699350575366     0.0906602845002682
\end{verbatim}
Then --- variance-based bounds for principal components:
\begin{minted}[mathescape, linenos, texcomments]{r}
res.hyperrectangle.bounds.KS.AD.pc2 <- calculate.hyperrectangle.bounds(  
    alpha = 0.05,
    gamma.left  = 0,
    gamma.right = 10,
    w = statistics.H0.p.comp$sdev^2 / sum(statistics.H0.p.comp$sdev^2),  
    statistics.H0.p.comp$scores
)
res.hyperrectangle.bounds.KS.AD.pc2
\end{minted}

\begin{verbatim}
$left.bounds   -0.767006622776559   -0.306994410363725
$right.bounds   2.11932210770307     0.138613994850501
\end{verbatim}

\begin{figure}[H]
\centering
\begin{minipage}{.5\textwidth}
  \centering
  \includegraphics[width=.9\linewidth]{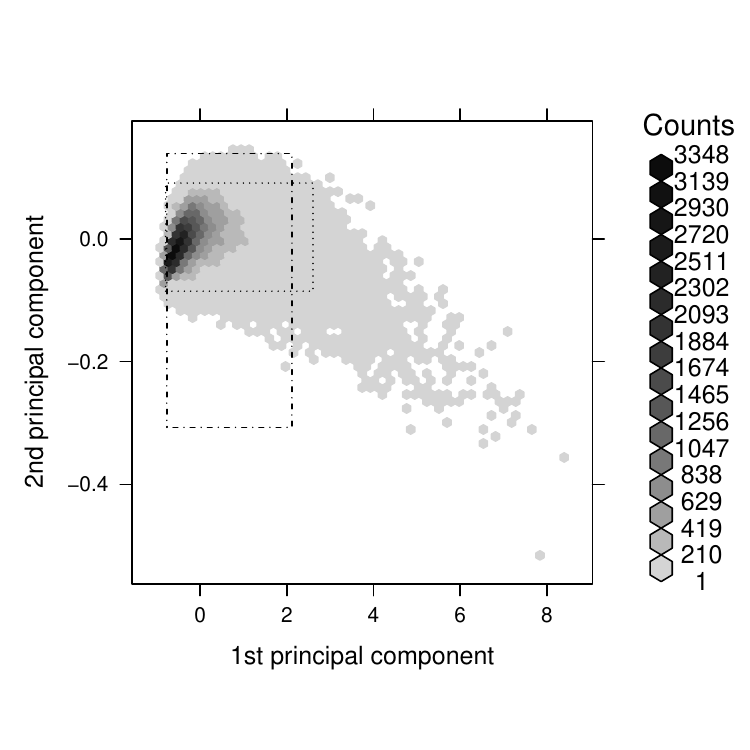}
  \captionof{figure}{Rotated joint distribution.}
  \label{fig:test1_2}
\end{minipage}%
\begin{minipage}{.5\textwidth}
  \centering
  \includegraphics[width=.9\linewidth]{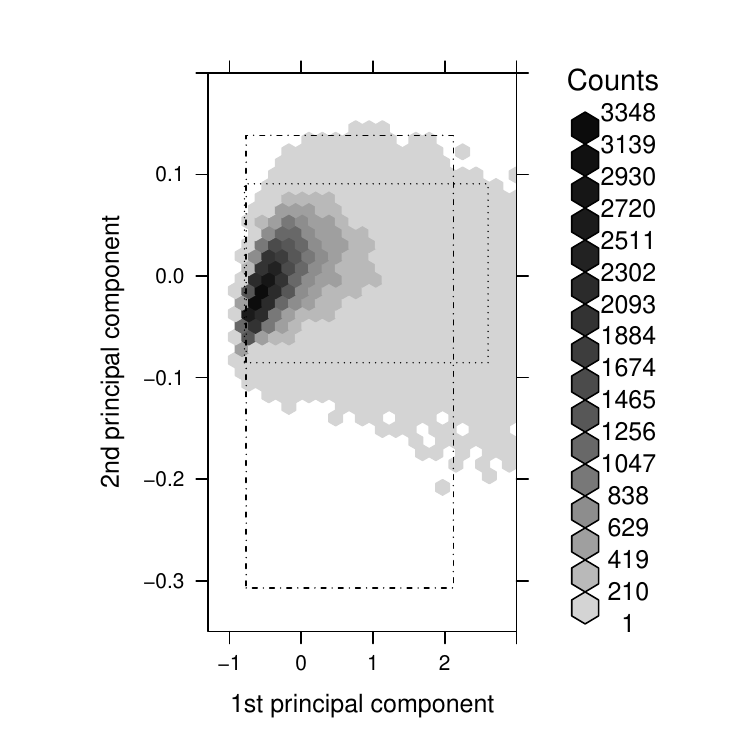}
  \captionof{figure}{Zoomed-in version of figure~\ref{fig:test1_2}.}
  \label{fig:test2_2}
\end{minipage}
\end{figure}

\noindent
Finally, we plot the rotated joint distribution and both bounds:

\begin{minted}[mathescape, linenos, texcomments]{r}
hexbinplot(
    y ~ x,
    data = data.frame(
        x = statistics.H0.p.comp$scores[,1],
        y = statistics.H0.p.comp$scores[,2]
    ),
    xbins = 50,
    panel = function(x, y, ...) {
        panel.hexbinplot(x, y, ...)
        plot.confidence.rectangle(res.hyperrectangle.bounds.KS.AD.pc1, 3)
        plot.confidence.rectangle(res.hyperrectangle.bounds.KS.AD.pc2, 4)
    },
    xlab = "1st principal component", ylab = "2nd principal component",
    # xlim = c(-1.3, 3), ylim = c(-0.35, 0.2)
)
\end{minted}
\noindent
The resulting plot and its zoomed-in version are presented on figures~\ref{fig:test1_2} and \ref{fig:test2_2}.

\noindent
Finally, we check the power of principal component based bounds:
\begin{minted}[mathescape, linenos, texcomments]{r}
p.comp.H1 <- predict(statistics.H0.p.comp, statistics.H1)
check.hyperrectangle.bounds(p.comp.H1, res.hyperrectangle.bounds.KS.AD.pc1)
check.hyperrectangle.bounds(p.comp.H1, res.hyperrectangle.bounds.KS.AD.pc2)
\end{minted}

\begin{verbatim}
0.33121
0.2314
\end{verbatim}

\noindent
We see that rotation does not help much, HDR based test is still the best.

\newpage

\section{Nearest neighbors test based on joint distribution of uniform order statistics; uniform vs beta case}
\label{sec:hdr_uniform_order_stat_test}

As we mentioned before in subsection~\ref{section:2dplots:uniform:hexplom}, for uniform samples joint distribution of order statistics is uniform in simplex $\Delta_n$ (defined by formula~\eqref{eq:simplex}). In this section we will show, that the approach based on nearest neighbors finds those parts of simplex's $\Delta_n$ border, which have the lowest values of local density. Interestingly, this produces a decent test. 

\subsection{Libraries to find nearest neighbors}
Throughout this chapter we will be using the \texttt{FNN} library to find nearest neighbors:
\begin{minted}[mathescape, linenos, texcomments]{r}
library(FNN)  # see \cite{FNN_package}
\end{minted}

\subsection{Sample size and number of samples}
Also, for this chapter we redefine the the sample size and number of samples:
\begin{minted}[mathescape, linenos]{r}
n <- 10      # sample size
m <- 100000  # number of samples
\end{minted}
We had to lower the number of samples since computational complexity of nearest neighbors approach depends on $m$ quadratically.

The default value of k, number of nearest neighbors to compare with, is defined as follows: 
\begin{minted}[mathescape, linenos]{r}
k = floor(sqrt(m))
k
\end{minted}

\begin{verbatim}
316
\end{verbatim}

\subsection{Simple demonstration on how to perform a nearest neighbor test}
To perform a test based on nearest neighbors, we do the following steps. 

First we generate the $H_0$ statistics matrix:
\begin{minted}[mathescape, linenos]{r}
sorted.samples.H0 <- get.sorted.samples.cpp(runif, n, m)
\end{minted}
Then we compute the distribution of sparsity estimates using the statistics matrix rows in the case when $H_0$ is true:
\begin{minted}[mathescape, linenos]{r}
mat.H0 <- knnx.dist(data = sorted.samples.H0, query = sorted.samples.H0, k = k)
kNN.sum.dists.H0 <- rowSums(mat.H0)
\end{minted}
For this distribution we determine $0.95$ quantile:
\begin{minted}[mathescape, linenos]{r}
q <- quantile(kNN.sum.dists.H0, 0.95)
q
\end{minted}

\begin{verbatim}
95%: 54.516334824856
\end{verbatim}

\noindent
Now, if we want to estimate power (or type 1 error) for some particular $H_1$ (that coincides with $H_0$, if we estimate type 1 error), we simulate statistics matrix for this $H_1$, compute sparsity estimate for each row of $H_1$ statistics matrix, and then compute the percentage of those rows that have sparsity higher than q. 

For type 1 error we do:

\begin{minted}[mathescape, linenos]{r}
sorted.samples.H1 <- get.sorted.samples.cpp(function(x) rbeta(x, 1, 1), n, m)
mat.H1 <- knnx.dist(data = sorted.samples.H0, query = sorted.samples.H1, k = k)
kNN.sum.dists.H1 <- rowSums(mat.H1)
sum(kNN.sum.dists.H1 > q) / m
\end{minted}

\begin{verbatim}
0.05151
\end{verbatim}

\noindent
For arbitrary $H_1$ we do:

\begin{minted}[mathescape, linenos]{r}
sorted.samples.H1 <- get.sorted.samples.cpp(function(x) rbeta(x, 0.5, 0.5), n, m)
mat.H1 <- knnx.dist(data = sorted.samples.H0, query = sorted.samples.H1, k = k)
kNN.sum.dists.H1 <- rowSums(mat.H1)
sum(kNN.sum.dists.H1 > q) / m
\end{minted}

\begin{verbatim}
0.37179
\end{verbatim}

\subsection{Function to calculate rejections for nearest neighbors tests \textasteriskcentered}

Let us write a function to calculate rejections in a loop for a given list of alternatives:

\begin{minted}[mathescape, linenos]{r}
calc.rejections.NN.tests <- function(
    sample.generation.function.H0,
    sample.generation.functions.H1,
    get.stat.matrix,
    alpha,
    h,
    n, m
)
{
    k <- floor(sqrt(m))
    s <- length(sample.generation.functions.H1)
    res.NN  <- matrix(0, h, s)
    for(i in 1:h)
    {
        print(paste("iteration", i))
        flush.console()
        
        stat.matrix.H0 <- get.stat.matrix(sample.generation.function.H0, n, m)
        
        print("calculate NN")
        flush.console()

        kNN.dists.H0 <- knnx.dist(data = stat.matrix.H0, query = stat.matrix.H0, k = k)
        kNN.sum.dists.H0 <- rowSums(kNN.dists.H0)
        q.H0 <- quantile(kNN.sum.dists.H0, alpha)
                                              
        for(j in 1:s){
            print(paste("H1", j))
            flush.console()
            stat.matrix.H1 <- get.stat.matrix(sample.generation.functions.H1[[j]], n, m)
            kNN.dists.H1 <- knnx.dist(data = stat.matrix.H0, query = stat.matrix.H1, k = k)
            kNN.sum.dists.H1 <- rowSums(kNN.dists.H1)
                                                   
            res.NN[i,j]  <- sum(kNN.sum.dists.H1 > q.H0) / m
        }
        print(paste("iteration", i, "done"))
        flush.console()
    }
    list(
        res.NN  = res.NN
    )
}
\end{minted}
The corresponding function to create power tables is as follows:
\begin{minted}[mathescape, linenos]{r}
create.rejection.table.kNN <- function(res.rejections)
    list(
        res.kNN.means = round(apply(res.rejections$res.NN, 2, mean), 4),
        res.kNN.sds = formatC(apply(res.rejections$res.NN, 2, sd  ), format = "e", digits = 1)
    )
\end{minted}

\subsection{Calculate rejections for nearest neighbor test in the case of uniform order statistics}
Now to the power tables. In this section we research ``uniform $H_0$ vs beta $H_1$'' case:
\begin{minted}[mathescape, linenos]{r}
res.uniform.kNN <- calc.rejections.NN.tests(
    sample.generation.function.H0  = runif,
    sample.generation.functions.H1 = c(
        lapply(1:10, function(x) function(nsamples) rbeta(nsamples, x, x)),
        lapply(rev(seq(0.1, 1.0, 0.1)), function(x) function(nsamples) rbeta(nsamples, x, x)),
        lapply(rev(seq(0.1, 1.0, 0.1)), function(x) function(nsamples) rbeta(nsamples, 1, x)),
        lapply(1:10, function(x) function(nsamples) rbeta(nsamples, 1, x))
    ),
    get.stat.matrix = get.sorted.samples.cpp,
    alpha = 0.95,
    h = 10,
    n = n, m = m
)
\end{minted}
The tables are as follows:
\begin{minted}[mathescape, linenos]{r}
res.uniform.kNN.table <- create.rejection.table.kNN(res.uniform.kNN)
res.uniform.kNN.table
\end{minted}

\begin{table}[H]
\centering

\begin{minipage}{.48\textwidth}
\centering

\begin{tabular}{ |l|p{1.8cm}||>{\columncolor[gray]{0.9}}p{1cm}|p{1.1cm}| } 
 \hline
  $H_0$ & $H_1$ & \multicolumn{2}{c|}{OS(U).kNN}  \\
  \hline
 &  & mean & sd  \\
 \hline
 \hline
 
$\mathrm{U}(0, 1)$ & $\mathrm{Beta}(1, 1)$ & 0.0521 & 8.5e-04\\
$\mathrm{U}(0, 1)$ & $\mathrm{Beta}(2, 2)$ & 0.0723 & 1.9e-03\\
$\mathrm{U}(0, 1)$ & $\mathrm{Beta}(3, 3)$ & 0.1799 & 5.1e-03\\
$\mathrm{U}(0, 1)$ & $\mathrm{Beta}(4, 4)$ & 0.3239 & 9.5e-03\\
$\mathrm{U}(0, 1)$ & $\mathrm{Beta}(5, 5)$ & 0.4722 & 1.2e-02\\
$\mathrm{U}(0, 1)$ & $\mathrm{Beta}(6, 6)$ & 0.607 & 1.4e-02\\
$\mathrm{U}(0, 1)$ & $\mathrm{Beta}(7, 7)$ & 0.7175 & 1.3e-02\\
$\mathrm{U}(0, 1)$ & $\mathrm{Beta}(8, 8)$ & 0.8033 & 1.2e-02\\
$\mathrm{U}(0, 1)$ & $\mathrm{Beta}(9, 9)$ & 0.8663 & 1.0e-02\\
$\mathrm{U}(0, 1)$ & $\mathrm{Beta}(10, 10)$ & 0.9104 & 8.6e-03\\

 \hline
 
 \end{tabular}
\caption{Test based on HDR for uniform order statistics; $\mathrm{U}(0,1)$ vs $\mathrm{Beta}(a,a)$; $a \geq 1$; alternatives have modes in the center, see figure~\ref{figure:beta:aa}.}
\label{table:beta:aa:uniform:kNN}

\end{minipage}
\hfill
\begin{minipage}{.48\textwidth}

\centering

\begin{tabular}{ |l|p{2cm}||>{\columncolor[gray]{0.9}}p{1cm}|p{1.1cm}| } 
 \hline
  $H_0$ & $H_1$ & \multicolumn{2}{c|}{OS(U).kNN}  \\
  \hline
 &  & mean & sd  \\
 \hline

$\mathrm{U}(0, 1)$ & $\mathrm{Beta}(1, 1)$ & 0.0518 & 6.1e-04\\
$\mathrm{U}(0, 1)$ & $\mathrm{Beta}(0.9, 0.9)$ & 0.0691 & 6.8e-04\\
$\mathrm{U}(0, 1)$ & $\mathrm{Beta}(0.8, 0.8)$ & 0.1004 & 9.0e-04\\
$\mathrm{U}(0, 1)$ & $\mathrm{Beta}(0.7, 0.7)$ & 0.1537 & 2.1e-03\\
$\mathrm{U}(0, 1)$ & $\mathrm{Beta}(0.6, 0.6)$ & 0.2409 & 2.2e-03\\
$\mathrm{U}(0, 1)$ & $\mathrm{Beta}(0.5, 0.5)$ & 0.3762 & 3.2e-03\\
$\mathrm{U}(0, 1)$ & $\mathrm{Beta}(0.4, 0.4)$ & 0.5628 & 3.1e-03\\
$\mathrm{U}(0, 1)$ & $\mathrm{Beta}(0.3, 0.3)$ & 0.7738 & 2.4e-03\\
$\mathrm{U}(0, 1)$ & $\mathrm{Beta}(0.2, 0.2)$ & 0.9395 & 1.0e-03\\
$\mathrm{U}(0, 1)$ & $\mathrm{Beta}(0.1, 0.1)$ & 0.9972 & 1.1e-04\\

 \hline
  \end{tabular}
\caption{Test based on HDR for uniform order statistics; $\mathrm{U}(0,1)$ vs $\mathrm{Beta}(a,a)$; $0 < a \leq 1$; alternatives have modes on both sides, see figure~\ref{figure:beta:0a0a}.}
\label{table:beta:0a0a:uniform:kNN}

\end{minipage}

\end{table}

\begin{table}[H]
\centering

\begin{minipage}{.48\textwidth}
\centering

\begin{tabular}{ |l|p{1.8cm}||>{\columncolor[gray]{0.9}}p{1cm}|p{1.1cm}| } 
 \hline
  $H_0$ & $H_1$ & \multicolumn{2}{c|}{OS(U).kNN}  \\
  \hline
 &  & mean & sd  \\
 \hline
 
$\mathrm{U}(0, 1)$ & $\mathrm{Beta}(1, 1)$ & 0.0526 & 9.6e-04\\
$\mathrm{U}(0, 1)$ & $\mathrm{Beta}(1, 0.9)$ & 0.0649 & 1.1e-03\\
$\mathrm{U}(0, 1)$ & $\mathrm{Beta}(1, 0.8)$ & 0.0993 & 1.9e-03\\
$\mathrm{U}(0, 1)$ & $\mathrm{Beta}(1, 0.7)$ & 0.1676 & 2.6e-03\\
$\mathrm{U}(0, 1)$ & $\mathrm{Beta}(1, 0.6)$ & 0.2865 & 3.1e-03\\
$\mathrm{U}(0, 1)$ & $\mathrm{Beta}(1, 0.5)$ & 0.4661 & 3.3e-03\\
$\mathrm{U}(0, 1)$ & $\mathrm{Beta}(1, 0.4)$ & 0.6848 & 4.8e-03\\
$\mathrm{U}(0, 1)$ & $\mathrm{Beta}(1, 0.3)$ & 0.8798 & 2.3e-03\\
$\mathrm{U}(0, 1)$ & $\mathrm{Beta}(1, 0.2)$ & 0.9805 & 8.2e-04\\
$\mathrm{U}(0, 1)$ & $\mathrm{Beta}(1, 0.1)$ & 0.9997 & 4.3e-05\\

  \hline
  \end{tabular}
\caption{Test based on HDR for uniform order statistics; $\mathrm{U}(0,1)$ vs $\mathrm{Beta}(1,a)$; $0 < a \leq 1$; alternatives have sharp modes on the right, see figure~\ref{figure:beta:10a}.}
\label{table:beta:10a:uniform:kNN}

\end{minipage}
\hfill
\begin{minipage}{.48\textwidth}

\centering

\begin{tabular}{ |l|p{2cm}||>{\columncolor[gray]{0.9}}p{1cm}|p{1.1cm}| } 
 \hline
  $H_0$ & $H_1$ & \multicolumn{2}{c|}{OS(U).kNN}  \\
  \hline
 &  & mean & sd  \\
 \hline

$\mathrm{U}(0, 1)$ & $\mathrm{Beta}(1, 1)$ & 0.0525 & 6.6e-04\\
$\mathrm{U}(0, 1)$ & $\mathrm{Beta}(1, 2)$ & 0.2562 & 4.1e-03\\
$\mathrm{U}(0, 1)$ & $\mathrm{Beta}(1, 3)$ & 0.6468 & 8.0e-03\\
$\mathrm{U}(0, 1)$ & $\mathrm{Beta}(1, 4)$ & 0.8903 & 6.0e-03\\
$\mathrm{U}(0, 1)$ & $\mathrm{Beta}(1, 5)$ & 0.9748 & 2.4e-03\\
$\mathrm{U}(0, 1)$ & $\mathrm{Beta}(1, 6)$ & 0.9954 & 5.9e-04\\
$\mathrm{U}(0, 1)$ & $\mathrm{Beta}(1, 7)$ & 0.9993 & 1.1e-04\\
$\mathrm{U}(0, 1)$ & $\mathrm{Beta}(1, 8)$ & 0.9999 & 2.9e-05\\
$\mathrm{U}(0, 1)$ & $\mathrm{Beta}(1, 9)$ & 1 & 1.3e-05\\
$\mathrm{U}(0, 1)$ & $\mathrm{Beta}(1, 10)$ & 1 & 4.2e-06\\
 
            \hline
\end{tabular}
\caption{Test based on HDR for uniform order statistics; $\mathrm{U}(0,1)$ vs $\mathrm{Beta}(1,a)$; $a \geq 1$; alternatives have sloping modes on the left, see figure~\ref{figure:beta:1a}.}
\label{table:beta:1a:uniform:kNN}

\end{minipage}

\end{table}

\noindent
Tables~\ref{table:beta:aa:uniform:kNN},~\ref{table:beta:0a0a:uniform:kNN},~\ref{table:beta:10a:uniform:kNN},~\ref{table:beta:1a:uniform:kNN} should be compared with tables from sections~\ref{section:hyperrectabgle_order_stats_uniform_full},~\ref{section:uniform:via:normal},~\ref{section:classical_tests_uniform},~\ref{sec:zhang_uniform}, and \ref{sec:2022verticals}. We denoted the newly introduced test as ``OS(U).kNN''\footnote{Which stands for ``ordered samples (uniform), k-nearest neighbors''}. Let us focus on comparison between ``OS(U)'', ``PC2(U)'' and ``OS(U).kNN'' tests since all of them are based on order statistics of uniform distribution. In tables~\ref{table:beta:aa:uniform:kNN},~\ref{table:beta:0a0a:uniform:kNN},~\ref{table:beta:10a:uniform:kNN} we see that the performance of ``OS(U).kNN'' test for the corresponding alternatives is better than the performance of ``PC2(U)'' test. The performance of ``OS(U).kNN'' for the alternatives from table~\ref{table:beta:1a:uniform:kNN} is a bit worse compared to ``PC2(U)''. The performance of ``OS(U)'' test is slightly better than the performance of ``OS(U).kNN'' test for the alternatives from ~\ref{table:beta:0a0a:uniform:kNN},~\ref{table:beta:10a:uniform:kNN},~\ref{table:beta:1a:uniform:kNN} tables, but for the alternatives from table~\ref{table:beta:aa:uniform:kNN} the performance of ``OS(U)'' test is almost nonexistent, and the performance of ``OS(U).kNN'' for this case is good. So far ``OS(U).kNN'' looks like a better version of ``PC2(U)''.

Ok, and what about a beta grid plot? We do:

\begin{minted}[mathescape, linenos, texcomments]{r}
res.uniform.beta2dgrid.kNN <- calc.rejections.NN.tests(
    sample.generation.function.H0  = runif,
    sample.generation.functions.H1 = apply(
        beta.parameters.grid,  # see formula \eqref{eq:beta_grid} 
        1, 
        function(par) function(nsamples) rbeta(nsamples, par[1], par[2])
    ),
    get.stat.matrix = get.sorted.samples.cpp,
    alpha = 0.95,
    h = 10,
    n = n, m = m
)

res.uniform.beta2dgrid.kNN.table <- create.rejection.table.kNN(res.uniform.beta2dgrid.kNN)
res.uniform.beta2dgrid.kNN.table

plot.powers.on.grid(beta.parameters.grid, res.uniform.beta2dgrid.kNN.table$res.kNN.means)
\end{minted}

\begin{figure}[H]
\centering
\includegraphics[width=16cm]{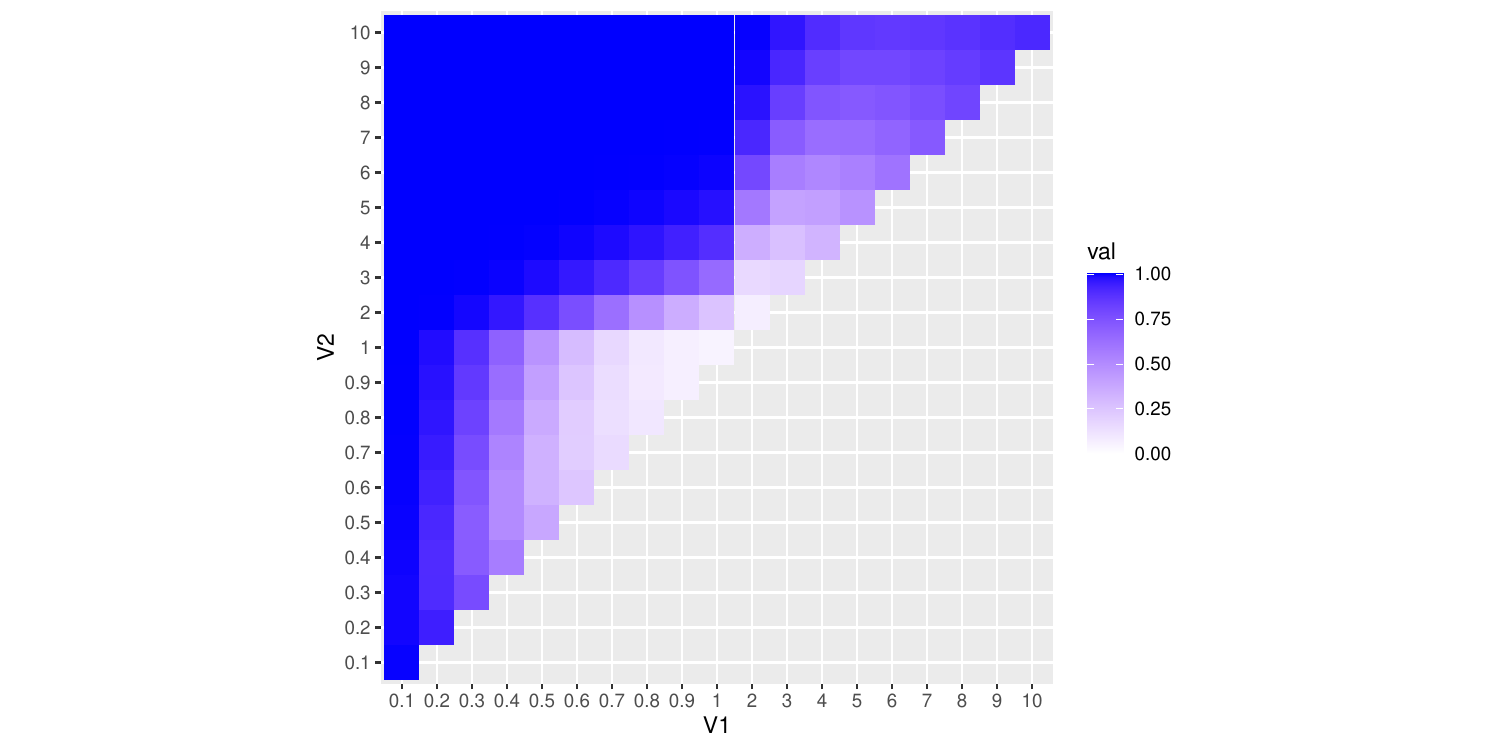}
\caption{OS(U).kNN}
\label{figure:osuknn}
\end{figure}

\noindent
Figure~\ref{figure:osuknn} should be compared to similar figures from sections~\ref{section:hyperrectabgle_order_stats_uniform_full},~\ref{section:uniform:via:normal},~\ref{section:classical_tests_uniform},~\ref{sec:zhang_uniform}, \ref{sec:2022verticals}, \ref{section:os_pairwise_dist_uniform}, and \ref{section:os_pairwise_dist_uniform_via_normal}. Figure~\ref{figure:osuknn} looks very similar to figure~\ref{figure:beta_grid:pc2u} for ``PC2(U)'' test with a bit better performance in the bottom left triangle.

\subsection{Uniform order statistics geometry}
To better understand how the proposed test works, let us plot order statistics for uniform samples of size $n = 3$ and mark those points form $H_0$ distribution, that are rejected by the test.

\begin{minted}[mathescape, linenos, texcomments]{r}
sorted.samples <- get.sorted.samples.cpp(runif, 3, 10000)

mat <- knnx.dist(data = sorted.samples, query = sorted.samples, k = round(sqrt(10000)))
v <- rowSums(mat)
filter <- v > quantile(v, 0.95)

options(rgl.useNULL = TRUE)
library(rgl)

plot3d(sorted.samples[,1], sorted.samples[,2], sorted.samples[,3])
points3d(sorted.samples[filter,1], sorted.samples[filter,2], sorted.samples[filter,3], col = "red", size = 10)

rglwidget()
\end{minted}

\begin{figure}[H]
\centering
\begin{minipage}{.33\textwidth}
  \centering
  \includegraphics[width=.99\linewidth]{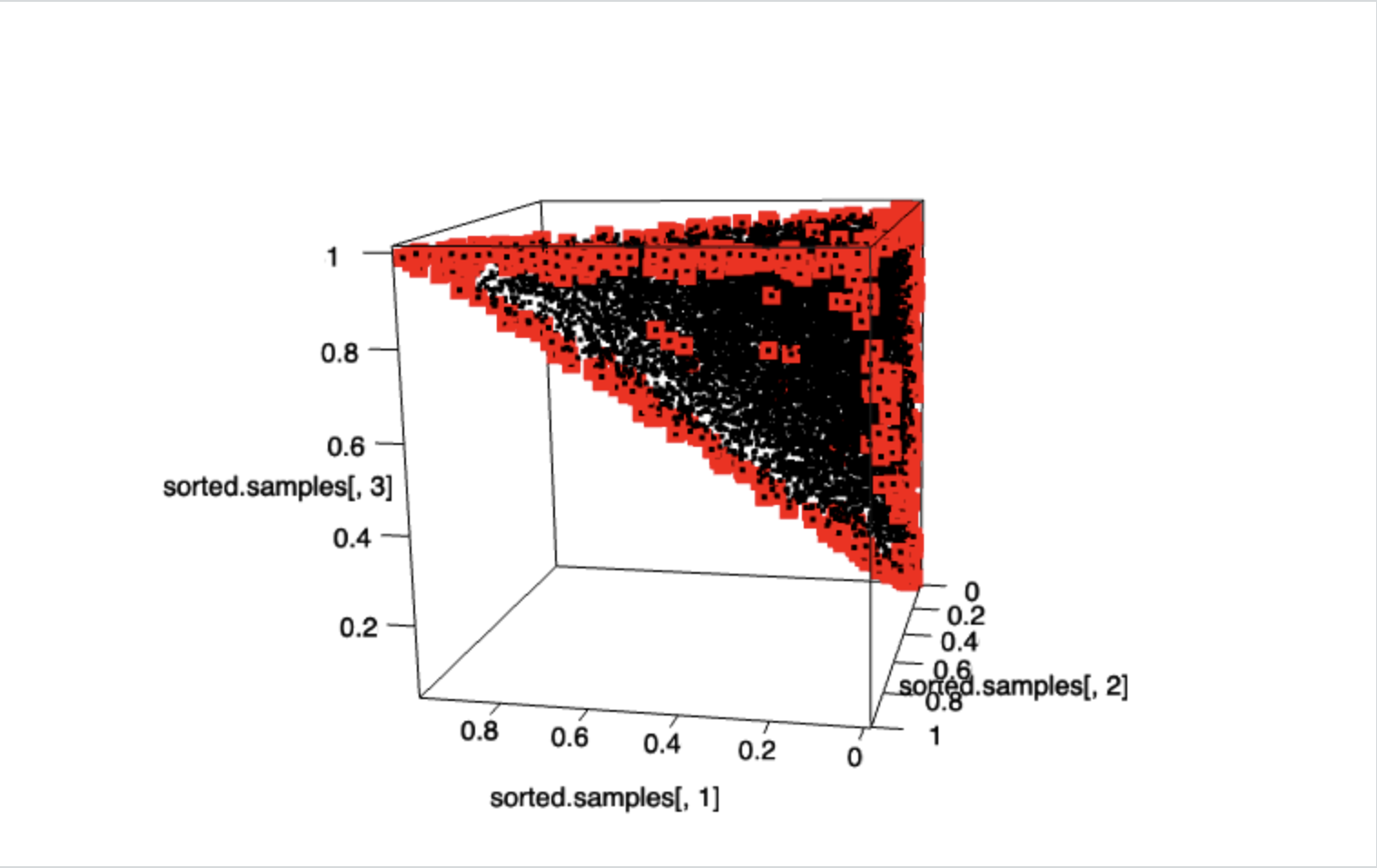}
\end{minipage}%
\begin{minipage}{.33\textwidth}
  \centering
  \includegraphics[width=.99\linewidth]{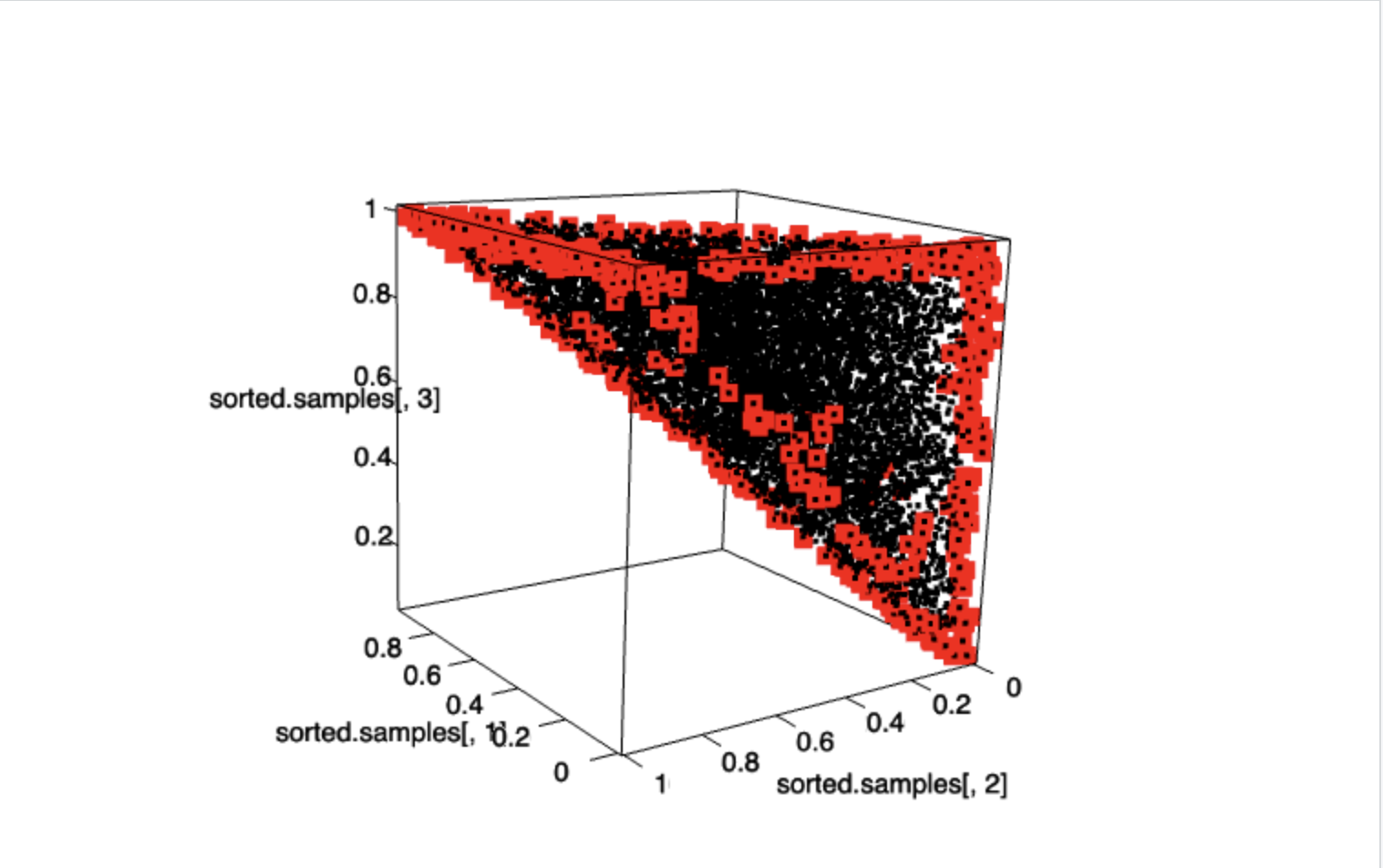}
\end{minipage}%
\begin{minipage}{.33\textwidth}
  \centering
  \includegraphics[width=.99\linewidth]{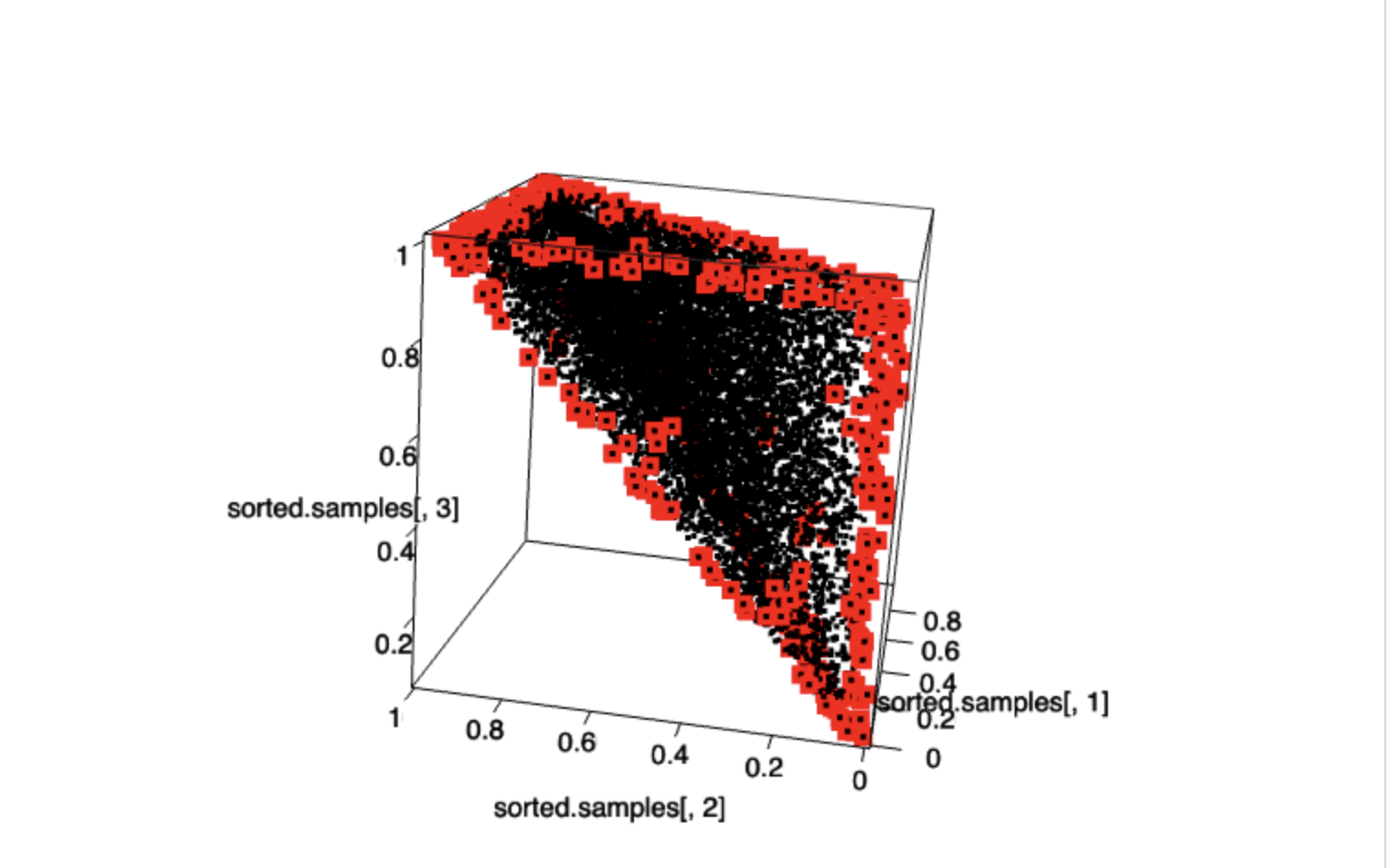}
\end{minipage}
\end{figure}

\begin{figure}[H]
\centering
\begin{minipage}{.33\textwidth}
  \centering
  \includegraphics[width=.99\linewidth]{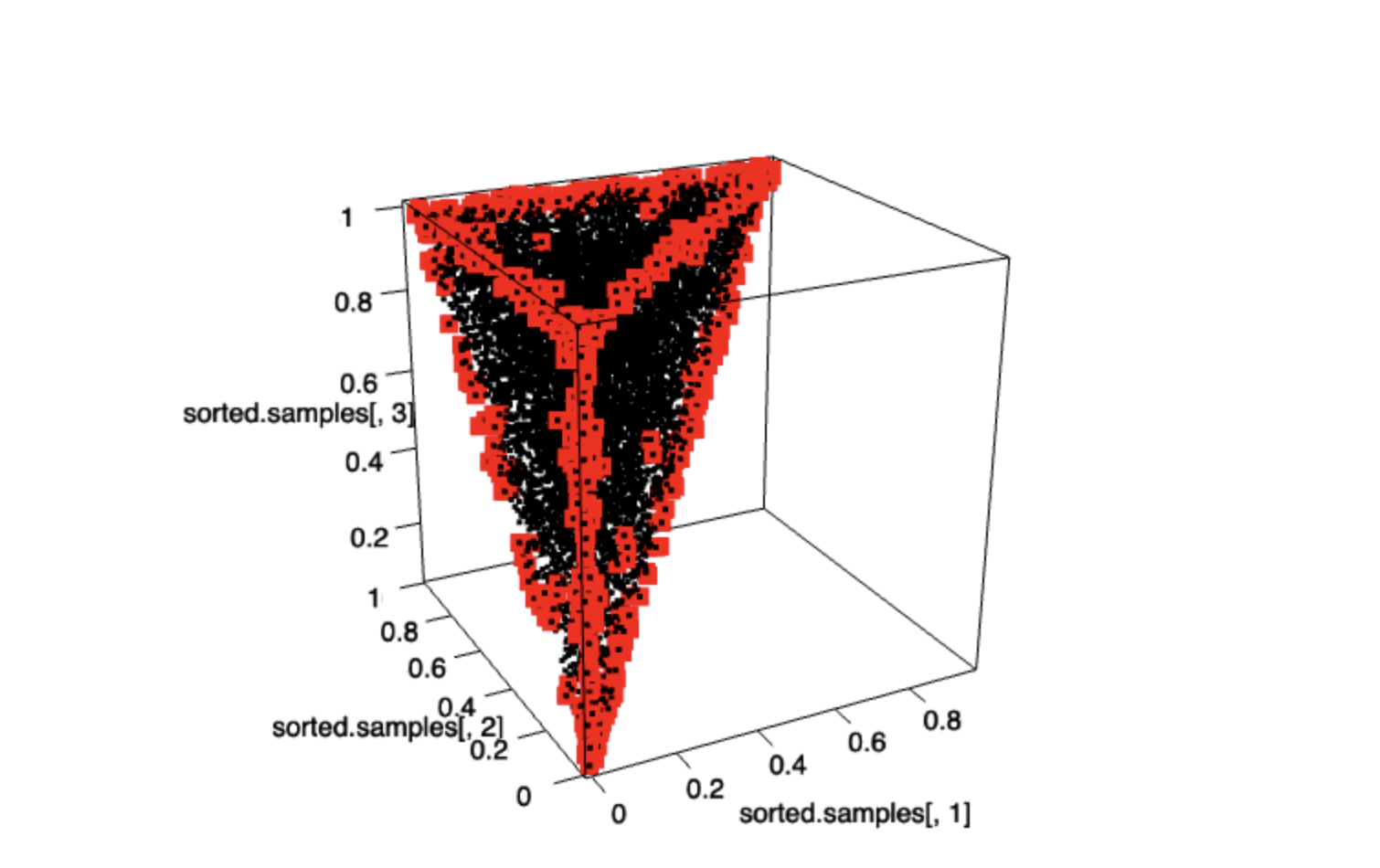}
\end{minipage}%
\begin{minipage}{.33\textwidth}
  \centering
  \includegraphics[width=.99\linewidth]{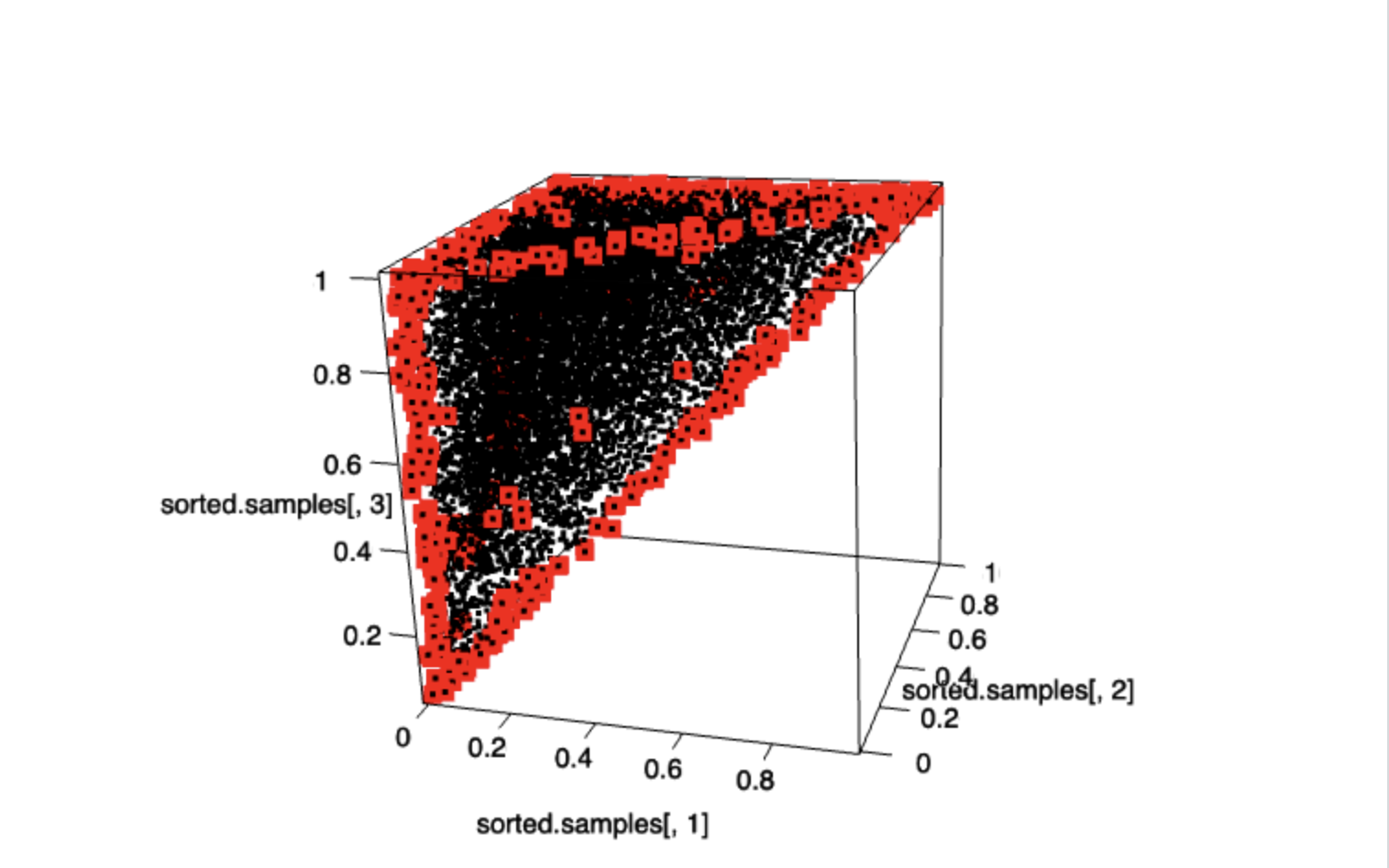}
\end{minipage}%
\begin{minipage}{.33\textwidth}
  \centering
  \includegraphics[width=.99\linewidth]{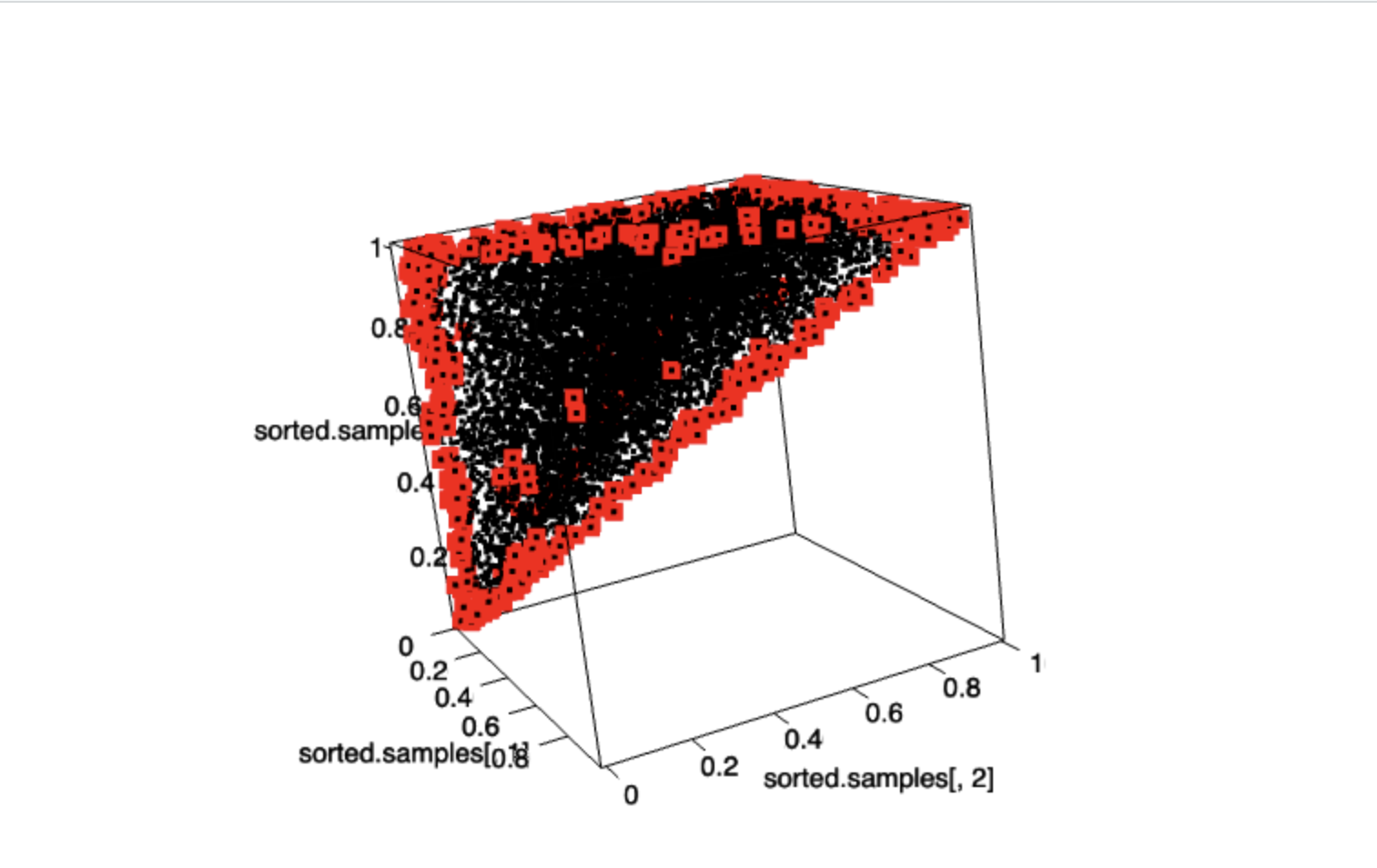}
\end{minipage}
\end{figure}

\begin{figure}[H]
\centering
\begin{minipage}{.33\textwidth}
  \centering
  \includegraphics[width=.99\linewidth]{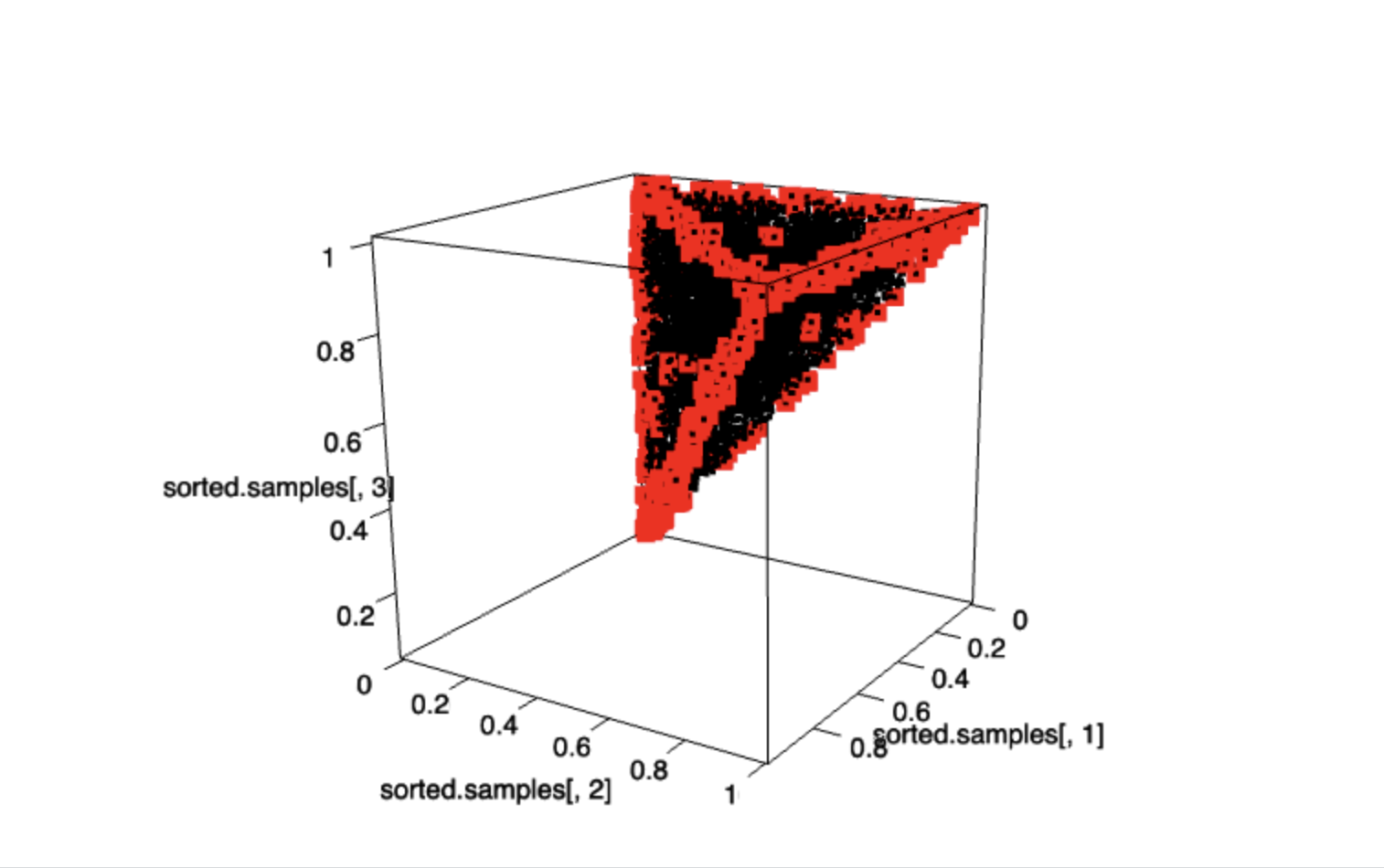}
\end{minipage}%
\begin{minipage}{.33\textwidth}
  \centering
  \includegraphics[width=.99\linewidth]{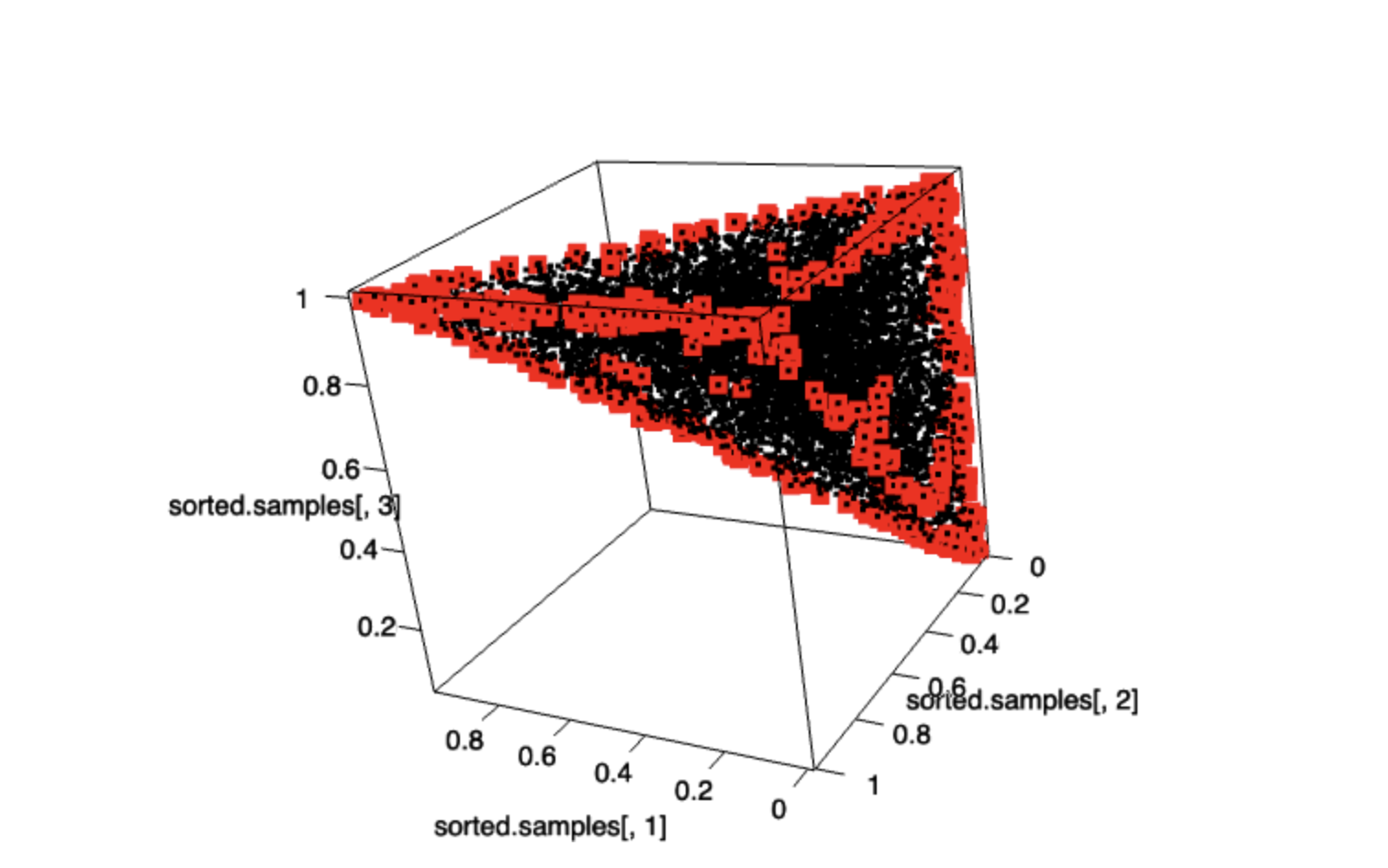}
\end{minipage}%
\begin{minipage}{.33\textwidth}
  \centering
  \includegraphics[width=.99\linewidth]{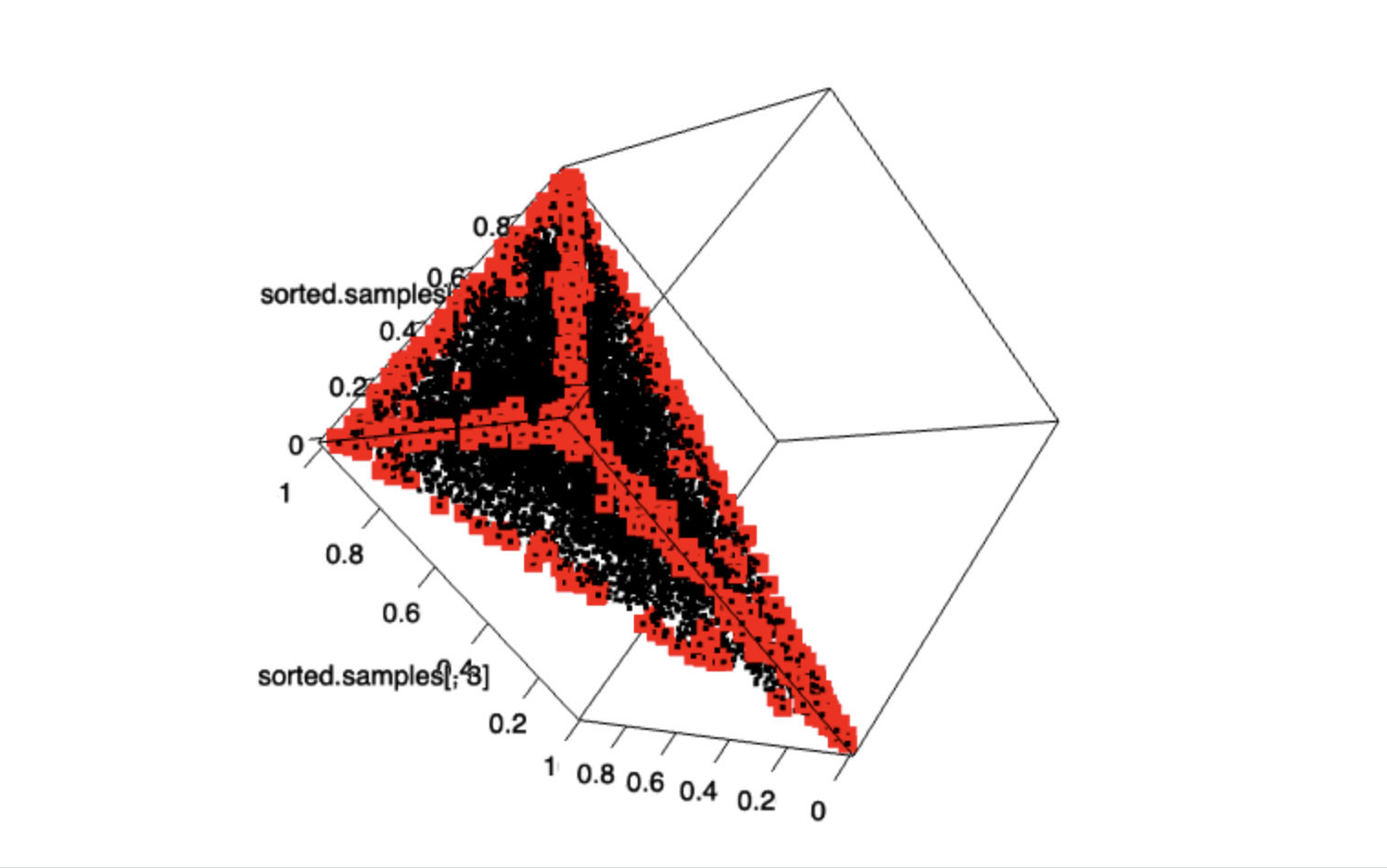}
\end{minipage}
\end{figure}

\noindent
We see that the nearest neighbors algorithm just finds the border of the simplex $\Delta_3$, where the local density values are the lowest.

\newpage
\section{Nearest neighbors test based on joint distribution of normal order statistics; uniform vs beta case (uniform via normal)}
\label{section:normal:nn:uniform:via:normal}

Now let us construct nearest neighbors test based on joint distribution of normal order statistics and apply it to check uniformity according to the methodology described in section~\ref{section:how:to:use:transforms:general} and subsection~\ref{section:how:to:use:normal:for:arbitrary}. We conduct simulations as follows:

\begin{minted}[mathescape, linenos]{r}
res.uniform.via.normal.kNN <- calc.rejections.NN.tests(
    sample.generation.function.H0  = rnorm,
    sample.generation.functions.H1 = c(
        lapply(1:10, function(x) function(nsamples) handle.infs(qnorm(rbeta(nsamples, x, x)))),
        lapply(
            rev(seq(0.1, 1.0, 0.1)), 
            function(x) function(nsamples) handle.infs(qnorm(rbeta(nsamples, x, x)))
        ),
        lapply(
            rev(seq(0.1, 1.0, 0.1)), 
            function(x) function(nsamples) handle.infs(qnorm(rbeta(nsamples, 1, x)))
        ),
        lapply(1:10, function(x) function(nsamples) handle.infs(qnorm(rbeta(nsamples, 1, x))))
    ),
    get.stat.matrix = get.sorted.samples.cpp,
    alpha = 0.95,
    h = 10,
    n = n, m = m
)
\end{minted}
Tables are created as follows:
\begin{minted}[mathescape, linenos]{r}
res.uniform.via.normal.kNN.table <- create.rejection.table.kNN(res.uniform.via.normal.kNN)
res.uniform.via.normal.kNN.table
\end{minted}
The new test is denoted as ``OS(N).kNN''\footnote{Which stands for ``order statistics (normal), k-nearest neighbors''.}.

\begin{table}[H]
\centering

\begin{minipage}{.48\textwidth}

\centering

\begin{tabular}{ |l|p{1.8cm}||>{\columncolor[gray]{0.9}}p{1cm}|p{1.1cm}| } 
 \hline
  $H_0$ & $H_1$ & \multicolumn{2}{c|}{OS(N).kNN}  \\
  \hline
 &  & mean & sd  \\
 \hline
 \hline
 
$\mathrm{U}(0, 1)$ & $\mathrm{Beta}(1, 1)$ & 0.0511 & 1.1e-03\\
$\mathrm{U}(0, 1)$ & $\mathrm{Beta}(2, 2)$ & 3e-04 & 7.0e-05\\
$\mathrm{U}(0, 1)$ & $\mathrm{Beta}(3, 3)$ & 4e-04 & 6.3e-05\\
$\mathrm{U}(0, 1)$ & $\mathrm{Beta}(4, 4)$ & 9e-04 & 1.6e-04\\
$\mathrm{U}(0, 1)$ & $\mathrm{Beta}(5, 5)$ & 0.0021 & 4.5e-04\\
$\mathrm{U}(0, 1)$ & $\mathrm{Beta}(6, 6)$ & 0.0041 & 8.9e-04\\
$\mathrm{U}(0, 1)$ & $\mathrm{Beta}(7, 7)$ & 0.0074 & 1.6e-03\\
$\mathrm{U}(0, 1)$ & $\mathrm{Beta}(8, 8)$ & 0.012 & 2.6e-03\\
$\mathrm{U}(0, 1)$ & $\mathrm{Beta}(9, 9)$ & 0.018 & 3.6e-03\\
$\mathrm{U}(0, 1)$ & $\mathrm{Beta}(10, 10)$ & 0.0259 & 5.3e-03\\

 \hline

 \end{tabular}
 \caption{Test based on HDR for normal order statistics; $\mathrm{U}(0,1)$ vs $\mathrm{Beta}(a,a)$; $a \geq 1$; alternatives have modes in the center, see figure~\ref{figure:beta:aa}.}
\label{table:beta:aa:uniform_via_normal:kNN}

\end{minipage}
\hfill
\begin{minipage}{.48\textwidth}

\centering

\begin{tabular}{ |l|p{2cm}||>{\columncolor[gray]{0.9}}p{1cm}|p{1.1cm}| } 
 \hline
  $H_0$ & $H_1$ & \multicolumn{2}{c|}{OS(N).kNN}  \\
  \hline
 &  & mean & sd  \\
 \hline

$\mathrm{U}(0, 1)$ & $\mathrm{Beta}(1, 1)$ & 0.0513 & 9.3e-04\\
$\mathrm{U}(0, 1)$ & $\mathrm{Beta}(0.9, 0.9)$ & 0.0932 & 1.1e-03\\
$\mathrm{U}(0, 1)$ & $\mathrm{Beta}(0.8, 0.8)$ & 0.1638 & 1.3e-03\\
$\mathrm{U}(0, 1)$ & $\mathrm{Beta}(0.7, 0.7)$ & 0.2752 & 2.3e-03\\
$\mathrm{U}(0, 1)$ & $\mathrm{Beta}(0.6, 0.6)$ & 0.4341 & 2.1e-03\\
$\mathrm{U}(0, 1)$ & $\mathrm{Beta}(0.5, 0.5)$ & 0.6264 & 1.8e-03\\
$\mathrm{U}(0, 1)$ & $\mathrm{Beta}(0.4, 0.4)$ & 0.8143 & 1.7e-03\\
$\mathrm{U}(0, 1)$ & $\mathrm{Beta}(0.3, 0.3)$ & 0.9451 & 8.8e-04\\
$\mathrm{U}(0, 1)$ & $\mathrm{Beta}(0.2, 0.2)$ & 0.9942 & 2.3e-04\\
$\mathrm{U}(0, 1)$ & $\mathrm{Beta}(0.1, 0.1)$ & 1 & 2.1e-05\\

 \hline
 
 \end{tabular}
\caption{Test based on HDR for normal order statistics; $\mathrm{U}(0,1)$ vs $\mathrm{Beta}(a,a)$; $0 < a \leq 1$; alternatives have modes on both sides, see figure~\ref{figure:beta:0a0a}.}
\label{table:beta:0a0a:uniform_via_normal:kNN}

\end{minipage}

\end{table}

\noindent
In table~\ref{table:beta:aa:uniform_via_normal:kNN} we immediately see that ``OS(N).kNN'' test lacks power for symmetric alternatives concentrated in the center. Table~\ref{table:beta:0a0a:uniform_via_normal:kNN} looks very good.

\begin{table}[H]
\centering

\begin{minipage}{.48\textwidth}

\centering

\begin{tabular}{ |l|p{2cm}||>{\columncolor[gray]{0.9}}p{1cm}|p{1.1cm}| } 
 \hline
  $H_0$ & $H_1$ & \multicolumn{2}{c|}{OS(N).kNN}  \\
  \hline
 &  & mean & sd  \\
 \hline

$\mathrm{U}(0, 1)$ & $\mathrm{Beta}(1, 1)$ & 0.0514 & 8.6e-04\\
$\mathrm{U}(0, 1)$ & $\mathrm{Beta}(1, 0.9)$ & 0.0749 & 1.2e-03\\
$\mathrm{U}(0, 1)$ & $\mathrm{Beta}(1, 0.8)$ & 0.1227 & 1.4e-03\\
$\mathrm{U}(0, 1)$ & $\mathrm{Beta}(1, 0.7)$ & 0.2116 & 2.3e-03\\
$\mathrm{U}(0, 1)$ & $\mathrm{Beta}(1, 0.6)$ & 0.3587 & 3.2e-03\\
$\mathrm{U}(0, 1)$ & $\mathrm{Beta}(1, 0.5)$ & 0.5645 & 3.4e-03\\
$\mathrm{U}(0, 1)$ & $\mathrm{Beta}(1, 0.4)$ & 0.7824 & 2.5e-03\\
$\mathrm{U}(0, 1)$ & $\mathrm{Beta}(1, 0.3)$ & 0.9388 & 1.4e-03\\
$\mathrm{U}(0, 1)$ & $\mathrm{Beta}(1, 0.2)$ & 0.9942 & 3.2e-04\\
$\mathrm{U}(0, 1)$ & $\mathrm{Beta}(1, 0.1)$ & 1 & 1.7e-05\\

 \hline
 
\end{tabular}
\caption{Test based on HDR for normal order statistics; $\mathrm{U}(0,1)$ vs $\mathrm{Beta}(1,a)$; $0 < a \leq 1$; alternatives have sharp modes on the right, see figure~\ref{figure:beta:10a}.}
\label{table:beta:10a:uniform_via_normal:kNN}
 
\end{minipage}
\hfill
\begin{minipage}{.48\textwidth}

\centering

\begin{tabular}{ |l|p{2cm}||>{\columncolor[gray]{0.9}}p{1cm}|p{1.1cm}| } 
 \hline
  $H_0$ & $H_1$ & \multicolumn{2}{c|}{OS(N).kNN}  \\
  \hline
 &  & mean & sd  \\
 \hline

$\mathrm{U}(0, 1)$ & $\mathrm{Beta}(1, 1)$ & 0.0514 & 9.9e-04\\
$\mathrm{U}(0, 1)$ & $\mathrm{Beta}(1, 2)$ & 0.1192 & 3.2e-03\\
$\mathrm{U}(0, 1)$ & $\mathrm{Beta}(1, 3)$ & 0.3955 & 7.5e-03\\
$\mathrm{U}(0, 1)$ & $\mathrm{Beta}(1, 4)$ & 0.6977 & 8.4e-03\\
$\mathrm{U}(0, 1)$ & $\mathrm{Beta}(1, 5)$ & 0.8849 & 5.1e-03\\
$\mathrm{U}(0, 1)$ & $\mathrm{Beta}(1, 6)$ & 0.9641 & 2.2e-03\\
$\mathrm{U}(0, 1)$ & $\mathrm{Beta}(1, 7)$ & 0.9907 & 9.9e-04\\
$\mathrm{U}(0, 1)$ & $\mathrm{Beta}(1, 8)$ & 0.9978 & 2.4e-04\\
$\mathrm{U}(0, 1)$ & $\mathrm{Beta}(1, 9)$ & 0.9996 & 6.0e-05\\
$\mathrm{U}(0, 1)$ & $\mathrm{Beta}(1, 10)$ & 0.9999 & 3.3e-05\\
 
             \hline
\end{tabular}
\caption{Test based on HDR for normal order statistics; $\mathrm{U}(0,1)$ vs $\mathrm{Beta}(1,a)$; $a \geq 1$; alternatives have sloping modes on the left, see figure~\ref{figure:beta:1a}.}
\label{table:beta:1a:uniform_via_normal:kNN}

\end{minipage}

\end{table}

\noindent
Tables~\ref{table:beta:10a:uniform_via_normal:kNN} and \ref{table:beta:1a:uniform_via_normal:kNN} are ok. 
And the beta grid plot: 

\begin{minted}[mathescape, linenos, texcomments]{r}
res.unif.via.normal.beta2dgrid.kNN <- calc.rejections.NN.tests(
    sample.generation.function.H0  = rnorm,
    sample.generation.functions.H1 = apply(
        beta.parameters.grid,  # see formula \eqref{eq:beta_grid} 
        1, 
        function(par) function(nsamples) handle.infs(qnorm(rbeta(nsamples, par[1], par[2])))
    ),
    get.stat.matrix = get.sorted.samples.cpp,
    alpha = 0.95,
    h = 10,
    n = n, m = m
)

res.unif.via.normal.beta2dgrid.kNN.table <- create.rejection.table.kNN(res.unif.via.normal.beta2dgrid.kNN)
res.unif.via.normal.beta2dgrid.kNN.table

plot.powers.on.grid(beta.parameters.grid, res.unif.via.normal.beta2dgrid.kNN.table$res.kNN.means)
\end{minted}

\begin{figure}[H]
\centering
\includegraphics[width=16cm]{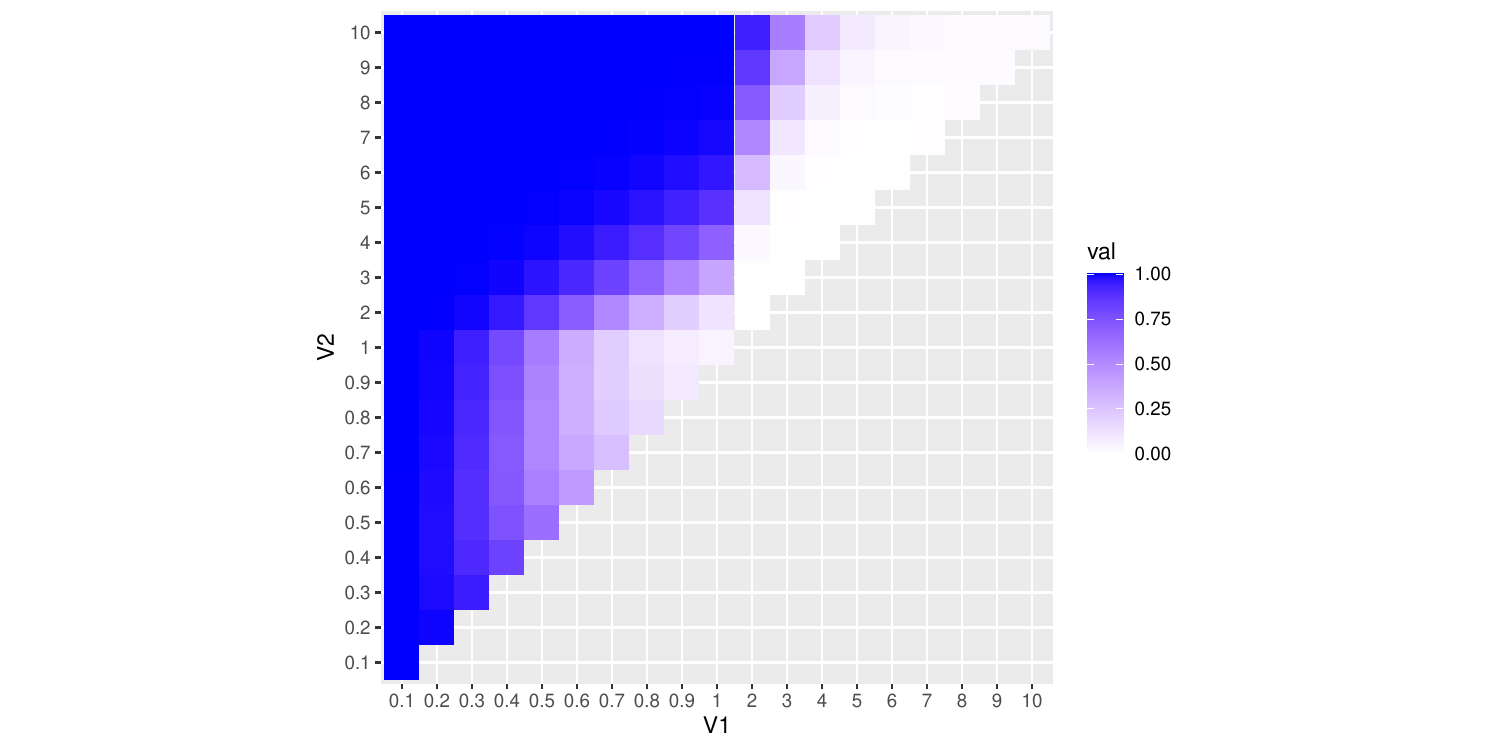}
\caption{OS(N).kNN}
\label{figure:osnknn}
\end{figure}

\noindent
On figure~\ref{figure:osnknn} we see that ``OS(N).kNN'' test has very small power for $\mathrm{Beta}(a,b)$, $a, b > 0$ alternatives, no matter symmetric or asymmetric.

\subsection{Normal order statistics geometry}

Let us plot normal order statistics vectors for samples of size 3 and mark those vectors that are rejected according to nearest neighbors test: 

\begin{minted}[mathescape, linenos, texcomments]{r}
sorted.samples <- get.sorted.samples.cpp(rnorm, 3, 10000)

mat <- knnx.dist(data = sorted.samples, query = sorted.samples, k = round(sqrt(10000)))
v <- rowSums(mat)
filter <- v > quantile(v, 0.95)

options(rgl.useNULL = TRUE)
library(rgl)

plot3d(sorted.samples[,1], sorted.samples[,2], sorted.samples[,3])
points3d(sorted.samples[filter,1], sorted.samples[filter,2], sorted.samples[filter,3], col = "red", size = 10)

rglwidget()
\end{minted}

\begin{figure}[H]
\centering
\begin{minipage}{.33\textwidth}
  \centering
  \includegraphics[width=.99\linewidth]{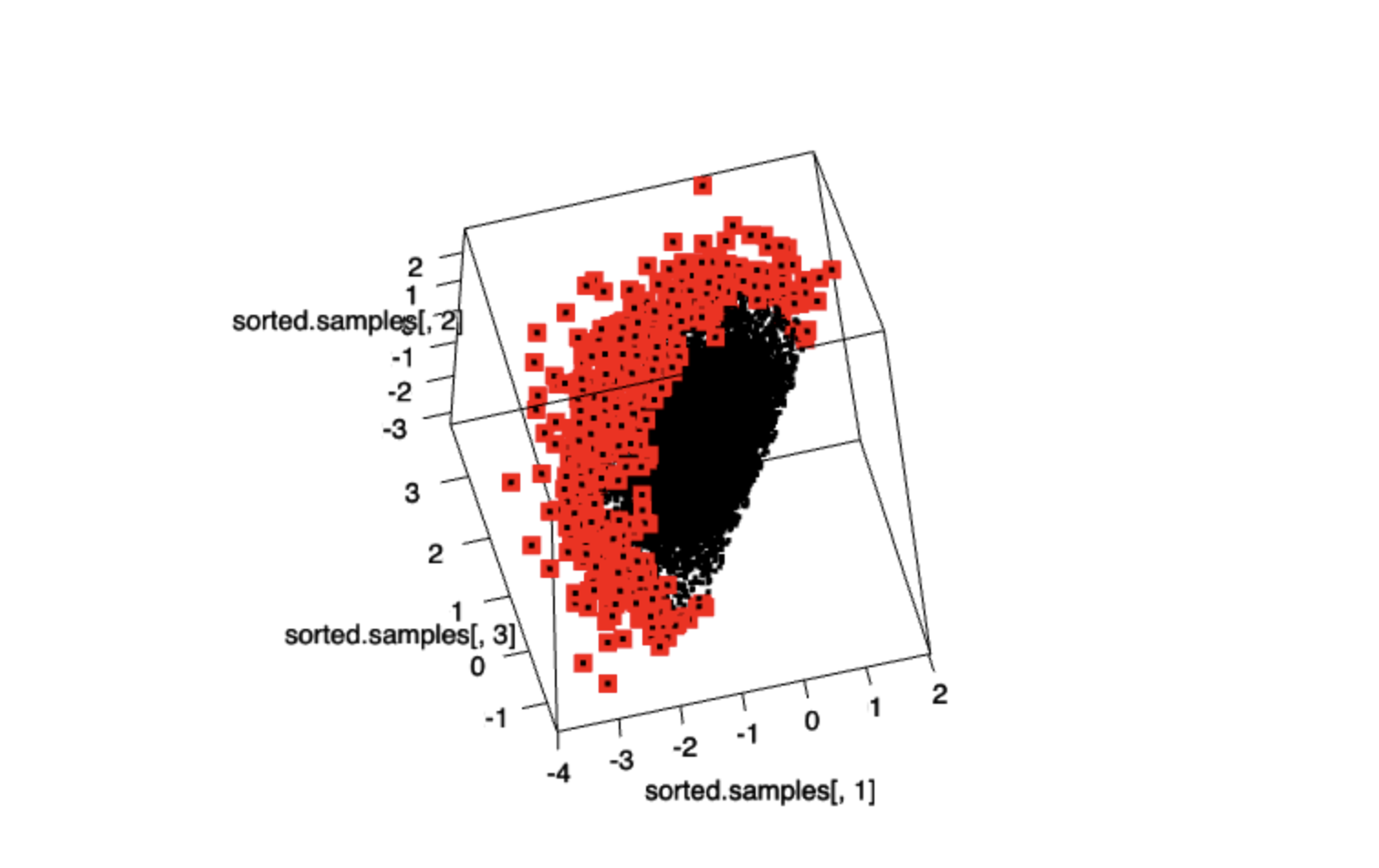}
\end{minipage}%
\begin{minipage}{.33\textwidth}
  \centering
  \includegraphics[width=.99\linewidth]{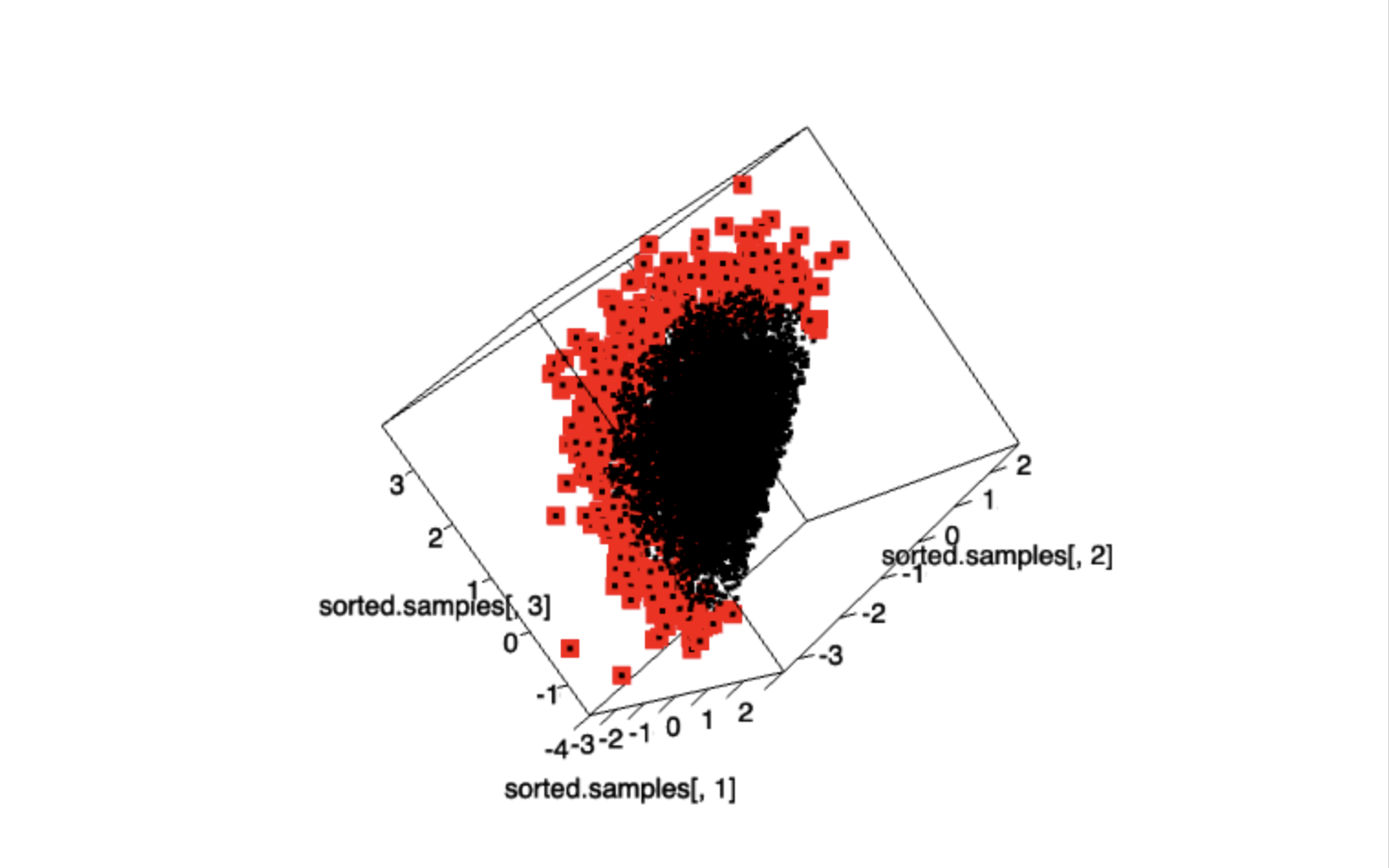}
\end{minipage}%
\begin{minipage}{.33\textwidth}
  \centering
  \includegraphics[width=.99\linewidth]{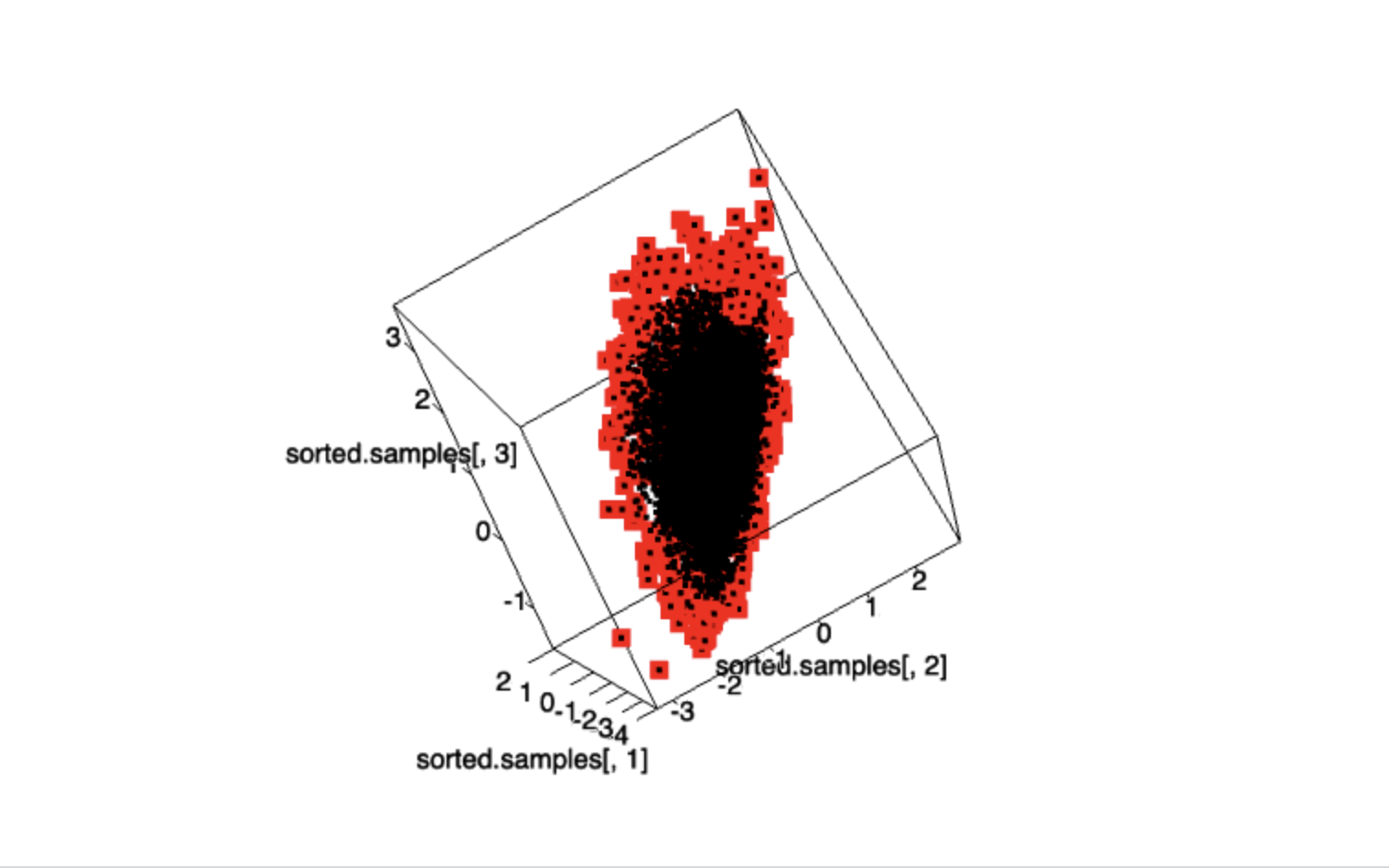}
\end{minipage}
\end{figure}

\begin{figure}[H]
\centering
\begin{minipage}{.33\textwidth}
  \centering
  \includegraphics[width=.99\linewidth]{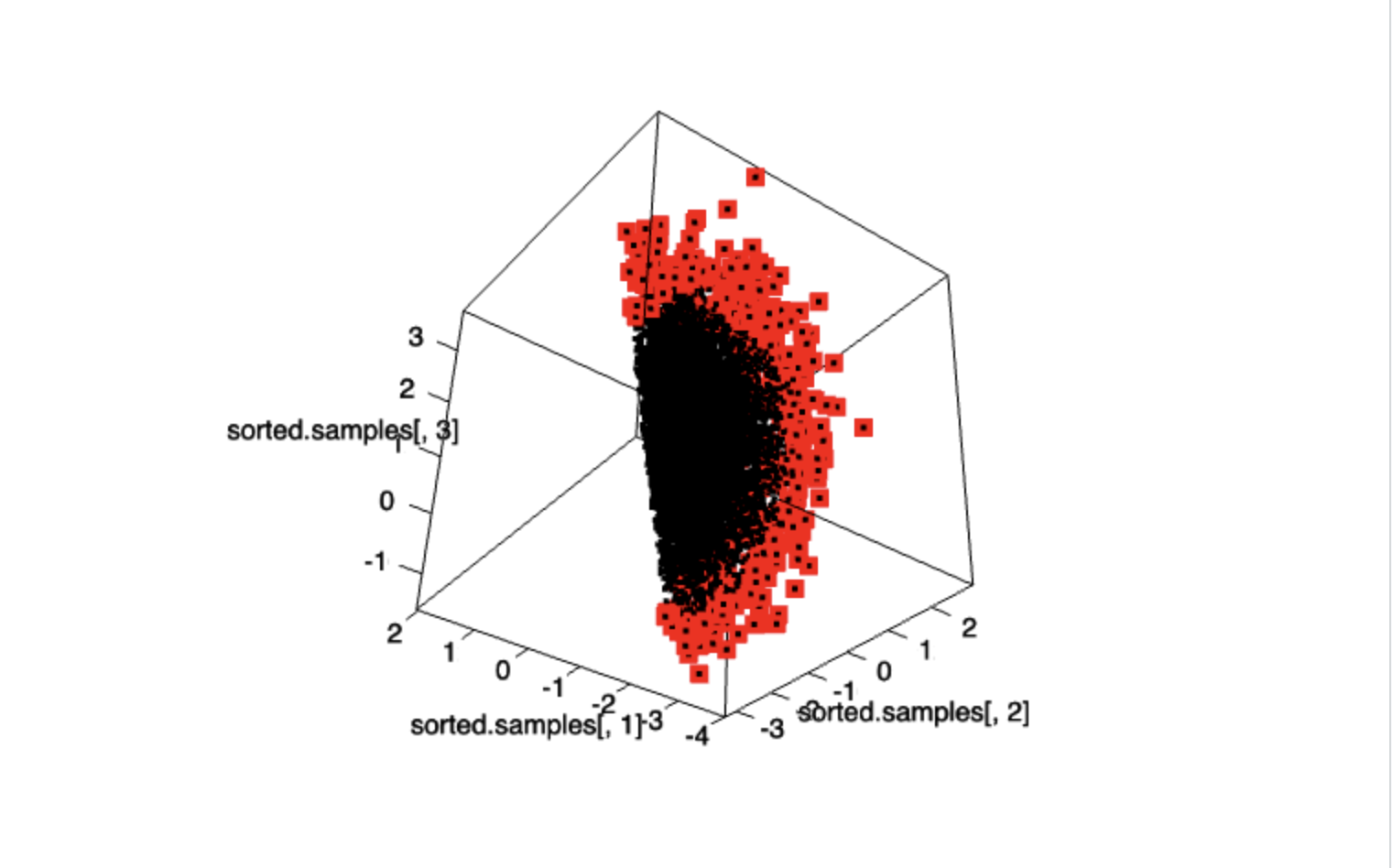}
\end{minipage}%
\begin{minipage}{.33\textwidth}
  \centering
  \includegraphics[width=.99\linewidth]{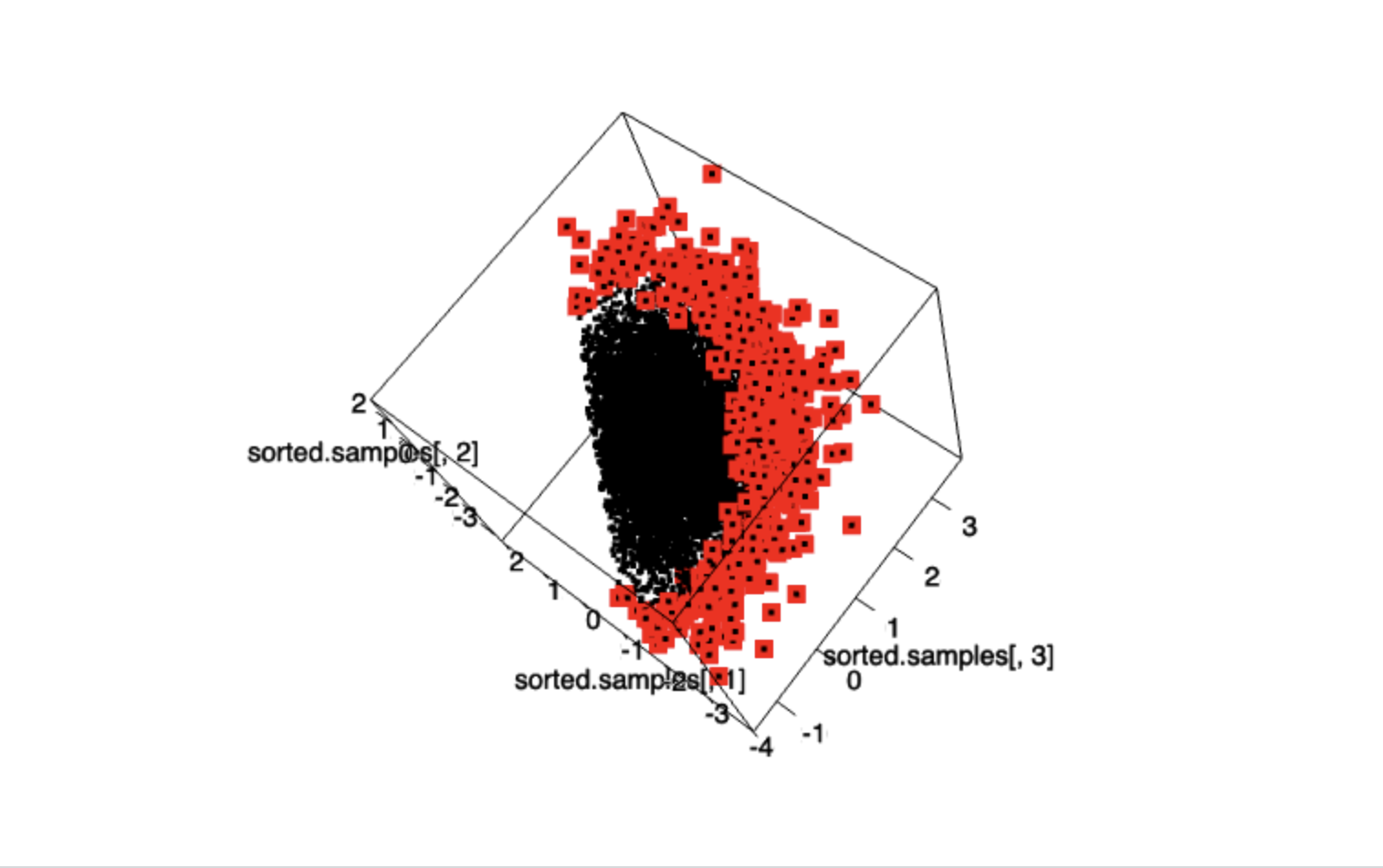}
\end{minipage}%
\begin{minipage}{.33\textwidth}
  \centering
  \includegraphics[width=.99\linewidth]{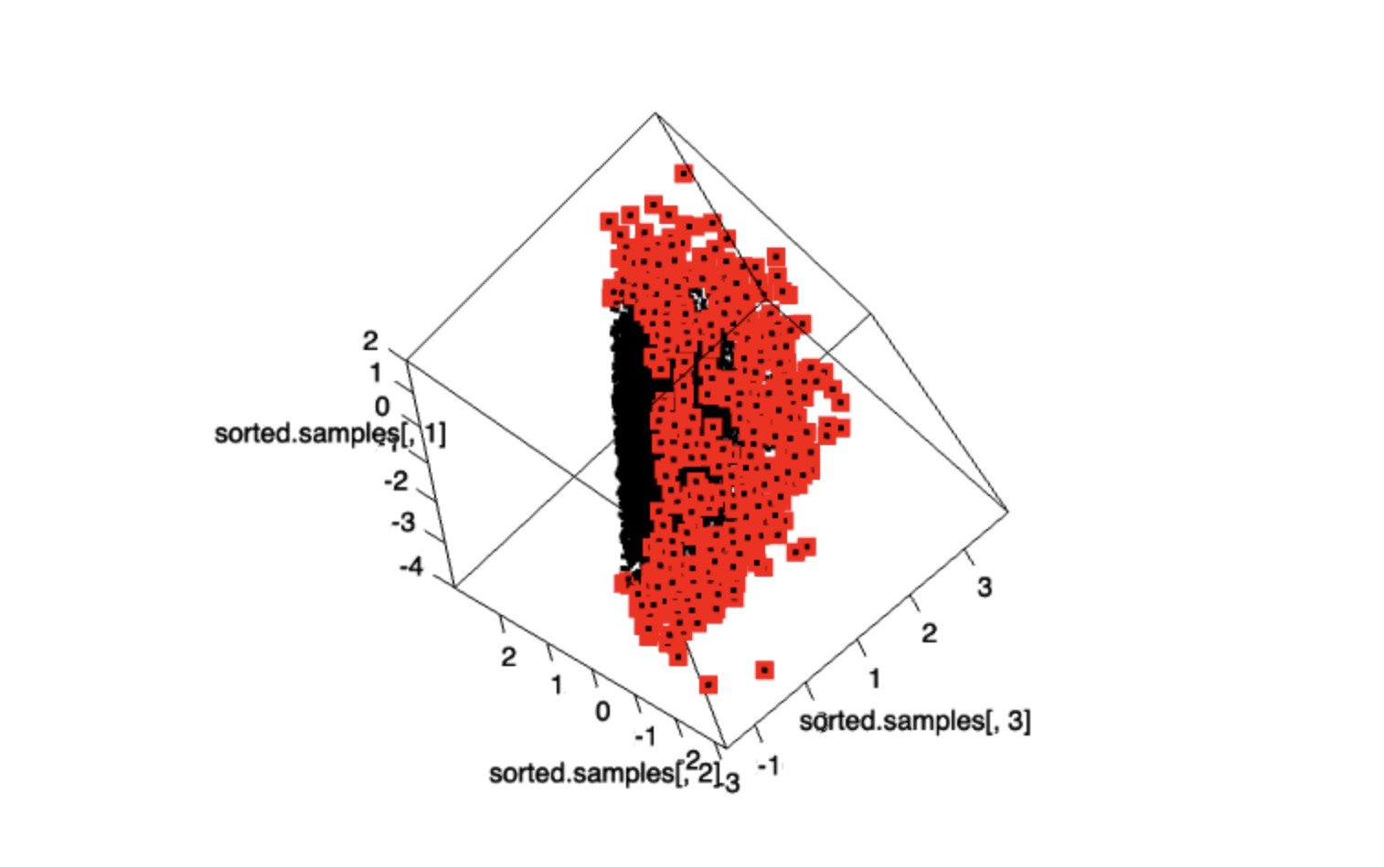}
\end{minipage}
\end{figure}

\begin{figure}[H]
\centering
\begin{minipage}{.33\textwidth}
  \centering
  \includegraphics[width=.99\linewidth]{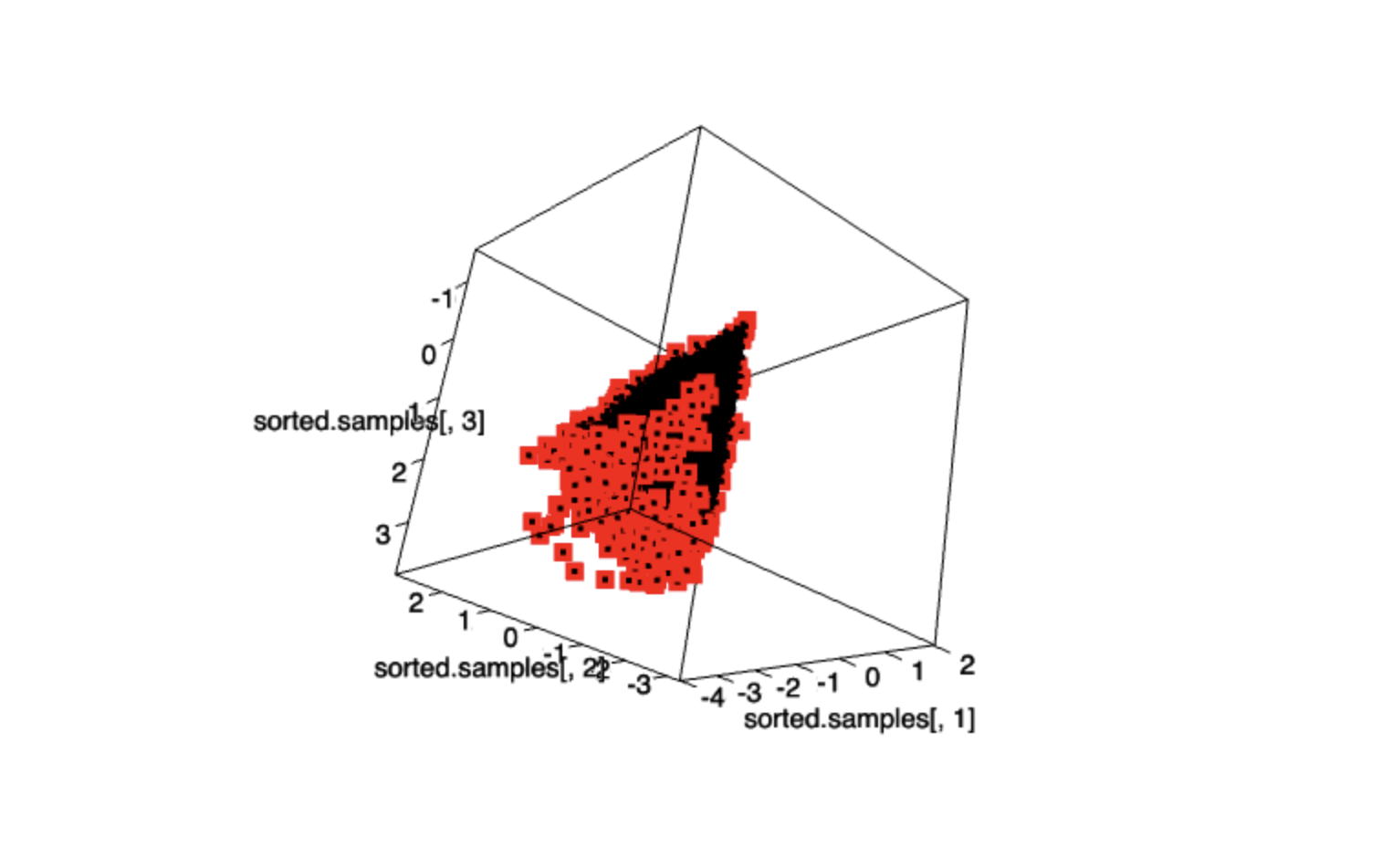}
\end{minipage}%
\begin{minipage}{.33\textwidth}
  \centering
  \includegraphics[width=.99\linewidth]{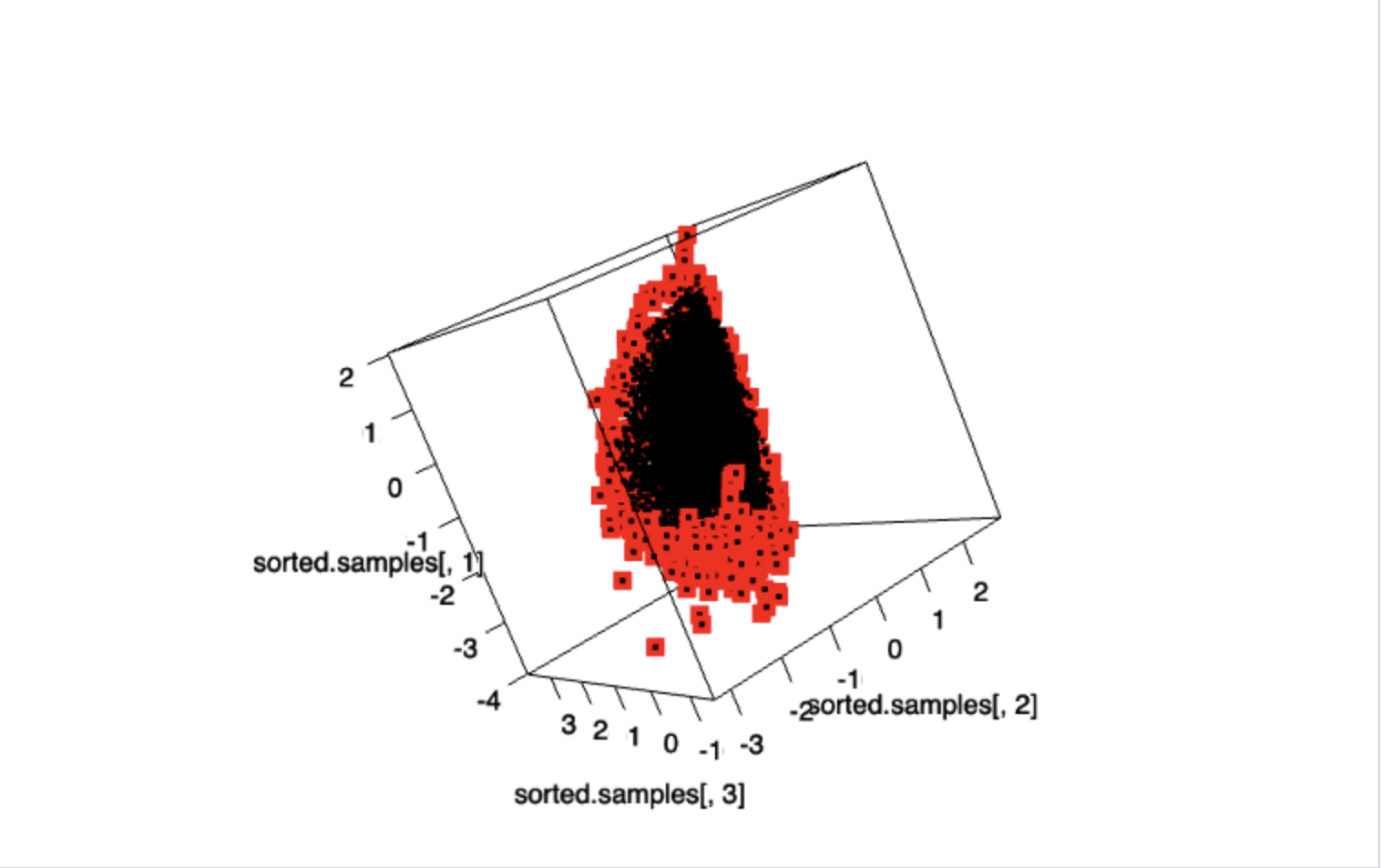}
\end{minipage}%
\begin{minipage}{.33\textwidth}
  \centering
  \includegraphics[width=.99\linewidth]{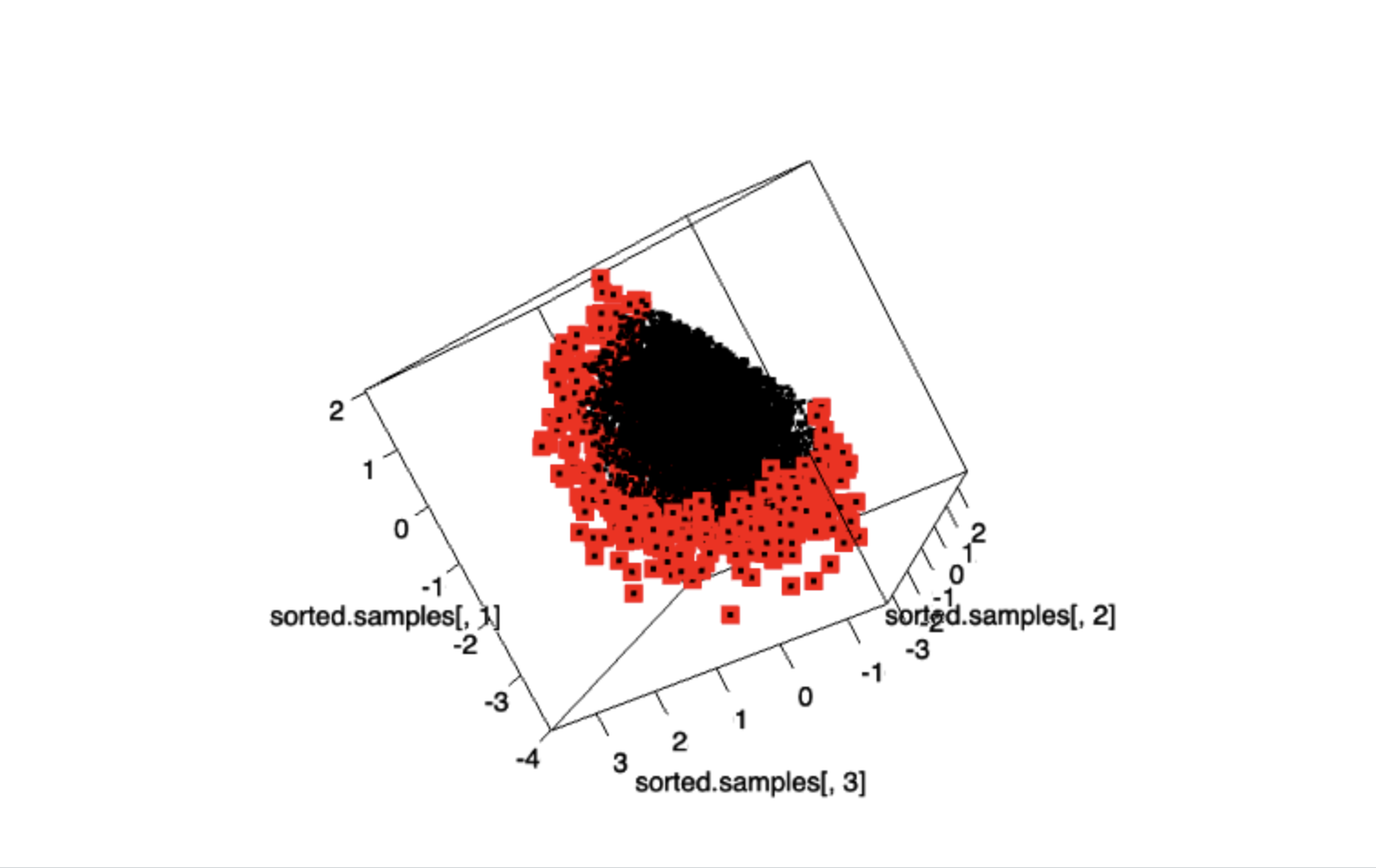}
\end{minipage}
\end{figure}

\newpage
\section{Nearest neighbors test based on joint distribution of normal order statistics; normal $H_0$}
\label{sec:hdr_normal_order_stats_test}

In section~\ref{section:normal:nn:uniform:via:normal} we introduced a nearest neighbor test based on normal order statistics. Its performance for testing uniformity was quite bad. Let us estimate this test's performance for checking normality. We start with ``normal vs normal'' case:

\begin{minted}[mathescape, linenos]{r}
res.normal.kNN <- calc.rejections.NN.tests(
    sample.generation.function.H0  = rnorm,
    sample.generation.functions.H1 = c(
        lapply(c(0.05, 0.1, 0.15), function(sd)   function(nsamples) rnorm(nsamples, 0,    sd)), 
        lapply(seq(0.2, 0.9, 0.1), function(sd)   function(nsamples) rnorm(nsamples, 0,    sd)),
        rnorm,
        lapply(seq(1.1, 2.4, 0.1), function(sd)   function(nsamples) rnorm(nsamples, 0,    sd)),
        lapply(seq(2.5,   4, 0.5), function(sd)   function(nsamples) rnorm(nsamples, 0,    sd)),
        lapply(seq(0.1,   2, 0.1), function(mean) function(nsamples) rnorm(nsamples, mean, 1 ))
    ),
    get.stat.matrix = get.sorted.samples.cpp,
    alpha = 0.95, h = 10, n = n, m = m
)
\end{minted}

\begin{table}[H]
\centering

\begin{minipage}[t]{.48\textwidth}
\vspace{0pt}

\centering

\begin{tabular}{ |l|p{1.6cm}||>{\columncolor[gray]{0.9}}p{1cm}|p{1.1cm}| } 
 \hline
  $H_0$ & $H_1$ & \multicolumn{2}{c|}{OS(N).kNN}  \\
  \hline
 &  & mean & sd  \\
 \hline
 \hline
 
 $\mathcal{N}(0, 1)$ & $\mathrm{N}(0,0.05^2)$ & 1 & 0.0e+00\\
$\mathcal{N}(0, 1)$ & $\mathrm{N}(0,0.1^2)$ & 0.984 & 4.1e-03\\
$\mathcal{N}(0, 1)$ & $\mathrm{N}(0,0.15^2)$ & 0.5853 & 2.9e-02\\
$\mathcal{N}(0, 1)$ & $\mathrm{N}(0,0.2^2)$ & 0.1907 & 1.6e-02\\
$\mathcal{N}(0, 1)$ & $\mathrm{N}(0,0.3^2)$ & 0.0171 & 1.6e-03\\
$\mathcal{N}(0, 1)$ & $\mathrm{N}(0,0.4^2)$ & 0.0024 & 2.3e-04\\
$\mathcal{N}(0, 1)$ & $\mathrm{N}(0,0.5^2)$ & 5e-04 & 9.0e-05\\
$\mathcal{N}(0, 1)$ & $\mathrm{N}(0,0.6^2)$ & 2e-04 & 5.3e-05\\
$\mathcal{N}(0, 1)$ & $\mathrm{N}(0,0.7^2)$ & 4e-04 & 9.0e-05\\
$\mathcal{N}(0, 1)$ & $\mathrm{N}(0,0.8^2)$ & 0.0027 & 1.8e-04\\
$\mathcal{N}(0, 1)$ & $\mathrm{N}(0,0.9^2)$ & 0.0144 & 4.5e-04\\
$\mathcal{N}(0, 1)$ & $\mathrm{N}(0,1^2)$ & 0.0514 & 7.0e-04\\
$\mathcal{N}(0, 1)$ & $\mathrm{N}(0,1.1^2)$ & 0.1228 & 1.0e-03\\
$\mathcal{N}(0, 1)$ & $\mathrm{N}(0,1.2^2)$ & 0.2274 & 2.0e-03\\
$\mathcal{N}(0, 1)$ & $\mathrm{N}(0,1.3^2)$ & 0.3515 & 2.7e-03\\
$\mathcal{N}(0, 1)$ & $\mathrm{N}(0,1.4^2)$ & 0.4756 & 1.6e-03\\
$\mathcal{N}(0, 1)$ & $\mathrm{N}(0,1.5^2)$ & 0.5911 & 2.9e-03\\
$\mathcal{N}(0, 1)$ & $\mathrm{N}(0,1.6^2)$ & 0.6873 & 2.0e-03\\
$\mathcal{N}(0, 1)$ & $\mathrm{N}(0,1.7^2)$ & 0.7658 & 2.1e-03\\
$\mathcal{N}(0, 1)$ & $\mathrm{N}(0,1.8^2)$ & 0.8264 & 1.4e-03\\
$\mathcal{N}(0, 1)$ & $\mathrm{N}(0,1.9^2)$ & 0.8725 & 1.1e-03\\
$\mathcal{N}(0, 1)$ & $\mathrm{N}(0,2^2)$ & 0.9061 & 1.3e-03\\
$\mathcal{N}(0, 1)$ & $\mathrm{N}(0,2.1^2)$ & 0.9315 & 7.8e-04\\
$\mathcal{N}(0, 1)$ & $\mathrm{N}(0,2.2^2)$ & 0.9501 & 6.9e-04\\
$\mathcal{N}(0, 1)$ & $\mathrm{N}(0,2.3^2)$ & 0.9629 & 7.3e-04\\
$\mathcal{N}(0, 1)$ & $\mathrm{N}(0,2.4^2)$ & 0.9726 & 5.3e-04\\
$\mathcal{N}(0, 1)$ & $\mathrm{N}(0,2.5^2)$ & 0.9799 & 5.5e-04\\
$\mathcal{N}(0, 1)$ & $\mathrm{N}(0,3^2)$ & 0.9953 & 2.2e-04\\
$\mathcal{N}(0, 1)$ & $\mathrm{N}(0,3.5^2)$ & 0.9986 & 1.5e-04\\
$\mathcal{N}(0, 1)$ & $\mathrm{N}(0,4^2)$ & 0.9996 & 7.9e-05\\
 
              \hline
\end{tabular}
\caption{$H_0$ --- standard normal, $H_1$ --- normals with different scale.}
\label{table:osnknn:normal:scale}

\end{minipage}
\hfill
\begin{minipage}[t]{.48\textwidth}
\vspace{120pt}

\centering
\begin{tabular}{ |l|p{1.6cm}||>{\columncolor[gray]{0.9}}p{1cm}|p{1.1cm}| } 
 \hline
  $H_0$ & $H_1$ & \multicolumn{2}{c|}{OS(N).kNN}  \\
  \hline
 &  & mean & sd  \\
 \hline
 \hline
 
 $\mathcal{N}(0, 1)$ & $\mathrm{N}(0.1, 1)$ & 0.0547 & 5.4e-04\\
$\mathcal{N}(0, 1)$ & $\mathrm{N}(0.2, 1)$ & 0.0684 & 8.7e-04\\
$\mathcal{N}(0, 1)$ & $\mathrm{N}(0.3, 1)$ & 0.093 & 2.0e-03\\
$\mathcal{N}(0, 1)$ & $\mathrm{N}(0.4, 1)$ & 0.1316 & 2.1e-03\\
$\mathcal{N}(0, 1)$ & $\mathrm{N}(0.5, 1)$ & 0.1868 & 3.2e-03\\
$\mathcal{N}(0, 1)$ & $\mathrm{N}(0.6, 1)$ & 0.2606 & 3.3e-03\\
$\mathcal{N}(0, 1)$ & $\mathrm{N}(0.7, 1)$ & 0.3529 & 4.7e-03\\
$\mathcal{N}(0, 1)$ & $\mathrm{N}(0.8, 1)$ & 0.4585 & 4.4e-03\\
$\mathcal{N}(0, 1)$ & $\mathrm{N}(0.9, 1)$ & 0.5675 & 5.3e-03\\
$\mathcal{N}(0, 1)$ & $\mathrm{N}(1, 1)$ & 0.6761 & 5.0e-03\\
$\mathcal{N}(0, 1)$ & $\mathrm{N}(1.1, 1)$ & 0.7715 & 4.3e-03\\
$\mathcal{N}(0, 1)$ & $\mathrm{N}(1.2, 1)$ & 0.8502 & 3.6e-03\\
$\mathcal{N}(0, 1)$ & $\mathrm{N}(1.3, 1)$ & 0.9089 & 2.5e-03\\
$\mathcal{N}(0, 1)$ & $\mathrm{N}(1.4, 1)$ & 0.9487 & 2.3e-03\\
$\mathcal{N}(0, 1)$ & $\mathrm{N}(1.5, 1)$ & 0.973 & 1.4e-03\\
$\mathcal{N}(0, 1)$ & $\mathrm{N}(1.6, 1)$ & 0.987 & 8.3e-04\\
$\mathcal{N}(0, 1)$ & $\mathrm{N}(1.7, 1)$ & 0.9943 & 4.2e-04\\
$\mathcal{N}(0, 1)$ & $\mathrm{N}(1.8, 1)$ & 0.9978 & 1.3e-04\\
$\mathcal{N}(0, 1)$ & $\mathrm{N}(1.9, 1)$ & 0.9991 & 1.1e-04\\
$\mathcal{N}(0, 1)$ & $\mathrm{N}(2, 1)$ & 0.9997 & 3.6e-05\\
 
 \hline
  \end{tabular}
\caption{$H_0$ --- standard normal, $H_1$ --- normals with different location.}
\label{table:osnknn:normal:location}

\end{minipage}

\end{table}

\noindent
In table~\ref{table:osnknn:normal:scale} we see that for $N(0, \sigma^2)$ alternatives ``OS(N).kNN'' test has excellent performance for $\sigma > 1$ and moderate performance for $\sigma < 1$ (still test reaches power of 1 for very small $\sigma$'s). In table~\ref{table:osnknn:normal:location} we see that for $N(\mu, 1)$ alternatives ``OS(N).kNN'' test performance has average performance.
 
 Now to Cauchy, gamma, and Student alternatives:
 
\begin{minted}[mathescape, linenos]{r}
res.normal.vs.Cauchy.gamma.t.kNN <- calc.rejections.NN.tests(
    sample.generation.function.H0  = rnorm,
    sample.generation.functions.H1 = c(
        # Cauchy alternatives
        lapply(seq(0.01, 0.09, 0.01), function(scale) function(nsamples) rcauchy(nsamples, 0, scale)),
        lapply(seq(0.1,  2,    0.1 ), function(scale) function(nsamples) rcauchy(nsamples, 0, scale)),
        # gamma alternatives
        lapply(
            c(seq(0.1, 0.9, 0.1), 1:5, 10, 100, 1000), 
            function(shape) 
                function(nsamples) 
                    (rgamma(nsamples, shape = shape, rate = 1) - shape) / sqrt(shape)
        ),
        # t alternatives
        lapply(seq(1,5), function(df) function(nsamples) rt(nsamples, df = df))
    ),
    get.stat.matrix = get.sorted.samples.cpp,
    alpha = 0.95, h = 10, n = n, m = m
)
\end{minted}

\begin{table}[H]
\centering

\begin{minipage}[t]{.48\textwidth}
\vspace{0pt}

\centering

\begin{tabular}{ |l|p{2.3cm}||>{\columncolor[gray]{0.9}}p{1cm}|p{1.1cm}| } 
 \hline
  $H_0$ & $H_1$ & \multicolumn{2}{c|}{OS(N).kNN}  \\
  \hline
 &  & mean & sd  \\
 \hline
 \hline
 
 $\mathcal{N}(0, 1)$ & $\mathrm{Cauchy}(0, 0.01)$ & 0.9892 & 9.6e-04\\
$\mathcal{N}(0, 1)$ & $\mathrm{Cauchy}(0, 0.02)$ & 0.9431 & 4.5e-03\\
$\mathcal{N}(0, 1)$ & $\mathrm{Cauchy}(0, 0.03)$ & 0.8644 & 9.3e-03\\
$\mathcal{N}(0, 1)$ & $\mathrm{Cauchy}(0, 0.04)$ & 0.7708 & 1.3e-02\\
$\mathcal{N}(0, 1)$ & $\mathrm{Cauchy}(0, 0.05)$ & 0.6775 & 1.5e-02\\
$\mathcal{N}(0, 1)$ & $\mathrm{Cauchy}(0, 0.06)$ & 0.592 & 1.7e-02\\
$\mathcal{N}(0, 1)$ & $\mathrm{Cauchy}(0, 0.07)$ & 0.521 & 1.6e-02\\
$\mathcal{N}(0, 1)$ & $\mathrm{Cauchy}(0, 0.08)$ & 0.4658 & 1.5e-02\\
$\mathcal{N}(0, 1)$ & $\mathrm{Cauchy}(0, 0.09)$ & 0.4226 & 1.3e-02\\
$\mathcal{N}(0, 1)$ & $\mathrm{Cauchy}(0, 0.1)$ & 0.3917 & 1.1e-02\\
$\mathcal{N}(0, 1)$ & $\mathrm{Cauchy}(0, 0.2)$ & 0.3851 & 1.9e-03\\
$\mathcal{N}(0, 1)$ & $\mathrm{Cauchy}(0, 0.3)$ & 0.4887 & 1.6e-03\\
$\mathcal{N}(0, 1)$ & $\mathrm{Cauchy}(0, 0.4)$ & 0.5865 & 1.1e-03\\
$\mathcal{N}(0, 1)$ & $\mathrm{Cauchy}(0, 0.5)$ & 0.6719 & 2.2e-03\\
$\mathcal{N}(0, 1)$ & $\mathrm{Cauchy}(0, 0.6)$ & 0.7429 & 1.4e-03\\
$\mathcal{N}(0, 1)$ & $\mathrm{Cauchy}(0, 0.7)$ & 0.8014 & 1.4e-03\\
$\mathcal{N}(0, 1)$ & $\mathrm{Cauchy}(0, 0.8)$ & 0.8487 & 8.0e-04\\
$\mathcal{N}(0, 1)$ & $\mathrm{Cauchy}(0, 0.9)$ & 0.8859 & 1.2e-03\\
$\mathcal{N}(0, 1)$ & $\mathrm{Cauchy}(0, 1)$ & 0.9149 & 9.2e-04\\
$\mathcal{N}(0, 1)$ & $\mathrm{Cauchy}(0, 1.1)$ & 0.9364 & 9.4e-04\\
$\mathcal{N}(0, 1)$ & $\mathrm{Cauchy}(0, 1.2)$ & 0.9531 & 6.6e-04\\
$\mathcal{N}(0, 1)$ & $\mathrm{Cauchy}(0, 1.3)$ & 0.9656 & 4.8e-04\\
$\mathcal{N}(0, 1)$ & $\mathrm{Cauchy}(0, 1.4)$ & 0.9743 & 7.0e-04\\
$\mathcal{N}(0, 1)$ & $\mathrm{Cauchy}(0, 1.5)$ & 0.9811 & 5.8e-04\\
$\mathcal{N}(0, 1)$ & $\mathrm{Cauchy}(0, 1.6)$ & 0.9863 & 4.9e-04\\
$\mathcal{N}(0, 1)$ & $\mathrm{Cauchy}(0, 1.7)$ & 0.9897 & 2.3e-04\\
$\mathcal{N}(0, 1)$ & $\mathrm{Cauchy}(0, 1.8)$ & 0.9924 & 2.3e-04\\
$\mathcal{N}(0, 1)$ & $\mathrm{Cauchy}(0, 1.9)$ & 0.9944 & 3.4e-04\\
$\mathcal{N}(0, 1)$ & $\mathrm{Cauchy}(0, 2)$ & 0.9957 & 2.0e-04\\
 
                \hline
\end{tabular}

\caption{Cauchy alternatives.}
\label{table:lab03}

\end{minipage}
\hfill
\begin{minipage}[t]{.48\textwidth}
\vspace{0pt}
\centering

\begin{tabular}{ |l|p{1.2cm}||>{\columncolor[gray]{0.9}}p{1cm}|p{1.1cm}| } 
 \hline
  $H_0$ & $H_1$ & \multicolumn{2}{c|}{OS(N).kNN}  \\
  \hline
 &  & mean & sd  \\
 \hline
 \hline
 
 $\mathcal{N}(0, 1)$ & $\mathrm{G}(0.1)$ & 0.7709 & 6.5e-03\\
$\mathcal{N}(0, 1)$ & $\mathrm{G}(0.2)$ & 0.5441 & 6.2e-03\\
$\mathcal{N}(0, 1)$ & $\mathrm{G}(0.3)$ & 0.4159 & 5.2e-03\\
$\mathcal{N}(0, 1)$ & $\mathrm{G}(0.4)$ & 0.3424 & 3.7e-03\\
$\mathcal{N}(0, 1)$ & $\mathrm{G}(0.5)$ & 0.2947 & 2.8e-03\\
$\mathcal{N}(0, 1)$ & $\mathrm{G}(0.6)$ & 0.2636 & 2.0e-03\\
$\mathcal{N}(0, 1)$ & $\mathrm{G}(0.7)$ & 0.2405 & 1.6e-03\\
$\mathcal{N}(0, 1)$ & $\mathrm{G}(0.8)$ & 0.2232 & 1.8e-03\\
$\mathcal{N}(0, 1)$ & $\mathrm{G}(0.9)$ & 0.21 & 2.3e-03\\
$\mathcal{N}(0, 1)$ & $\mathrm{G}(1)$ & 0.1979 & 1.1e-03\\
$\mathcal{N}(0, 1)$ & $\mathrm{G}(2)$ & 0.1417 & 1.5e-03\\
$\mathcal{N}(0, 1)$ & $\mathrm{G}(3)$ & 0.1193 & 1.1e-03\\
$\mathcal{N}(0, 1)$ & $\mathrm{G}(4)$ & 0.106 & 1.1e-03\\
$\mathcal{N}(0, 1)$ & $\mathrm{G}(5)$ & 0.0977 & 1.3e-03\\
$\mathcal{N}(0, 1)$ & $\mathrm{G}(10)$ & 0.078 & 1.5e-03\\
$\mathcal{N}(0, 1)$ & $\mathrm{G}(100)$ & 0.0542 & 9.0e-04\\
$\mathcal{N}(0, 1)$ & $\mathrm{G}(1000)$ & 0.0515 & 6.8e-04\\
 
                 \hline
  \end{tabular}
\caption{Gamma alternatives.}
\label{table:lab04}

\vspace{32pt}

\begin{tabular}{ |l|p{1.2cm}||>{\columncolor[gray]{0.9}}p{1cm}|p{1.1cm}| } 
 \hline
  $H_0$ & $H_1$ & \multicolumn{2}{c|}{OS(N).kNN}  \\
  \hline
 &  & mean & sd  \\
 \hline
 \hline
 
$\mathcal{N}(0, 1)$ & $\mathrm{t}(1)$ & 0.9142 & 9.1e-04\\
$\mathcal{N}(0, 1)$ & $\mathrm{t}(2)$ & 0.6746 & 2.1e-03\\
$\mathcal{N}(0, 1)$ & $\mathrm{t}(3)$ & 0.4967 & 1.6e-03\\
$\mathcal{N}(0, 1)$ & $\mathrm{t}(4)$ & 0.3832 & 2.2e-03\\
$\mathcal{N}(0, 1)$ & $\mathrm{t}(5)$ & 0.31 & 2.5e-03\\
 
                 \hline
\end{tabular}
\caption{Student alternatives.}
\label{table:lab05}

\end{minipage}

\end{table}

\noindent
All in all, we see that this test is best at nothing.

\newpage
\section{Nearest neighbors test based on joint distribution of uniform order statistics; normal $H_0$ (normal via uniform)}

Finally, to conclude the comparison between ``OS(U).kNN'' and ``OS(N).kNN'' tests, let us research how ``OS(U).kNN'' test handles normality checks. For normal alternatives we have:

\begin{minted}[mathescape, linenos]{r}
res.normal.via.uniform.kNN <- calc.rejections.NN.tests(
    sample.generation.function.H0  = runif,
    sample.generation.functions.H1 = c(
        lapply(c(0.05, 0.1, 0.15), function(sd)   function(nsamples) pnorm(rnorm(nsamples, 0,    sd), 0, 1)), 
        lapply(seq(0.2, 0.9, 0.1), function(sd)   function(nsamples) pnorm(rnorm(nsamples, 0,    sd), 0, 1)),
                                                  function(nsamples) pnorm(rnorm(nsamples          )       ),
        lapply(seq(1.1, 2.4, 0.1), function(sd)   function(nsamples) pnorm(rnorm(nsamples, 0,    sd), 0, 1)),
        lapply(seq(2.5,   4, 0.5), function(sd)   function(nsamples) pnorm(rnorm(nsamples, 0,    sd), 0, 1)),
        lapply(seq(0.1,   2, 0.1), function(mean) function(nsamples) pnorm(rnorm(nsamples, mean, 1 ), 0, 1))
    ),
    get.stat.matrix = get.sorted.samples.cpp,
    alpha = 0.95, h = 10, n = n, m = m
)
\end{minted}

\begin{table}[H]
\centering

\begin{minipage}[t]{.48\textwidth}

\vspace{0pt}

\centering

\begin{tabular}{ |l|p{1.6cm}||>{\columncolor[gray]{0.9}}p{1cm}|p{1.1cm}| } 
 \hline
  $H_0$ & $H_1$ & \multicolumn{2}{c|}{OS(U).kNN}  \\
  \hline
 &  & mean & sd  \\
 \hline
 \hline
 
 $\mathcal{N}(0, 1)$ & $\mathrm{N}(0,0.05^2)$ & 1 & 0.0e+00\\
$\mathcal{N}(0, 1)$ & $\mathrm{N}(0,0.1^2)$ & 1 & 0.0e+00\\
$\mathcal{N}(0, 1)$ & $\mathrm{N}(0,0.15^2)$ & 1 & 0.0e+00\\
$\mathcal{N}(0, 1)$ & $\mathrm{N}(0,0.2^2)$ & 0.9992 & 2.7e-04\\
$\mathcal{N}(0, 1)$ & $\mathrm{N}(0,0.3^2)$ & 0.8672 & 1.3e-02\\
$\mathcal{N}(0, 1)$ & $\mathrm{N}(0,0.4^2)$ & 0.4977 & 1.6e-02\\
$\mathcal{N}(0, 1)$ & $\mathrm{N}(0,0.5^2)$ & 0.2358 & 7.4e-03\\
$\mathcal{N}(0, 1)$ & $\mathrm{N}(0,0.6^2)$ & 0.1125 & 3.5e-03\\
$\mathcal{N}(0, 1)$ & $\mathrm{N}(0,0.7^2)$ & 0.0606 & 1.9e-03\\
$\mathcal{N}(0, 1)$ & $\mathrm{N}(0,0.8^2)$ & 0.0404 & 9.7e-04\\
$\mathcal{N}(0, 1)$ & $\mathrm{N}(0,0.9^2)$ & 0.0387 & 9.9e-04\\
$\mathcal{N}(0, 1)$ & $\mathrm{N}(0,1^2)$ & 0.0519 & 1.3e-03\\
$\mathcal{N}(0, 1)$ & $\mathrm{N}(0,1.1^2)$ & 0.081 & 1.8e-03\\
$\mathcal{N}(0, 1)$ & $\mathrm{N}(0,1.2^2)$ & 0.1281 & 1.9e-03\\
$\mathcal{N}(0, 1)$ & $\mathrm{N}(0,1.3^2)$ & 0.1911 & 2.3e-03\\
$\mathcal{N}(0, 1)$ & $\mathrm{N}(0,1.4^2)$ & 0.2631 & 3.1e-03\\
$\mathcal{N}(0, 1)$ & $\mathrm{N}(0,1.5^2)$ & 0.3398 & 2.6e-03\\
$\mathcal{N}(0, 1)$ & $\mathrm{N}(0,1.6^2)$ & 0.4187 & 2.7e-03\\
$\mathcal{N}(0, 1)$ & $\mathrm{N}(0,1.7^2)$ & 0.4938 & 2.9e-03\\
$\mathcal{N}(0, 1)$ & $\mathrm{N}(0,1.8^2)$ & 0.5636 & 3.3e-03\\
$\mathcal{N}(0, 1)$ & $\mathrm{N}(0,1.9^2)$ & 0.6262 & 3.5e-03\\
$\mathcal{N}(0, 1)$ & $\mathrm{N}(0,2^2)$ & 0.6815 & 2.9e-03\\
$\mathcal{N}(0, 1)$ & $\mathrm{N}(0,2.1^2)$ & 0.7313 & 3.3e-03\\
$\mathcal{N}(0, 1)$ & $\mathrm{N}(0,2.2^2)$ & 0.7719 & 2.6e-03\\
$\mathcal{N}(0, 1)$ & $\mathrm{N}(0,2.3^2)$ & 0.8077 & 2.9e-03\\
$\mathcal{N}(0, 1)$ & $\mathrm{N}(0,2.4^2)$ & 0.8379 & 2.2e-03\\
$\mathcal{N}(0, 1)$ & $\mathrm{N}(0,2.5^2)$ & 0.8629 & 2.1e-03\\
$\mathcal{N}(0, 1)$ & $\mathrm{N}(0,3^2)$ & 0.9396 & 1.2e-03\\
$\mathcal{N}(0, 1)$ & $\mathrm{N}(0,3.5^2)$ & 0.9721 & 6.0e-04\\
$\mathcal{N}(0, 1)$ & $\mathrm{N}(0,4^2)$ & 0.9862 & 5.5e-04\\
 
                  \hline
\end{tabular}
\caption{$N(0, \sigma^2)$, $\sigma > 0$ alternatives.}
\label{table:osuknn:normal:scale}

\end{minipage}
\hfill
\begin{minipage}[t]{.48\textwidth}
\vspace{120pt}

\centering

\begin{tabular}{ |l|p{1.6cm}||>{\columncolor[gray]{0.9}}p{1cm}|p{1.1cm}| } 
 \hline
  $H_0$ & $H_1$ & \multicolumn{2}{c|}{OS(U).kNN}  \\
  \hline
 &  & mean & sd  \\
 \hline
 \hline

$\mathcal{N}(0, 1)$ & $\mathrm{N}(0.1, 1)$ & 0.0579 & 5.5e-04\\
$\mathcal{N}(0, 1)$ & $\mathrm{N}(0.2, 1)$ & 0.0763 & 1.3e-03\\
$\mathcal{N}(0, 1)$ & $\mathrm{N}(0.3, 1)$ & 0.1093 & 1.8e-03\\
$\mathcal{N}(0, 1)$ & $\mathrm{N}(0.4, 1)$ & 0.1598 & 1.8e-03\\
$\mathcal{N}(0, 1)$ & $\mathrm{N}(0.5, 1)$ & 0.2293 & 2.3e-03\\
$\mathcal{N}(0, 1)$ & $\mathrm{N}(0.6, 1)$ & 0.317 & 3.6e-03\\
$\mathcal{N}(0, 1)$ & $\mathrm{N}(0.7, 1)$ & 0.4194 & 4.5e-03\\
$\mathcal{N}(0, 1)$ & $\mathrm{N}(0.8, 1)$ & 0.5302 & 5.3e-03\\
$\mathcal{N}(0, 1)$ & $\mathrm{N}(0.9, 1)$ & 0.6401 & 5.4e-03\\
$\mathcal{N}(0, 1)$ & $\mathrm{N}(1, 1)$ & 0.7394 & 4.9e-03\\
$\mathcal{N}(0, 1)$ & $\mathrm{N}(1.1, 1)$ & 0.8236 & 3.9e-03\\
$\mathcal{N}(0, 1)$ & $\mathrm{N}(1.2, 1)$ & 0.8898 & 3.2e-03\\
$\mathcal{N}(0, 1)$ & $\mathrm{N}(1.3, 1)$ & 0.9353 & 2.1e-03\\
$\mathcal{N}(0, 1)$ & $\mathrm{N}(1.4, 1)$ & 0.9646 & 1.6e-03\\
$\mathcal{N}(0, 1)$ & $\mathrm{N}(1.5, 1)$ & 0.9819 & 9.8e-04\\
$\mathcal{N}(0, 1)$ & $\mathrm{N}(1.6, 1)$ & 0.9917 & 4.5e-04\\
$\mathcal{N}(0, 1)$ & $\mathrm{N}(1.7, 1)$ & 0.9964 & 2.0e-04\\
$\mathcal{N}(0, 1)$ & $\mathrm{N}(1.8, 1)$ & 0.9986 & 1.3e-04\\
$\mathcal{N}(0, 1)$ & $\mathrm{N}(1.9, 1)$ & 0.9995 & 5.1e-05\\
$\mathcal{N}(0, 1)$ & $\mathrm{N}(2, 1)$ & 0.9998 & 4.9e-05\\

                  \hline
\end{tabular}
\caption{$N(\mu, 1)$ alternatives.}
\label{table:osuknn:normal:location}

\end{minipage}

\end{table}

\noindent
In table~\ref{table:osuknn:normal:scale} we see that for $N(0,\sigma^2)$ alternatives for $\sigma < 1$ ``OS(U).kNN'' test performs significantly better than ``OS(N).kNN'' test, and for $\sigma > 1$ ``OS(U).kNN'' test is worse than ``OS(N).kNN'' test. In table~\ref{table:osuknn:normal:location} we see that for $N(\mu, 1)$ alternatives ``OS(U).kNN'' is again better than ``OS(N).kNN''.

Now to Cauchy, gamma, and Student alternatives:

\begin{minted}[mathescape, linenos]{r}
res.normal.via.uniform.vs.Cauchy.kNN <- calc.rejections.NN.tests(
    sample.generation.function.H0  = runif,
    sample.generation.functions.H1 = c(
        # Cauchy
        lapply(seq(0.01, 0.09, 0.01), function(scale) function(nsamples) pnorm(rcauchy(nsamples, 0, scale))),
        lapply(seq(0.1,  2,    0.1 ), function(scale) function(nsamples) pnorm(rcauchy(nsamples, 0, scale))),
        # gamma
        lapply(
            c(seq(0.1, 0.9, 0.1), 1:5, 10, 100, 1000), 
            function(shape) 
                function(nsamples) pnorm( (rgamma(nsamples, shape = shape, rate = 1) - shape) / sqrt(shape) )
        ),
        # Student
        lapply(seq(1,5), function(df) function(nsamples) pnorm(rt(nsamples, df = df)))
    ),
    get.stat.matrix = get.sorted.samples.cpp,
    alpha = 0.95, h = 10, n = n, m = m
)
\end{minted}

\begin{table}[H]
\centering

\begin{minipage}[t]{.48\textwidth}
\vspace{0pt}

\centering

\begin{tabular}{ |l|p{2.3cm}||>{\columncolor[gray]{0.9}}p{1cm}|p{1.1cm}| } 
 \hline
  $H_0$ & $H_1$ & \multicolumn{2}{c|}{OS(U).kNN}  \\
  \hline
 &  & mean & sd  \\
 \hline
 \hline
 
 $\mathcal{N}(0, 1)$ & $\mathrm{Cauchy}(0, 0.01)$ & 1 & 4.2e-06\\
$\mathcal{N}(0, 1)$ & $\mathrm{Cauchy}(0, 0.02)$ & 1 & 1.3e-05\\
$\mathcal{N}(0, 1)$ & $\mathrm{Cauchy}(0, 0.03)$ & 0.9997 & 2.7e-05\\
$\mathcal{N}(0, 1)$ & $\mathrm{Cauchy}(0, 0.04)$ & 0.999 & 1.1e-04\\
$\mathcal{N}(0, 1)$ & $\mathrm{Cauchy}(0, 0.05)$ & 0.9968 & 1.7e-04\\
$\mathcal{N}(0, 1)$ & $\mathrm{Cauchy}(0, 0.06)$ & 0.9923 & 3.1e-04\\
$\mathcal{N}(0, 1)$ & $\mathrm{Cauchy}(0, 0.07)$ & 0.9847 & 4.2e-04\\
$\mathcal{N}(0, 1)$ & $\mathrm{Cauchy}(0, 0.08)$ & 0.9736 & 8.3e-04\\
$\mathcal{N}(0, 1)$ & $\mathrm{Cauchy}(0, 0.09)$ & 0.9585 & 7.2e-04\\
$\mathcal{N}(0, 1)$ & $\mathrm{Cauchy}(0, 0.1)$ & 0.939 & 1.0e-03\\
$\mathcal{N}(0, 1)$ & $\mathrm{Cauchy}(0, 0.2)$ & 0.6112 & 3.5e-03\\
$\mathcal{N}(0, 1)$ & $\mathrm{Cauchy}(0, 0.3)$ & 0.3483 & 2.4e-03\\
$\mathcal{N}(0, 1)$ & $\mathrm{Cauchy}(0, 0.4)$ & 0.2372 & 2.3e-03\\
$\mathcal{N}(0, 1)$ & $\mathrm{Cauchy}(0, 0.5)$ & 0.2173 & 2.1e-03\\
$\mathcal{N}(0, 1)$ & $\mathrm{Cauchy}(0, 0.6)$ & 0.2459 & 2.5e-03\\
$\mathcal{N}(0, 1)$ & $\mathrm{Cauchy}(0, 0.7)$ & 0.2992 & 1.8e-03\\
$\mathcal{N}(0, 1)$ & $\mathrm{Cauchy}(0, 0.8)$ & 0.3652 & 2.7e-03\\
$\mathcal{N}(0, 1)$ & $\mathrm{Cauchy}(0, 0.9)$ & 0.4389 & 2.4e-03\\
$\mathcal{N}(0, 1)$ & $\mathrm{Cauchy}(0, 1)$ & 0.5117 & 2.7e-03\\
$\mathcal{N}(0, 1)$ & $\mathrm{Cauchy}(0, 1.1)$ & 0.5802 & 2.8e-03\\
$\mathcal{N}(0, 1)$ & $\mathrm{Cauchy}(0, 1.2)$ & 0.6437 & 2.3e-03\\
$\mathcal{N}(0, 1)$ & $\mathrm{Cauchy}(0, 1.3)$ & 0.7006 & 1.7e-03\\
$\mathcal{N}(0, 1)$ & $\mathrm{Cauchy}(0, 1.4)$ & 0.7494 & 2.6e-03\\
$\mathcal{N}(0, 1)$ & $\mathrm{Cauchy}(0, 1.5)$ & 0.7906 & 2.1e-03\\
$\mathcal{N}(0, 1)$ & $\mathrm{Cauchy}(0, 1.6)$ & 0.8258 & 1.9e-03\\
$\mathcal{N}(0, 1)$ & $\mathrm{Cauchy}(0, 1.7)$ & 0.8564 & 1.8e-03\\
$\mathcal{N}(0, 1)$ & $\mathrm{Cauchy}(0, 1.8)$ & 0.881 & 1.1e-03\\
$\mathcal{N}(0, 1)$ & $\mathrm{Cauchy}(0, 1.9)$ & 0.9013 & 1.2e-03\\
$\mathcal{N}(0, 1)$ & $\mathrm{Cauchy}(0, 2)$ & 0.9175 & 1.1e-03\\
 
                  \hline
\end{tabular}
\caption{Cauchy alternatives.}

\end{minipage}
\hfill
\begin{minipage}[t]{.48\textwidth}
\vspace{0pt}
\centering

\begin{tabular}{ |l|p{1.2cm}||>{\columncolor[gray]{0.9}}p{1cm}|p{1.1cm}| } 
 \hline
  $H_0$ & $H_1$ & \multicolumn{2}{c|}{OS(U).kNN}  \\
  \hline
 &  & mean & sd  \\
 \hline
 \hline
 
 $\mathcal{N}(0, 1)$ & $\mathrm{G}(0.1)$ & 0.9974 & 3.0e-04\\
$\mathcal{N}(0, 1)$ & $\mathrm{G}(0.2)$ & 0.9574 & 1.7e-03\\
$\mathcal{N}(0, 1)$ & $\mathrm{G}(0.3)$ & 0.8552 & 1.8e-03\\
$\mathcal{N}(0, 1)$ & $\mathrm{G}(0.4)$ & 0.722 & 5.3e-03\\
$\mathcal{N}(0, 1)$ & $\mathrm{G}(0.5)$ & 0.5938 & 5.8e-03\\
$\mathcal{N}(0, 1)$ & $\mathrm{G}(0.6)$ & 0.4833 & 6.3e-03\\
$\mathcal{N}(0, 1)$ & $\mathrm{G}(0.7)$ & 0.3981 & 6.2e-03\\
$\mathcal{N}(0, 1)$ & $\mathrm{G}(0.8)$ & 0.331 & 6.8e-03\\
$\mathcal{N}(0, 1)$ & $\mathrm{G}(0.9)$ & 0.2813 & 6.2e-03\\
$\mathcal{N}(0, 1)$ & $\mathrm{G}(1)$ & 0.2421 & 5.6e-03\\
$\mathcal{N}(0, 1)$ & $\mathrm{G}(2)$ & 0.1065 & 2.5e-03\\
$\mathcal{N}(0, 1)$ & $\mathrm{G}(3)$ & 0.0799 & 1.8e-03\\
$\mathcal{N}(0, 1)$ & $\mathrm{G}(4)$ & 0.0705 & 1.9e-03\\
$\mathcal{N}(0, 1)$ & $\mathrm{G}(5)$ & 0.0653 & 1.7e-03\\
$\mathcal{N}(0, 1)$ & $\mathrm{G}(10)$ & 0.0569 & 1.4e-03\\
$\mathcal{N}(0, 1)$ & $\mathrm{G}(100)$ & 0.0521 & 1.2e-03\\
$\mathcal{N}(0, 1)$ & $\mathrm{G}(1000)$ & 0.0518 & 7.6e-04\\
 
                   \hline
\end{tabular}
\caption{Gamma alternatives.}

\vspace{32pt}

\begin{tabular}{ |l|p{1.2cm}||>{\columncolor[gray]{0.9}}p{1cm}|p{1.1cm}| } 
 \hline
  $H_0$ & $H_1$ & \multicolumn{2}{c|}{OS(U).kNN}  \\
  \hline
 &  & mean & sd  \\
 \hline
 \hline

$\mathcal{N}(0, 1)$ & $\mathrm{t}(1)$ & 0.5099 & 2.1e-03\\
$\mathcal{N}(0, 1)$ & $\mathrm{t}(2)$ & 0.2502 & 2.6e-03\\
$\mathcal{N}(0, 1)$ & $\mathrm{t}(3)$ & 0.1637 & 2.0e-03\\
$\mathcal{N}(0, 1)$ & $\mathrm{t}(4)$ & 0.1273 & 1.3e-03\\
$\mathcal{N}(0, 1)$ & $\mathrm{t}(5)$ & 0.1076 & 1.0e-03\\

                   \hline
\end{tabular}
\caption{Student alternatives.}

\end{minipage}

\end{table}

\noindent
For Cauchy alternatives ``OS(U).kNN'' has average performance, for gamma alternatives --- good performance, and for Student alternatives --- bad performance.

\newpage

\section{Goodness of fit tests based on joint distribution of four empirical moments}
\label{section:4moments}

Let us try to construct goodness of fit tests based on joint distribution of sample mean, sample variance, sample skewness, and sample kurtosis of standard normal samples of size $n = 10$. We simulate $m$ samples of size $n$, compute sample moments for each sample, and then determine a highest density region for joint distribution of sample moments:

\begin{minted}[mathescape, linenos]{r}
res.normal.4moments.kNN <- calc.rejections.NN.tests(
    sample.generation.function.H0  = rnorm,
    sample.generation.functions.H1 = c(
        # normal, scale
        lapply(c(0.05, 0.1, 0.15), function(sd)   function(nsamples) rnorm(nsamples, 0,    sd)), 
        lapply(seq(0.2, 0.9, 0.1), function(sd)   function(nsamples) rnorm(nsamples, 0,    sd)),
        rnorm,
        lapply(seq(1.1, 2.4, 0.1), function(sd)   function(nsamples) rnorm(nsamples, 0,    sd)),
        lapply(seq(2.5,   4, 0.5), function(sd)   function(nsamples) rnorm(nsamples, 0,    sd)),
        # normal, location
        lapply(seq(0.1,   2, 0.1), function(mean) function(nsamples) rnorm(nsamples, mean, 1 )),
        # Cauchy
        lapply(seq(0.01, 0.09, 0.01), function(scale) function(nsamples) rcauchy(nsamples, 0, scale)),
        lapply(seq(0.1,  2,    0.1 ), function(scale) function(nsamples) rcauchy(nsamples, 0, scale)),
        # gamma
        lapply(
            c(seq(0.1, 0.9, 0.1), 1:5, 10, 100, 1000), 
            function(shape) 
                function(nsamples) (rgamma(nsamples, shape = shape, rate = 1) - shape) / sqrt(shape)
        ),
        # Student
        lapply(seq(1,5), function(df) function(nsamples) rt(nsamples, df = df))
    ),
    get.stat.matrix = function(sample.generation.function, n, m){
        samples <- sample.generation.function(n * m)
        dim(samples) <- c(m, n)
        
        statistics <- cbind(
            apply(samples, 1, mean),
            apply(samples, 1, var),
            apply(samples, 1, e1071::skewness),
            apply(samples, 1, e1071::kurtosis)
        )
    },
    alpha = 0.95,
    h = 10,
    n = n, m = m
)
\end{minted}

\begin{table}[H]
\centering

\begin{minipage}[t]{.48\textwidth}
\vspace{0pt}

\centering

\begin{tabular}{ |l|p{1.6cm}||>{\columncolor[gray]{0.9}}p{1cm}|p{1.1cm}| } 
 \hline
  $H_0$ & $H_1$ & \multicolumn{2}{c|}{4M.kNN}  \\
  \hline
 &  & mean & sd  \\
 \hline
 \hline
 
 $\mathcal{N}(0, 1)$ & $\mathrm{N}(0,0.05^2)$ & 0.2998 & 1.5e-02\\
$\mathcal{N}(0, 1)$ & $\mathrm{N}(0,0.1^2)$ & 0.2786 & 1.4e-02\\
$\mathcal{N}(0, 1)$ & $\mathrm{N}(0,0.15^2)$ & 0.2476 & 1.2e-02\\
$\mathcal{N}(0, 1)$ & $\mathrm{N}(0,0.2^2)$ & 0.2126 & 8.7e-03\\
$\mathcal{N}(0, 1)$ & $\mathrm{N}(0,0.3^2)$ & 0.1459 & 4.2e-03\\
$\mathcal{N}(0, 1)$ & $\mathrm{N}(0,0.4^2)$ & 0.0951 & 3.0e-03\\
$\mathcal{N}(0, 1)$ & $\mathrm{N}(0,0.5^2)$ & 0.0614 & 1.6e-03\\
$\mathcal{N}(0, 1)$ & $\mathrm{N}(0,0.6^2)$ & 0.0417 & 1.2e-03\\
$\mathcal{N}(0, 1)$ & $\mathrm{N}(0,0.7^2)$ & 0.0305 & 8.5e-04\\
$\mathcal{N}(0, 1)$ & $\mathrm{N}(0,0.8^2)$ & 0.0266 & 7.9e-04\\
$\mathcal{N}(0, 1)$ & $\mathrm{N}(0,0.9^2)$ & 0.0309 & 6.5e-04\\
$\mathcal{N}(0, 1)$ & $\mathrm{N}(0,1^2)$ & 0.0512 & 1.2e-03\\
$\mathcal{N}(0, 1)$ & $\mathrm{N}(0,1.1^2)$ & 0.0946 & 1.2e-03\\
$\mathcal{N}(0, 1)$ & $\mathrm{N}(0,1.2^2)$ & 0.1701 & 1.7e-03\\
$\mathcal{N}(0, 1)$ & $\mathrm{N}(0,1.3^2)$ & 0.2675 & 3.1e-03\\
$\mathcal{N}(0, 1)$ & $\mathrm{N}(0,1.4^2)$ & 0.3799 & 2.9e-03\\
$\mathcal{N}(0, 1)$ & $\mathrm{N}(0,1.5^2)$ & 0.4919 & 2.6e-03\\
$\mathcal{N}(0, 1)$ & $\mathrm{N}(0,1.6^2)$ & 0.5946 & 3.1e-03\\
$\mathcal{N}(0, 1)$ & $\mathrm{N}(0,1.7^2)$ & 0.6843 & 2.4e-03\\
$\mathcal{N}(0, 1)$ & $\mathrm{N}(0,1.8^2)$ & 0.758 & 2.5e-03\\
$\mathcal{N}(0, 1)$ & $\mathrm{N}(0,1.9^2)$ & 0.8169 & 2.4e-03\\
$\mathcal{N}(0, 1)$ & $\mathrm{N}(0,2^2)$ & 0.8611 & 1.8e-03\\
$\mathcal{N}(0, 1)$ & $\mathrm{N}(0,2.1^2)$ & 0.8955 & 9.6e-04\\
$\mathcal{N}(0, 1)$ & $\mathrm{N}(0,2.2^2)$ & 0.922 & 1.2e-03\\
$\mathcal{N}(0, 1)$ & $\mathrm{N}(0,2.3^2)$ & 0.9413 & 1.3e-03\\
$\mathcal{N}(0, 1)$ & $\mathrm{N}(0,2.4^2)$ & 0.9561 & 7.6e-04\\
$\mathcal{N}(0, 1)$ & $\mathrm{N}(0,2.5^2)$ & 0.9674 & 7.2e-04\\
$\mathcal{N}(0, 1)$ & $\mathrm{N}(0,3^2)$ & 0.9917 & 3.6e-04\\
$\mathcal{N}(0, 1)$ & $\mathrm{N}(0,3.5^2)$ & 0.9977 & 1.5e-04\\
$\mathcal{N}(0, 1)$ & $\mathrm{N}(0,4^2)$ & 0.9993 & 9.7e-05\\

                    \hline
\end{tabular}
\caption{$N(0,\sigma^2)$ alternatives.}
\label{table:4mknn:normal:scale}

\end{minipage}
\hfill
\begin{minipage}[t]{.48\textwidth}
\vspace{120pt}

\centering

\begin{tabular}{ |l|p{1.6cm}||>{\columncolor[gray]{0.9}}p{1cm}|p{1.1cm}| } 
 \hline
  $H_0$ & $H_1$ & \multicolumn{2}{c|}{4M.kNN}  \\
  \hline
 &  & mean & sd  \\
 \hline
 \hline
 
 $\mathcal{N}(0, 1)$ & $\mathrm{N}(0.1, 1)$ & 0.0549 & 1.2e-03\\
$\mathcal{N}(0, 1)$ & $\mathrm{N}(0.2, 1)$ & 0.0672 & 9.6e-04\\
$\mathcal{N}(0, 1)$ & $\mathrm{N}(0.3, 1)$ & 0.0905 & 1.3e-03\\
$\mathcal{N}(0, 1)$ & $\mathrm{N}(0.4, 1)$ & 0.1275 & 2.1e-03\\
$\mathcal{N}(0, 1)$ & $\mathrm{N}(0.5, 1)$ & 0.1811 & 3.1e-03\\
$\mathcal{N}(0, 1)$ & $\mathrm{N}(0.6, 1)$ & 0.2516 & 3.5e-03\\
$\mathcal{N}(0, 1)$ & $\mathrm{N}(0.7, 1)$ & 0.3406 & 3.8e-03\\
$\mathcal{N}(0, 1)$ & $\mathrm{N}(0.8, 1)$ & 0.4436 & 5.4e-03\\
$\mathcal{N}(0, 1)$ & $\mathrm{N}(0.9, 1)$ & 0.5525 & 4.8e-03\\
$\mathcal{N}(0, 1)$ & $\mathrm{N}(1, 1)$ & 0.6603 & 4.8e-03\\
$\mathcal{N}(0, 1)$ & $\mathrm{N}(1.1, 1)$ & 0.7575 & 5.1e-03\\
$\mathcal{N}(0, 1)$ & $\mathrm{N}(1.2, 1)$ & 0.8391 & 3.8e-03\\
$\mathcal{N}(0, 1)$ & $\mathrm{N}(1.3, 1)$ & 0.8996 & 3.1e-03\\
$\mathcal{N}(0, 1)$ & $\mathrm{N}(1.4, 1)$ & 0.9426 & 1.9e-03\\
$\mathcal{N}(0, 1)$ & $\mathrm{N}(1.5, 1)$ & 0.9695 & 1.4e-03\\
$\mathcal{N}(0, 1)$ & $\mathrm{N}(1.6, 1)$ & 0.985 & 6.6e-04\\
$\mathcal{N}(0, 1)$ & $\mathrm{N}(1.7, 1)$ & 0.9932 & 6.8e-04\\
$\mathcal{N}(0, 1)$ & $\mathrm{N}(1.8, 1)$ & 0.9973 & 2.0e-04\\
$\mathcal{N}(0, 1)$ & $\mathrm{N}(1.9, 1)$ & 0.999 & 1.6e-04\\
$\mathcal{N}(0, 1)$ & $\mathrm{N}(2, 1)$ & 0.9996 & 7.1e-05\\
 
                     \hline
\end{tabular}

\caption{$N(\mu,1)$ alternatives.}

\end{minipage}

\end{table}

\noindent
We call the new moment based test as ``4M.kNN'' test. In table~\ref{table:4mknn:normal:scale} we see that ``4M.kNN'' does not perform well for $N(0, \sigma^2)$ alternatives if $\sigma < 1$. In table~\ref{table:4mknn:normal:scale} we see that the performance for $N(\mu, 1)$ alternatives is average.

\begin{table}[H]
\centering

\begin{minipage}[t]{.48\textwidth}
\vspace{0pt}

\centering
\begin{tabular}{ |l|p{2.3cm}||>{\columncolor[gray]{0.9}}p{1cm}|p{1.1cm}| } 
 \hline
  $H_0$ & $H_1$ & \multicolumn{2}{c|}{4M.kNN}  \\
  \hline
 &  & mean & sd  \\
 \hline
 \hline
 
 $\mathcal{N}(0, 1)$ & $\mathrm{Cauchy}(0, 0.01)$ & 0.8429 & 7.4e-03\\
$\mathcal{N}(0, 1)$ & $\mathrm{Cauchy}(0, 0.02)$ & 0.8392 & 7.1e-03\\
$\mathcal{N}(0, 1)$ & $\mathrm{Cauchy}(0, 0.03)$ & 0.8335 & 6.9e-03\\
$\mathcal{N}(0, 1)$ & $\mathrm{Cauchy}(0, 0.04)$ & 0.8238 & 7.1e-03\\
$\mathcal{N}(0, 1)$ & $\mathrm{Cauchy}(0, 0.05)$ & 0.8131 & 7.2e-03\\
$\mathcal{N}(0, 1)$ & $\mathrm{Cauchy}(0, 0.06)$ & 0.8005 & 7.1e-03\\
$\mathcal{N}(0, 1)$ & $\mathrm{Cauchy}(0, 0.07)$ & 0.7876 & 7.5e-03\\
$\mathcal{N}(0, 1)$ & $\mathrm{Cauchy}(0, 0.08)$ & 0.7728 & 7.1e-03\\
$\mathcal{N}(0, 1)$ & $\mathrm{Cauchy}(0, 0.09)$ & 0.7588 & 6.4e-03\\
$\mathcal{N}(0, 1)$ & $\mathrm{Cauchy}(0, 0.1)$ & 0.7444 & 5.3e-03\\
$\mathcal{N}(0, 1)$ & $\mathrm{Cauchy}(0, 0.2)$ & 0.6376 & 3.1e-03\\
$\mathcal{N}(0, 1)$ & $\mathrm{Cauchy}(0, 0.3)$ & 0.6242 & 1.6e-03\\
$\mathcal{N}(0, 1)$ & $\mathrm{Cauchy}(0, 0.4)$ & 0.6588 & 1.6e-03\\
$\mathcal{N}(0, 1)$ & $\mathrm{Cauchy}(0, 0.5)$ & 0.7088 & 1.2e-03\\
$\mathcal{N}(0, 1)$ & $\mathrm{Cauchy}(0, 0.6)$ & 0.7611 & 1.4e-03\\
$\mathcal{N}(0, 1)$ & $\mathrm{Cauchy}(0, 0.7)$ & 0.808 & 1.8e-03\\
$\mathcal{N}(0, 1)$ & $\mathrm{Cauchy}(0, 0.8)$ & 0.8485 & 1.4e-03\\
$\mathcal{N}(0, 1)$ & $\mathrm{Cauchy}(0, 0.9)$ & 0.8806 & 9.7e-04\\
$\mathcal{N}(0, 1)$ & $\mathrm{Cauchy}(0, 1)$ & 0.9072 & 9.8e-04\\
$\mathcal{N}(0, 1)$ & $\mathrm{Cauchy}(0, 1.1)$ & 0.9286 & 1.3e-03\\
$\mathcal{N}(0, 1)$ & $\mathrm{Cauchy}(0, 1.2)$ & 0.9455 & 7.6e-04\\
$\mathcal{N}(0, 1)$ & $\mathrm{Cauchy}(0, 1.3)$ & 0.9586 & 5.9e-04\\
$\mathcal{N}(0, 1)$ & $\mathrm{Cauchy}(0, 1.4)$ & 0.9687 & 4.9e-04\\
$\mathcal{N}(0, 1)$ & $\mathrm{Cauchy}(0, 1.5)$ & 0.9764 & 5.2e-04\\
$\mathcal{N}(0, 1)$ & $\mathrm{Cauchy}(0, 1.6)$ & 0.982 & 4.6e-04\\
$\mathcal{N}(0, 1)$ & $\mathrm{Cauchy}(0, 1.7)$ & 0.9865 & 3.2e-04\\
$\mathcal{N}(0, 1)$ & $\mathrm{Cauchy}(0, 1.8)$ & 0.9895 & 3.4e-04\\
$\mathcal{N}(0, 1)$ & $\mathrm{Cauchy}(0, 1.9)$ & 0.9922 & 1.9e-04\\
$\mathcal{N}(0, 1)$ & $\mathrm{Cauchy}(0, 2)$ & 0.9941 & 1.6e-04\\
 
                      \hline
\end{tabular}
\caption{Cauchy alternatives.}

\end{minipage}
\hfill
\begin{minipage}[t]{.48\textwidth}
\vspace{0pt}
\centering

\begin{tabular}{ |l|p{1.2cm}||>{\columncolor[gray]{0.9}}p{1cm}|p{1.1cm}| } 
 \hline
  $H_0$ & $H_1$ & \multicolumn{2}{c|}{4M.kNN}  \\
  \hline
 &  & mean & sd  \\
 \hline
 \hline
 
 $\mathcal{N}(0, 1)$ & $\mathrm{G}(0.1)$ & 0.9329 & 2.1e-03\\
$\mathcal{N}(0, 1)$ & $\mathrm{G}(0.2)$ & 0.7794 & 3.9e-03\\
$\mathcal{N}(0, 1)$ & $\mathrm{G}(0.3)$ & 0.6397 & 2.9e-03\\
$\mathcal{N}(0, 1)$ & $\mathrm{G}(0.4)$ & 0.5346 & 2.5e-03\\
$\mathcal{N}(0, 1)$ & $\mathrm{G}(0.5)$ & 0.4565 & 2.0e-03\\
$\mathcal{N}(0, 1)$ & $\mathrm{G}(0.6)$ & 0.4003 & 2.4e-03\\
$\mathcal{N}(0, 1)$ & $\mathrm{G}(0.7)$ & 0.357 & 1.9e-03\\
$\mathcal{N}(0, 1)$ & $\mathrm{G}(0.8)$ & 0.3237 & 2.7e-03\\
$\mathcal{N}(0, 1)$ & $\mathrm{G}(0.9)$ & 0.2979 & 3.1e-03\\
$\mathcal{N}(0, 1)$ & $\mathrm{G}(1)$ & 0.2759 & 3.1e-03\\
$\mathcal{N}(0, 1)$ & $\mathrm{G}(2)$ & 0.1781 & 1.4e-03\\
$\mathcal{N}(0, 1)$ & $\mathrm{G}(3)$ & 0.143 & 1.9e-03\\
$\mathcal{N}(0, 1)$ & $\mathrm{G}(4)$ & 0.1239 & 1.3e-03\\
$\mathcal{N}(0, 1)$ & $\mathrm{G}(5)$ & 0.1116 & 1.2e-03\\
$\mathcal{N}(0, 1)$ & $\mathrm{G}(10)$ & 0.0851 & 1.1e-03\\
$\mathcal{N}(0, 1)$ & $\mathrm{G}(100)$ & 0.0545 & 1.2e-03\\
$\mathcal{N}(0, 1)$ & $\mathrm{G}(1000)$ & 0.0513 & 8.4e-04\\
 
                        \hline
\end{tabular}
\caption{Gamma alternatives.}

\vspace{32pt}

\begin{tabular}{ |l|p{1.2cm}||>{\columncolor[gray]{0.9}}p{1cm}|p{1.1cm}| } 
 \hline
  $H_0$ & $H_1$ & \multicolumn{2}{c|}{4M.kNN}  \\
  \hline
 &  & mean & sd  \\
 \hline
 \hline
 
 $\mathcal{N}(0, 1)$ & $\mathrm{t}(1)$ & 0.9079 & 7.2e-04\\
$\mathcal{N}(0, 1)$ & $\mathrm{t}(2)$ & 0.6561 & 1.2e-03\\
$\mathcal{N}(0, 1)$ & $\mathrm{t}(3)$ & 0.475 & 2.2e-03\\
$\mathcal{N}(0, 1)$ & $\mathrm{t}(4)$ & 0.3623 & 2.1e-03\\
$\mathcal{N}(0, 1)$ & $\mathrm{t}(5)$ & 0.2908 & 1.8e-03\\
 
                         \hline
\end{tabular}
\caption{Student alternatives.}

\end{minipage}

\end{table}

\noindent
In the tables above we see that ``4M.kNN'' test is good for gamma and Student alternatives, but in case of Cauchy alternatives with small scale this test does not reach the power of 1.

In the next section we'll try to fix ``4M.kNN'' test's performance for normal and Cauchy alternatives with small scale parameters by introducing weights.

\newpage
\section{Goodness of fit tests based on joint distribution of four scaled empirical moments }

In section~\ref{section:4moments} we introduced a test based on joint distribution of four sample moments. The problem is that the moments do have pretty different scale. In this section we will try almost the same test, but we will divide each sample moment on its asymptotic variance:

\begin{minted}[mathescape, linenos]{r}
res.normal.4moments.w.kNN <- calc.rejections.NN.tests(
    sample.generation.function.H0  = rnorm,
    sample.generation.functions.H1 = c(
        # noemal, scale
        lapply(c(0.05, 0.1, 0.15), function(sd)   function(nsamples) rnorm(nsamples, 0,    sd)), 
        lapply(seq(0.2, 0.9, 0.1), function(sd)   function(nsamples) rnorm(nsamples, 0,    sd)),
        rnorm,
        lapply(seq(1.1, 2.4, 0.1), function(sd)   function(nsamples) rnorm(nsamples, 0,    sd)),
        lapply(seq(2.5,   4, 0.5), function(sd)   function(nsamples) rnorm(nsamples, 0,    sd)),
        # normal, location
        lapply(seq(0.1,   2, 0.1), function(mean) function(nsamples) rnorm(nsamples, mean, 1 )),
        # Cauchy
        lapply(seq(0.01, 0.09, 0.01), function(scale) function(nsamples) rcauchy(nsamples, 0, scale)),
        lapply(seq(0.1,  2,    0.1 ), function(scale) function(nsamples) rcauchy(nsamples, 0, scale)),
        # gamma
        lapply(
            c(seq(0.1, 0.9, 0.1), 1:5, 10, 100, 1000), 
            function(shape) 
                function(nsamples) (rgamma(nsamples, shape = shape, rate = 1) - shape) / sqrt(shape)
        ),
        # Student
        lapply(seq(1,5), function(df) function(nsamples) rt(nsamples, df = df))
    ),
    get.stat.matrix = function(sample.generation.function, n, m){
        samples <- sample.generation.function(n * m)
        dim(samples) <- c(m, n)
        
        statistics <- cbind(
            apply(samples, 1, mean),
            apply(samples, 1, var) / 2,
            apply(samples, 1, e1071::skewness) / 6,
            apply(samples, 1, e1071::kurtosis) / 24
        )
    },
    alpha = 0.95,
    h = 10,
    n = n, m = m
)
\end{minted}

\begin{table}[H]
\centering

\begin{minipage}[t]{.48\textwidth}
\vspace{0pt}

\centering

\begin{tabular}{ |l|p{1.6cm}||>{\columncolor[gray]{0.9}}p{1cm}|p{1.1cm}| } 
 \hline
  $H_0$ & $H_1$ & \multicolumn{2}{c|}{4MW.kNN}  \\
  \hline
 &  & mean & sd  \\
 \hline
 \hline
 
 $\mathcal{N}(0, 1)$ & $\mathrm{N}(0,0.05^2)$ & 1 & 0.0e+00\\
$\mathcal{N}(0, 1)$ & $\mathrm{N}(0,0.1^2)$ & 1 & 4.1e-05\\
$\mathcal{N}(0, 1)$ & $\mathrm{N}(0,0.15^2)$ & 0.9752 & 1.2e-02\\
$\mathcal{N}(0, 1)$ & $\mathrm{N}(0,0.2^2)$ & 0.7929 & 3.5e-02\\
$\mathcal{N}(0, 1)$ & $\mathrm{N}(0,0.3^2)$ & 0.3395 & 1.8e-02\\
$\mathcal{N}(0, 1)$ & $\mathrm{N}(0,0.4^2)$ & 0.1412 & 5.5e-03\\
$\mathcal{N}(0, 1)$ & $\mathrm{N}(0,0.5^2)$ & 0.067 & 2.1e-03\\
$\mathcal{N}(0, 1)$ & $\mathrm{N}(0,0.6^2)$ & 0.0365 & 1.2e-03\\
$\mathcal{N}(0, 1)$ & $\mathrm{N}(0,0.7^2)$ & 0.024 & 9.6e-04\\
$\mathcal{N}(0, 1)$ & $\mathrm{N}(0,0.8^2)$ & 0.0202 & 7.1e-04\\
$\mathcal{N}(0, 1)$ & $\mathrm{N}(0,0.9^2)$ & 0.0263 & 9.6e-04\\
$\mathcal{N}(0, 1)$ & $\mathrm{N}(0,1^2)$ & 0.0504 & 1.2e-03\\
$\mathcal{N}(0, 1)$ & $\mathrm{N}(0,1.1^2)$ & 0.104 & 1.2e-03\\
$\mathcal{N}(0, 1)$ & $\mathrm{N}(0,1.2^2)$ & 0.1902 & 2.0e-03\\
$\mathcal{N}(0, 1)$ & $\mathrm{N}(0,1.3^2)$ & 0.2991 & 2.3e-03\\
$\mathcal{N}(0, 1)$ & $\mathrm{N}(0,1.4^2)$ & 0.4177 & 2.8e-03\\
$\mathcal{N}(0, 1)$ & $\mathrm{N}(0,1.5^2)$ & 0.5326 & 3.0e-03\\
$\mathcal{N}(0, 1)$ & $\mathrm{N}(0,1.6^2)$ & 0.6345 & 1.8e-03\\
$\mathcal{N}(0, 1)$ & $\mathrm{N}(0,1.7^2)$ & 0.7205 & 1.4e-03\\
$\mathcal{N}(0, 1)$ & $\mathrm{N}(0,1.8^2)$ & 0.788 & 1.3e-03\\
$\mathcal{N}(0, 1)$ & $\mathrm{N}(0,1.9^2)$ & 0.8413 & 1.7e-03\\
$\mathcal{N}(0, 1)$ & $\mathrm{N}(0,2^2)$ & 0.8822 & 1.4e-03\\
$\mathcal{N}(0, 1)$ & $\mathrm{N}(0,2.1^2)$ & 0.9128 & 9.2e-04\\
$\mathcal{N}(0, 1)$ & $\mathrm{N}(0,2.2^2)$ & 0.935 & 7.4e-04\\
$\mathcal{N}(0, 1)$ & $\mathrm{N}(0,2.3^2)$ & 0.9521 & 9.7e-04\\
$\mathcal{N}(0, 1)$ & $\mathrm{N}(0,2.4^2)$ & 0.9643 & 7.8e-04\\
$\mathcal{N}(0, 1)$ & $\mathrm{N}(0,2.5^2)$ & 0.9732 & 7.1e-04\\
$\mathcal{N}(0, 1)$ & $\mathrm{N}(0,3^2)$ & 0.9934 & 3.0e-04\\
$\mathcal{N}(0, 1)$ & $\mathrm{N}(0,3.5^2)$ & 0.9981 & 7.6e-05\\
$\mathcal{N}(0, 1)$ & $\mathrm{N}(0,4^2)$ & 0.9994 & 7.9e-05\\
 
                       \hline
\end{tabular}
\caption{$N(0,\sigma^2)$ alternatives.}

\end{minipage}
\hfill
\begin{minipage}[t]{.48\textwidth}
\vspace{120pt}

\centering

\begin{tabular}{ |l|p{1.6cm}||>{\columncolor[gray]{0.9}}p{1cm}|p{1.1cm}| } 
 \hline
  $H_0$ & $H_1$ & \multicolumn{2}{c|}{4MW.kNN}  \\
  \hline
 &  & mean & sd  \\
 \hline
 \hline
 
 $\mathcal{N}(0, 1)$ & $\mathrm{N}(0.1, 1)$ & 0.0563 & 1.3e-03\\
$\mathcal{N}(0, 1)$ & $\mathrm{N}(0.2, 1)$ & 0.0762 & 1.6e-03\\
$\mathcal{N}(0, 1)$ & $\mathrm{N}(0.3, 1)$ & 0.1112 & 2.0e-03\\
$\mathcal{N}(0, 1)$ & $\mathrm{N}(0.4, 1)$ & 0.1635 & 2.4e-03\\
$\mathcal{N}(0, 1)$ & $\mathrm{N}(0.5, 1)$ & 0.2373 & 3.2e-03\\
$\mathcal{N}(0, 1)$ & $\mathrm{N}(0.6, 1)$ & 0.3309 & 3.3e-03\\
$\mathcal{N}(0, 1)$ & $\mathrm{N}(0.7, 1)$ & 0.4397 & 3.9e-03\\
$\mathcal{N}(0, 1)$ & $\mathrm{N}(0.8, 1)$ & 0.5546 & 5.1e-03\\
$\mathcal{N}(0, 1)$ & $\mathrm{N}(0.9, 1)$ & 0.668 & 4.1e-03\\
$\mathcal{N}(0, 1)$ & $\mathrm{N}(1, 1)$ & 0.7681 & 3.9e-03\\
$\mathcal{N}(0, 1)$ & $\mathrm{N}(1.1, 1)$ & 0.8485 & 3.3e-03\\
$\mathcal{N}(0, 1)$ & $\mathrm{N}(1.2, 1)$ & 0.9086 & 1.7e-03\\
$\mathcal{N}(0, 1)$ & $\mathrm{N}(1.3, 1)$ & 0.9495 & 1.2e-03\\
$\mathcal{N}(0, 1)$ & $\mathrm{N}(1.4, 1)$ & 0.974 & 1.0e-03\\
$\mathcal{N}(0, 1)$ & $\mathrm{N}(1.5, 1)$ & 0.9879 & 5.3e-04\\
$\mathcal{N}(0, 1)$ & $\mathrm{N}(1.6, 1)$ & 0.9947 & 2.7e-04\\
$\mathcal{N}(0, 1)$ & $\mathrm{N}(1.7, 1)$ & 0.9979 & 1.6e-04\\
$\mathcal{N}(0, 1)$ & $\mathrm{N}(1.8, 1)$ & 0.9992 & 1.1e-04\\
$\mathcal{N}(0, 1)$ & $\mathrm{N}(1.9, 1)$ & 0.9997 & 5.7e-05\\
$\mathcal{N}(0, 1)$ & $\mathrm{N}(2, 1)$ & 0.9999 & 3.2e-05\\
 
                        \hline
\end{tabular}
\caption{$N(\mu, 1)$ alternatives.}

\end{minipage}

\end{table}

\noindent
The scaled moments test is denoted as ``4MW.kNN''. We see that moment scaling fixed the ``4MW.kNN'' test's behavior for $N(0,\sigma^2)$, $\sigma < 1$ alternatives compared to ``4M.kNN'' test. Moreover, the performance of ``4MW.kNN'' test for detecting location shifts is better than the performance of ``4M.kNN'' test.

\begin{table}[H]
\centering

\begin{minipage}[t]{.48\textwidth}
\vspace{0pt}

\centering
\begin{tabular}{ |l|p{2.3cm}||>{\columncolor[gray]{0.9}}p{1cm}|p{1.1cm}| } 
 \hline
  $H_0$ & $H_1$ & \multicolumn{2}{c|}{4MW.kNN}  \\
  \hline
 &  & mean & sd  \\
 \hline
 \hline

$\mathcal{N}(0, 1)$ & $\mathrm{Cauchy}(0, 0.01)$ & 0.9992 & 9.2e-05\\
$\mathcal{N}(0, 1)$ & $\mathrm{Cauchy}(0, 0.02)$ & 0.9962 & 2.8e-04\\
$\mathcal{N}(0, 1)$ & $\mathrm{Cauchy}(0, 0.03)$ & 0.9897 & 3.1e-04\\
$\mathcal{N}(0, 1)$ & $\mathrm{Cauchy}(0, 0.04)$ & 0.9788 & 7.5e-04\\
$\mathcal{N}(0, 1)$ & $\mathrm{Cauchy}(0, 0.05)$ & 0.963 & 1.6e-03\\
$\mathcal{N}(0, 1)$ & $\mathrm{Cauchy}(0, 0.06)$ & 0.9428 & 2.0e-03\\
$\mathcal{N}(0, 1)$ & $\mathrm{Cauchy}(0, 0.07)$ & 0.9184 & 2.7e-03\\
$\mathcal{N}(0, 1)$ & $\mathrm{Cauchy}(0, 0.08)$ & 0.8916 & 3.9e-03\\
$\mathcal{N}(0, 1)$ & $\mathrm{Cauchy}(0, 0.09)$ & 0.8628 & 4.0e-03\\
$\mathcal{N}(0, 1)$ & $\mathrm{Cauchy}(0, 0.1)$ & 0.8329 & 4.1e-03\\
$\mathcal{N}(0, 1)$ & $\mathrm{Cauchy}(0, 0.2)$ & 0.6178 & 2.9e-03\\
$\mathcal{N}(0, 1)$ & $\mathrm{Cauchy}(0, 0.3)$ & 0.5865 & 1.7e-03\\
$\mathcal{N}(0, 1)$ & $\mathrm{Cauchy}(0, 0.4)$ & 0.628 & 1.2e-03\\
$\mathcal{N}(0, 1)$ & $\mathrm{Cauchy}(0, 0.5)$ & 0.6879 & 1.6e-03\\
$\mathcal{N}(0, 1)$ & $\mathrm{Cauchy}(0, 0.6)$ & 0.7474 & 1.1e-03\\
$\mathcal{N}(0, 1)$ & $\mathrm{Cauchy}(0, 0.7)$ & 0.7998 & 1.7e-03\\
$\mathcal{N}(0, 1)$ & $\mathrm{Cauchy}(0, 0.8)$ & 0.8445 & 1.7e-03\\
$\mathcal{N}(0, 1)$ & $\mathrm{Cauchy}(0, 0.9)$ & 0.8803 & 1.1e-03\\
$\mathcal{N}(0, 1)$ & $\mathrm{Cauchy}(0, 1)$ & 0.9085 & 7.7e-04\\
$\mathcal{N}(0, 1)$ & $\mathrm{Cauchy}(0, 1.1)$ & 0.9306 & 4.3e-04\\
$\mathcal{N}(0, 1)$ & $\mathrm{Cauchy}(0, 1.2)$ & 0.9474 & 6.7e-04\\
$\mathcal{N}(0, 1)$ & $\mathrm{Cauchy}(0, 1.3)$ & 0.9606 & 5.2e-04\\
$\mathcal{N}(0, 1)$ & $\mathrm{Cauchy}(0, 1.4)$ & 0.9708 & 5.4e-04\\
$\mathcal{N}(0, 1)$ & $\mathrm{Cauchy}(0, 1.5)$ & 0.9781 & 4.0e-04\\
$\mathcal{N}(0, 1)$ & $\mathrm{Cauchy}(0, 1.6)$ & 0.9838 & 4.3e-04\\
$\mathcal{N}(0, 1)$ & $\mathrm{Cauchy}(0, 1.7)$ & 0.9879 & 2.0e-04\\
$\mathcal{N}(0, 1)$ & $\mathrm{Cauchy}(0, 1.8)$ & 0.9907 & 2.9e-04\\
$\mathcal{N}(0, 1)$ & $\mathrm{Cauchy}(0, 1.9)$ & 0.9932 & 3.2e-04\\
$\mathcal{N}(0, 1)$ & $\mathrm{Cauchy}(0, 2)$ & 0.9948 & 1.6e-04\\

                        \hline
\end{tabular}
\caption{Cauchy alternatives.}

\end{minipage}
\hfill
\begin{minipage}[t]{.48\textwidth}
\vspace{0pt}
\centering

\begin{tabular}{ |l|p{1.2cm}||>{\columncolor[gray]{0.9}}p{1cm}|p{1.1cm}| } 
 \hline
  $H_0$ & $H_1$ & \multicolumn{2}{c|}{4MW.kNN}  \\
  \hline
 &  & mean & sd  \\
 \hline
 \hline
 
 $\mathcal{N}(0, 1)$ & $\mathrm{G}(0.1)$ & 0.9034 & 2.5e-03\\
$\mathcal{N}(0, 1)$ & $\mathrm{G}(0.2)$ & 0.7548 & 4.8e-03\\
$\mathcal{N}(0, 1)$ & $\mathrm{G}(0.3)$ & 0.6298 & 7.2e-03\\
$\mathcal{N}(0, 1)$ & $\mathrm{G}(0.4)$ & 0.5344 & 6.1e-03\\
$\mathcal{N}(0, 1)$ & $\mathrm{G}(0.5)$ & 0.4626 & 7.3e-03\\
$\mathcal{N}(0, 1)$ & $\mathrm{G}(0.6)$ & 0.409 & 5.8e-03\\
$\mathcal{N}(0, 1)$ & $\mathrm{G}(0.7)$ & 0.365 & 5.6e-03\\
$\mathcal{N}(0, 1)$ & $\mathrm{G}(0.8)$ & 0.3325 & 4.0e-03\\
$\mathcal{N}(0, 1)$ & $\mathrm{G}(0.9)$ & 0.3054 & 3.4e-03\\
$\mathcal{N}(0, 1)$ & $\mathrm{G}(1)$ & 0.2836 & 3.5e-03\\
$\mathcal{N}(0, 1)$ & $\mathrm{G}(2)$ & 0.1835 & 1.9e-03\\
$\mathcal{N}(0, 1)$ & $\mathrm{G}(3)$ & 0.1467 & 1.9e-03\\
$\mathcal{N}(0, 1)$ & $\mathrm{G}(4)$ & 0.1275 & 2.6e-03\\
$\mathcal{N}(0, 1)$ & $\mathrm{G}(5)$ & 0.1158 & 1.2e-03\\
$\mathcal{N}(0, 1)$ & $\mathrm{G}(10)$ & 0.0873 & 1.5e-03\\
$\mathcal{N}(0, 1)$ & $\mathrm{G}(100)$ & 0.0554 & 1.2e-03\\
$\mathcal{N}(0, 1)$ & $\mathrm{G}(1000)$ & 0.0509 & 1.2e-03\\
 
                         \hline
\end{tabular}
\caption{Gamma alternatives.}

\vspace{32pt}

\begin{tabular}{ |l|p{1.2cm}||>{\columncolor[gray]{0.9}}p{1cm}|p{1.1cm}| } 
 \hline
  $H_0$ & $H_1$ & \multicolumn{2}{c|}{4MW.kNN}  \\
  \hline
 &  & mean & sd  \\
 \hline
 \hline
 
$\mathcal{N}(0, 1)$ & $\mathrm{t}(1)$ & 0.9084 & 9.1e-04\\
$\mathcal{N}(0, 1)$ & $\mathrm{t}(2)$ & 0.6585 & 2.3e-03\\
$\mathcal{N}(0, 1)$ & $\mathrm{t}(3)$ & 0.4773 & 2.1e-03\\
$\mathcal{N}(0, 1)$ & $\mathrm{t}(4)$ & 0.3647 & 2.2e-03\\
$\mathcal{N}(0, 1)$ & $\mathrm{t}(5)$ & 0.294 & 1.9e-03\\
 
                          \hline
\end{tabular}
\caption{Student moments.}

\end{minipage}

\end{table}

\noindent
We see that the behavior for Cauchy alternatives with small scales is also fixed, while powers for gamma and Student alternatives remain good.  
We can conclude that the scaled moments based test is a good balanced test for checking standard normality.

\newpage
\section{Goodness of fit test based on joint distribution of three classical distances}

In this section we construct a goodness of fit test based on joint distribution of Kolmogorov--Smirnov, Cramer--von Mises and Anderson--Darling distances.

\begin{minted}[mathescape, linenos]{r}
res.uniform.vs.beta.3classic.kNN <- calc.rejections.NN.tests(
    sample.generation.function.H0  = runif,
    sample.generation.functions.H1 = c(
        lapply(1:10, function(x) function(nsamples) rbeta(nsamples, x, x)),
        lapply(c(1,rev(seq(0.1, 0.9, 0.1))), function(x) function(nsamples) rbeta(nsamples, x, x)),
        lapply(c(1,rev(seq(0.1, 0.9, 0.1))), function(x) function(nsamples) rbeta(nsamples, 1, x)),
        lapply(1:10, function(x) function(nsamples) rbeta(nsamples, 1, x))
    ),
    get.stat.matrix = function(sample.generation.function, n, m){
        sorted.samples <- get.sorted.samples.cpp(sample.generation.function, n, m)
        classicalDistances(sorted.samples, punif)
    },
    alpha = 0.95,
    h = 10,
    n = n, m = m
)
res.uniform.vs.beta.3classic.kNN.table <- create.rejection.table.kNN(res.uniform.vs.beta.3classic.kNN)
res.uniform.vs.beta.3classic.kNN.table
\end{minted}

\begin{table}[H]
\centering

\begin{minipage}[t]{.48\textwidth}
\vspace{0pt}

\centering

\begin{tabular}{ |l|p{1.8cm}||>{\columncolor[gray]{0.9}}p{1cm}|p{1.1cm}| } 
 \hline
  $H_0$ & $H_1$ & \multicolumn{2}{c|}{3DIST.kNN}  \\
  \hline
 
 $\mathrm{U}(0, 1)$ & $\mathrm{Beta}(1, 1)$ & 0.0504 & 1.2e-03\\
$\mathrm{U}(0, 1)$ & $\mathrm{Beta}(2, 2)$ & 0.033 & 1.4e-03\\
$\mathrm{U}(0, 1)$ & $\mathrm{Beta}(3, 3)$ & 0.093 & 1.8e-03\\
$\mathrm{U}(0, 1)$ & $\mathrm{Beta}(4, 4)$ & 0.1845 & 3.2e-03\\
$\mathrm{U}(0, 1)$ & $\mathrm{Beta}(5, 5)$ & 0.2935 & 4.7e-03\\
$\mathrm{U}(0, 1)$ & $\mathrm{Beta}(6, 6)$ & 0.4088 & 6.9e-03\\
$\mathrm{U}(0, 1)$ & $\mathrm{Beta}(7, 7)$ & 0.5208 & 6.6e-03\\
$\mathrm{U}(0, 1)$ & $\mathrm{Beta}(8, 8)$ & 0.6194 & 7.2e-03\\
$\mathrm{U}(0, 1)$ & $\mathrm{Beta}(9, 9)$ & 0.7056 & 9.0e-03\\
$\mathrm{U}(0, 1)$ & $\mathrm{Beta}(10, 10)$ & 0.7774 & 7.6e-03\\

\hline

 \end{tabular}
\caption{Test based on KS, CvM and AD distances; $\mathrm{U}(0,1)$ vs $\mathrm{Beta}(a,a)$; $a \geq 1$; alternatives have modes in the center, see figure~\ref{figure:beta:aa}.}
\label{table:beta:aa:3dist:kNN}

\end{minipage}
\hfill
\begin{minipage}[t]{.48\textwidth}
\vspace{0pt}

\centering

\begin{tabular}{ |l|p{2cm}||>{\columncolor[gray]{0.9}}p{1cm}|p{1.22cm}| } 
 \hline
  $H_0$ & $H_1$ & \multicolumn{2}{c|}{3DIST.kNN}  \\
  \hline

$\mathrm{U}(0, 1)$ & $\mathrm{Beta}(1, 1)$ & 0.0509 & 7.7e-04\\
$\mathrm{U}(0, 1)$ & $\mathrm{Beta}(0.9, 0.9)$ & 0.0764 & 9.7e-04\\
$\mathrm{U}(0, 1)$ & $\mathrm{Beta}(0.8, 0.8)$ & 0.1213 & 9.1e-04\\
$\mathrm{U}(0, 1)$ & $\mathrm{Beta}(0.7, 0.7)$ & 0.1972 & 1.7e-03\\
$\mathrm{U}(0, 1)$ & $\mathrm{Beta}(0.6, 0.6)$ & 0.3156 & 2.2e-03\\
$\mathrm{U}(0, 1)$ & $\mathrm{Beta}(0.5, 0.5)$ & 0.4886 & 2.5e-03\\
$\mathrm{U}(0, 1)$ & $\mathrm{Beta}(0.4, 0.4)$ & 0.6977 & 2.3e-03\\
$\mathrm{U}(0, 1)$ & $\mathrm{Beta}(0.3, 0.3)$ & 0.887 & 2.1e-03\\
$\mathrm{U}(0, 1)$ & $\mathrm{Beta}(0.2, 0.2)$ & 0.9837 & 4.6e-04\\
$\mathrm{U}(0, 1)$ & $\mathrm{Beta}(0.1, 0.1)$ & 0.9998 & 2.9e-05\\

\hline

  \end{tabular}
\caption{Test based on KS, CvM and AD distances; $\mathrm{U}(0,1)$ vs $\mathrm{Beta}(a,a)$; $0 < a \leq 1$; alternatives have modes on both sides, see figure~\ref{figure:beta:0a0a}.}
\label{table:beta:0a0a:3dist:kNN}

\end{minipage}

\end{table}

\begin{table}[H]
\centering

\begin{minipage}[t]{.48\textwidth}
\vspace{0pt}

\centering

\begin{tabular}{ |l|p{1.8cm}||>{\columncolor[gray]{0.9}}p{1cm}|p{1.1cm}| } 
 \hline
  $H_0$ & $H_1$ & \multicolumn{2}{c|}{3DIST.kNN}  \\
  \hline
 &  & mean & sd  \\
 \hline

$\mathrm{U}(0, 1)$ & $\mathrm{Beta}(1, 1)$ & 0.0504 & 7.6e-04\\
$\mathrm{U}(0, 1)$ & $\mathrm{Beta}(1, 0.9)$ & 0.0708 & 9.8e-04\\
$\mathrm{U}(0, 1)$ & $\mathrm{Beta}(1, 0.8)$ & 0.1215 & 1.1e-03\\
$\mathrm{U}(0, 1)$ & $\mathrm{Beta}(1, 0.7)$ & 0.2213 & 2.1e-03\\
$\mathrm{U}(0, 1)$ & $\mathrm{Beta}(1, 0.6)$ & 0.3806 & 2.3e-03\\
$\mathrm{U}(0, 1)$ & $\mathrm{Beta}(1, 0.5)$ & 0.5927 & 2.1e-03\\
$\mathrm{U}(0, 1)$ & $\mathrm{Beta}(1, 0.4)$ & 0.8065 & 1.3e-03\\
$\mathrm{U}(0, 1)$ & $\mathrm{Beta}(1, 0.3)$ & 0.9478 & 9.8e-04\\
$\mathrm{U}(0, 1)$ & $\mathrm{Beta}(1, 0.2)$ & 0.9953 & 1.5e-04\\
$\mathrm{U}(0, 1)$ & $\mathrm{Beta}(1, 0.1)$ & 1 & 2.4e-05\\

  \hline
  \end{tabular}
\caption{Test based on KS, CvM and AD distances; $\mathrm{U}(0,1)$ vs $\mathrm{Beta}(1,a)$; $0 < a \leq 1$; alternatives have sharp modes on the right, see figure~\ref{figure:beta:10a}.}
\label{table:beta:10a:3dist:kNN}

\end{minipage}
\hfill
\begin{minipage}[t]{.48\textwidth}
\vspace{0pt}

\centering

\begin{tabular}{ |l|p{2cm}||>{\columncolor[gray]{0.9}}p{1cm}|p{1.22cm}| } 
 \hline
  $H_0$ & $H_1$ & \multicolumn{2}{c|}{3DIST.kNN}  \\
  \hline
 &  & mean & sd  \\
 \hline

$\mathrm{U}(0, 1)$ & $\mathrm{Beta}(1, 1)$ & 0.0501 & 7.2e-04\\
$\mathrm{U}(0, 1)$ & $\mathrm{Beta}(1, 2)$ & 0.3366 & 4.4e-03\\
$\mathrm{U}(0, 1)$ & $\mathrm{Beta}(1, 3)$ & 0.7745 & 5.1e-03\\
$\mathrm{U}(0, 1)$ & $\mathrm{Beta}(1, 4)$ & 0.9558 & 1.6e-03\\
$\mathrm{U}(0, 1)$ & $\mathrm{Beta}(1, 5)$ & 0.9942 & 3.8e-04\\
$\mathrm{U}(0, 1)$ & $\mathrm{Beta}(1, 6)$ & 0.9994 & 7.9e-05\\
$\mathrm{U}(0, 1)$ & $\mathrm{Beta}(1, 7)$ & 1 & 2.5e-05\\
$\mathrm{U}(0, 1)$ & $\mathrm{Beta}(1, 8)$ & 1 & 7.0e-06\\
$\mathrm{U}(0, 1)$ & $\mathrm{Beta}(1, 9)$ & 1 & 0.0e+00\\
$\mathrm{U}(0, 1)$ & $\mathrm{Beta}(1, 10)$ & 1 & 0.0e+00\\
 
                           \hline
\end{tabular}

\caption{Test based on KS, CvM and AD distances; $\mathrm{U}(0,1)$ vs $\mathrm{Beta}(1,a)$; $a \geq 1$; alternatives have sloping modes on the left, see figure~\ref{figure:beta:1a}.}
\label{table:beta:1a:3dist:kNN}

\end{minipage}

\end{table}

\noindent
The test is denoted as ``3DIST.kNN''. The results for beta grid are as follows:

\begin{minted}[mathescape, linenos, texcomments]{r}
res.u.vs.b.2dgrid.3classic.kNN <- calc.rejections.NN.tests(
    sample.generation.function.H0  = runif,
    sample.generation.functions.H1 = apply(
        beta.parameters.grid,  # see formula \eqref{eq:beta_grid} 
        1, 
        function(par) function(nsamples) rbeta(nsamples, par[1], par[2])
    ),
    get.stat.matrix = function(sample.generation.function, n, m){
        sorted.samples <- get.sorted.samples.cpp(sample.generation.function, n, m)
        handle.infs(classicalDistances(sorted.samples, punif))
    },
    alpha = 0.95,
    h = 10,
    n = n, m = m
)

res.u.vs.b.2dgrid.3classic.kNN.table <- create.rejection.table.kNN(res.u.vs.b.2dgrid.3classic.kNN)
res.u.vs.b.2dgrid.3classic.kNN.table

plot.powers.on.grid(beta.parameters.grid, res.u.vs.b.2dgrid.3classic.kNN.table$res.kNN.means)
\end{minted}

\begin{figure}[H]
\centering
\includegraphics[width=16cm]{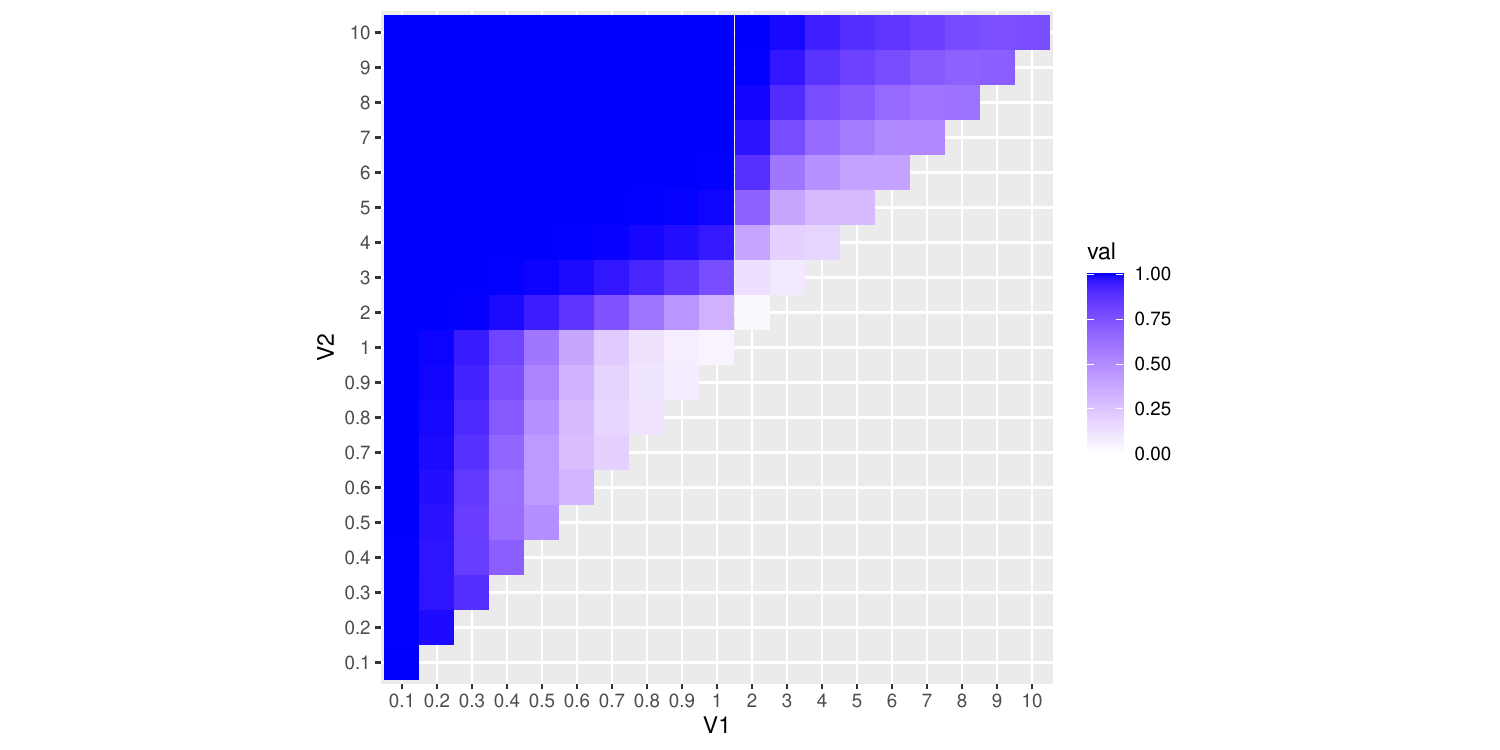}
\caption{3DIST.kNN}
\label{figure:3dist:2dgrid}
\end{figure}

\noindent
Figure~\ref{figure:3dist:2dgrid} for ``3DIST.kNN'' is very similar to the corresponding figure for $Z_C$ test. The interesting thing is that the test based on joint distribution of 3 classical distances performs better than each of the classical tests individually.

\newpage
\section{Sneak peek into two sample case}

How to generalize the ideas from the present work to two sample scenario? The logical way is to try our luck with permutation tests, since they also have the stage of $H_0$ simulation, just as the goodness of fit tests we discussed so far.  

\subsection{Permutation tests based on joint distribution of multiple sample statistics}

Suppose that we have two samples, 
\begin{align}
X = x_1, \dots, x_{n_1} \sim F \text{ and } Y = y_1, \dots, y_{n_2} \sim G,
\end{align}
from absolutely continuous distributions with cumulative distribution functions $F$ and $G$. We want to test the hypothesis 
\begin{align}
H_0 : F = G;
\end{align}
against the alternative
\begin{align}
H_1 : F \neq G.
\end{align}
A permutation test based on HDR for some joint density of multiple statistics consists of the following steps:
\begin{enumerate}
\item We combine the given two samples into the single sample 
\begin{align}
\label{eq:Z}
Z = z_1, \dots, z_n = x_1, \dots, x_{n_1}, y_1, \dots, y_{n_2},
\end{align}
where $n = n_1 + n_2$.
\item 
Then we simulate $m$ permutations of $z_1, \dots, z_m$:
\begin{align}
\begin{bmatrix}
Z_1^* \\
Z_2^* \\
\vdots \\
Z_m^*
\end{bmatrix} 
=
\begin{bmatrix}
z_{\sigma_1(1)} & z_{\sigma_1(2)} & \dots & z_{{\sigma_1(n)}}\\
z_{\sigma_2(1)} & z_{\sigma_2(2)} & \dots & z_{{\sigma_2(n)}}\\
\vdots & \vdots & \ddots & \vdots \\
z_{\sigma_m(1)} & z_{\sigma_m(2)} & \dots & z_{{\sigma_m(n)}}
\end{bmatrix},
\end{align}
where $m \leq n!$, $\sigma_1, \dots, \sigma_m \in S_n$, and $S_n$ is the symmetric group on $\left\{ 1,\dots,n \right\}$.

\item
Then we divide each $Z_i^*$, $i = 1,\dots,n$ into two parts. 

First $n_1$ elements of $Z_i^*$ form $X_i^*$, the simulated sample from unknown distribution $F$:
\begin{align}
\begin{bmatrix}
X_1^* \\
X_2^* \\
\vdots \\
X_m^*
\end{bmatrix} 
=
\begin{bmatrix}
x^*_{1,1} & x^*_{1,2} & \dots & x^*_{1,n_1}\\
x^*_{2,1} & x^*_{2,2} & \dots & x^*_{2,n_1}\\
\vdots & \vdots & \ddots & \vdots \\
x^*_{m,1} & x^*_{m,2} & \dots & x^*_{m,n_1}
\end{bmatrix} 
=
\begin{bmatrix}
z_{\sigma_1(1)} & z_{\sigma_1(2)} & \dots & z_{{\sigma_1(n_1)}}\\
z_{\sigma_2(1)} & z_{\sigma_2(2)} & \dots & z_{{\sigma_2(n_1)}}\\
\vdots & \vdots & \ddots & \vdots \\
z_{\sigma_m(1)} & z_{\sigma_m(2)} & \dots & z_{{\sigma_m(n_1)}}
\end{bmatrix}.
\end{align}
Last $n_2$ elements of $Z_i^*$ form $Y_i^*$, the simulated sample from unknown distribution $G$:
\begin{align}
\begin{bmatrix}
Y_1^* \\
Y_2^* \\
\vdots \\
Y_m^*
\end{bmatrix} 
=
\begin{bmatrix}
y^*_{1,1} & y^*_{1,2} & \dots & y^*_{1,n_2}\\
y^*_{2,1} & y^*_{2,2} & \dots & y^*_{2,n_2}\\
\vdots & \vdots & \ddots & \vdots \\
y^*_{m,1} & y^*_{m,2} & \dots & y^*_{m,n_2}
\end{bmatrix} 
=
\begin{bmatrix}
z_{\sigma_1(n_1+1)} & z_{\sigma_1(n_1+2)} & \dots & z_{{\sigma_1(n_1+n_2)}}\\
z_{\sigma_2(n_1+1)} & z_{\sigma_2(n_1+2)} & \dots & z_{{\sigma_2(n_1+n_2)}}\\
\vdots & \vdots & \ddots & \vdots \\
z_{\sigma_m(n_1+1)} & z_{\sigma_m(n_1+2)} & \dots & z_{{\sigma_m(n_1+n_2)}}
\end{bmatrix}.
\end{align}

\item
Then we choose $k$ statistics 
\begin{align}
T_j : \mathbb{R}^{n_1} \times \mathbb{R}^{n_2} \rightarrow \mathbb{R}; \qquad j = 1, \dots, k;
\end{align}
on the basis of which we will decide on the truth of the null hypothesis,
and form a table
\begin{align}
\label{eq:stat:table:permut}
\begin{bmatrix}
T_{1} (X_1^*, Y_1^*) & T_{2} (X_1^*, Y_1^*) & \dots & T_{k}(X_1^*,Y_1^*)\\
T_{1} (X_2^*, Y_2^*) & T_{2} (X_2^*, Y_2^*) & \dots & T_{k}(X_2^*,Y_2^*)\\
\vdots & \vdots & \ddots & \vdots \\
T_{1} (X_m^*,Y_m^*) & T_{2} (X_m^*, Y_m^*) & \dots & T_{k} (X_m^*,Y_m^*)
\end{bmatrix} .
\end{align}
\item 
Rows of table~\eqref{eq:stat:table:permut} form $\mathcal{P}_{T_1,\dots,T_k}(Z)$, the empirical joint distribution of statistics $T_1, \dots, T_k$. If the null hypothesis is true, we expect the vector 
\begin{align}
[T_1(X, Y), \dots, T_k(X, Y)]
\end{align}
to be in the highest density region of $\mathcal{P}_{T_1,\dots,T_k}(Z)$. If alternative hypothesis is true, we expect this vector to be out of the highest density region.
\end{enumerate}

\noindent
In a sense, if $n_2 \rightarrow \infty$, then permutation test becomes a goodness of fit test. 

Indeed, if $n_2$ is much larger than $n_1$, then
\begin{enumerate}
\item
the effect of $x_1,\dots,x_{n_1}$ points in Z is negligible;
\item
$G_{n_2}$, the empirical cumulative distribution function of sample $Y$, is very close to $G$;
\item
$X_i^*$ will be almost the samples of size $n_1$ from $G$;
\item
$Y_i^*$ will be the samples with eCDFs that are very close to $G_{n_2}$ (and consequently very close to $G$). 
\end{enumerate}
So we can say that the proposed permutation approach is the direct generalization of the discussed goodness of fit approach to two sample case.

\subsection{Notes on power estimation}
Let us discuss how to measure power (and type 1 error) of a permutation test for some given fixed $n_1$, $n_2$, $F$ and $G$: 
\begin{enumerate}
\item Firstly, we simulate $v$ samples $X_1, \dots, X_v$ of size $n_1$ from $F$, and $v$ samples $Y_1, \dots, Y_v$ of size $n_2$ from $G$.
\item Then we create $v$ combined samples $Z_1, \dots, Z_v$, each $Z_i$ comprised of the elements from $X_i$ and $Y_i$.
\item
Then we simulate 
\begin{align*}
\begin{bmatrix}
X_{[i];1}^* \\
X_{[i];2}^* \\
\vdots \\
X_{[i];m}^*
\end{bmatrix} 
\text{ and }
\begin{bmatrix}
Y_{[i];1}^* \\
Y_{[i];2}^* \\
\vdots \\
Y_{[i];m}^*
\end{bmatrix}
 ;
 \qquad
 i = 1, \dots, v;
\end{align*}
based on permutations of corresponding $Z_i$, $i = 1,\dots,v$.
\item
Then we compute tables
\begin{align*}
\begin{bmatrix}
T_1 (X_{[i];1}^*, Y_{[i];1}^*) & T_2 (X_{[i];1}^*, Y_{[i];1}^*) & \dots & T_k(X_{[i];1}^*,Y_{[i];1}^*)\\
T_1 (X_{[i];2}^*, Y_{[i];2}^*) & T_2 (X_{[i];2}^*, Y_{[i];2}^*) & \dots & T_k(X_{[i];2}^*,Y_{[i];2}^*)\\
\vdots & \vdots & \ddots & \vdots \\
T_1 (X_{[i];m}^*,Y_{[i];m}^*) & T_2 (X_{[i];m}^*, Y_{[i];m}^*) & \dots & T_k (X_{[i];m}^*,Y_{[i];m}^*)
\end{bmatrix};
\qquad i = 1,\dots,v; 
\end{align*}
to determine empirical joint statistics distributions $\mathcal{P}_{T_1,\dots,T_k} (Z_i)$ and corresponding highest density regions $\mathrm{HDR}\left( \mathcal{P}_{T_1,\dots,T_k} (Z_i) \right)$ for them.
\item
Finally, we compute the power estimate
\begin{align}
\label{eq:permutation:power}
\Big\{ 
[T_1(X_i,Y_i), \dots, T_k(X_i,Y_i)] \notin \mathrm{HDR}\left( \mathcal{P}_{T_1,\dots,T_k} (Z_i) \right) \: \Big| \: i = 1,\dots,v
\Big\} / v.
\end{align}
\end{enumerate}
If $F = G$, then~\eqref{eq:permutation:power} is a type 1 error estimate.

\subsection{Permutation test based on joint distribution of KS and AD distances}

In this subsection we construct the permutation test based on joint distribution of two sample Kolmogorov--Smirnov and Anderson--Darling distances.

Let 
\[
F_{n_1} (x; X) = \frac{1}{n_1}\sum_{i=1}^{n_1} \mathbf{1}\{x_i \le x\}, 
\qquad
G_{n_2} (x; Y) = \frac{1}{n_2}\sum_{i=1}^{n_2} \mathbf{1}\{y_i \le x\};
\]
be the empirical cumulative distribution functions for samples $X$ and $Y$. Kolmogorov--Smirnov two sample distance is as follows:
\[
D_{KS} (X, Y)
=
\sup_{x \in \mathbb{R}}
\left|
F_{n_1}(x;X)
-
G_{n_2}(x;Y)
\right|.
\]
The computational formula is
\[
D_{KS} (X, Y)
=
\max_{1 \le j \le n_1+n_2}
\left|
F_{n_1}(z_{(j)};X)
-
G_{n_2}( z_{(j)};Y)
\right|,
\]
where $z_{(j)}$, $j = 1,\dots,n_1+n_2$, are order statistics of combined sample $Z$ defined in \eqref{eq:Z}.

We implement two sample KS distance like this:
\begin{minted}[mathescape, linenos, texcomments]{r}
distance.KS.2samples <- function(x, y) {
    x <- sort(x)
    y <- sort(y)

    z <- sort(unique(c(x, y)))

    Fx <- findInterval(z, x) / length(x)
    Gy <- findInterval(z, y) / length(y)

    max(abs(Fx - Gy))
}
\end{minted}

Let further
\[
H_N(x)=\frac{n_1 F_{n_1}(x;X) + n_2 G_{n_2}(x;Y)}{N},
\qquad N=n_1 + n_2.
\]
The two-sample Anderson--Darling distance is denoted as 
\[
D_{AD} (X,Y)
=
\frac{n_1 n_2}{N}
\int_{-\infty}^{\infty}
\frac{\bigl(F_{n_1}(x;X)-G_{n_2}(x;Y)\bigr)^2}
{H_N(x)\bigl(1-H_N(x)\bigr)}
\,dH_N(x).
\]

If 
\[
M_i
=
\#\{x_j : x_j \le z_{(i)} \: | \: j = 1,\dots,n_1 \},
\qquad i=1,\dots,N;
\]
then among the first $i$ observations in combined sample $Z$, $M_i$ come from sample $X$ and $i-M_i$ come from sample $Y$. Hence
\[
F_{n_1} \bigl(z_{(i)}\bigr)=\frac{M_i}{n_1},
\qquad
G_{n_2}\bigl(z_{(i)}\bigr)=\frac{i-M_i}{n_2}.
\]
So a computational form of the two-sample Anderson--Darling distance is
\[
D_{AD} (X,Y)
=
\frac{n_1 n_2}{N^2}
\sum_{i=1}^{N-1}
\frac{
\left(
\frac{M_i}{n_1}
-
\frac{i-M_i}{n_2}
\right)^2
}{
\frac{i}{N}\left(1-\frac{i}{N}\right)
}.
\]
We implement Anderson--Darling two sample distance as follows:
\begin{minted}[mathescape, linenos, texcomments]{r}
distance.AD.2samples <- function(x, y) {
    x <- sort(x)
    y <- sort(y)
    
    n1 <- length(x)
    n2 <- length(y)
    N <- n1 + n2
    
    z <- sort(c(x, y))
    
    i <- seq_len(N - 1)
    
    M_i <- findInterval(z[-N], x) 
    H_i <- i / N
    
    F_i <- M_i / n1
    G_i <- (i - M_i) / n2
    
    (n1 * n2 / N) * sum(
        (F_i - G_i)^2 / (H_i * (1 - H_i))
    ) / N
}
\end{minted}
Finally, we implement the power estimate for permutation test based on joint distribution of two sample KS and AD distances as follows:
\begin{minted}[mathescape, linenos, texcomments]{r}
KS.AD.2sample.permutation.test <- function(
    sample.1.generation.function,
    sample.2.generation.function,
    num.iter = 100,
    m = 10000
)
{
    power_array_1 <- numeric(num.iter)
    power_array_2 <- numeric(num.iter)
    power_array_combined <- numeric(num.iter)
    
    k = round(sqrt(m))
    
    for(j in 1:num.iter)
    {
        sample.1 <- sample.1.generation.function()
        sample.2 <- sample.2.generation.function()
      
        d1 <- distance.KS.2samples(sample.1, sample.2)
        d2 <- distance.AD.2samples(sample.1, sample.2)
    
        combined.sample <- c(sample.1, sample.2)
        res1 <- numeric(m)
        res2 <- numeric(m)
        for(i in 1:m)
        {
            indices <- sample.int(length(combined.sample), size = length(sample.1))
            sample1_ <- combined.sample[indices]
            sample2_ <- combined.sample[-indices]
            res1[i] <- distance.KS.2samples(sample1_, sample2_)
            res2[i] <- distance.AD.2samples(sample1_, sample2_)
        }
      
        joint.distr.2dists <- cbind(res1, res2)
        mat <- knnx.dist(data = joint.distr.2dists, query = joint.distr.2dists, k = k)
        kNN.sum.dists <- rowSums(mat)
    
        d.j <- sum(knnx.dist(data = joint.distr.2dists, query = cbind(d1, d2), k = k))
        
        power_array_1[j]        <- d1  > quantile(res1, 0.95)
        power_array_2[j]        <- d2  > quantile(res2, 0.95)
        power_array_combined[j] <- d.j > quantile(kNN.sum.dists, 0.95)
        print(j)
        flush.console()
    }
    list(
        KS = sum(power_array_1) / length(power_array_1),
        AD = sum(power_array_2) / length(power_array_2),
        joint = sum(power_array_combined) / length(power_array_combined)
    )
}
\end{minted}
Let us achieve some quick perspective about the introduced test's power.

First, type 1 error check: 
\begin{minted}[mathescape, linenos, texcomments]{r}
KS.AD.2sample.permutation.test(function() rnorm(30, 0, 1), function() rnorm(42, 0, 1))
\end{minted}
\begin{verbatim}
$KS    0.02
$AD    0.04
$joint 0.04
\end{verbatim}
We see, that the values are quite rough. To get more accurate results, one might increase \texttt{m} and \texttt{num.iter}.

Now to the powers:
\begin{minted}[mathescape, linenos, texcomments]{r}
KS.AD.2sample.permutation.test(function() rnorm(30, 0, 1), function() rnorm(42, 0, 1.5))
\end{minted}
\begin{verbatim}
$KS    0.1
$AD    0.14
$joint 0.13
\end{verbatim}

\begin{minted}[mathescape, linenos, texcomments]{r}
KS.AD.2sample.permutation.test(function() rnorm(30, 0, 1), function() rnorm(42, 0, 2))
\end{minted}
\begin{verbatim}
$KS    0.28
$AD    0.42
$joint 0.39
\end{verbatim}

\begin{minted}[mathescape, linenos, texcomments]{r}
KS.AD.2sample.permutation.test(function() rnorm(30, 0, 1), function() rnorm(42, 0, 0.5))
\end{minted}
\begin{verbatim}
$KS    0.35
$AD    0.6
$joint 0.61
\end{verbatim}

\begin{minted}[mathescape, linenos, texcomments]{r}
KS.AD.2sample.permutation.test(function() rnorm(30, 0, 1), function() rnorm(42, 0.5, 1))
\end{minted}
\begin{verbatim}
$KS    0.4
$AD    0.46
$joint 0.48
\end{verbatim}

\begin{minted}[mathescape, linenos, texcomments]{r}
KS.AD.2sample.permutation.test(function() rnorm(30, 0, 1), function() rt(42, df = 2))
\end{minted}
\begin{verbatim}
$KS    0.04
$AD    0.09
$joint 0.08
\end{verbatim}
Well it seems that adding a second distance doesn't bring anything new to the table compared to using just AD distance. Two distance test is sometimes performs even worse than raw AD test. 

\subsection{Permutation test based on joint distribution of mean, variance, skewness and kurtosis}

In this subsection we propose a permutation test based on joint distribution of mean, variance, skewness and kurtosis of the smaller sample.

The implementation is as follows:

\begin{minted}[mathescape, linenos, texcomments]{r}
f.4moments.2sample.permutation.test <- function(
    sample.1.generation.function,
    sample.2.generation.function,
    num.iter = 100,
    m = 10000
)
{
    power_array_combined <- numeric(num.iter)
    k = round(sqrt(m))
    for(j in 1:num.iter)
    {
        sample.1 <- sample.1.generation.function()
        sample.2 <- sample.2.generation.function()
        
        d1 <- mean(sample.1)
        d2 <- var(sample.1)
        d3 <- e1071::skewness(sample.1)
        d4 <- e1071::kurtosis(sample.1)
        
        combined.sample <- c(sample.1, sample.2)
        res1 <- numeric(m)
        res2 <- numeric(m)
        res3 <- numeric(m)
        res4 <- numeric(m)
        for(i in 1:m)
        {
            indices <- sample.int(length(combined.sample), size = length(sample.1))
            sample1_ <- combined.sample[indices]
            sample2_ <- combined.sample[-indices]
            res1[i] <- mean(sample1_)
            res2[i] <- var(sample1_)
            res3[i] <- e1071::skewness(sample1_)
            res4[i] <- e1071::kurtosis(sample1_)
        }
      
        joint.distr.2dists <- cbind(res1, res2 / 2, res3 / 6, res4 / 24)
        mat <- knnx.dist(data = joint.distr.2dists, query = joint.distr.2dists, k = k)
        kNN.sum.dists <- rowSums(mat)
    
        d.j <- sum(knnx.dist(data = joint.distr.2dists, query = cbind(d1, d2 / 2, d3 / 6, d4 / 24), k = k))
    
        power_array_combined[j] <- d.j > quantile(kNN.sum.dists, 0.95)
        
        print(j)
        flush.console()
    }
    sum(power_array_combined) / length(power_array_combined)
}
\end{minted}
Type 1 error check:
\begin{minted}[mathescape, linenos, texcomments]{r}
f.4moments.2sample.permutation.test(function() rnorm(30, 0, 1), function() rnorm(42, 0, 1))
\end{minted}
\begin{verbatim}
0.05
\end{verbatim}
And the powers:
\begin{minted}[mathescape, linenos, texcomments]{r}
f.4moments.2sample.permutation.test(function() rnorm(30, 0, 1), function() rnorm(42, 0, 1.5))
\end{minted}
\begin{verbatim}
0.49
\end{verbatim}
\begin{minted}[mathescape, linenos, texcomments]{r}
f.4moments.2sample.permutation.test(function() rnorm(30, 0, 1), function() rnorm(42, 0, 2))
\end{minted}
\begin{verbatim}
0.85
\end{verbatim}
\begin{minted}[mathescape, linenos, texcomments]{r}
f.4moments.2sample.permutation.test(function() rnorm(30, 0, 1), function() rnorm(42, 0, 0.5))
\end{minted}
\begin{verbatim}
0.77
\end{verbatim}
\begin{minted}[mathescape, linenos, texcomments]{r}
f.4moments.2sample.permutation.test(function() rnorm(30, 0, 1), function() rnorm(42, 0.5, 1))
\end{minted}
\begin{verbatim}
0.42
\end{verbatim}
\begin{minted}[mathescape, linenos, texcomments]{r}
f.4moments.2sample.permutation.test(function() rnorm(30, 0, 1), function() rt(42, df = 2))
\end{minted}
\begin{verbatim}
0.16
\end{verbatim}
This test looks very good. 

\subsection{For comparison: energy distance permutation test}
Energy distance is quite a popular option for permutation tests. Let us implement a permutation test based on energy distance and compare it with the test from the previous subsection.

Energy distance between two samples is defined as
$$
D_E(X,Y)
=
\frac{2}{n_1 n_2}\sum_{i=1}^{n_1}\sum_{j=1}^{n_2}|x_i-y_j|
-
\frac{1}{n_1^2}\sum_{i=1}^{n_1}\sum_{j=1}^{n_1}|x_i-x_j|
-
\frac{1}{n_2^2}\sum_{i=1}^{n_2}\sum_{j=1}^{n_2}|y_i-y_j|
$$
We implement it like this
\begin{minted}[mathescape, linenos, texcomments]{r}
energy_distance_squared <- function(x, y) {
  x <- as.matrix(x)
  y <- as.matrix(y)

  dxy <- as.matrix(dist(rbind(x, y)))[
    1:nrow(x),
    (nrow(x) + 1):(nrow(x) + nrow(y))
  ]

  dxx <- as.matrix(dist(x))
  dyy <- as.matrix(dist(y))

  2 * mean(dxy) - mean(dxx) - mean(dyy)
}
\end{minted}
The permutation test based on energy distance is as follows:
\begin{minted}[mathescape, linenos, texcomments]{r}
energy.distance.permutation.test <- function(
    sample.1.generation.function,
    sample.2.generation.function,
    num.iter = 100,
    m = 10000
)
{
    power_array_combined <- numeric(num.iter)
    for(j in 1:num.iter)
    {
        sample.1 <- sample.1.generation.function()
        sample.2 <- sample.2.generation.function()
        
        d <- energy_distance_squared(sample.1, sample.2)
        
        combined.sample <- c(sample.1, sample.2)
        res <- numeric(m)
        for(i in 1:m)
        {
            indices <- sample.int(length(combined.sample), size = length(sample.1))
            sample1_ <- combined.sample[indices]
            sample2_ <- combined.sample[-indices]
            res[i] <- energy_distance_squared(sample1_, sample2_)
        }
    
        power_array_combined[j] <- d > quantile(res, 0.95)
        
        print(j)
        flush.console()
    }
    sum(power_array_combined) / length(power_array_combined)
}
\end{minted}
Type 1 error check:
\begin{minted}[mathescape, linenos, texcomments]{r}
energy.distance.permutation.test(function() rnorm(30, 0, 1), function() rnorm(42, 0, 1))
\end{minted}
\begin{verbatim}
0.07
\end{verbatim}

\begin{minted}[mathescape, linenos, texcomments]{r}
energy.distance.permutation.test(function() rnorm(30, 0, 1), function() rnorm(42, 0, 1.5))
\end{minted}
\begin{verbatim}
0.09
\end{verbatim}

\begin{minted}[mathescape, linenos, texcomments]{r}
energy.distance.permutation.test(function() rnorm(30, 0, 1), function() rnorm(42, 0, 2))
\end{minted}
\begin{verbatim}
0.41
\end{verbatim}

\begin{minted}[mathescape, linenos, texcomments]{r}
energy.distance.permutation.test(function() rnorm(30, 0, 1), function() rnorm(42, 0, 0.5))
\end{minted}
\begin{verbatim}
0.61
\end{verbatim}

\begin{minted}[mathescape, linenos, texcomments]{r}
energy.distance.permutation.test(function() rnorm(30, 0, 1), function() rnorm(42, 0.5, 1))
\end{minted}
\begin{verbatim}
0.49
\end{verbatim}

\begin{minted}[mathescape, linenos, texcomments]{r}
energy.distance.permutation.test(function() rnorm(30, 0, 1), function() rt(42, df = 2))
\end{minted}
\begin{verbatim}
0.09
\end{verbatim}

\noindent
We see that the permutation test based on mean, variance, skewness and kurtosis is better than the test based on energy distance for every checked situation except the difference in means. 

\bibliography{literature2}{}
\bibliographystyle{plainurl}

\end{document}